\documentclass[structabstract]{aa}

\usepackage{amsmath}
\usepackage{color}
\usepackage{subfigure}
\usepackage{morefloats}
\usepackage{epsf}
\usepackage{graphics}
\usepackage{color} 
\usepackage[section] {placeins}
\usepackage{array}
\usepackage{lscape}

\usepackage[ansinew]{inputenc}
\usepackage{textcomp}
\usepackage{bm}
\usepackage{rotating}
\usepackage{amssymb}

\usepackage{graphicx}
\usepackage{txfonts}
\usepackage{natbib}
\usepackage{times}

\usepackage{longtable}

\newcommand{\aox}{\alpha_{\rm ox}}

\newcommand{\kbol}{k_{\rm bol}}

\newcommand{\Lhard}{\rm{L_{[2-10]keV}}}

 \newcommand{\Ldisc}{\rm L_{\rm disc}}

\def\lappr{\lower 3pt\hbox{$\buildrel < \over \sim\;$}}
\def\llappr{\lower 3pt\hbox{$\buildrel > \over \sim\;$}}

\begin{document}

\title{The optical-UV spectral energy distribution  of the unabsorbed AGN population in the   XMM-\emph{Newton} Bright Serendipitous Survey  \thanks{The XMM-\emph{Newton} Bright Serendipitous Survey is part of the follow-up program being conducted by the XMM-\emph{Newton} Survey Science Centre (SSC), http://xmmssc-www.star.le.ac.uk/.}}
\titlerunning{The optical-UV  spectral energy distribution of type 1 AGN in the XBS}

   \author{E. Marchese$^{1}$\thanks{elena.marchese@brera.inaf.it}, R. Della Ceca$^{1}$\thanks{roberto.dellaceca@brera.inaf.it}, A. Caccianiga$^{1}$, P. Severgnini$^{1}$, A. Corral$^{1}$ and
          R. Fanali$^{2}$.
       }

   \institute{$^{1}$INAF-Osservatorio Astronomico di Brera, via Brera 28, 20121, Milano, Italy\\
              $^{2}$ Universit\`a Milano Bicocca, Dipartimento di Fisica G. Occhialini, Piazza della Scienza 3, 20126, Milano, Italy\\
}

\authorrunning{E. Marchese et al.}

   \date{Revised version, \today}

\abstract
{Active galactic nuclei (AGN) emit radiation over a wide range of wavelengths, with a peak of emission in the far-UV region of the electromagnetic spectrum, a spectral region that is historically difficult to observe.}
{Using  optical, \emph{GALEX} UV, and XMM-\emph{Newton} data we derive the spectral energy distribution (SED) from the optical/UV to X-ray regime of a sizeable sample of AGN. The principal motivation is to investigate the relationship between the optical/UV emission and the X-ray emission and  provide bolometric corrections to the hard X-ray (2-10 keV) energy range, $k_{bol}$, the latter being a fundamental parameter  in current physical cosmology.}
{We construct and study  the X-ray to optical SED of a sample of 195 X-ray selected Type 1 AGN belonging to  the XMM-\emph{Newton} bright serendipitous survey (XBS).
The optical-UV luminosity was computed using data from the \emph{Sloan Digital Sky Survey} (SDSS), from our own dedicated optical spectroscopy  and the satellite \emph{Galaxy Evolution Explorer} (\emph{GALEX}), while the X-ray luminosity was computed using  XMM-\emph{Newton} data.
%In particolar, we calculate bolometric corrections for a sample of sources  
Because it covers a wide range of redshift ($0.03<z\lesssim 2.2$), X-ray luminosities ($41.8<logL_{[2-10]keV}<45.5$ erg/s) and because it is composed of ``bright objects'', this sample is ideal for this kind of investigation.}
{ We confirm   a highly significant correlation between the accretion disc luminosity $L_{disc}$  and the hard X-ray luminosity $L_{[2-10]keV}$, in the form $L_{disc} \propto L_{[2-10] keV}^{\beta}$, where $\beta =1.18 \pm 0.05$.  We find a very shallow dependence of  $k_{bol}$ on the X-ray luminosity with respect to the broad distribution of values of   $k_{bol}$.
 We find a correlation between $k_{bol}$ and the hard X-ray photon index $\Gamma_{2-10 keV}$ and a tight correlation between the optical-to-X-ray spectral index $\aox$ and $k_{bol}$, so we conclude that both  $\Gamma_{2-10 keV}$ and  $\aox$ can be used as a proxy for $k_{bol}$.    }
{}

 \maketitle

\section{Introduction}
A large fraction of the active galactic nuclei (AGN) bolometric luminosity  is
emitted in a strong, broad feature that begins to dominate the spectral energy distribution
(SED) at the bluest optical wavelengths, and appears to extend short-wards of the current limits of UV satellite data ($\sim$ 100 $\AA$). This feature of the continuum, also known as the Big Blue Bump (\citealt{Sanders89}), 
is most likely   thermal emission arising from a geometrically thin, optically thick accretion disc \citep{Shields78,Malkan82,Ward87}. Another large fraction of the total luminosity in AGN is also emitted in the X-ray band, which is more likely arising from the inverse Compton scattering of the disc's photons  by a  corona of hot plasma surrounding the central regions of the disc, therefore X-ray and optical/UV observations are  critical probes of the physics of the innermost regions of AGN, and investigating the relationship between the UV and X-ray emission is an important step towards  better understanding  the physics involved. Indeed the study of the correlation between X-ray and UV luminosities has in the past been  the subject of many works on optically or X-ray selected samples of AGN  (see section \ref{discussion}).

In this paper we investigate the SEDs of a sample of 195 AGNs belonging to the XMM-\emph{Newton} Bright Serendipitous Survey (XBS, \citealt{DellaCeca1}) having UV observations from the satellite \emph{Galaxy Evolution Explorer} (\emph{GALEX}), optical magnitudes from 
the \emph{Sloan Digital Sky Survey, SDSS} (complemented with optical data reported in Caccianiga et al. 2008)
and X-ray observations from   \emph{XMM-Newton} (see \citealt{Corral11}, hereafter referred as C11).
The main goals of this work are: \\
a) to derive accretion disc luminosities for a significant and representative sample of AGN;\\
b) to investigate the correlations between the accretion disc   and the X-ray luminosity;\\
c) to evaluate the bolometric luminosity (computed as the contribution of X-ray, UV and optical emission) and thus the bolometric correction to the hard X-ray (2-10 keV), defined as
 \begin{equation}
 k_{bol}=\frac{L_{bol}}{L_{[2-10]keV}}
 \label{kbol}
 \end{equation}
 for a sample of AGN spanning a wide range in X-ray luminosities and redshifts.  We stress that  the infrared
  emission is not taken into account in the computation of the bolometric luminosity, since it is known to be re-processed emission mainly from the ultraviolet (see \citealt{Antonucci93}); indeed its inclusion would mean  counting part of the emission twice, overestimating the derived bolometric luminosities.

The paper is organised as follows: in section \ref{data} we discuss the sample selection and the procedure used in cross-matching the sources in the XBS database with the \emph{GALEX} and SDSS catalogues; section \ref{SED}  covers the construction of reliable SEDs for each source in the sample, taking into account intrinsic extinction and host galaxy contamination to the observed emission, as well as the emission lines contribution and absorption from Lyman $\alpha$ systems along the line of sight. In section \ref{correlations} we describe the results obtained in this work, focusing on the correlation analysis between  the accretion disc luminosity and the X-ray luminosity, and  the relation between  bolometric correction and the X-ray luminosity. 
Our results are discussed within the context of previous works in section \ref{discussion} while in
section \ref{summary} summary and conclusions are presented. 
As done for other papers on the XBS survey we assume here  the cosmological model $H_0=65km \ s^{-1} \ Mpc^{-1}$, $\Omega_{\lambda}=0.7$ and $\Omega_{M}=0.3$ throughout this paper.

\section{Data sources}
\label{data}

\subsection{XMM-Newton bright serendipitous survey}
\label{xbs}
The \emph{XMM-Newton} bright serendipitous survey (XBS survey) is a wide-angle ($\sim $28 sq. deg) high
Galactic latitude ($\mid b \mid >20^{\circ }$) survey based on the XMM-\emph{Newton}
archival data. It is composed of two flux-limited serendipitous samples of X-ray selected sources: the XMM bright source sample (BSS, 0.5-4.5 keV band,  389  sources) and the XMM Hard bright source sample (HBSS, 4.5-7.5 keV band, 67 sources, with 56 sources in common with the BSS sample), having a flux limit of $\sim  7 \times 10^{-14} \rm erg \ cm^{-2} \ s^{-1}$ in both energy selection bands.
%having an EPICMOS2 count rate limit, corrected for vignetting, of $10^{-2}$ cts/s ($2 \times 10^{-3}$ cts/s) in the 0.5-4.5 keV (4.5-7.5 keV) energy band; the flux limit is $\sim  7 \times 10^{-14} erg \ cm^{-2} \ s^{-1}$ in both energy selection bands.
The details on the XMM-Newton fields selection strategy and the source selection criteria 
of the XMM BSS and HBSS samples are discussed in \cite{DellaCeca2} and \cite{DellaCeca3}.

To date, the spectroscopic identification level has reached 93\% and 97\% for the BSS and the HBSS samples, respectively. The current classification of the XBS sample is as follows: 305 AGN (including 5 BL Lacs), 8 clusters of galaxies, 2 normal galaxies and 58 X-ray emitting stars (\citealp{Caccianiga2,Corral11}).
The large majority of the still unidentified objects are expected to be absorbed AGN and BL Lac objects, so the sample of type 1 AGN in the XBS can be considered complete at a confidence level approaching 100\%.
The analysis of the optical data, along with the relevant classification scheme and the optical properties of the extragalactic sources are presented in \cite{Caccianiga3} and \cite{Caccianiga2}; the optical and X-ray properties of
the galactic population are discussed in \cite{Lopez07}.\\

The availability of good XMM-\emph{Newton} data for the sources in the XBS sample, spanning the energy range between 
$\sim0.3$ and $\sim10$ keV, allowed us to perform a reliable X-ray spectral analysis for almost every AGN of the sample.
% The results on  different sub-samples of sources have been already presented in literature (Severgnini et al. 2003, Caccianiga et al. 2004, Caccianiga et al. 2007, Galbiati et al. 2005, Della Ceca et al. 2008). 
The X-ray spectral analysis of the complete XBS AGN sample is presented in C11, which provide reliable X-ray photon indices,  intrinsic column densities $N_{\rm H}$, and  X-ray luminosities, which are necessary to derive  bolometric luminosities.\\

The following work  is focused on type 1 AGN, in order to limit uncertainties due to obscuration in the determination of the intrinsic SED shape. Furthermore, in the source selection, we applied a cut in intrinsic column density $N_{\rm H}$, selecting only type 1 AGN with $N_{\rm H}<4 \times 10^{21}  \ \rm cm^{-2}$, resulting in a  sample of 262 sources (7 AGN have been excluded). From this sample we   excluded 14 sources, classified as radio loud AGN (\citealt{Galbiati05}), because we do not know the orientation of the relativistic jet respect to the line of sight, and thus we can not quantify  its contamination on the X-ray observed spectrum  (\citealt{Zamorani81,Wilkes87,Galbiati05}). We note that selecting X-ray unabsorbed ($N_{\rm H}<4 \times 10^{21}  \ \rm cm^{-2}$) AGN also mitigate the possible ``contamination" due to broad absorption line quasars (BALQSOs). Indeed BALQSOs seem to be characterised by the same intrinsic underlying X-ray continuum as the majority of the AGN population but their X-ray emission is often depressed by large amounts of intrinsic absorption (\citealt{Streblyanska10,Giustini08,Gallagher06}). An X-ray to optical investigation of optically selected samples of Type 1 AGN  without taking into account this contamination could lead to misleading results, since BALQSOs can cause an ``artificial" steepening of the optical to X-ray correlation. To our knowledge there are no BALQSOs in our type 1 AGN  sample.  \\
Thus the starting sample of our analysis is composed of 248 X-ray selected type 1 AGN.   We remark that the exclusion of obscured sources from our analysis does not necessarily mean that the conclusions we will outline further on are valid for unobscured sources only.  Our results are still applicable to obscured AGNs if the obscuration is a line-of-sight orientation effect and does not affect the emission process at work in AGN.

\subsection{Cross correlation with \emph{GALEX}}
\label{galex}
The \emph{Galaxy Evolution Explorer} satellite is  performing the first large-scale
UV imaging survey (\citealt{Martin05,Morrissey07}). Most images are taken simultaneously in two broad
bands, the near UV (NUV, $\sim$ 1770 - 2850$\AA$) and the far UV
(FUV, $\sim$1350 - 1780$\AA$) at a resolution of $\sim$5'' full width at
half maximum (FWHM). Three nested \emph{GALEX} imaging surveys
have been defined: the All-Sky Survey (AIS) expected to cover
a large fraction ($\sim 85$\%) of the high Galactic latitude ($\mid b \mid >20^{\circ }$) sky 
to $m_{AB} \sim 21$, the Medium Imaging Survey (MIS) reaching
$m_{AB} \sim 23$ on 1000 deg$^{2}$, and the Deep Imaging Survey (DIS)
extending to $m_{AB} \sim 25$ on 80 deg$^2$. These main surveys are
complemented by guest investigator programs. \\
Here we used the \emph{GALEX} data from the officially distributed Data Release 4 (GR4), which 
has been homogeneously reduced and analysed by a dedicated software pipeline.
A previous version of this pipeline used for the earlier GR3 data release is described in detail by \cite{Morrissey07}. For details about the changes between GR3 and GR4  and on the \emph{GALEX} mission see respectively  http://galex.stsci.edu/GR4/ and  http://www.galex.caltech.edu/.

\subsubsection{\emph{GALEX} detections}

The cross correlation of the 248 type 1 XBS AGN    with the \emph{GALEX} catalogue GR4 was performed by using the coordinates of the optical counterparts of the X-ray sources (reported in \citealt{Caccianiga2}), with an impact parameter of $2.6^{\prime\prime}$.
This latter value was derived by \cite{Trammell07} cross-correlating a sample of 6371 quasar from the SDSS with \emph{GALEX}; they find that 99\% of the matches is recovered with a search radius of 2.6''.\\

The cross-correlation produced (multiple or single) matches for 182 X-ray sources.
We now describe how these   matches  were analysed.   In the case   of multiple matches (115 X-ray sources) the duplicates were removed with the following procedure. If two \emph{GALEX} sources were within 2.6'', but had the same ``photoextractid'' (i.e. they were both from the same observation) they were  considered as two independent sources. In these cases (6 matches:  XBSJ003418.9-115940, XBSJ012000.0-110429, XBSJ120359.1+443715, XBSJ120413.7+443149, XBSJ141809.1+250040, XBSJ163309.8+571039) the brightest source was selected as the best candidate to be the counterpart of the XBS source.  Otherwise, if the multiple matches were from different observations, they were assumed to be multiple observations of the same source.  In these cases the observation with the longest exposure time was retained. In the cases of almost equal exposure times we chose the observation where the source was closer to the  centre of the field of view (that is the source with the smallest ``fov-radius'' from the ``photoobjall'' table), as generally the photometric quality is   better in the central part of the field (\citealt{Bianchi10a}). All the  selected sources, with the exception of two (XBSJ210355.3-121858, XBSJ165406.6+142123) have  photometric errors (based on Poissonian source counts statistics, see also section \ref{phot_err}) on the UV magnitudes lower than 0.5 mag, in agreement with the selection procedure used by \cite{Bianchi10a}. The two exceptions reported above, having respectively errors on NUV magnitudes of 0.56 mag and 0.58 mag, were still considered in our analysis.\\

 From this sample of 182 type 1 AGN we excluded two sources
(XBSJ031851.9-441815 and XBSJ062134.8-643150) from the analysis,  
due to huge (up to a factor 1000) uncertainties in the estimate of the UV-optical fluxes,
once the corrections discussed in section \ref{SED} were taken into
account. 
This procedure allowed us to define  180 X-ray/UV matches, from the original sample of 248 AGN;   about $ 98\%$ of the sources have at least a NUV detection, $\sim 2\%$ were detected in the FUV band only, and about 75\% of these sources were detected both in the NUV and in the FUV band.\\

In order to estimate the reliability of these  matches we   computed the expected number of random matches within 2.6''. This was done by performing the same cross-correlation after shifting  the X-ray catalogue (along the right ascension and the declination) so that only chance coincidences are expected. To account for the non-uniform distribution of the sources in the sky we repeated several times the cross-correlation with different offsets and computed the average number of random sources found in each cross-correlation. Thus we derived that the probability of finding a \emph{GALEX} UV spurious source within 2.6'' is 0.015, which correspond to  $\sim 4$ chance matches over  the original sample of 248 sources.\\

\subsubsection{Upper limits on \emph{GALEX} fluxes}
\label{galex_upper_lim}

In this section we will discuss how the remaining 66  sources (for which we do not have \emph{GALEX} photometric data) were treated. 
Forty six of these 66 sources  ($\sim 18.5 $\% of the original sample of 248 sources) do not fall in the area of sky covered by \emph{GALEX} (Data Release 4), thus the lack of UV data does not depend on the properties of the sources in this sub-sample.
It is worth noting that, as the sources contained in the XBS catalogue have a serendipitous distribution in the sky 
(at $\mid b \mid >20^{\circ}$), we expect that a fraction of them will be ``not covered" with \emph{GALEX} data since the AIS survey is not covering the whole sky. Indeed,  the total sky area at $\mid b \mid >20^{\circ}$ covered from AIS 
(\emph{GALEX} DR4)  is $\sim 23,000$ square degrees (Karl Forster, mission planner, private communication), i.e. $\sim 85\%$ of the sky at 
$\mid b \mid >20^{\circ}$. Therefore the fact that $18.5 $\% of the original sample of sources is not covered by \emph{GALEX} is fully consistent with the ``missing"  \emph{GALEX} coverage of the $\mid b \mid >20^{\circ}$ sky ($\sim 15\%$).
 
The remaining 20 sources  are indeed located in sky regions covered by \emph{GALEX} observations. A visual inspection of the regions near these sources lead to the following conclusions: 3 sources (XBSJ002707.5+170748, XBSJ022253.0-044515, XBSJ220320.8+184930) were not detected since each of them lies in the wings of a very bright UV source in the sky, while 17 sources were simply too faint to be detected with the exposure time of the \emph{GALEX} data along their line of sight. The former three sources were   removed from the sample, while for the latter 17 sources we decided to use as an upper limit on the  NUV flux (since 98\% of the sources have at least a NUV detection) the one corresponding to the faintest detected source (with photometric error $<$0.5 mag, see \citealt{Bianchi10a}) within $\sim 2$ arcmin from the position of our undetected target. In order to investigate if these sources have different properties from the detected ones we compare (see figure \ref{fig:mr_muv}) the optical r magnitude   against the magnitude in the NUV band for the final sample of \emph{GALEX} detected sources (black circles) and the sources with \emph{GALEX} upper limits (red crosses, in the electronic form only); as expected, the latter ones are on average fainter than the detections in the  r magnitude too. We also investigated the distribution in the $L_{[2-10]keV}\rm-z  $ plane of these two samples (see figure \ref{fig:lx_z}). It is evident that the undetected sources do not show any difference in the X-ray luminosity distribution as compared to the detected ones.\\
 From this  sub-sample  of 17 source we excluded from the following analysis 2 sources (XBSJ014109.9-675639 and XBSJ050453.4-284532): the reason is that only one optical flux (in the red band) is available  and it corresponds to an intrinsic  rest-frame $\lambda > 5000 \AA$, a wavelength range we did not consider  in the SED fitting (see section \ref{lambda} for further details).
At this stage, the sample is composed of 195 sources: 180 \emph{GALEX} detections and 15 sources with \emph{GALEX} upper limits.

\subsection{\label{sdss}Cross correlation with the \emph{Sloan Digital Sky Survey}}
 All the sources of the XBS sample have a measured optical magnitude (mainly in the red optical band, \citealt{Caccianiga2}). In order to have more reliable SEDs we need more than one optical point, preferably of the same epoch, to avoid variability effects. Thus we searched for optical information from  the \emph{Sloan Digital Sky Survey} (SDSS). This is one of the most ambitious  surveys, which over eight years of operations (SDSS-I, 2000-2005; SDSS-II, 2005-2008) obtained deep, multi-colour images covering more than a quarter of the sky and created three-dimensional maps containing more than 930,000 galaxies and more than 120,000 quasars. The SDSS uses a dedicated 2.5-metre wide-angle optical  telescope at Apache Point Observatory (New Mexico), and takes images using photometric system of five filters (named u, g, r, i and z). For more information see http://www.sdss.org).\\

We cross-matched the optical positions of the resulting   sample of 195 sources (including both the 180 XBS-\emph{GALEX} detected sources and the 15 sources with \emph{GALEX} upper limits) with the  \emph{SDSS} DR7 catalogue. 
This has produced 101 matches, i.e. 101 sources with a measured fluxes in the X-ray, UV (detection or upper limit) and at least one of the optical SDSS (u,g,r,i,z) bands; for the remaining 94 sources we used the optical magnitudes reported in \cite{Caccianiga2}.\\

\begin{figure}[htbp]
    \centering
    \includegraphics[height=8cm, width=8cm]{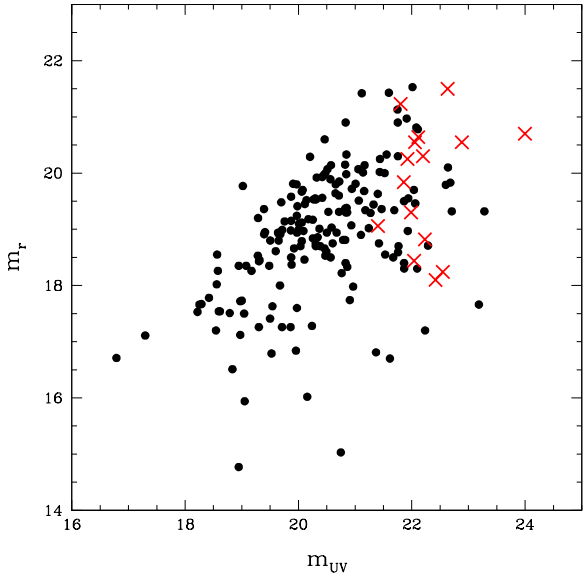}
  \caption{Distribution in the $m_{UV} - m_r $ plane of the final sample of  type 1 AGN analysed. The red crosses  represent the  sources with UV upper limits, the black circles are the  sources detected by \emph{GALEX}. }
    \label{fig:mr_muv}
  \end{figure}

\begin{figure}[htbp]
  \centering
   \includegraphics[height=8cm, width=8cm]{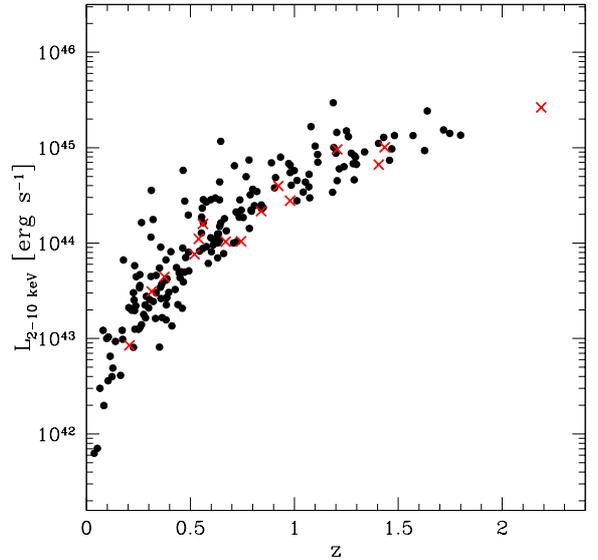}
 \caption{Distribution in the $L_X-z$ plane of the final sample of  type 1 AGN. Black  circles:\emph{GALEX} detections; red crosses:upper limits.}
   \label{fig:lx_z}
 \end{figure}

To summarise, the final sample of type 1 unabsorbed AGN ($N_{\rm{H}}<4 \times 10^{21} \ cm^{-2}$) used in the analysis reported here is composed of 195 sources; their distribution in the $L_X-z$ and in the $ m_{UV}-m_r$ planes are shown in Figs.  \ref{fig:mr_muv} and \ref{fig:lx_z}. The sample covers a range of redshift between 0.03 and $\sim $2 and a wide range of X-ray luminosities, i.e. 
from $6\times 10^{41}$ to $3\times 10^{45}$ erg/s. Besides the X-ray information (e.g. spectral slope, flux), which are available for all the sources, we have a) optical magnitudes (from SDSS and/or from the XBS project) for all the  195 sources analysed and b) \emph{GALEX} detections for 180 sources and \emph{GALEX} upper limits for 15 sources.

\section{\label{SED} Construction of SEDs}

\subsection{Exclusion of data at $\lambda_{rest}>5000 \ \AA$}
\label{lambda}
In order to study the  intrinsic emission produced by the accretion disc it is necessary to consider wavelength ranges which are free from contamination due to different spectral components (e.g. stellar emission from the host galaxy). \cite{Vandenberk01}, studying the composite spectra of a sample of SDSS quasars, showed that the slope of the continuum changes abruptly  at a rest-frame wavelength of about 5000 $\AA$, becoming steeper (optical spectral index $\alpha_{\nu}$ changing from  -0.44 to -2.45) at longer
rest-frame wavelengths
%untill  8555 \AA \ 
(see Figure 5  in \citealt{Vandenberk01}). The authors hypothesise that this behaviour can be partly due to the host-galaxy starlight contamination, as suggested from the presence of stellar absorption lines in the composite spectrum. Nevertheless, they also theorise that this contribution of emission at wavelengths beyond 5000 $\AA$ \ could   be caused by an intrinsic change in the quasar continuum (e.g. emission from hot dust).
As the emission from the accretion disc is concentrated at wavelengths lower than 5000 $\AA$ \ (we expect a peak of emission in the Far-UV), we did not consider in the fit of the accretion disc emission  
the  rest-frame wavelengths longer than 5000 $\AA$. For this reason two sources (XBSJ014109.9-675639 and XBSJ050453.4-284532) having only \emph{GALEX} upper limits and optical magnitudes corresponding  to a rest-frame wavelength $\lambda > 5000 \ \AA$ \ were already from our analysis (see section \ref{galex_upper_lim}). 
However, since  a contamination from the host-galaxy (although weaker)  is expected  also for 
wavelength below 5000  $\AA$ \ we have developed a method (see section \ref{emiss_host}) to take into account this 
possible contamination.

\subsection{\label{corrections} Corrections to measured fluxes}
The radiation emitted from an astronomical source has obviously undergone various interactions during its journey, depending on different factors (redshift, environment etc..). For this reason, in order to reliably study the intrinsic luminosity of our sources, the potential effects  on the primary radiation have to be considered. First of all, Galactic reddening: for our sample  this effect was estimated from the extinction law computed by \cite{Allen76}, with $R_V=3.1$, adopting the values of the Galactic colour excess $E_{B-V}$ available for each source in the \emph{GALEX} database.

\subsubsection{\label{Ly} Lyman $\alpha$ forest}
Even if the large majority (97\%) of the sources in our sample lies at z$<$1.6, we decided to account for the potential attenuation given by absorption of neutral hydrogen in intervening Lyman-$\alpha$ absorption systems. For our sample this is possible only adopting a statistical approach, to produce an estimate of the attenuation $\tau_{eff}= -ln < e^{-\tau} >$, where the average is taken over all possible lines of sights.
 In order to account for this effect  we considered the results reported in \cite{Ghisellini10}, where a relation between the attenuation and redshift has been estimated for the six central wavelengths  of the UV filters of the instrument UVOT on board of the satellite \emph{Swift}.
We extrapolated the average attenuation corresponding to the effective wavelengths of the \emph{GALEX} filters FUV and NUV and we found that for $z<1$ ($\sim 75 \%$ of the sources) the attenuation produces a flux reduction lower than 15\% (5\%) in the FUV (NUV) band while, for $1<z<1.6$ ($\sim 22 \%$ of the sources) the reduction in flux is between 15-50\% (5-20\%) in the FUV (NUV) band. Only 3\% of our sources have $1.6<z<2.3$, implying corrections between 50-80\% (20-70\%) in the FUV (NUV). We applied these corrections  to the UV fluxes of the whole sample of 195 sources with \emph{GALEX} detections, even if they are negligible for $\sim 75 \%$ of them.

\subsubsection{\label{emiss_host} Host Galaxy contamination}
\begin{figure}[htbp]
   \centering
   \includegraphics[height=8cm, width=8cm]{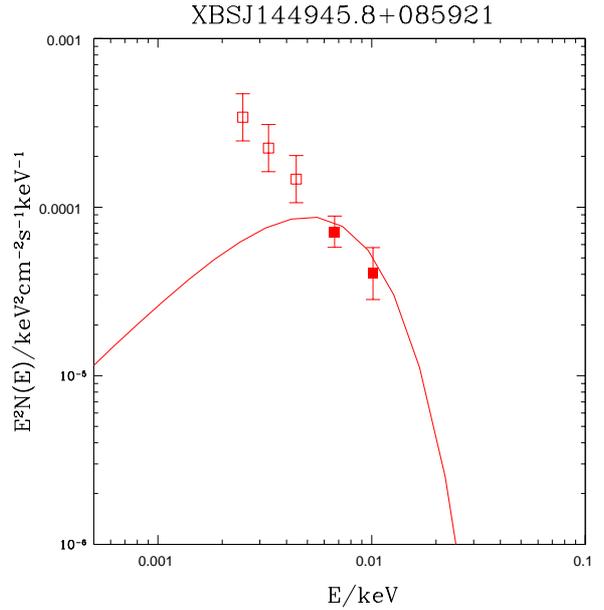}
 \caption{Example of the rest-frame SED of one source showing  the optical fluxes (empty squares) contaminated by the emission contribution of the host galaxy. The solid line represents the accretion disc emission model described in section \ref{model}.}
   \label{ESEMPIO}
 \end{figure}

A first look at the SEDs obtained after these first corrections to the observed fluxes highlighted out that $\sim$ 20\% of the sources shows a steepening of the SED at the optical wavelengths, in evident disagreement with the expected accretion disc emission continuum (see figure \ref{ESEMPIO}). 

We also noted that all these sources are characterised by a value of the 4000 $\AA$ \ break\footnote{The 4000 $\AA$ \ break is defined as $\Delta=\frac{F^+-F^-}{F^+}$ where $F^+$ and $F^-$ represent the mean value of the flux density (expressed per unit frequency) in the region 4050-4250 $\AA$ \ and 3750-3950 $\AA$ \ (in the source's rest frame) respectively.} (also known as {\it Calcium break} $\Delta$) higher than $\sim$17\%. This strongly suggests a not negligible contribution of the stellar emission of the host galaxy to the total observed continuum.\\
In order to subtract this contribution and to constrain the SED shape at optical wavelengths, we assumed a first approximation model of  a host galaxy composed by a step function having a {\it Calcium break} of 50\%. This value was chosen as an intermediate value between 45\% and 55\%, enclosing the  4000 $\AA$ \ breaks computed using the templates reported in \cite{Polletta07}, for  elliptical galaxies and spirals S0, Sa, Sb. The  AGN is instead represented by a power law with a mean spectral index $\alpha_{\nu}=-0.44$ in the wavelength region 1500--5000 $\AA$ \ (\citealt{Vandenberk01}). Superimposing these two components we derived an empirical relation between the observed 4000 $\AA$  \ break and the relative normalization of the AGN component with respect to the host galaxy component. This procedure  allowed us to derive an empirical correction for the contamination of the host-galaxy for all the sources with measured values of the 4000 $\AA$ \ break.

\begin{figure}[htbp]
  \centering
  \includegraphics[height=8cm, width=8cm]{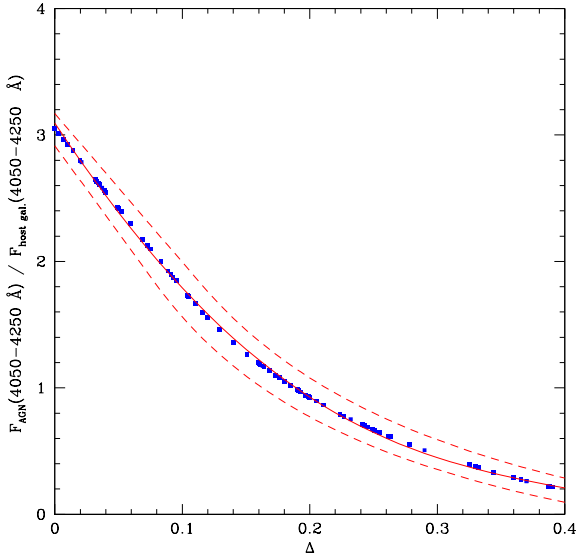}
\caption{Plot of the ratio between the fluxes of AGN and host galaxy in the wavelength range 4050-4250 $\AA$ \   as a function of the intensity of the Calcium break $\Delta$ for the sources in our sample.  The relations obtained assuming a host galaxy  Calcium break of respectively 45$\%$ or 55$\%$ are shown as dashed lines.}
  \label{ratio_norm_break}
\end{figure}

In figure \ref{ratio_norm_break} we report the ratio between the fluxes of AGN and host galaxy in the wavelength range 4050-4250 $\AA$ \ as a function of the intensity of the Calcium break $\Delta$ measured for the sources in our sample (blue points).

The data points have been fitted by a best fit relation described by the  polynomial:
\begin{equation}
y=a+bx+cx^2+dx^3+ex^4
\end{equation}
where $y=\frac{Flux_{AGN}(4050-4250 \AA)}{Flux_{gal}(4050-4250 \AA)}$, $x=\Delta$, a=3.09, b=-15.23, c=20.96, d=12.34 and e=-36.64.
  This plot and the relative best fit relation represent a tool to derive an  estimate of the contribution of the host galaxy   to the AGN emission, once the intensity of the 4000 $\AA$ \ break is known. We also show the expected relation when the intensity of the host-galaxy 4000 $\AA$ \ break takes values in the range between 45\% (lower dashed line) and 55\% (upper dashed  line). A variation from 45\% to 55\% of  the assumed host-galaxy Calcium break    implies a variation on the accretion disc luminosity of $\sim 45\%$ at 2$\sigma$, with a mean variation of   $\sim$14\%. Observing the empirical relations plotted in figure  \ref{ratio_norm_break} we can infer that the relative optical flux ratio AGN/host galaxy changes  from $\sim 15\%$ (for observed $\Delta$ less than $\sim$15\%) up to a maximum of 60\% (for observed $\Delta \sim 40\%$).  We also tested that the results obtained in this work (discussed in section \ref{correlations}) were not statistically affected by a   change in the range 45\%--55\%  of our assumed $\Delta_{galaxy}=50\%$.\\
  Furthermore we also checked if a variation of the    AGN  spectral index could lead to different results: considering the uncertainty of $\approx 0.1$ on the spectral index, reported by  \cite{Vandenberk01}, we found that the accretion disc luminosities  are affected by less then 1\%.

\subsubsection{\label{intrinsic} Intrinsic extinction}
The shape of the Big Blue Bump in the optical-UV region of AGN spectra is notoriously prone to effects of reddening: it is thus of great importance to account for intrinsic extinction (at the source redshift) at these wavelengths, in order to constrain the effective energy output.  The effect is stronger in type 2 objects but a weaker effect is likely to be present in type 1 objects as well. These corrections are not trivial, as the exact shape of the extinction curve in the far-UV for AGN is still a matter of debate. 

We adopted the results of Gaskell \& Benker (2007), who derived a mean extinction curve for 17 AGN with data from \emph{FUSE} and \emph{HST}.
 
We also  verified ``a posteriori''  that these corrections are in good agreement with the expected SED shapes.\\
The extinction curve obtained by \cite{Gaskell07} is only valid in the range 1216 $\AA$--6565 $\AA$. To apply the intrinsic absorption correction below 1216 $\AA$, we extrapolated this curve by preserving the flat trend that characterises the UV emission at short wavelengths.\\
The intrinsic colour excess $E_{B-V}$ for each source was derived from the values of intrinsic column density $N_{\rm H}$ measured from the X-ray spectral analysis, assuming a Galactic gas-to-dust ratio $N_{\rm H}/E_{B-V}=4.8 \times 10^{21} \rm cm^{-2}mag^{-1}$ (\citealt{Bohlin78}).  We recall that there are claims in the literature of a non-Galactic gas-to-dust ratio for some AGN (e.g. \citealt{Maiolino01}), thus there could be objects where this assumption cannot be fulfilled. However, from the analysis of the optical and X-ray spectra of the AGN of the XBS sample, we found  good consistency between the optical and the X-ray classifications, with very few exceptions (\citealt{Caccianiga1,Corral11}) thus supporting the idea of using the Galactic gas-to-dust ratio. Despite having reliable X-ray spectral information,  we only have  upper limits on $N_{\rm H}$ for $\sim$68$\%$ of the sources in the sample.
 
To derive the best guess for the intrinsic $N_{\rm H}$ for each source we made use of the survival analysis, an efficient tool to work with censored data (Isobe et al. 1986, 1990). 
By using this statistical approach we could estimate a cumulative and a differential distribution of $N_{\rm H}$ for the whole sample of AGN.
 
The differential distribution   was then fitted with a gaussian curve, deriving values for the mean and the sigma of the $N_{\rm H}$ distribution. Thus, for each  source with $N_{\rm H}$ upper limits, we produced a series of 100 random values between the minimum $N_{\rm H}^{min}$ of the distribution and the upper limit relative to the individual source. Finally, we adopted the average of these 100 values  as the estimated $N_{\rm H}^{estim}$ of the source; the $1 \sigma$ confidence interval on this average encloses the 68\% of the values of $N_{\rm H}$ computed for each source around $N_{\rm H}^{estim}$.  The derived  $N_{\rm H}^{estim}$ (and the relative 68\% confidence error) for each source was used to derive the corresponding $E_{B-V}$ using the equation reported above.

\subsubsection{\label{lines} Emission lines contribution}
The presence of broad emission lines within the bandpass of a given filter contributes significantly to the observed photometric magnitudes. 

It is thus important  to correct for the presence of the emission lines to obtain magnitudes closer to the continuum emission alone.

The ``redshift dependent" magnitude correction is given by (``Allen's Astrophysical Quantities'', 1990):
\begin{equation}
\Delta m  =2.5 log_{10} ( 1+ (EW)_e (1+z) \frac { R_m(\lambda )} {\int R_m (\lambda) d \lambda  } )
\end{equation}
where $(EW)_e $ is the rest-frame equivalent width of the emission line, $\lambda =\lambda_e(1+z)$ is the observed wavelength of the line and  $R_m$ is the response of the filter (  $\AA^{-1}$)\footnote{http://svo.cab.inta-csic.es/theory/filters/}.

Only the most prominent AGN emission lines in the wavelengths of interest were considered: Ly$\alpha$+NV, CIV, MgII, OIV+Ly$\beta$ and CIII+SiIII.
The rest frame equivalent widths assumed to derive these corrections were taken from   \cite{Telfer02}, and these equivalent widths were based on the investigations of 184 QSOs at $z>0.33$ having \emph{HST} spectra. The line that produces the higher contamination to the measured continuum is the Ly$\alpha$ emission line. This line contribution to the computed magnitudes can amount to 0.4-0.5 mag in the FUV, for objects in the 0.1-0.4 redshift range, and to 0.2-0.3 mag in the NUV, for objects in the 0.6-1.2 redshift range. The equivalent width of the Ly$\alpha$ line used here is in very good agreement with the average  one computed from the composite quasar spectra of the SDSS (\citealt{Vandenberk01}).
 Another significant contribution is produced by the CIV emission line (e.g. $\sim 0.2$ mag for sources at $1.1<z<1.5$ in the $u$ band of the SDSS). In this case the equivalent
width of the CIV emission line considered here (54 $\AA$)
is quite different from the one computed by \cite{Vandenberk01} (24 $\AA$); we stress that in the last case the variation in
magnitude would be  $\sim 0.1$ mag. 

\subsection{\label{errors} Errors on fluxes}
\subsubsection{\label{phot_err}\emph{GALEX} Photometric errors}
The GR4 UV source extractor of the \emph{GALEX} image processing pipeline reports photometric
uncertainties for each object by assuming the observations are  limited by Poisson noise. Therefore, the magnitude errors given in the
GR4 object tables do not take additional sources of noise into account,  including unknown
variances of the detector background level and flat-field maps, or any other systematic errors
present in the data. To account for these errors we referred to the work of \cite{Trammell07}. Assuming that the bulk of the stars observed in a \emph{GALEX} field are
non-variable, \cite{Trammell07} used  fields with multi-epoch observations as a   tool for analysing the
repeatability of the \emph{GALEX} UV photometry for the same objects, using a large
number of objects to empirically estimate the true photometric uncertainties. They  derived an empirical relation between the NUV (FUV) magnitude and the corresponding true photometric error for both the AIS and DIS surveys. Applying this empirical relation to the \emph{GALEX} magnitudes of our   sources  we  find that the actual
photometric errors are  on average  $\sim 0.2$ mag (in the FUV band) and  $\sim0.11 $ (in the NUV band) larger than the Poissonian errors quoted in the GR4 tables.

\subsubsection{Errors due to long-term variability}
To account for the the lack of simultaneity in the optical and UV data it is necessary to have an estimate of the average uncertainties on optical and UV fluxes, due to long term variability. \cite{Devries05}, in a study on optical long term variability of a sample of 41,391 quasars, derived a distribution of the quasar variability as a function of the time-lag between observations. In this work  \cite{Devries05} found that in a time-lag of   years the magnitude difference given by long term variability is $\sim 0.35 $ magnitudes; consequently,  we associated this value to the long term variability for each source.

\subsubsection{Total errors}

The asymmetric $1\sigma$ total errors on the corrected fluxes that were adopted  hereafter are given by the quadratic sum of the $1\sigma$ errors due to long term variability, the $1\sigma$ errors due to the photometric errors, and  $1\sigma$ errors related to the corrections for the intrinsic extinction.\\

\subsection{\label{model} The accretion disc emission model}

We described the optical-UV data points with a simple multicolour disc model (\textsc{diskpn} in the \textsc{xspec12} package, \citealt{Arnaud96}). The parameters of this model are  $kT_{max}$ (maximum temperature of the accretion disc), $R_{in}$ (inner radius of the accretion disc), and the normalization K (for details on this model see \citealt{Gierlinski99}). In our study the inner radius $R_{in}$ was set at 6.0 gravitational radii, and the normalization K was left as a free parameter. We created a   grid of models corresponding to  values  of $kT_{max}$
in the range between $kT\approx 1$ eV and  $kT\approx 10 $ eV, and we  fitted these models to the photometric data  with our routines.

For the 15 sources with  only an upper limit from \emph{GALEX}, we assumed a fixed average temperature of the accretion disc ($kT_{max}$ $\approx $4 eV, as derived from the best-fit temperature computed for the 180 sources with detected \emph{GALEX} fluxes) or a lower value of $T_{max}$, in case of inconsistency of the UV upper limits with the   fit with $kT_{max}$ $\approx $4 eV. In these cases the SED normalization was determined using only the optical data.
 
\subsection{\label{sed_final} Resulting SEDs}

One example of the optical-UV-Xray spectral energy distribution obtained applying the correction discussed so far is shown in figure \ref{sed}. The SEDs for  all the remaining 194 objects are reported in the appendix.  We also show  the best-fit power-law model in the X-ray energy range 2-10 keV obtained from the X-ray spectral analysis (for more details see C11).

 The optical/UV SED was carried out using the model and the  grid of $kT_{max}$ quoted in section \ref{model} leaving the normalization as free parameter. 
For each $kT_{max}$ the best fit normalization was established by calculating an asymmetric weighted mean that uses the 
asymmetric errors as weights in the computation.
Finally for each source, we chose the model whose $T_{max}$ minimises the value of the reduced $\chi ^2$.
Given this best fit $T_{max}$, the asymmetric range of $1\sigma$ errors on the disc luminosity reported hereafter is given by the normalization range corresponding to $\chi^{2}_{min} \pm 1$.

 \begin{figure}[htbp]
  \centering
 \includegraphics[height=7.5cm, width=8.6cm]{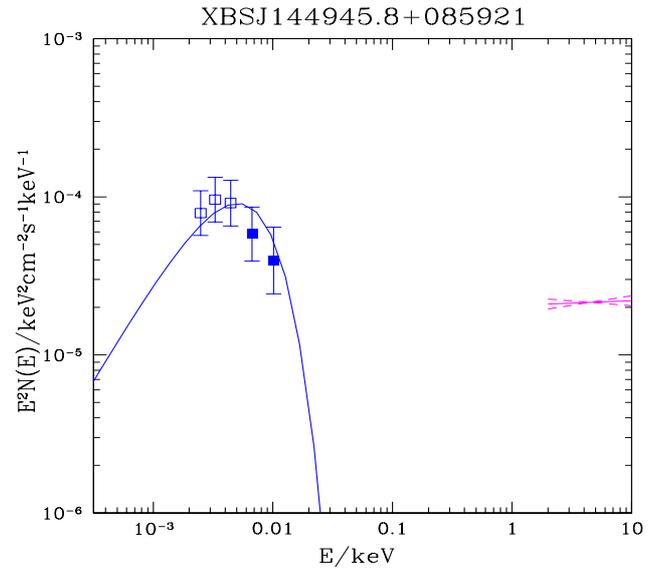} 
\caption{Example of the optical-UV-Xray spectral energy distribution obtained by applying the correction discussed so far. The filled blue squares are the fluxes in the \emph{GALEX} NUV/FUV bands, while the empty squares represent the optical data. The fit used the model quoted in section \ref{model} (blue curve in the electronic form).
The magenta curve  is  the best-fit power-law model in the X-ray energy range 2-10 keV   obtained from the X-ray spectral analysis (see C11) The dashed magenta lines represent the errors on the best-fit model of the X-ray data, given by the errors on the spectral index $\Gamma$.}
\label{sed}
 \end{figure}

\subsection{\label{Lbol_calc} Bolometric luminosities}

\begin{figure}[htbp]
  \centering
 \includegraphics[height=7.5cm, width=8.6cm]{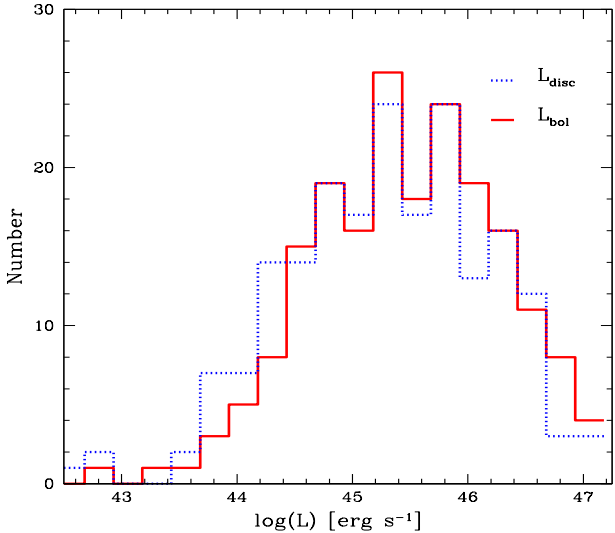}
\caption{ Distribution of the AGN accretion disc luminosities (dotted blue line), computed  by integrating over the optical-UV continuum spectra of each source and distribution of the bolometric luminosities (solid red line), obtained as a sum of the accretion disc luminosities and the  0.1-500 keV X-ray luminosities. }
\label{isto_Ld}
 \end{figure}

In  figure  \ref{isto_Ld} we report the distribution of the accretion disc  luminosities, which were computed  by integrating over the optical-UV continuum spectra of each source. The accretion disc luminosity covers a luminosity range $42.4<log(L_{disc})< 47.3 $ \ erg/s with a median value  $log(L_{disc})^{med}\cong45.4 $ erg/s.\\
The bolometric luminosities were obtained as the sum of the accretion disc luminosity and the  0.1-500 keV X-ray luminosity, and the distribution is reported in figure \ref{isto_Ld}. We computed this with the photon indices $\Gamma$  available from the X-ray spectral analysis (C11) and introduced an exponential cut-off at 200 keV (\citealt{Dadina08}), $F(E)\propto E^{-\Gamma}e^{-E/200}$. The values of $L_{[2-10] keV}$ and $\Gamma$   used hereafter are reported in Table \ref{tab_lunga}.\\
Since the largest part of the bolometric luminosity is given by the accretion disc luminosity we obtained a very similar distribution, covering the range $42.8<log(L_{bol})< 47.3 \ \rm erg/s$ with  median value $log(L_{bol})^{med}\cong45.5 \ \rm erg/s$.

%#################################################################################################################

\section{Results \label{correlations}}

We derived the accretion disc luminosities, the X-ray luminosities, and the bolometric luminosities for a 
significant sample of type 1 AGN covering a wide range of redshift ($0.03<z\lesssim 2.2$) and a wide range of X-ray luminosities ($41.8<logL_{[2-10]keV}<45.5$  erg/s). The next step is to investigate the correlations between  
$\Ldisc$, 
$\Lhard$,  
and the bolometric correction $\kbol = \frac{L_{bol}}{L_{[2-10]keV}}$.

We   computed the slope of the relation between two linearly correlated variables using two different methods\\
\begin{enumerate}
\item A linear least squares regression method choosing:\\
	   a) X as independent variable (LSQ($y\mid x$))\\
	   b) Y as independent variable (LSQ($x\mid y$));
\item  A symmetric approach (i.e. without fixing one of the variables as independent and the other as dependent) that computes the bisector between the two  least squares regression lines obtained by interchanging the choice of the independent variable (hereafter the ``bisector method",  \citealt{Isobe90}).
\end{enumerate}

In the following we   use the bisector method in those cases where the choice of the independent variable is  not obvious, and otherwise   the linear least square regression of the dependent variable Y against the independent variable X (LSQ($y\mid x$)).\\	   

The study of correlations between outputs at different wavelengths is better done by directly comparing   luminosities rather than fluxes, since any correlation in the luminosity space will be distorted in the flux space unless the luminosities are linearly correlated (\citealt{Feigelson83,Padovani92}).
However, the use of luminosities instead of fluxes always introduces a redshift bias in flux limited samples, as luminosities are strongly correlated with redshift. It is therefore crucial to estimate the influence of this effect on the correlations in order to draw reliable conclusions on the true physical relationship between two redshift-dependent variables.\\
 A way of dealing with this problem is via a partial-correlation analysis, i.e. computing the correlation between two variables (e.g. $\Ldisc$,  $\Lhard$) and checking the effect of additional parameters (in this case the redshift) that the two variables depend on.
The procedure used here is based on the Spearman rank-order correlation coefficient $r_{S,z}$ (and the two-sided probability  of no correlation P, computed given the degrees of freedom and the t-value) as modified by \cite{Kendall79} and \cite{Padovani92}, in order to take   the dependence on z into account. 
We assumed the correlation to be marginally significant if the probability P of no correlation is less than $10\%$, significant if  $P\leq5\%$, and highly significant if $P\leq 1\%$.

Finally, to check the stability of our results, the analysis was done   for the main sample of 195 sources and for a sub-sample of  176 sources  (\emph{Low-err} sample hereafter) obtained by excluding the 19 sources with the highest uncertainties (i.e. those sources having an uncertainty (1$\sigma$) on $k_{bol}$ of more than a factor 1.5).

\subsection{$L_{[2-10]keV}-L_{disc}$}
\label{LD_LX}

The  of the correlation between X-ray and UV luminosities has been, in the past, the subject of many works on optically or X-ray selected samples of AGNs (see Sect. \ref{Lx_Ld_disc} for an account of these works). For our sources, the Spearman correlation coefficient is $r_{S,z} =0.34$, giving a highly significant correlation ($P<10^{-3}$) according to our criteria, and a similar highly significant correlation is obtained for the sub-sample \emph{Low-err}. 
 
\begin{figure}[htbp]
  \centering
 \includegraphics[height=6.8cm, width=8cm]{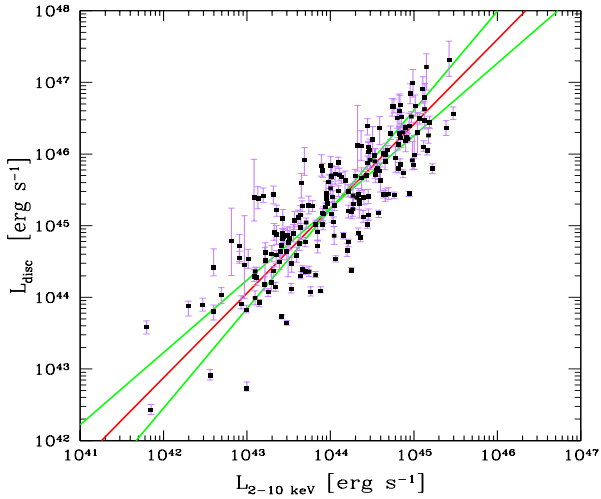}
\caption{Plot of $L_{disc} $ vs.  X-ray luminosity $L_{[2-10] keV}$ for our sample of sources. The two green lines represent the best-fit regression lines, given by Eqs. \ref{eq1_LDLX} and \ref{eq2_LDLX},  obtained by interchanging the independent and dependent variable; the red line is the bisector of these two lines, parametrised by Eq. \ref{eq3_LDLX}.}
\label{Ld_Lx}
 \end{figure}

Treating $L_{[2-10]keV}$ as the independent variable, we found
\begin{equation}
logL_{disc}=(1.009\pm0.05)logL_{[2-10]keV}+ 0.84 ,
\label{eq1_LDLX}
\end{equation}
and while  treating $L_{disc}$ as the independent variable  we found
\begin{equation}
logL_{disc}=(1.38\pm0.06)logL_{[2-10]keV}- 15.71  .
\label{eq2_LDLX}
\end{equation}
Computing the bisector of the two regression lines as described by \cite{Isobe90} the best-fit relation is

\begin{equation}
logL_{disc}=(1.18\pm 0.05)logL_{[2-10]keV} -6.68 .
\label{eq3_LDLX}
\end{equation}
 In Figure \ref{Ld_Lx} we display the  relations  \ref{eq1_LDLX},  \ref{eq2_LDLX}, and  \ref{eq3_LDLX},  between  $L_{[2-10]keV}$ and   $L_{disc}$.\\
 A similar relation was obtained for the \emph{Low err} sub-sample. Using the ``bisector method" we obtain $logL_{disc}=(1.20\pm 0.05)logL_{[2-10]keV} -7.65$. \\

It is worth noting that the correlation between the disc and the X-ray luminosity has often been  
studied in the literature by using the monochromatic optical luminosity, $L_{2500\AA }$, and the monochromatic X-ray luminosity at 2 keV, $L_{2 keV}$. For completeness, we also adopted this approach. For each source, $L_{2500\AA }$ was computed using the intrinsic best-fit SED discussed in Section 3.5, while $L_{2 keV}$ was derived from the intrinsic X-ray spectrum  (C11). 
If we use these two quantities for our sample we obtain

\begin{equation}
logL_{2500\AA }=(1.05\pm0.05)logL_{2 keV}+ 2.20
\label{eq1_4_LDLX}
\end{equation}
if treating $L_{2 keV}$ as the independent variable,

\begin{equation}
logL_{2500\AA }=(1.33\pm0.05)logL_{2 keV}- 5.09
\label{eq5_LDLX}
 \end{equation}
when treating $L_{2500\AA }$ as the independent variable, and

\begin{equation}
logL_{2500\AA }=(1.18\pm 0.05)logL_{2 keV} -1.20
\label{eq6_LDLX}
\end{equation}
using the ``bisector" method. The derived slopes are in very good agreement with those derived using $logL_{disc}$ and $logL_{[2-10]keV}$, 
confirming the equivalence of the two approaches.

Finally, we  also tried to evaluate whether $\beta$ ($L_{disc}\propto L_{[2-10]keV}^{\beta}$) could change as a function of the redshift, by splitting the sample in two sub-samples containing almost the same number of sources (having z above and below 0.6, respectively). We found marginal ($2\sigma$) evidence of an evolution of the slope with redshift, because it is steeper for the high z sub-sample.
%we do not have compelling evidence of differences between the slopes of the two subsamples, which are consistent within $\sim 2\sigma$.

\subsection{$k_{bol}-L_{[2-10] keV}$}
 
\begin{figure}[htbp]
\centering
 \includegraphics[height=6.8cm, width=8cm]{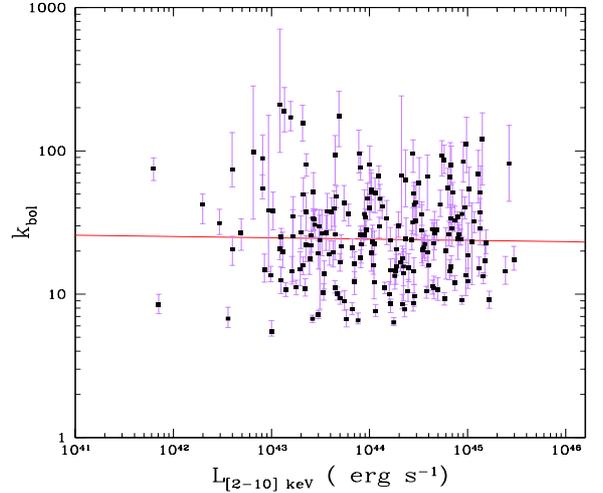}
\caption{Plot of bolometric correction vs. X-ray luminosity $L_{[2-10] keV}$ for our
sample of  sources. The red line represents the best fit relation obtained assuming  $L_{[2-10] keV}$ as independent variable. }
\label{fig:bc_Lx}
 \end{figure}

A fundamental parameter for cosmological studies of AGN is the bolometric correction to the hard X-ray (2-10 keV), $k_{bol}$, defined in equation \ref{kbol}. In Figure \ref{fig:bc_Lx} we display $k_{bol}$ against the 2-10 keV X-ray luminosity. Since the dependence of $L_{disc}$ from $L_{[2-10]keV}$ is  very close to (although statistically different from) linear,  the ``slope" in the correlation between $k_{bol}$ and $L_{[2-10]keV}$ (that was found to be highly significant when the redshift is considered) is expected to be close to 0. Indeed, using the data shown in figure \ref{fig:bc_Lx} we derived a best fit relation 
(assuming this time $L_{[2-10]keV}$ as the independent variable since $k_{bol}$ and $L_{[2-10]keV}$ are related, see Eq.
\ref{kbol}) with a flat slope\footnote { This slope is very close to what is expected  from the relationship $L_{disc}\propto L_{[2-10]keV}^{\beta}$ when we apply the {\it same} fitting method (i.e. assuming $L_{[2-10]keV}$ as the independent variable) as requested from a statistical point of view (see \citealt{Andreon10} for a discussion of this point). In this case, if ${\beta}=1.009$ (see Eq. \ref{eq1_LDLX}), and if we assume that $L_{bol}\sim L_{disc}$, we expect $logk_{bol}\propto 0.01 \times logL_{X}$, well within the errors (0.04) on the slope derived from the fit ($\beta=-0.01$).   }, $logk_{bol}\propto -(0.01\pm0.04) \times logL_{X}$

\begin{figure}[htbp]
  \centering
 \includegraphics[height=7.2cm, width=8cm]{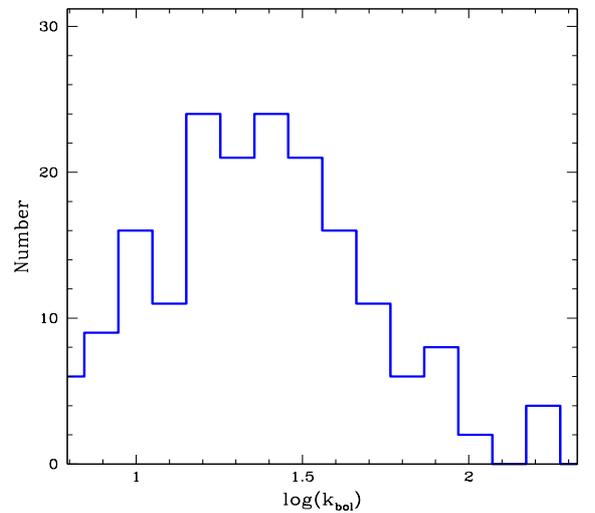}
\caption{Histogram of the bolometric corrections for the AGN population
(obtained using the 
\emph{Low-err} sample composed of 176 objects).}
\label{fig:isto_bc}   
 \end{figure}

This  flat slope (figure \ref{fig:bc_Lx}) implies a variation of less than $10\%$ in $k_{bol}$ over about five order of magnitude in  luminosities. 
This said, and given the very large dispersion observed, $k_{bol}$ could be therefore considered practically independent from $L_{[2-10]keV}$. It is   worth noting that the bulk of this large dispersion is intrinsic; i.e., the measurement errors do not constitute the dominant source of dispersion. 
In figure \ref{fig:isto_bc} we report the histogram of the bolometric corrections   derived using the \emph{Low-err} sample (176 objects),  in order to give a ``representative" distribution of $k_{bol}$ for the AGN population.

 \begin{table}
\begin{center}
\begin{tabular}{cccc}
sample & Correlation & $r_{S,z}$ & P \\
\hline
\hline
Main sample(195) & $L_{disc}-L_{[2-10]keV}$ & 0.34 & $<10^{-3}$ \\
 & $k_{bol}-L_{[2-10]keV}$ & -0.29 & $<10^{-6}$ \\
 \hline
$Low-err$(176) & $L_{disc}-L_{[2-10]keV}$  & 0.35 &  $<10^{-4}$ \\
& $k_{bol}-L_{[2-10]keV}$  & -0.23 &  $<10^{-3}$ \\
\end{tabular}
\caption[]{Spearman rank correlation coefficients ($r_{S,z}$) and probabilities of null
correlation (P) for the samples considered in this work (in parenthesis the number of
sources in each sample).}
\label{corr_table}
\end{center}
\end{table}
    
\section{Discussion}
\label{discussion}

\subsection{The X-ray to Optical/UV correlation}
\label{Lx_Ld_disc}

The study of the correlation between X-ray and UV luminosities has been subject of many works on optically or X-ray selected samples of AGNs. This correlation is found to be $L_{UV}\propto L_X^{\beta}$, with $\beta$ ranging, 
in most studies, from 1.2 to 1.6 (\citealt{Avni82,Avni86,Kriss85,Anderson87,Wilkes94,Yuan98,Vignali03,Strateva05,Steffen06,Just07,Young09,Lusso10,Grupe10,Stalin10}).
This correlation provides a strong constraint on the physical processes at work, since it tells us that the fraction of power in the accretion disc corona, emitted in the X-rays, decreases/increases as a function of the accretion disc power (mainly emitted in the UV).

A direct comparison between our results and previous works is not very straightforward since different 
samples (each one with its selection effects that should be carefully evaluated) and different fitting methods 
(LSQ($y\mid x$)), LSQ($x\mid y$), or else the bisector method, have been used in the past. Furthermore, as also 
stressed by \cite{Green09},  different regression methods can yield very 
different results in samples with large dispersions. 

As discussed in \cite{Isobe90}  when the 
physics (or the question to be solved) does  not clearly indicate which variable depends on the other,
a symmetric approach (such as the bisector method) should provide the best guess for the {\it intrinsic} relationship between the two variables (e.g. the X-ray and the disc luminosities in this case). 
Using the bisector method we found a slope that   is different from linear at $\sim 3.6\sigma$, i.e. $L_{Disc} \propto L_{[2-10] keV}^{(1.18\pm 0.05)}$, and  this result can now  be  compared with those obtained from other authors using a similar symmetric approach.

Statistically we found that   our best-fit slope differs $\sim 2.8 \sigma$   from the slope obtained by Lusso et al. 
(2010, $\beta = 1.32\pm 0.04$), $4.6 \sigma$   from Just et al. (2007, $\beta = 1.41\pm 0.01$), and $5.6 \sigma$  from Green et al. (2009, $\beta = 0.90\pm 0.02$).  However, there are a number of differences between our analysis and those reported above that can alleviate these differences. 

First of all, as mentioned above, we are dealing with a statistically representative sample of X-ray selected, \emph{spectroscopically confirmed}, Type 1 AGN. For all the studied objects we have an X-ray spectrum allowing us to measure the intrinsic X-ray luminosity; UV data from \emph{GALEX} are available for about 93\% of the sample.
Finally we   also tried to  account for the factors contaminating or absorbing the primary radiation emitted from the AGN.  

For the optically selected sample of SDSS AGN used in \cite{Green09}, the result 
quoted above refers to what is obtained with their ``Main" sample ($\sim$ 2300 sources) having an X-ray detection fraction of only about 50\%. When using their ``zLxBox" sample, with a detection fraction of 100\%, \cite{Green09} obtain $\beta = 1.19\pm 0.02$, in excellent agreement with our results. \cite{Just07} combine a sample of 26 optically selected, X-ray-observed AGN, at high z (between 1.5 and 4.5) with an homogeneous sample of 333 optically selected AGN from \cite{Steffen06}. Most of the Type 1 unabsorbed AGN have been classified as such using only photometric data. Finally \cite{Lusso10} use a sample of X-ray selected Type 1 AGN from the COSMOS survey (545 objects) with about 40\% of the objects with a photometric redshift and a classification based on their multi-band SED. All   considered, it is difficult to say whether these differences in sample selections  have any influence on the best-fit relationship.   We aimed at obtaining the most accurate result by adopting a new large (195 objects), statistically complete sample of X-ray selected Type 1 AGN, spanning a wide range of redshift ($0.03<z\lesssim 2.2$) and X-ray luminosities ($41.8<logL_{[2-10]keV}<45.5$ erg/s),  with complete spectroscopic identifications and accurate X-ray spectral analysis (C11);
furthermore, we also accounted for the factors contaminating or absorbing the primary radiation emitted from the AGN.  Because statistically representative and composed of ``X-ray bright" sources, this sample is ideal for this kind of investigation.

\subsection{The bolometric correction, $k_{bol}$}
\label{bol_corr}

The bolometric correction $k_{bol}$, used to derive the bolometric luminosities (laborious to obtain) from the measured X-ray luminosity, 
is a fundamental parameter for much important research in current physical cosmology, 
including the study of the accretion rate (e.g., \citealt{Marconi04}), 
  measurement of SMBH densities in the universe (e.g., \citealt{Marconi04}), 
 estimation of the active accretion lifetimes or duty cycles (e.g., \citealt{Hopkins05,Adelberger05}), and  the conversion the energy density of the X-ray background (which is produced by the 
integrated X-ray emission of AGN) into a mass density of SMBHs in the local Universe (\citealt{Soltan82}). 

In section 4.2 we  discussed our $k_{bol}$ values, where we had  a very shallow dependence 
with the X-ray luminosity if compared with its very broad distribution (ranging from $k_{bol}\sim 5$ up to $k_{bol}\sim $ few hundred).  Similar results were obtained by \cite{Vasudevan07} with a sample of 54 nearby AGN, where they also found a significant spread in $k_{bol} $ (ranging from $\sim 5$ up to $\sim 100$ when Radio Loud objects are removed). It is clear that this rather broad and ``flat" distribution implies a very large distribution in the SED of AGN, preventing the use of a single value for the total population. Given the importance of $k_{bol}$ in a cosmological context, many authors investigated the possibility of deriving it from correlations with other simpler observables   to compute. After already showing that we cannot use the X-ray luminosities, we discuss below two other    observables recently proposed in literature. 

\vskip 0.2truecm

In a recent paper, \cite{Lusso10} have reported a very tight correlation between $k_{bol}$ and the   X-ray to optical-UV luminosity ratio  $\aox$
\footnote{
Usually the X-ray to optical-UV luminosity ratio in AGN is parametrised by the optical-to-X-ray spectral index
\begin{equation}
\alpha_{ox}=-0.384 log\left[\frac{L_{2 keV}}{L_{2500\AA } }\right] .  
\end{equation} 
For each source $L_{2500\AA }$ was computed using the intrinsic best-fit SED discussed in Section 3.5, while $L_{2 keV}$ was derived from the intrinsic X-ray spectrum (C11).}, which is expected since both quantities are sensitive to the strength of the optical/UV part of the AGN spectrum compared with the X-ray part . However,  to compute $\aox$  ideally, only two measurements are needed, one in a proximity of 2500 $\AA$ \ and another of 2 keV
(both rest frame). Because of the good correlation between $k_{bol}$ and $\aox$, \cite{Lusso10} state that $\aox$ can be used as an accurate estimate ($\sim 20\%$ at $1\sigma$) of $k_{bol}$. After studying  an independent sample of type 1 AGN using a different analysis to derive the bolometric output, we can now test the validity of this suggestion. 

\begin{figure}[htbp]
  \centering
 \includegraphics[height=6.8cm, width=8cm]{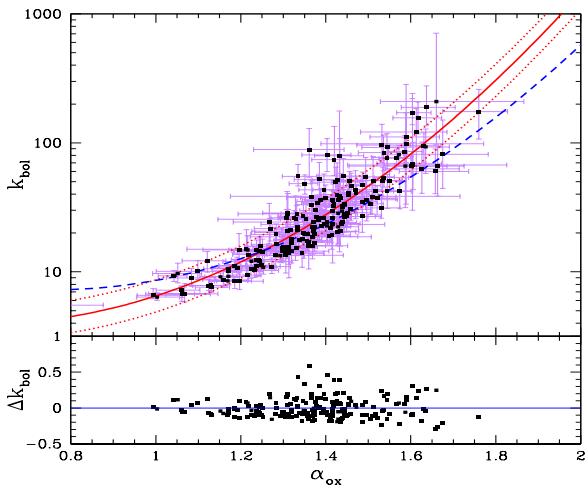}
\caption{Upper panel: plot of $k_{bol}$ against $\aox$ for our sample of Type 1 AGN. 
The solid line and dotted lines  represent our best fit and  the range of variation with respect to the best-fit, including 68\% of the sources. The dashed line represent the best fit relation computed by \cite{Lusso10}; in the bottom panel we show the residuals with respect to our best-fit relationship. }
\label{alfa_bc}
 \end{figure}

In figure \ref{alfa_bc} we show $k_{bol}$ against $\aox$ for our sample of sources, where a tight correlation is clearly 
present. However we found a steeper correlation with respect to the best fit in \cite{Lusso10}. Fitting our data with a quadratic polynomial we found the following best-fit relation:
 
\begin{equation}
logk_{bol}=1.05 - 1.52 \alpha_{ox} + 1.29 \alpha_{ox}^2
\label{eq11}
\end{equation}
(see figure \ref{alfa_bc} upper panel). In the bottom panel we also show the residual $\Delta k_{bol}$ from our best-fit relation: about 68\% ($\simeq 1\sigma$) of the objects are contained within 
$14\%$ of the best-fit relationship while the maximum deviation of the data points from the expected relation is 
$\sim 60\%$. Provided that measurements around $2500 \ \AA \ $ and around 2 keV (rest frame) are available and \emph{that all the   discussed corrections to these measurements are applied}, we confirm that $\aox$ could be used as a proxy of $k_{bol}$.  In particular, using the relation (\ref{eq11}) the value of $k_{bol}$ can be 
estimated with a  mean accuracy of 0.14 dex ($\sim 30\%$).
 
\vskip 0.3truecm

\begin{figure}[htbp]
  \centering
 \includegraphics[height=6.8cm, width=8cm]{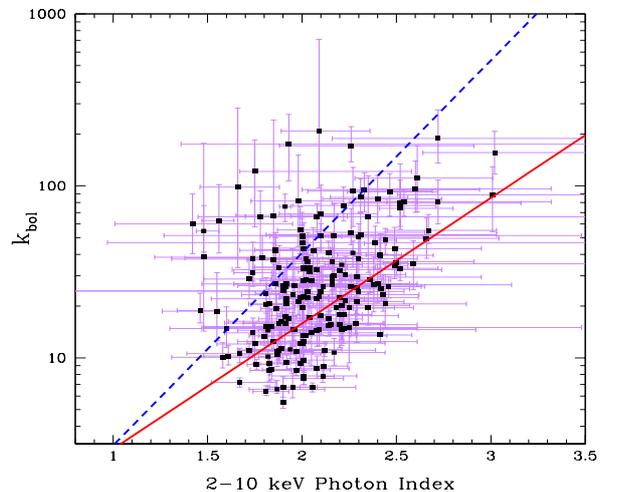}
\caption{Plot of  $k_{bol}$ vs. the 2-10 keV photon index $\Gamma_{2-10 keV}$. We show our best-fit relation (red solid line)  and the best-fit relation computed by Zhou and Zhao (2010) (blue dashed line).}
\label{g_bc}
\end{figure}

A correlation between the hard X-ray photon index $\Gamma_{2-10 keV}$ and the bolometric correction has been recently 
proposed by \cite{Zhou10} based on the data on 29 optically selected low-redshift ($z<0.33$) AGN. They propose that the rest frame $\Gamma_{2-10 keV}$ can be used as a proxy for the bolometric correction with a mean uncertainty of a factor 2-3. The AGN sample discussed here is ideal for testing this correlation, since a detailed X-ray spectral analysis was carried out for all the sources in the sample (see C11). In figure \ref{g_bc} we report the relation between the hard X-ray photon index $\Gamma_{2-10 keV}$ and the bolometric correction for our sample of AGN.

A highly significant correlation between the two quantities is clearly present ($r_S =0.37, P<10^{-6}$), confirming the results of \cite{Zhou10}; for comparison, we also show in figure \ref{g_bc} the best-fit relationship reported by \cite{Zhou10}, obtained using a symmetric approach ($log k_{bol}=(1.12 \pm 0.30)\Gamma_{2-10 keV} -(0.63 \pm 0.53)$). For our sample, the best-fit relation, obtained assuming $\Gamma_{2-10 keV}$ as independent variable, is
\begin{equation}
log k_{bol}=(0.73 \pm 0.04)\Gamma_{2-10 keV} -(0.26\pm 0.07).
\label{eq_g_k}
\end{equation}
 The divergence  between the best-fit relation derived for our data set and the \cite{Zhou10} relationship can be attributed to the different statistical approach   and to selection effects arising when comparing X-ray and optically selected samples.  We conclude that   equation (\ref{eq_g_k}) can be adopted to derive $k_{bol}$ from $\Gamma_{2-10 keV}$ with a mean error of $\sim 0.34$ dex (i.e $\sim 80\%$).\\ The physical justification of the observed $ k_{bol}-\Gamma_{2-10 keV}$ relation could 
be related to the common dependence of these two quantities to the same 
physical parameter, namely the Eddington ratio ($\lambda_{Edd}$). Indeed, evidence 
for a direct dependence of both $k_{bol}$ and $\Gamma$  on $\lambda_{Edd}$ have recently been  
found by different authors (e.g. \citealt{Vasudevan09,Risaliti09,Lusso10,Shemmer09,Grupe10,Caccianiga4,Grupe11}). This common dependence 
on  $\lambda_{Edd}$ may naturally lead to a mutual correlation between these two 
quantities. A more detailed investigation of these correlations, using the 
AGN of the XBS, is currently in progress.

\section{Summary and conclusions}
\label{summary}  

In this paper we have presented the analysis of the optical-UV-X-ray SEDs of a complete and representative sample of 195 X-ray selected, spectroscopically identified, type 1 AGN, with intrinsic absorbing column densities $N_{\rm H}<4 \times 10^{21}  \ cm^{-2}$. The adopted sample, spanning a wide range of redshift $0.03 <z\lesssim 2.2$ and X-ray luminosity, $6 \times 10^{41}$ erg/s $< L_{[2-10]keV}<$ $3 \times 10^{45}$ erg/s, is composed of relatively bright AGN and thus ideal for this study. The optical-UV SED was investigated using data from the \emph{Sloan Digital Sky Survey} (SDSS), from our own dedicated optical spectroscopy and photometry and from the satellite \emph{Galaxy Evolution Explorer} (\emph{GALEX}), while the X-ray part was studied using data from XMM-\emph{Newton}.

While the X-ray spectra was presented and discussed in C11, we  derived here the intrinsic optical-UV SED, trying to take  all the factors contaminating or absorbing the primary optical-UV radiation emitted from the AGN into account. In particular we   applied correction for 
\begin{itemize}
\item the Galactic extinction;
\item the potential attenuation given by absorption of neutral hydrogen in intervening Lyman-$\alpha$ absorption systems; 
\item the contamination from the stellar emission of the host galaxy of the AGN;
\item the intrinsic extinction, using the mean extinction curve computed by \cite{Gaskell07}, a galactic gas-to-dust ratio, and the measured values (or upper limits) of intrinsic column density $N_{\rm H}$ available from the X-ray spectral analysis;
\item the contribution of the emission lines Lyman $\alpha$+NV, CIV, MgII, OIV+Ly$\beta$ and CIII+SiIII.
\end{itemize}
The intrinsic optical-UV radiation was thus fitted using an accretion disc emission model.

We investigated here the correlations between the physical parameters inferred from the optical to X-ray SEDs, such as accretion disc luminosity, bolometric luminosity, bolometric correction, and X-ray photon index.
 We recall that our results are  also applicable to Type 2 (obscured) AGNs if the obscuration is a line-of-sight orientation effect and does not affect the intrinsic energy generation mechanisms within AGNs (i.e. if the unified model of AGNs is valid).
The main results of this work are the following 
\begin{enumerate}

\item Using a symmetric fitting approach (the bisector method) we found a highly significant {\it intrinsic} correlation between the accretion disc luminosity $L_{disc}$ and the X-ray luminosity $L_{[2-10] keV}$, in the form $L_{disc} \propto L_{[2-10] keV}^{\beta}$, where $\beta =1.18 \pm 0.05$. The non-linearity of this relation implies  that the fraction of X-ray emission decreases with increasing accretion disc luminosity. We also found marginal evidence  that this relationship depends on redshift;

\item We found a very shallow dependence of the bolometric correction on the X-ray luminosity if compared with its very broad distribution (ranging from $k_{bol}\sim 5$ up to $k_{bol}\sim $ few hundred). This rather broad and ``flat" distribution ( mainly intrinsic, i.e. it is not caused by the measurement errors)   implies a very large distribution of the SED in AGN, preventing the use of a single value for the total AGN family. Furthermore, this  implies that the X-ray luminosities are not a useful proxy to derive bolometric corrections;

\item We confirm a tight correlation between $k_{bol}$ and $\alpha_{ox}$ and a correlation between $k_{bol}$ and 
the hard-X-ray photon index. The first correlation could be used to derive a relatively good
estimate of $k_{bol}$,  with a mean error of $ \sim 30\% $ (providing that all the main corrections 
discussed here are applied). The  second
correlation, despite its higher dispersion, can  also be adopted (with a mean error of $\sim 80\%$) to estimate $k_{bol}$ once the
photon index is known.  

\end{enumerate}

A correlation analysis of $L_{disc}$, $L_{[2-10]keV}$, and $k_{bol}$ with the physical parameters
of the central supermassive black-hole (e.g. black hole mass and Eddington ratio) for the AGN sample
belonging to the XBS survey will be reported in a forthcoming paper. Expanding the analysis to include these parameters may provide a fundamental tool for constraining AGN models (see e.g. \citealt{Sobolewska04a,Sobolewska04b}).

\acknowledgement{We thank Luciana Bianchi for help with the \emph{GALEX} data, Stefano Andreon for the useful suggestions about the statistical issues, Valentina Braito, Cristian Vignali and the anonymous referee for many 
useful suggestions and comments that significantly improved the paper. The authors acknowledge financial support from ASI (grant n. I/088/06/0, COFIS contract
and ASI-INAF grant n. I/009/10/0).}

\newpage

\onecolumn

\renewcommand{\arraystretch}{2} 
\renewcommand{\thefootnote}{\alph{footnote}}
\clearpage

\begin{longtable}[|c|]{|p{2.5cm}|p{0.95cm}|p{1.1cm}|p{1.5cm}|p{1.1cm}|p{0.9cm}|p{1cm}|p{1cm}|}  

\caption{\label{tab_lunga} Main properties of the sample of 195 type 1 AGN analysed in this work.}\\
\hline 

 \multicolumn{1}{|p{2.5cm}|}{name XBSJ} &
  \multicolumn{1}{|p{0.95cm}|}{z} &
  \multicolumn{1}{|p{1.1cm}|}{$logL_{disc}  \ ^{\mathrm{a}}$} &
  \multicolumn{1}{|p{1.65cm}|}{$logL_{[2-10]\rm keV}  \ ^{\mathrm{a,b}}$} &
  \multicolumn{1}{|p{1.1cm}|}{$logL_{bol} \ ^{\mathrm{a}}$} &
  \multicolumn{1}{|p{0.9cm}|}{$\Gamma \ ^{\mathrm{b}} $} &
  \multicolumn{1}{|p{1cm}|}{$\alpha_{ox}$} & 
  \multicolumn{1}{|p{1cm}|}{$log k_{bol}$}     \\
  \hline
  \hline

\endfirsthead
\multicolumn{1}{l}{{Continued on Next Page\ldots}} \\
\endfoot
\hline
\endfoot
\hline 
	 \multicolumn{1}{|p{2.5cm}|}{name XBSJ} &
  \multicolumn{1}{|p{0.95cm}|}{z} &
  \multicolumn{1}{|p{1.1cm}|}{$logL_{disc}  \ ^{\mathrm{a}}$} &
  \multicolumn{1}{|p{1.65cm}|}{$logL_{[2-10]\rm keV}  \ ^{\mathrm{a,b}}$} &
  \multicolumn{1}{|p{1.1cm}|}{$logL_{bol} \ ^{\mathrm{a}}$} &
  \multicolumn{1}{|p{0.9cm}|}{$\Gamma \ ^{\mathrm{b}} $} &
  \multicolumn{1}{|p{1cm}|}{$\alpha_{ox}$} & 
   \multicolumn{1}{|p{1cm}|}{$log k_{bol}$}     \\
  \hline
\hline
\endhead
\hline
\multicolumn{3}{c}{}\\
%\caption{}

\endlastfoot
000027.7-250442 &  0.34 & $ 44.64^{+0.09}_{-0.11} $  &  43.48 & $ 44.76^{+0.09}_{-0.11} $ & $ 1.87^{+0.09}_{-0.08} $ & $  1.41^{+0.04}_{-0.06} $ & $ 1.29^{+0.09}_{-0.11} $   \\ 
000031.7-245502 &  0.28 & $ 44.52^{+0.18}_{-0.32} $  &  43.22 & $ 44.62^{+0.18}_{-0.32} $ & $ 2.29^{+0.14}_{-0.13} $ & $  1.33^{+0.09}_{-0.18} $ & $ 1.41^{+0.18}_{-0.32} $   \\ 
000102.4-245850 &  0.43 & $ 44.36^{+0.08}_{-0.07} $  &  43.74 & $ 44.70^{+0.07}_{-0.06} $ & $ 2.12^{+0.13}_{-0.12} $ & $  1.08^{+0.06}_{-0.06} $ & $ 0.95^{+0.07}_{-0.06} $   \\ 
001831.6+162925 &  0.55 & $ 45.71^{+0.10}_{-0.08} $  &  44.11 & $ 45.78^{+0.10}_{-0.08} $ & $ 2.39^{+0.06}_{-0.06} $ & $  1.46^{+0.04}_{-0.04} $ & $ 1.67^{+0.10}_{-0.08} $   \\ 
002618.5+105019 &  0.47 & $ 45.87^{+0.10}_{-0.08} $  &  44.44 & $ 45.94^{+0.10}_{-0.08} $ & $ 2.04^{+0.06}_{-0.06} $ & $  1.44^{+0.04}_{-0.04} $ & $ 1.50^{+0.10}_{-0.08} $   \\ 
002637.4+165953 &  0.55 & $ 45.31^{+0.08}_{-0.07} $  &  44.27 & $ 45.47^{+0.08}_{-0.06} $ & $ 2.15^{+0.07}_{-0.05} $ & $  1.30^{+0.04}_{-0.04} $ & $ 1.20^{+0.08}_{-0.06} $   \\ 
003315.5-120700 &  1.21 & $ 45.91^{+0.26}_{-0.15} $  &  44.98 & $ 46.11^{+0.25}_{-0.16} $ & $ 2.01^{+0.28}_{-0.16} $ & $  1.31^{+0.14}_{-0.10} $ & $ 1.13^{+0.25}_{-0.16} $   \\ 
003316.0-120456 &  0.66 & $ 45.84^{+0.16}_{-0.14} $  &  43.89 & $ 45.89^{+0.16}_{-0.14} $ & $ 2.60^{+0.72}_{-0.29} $ & $  1.54^{+0.07}_{-0.06} $ & $ 2.00^{+0.16}_{-0.14} $   \\ 
003418.9-115940 &  0.85 & $ 45.68^{+0.15}_{-0.17} $  &  44.37 & $ 45.78^{+0.15}_{-0.17} $ & $ 2.10^{+0.44}_{-0.26} $ & $  1.28^{+0.07}_{-0.08} $ & $ 1.40^{+0.15}_{-0.17} $   \\ 
005009.9-515934 &  0.61 & $ 45.06^{+0.08}_{-0.06} $  &  44.04 & $ 45.24^{+0.08}_{-0.06} $ & $ 2.28^{+0.15}_{-0.13} $ & $  1.25^{+0.04}_{-0.04} $ & $ 1.21^{+0.08}_{-0.06} $   \\ 
005031.1-520012 &  0.46 & $ 45.45^{+0.17}_{-0.17} $  &  43.95 & $ 45.51^{+0.17}_{-0.17} $ & $ 2.03^{+0.35}_{-0.19} $ & $  1.50^{+0.07}_{-0.08} $ & $ 1.56^{+0.17}_{-0.17} $   \\ 
005032.3-521543 &  1.22 & $ 45.92^{+0.16}_{-0.20} $  &  44.78 & $ 46.06^{+0.16}_{-0.20} $ & $ 2.21^{+0.36}_{-0.25} $ & $  1.34^{+0.08}_{-0.12} $ & $ 1.29^{+0.16}_{-0.20} $   \\ 
010421.4-061418 &  0.52 & $ 44.09^{+0.06}_{-0.02} $  &  43.88 & $ 44.70^{+0.05}_{-0.02} $ & $ 1.87^{+0.25}_{-0.15} $ & $  0.99^{+0.07}_{-0.03} $ & $ 0.82^{+0.05}_{-0.02} $   \\ 
010432.8-583712 &  1.64 & $ 46.37^{+0.11}_{-0.10} $  &  45.38 & $ 46.56^{+0.10}_{-0.09} $ & $ 1.95^{+0.08}_{-0.06} $ & $  1.25^{+0.06}_{-0.06} $ & $ 1.17^{+0.10}_{-0.09} $   \\ 
010701.5-172748 &  0.89 & $ 46.23^{+0.15}_{-0.13} $  &  44.84 & $ 46.31^{+0.15}_{-0.12} $ & $ 2.02^{+0.35}_{-0.19} $ & $  1.32^{+0.07}_{-0.06} $ & $ 1.47^{+0.15}_{-0.12} $   \\ 
010747.2-172044 &  0.98 & $ 46.67^{+0.11}_{-0.14} $  &  44.74 & $ 46.70^{+0.11}_{-0.14} $ & $ 2.47^{+0.25}_{-0.22} $ & $  1.54^{+0.04}_{-0.06} $ & $ 1.96^{+0.11}_{-0.14} $   \\ 
012000.0-110429 &  0.35 & $ 44.71^{+0.16}_{-0.21} $  &  42.91 & $ 44.86^{+0.16}_{-0.21} $ & $ 3.01^{+3.62}_{-0.50} $ & $  1.36^{+0.08}_{-0.13} $ & $ 1.95^{+0.16}_{-0.21} $   \\ 
012025.2-105441 &  1.34 & $ 46.81^{+0.14}_{-0.14} $  &  44.96 & $ 46.85^{+0.14}_{-0.14} $ & $ 2.40^{+0.34}_{-0.29} $ & $  1.56^{+0.06}_{-0.06} $ & $ 1.90^{+0.14}_{-0.14} $   \\ 
012119.9-110418 &  0.20 & $ 44.91^{+0.12}_{-0.12} $  &  43.32 & $ 45.02^{+0.12}_{-0.12} $ & $ 2.66^{+0.38}_{-0.23} $ & $  1.42^{+0.06}_{-0.06} $ & $ 1.70^{+0.12}_{-0.12} $   \\ 
012540.2+015752 &  0.12 & $ 43.80^{+0.09}_{-0.12} $  &  42.60 & $ 43.92^{+0.09}_{-0.11} $ & $ 1.83^{+0.13}_{-0.11} $ & $  1.43^{+0.04}_{-0.06} $ & $ 1.31^{+0.09}_{-0.11} $   \\ 
013204.9-400050 &  0.45 & $ 45.15^{+0.12}_{-0.12} $  &  43.64 & $ 45.24^{+0.12}_{-0.12} $ & $ 2.42^{+0.28}_{-0.23} $ & $  1.45^{+0.06}_{-0.06} $ & $ 1.60^{+0.12}_{-0.12} $   \\ 
013944.0-674909 &  0.10 & $ 42.91^{+0.08}_{-0.06} $  &  42.56 & $ 43.39^{+0.08}_{-0.06} $ & $ 1.95^{+0.13}_{-0.12} $ & $  1.07^{+0.08}_{-0.08} $ & $ 0.83^{+0.08}_{-0.06} $   \\ 
014227.0+133453 &  0.28 & $ 44.07^{+0.05}_{-0.06} $  &  43.25 & $ 44.30^{+0.05}_{-0.05} $ & $ 1.94^{+0.27}_{-0.23} $ & $  1.19^{+0.03}_{-0.04} $ & $ 1.05^{+0.05}_{-0.05} $   \\ 
014251.5+133352 &  1.07 & $ 46.29^{+0.10}_{-0.09} $  &  44.72 & $ 46.35^{+0.10}_{-0.08} $ & $ 1.86^{+0.24}_{-0.19} $ & $  1.52^{+0.04}_{-0.04} $ & $ 1.62^{+0.10}_{-0.08} $   \\ 
015957.5+003309 &  0.31 & $ 44.53^{+0.05}_{-0.04} $  &  44.06 & $ 44.94^{+0.05}_{-0.04} $ & $ 2.01^{+0.21}_{-0.11} $ & $  1.13^{+0.04}_{-0.04} $ & $ 0.88^{+0.05}_{-0.04} $   \\ 
020029.0+002846 &  0.17 & $ 43.83^{+0.06}_{-0.05} $  &  42.99 & $ 44.13^{+0.06}_{-0.05} $ & $ 2.42^{+0.17}_{-0.16} $ & $  1.22^{+0.04}_{-0.04} $ & $ 1.13^{+0.06}_{-0.05} $   \\ 
021808.3-045845 &  0.71 & $ 46.15^{+0.10}_{-0.08} $  &  44.81 & $ 46.23^{+0.10}_{-0.08} $ & $ 1.91^{+0.07}_{-0.05} $ & $  1.42^{+0.04}_{-0.04} $ & $ 1.42^{+0.09}_{-0.08} $   \\ 
021817.4-045113 &  1.08 & $ 45.78^{+0.06}_{-0.07} $  &  45.22 & $ 46.18^{+0.05}_{-0.06} $ & $ 1.83^{+0.07}_{-0.05} $ & $  1.16^{+0.05}_{-0.07} $ & $ 0.96^{+0.05}_{-0.06} $   \\ 
021820.6-050427 &  0.65 & $ 45.49^{+0.07}_{-0.08} $  &  44.21 & $ 45.59^{+0.06}_{-0.07} $ & $ 1.81^{+0.07}_{-0.06} $ & $  1.44^{+0.03}_{-0.04} $ & $ 1.38^{+0.06}_{-0.07} $   \\ 
021923.2-045148 &  0.63 & $ 45.53^{+0.10}_{-0.08} $  &  44.00 & $ 45.61^{+0.10}_{-0.08} $ & $ 2.41^{+0.12}_{-0.07} $ & $  1.43^{+0.04}_{-0.04} $ & $ 1.61^{+0.10}_{-0.08} $   \\ 
023459.7-294436 &  0.45 & $ 45.92^{+0.17}_{-0.21} $  &  43.69 & $ 45.93^{+0.17}_{-0.21} $ & $ 1.93^{+1.83}_{-0.57} $ & $  1.76^{+0.07}_{-0.08} $ & $ 2.24^{+0.17}_{-0.21} $   \\ 
024200.9+000020 &  1.11 & $ 46.21^{+0.07}_{-0.04} $  &  44.93 & $ 46.31^{+0.07}_{-0.04} $ & $ 2.03^{+0.08}_{-0.07} $ & $  1.44^{+0.03}_{-0.02} $ & $ 1.38^{+0.07}_{-0.04} $   \\ 
024204.7+000814 &  0.38 & $ 45.42^{+0.11}_{-0.09} $  &  43.20 & $ 45.43^{+0.11}_{-0.09} $ & $ 2.26^{+0.65}_{-0.38} $ & $  1.61^{+0.04}_{-0.04} $ & $ 2.24^{+0.11}_{-0.09} $   \\ 
024207.3+000037 &  0.38 & $ 44.83^{+0.06}_{-0.07} $  &  43.43 & $ 44.95^{+0.06}_{-0.07} $ & $ 2.52^{+0.20}_{-0.14} $ & $  1.37^{+0.03}_{-0.04} $ & $ 1.52^{+0.06}_{-0.07} $   \\ 
024325.6-000413 &  0.36 & $ 44.43^{+0.11}_{-0.11} $  &  43.54 & $ 44.66^{+0.10}_{-0.11} $ & $ 1.74^{+0.28}_{-0.15} $ & $  1.33^{+0.06}_{-0.08} $ & $ 1.12^{+0.10}_{-0.10} $   \\ 
025606.1+001635 &  0.63 & $ 45.29^{+0.09}_{-0.07} $  &  44.05 & $ 45.40^{+0.09}_{-0.07} $ & $ 2.20^{+0.41}_{-0.21} $ & $  1.37^{+0.04}_{-0.04} $ & $ 1.35^{+0.09}_{-0.07} $   \\ 
025645.4+000031 &  0.36 & $ 43.73^{+0.03}_{-0.03} $  &  43.42 & $ 44.25^{+0.03}_{-0.03} $ & $ 2.06^{+0.23}_{-0.20} $ & $  1.06^{+0.03}_{-0.04} $ & $ 0.83^{+0.03}_{-0.03} $   \\ 
030206.8-000121 &  0.64 & $ 45.44^{+0.05}_{-0.06} $  &  44.64 & $ 45.70^{+0.05}_{-0.05} $ & $ 1.89^{+0.05}_{-0.05} $ & $  1.22^{+0.03}_{-0.04} $ & $ 1.05^{+0.05}_{-0.05} $   \\ 
031015.5-765131 &  1.19 & $ 46.54^{+0.10}_{-0.08} $  &  45.47 & $ 46.70^{+0.09}_{-0.07} $ & $ 1.91^{+0.04}_{-0.04} $ & $  1.35^{+0.05}_{-0.04} $ & $ 1.23^{+0.09}_{-0.07} $   \\ 
031311.7-765428 &  1.27 & $ 46.37^{+0.18}_{-0.18} $  &  44.94 & $ 46.49^{+0.16}_{-0.16} $ & $ 2.16^{+0.75}_{-0.16} $ & $  1.42^{+0.08}_{-0.08} $ & $ 1.54^{+0.16}_{-0.16} $   \\ 
031401.3-545959 &  0.84 & $ 45.41^{+0.11}_{-0.10} $  &  44.40 & $ 45.59^{+0.10}_{-0.10} $ & $ 1.84^{+0.35}_{-0.36} $ & $  1.31^{+0.06}_{-0.06} $ & $ 1.19^{+0.10}_{-0.10} $   \\ 
031549.4-551811 &  0.81 & $ 45.15^{+0.09}_{-0.08} $  &  44.39 & $ 45.43^{+0.08}_{-0.07} $ & $ 1.87^{+0.23}_{-0.21} $ & $  1.23^{+0.06}_{-0.06} $ & $ 1.03^{+0.08}_{-0.07} $   \\ 
031859.2-441627 &  0.14 & $ 44.46^{+0.68}_{-0.47} $  &  42.97 & $ 44.56^{+0.66}_{-0.43} $ & $ 1.48^{+0.31}_{-0.26} $ & $  1.43^{+0.28}_{-0.26} $ & $ 1.59^{+0.66}_{-0.43} $   \\ 
033208.7-274735 &  0.54 & $ 45.17^{+0.10}_{-0.14} $  &  43.92 & $ 45.27^{+0.10}_{-0.13} $ & $ 1.99^{+0.15}_{-0.11} $ & $  1.43^{+0.05}_{-0.07} $ & $ 1.35^{+0.10}_{-0.13} $   \\ 
033506.0-255619 &  1.43 & $ 46.92^{+0.17}_{-0.20} $  &  45.11 & $ 46.95^{+0.17}_{-0.20} $ & $ 2.10^{+0.30}_{-0.22} $ & $  1.61^{+0.07}_{-0.08} $ & $ 1.85^{+0.17}_{-0.20} $   \\ 
033851.4-352646 &  1.07 & $ 46.33^{+0.15}_{-0.24} $  &  44.59 & $ 46.37^{+0.15}_{-0.23} $ & $ 1.78^{+0.08}_{-0.08} $ & $  1.62^{+0.06}_{-0.10} $ & $ 1.79^{+0.15}_{-0.23} $   \\ 
033912.1-352813 &  0.47 & $ 44.58^{+0.12}_{-0.09} $  &  43.59 & $ 44.87^{+0.10}_{-0.07} $ & $ 1.46^{+0.10}_{-0.12} $ & $  1.41^{+0.07}_{-0.06} $ & $ 1.28^{+0.10}_{-0.07} $   \\ 
033942.8-352411 &  1.04 & $ 45.97^{+0.09}_{-0.07} $  &  44.53 & $ 46.08^{+0.09}_{-0.07} $ & $ 2.50^{+0.06}_{-0.07} $ & $  1.38^{+0.04}_{-0.04} $ & $ 1.54^{+0.09}_{-0.07} $   \\ 
041108.1-711341 &  0.92 & $ 45.64^{+0.18}_{-0.17} $  &  44.60 & $ 45.80^{+0.18}_{-0.18} $ & $ 1.91^{+0.52}_{-0.32} $ & $  1.36^{+0.09}_{-0.10} $ & $ 1.20^{+0.18}_{-0.18} $   \\ 
043448.3-775329 &  0.10 & $ 42.73^{+0.08}_{-0.04} $  &  43.00 & $ 43.74^{+0.07}_{-0.04} $ & $ 1.90^{c} $ & $  0.70^{+0.18}_{-0.27} $ & $ 0.74^{+0.07}_{-0.04} $   \\ 
050446.3-283821 &  0.84 & $ 44.89^{+0.07}_{-0.02} $  &  44.33 & $ 45.26^{+0.07}_{-0.02} $ & $ 1.97^{+0.18}_{-0.14} $ & $  1.17^{+0.06}_{-0.01} $ & $ 0.93^{+0.07}_{-0.02} $   \\ 
050501.8-284149 &  0.26 & $ 44.28^{+0.14}_{-0.11} $  &  43.11 & $ 44.41^{+0.14}_{-0.11} $ & $ 2.18^{+0.09}_{-0.09} $ & $  1.35^{+0.07}_{-0.06} $ & $ 1.29^{+0.14}_{-0.11} $   \\ 
051651.9+794314 &  0.56 & $ 46.13^{+0.22}_{-0.20} $  &  44.37 & $ 46.18^{+0.21}_{-0.19} $ & $ 1.56^{+0.08}_{-0.09} $ & $  1.63^{+0.09}_{-0.08} $ & $ 1.81^{+0.21}_{-0.19} $   \\ 
051822.6+793208 &  0.05 & $ 42.43^{+0.07}_{-0.07} $  &  41.85 & $ 42.78^{+0.07}_{-0.06} $ & $ 1.83^{+0.12}_{-0.10} $ & $  1.20^{+0.06}_{-0.06} $ & $ 0.93^{+0.07}_{-0.06} $   \\ 
051955.5-455727 &  0.56 & $ 45.22^{+0.08}_{-0.10} $  &  44.21 & $ 45.39^{+0.08}_{-0.10} $ & $ 2.09^{+0.06}_{-0.06} $ & $  1.24^{+0.04}_{-0.06} $ & $ 1.18^{+0.08}_{-0.10} $   \\ 
052022.0-252309 &  0.75 & $ 45.33^{+0.16}_{-0.20} $  &  44.35 & $ 45.50^{+0.15}_{-0.20} $ & $ 2.05^{+0.38}_{-0.17} $ & $  1.32^{+0.08}_{-0.13} $ & $ 1.16^{+0.15}_{-0.20} $   \\ 
052116.2-252957 &  0.33 & $ 44.18^{+0.10}_{-0.09} $  &  43.21 & $ 44.37^{+0.10}_{-0.09} $ & $ 2.21^{+0.71}_{-0.44} $ & $  1.30^{+0.06}_{-0.06} $ & $ 1.16^{+0.10}_{-0.09} $   \\ 
052144.1-251518 &  0.32 & $ 44.49^{+0.09}_{-0.11} $  &  43.39 & $ 44.63^{+0.09}_{-0.11} $ & $ 2.10^{+0.41}_{-0.27} $ & $  1.34^{+0.04}_{-0.06} $ & $ 1.25^{+0.09}_{-0.11} $   \\ 
052543.6-334856 &  0.73 & $ 45.44^{+0.17}_{-0.14} $  &  44.27 & $ 45.60^{+0.17}_{-0.14} $ & $ 2.44^{+0.43}_{-0.45} $ & $  1.32^{+0.09}_{-0.08} $ & $ 1.33^{+0.17}_{-0.14} $   \\ 
065214.1+743230 &  0.62 & $ 46.07^{+0.15}_{-0.23} $  &  44.47 & $ 46.12^{+0.15}_{-0.23} $ & $ 2.01^{+0.16}_{-0.15} $ & $  1.48^{+0.06}_{-0.10} $ & $ 1.65^{+0.15}_{-0.23} $   \\ 
065400.0+742045 &  0.36 & $ 45.07^{+0.13}_{-0.13} $  &  43.57 & $ 45.15^{+0.13}_{-0.13} $ & $ 2.30^{+0.31}_{-0.19} $ & $  1.43^{+0.06}_{-0.06} $ & $ 1.58^{+0.13}_{-0.13} $   \\ 
074202.7+742625 &  0.60 & $ 44.98^{+0.07}_{-0.06} $  &  44.46 & $ 45.36^{+0.07}_{-0.06} $ & $ 2.01^{+0.16}_{-0.14} $ & $  1.06^{+0.06}_{-0.06} $ & $ 0.91^{+0.07}_{-0.06} $   \\ 
074312.1+742937 &  0.31 & $ 45.77^{+0.09}_{-0.08} $  &  44.55 & $ 45.88^{+0.09}_{-0.07} $ & $ 1.98^{+0.07}_{-0.07} $ & $  1.33^{+0.04}_{-0.04} $ & $ 1.33^{+0.09}_{-0.07} $   \\ 
074352.0+744258 &  0.80 & $ 45.82^{+0.09}_{-0.12} $  &  44.56 & $ 45.93^{+0.09}_{-0.12} $ & $ 2.03^{+0.11}_{-0.10} $ & $  1.38^{+0.04}_{-0.06} $ & $ 1.36^{+0.09}_{-0.12} $   \\ 
075117.9+180856 &  0.25 & $ 44.22^{+0.08}_{-0.07} $  &  43.53 & $ 44.58^{+0.07}_{-0.06} $ & $ 1.61^{+0.13}_{-0.13} $ & $  1.24^{+0.06}_{-0.06} $ & $ 1.05^{+0.07}_{-0.06} $   \\ 
080504.6+245156 &  0.98 & $ 45.02^{+0.08}_{-0.05} $  &  44.44 & $ 45.38^{+0.07}_{-0.06} $ & $ 2.08^{+0.17}_{-0.16} $ & $  1.16^{+0.06}_{-0.05} $ & $ 0.94^{+0.07}_{-0.06} $   \\ 
080608.1+244420 &  0.36 & $ 45.38^{+0.06}_{-0.07} $  &  43.96 & $ 45.49^{+0.06}_{-0.07} $ & $ 2.49^{+0.07}_{-0.05} $ & $  1.38^{+0.03}_{-0.04} $ & $ 1.53^{+0.06}_{-0.07} $   \\ 
083049.8+524908 &  1.20 & $ 45.45^{+0.04}_{-0.04} $  &  44.94 & $ 45.90^{+0.03}_{-0.03} $ & $ 1.76^{+0.06}_{-0.05} $ & $  1.15^{+0.03}_{-0.04} $ & $ 0.96^{+0.03}_{-0.03} $   \\ 
083737.0+255151 &  0.11 & $ 44.54^{+0.13}_{-0.13} $  &  43.02 & $ 44.60^{+0.13}_{-0.13} $ & $ 1.79^{+0.48}_{-0.41} $ & $  1.51^{+0.06}_{-0.06} $ & $ 1.58^{+0.13}_{-0.13} $   \\ 
083737.1+254751 &  0.08 & $ 44.29^{+0.09}_{-0.08} $  &  43.09 & $ 44.40^{+0.09}_{-0.07} $ & $ 1.92^{+0.14}_{-0.12} $ & $  1.37^{+0.04}_{-0.04} $ & $ 1.31^{+0.09}_{-0.07} $   \\ 
083838.6+253616 &  0.60 & $ 45.68^{+0.11}_{-0.09} $  &  43.91 & $ 45.72^{+0.11}_{-0.09} $ & $ 2.22^{+0.49}_{-0.27} $ & $  1.55^{+0.04}_{-0.04} $ & $ 1.81^{+0.11}_{-0.09} $   \\ 
083905.9+255010 &  0.25 & $ 43.99^{+0.10}_{-0.09} $  &  43.10 & $ 44.19^{+0.10}_{-0.09} $ & $ 2.01^{+0.60}_{-0.35} $ & $  1.24^{+0.06}_{-0.06} $ & $ 1.09^{+0.10}_{-0.09} $   \\ 
085530.7+585129 &  0.91 & $ 45.18^{+0.04}_{-0.06} $  &  44.58 & $ 45.60^{+0.03}_{-0.05} $ & $ 1.67^{+0.29}_{-0.28} $ & $  1.17^{+0.03}_{-0.06} $ & $ 1.02^{+0.03}_{-0.05} $   \\ 
094526.2-085006 &  0.31 & $ 44.48^{+0.15}_{-0.20} $  &  43.49 & $ 44.67^{+0.15}_{-0.21} $ & $ 2.25^{+1.23}_{-0.78} $ & $  1.21^{+0.08}_{-0.14} $ & $ 1.18^{+0.15}_{-0.21} $   \\ 
094548.3-084824 &  1.75 & $ 47.22^{+0.18}_{-0.32} $  &  45.15 & $ 47.24^{+0.18}_{-0.32} $ & $ 1.75^{+0.83}_{-0.08} $ & $  1.62^{+0.07}_{-0.13} $ & $ 2.09^{+0.18}_{-0.32} $   \\ 
095054.5+393924 &  1.30 & $ 46.28^{+0.10}_{-0.08} $  &  44.83 & $ 46.35^{+0.10}_{-0.08} $ & $ 2.01^{+0.38}_{-0.26} $ & $  1.42^{+0.04}_{-0.04} $ & $ 1.53^{+0.10}_{-0.08} $   \\ 
095309.7+013558 &  0.48 & $ 45.05^{+0.09}_{-0.12} $  &  43.85 & $ 45.17^{+0.09}_{-0.11} $ & $ 1.89^{+0.40}_{-0.28} $ & $  1.26^{+0.04}_{-0.06} $ & $ 1.32^{+0.09}_{-0.11} $   \\ 
095509.6+174124 &  1.29 & $ 46.24^{+0.07}_{-0.08} $  &  44.91 & $ 46.33^{+0.06}_{-0.08} $ & $ 1.90^{+0.14}_{-0.09} $ & $  1.31^{+0.03}_{-0.04} $ & $ 1.42^{+0.06}_{-0.08} $   \\ 
100100.0+252103 &  0.79 & $ 45.44^{+0.08}_{-0.07} $  &  44.33 & $ 45.59^{+0.08}_{-0.07} $ & $ 2.20^{+0.12}_{-0.07} $ & $  1.35^{+0.04}_{-0.04} $ & $ 1.25^{+0.08}_{-0.07} $   \\ 
100309.4+554135 &  0.67 & $ 45.67^{+0.07}_{-0.08} $  &  44.13 & $ 45.74^{+0.07}_{-0.08} $ & $ 2.27^{+0.12}_{-0.10} $ & $  1.45^{+0.03}_{-0.04} $ & $ 1.61^{+0.07}_{-0.08} $   \\ 
100828.8+535408 &  0.38 & $ 44.85^{+0.07}_{-0.08} $  &  43.43 & $ 44.92^{+0.07}_{-0.08} $ & $ 2.04^{+0.20}_{-0.15} $ & $  1.49^{+0.03}_{-0.04} $ & $ 1.49^{+0.07}_{-0.08} $   \\ 
100921.7+534926 &  0.39 & $ 44.75^{+0.08}_{-0.10} $  &  43.63 & $ 44.92^{+0.08}_{-0.10} $ & $ 2.35^{+0.14}_{-0.08} $ & $  1.28^{+0.04}_{-0.06} $ & $ 1.29^{+0.08}_{-0.10} $   \\ 
100926.5+533426 &  1.72 & $ 46.43^{+0.09}_{-0.08} $  &  45.19 & $ 46.54^{+0.09}_{-0.07} $ & $ 2.01^{+0.13}_{-0.12} $ & $  1.40^{+0.04}_{-0.04} $ & $ 1.36^{+0.09}_{-0.07} $   \\ 
101506.0+520157 &  0.61 & $ 45.60^{+0.10}_{-0.09} $  &  43.98 & $ 45.65^{+0.10}_{-0.09} $ & $ 2.00^{+1.57}_{-1.03} $ & $  1.49^{+0.04}_{-0.04} $ & $ 1.67^{+0.10}_{-0.09} $   \\ 
101838.0+411635 &  0.58 & $ 45.30^{+0.06}_{-0.07} $  &  43.96 & $ 45.41^{+0.06}_{-0.07} $ & $ 2.36^{+0.11}_{-0.10} $ & $  1.33^{+0.03}_{-0.04} $ & $ 1.45^{+0.06}_{-0.07} $   \\ 
101843.0+413515 &  0.08 & $ 43.88^{+0.07}_{-0.14} $  &  42.30 & $ 43.93^{+0.07}_{-0.13} $ & $ 1.86^{+0.20}_{-0.11} $ & $  1.57^{+0.03}_{-0.06} $ & $ 1.63^{+0.07}_{-0.13} $   \\ 
101850.5+411506 &  0.58 & $ 45.70^{+0.06}_{-0.07} $  &  44.43 & $ 45.81^{+0.06}_{-0.07} $ & $ 2.30^{+0.08}_{-0.05} $ & $  1.37^{+0.03}_{-0.04} $ & $ 1.38^{+0.06}_{-0.07} $   \\ 
101922.6+412049 &  0.24 & $ 44.30^{+0.06}_{-0.05} $  &  43.65 & $ 44.69^{+0.05}_{-0.04} $ & $ 2.12^{+0.27}_{-0.08} $ & $  1.19^{+0.04}_{-0.04} $ & $ 1.04^{+0.05}_{-0.04} $   \\ 
102412.3+042023 &  1.46 & $ 46.30^{+0.10}_{-0.08} $  &  44.87 & $ 46.38^{+0.10}_{-0.08} $ & $ 2.01^{+0.16}_{-0.10} $ & $  1.44^{+0.04}_{-0.04} $ & $ 1.52^{+0.10}_{-0.08} $   \\ 
103120.0+311404 &  1.19 & $ 45.85^{+0.10}_{-0.06} $  &  45.00 & $ 46.09^{+0.09}_{-0.05} $ & $ 1.85^{+0.19}_{-0.14} $ & $  1.24^{+0.06}_{-0.04} $ & $ 1.09^{+0.09}_{-0.05} $   \\ 
103154.1+310732 &  0.30 & $ 44.37^{+0.06}_{-0.07} $  &  43.32 & $ 44.52^{+0.06}_{-0.07} $ & $ 1.88^{+0.22}_{-0.19} $ & $  1.37^{+0.03}_{-0.04} $ & $ 1.20^{+0.06}_{-0.07} $   \\ 
103909.4+205222 &  0.98 & $ 45.80^{+0.08}_{-0.06} $  &  44.82 & $ 45.98^{+0.08}_{-0.06} $ & $ 1.96^{+0.26}_{-0.16} $ & $  1.31^{+0.04}_{-0.04} $ & $ 1.16^{+0.08}_{-0.06} $   \\ 
103932.7+205426 &  0.24 & $ 44.14^{+0.07}_{-0.06} $  &  43.34 & $ 44.38^{+0.07}_{-0.05} $ & $ 1.87^{+0.18}_{-0.15} $ & $  1.27^{+0.04}_{-0.04} $ & $ 1.04^{+0.07}_{-0.05} $   \\ 
103935.8+533036 &  0.23 & $ 44.65^{+0.09}_{-0.12} $  &  43.40 & $ 44.75^{+0.09}_{-0.12} $ & $ 2.08^{+0.25}_{-0.16} $ & $  1.33^{+0.04}_{-0.06} $ & $ 1.34^{+0.09}_{-0.12} $   \\ 
104026.9+204542 &  0.47 & $ 45.43^{+0.04}_{-0.05} $  &  44.76 & $ 45.73^{+0.04}_{-0.05} $ & $ 1.99^{+0.05}_{-0.05} $ & $  1.04^{+0.03}_{-0.04} $ & $ 0.97^{+0.04}_{-0.05} $   \\ 
104034.3+205110 &  0.67 & $ 45.70^{+0.14}_{-0.03} $  &  44.01 & $ 45.75^{+0.14}_{-0.05} $ & $ 2.26^{+0.39}_{-0.18} $ & $  1.54^{+0.06}_{-0.01} $ & $ 1.73^{+0.14}_{-0.05} $   \\ 
104425.0-013521 &  1.57 & $ 46.63^{+0.10}_{-0.13} $  &  45.13 & $ 46.70^{+0.10}_{-0.13} $ & $ 1.85^{+0.16}_{-0.14} $ & $  1.52^{+0.04}_{-0.06} $ & $ 1.58^{+0.10}_{-0.13} $   \\ 
104509.3-012442 &  0.47 & $ 44.72^{+0.06}_{-0.06} $  &  43.70 & $ 44.89^{+0.06}_{-0.06} $ & $ 2.14^{+0.18}_{-0.10} $ & $  1.30^{+0.03}_{-0.04} $ & $ 1.19^{+0.06}_{-0.06} $   \\ 
104522.1-012843 &  0.78 & $ 45.74^{+0.05}_{-0.06} $  &  44.87 & $ 45.96^{+0.05}_{-0.06} $ & $ 2.00^{+0.13}_{-0.03} $ & $  1.23^{+0.03}_{-0.04} $ & $ 1.09^{+0.05}_{-0.06} $   \\ 
104912.8+330459 &  0.23 & $ 43.64^{+0.03}_{-0.03} $  &  43.48 & $ 44.34^{+0.03}_{-0.03} $ & $ 1.67^{+0.19}_{-0.15} $ & $  1.06^{+0.04}_{-0.06} $ & $ 0.86^{+0.03}_{-0.03} $   \\ 
105014.9+331013 &  1.01 & $ 46.42^{+0.10}_{-0.13} $  &  44.44 & $ 46.45^{+0.10}_{-0.13} $ & $ 2.33^{+0.61}_{-0.33} $ & $  1.64^{+0.04}_{-0.05} $ & $ 2.01^{+0.10}_{-0.13} $   \\ 
105239.7+572431 &  1.11 & $ 46.51^{+0.07}_{-0.09} $  &  44.85 & $ 46.56^{+0.07}_{-0.09} $ & $ 2.10^{+0.04}_{-0.03} $ & $  1.55^{+0.03}_{-0.04} $ & $ 1.71^{+0.07}_{-0.09} $   \\ 
105316.9+573551 &  1.20 & $ 46.02^{+0.05}_{-0.06} $  &  45.16 & $ 46.27^{+0.05}_{-0.05} $ & $ 1.80^{+0.04}_{-0.03} $ & $  1.29^{+0.03}_{-0.04} $ & $ 1.11^{+0.05}_{-0.05} $   \\ 
105335.0+572540 &  0.78 & $ 44.86^{+0.04}_{-0.05} $  &  44.15 & $ 45.20^{+0.04}_{-0.04} $ & $ 1.72^{+0.08}_{-0.06} $ & $  1.24^{+0.03}_{-0.04} $ & $ 1.05^{+0.04}_{-0.04} $   \\ 
105339.7+573104 &  0.59 & $ 45.28^{+0.07}_{-0.08} $  &  43.79 & $ 45.35^{+0.07}_{-0.08} $ & $ 2.16^{+0.07}_{-0.07} $ & $  1.41^{+0.03}_{-0.04} $ & $ 1.56^{+0.07}_{-0.08} $   \\ 
105624.2-033522 &  0.63 & $ 45.45^{+0.07}_{-0.08} $  &  44.10 & $ 45.54^{+0.07}_{-0.08} $ & $ 2.16^{+0.15}_{-0.10} $ & $  1.42^{+0.03}_{-0.04} $ & $ 1.44^{+0.07}_{-0.08} $   \\ 
110652.0-182738 &  1.43 & $ 45.99^{+0.27}_{-0.21} $  &  45.01 & $ 46.28^{+0.22}_{-0.17} $ & $ 1.55^{+0.36}_{-0.29} $ & $  1.37^{+0.14}_{-0.15} $ & $ 1.27^{+0.22}_{-0.17} $   \\ 
111933.0+212756 &  0.28 & $ 44.59^{+0.06}_{-0.12} $  &  43.35 & $ 44.69^{+0.06}_{-0.12} $ & $ 1.92^{+0.52}_{-0.40} $ & $  1.41^{+0.03}_{-0.06} $ & $ 1.34^{+0.06}_{-0.12} $   \\ 
111942.1+211516 &  1.29 & $ 46.04^{+0.12}_{-0.16} $  &  44.66 & $ 46.13^{+0.11}_{-0.15} $ & $ 1.92^{+0.40}_{-0.28} $ & $  1.46^{+0.05}_{-0.08} $ & $ 1.46^{+0.11}_{-0.15} $   \\ 
112022.3+125252 &  0.41 & $ 45.03^{+0.06}_{-0.07} $  &  43.91 & $ 45.17^{+0.06}_{-0.07} $ & $ 2.22^{+0.15}_{-0.13} $ & $  1.30^{+0.03}_{-0.04} $ & $ 1.26^{+0.06}_{-0.07} $   \\ 
112046.7+125429 &  0.38 & $ 45.02^{+0.06}_{-0.07} $  &  43.83 & $ 45.15^{+0.06}_{-0.07} $ & $ 2.29^{+0.25}_{-0.15} $ & $  1.32^{+0.03}_{-0.04} $ & $ 1.33^{+0.06}_{-0.07} $   \\ 
113106.9+312518 &  1.48 & $ 46.47^{+0.12}_{-0.12} $  &  45.13 & $ 46.59^{+0.12}_{-0.11} $ & $ 1.72^{+0.27}_{-0.24} $ & $  1.45^{+0.06}_{-0.06} $ & $ 1.46^{+0.12}_{-0.11} $   \\ 
115317.9+364712 &  0.72 & $ 45.64^{+0.13}_{-0.14} $  &  44.02 & $ 45.69^{+0.13}_{-0.14} $ & $ 2.00^{+1.48}_{-0.73} $ & $  1.41^{+0.06}_{-0.06} $ & $ 1.67^{+0.13}_{-0.14} $   \\ 
120359.1+443715 &  0.64 & $ 45.40^{+0.09}_{-0.07} $  &  44.03 & $ 45.40^{+0.11}_{-0.10} $ & $ 2.43^{+0.19}_{-0.19} $ & $  1.40^{+0.04}_{-0.04} $ & $ 1.37^{+0.11}_{-0.10} $   \\ 
120413.7+443149 &  0.49 & $ 45.04^{+0.07}_{-0.08} $  &  43.71 & $ 45.04^{+0.08}_{-0.10} $ & $ 2.23^{+0.22}_{-0.14} $ & $  1.27^{+0.03}_{-0.04} $ & $ 1.33^{+0.08}_{-0.10} $   \\ 
123036.2+642531 &  0.74 & $ 45.16^{+0.17}_{-0.11} $  &  44.02 & $ 45.31^{+0.17}_{-0.12} $ & $ 2.25^{+0.34}_{-0.21} $ & $  1.35^{+0.08}_{-0.06} $ & $ 1.29^{+0.17}_{-0.12} $   \\ 
123116.5+641115 &  0.45 & $ 44.37^{+0.04}_{-0.05} $  &  43.70 & $ 44.67^{+0.04}_{-0.04} $ & $ 1.92^{+0.09}_{-0.08} $ & $  1.22^{+0.03}_{-0.04} $ & $ 0.98^{+0.04}_{-0.05} $   \\ 
123218.5+640311 &  1.01 & $ 45.43^{+0.09}_{-0.08} $  &  44.66 & $ 45.70^{+0.08}_{-0.08} $ & $ 1.88^{+0.25}_{-0.22} $ & $  1.24^{+0.06}_{-0.06} $ & $ 1.04^{+0.08}_{-0.08} $   \\ 
123759.6+621102 &  0.91 & $ 46.05^{+0.07}_{-0.08} $  &  44.69 & $ 46.14^{+0.07}_{-0.08} $ & $ 2.05^{+0.06}_{-0.06} $ & $  1.44^{+0.03}_{-0.04} $ & $ 1.45^{+0.07}_{-0.08} $   \\ 
123800.9+621338 &  0.44 & $ 45.22^{+0.07}_{-0.09} $  &  43.36 & $ 45.26^{+0.07}_{-0.09} $ & $ 2.54^{+0.07}_{-0.09} $ & $  1.57^{+0.03}_{-0.04} $ & $ 1.91^{+0.07}_{-0.09} $   \\ 
124214.1-112512 &  0.82 & $ 45.73^{+0.06}_{-0.12} $  &  44.54 & $ 45.86^{+0.06}_{-0.11} $ & $ 1.81^{+0.09}_{-0.08} $ & $  1.43^{+0.03}_{-0.06} $ & $ 1.32^{+0.06}_{-0.11} $   \\ 
124557.6+022659 &  0.71 & $ 45.81^{+0.13}_{-0.13} $  &  44.00 & $ 45.89^{+0.13}_{-0.13} $ & $ 2.72^{+0.60}_{-0.47} $ & $  1.39^{+0.06}_{-0.06} $ & $ 1.89^{+0.13}_{-0.13} $   \\ 
124607.6+022153 &  0.49 & $ 45.18^{+0.06}_{-0.07} $  &  43.90 & $ 45.32^{+0.06}_{-0.07} $ & $ 2.46^{+0.19}_{-0.14} $ & $  1.32^{+0.03}_{-0.04} $ & $ 1.41^{+0.06}_{-0.07} $   \\ 
124641.8+022412 &  0.93 & $ 46.37^{+0.04}_{-0.08} $  &  44.90 & $ 46.44^{+0.04}_{-0.08} $ & $ 2.21^{+0.11}_{-0.09} $ & $  1.49^{+0.02}_{-0.04} $ & $ 1.54^{+0.04}_{-0.08} $   \\ 
124647.9+020955 &  1.07 & $ 45.89^{+0.13}_{-0.13} $  &  44.47 & $ 45.96^{+0.13}_{-0.13} $ & $ 2.08^{+0.66}_{-0.45} $ & $  1.43^{+0.06}_{-0.06} $ & $ 1.49^{+0.13}_{-0.13} $   \\ 
124914.6-060910 &  1.63 & $ 46.51^{+0.07}_{-0.08} $  &  44.97 & $ 46.57^{+0.07}_{-0.08} $ & $ 2.14^{+0.13}_{-0.12} $ & $  1.47^{+0.03}_{-0.04} $ & $ 1.60^{+0.07}_{-0.08} $   \\ 
124949.4-060722 &  1.05 & $ 45.99^{+0.07}_{-0.08} $  &  44.64 & $ 46.08^{+0.07}_{-0.08} $ & $ 2.16^{+0.11}_{-0.10} $ & $  1.42^{+0.03}_{-0.04} $ & $ 1.44^{+0.07}_{-0.08} $   \\ 
130619.7-233857 &  0.35 & $ 45.29^{+0.12}_{-0.12} $  &  43.74 & $ 45.37^{+0.12}_{-0.12} $ & $ 2.49^{+0.46}_{-0.34} $ & $  1.40^{+0.06}_{-0.06} $ & $ 1.63^{+0.12}_{-0.12} $   \\ 
130658.1-234849 &  0.38 & $ 44.98^{+0.17}_{-0.17} $  &  43.65 & $ 45.07^{+0.17}_{-0.18} $ & $ 1.96^{+0.40}_{-0.30} $ & $  1.38^{+0.08}_{-0.08} $ & $ 1.42^{+0.17}_{-0.18} $   \\ 
132038.0+341124 &  0.06 & $ 43.89^{+0.10}_{-0.08} $  &  42.47 & $ 43.97^{+0.10}_{-0.08} $ & $ 1.74^{+0.10}_{-0.10} $ & $  1.53^{+0.04}_{-0.04} $ & $ 1.50^{+0.10}_{-0.08} $   \\ 
132101.6+340656 &  0.34 & $ 45.28^{+0.07}_{-0.08} $  &  43.66 & $ 45.35^{+0.07}_{-0.08} $ & $ 2.44^{+0.06}_{-0.06} $ & $  1.35^{+0.03}_{-0.04} $ & $ 1.68^{+0.07}_{-0.08} $   \\ 
133807.5+242411 &  0.63 & $ 45.88^{+0.07}_{-0.09} $  &  44.09 & $ 45.92^{+0.07}_{-0.09} $ & $ 2.08^{+0.16}_{-0.14} $ & $  1.60^{+0.03}_{-0.04} $ & $ 1.82^{+0.07}_{-0.09} $   \\ 
133942.6-315004 &  0.11 & $ 44.78^{+0.46}_{-0.48} $  &  42.81 & $ 44.81^{+0.46}_{-0.47} $ & $ 1.66^{+0.22}_{-0.20} $ & $  1.59^{+0.18}_{-0.20} $ & $ 1.99^{+0.46}_{-0.47} $   \\ 
134749.9+582111 &  0.65 & $ 46.50^{+0.07}_{-0.08} $  &  45.07 & $ 46.58^{+0.07}_{-0.08} $ & $ 2.20^{+0.03}_{-0.03} $ & $  1.42^{+0.03}_{-0.04} $ & $ 1.51^{+0.07}_{-0.08} $   \\ 
140100.0-110942 &  0.16 & $ 44.42^{+0.26}_{-0.12} $  &  42.60 & $ 44.47^{+0.26}_{-0.12} $ & $ 2.52^{+0.28}_{-0.11} $ & $  1.42^{+0.11}_{-0.05} $ & $ 1.87^{+0.26}_{-0.12} $   \\ 
140102.0-111224 &  0.04 & $ 43.59^{+0.08}_{-0.10} $  &  41.80 & $ 43.68^{+0.07}_{-0.09} $ & $ 1.91^{+0.03}_{-0.03} $ & $  1.57^{+0.03}_{-0.04} $ & $ 1.88^{+0.07}_{-0.09} $   \\ 
140113.4+024016 &  0.63 & $ 44.69^{+0.07}_{-0.09} $  &  43.85 & $ 44.92^{+0.07}_{-0.08} $ & $ 1.99^{+0.45}_{-0.21} $ & $  1.11^{+0.04}_{-0.06} $ & $ 1.07^{+0.07}_{-0.08} $   \\ 
140127.7+025605 &  0.26 & $ 44.77^{+0.04}_{-0.06} $  &  44.22 & $ 45.15^{+0.04}_{-0.06} $ & $ 1.84^{+0.10}_{-0.05} $ & $  1.19^{+0.03}_{-0.06} $ & $ 0.93^{+0.04}_{-0.06} $   \\ 
140921.1+261336 &  1.10 & $ 46.66^{+0.16}_{-0.13} $  &  45.02 & $ 46.75^{+0.15}_{-0.12} $ & $ 1.48^{+0.08}_{-0.04} $ & $  1.49^{+0.07}_{-0.06} $ & $ 1.74^{+0.15}_{-0.12} $   \\ 
141531.5+113156 &  0.26 & $ 44.40^{+0.05}_{-0.05} $  &  43.67 & $ 44.68^{+0.04}_{-0.05} $ & $ 1.82^{+0.06}_{-0.05} $ & $  1.17^{+0.03}_{-0.04} $ & $ 1.01^{+0.04}_{-0.05} $   \\ 
141722.6+251335 &  0.56 & $ 45.26^{+0.06}_{-0.08} $  &  43.94 & $ 45.36^{+0.06}_{-0.07} $ & $ 2.26^{+0.37}_{-0.17} $ & $  1.40^{+0.03}_{-0.04} $ & $ 1.42^{+0.06}_{-0.07} $   \\ 
141736.3+523028 &  0.99 & $ 45.75^{+0.12}_{-0.11} $  &  44.59 & $ 45.88^{+0.11}_{-0.11} $ & $ 2.00^{+0.07}_{-0.07} $ & $  1.37^{+0.06}_{-0.06} $ & $ 1.29^{+0.11}_{-0.11} $   \\ 
141809.1+250040 &  0.73 & $ 45.39^{+0.06}_{-0.07} $  &  44.31 & $ 45.54^{+0.06}_{-0.07} $ & $ 1.93^{+0.19}_{-0.16} $ & $  1.37^{+0.03}_{-0.04} $ & $ 1.22^{+0.06}_{-0.07} $   \\ 
144937.5+090826 &  1.26 & $ 46.11^{+0.08}_{-0.06} $  &  45.12 & $ 46.30^{+0.08}_{-0.06} $ & $ 1.81^{+0.11}_{-0.07} $ & $  1.33^{+0.04}_{-0.04} $ & $ 1.19^{+0.08}_{-0.06} $   \\ 
144945.8+085921 &  0.26 & $ 43.89^{+0.05}_{-0.05} $  &  43.15 & $ 44.16^{+0.04}_{-0.05} $ & $ 1.97^{+0.09}_{-0.09} $ & $  1.24^{+0.03}_{-0.04} $ & $ 1.02^{+0.04}_{-0.05} $   \\ 
150428.3+101856 &  1.00 & $ 46.66^{+0.11}_{-0.09} $  &  44.76 & $ 46.69^{+0.11}_{-0.09} $ & $ 2.31^{+0.25}_{-0.16} $ & $  1.59^{+0.04}_{-0.04} $ & $ 1.93^{+0.11}_{-0.09} $   \\ 
151815.0+060851 &  1.29 & $ 46.18^{+0.11}_{-0.15} $  &  44.90 & $ 46.29^{+0.11}_{-0.15} $ & $ 1.90^{c} $ & $  1.40^{+0.05}_{-0.08} $ & $ 1.39^{+0.11}_{-0.15} $   \\ 
153205.7-082952 &  1.24 & $ 46.48^{+0.14}_{-0.14} $  &  44.80 & $ 46.53^{+0.13}_{-0.14} $ & $ 1.99^{+0.13}_{-0.12} $ & $  1.43^{+0.06}_{-0.06} $ & $ 1.73^{+0.13}_{-0.14} $   \\ 
153419.0+011808 &  1.28 & $ 46.66^{+0.13}_{-0.14} $  &  44.83 & $ 46.71^{+0.13}_{-0.14} $ & $ 2.52^{+0.64}_{-0.36} $ & $  1.42^{+0.06}_{-0.06} $ & $ 1.88^{+0.13}_{-0.14} $   \\ 
153456.1+013033 &  0.31 & $ 45.58^{+0.19}_{-0.14} $  &  43.65 & $ 45.61^{+0.19}_{-0.14} $ & $ 2.27^{+0.42}_{-0.24} $ & $  1.57^{+0.08}_{-0.06} $ & $ 1.96^{+0.19}_{-0.14} $   \\ 
160706.6+075709 &  0.23 & $ 44.36^{+0.06}_{-0.07} $  &  43.10 & $ 44.50^{+0.06}_{-0.07} $ & $ 2.42^{+0.15}_{-0.14} $ & $  1.38^{+0.03}_{-0.04} $ & $ 1.40^{+0.06}_{-0.07} $   \\ 
160731.5+081202 &  0.23 & $ 44.55^{+0.09}_{-0.08} $  &  42.91 & $ 44.65^{+0.09}_{-0.08} $ & $ 2.67^{+0.36}_{-0.22} $ & $  1.33^{+0.04}_{-0.04} $ & $ 1.74^{+0.09}_{-0.08} $   \\ 
161544.2+121708 &  0.21 & $ 43.91^{+0.11}_{-0.07} $  &  42.93 & $ 44.10^{+0.11}_{-0.09} $ & $ 2.22^{+0.39}_{-0.19} $ & $  1.22^{+0.06}_{-0.04} $ & $ 1.17^{+0.11}_{-0.09} $   \\ 
161615.1+121353 &  0.84 & $ 44.82^{+0.03}_{-0.06} $  &  44.36 & $ 45.25^{+0.03}_{-0.05} $ & $ 2.01^{+0.28}_{-0.17} $ & $  1.13^{+0.03}_{-0.06} $ & $ 0.89^{+0.03}_{-0.05} $   \\ 
161825.4+124145 &  0.40 & $ 44.86^{+0.12}_{-0.08} $  &  43.48 & $ 44.96^{+0.12}_{-0.08} $ & $ 2.29^{+0.66}_{-0.45} $ & $  1.42^{+0.06}_{-0.04} $ & $ 1.47^{+0.12}_{-0.08} $   \\ 
162813.9+780342 &  0.64 & $ 46.23^{+0.11}_{-0.09} $  &  44.45 & $ 46.27^{+0.10}_{-0.09} $ & $ 2.30^{+0.34}_{-0.29} $ & $  1.43^{+0.04}_{-0.04} $ & $ 1.82^{+0.10}_{-0.09} $   \\ 
163309.8+571039 &  0.29 & $ 44.84^{+0.07}_{-0.08} $  &  43.44 & $ 44.93^{+0.07}_{-0.08} $ & $ 2.23^{+0.32}_{-0.18} $ & $  1.46^{+0.03}_{-0.04} $ & $ 1.48^{+0.07}_{-0.08} $   \\ 
163332.3+570520 &  0.39 & $ 45.03^{+0.13}_{-0.13} $  &  43.43 & $ 45.09^{+0.13}_{-0.13} $ & $ 2.31^{+0.72}_{-0.50} $ & $  1.36^{+0.06}_{-0.06} $ & $ 1.66^{+0.13}_{-0.13} $   \\ 
165406.6+142123 &  0.64 & $ 45.66^{+0.13}_{-0.13} $  &  44.18 & $ 45.72^{+0.13}_{-0.13} $ & $ 1.88^{+0.20}_{-0.14} $ & $  1.45^{+0.06}_{-0.06} $ & $ 1.55^{+0.13}_{-0.13} $   \\ 
165425.3+142159 &  0.18 & $ 44.31^{+0.05}_{-0.04} $  &  43.82 & $ 44.72^{+0.05}_{-0.04} $ & $ 2.11^{+0.06}_{-0.03} $ & $  1.12^{+0.04}_{-0.04} $ & $ 0.89^{+0.05}_{-0.04} $   \\ 
165448.5+141311 &  0.32 & $ 44.41^{+0.02}_{-0.02} $  &  44.25 & $ 45.06^{+0.02}_{-0.02} $ & $ 1.81^{+0.12}_{-0.07} $ & $  1.02^{+0.03}_{-0.04} $ & $ 0.81^{+0.02}_{-0.02} $   \\ 
165800.7+352333 &  0.13 & $ 44.04^{+0.10}_{-0.13} $  &  42.69 & $ 44.12^{+0.10}_{-0.12} $ & $ 1.86^{+0.75}_{-0.39} $ & $  1.49^{+0.04}_{-0.06} $ & $ 1.43^{+0.10}_{-0.12} $   \\ 
185518.7-462504 &  0.79 & $ 46.21^{+0.19}_{-0.19} $  &  44.51 & $ 46.30^{+0.17}_{-0.17} $ & $ 1.42^{+0.53}_{-0.41} $ & $  1.66^{+0.08}_{-0.08} $ & $ 1.79^{+0.17}_{-0.17} $   \\ 
185613.7-462239 &  0.77 & $ 45.46^{+0.09}_{-0.08} $  &  44.70 & $ 45.74^{+0.08}_{-0.08} $ & $ 2.17^{+0.27}_{-0.23} $ & $  1.20^{+0.06}_{-0.06} $ & $ 1.04^{+0.08}_{-0.08} $   \\ 
204159.2-321439 &  0.74 & $ 45.42^{+0.10}_{-0.10} $  &  44.45 & $ 45.62^{+0.10}_{-0.09} $ & $ 2.08^{+0.18}_{-0.11} $ & $  1.26^{+0.06}_{-0.06} $ & $ 1.17^{+0.10}_{-0.09} $   \\ 
204204.1-321601 &  0.38 & $ 44.87^{+0.13}_{-0.13} $  &  43.35 & $ 44.93^{+0.13}_{-0.13} $ & $ 2.02^{+0.38}_{-0.21} $ & $  1.37^{+0.06}_{-0.06} $ & $ 1.58^{+0.13}_{-0.13} $   \\ 
204208.2-323523 &  1.18 & $ 45.87^{+0.18}_{-0.23} $  &  44.53 & $ 45.96^{+0.17}_{-0.23} $ & $ 2.01^{+0.37}_{-0.23} $ & $  1.30^{+0.08}_{-0.12} $ & $ 1.43^{+0.17}_{-0.23} $   \\ 
204548.4-025234 &  2.19 & $ 47.31^{+0.27}_{-0.22} $  &  45.42 & $ 47.33^{+0.27}_{-0.27} $ & $ 1.98^{+0.24}_{-0.25} $ & $  1.68^{+0.11}_{-0.09} $ & $ 1.91^{+0.27}_{-0.27} $   \\ 
205635.7-044717 &  0.22 & $ 44.61^{+0.11}_{-0.11} $  &  43.30 & $ 44.73^{+0.11}_{-0.11} $ & $ 2.40^{+0.17}_{-0.14} $ & $  1.35^{+0.06}_{-0.06} $ & $ 1.43^{+0.11}_{-0.11} $   \\ 
205829.9-423634 &  0.23 & $ 44.08^{+0.07}_{-0.06} $  &  43.76 & $ 44.59^{+0.06}_{-0.06} $ & $ 1.90^{+0.09}_{-0.08} $ & $  1.06^{+0.07}_{-0.08} $ & $ 0.83^{+0.06}_{-0.06} $   \\ 
210325.4-112011 &  0.72 & $ 46.14^{+0.57}_{-0.36} $  &  44.33 & $ 46.18^{+0.57}_{-0.35} $ & $ 1.85^{+0.34}_{-0.20} $ & $  1.61^{+0.23}_{-0.15} $ & $ 1.85^{+0.57}_{-0.35} $   \\ 
210355.3-121858 &  0.79 & $ 45.39^{+0.13}_{-0.14} $  &  44.35 & $ 45.56^{+0.13}_{-0.14} $ & $ 2.20^{+0.41}_{-0.15} $ & $  1.32^{+0.07}_{-0.08} $ & $ 1.21^{+0.13}_{-0.14} $   \\ 
213002.3-153414 &  0.56 & $ 46.04^{+0.13}_{-0.14} $  &  44.46 & $ 46.09^{+0.13}_{-0.13} $ & $ 2.06^{+0.21}_{-0.19} $ & $  1.55^{+0.06}_{-0.06} $ & $ 1.64^{+0.13}_{-0.13} $   \\ 
213729.7-423601 &  0.66 & $ 45.21^{+0.13}_{-0.10} $  &  44.26 & $ 45.40^{+0.12}_{-0.09} $ & $ 2.02^{+0.43}_{-0.30} $ & $  1.29^{+0.07}_{-0.06} $ & $ 1.14^{+0.12}_{-0.09} $   \\ 
213733.2-434800 &  0.43 & $ 44.82^{+0.16}_{-0.16} $  &  43.51 & $ 44.94^{+0.16}_{-0.16} $ & $ 2.38^{+0.73}_{-0.50} $ & $  1.38^{+0.08}_{-0.08} $ & $ 1.42^{+0.16}_{-0.16} $   \\ 
213757.6-422334 &  0.36 & $ 44.63^{+0.14}_{-0.15} $  &  43.22 & $ 44.77^{+0.13}_{-0.15} $ & $ 2.59^{+2.19}_{-0.72} $ & $  1.42^{+0.07}_{-0.08} $ & $ 1.55^{+0.13}_{-0.15} $   \\ 
213824.0-423019 &  0.26 & $ 44.89^{+0.12}_{-0.12} $  &  43.56 & $ 44.98^{+0.12}_{-0.12} $ & $ 2.16^{+0.11}_{-0.16} $ & $  1.44^{+0.06}_{-0.06} $ & $ 1.43^{+0.12}_{-0.12} $   \\ 
213829.8-423958 &  1.47 & $ 47.00^{+0.19}_{-0.20} $  &  44.99 & $ 47.04^{+0.19}_{-0.20} $ & $ 2.61^{+0.40}_{-0.33} $ & $  1.59^{+0.08}_{-0.08} $ & $ 2.06^{+0.19}_{-0.20} $   \\ 
213852.2-434714 &  0.46 & $ 45.43^{+0.12}_{-0.12} $  &  43.32 & $ 45.51^{+0.12}_{-0.12} $ & $ 3.02^{+0.59}_{-0.38} $ & $  1.62^{+0.06}_{-0.06} $ & $ 2.19^{+0.12}_{-0.12} $   \\ 
214041.4-234720 &  0.49 & $ 45.69^{+0.10}_{-0.08} $  &  44.29 & $ 45.77^{+0.10}_{-0.08} $ & $ 2.17^{+0.09}_{-0.08} $ & $  1.38^{+0.04}_{-0.04} $ & $ 1.48^{+0.10}_{-0.08} $   \\ 
220446.8-014535 &  0.54 & $ 44.86^{+0.13}_{-0.11} $  &  44.04 & $ 45.13^{+0.12}_{-0.11} $ & $ 1.75^{+0.25}_{-0.20} $ & $  1.24^{+0.08}_{-0.08} $ & $ 1.08^{+0.12}_{-0.11} $   \\ 
221623.3-174317 &  0.75 & $ 45.29^{+0.14}_{-0.13} $  &  44.27 & $ 45.47^{+0.13}_{-0.12} $ & $ 1.82^{+0.25}_{-0.16} $ & $  1.34^{+0.07}_{-0.08} $ & $ 1.20^{+0.13}_{-0.12} $   \\ 
223547.9-255836 &  0.30 & $ 44.72^{+0.12}_{-0.12} $  &  43.41 & $ 44.82^{+0.12}_{-0.12} $ & $ 2.11^{+0.27}_{-0.25} $ & $  1.35^{+0.06}_{-0.06} $ & $ 1.41^{+0.12}_{-0.12} $   \\ 
223555.0-255833 &  1.80 & $ 46.80^{+0.19}_{-0.20} $  &  45.13 & $ 46.84^{+0.19}_{-0.19} $ & $ 2.17^{+0.15}_{-0.13} $ & $  1.51^{+0.08}_{-0.08} $ & $ 1.71^{+0.19}_{-0.19} $   \\ 
223949.8+080926 &  1.41 & $ 46.60^{+0.24}_{-0.32} $  &  44.82 & $ 46.64^{+0.24}_{-0.34} $ & $ 2.35^{+1.66}_{-0.90} $ & $  1.59^{+0.10}_{-0.14} $ & $ 1.82^{+0.24}_{-0.34} $   \\ 
224756.6-642721 &  0.60 & $ 45.73^{+0.11}_{-0.14} $  &  44.06 & $ 45.77^{+0.10}_{-0.14} $ & $ 2.00^{+0.21}_{-0.17} $ & $  1.54^{+0.04}_{-0.06} $ & $ 1.72^{+0.10}_{-0.14} $   \\ 
225025.1-643225 &  1.21 & $ 45.89^{+0.15}_{-0.12} $  &  44.65 & $ 46.00^{+0.14}_{-0.12} $ & $ 2.09^{+0.19}_{-0.14} $ & $  1.36^{+0.07}_{-0.06} $ & $ 1.35^{+0.14}_{-0.12} $   \\ 
225050.2-642900 &  1.25 & $ 46.23^{+0.11}_{-0.11} $  &  45.18 & $ 46.39^{+0.11}_{-0.10} $ & $ 2.04^{+0.07}_{-0.07} $ & $  1.35^{+0.06}_{-0.06} $ & $ 1.21^{+0.11}_{-0.10} $   \\ 
225118.0-175951 &  0.17 & $ 45.40^{+0.53}_{-0.33} $  &  43.09 & $ 45.41^{+0.53}_{-0.33} $ & $ 2.09^{+0.27}_{-0.20} $ & $  1.66^{+0.21}_{-0.13} $ & $ 2.32^{+0.53}_{-0.33} $   \\ 
230400.4-083755 &  0.41 & $ 45.38^{+0.17}_{-0.14} $  &  43.13 & $ 45.41^{+0.17}_{-0.14} $ & $ 2.72^{+0.90}_{-0.56} $ & $  1.64^{+0.07}_{-0.06} $ & $ 2.28^{+0.17}_{-0.14} $   \\ 
230434.1+122728 &  0.23 & $ 44.21^{+0.14}_{-0.13} $  &  43.29 & $ 44.46^{+0.13}_{-0.11} $ & $ 1.60^{+0.37}_{-0.30} $ & $  1.20^{+0.08}_{-0.08} $ & $ 1.17^{+0.13}_{-0.11} $   \\ 
230443.8+121636 &  1.40 & $ 46.30^{+0.27}_{-0.31} $  &  45.04 & $ 46.41^{+0.27}_{-0.30} $ & $ 1.95^{+0.29}_{-0.27} $ & $  1.41^{+0.12}_{-0.17} $ & $ 1.36^{+0.27}_{-0.30} $   \\ 
230459.6+121205 &  0.56 & $ 44.65^{+0.11}_{-0.07} $  &  44.20 & $ 45.20^{+0.09}_{-0.06} $ & $ 1.58^{+0.32}_{-0.27} $ & $  1.10^{+0.10}_{-0.09} $ & $ 1.00^{+0.09}_{-0.06} $   \\ 
231342.5-423210 &  0.97 & $ 45.85^{+0.08}_{-0.06} $  &  44.83 & $ 46.02^{+0.08}_{-0.06} $ & $ 2.14^{+0.13}_{-0.07} $ & $  1.30^{+0.04}_{-0.04} $ & $ 1.19^{+0.08}_{-0.06} $   \\ 
231601.7-424038 &  0.38 & $ 45.12^{+0.10}_{-0.13} $  &  43.61 & $ 45.19^{+0.10}_{-0.13} $ & $ 1.74^{+0.42}_{-0.43} $ & $  1.40^{+0.04}_{-0.06} $ & $ 1.57^{+0.10}_{-0.13} $   \\

\hline

\end{longtable}
\begin{list}{}{}
\item[$^{\mathrm{a}}$]The luminosity is expressed in units of $\rm erg \ s^{-1}$.
\item[$^{\mathrm{b}}$]For a detailed account of  the X-ray properties of all the sources see Corral et a. 2011 .
\item[$^{\mathrm{c}}$]Fixed parameter. \\
We assume the cosmological model $H_0=65km \ s^{-1} \ Mpc^{-1}$, $\Omega_{\lambda}=0.7$ and $\Omega_{M}=0.3$ throughout this paper.
\end{list}

\renewcommand{\arraystretch}{1}
\renewcommand{\thefootnote}{\arabic{footnote}}

\begin{appendix}
\onecolumn
  \section{}
We present here the optical-UV-X-ray SEDs obtained applying the correction discussed in section \ref{corrections}, for all the 195 sources analysed in this paper. The filled blue squares represent the fluxes in the \emph{GALEX} NUV/FUV bands, while the empty squares represent the optical data. The upper limits on the NUV/FUV fluxes are represented as downward arrays.  The fit has been done using the model quoted in section \ref{model} (blue curve).
We also show the slope of the best-fit model in the rest-frame energy range 1-10 keV (magenta curve)  obtained from the X-ray spectral analysis (for further details see Corral et al. 2011).  The dashed magenta lines represent the errors on the best-fit model of the X-ray data, given by the errors on the spectral index $\Gamma$.

%\newpage

\begin{figure*}
\label{all_seds}  
\centering
\subfigure{ 
  \includegraphics[height=5.6cm, width=6cm]{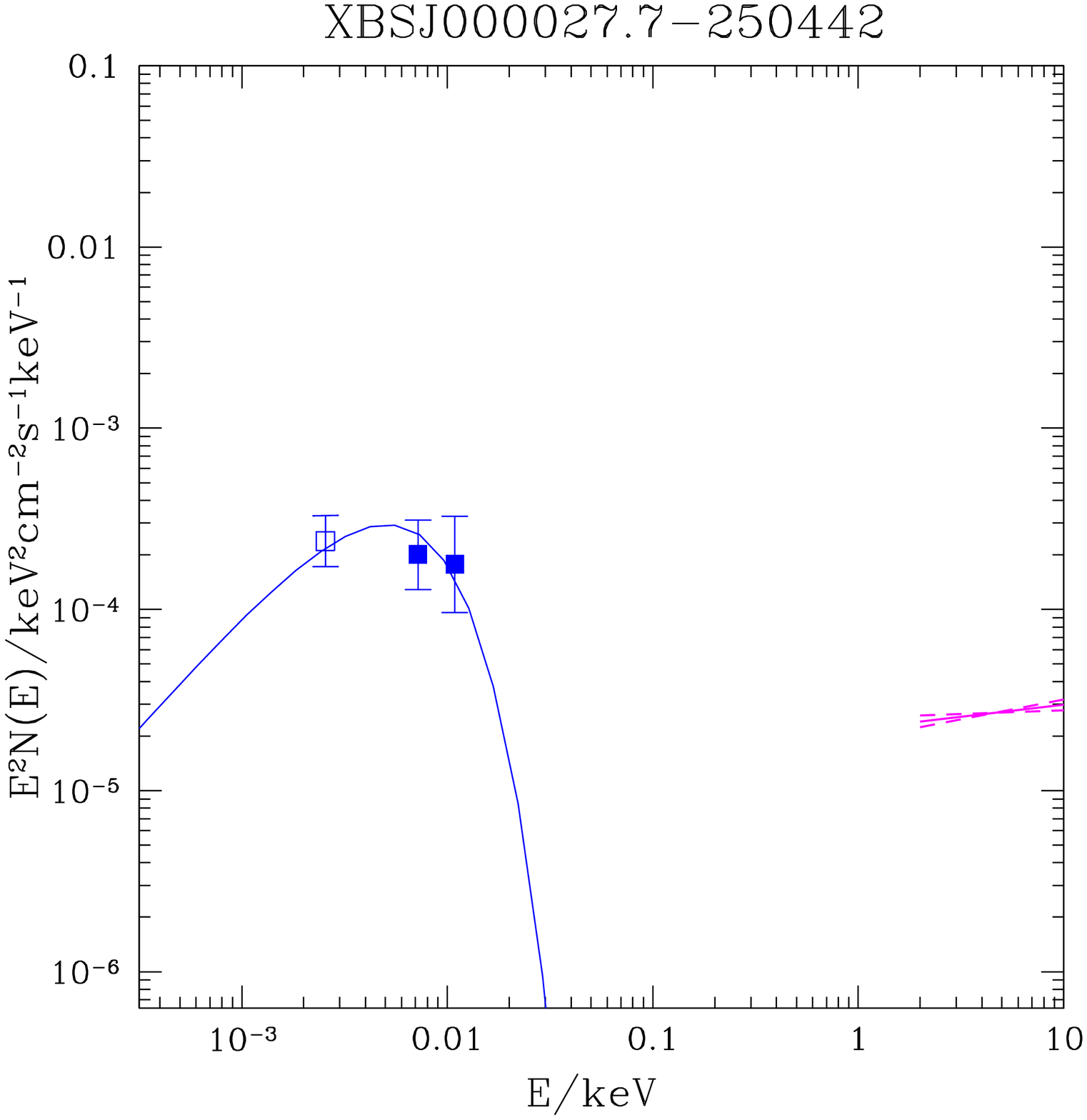}
  \includegraphics[height=5.6cm, width=6cm]{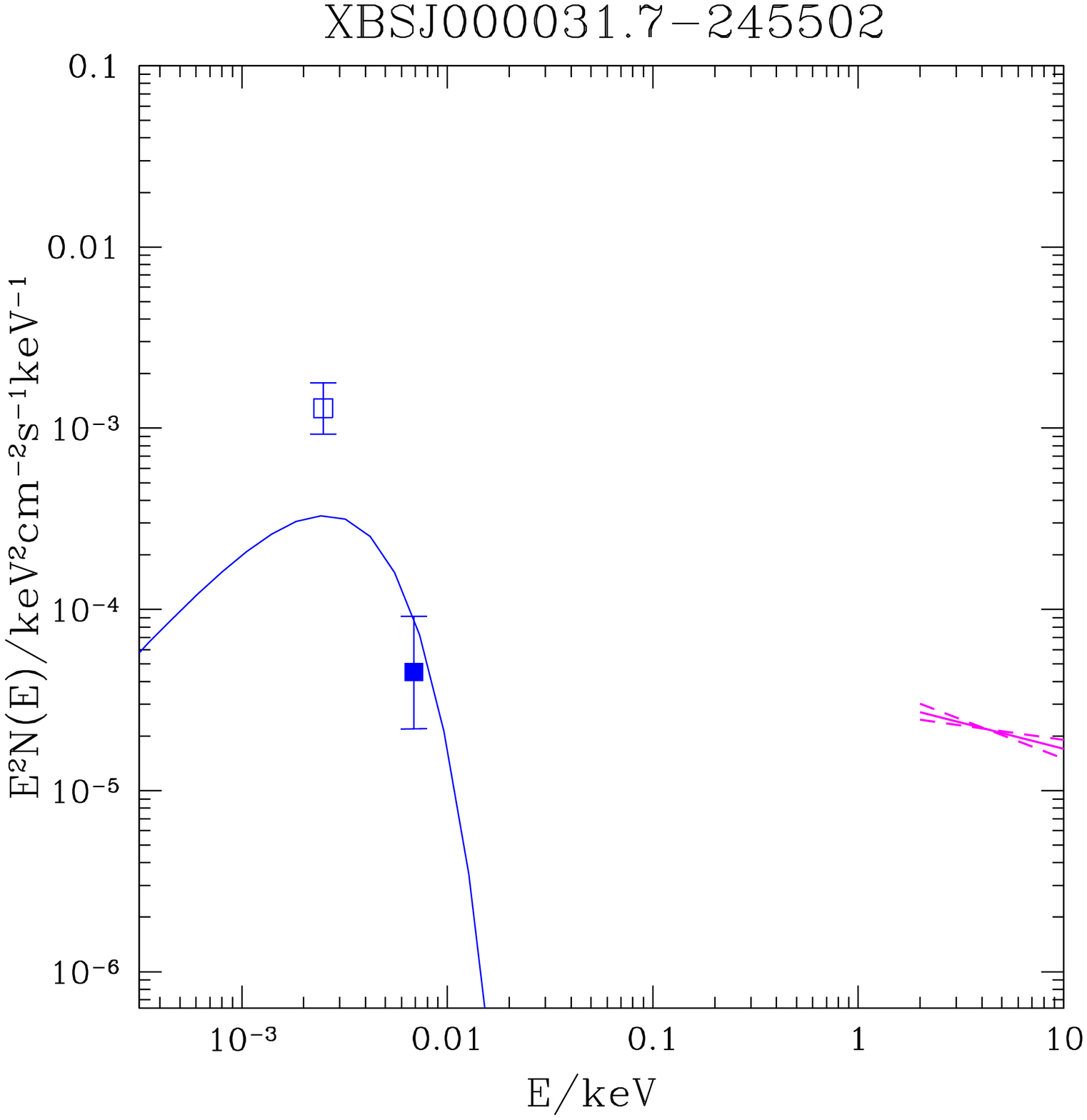}}      
  \subfigure{  
  \includegraphics[height=5.6cm, width=6cm]{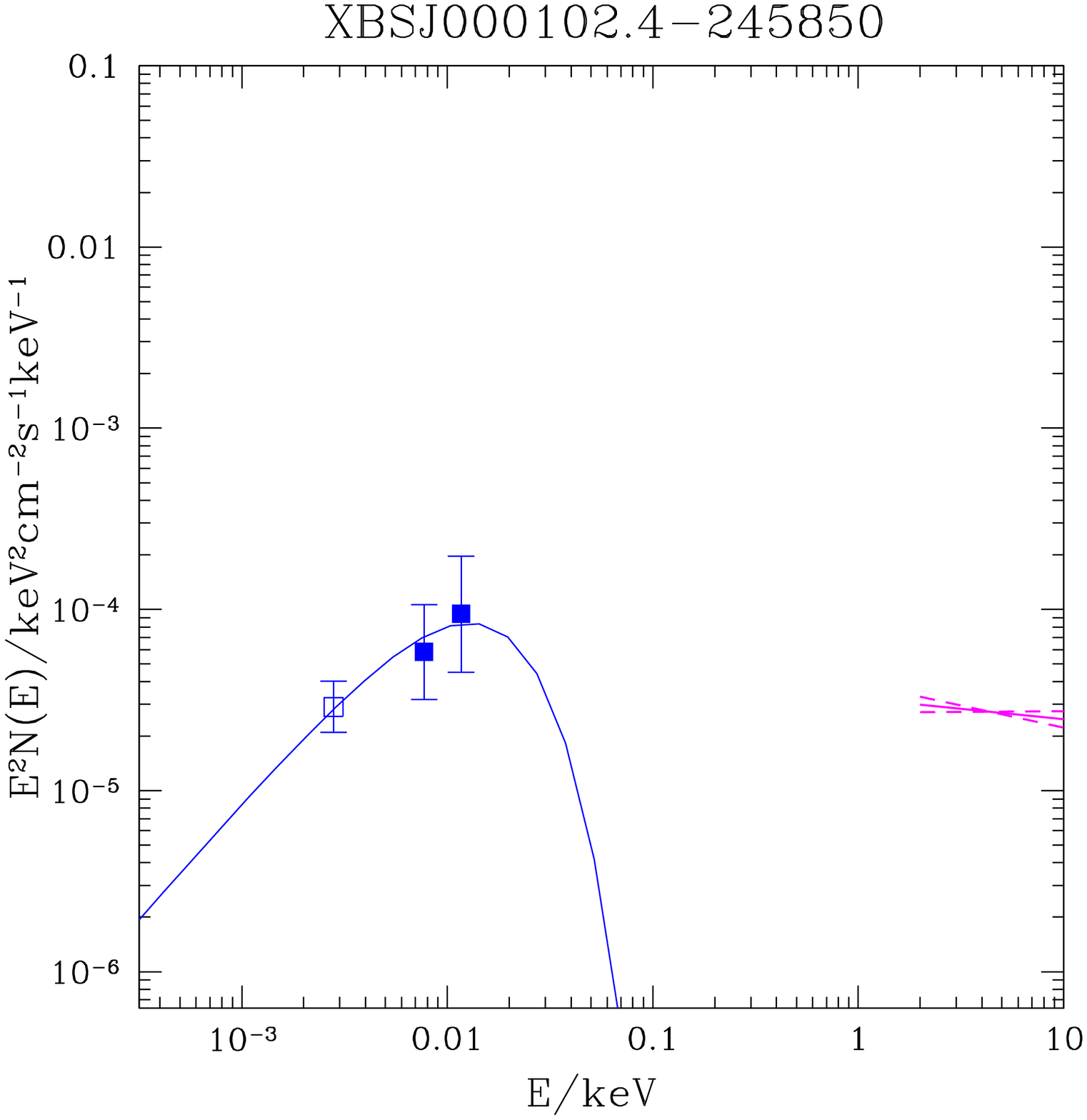}
  \includegraphics[height=5.6cm, width=6cm]{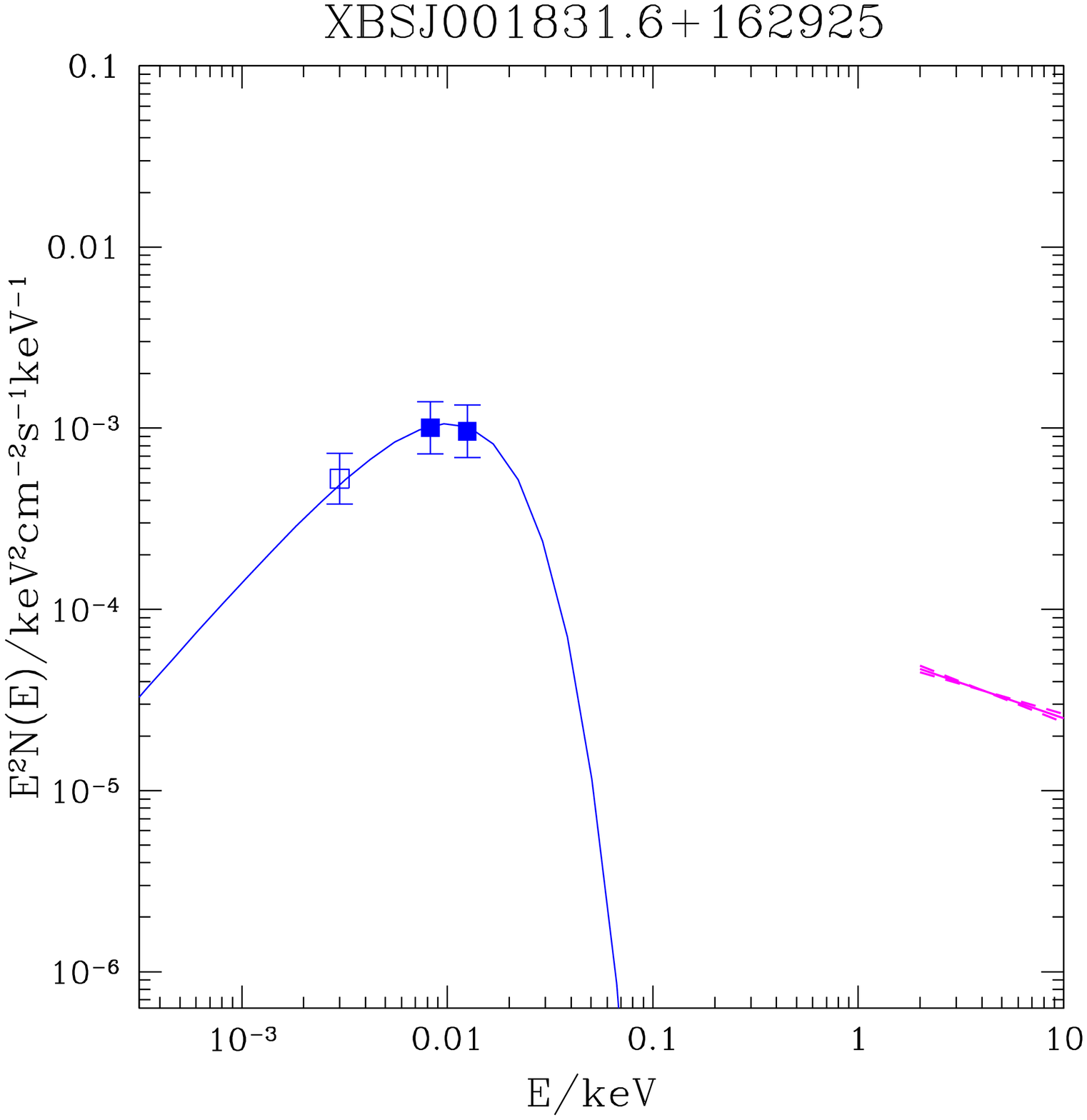}}  
 \end{figure*} 
 
 \begin{figure*}
 \centering
   \subfigure{ 
  \includegraphics[height=5.6cm, width=6cm]{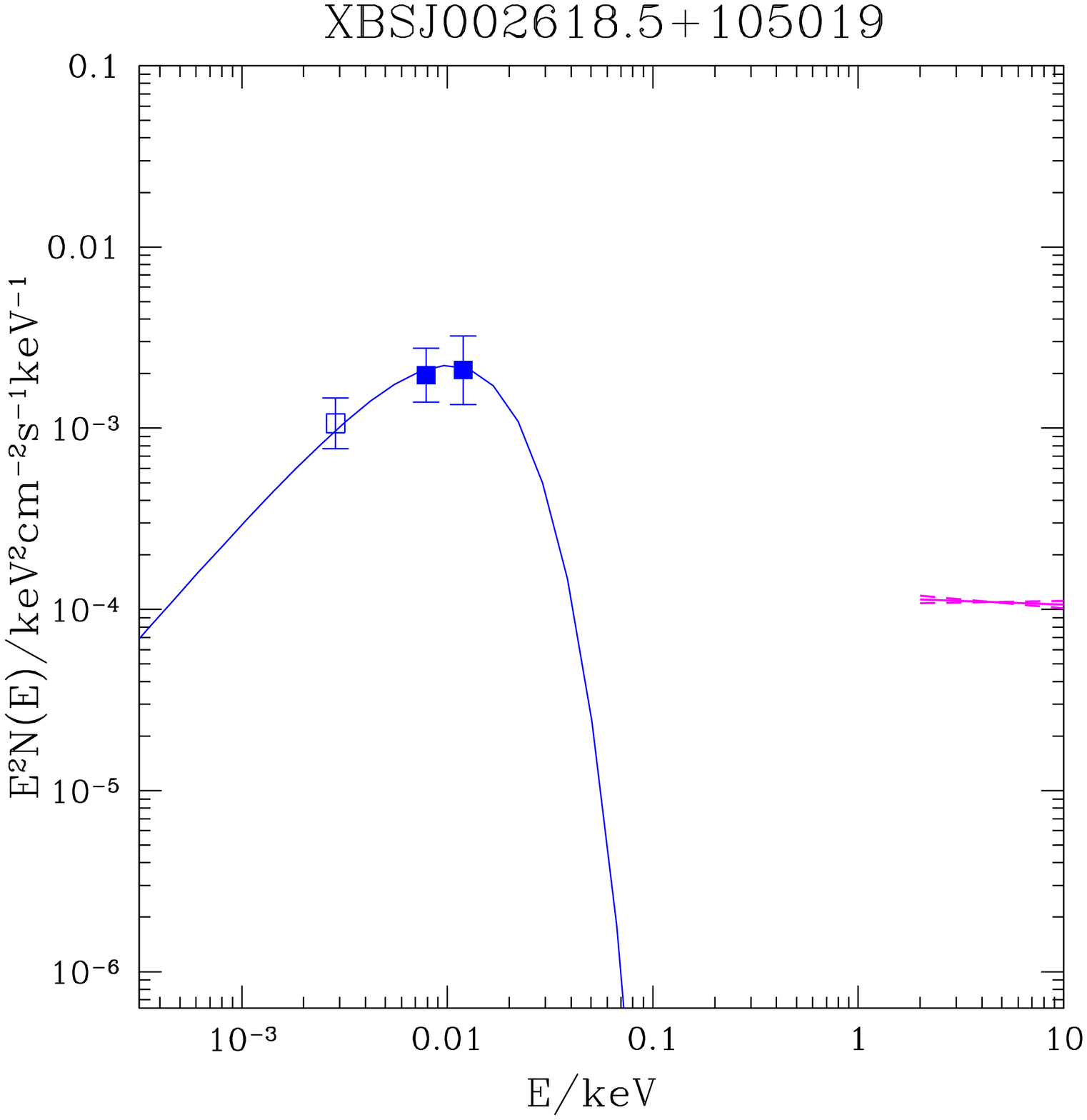}
  \includegraphics[height=5.6cm, width=6cm]{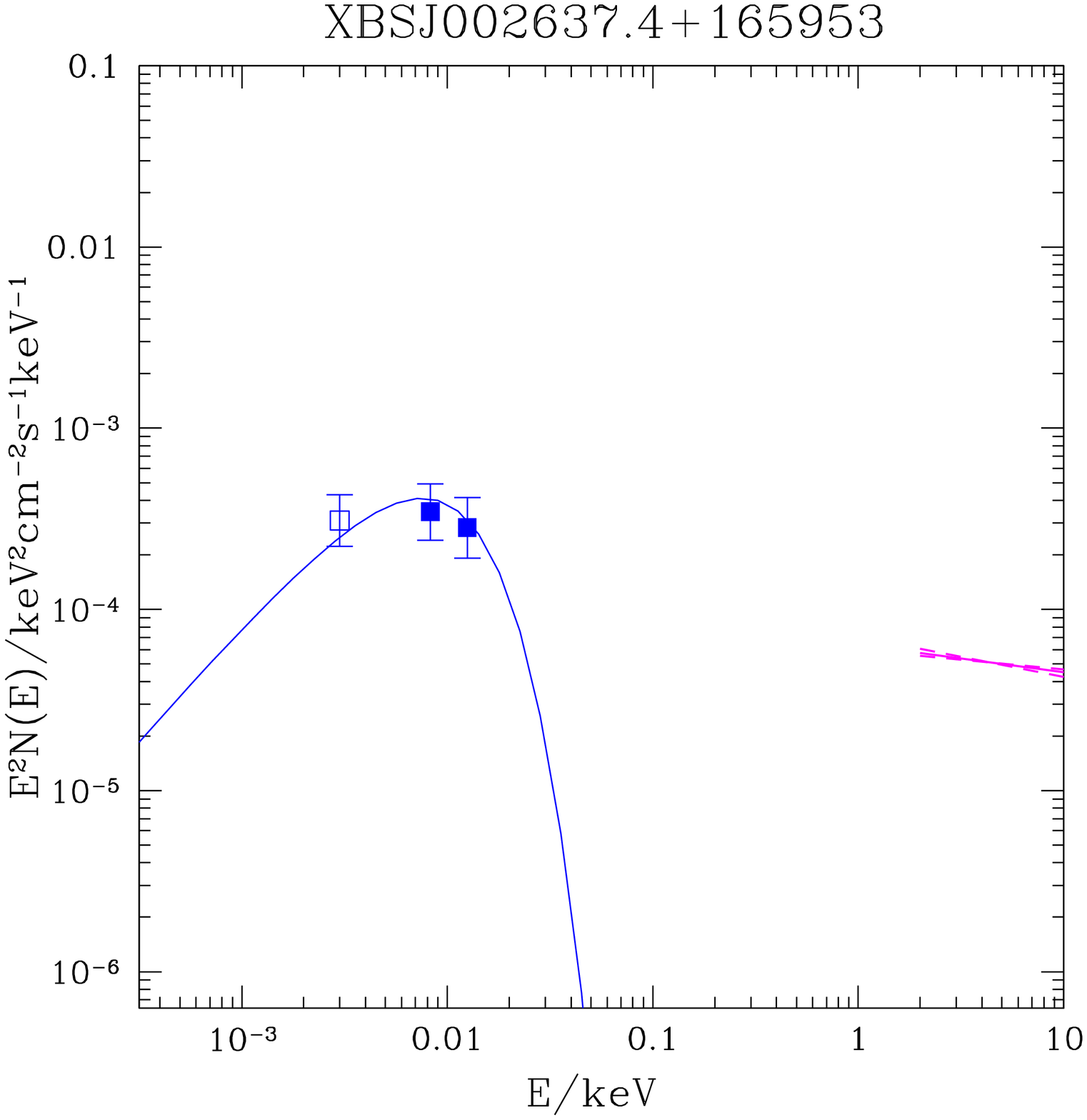}}     
  \subfigure{ 
  \includegraphics[height=5.6cm, width=6cm]{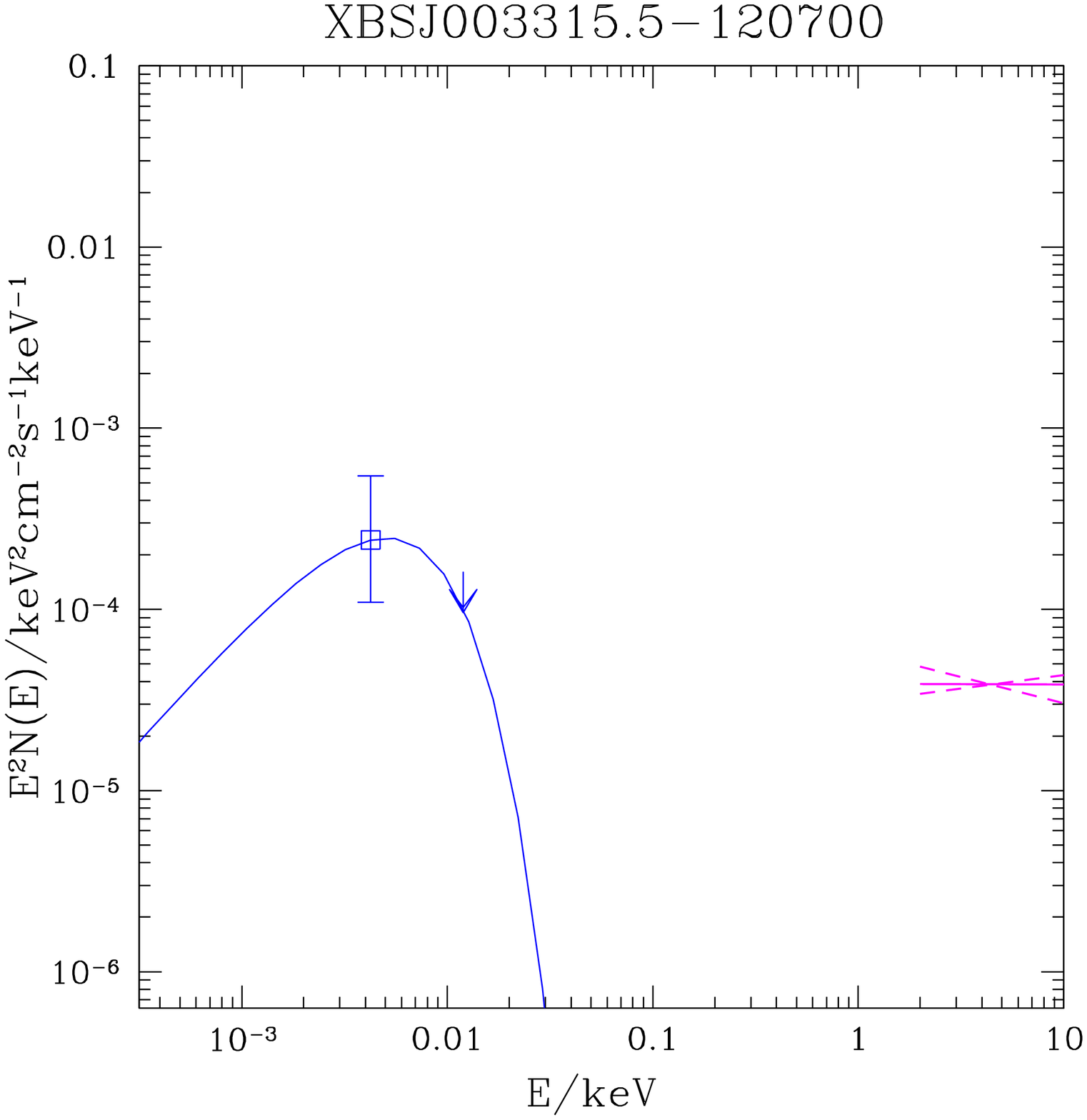}
  \includegraphics[height=5.6cm, width=6cm]{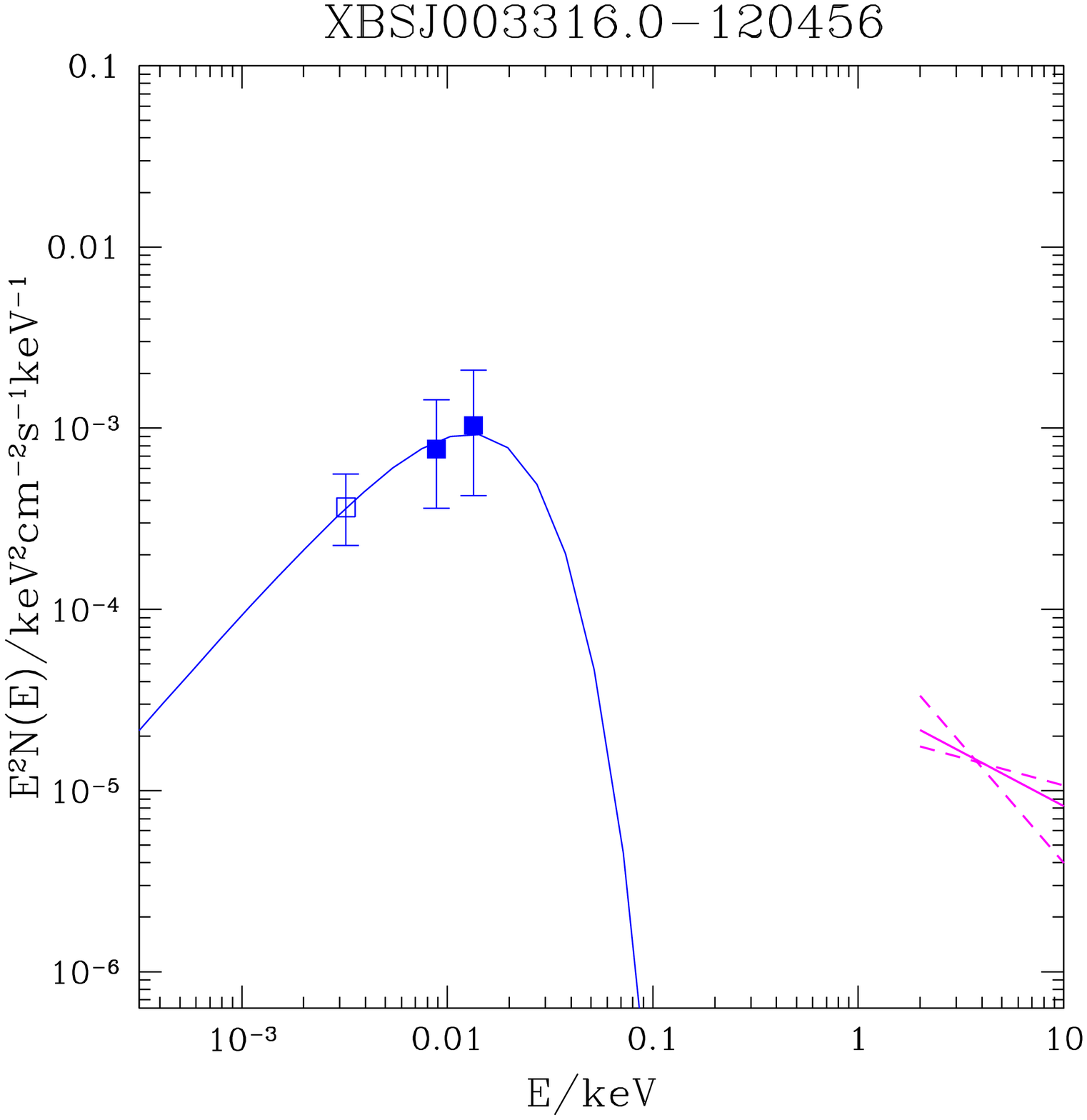}}
   \subfigure{ 
  \includegraphics[height=5.6cm, width=6cm]{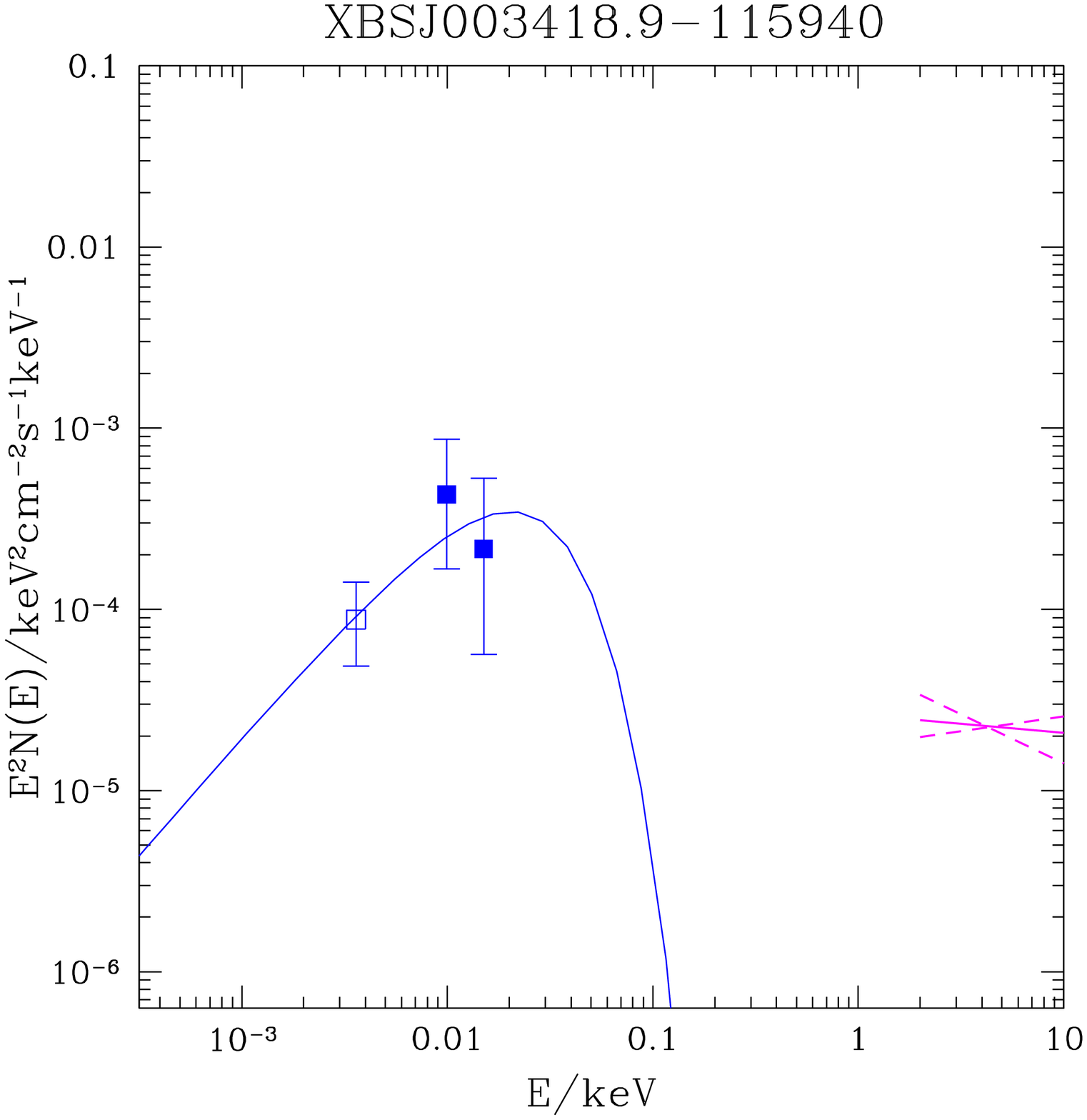}
  \includegraphics[height=5.6cm, width=6cm]{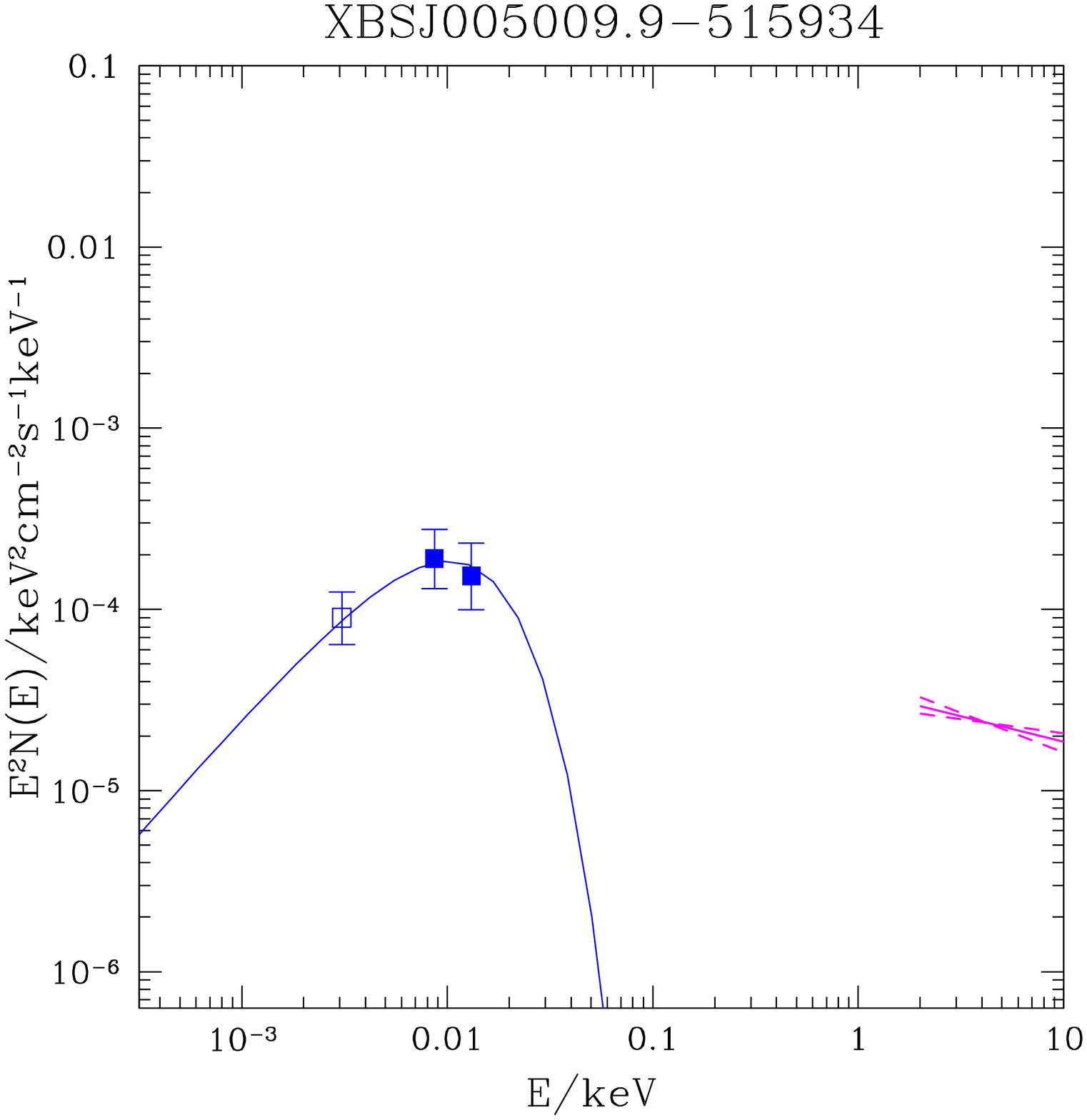}}      
\subfigure{ 
  \includegraphics[height=5.6cm, width=6cm]{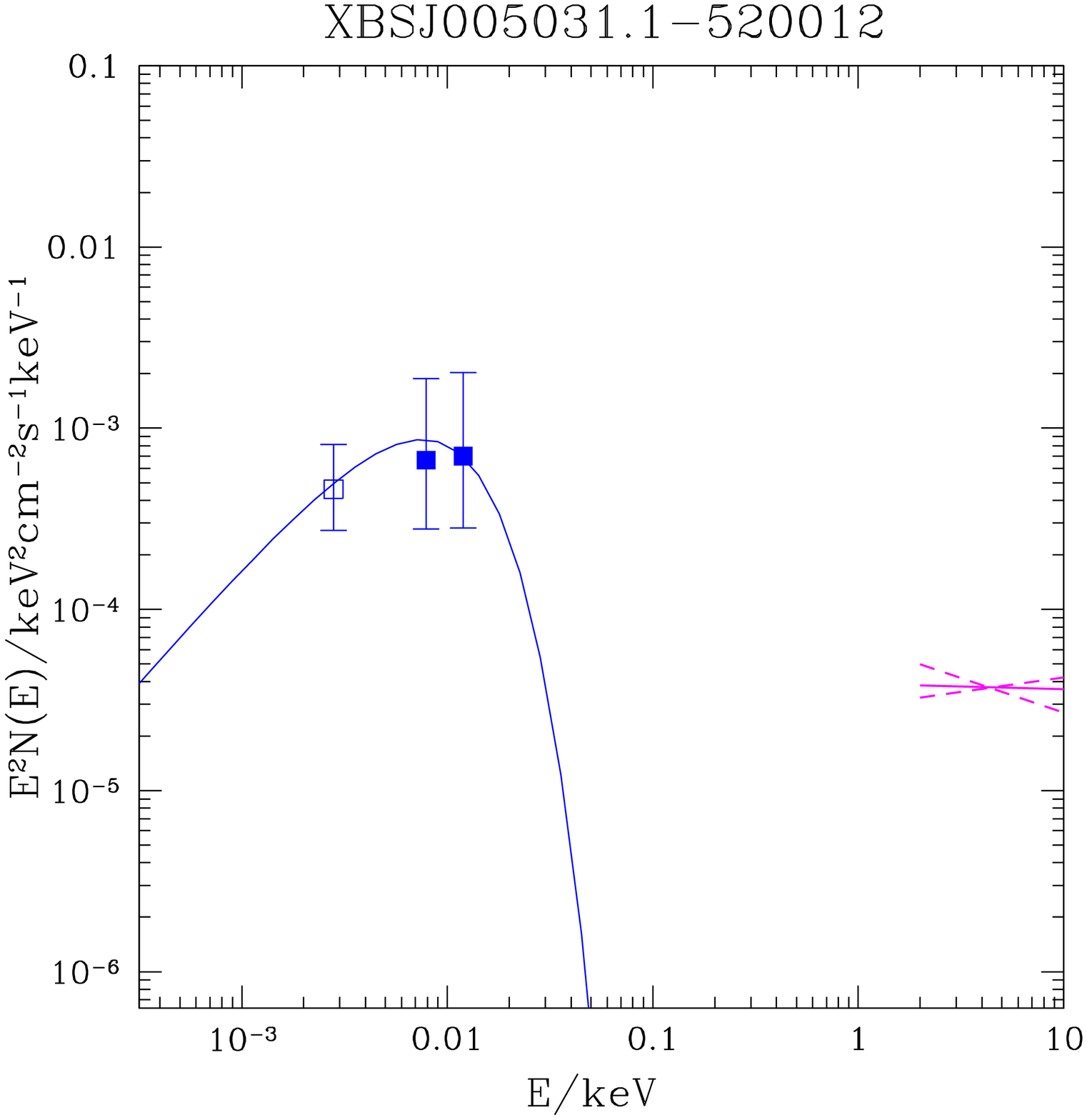}
  \includegraphics[height=5.6cm, width=6cm]{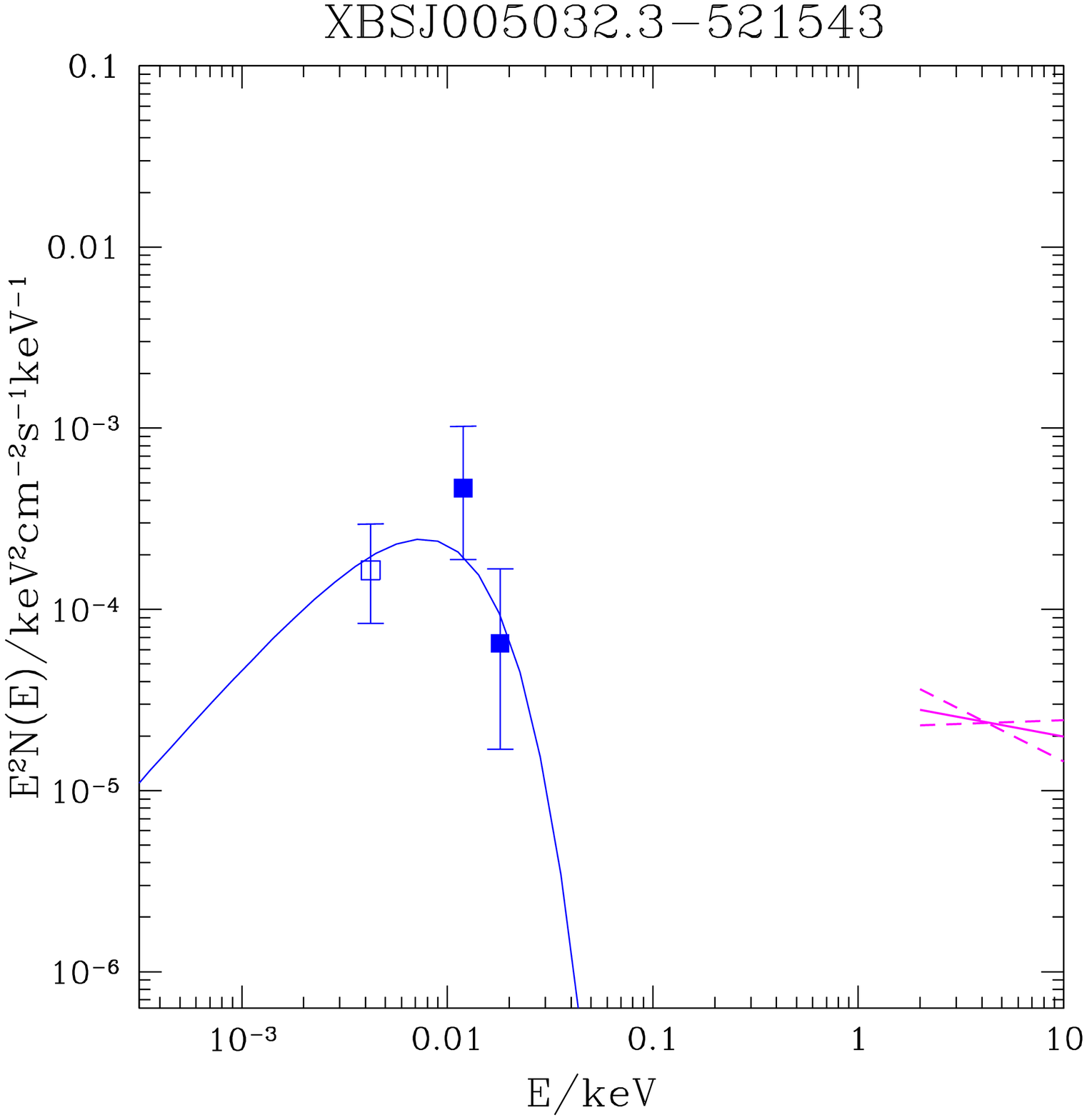}}   
 \end{figure*} 

 \FloatBarrier
 
\begin{figure*}
\centering
 \subfigure{ 
  \includegraphics[height=5.6cm, width=6cm]{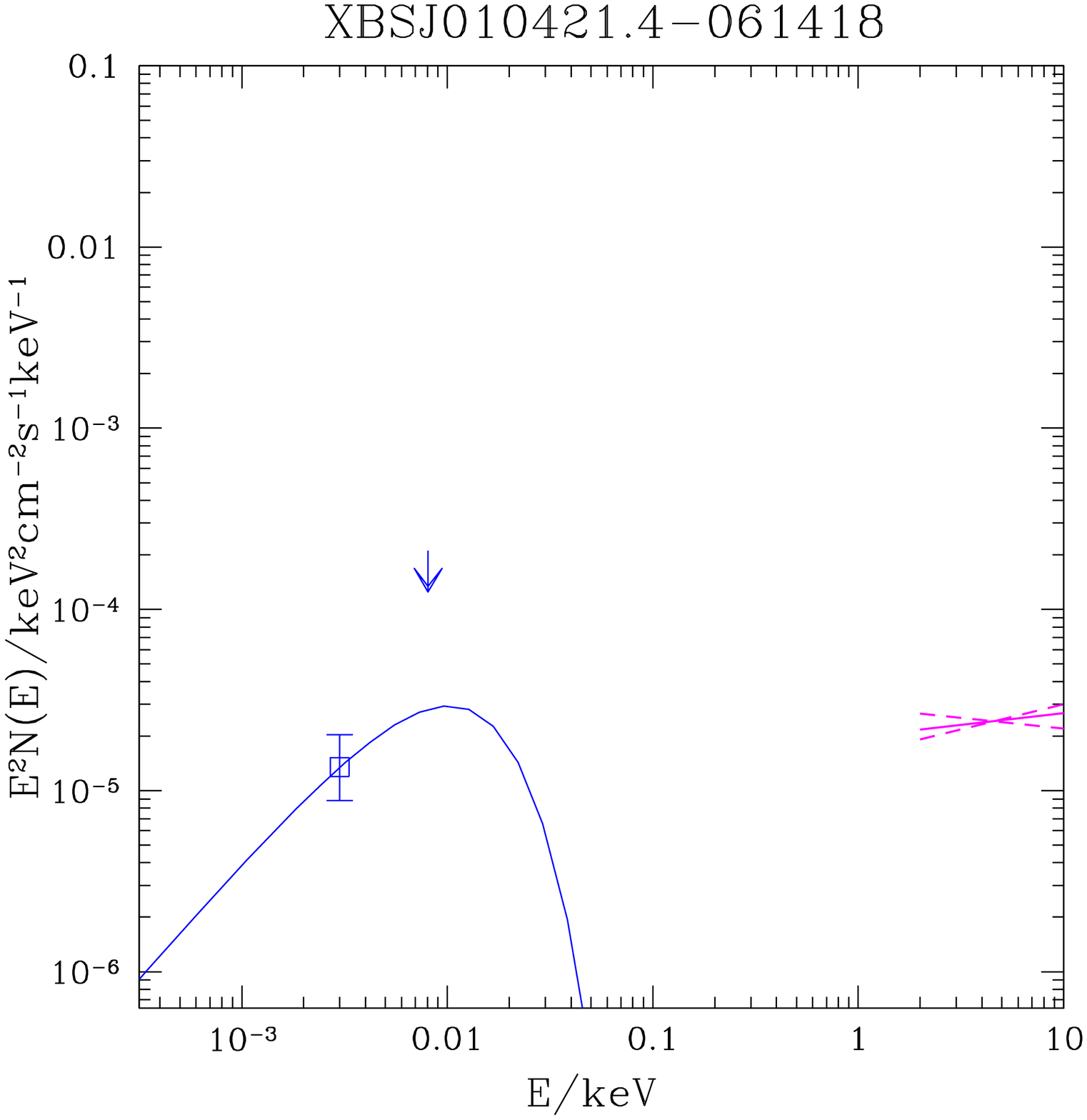}
  \includegraphics[height=5.6cm, width=6cm]{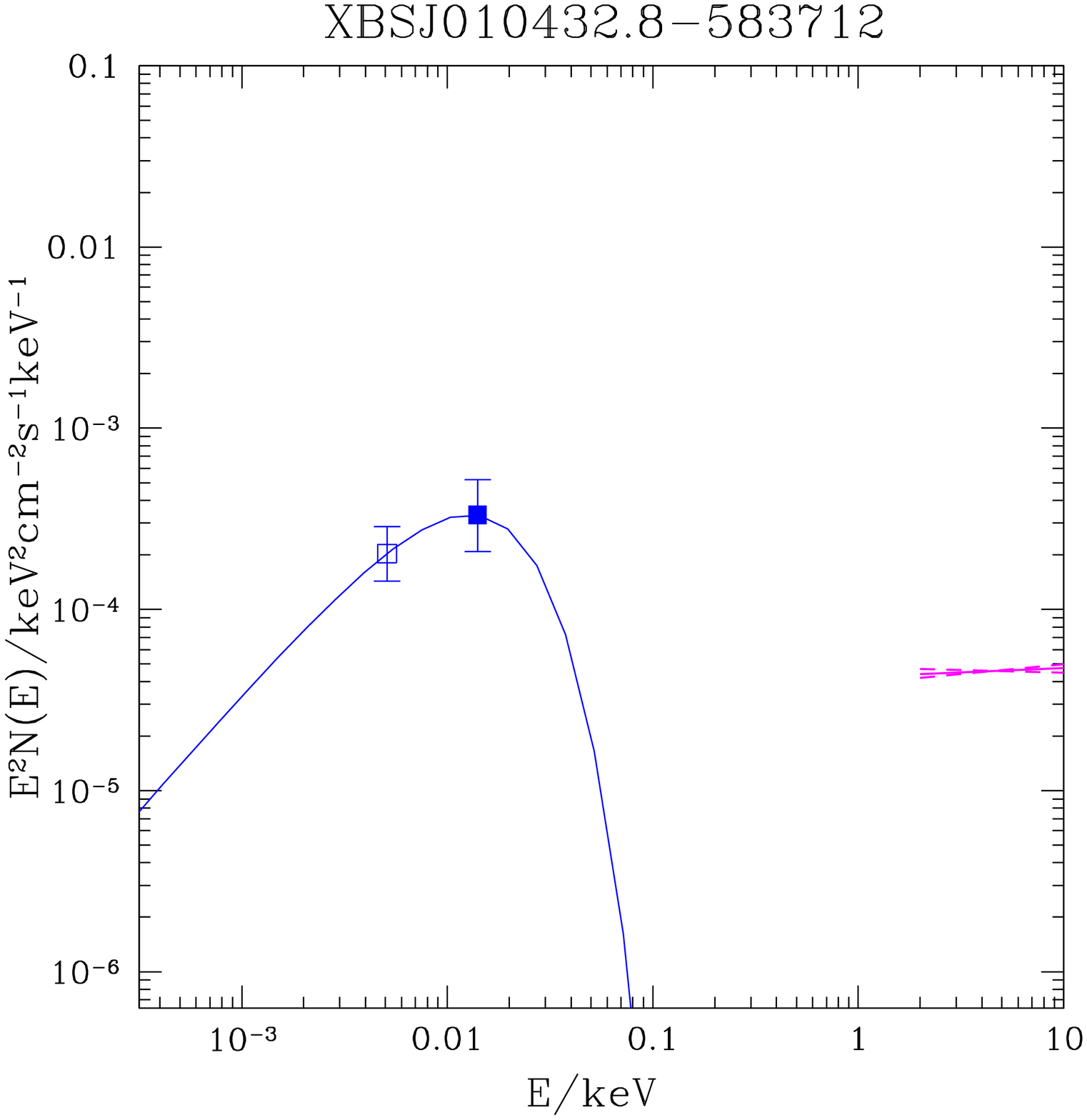}}   
   \subfigure{ 
  \includegraphics[height=5.6cm, width=6cm]{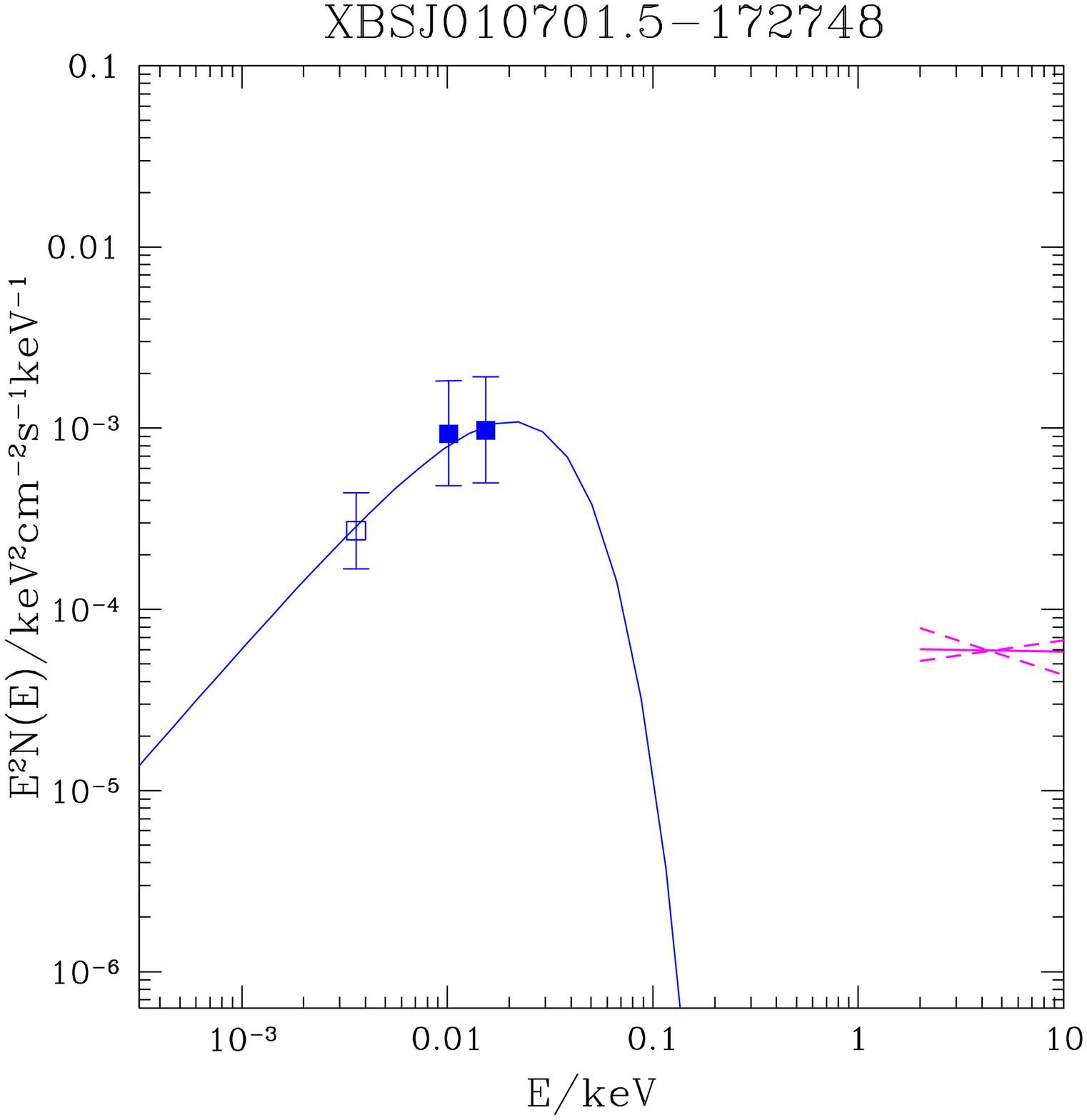}
  \includegraphics[height=5.6cm, width=6cm]{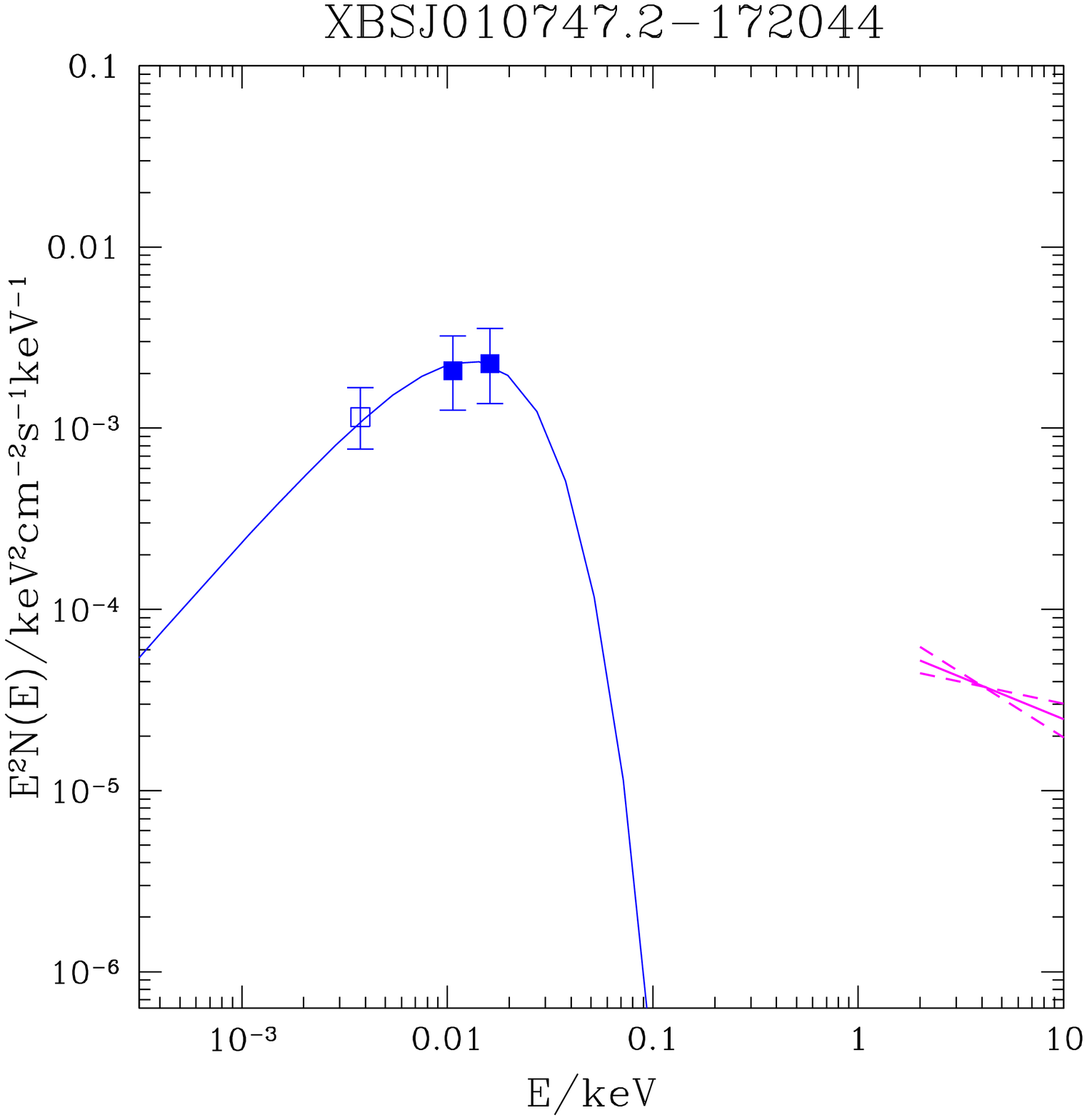}} 
    \subfigure{ 
  \includegraphics[height=5.6cm, width=6cm]{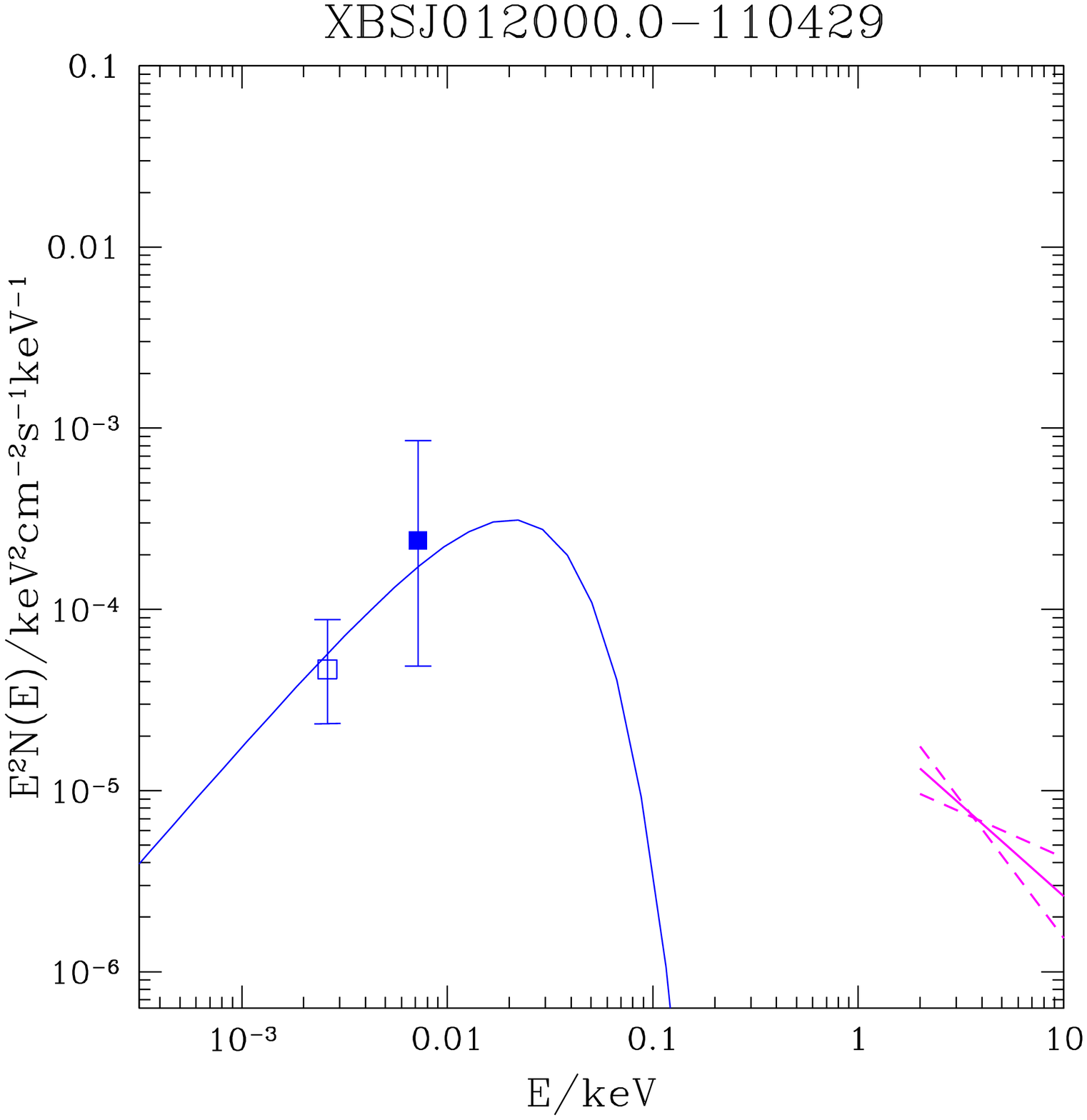}
  \includegraphics[height=5.6cm, width=6cm]{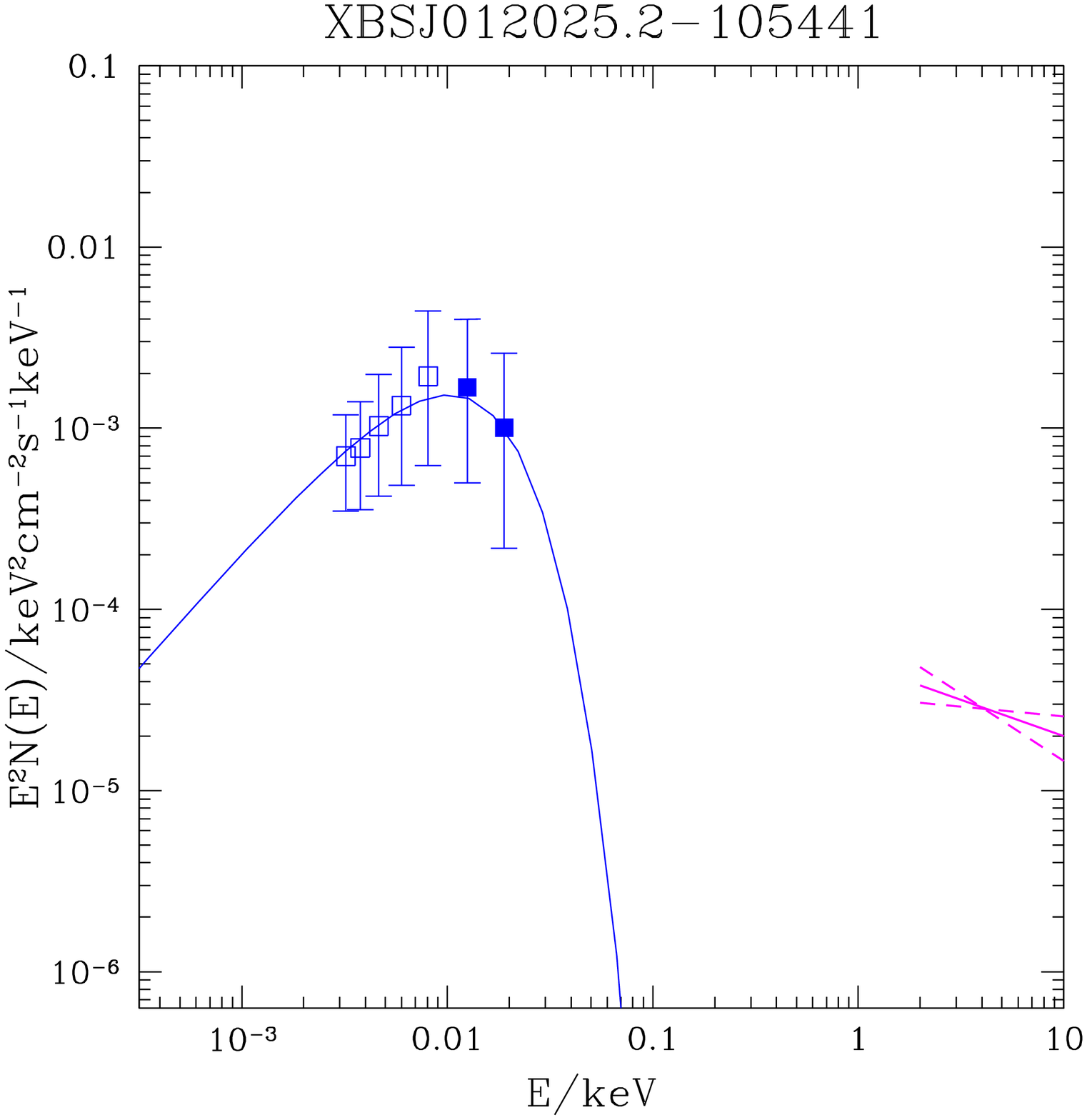}}    
  \subfigure{ 
  \includegraphics[height=5.6cm, width=6cm]{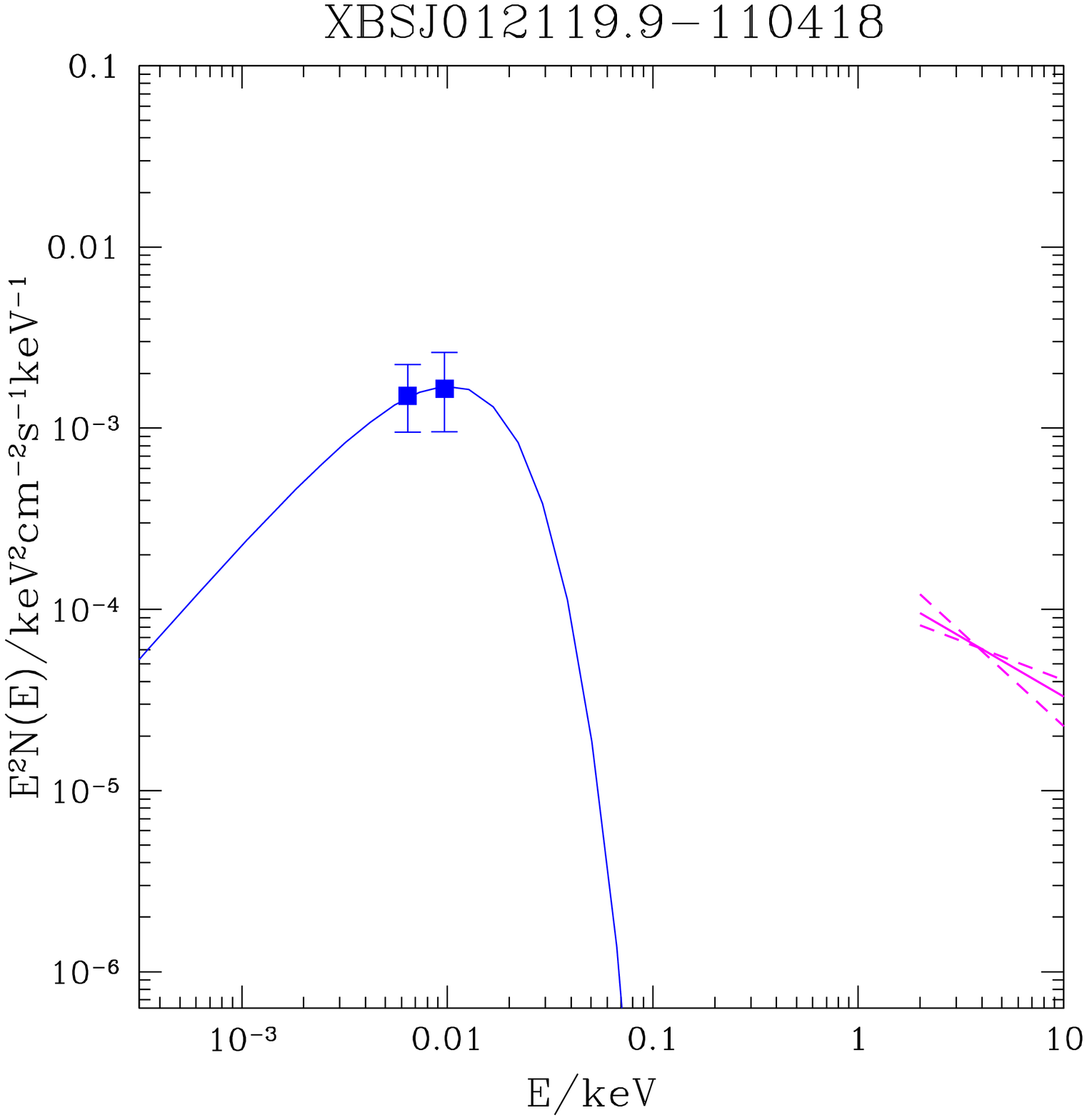}
  \includegraphics[height=5.6cm, width=6cm]{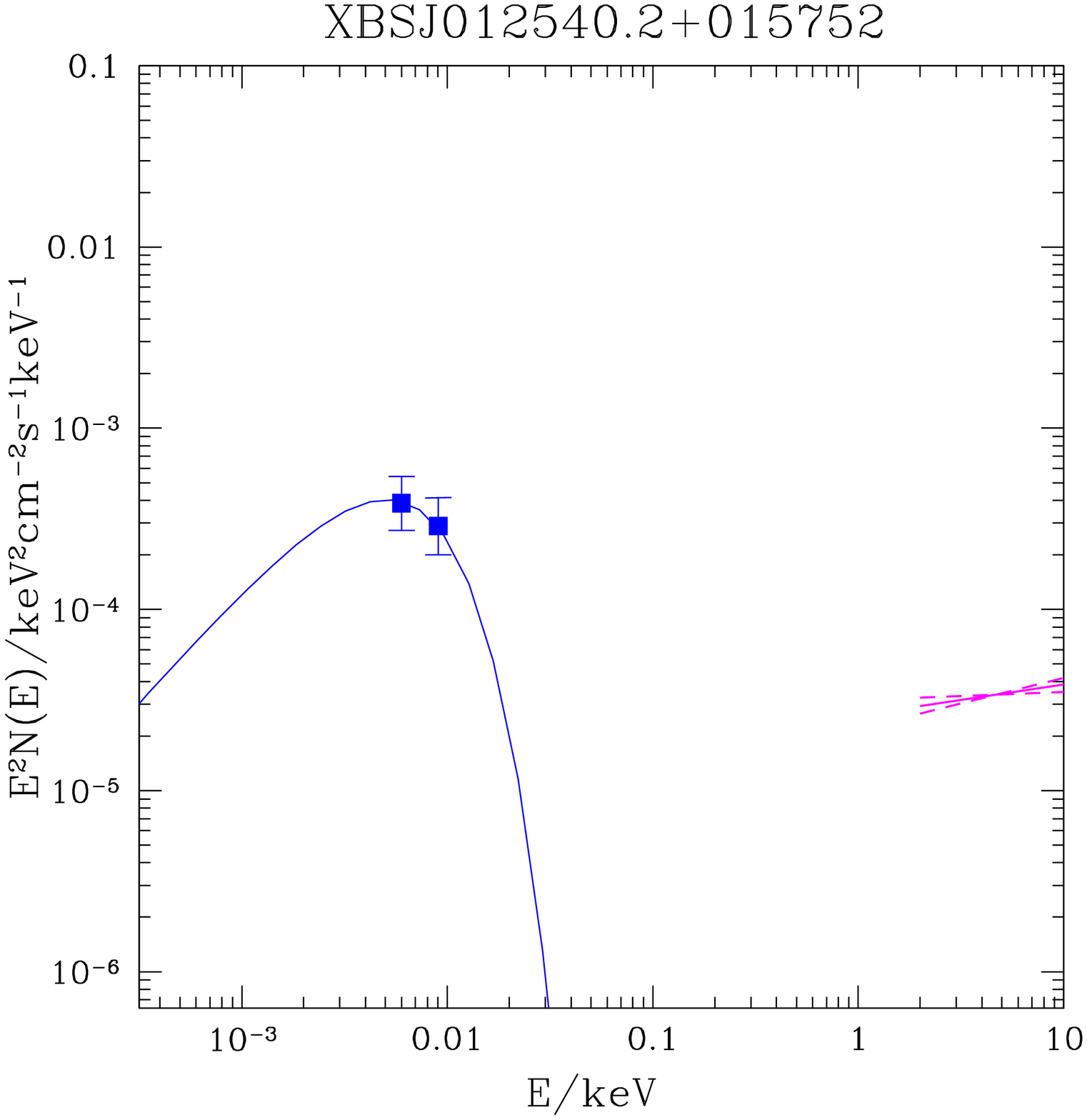}} 
  \end{figure*} 
  
 \begin{figure*}
\centering
 \subfigure{ 
  \includegraphics[height=5.6cm, width=6cm]{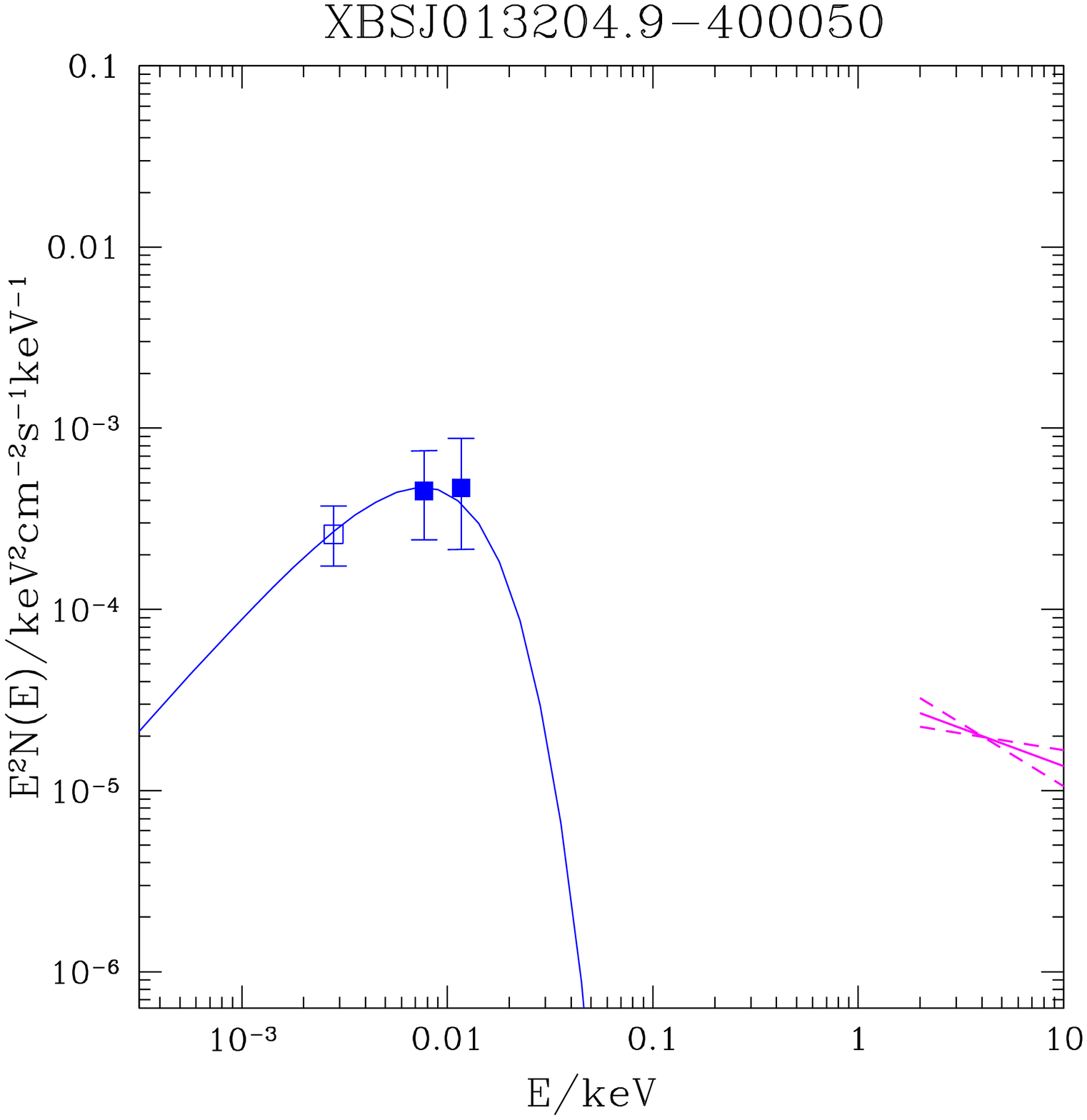}
  \includegraphics[height=5.6cm, width=6cm]{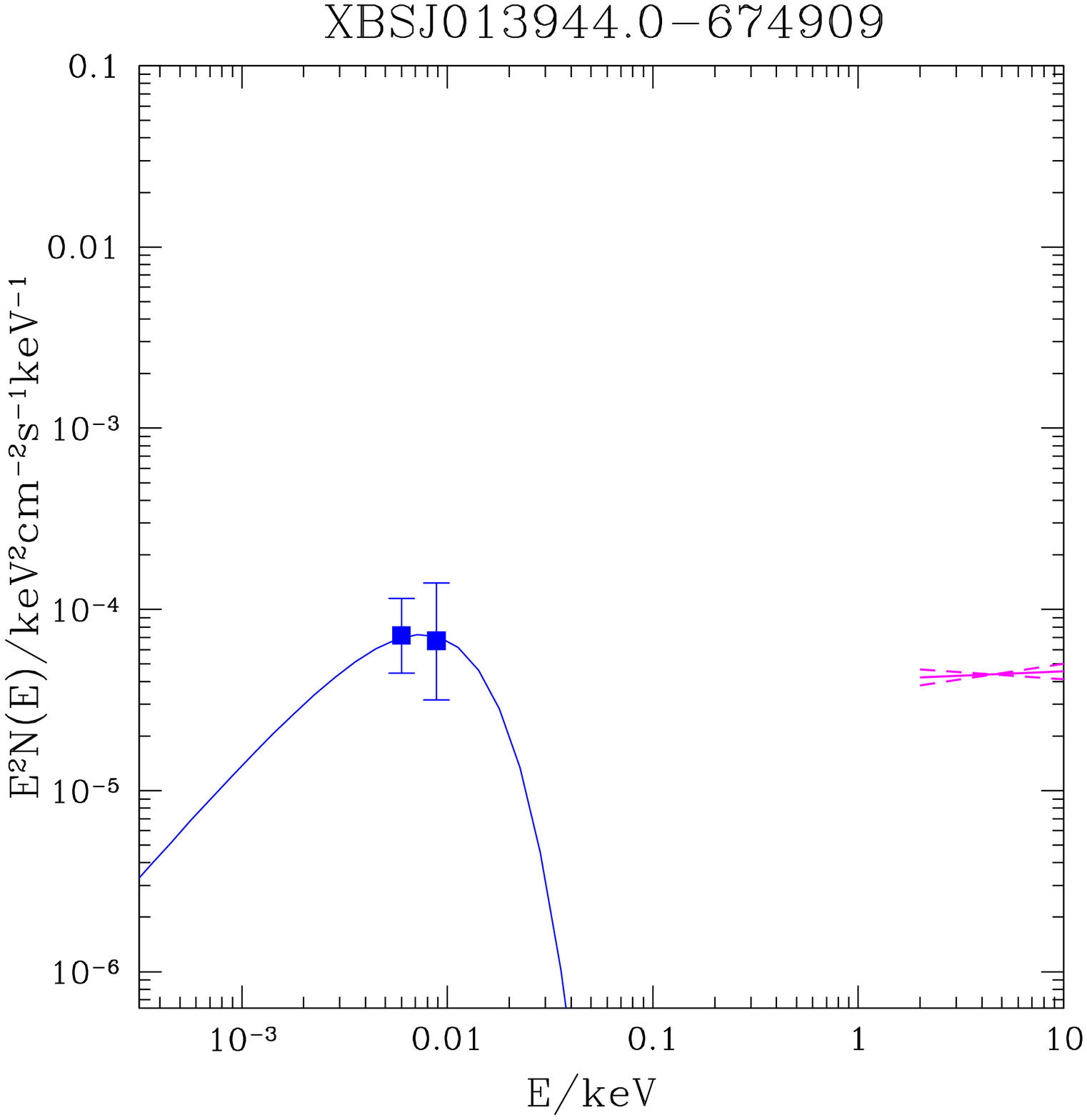}}    
  \subfigure{ 
  \includegraphics[height=5.6cm, width=6cm]{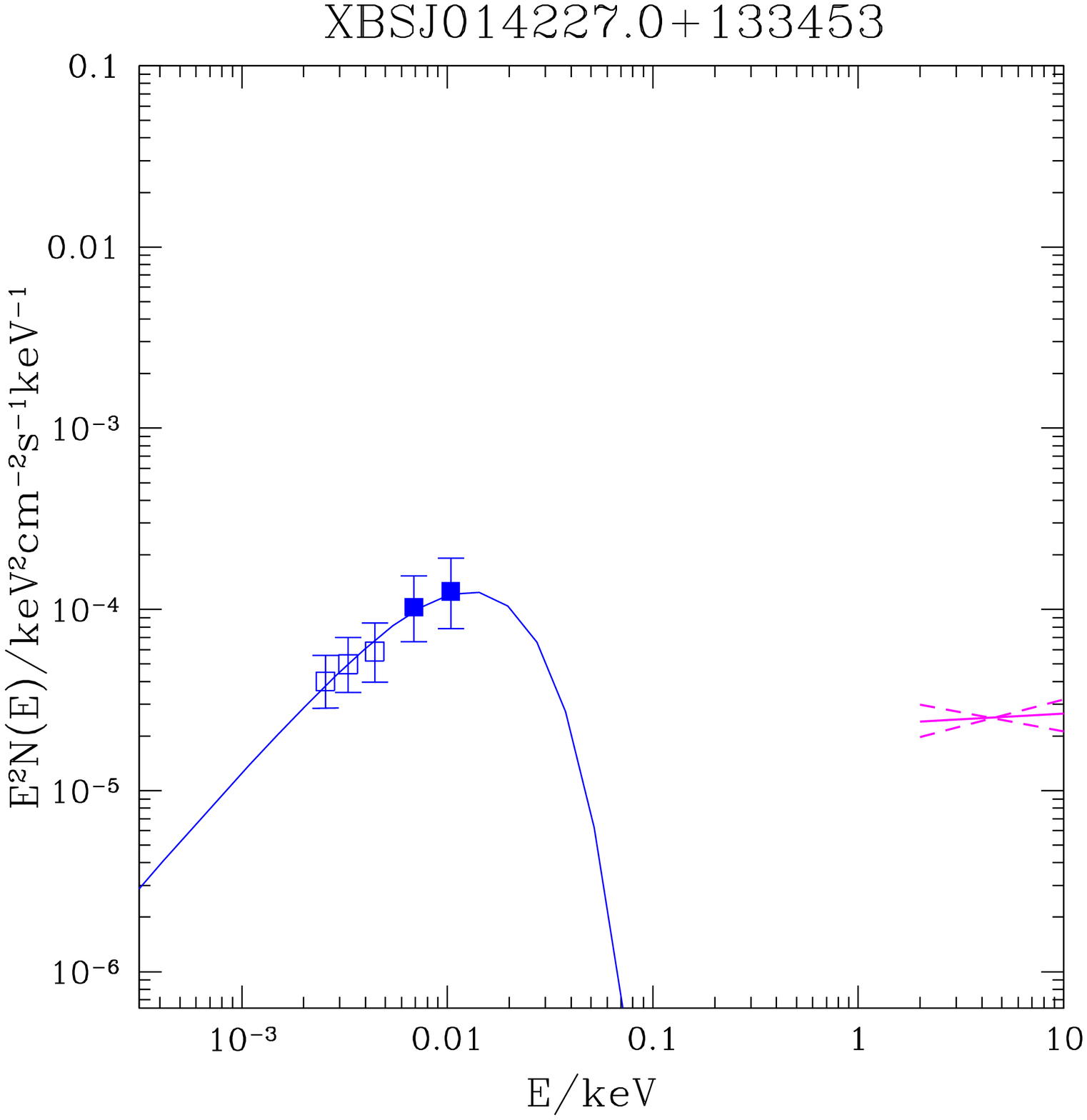}
  \includegraphics[height=5.6cm, width=6cm]{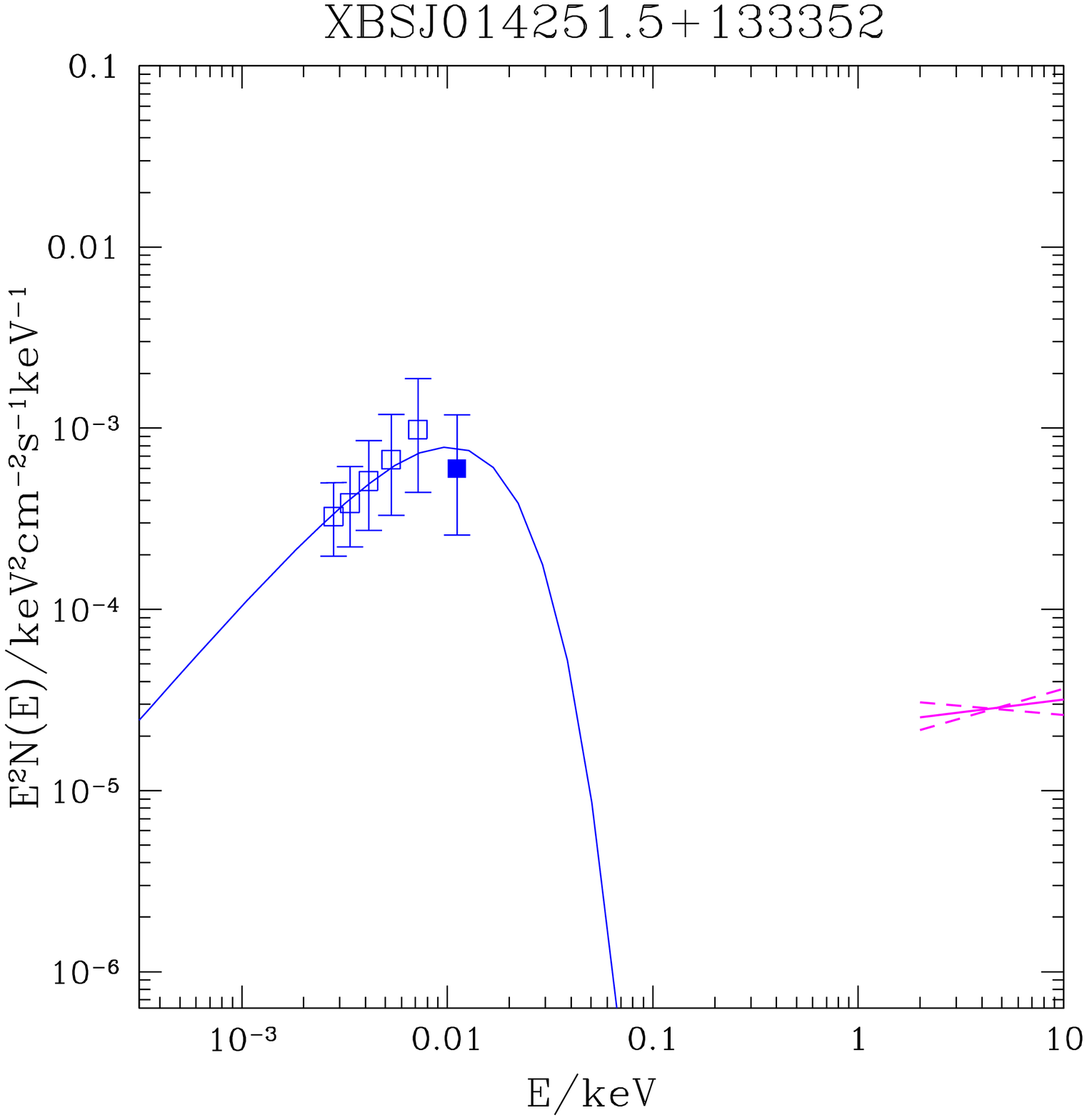}}   
  \subfigure{ 
  \includegraphics[height=5.6cm, width=6cm]{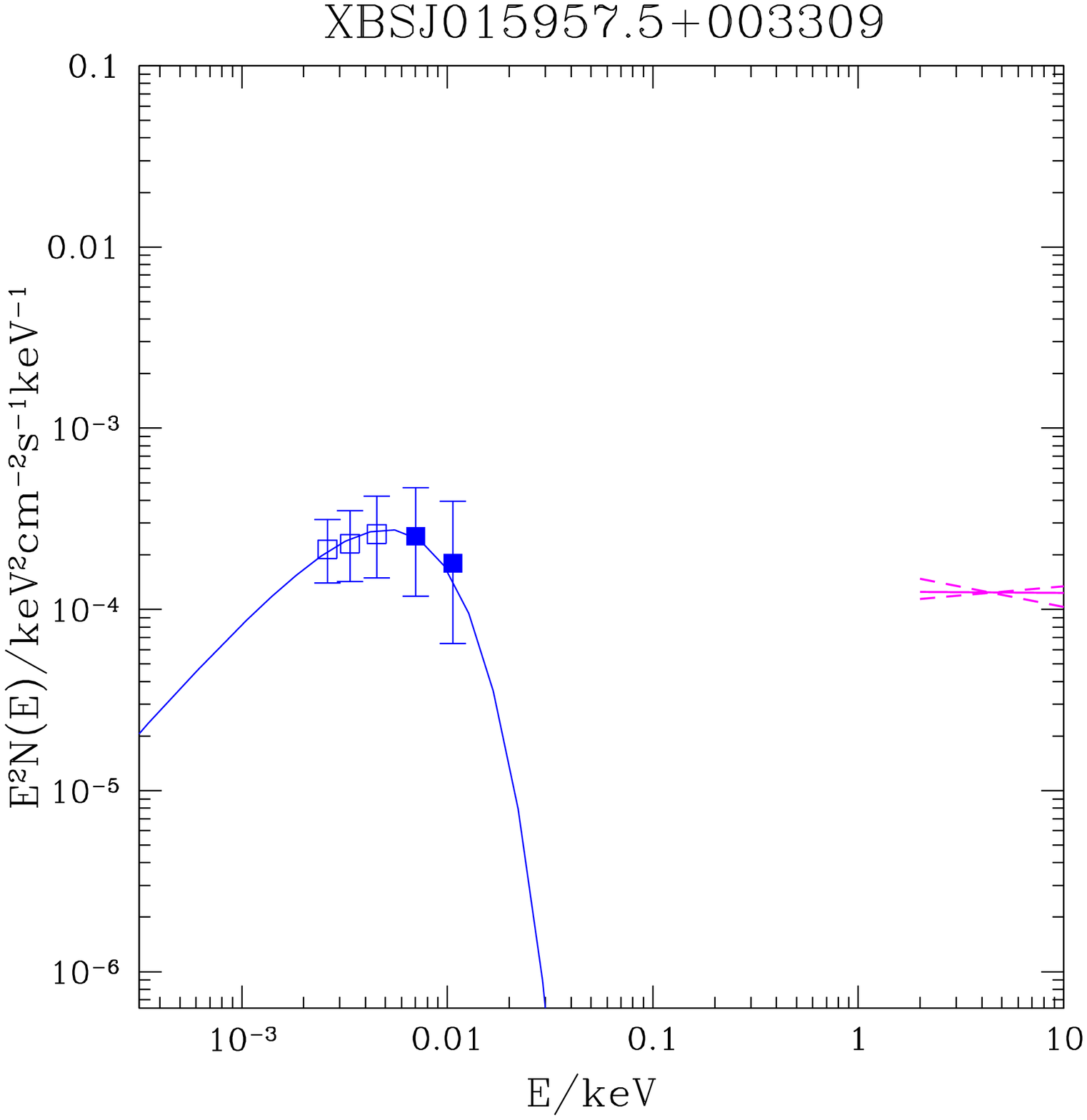}
  \includegraphics[height=5.6cm, width=6cm]{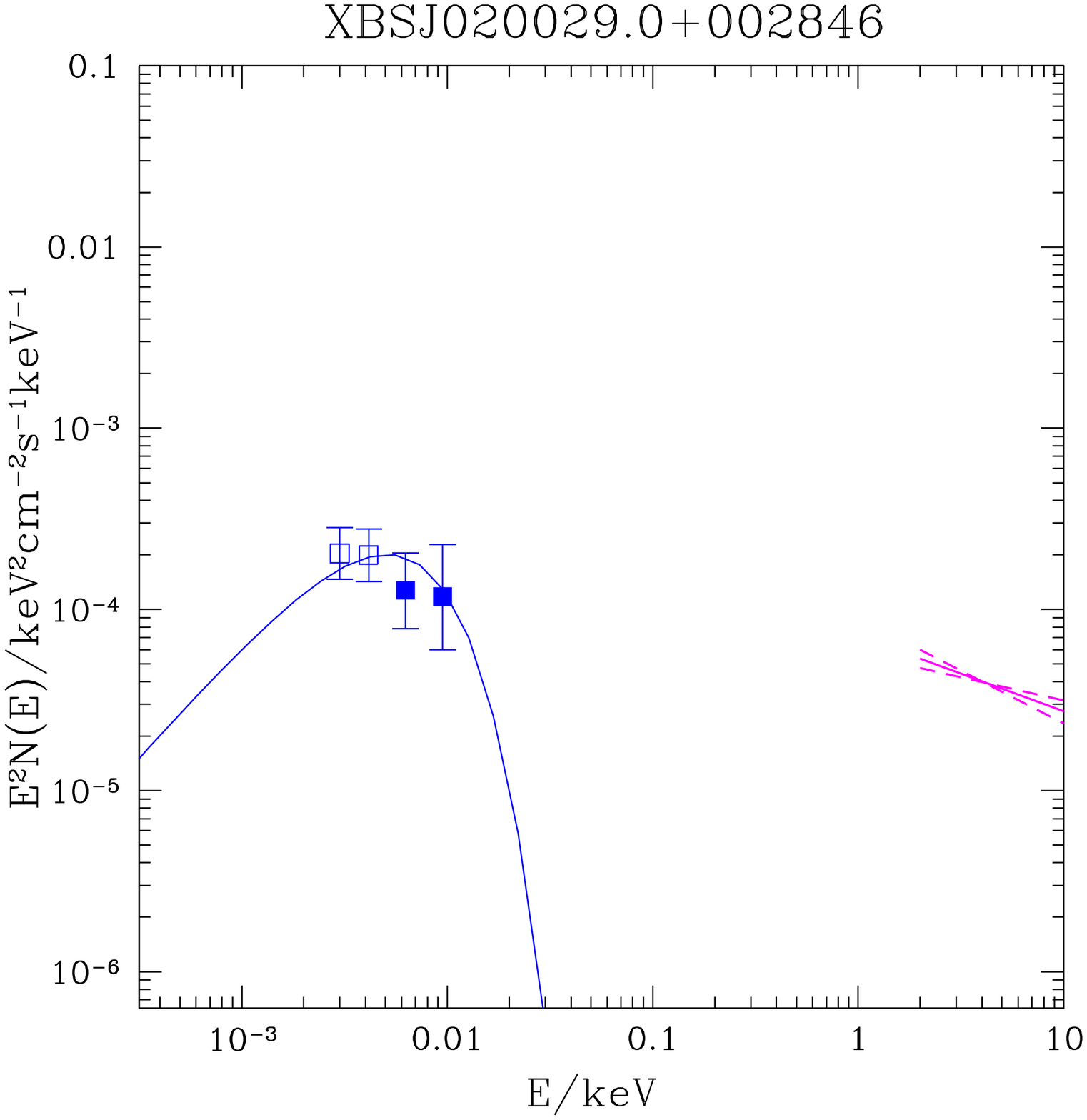}}     
 \subfigure{ 
  \includegraphics[height=5.6cm, width=6cm]{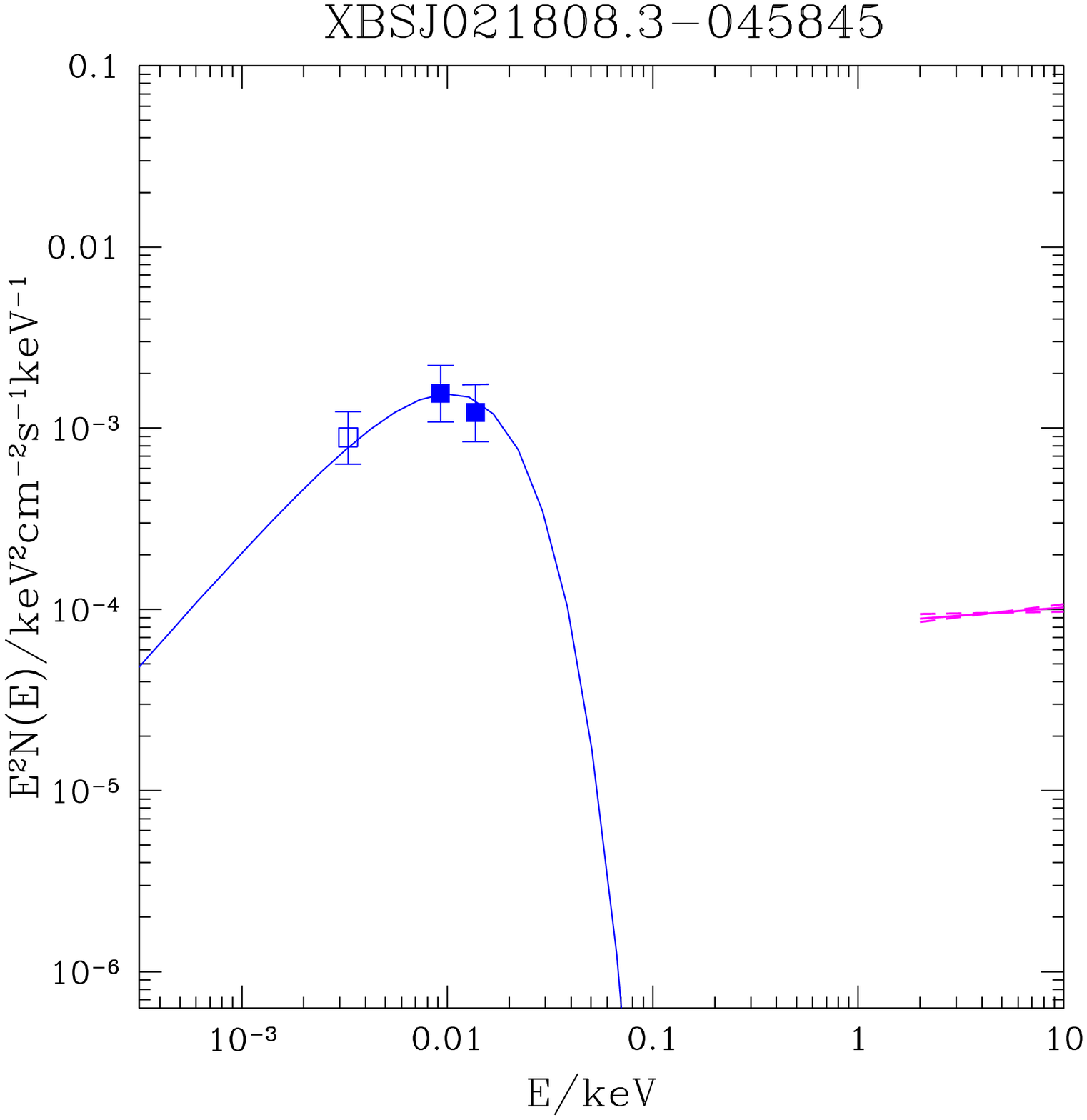}
  \includegraphics[height=5.6cm, width=6cm]{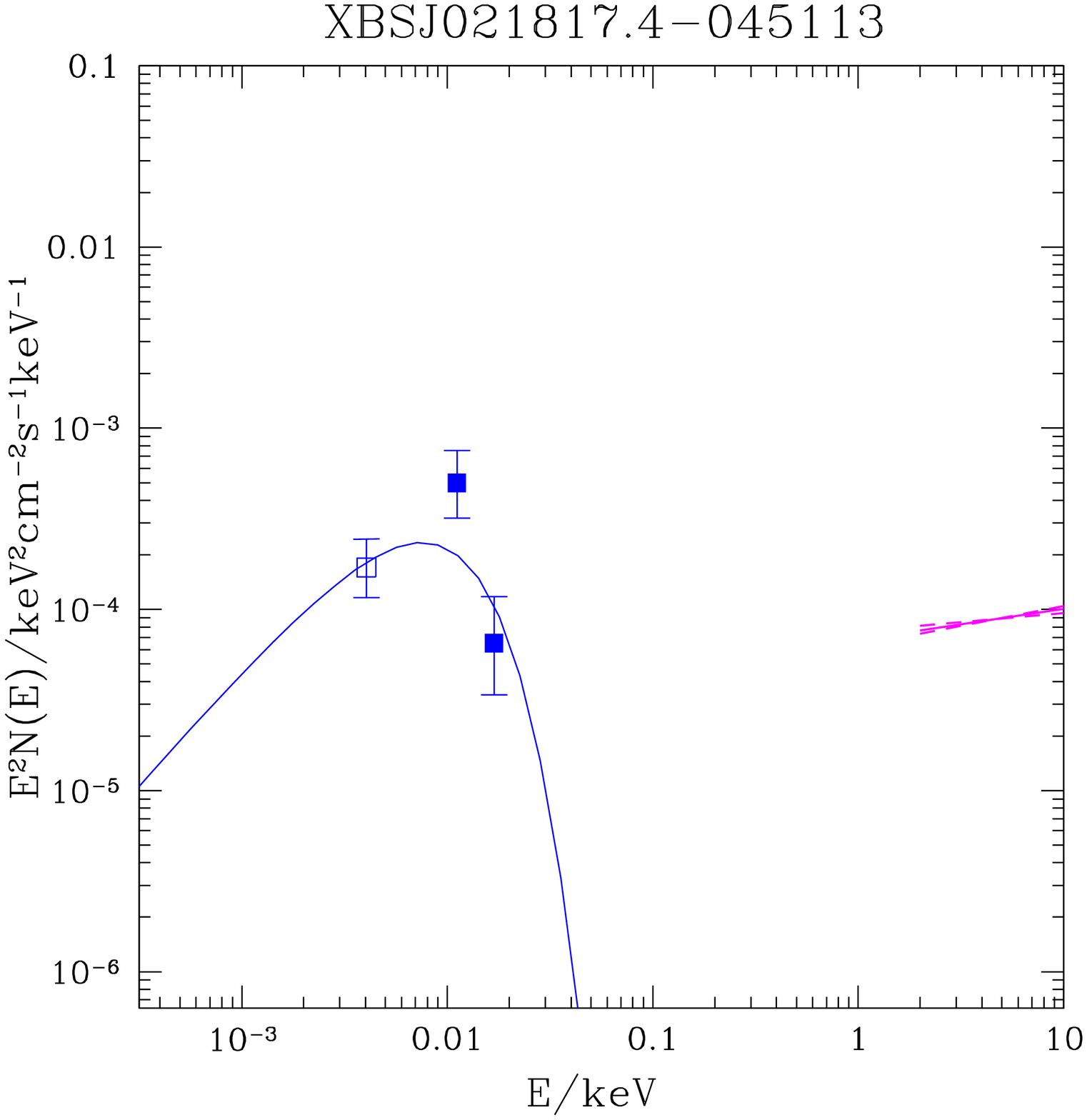}}  
  \end{figure*}  
   
 \FloatBarrier
   
  \begin{figure*}
\centering
 \subfigure{ 
  \includegraphics[height=5.6cm, width=6cm]{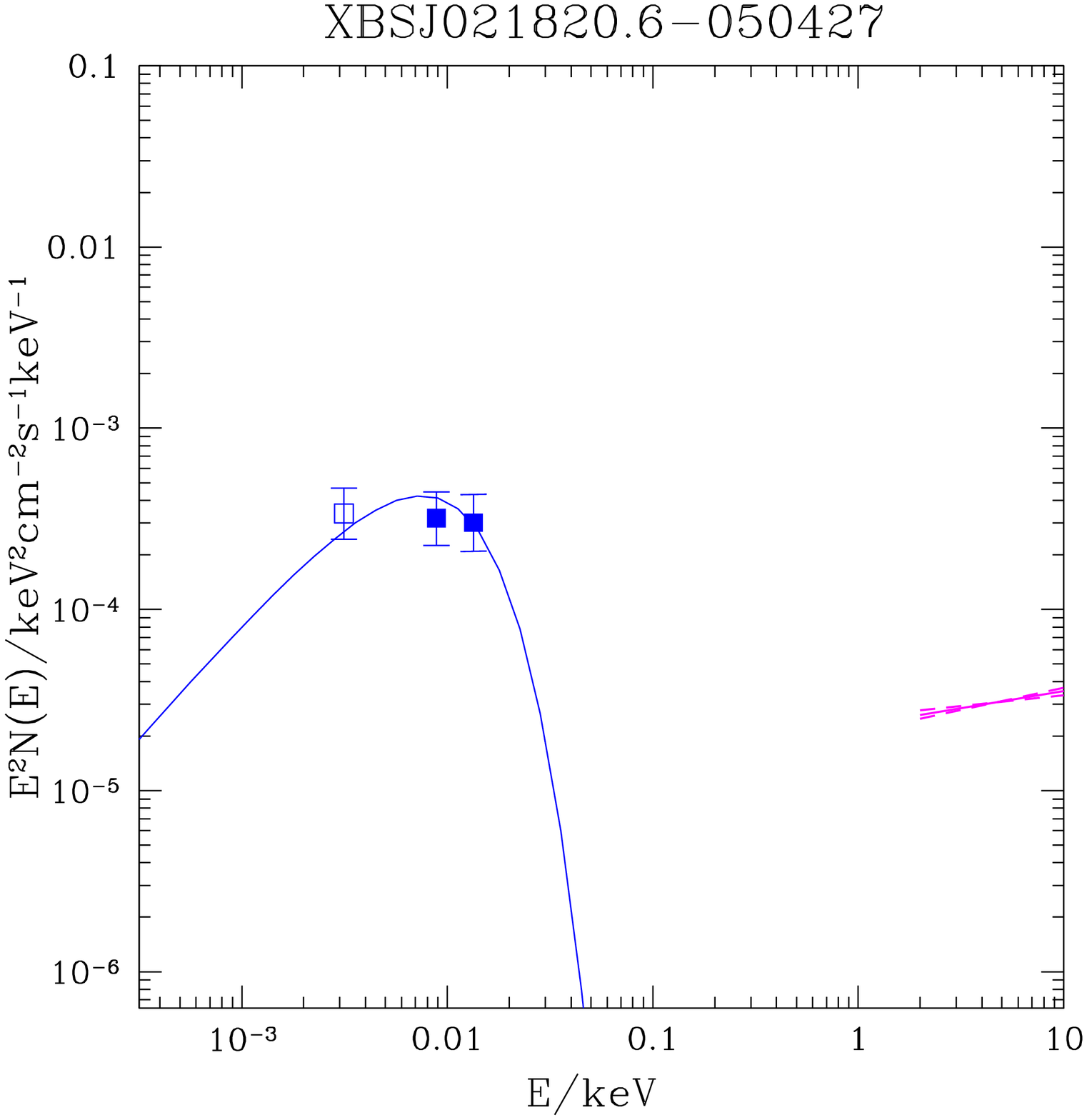}
  \includegraphics[height=5.6cm, width=6cm]{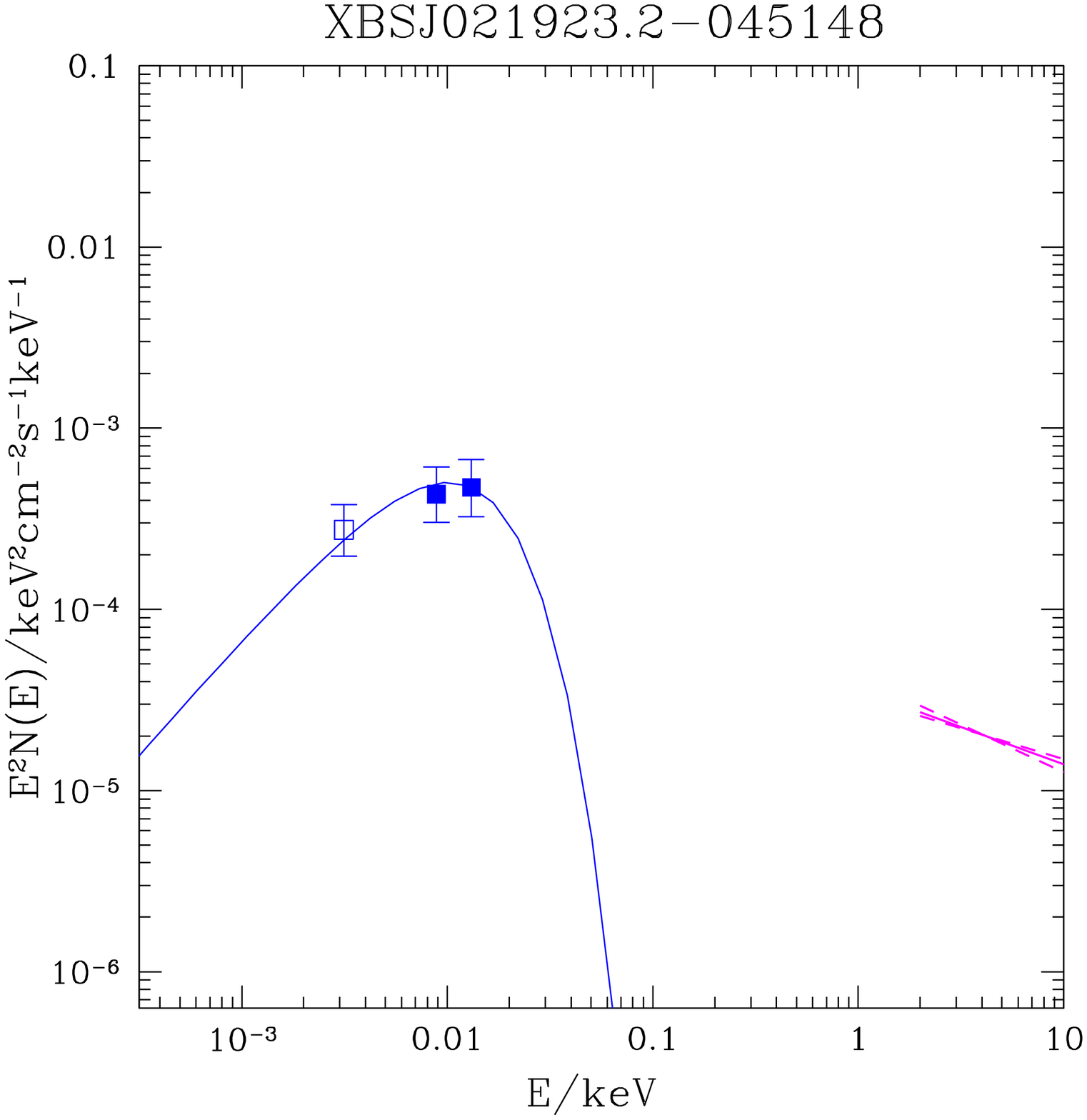}}     
 \subfigure{ 
  \includegraphics[height=5.6cm, width=6cm]{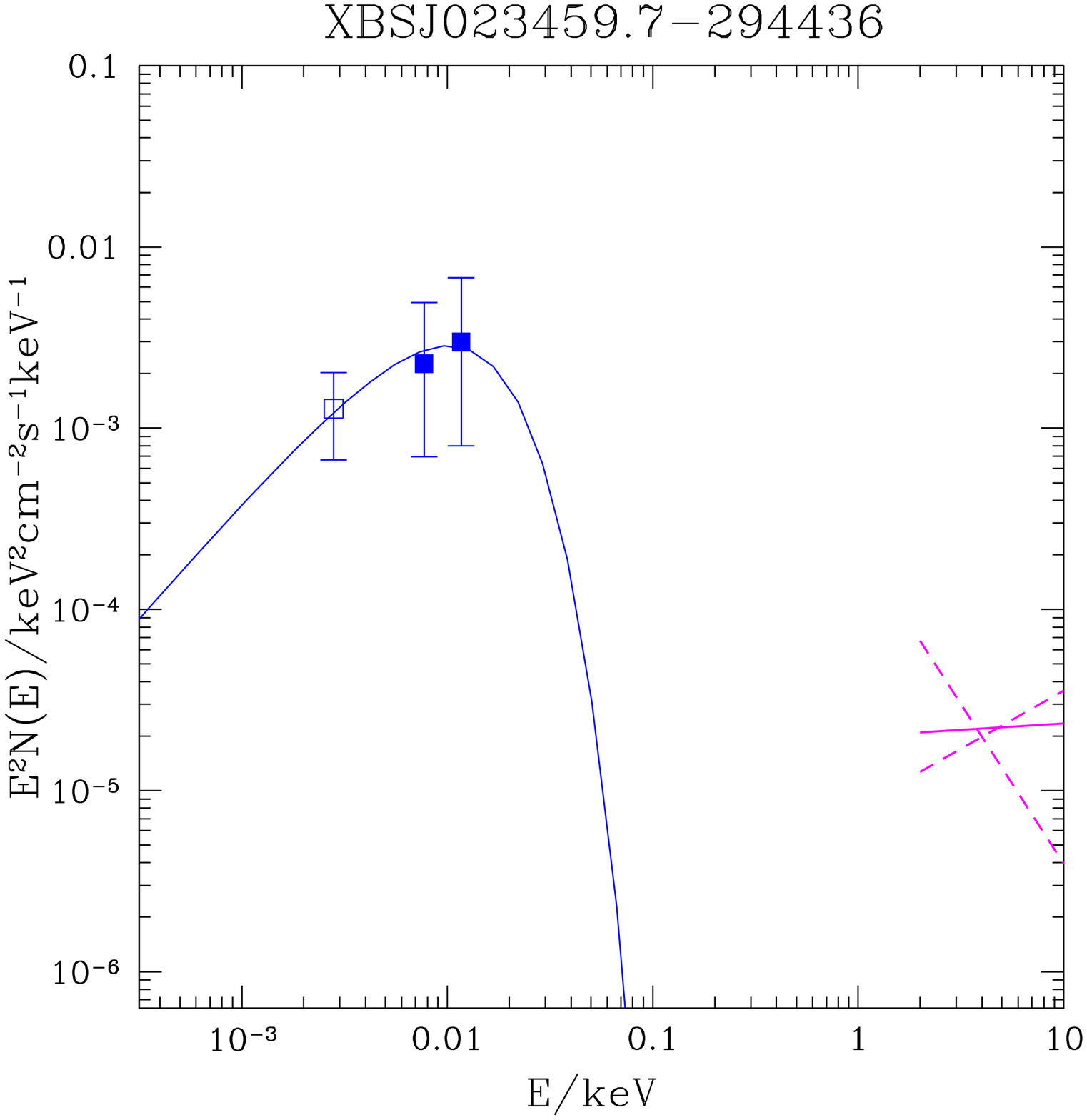}
  \includegraphics[height=5.6cm, width=6cm]{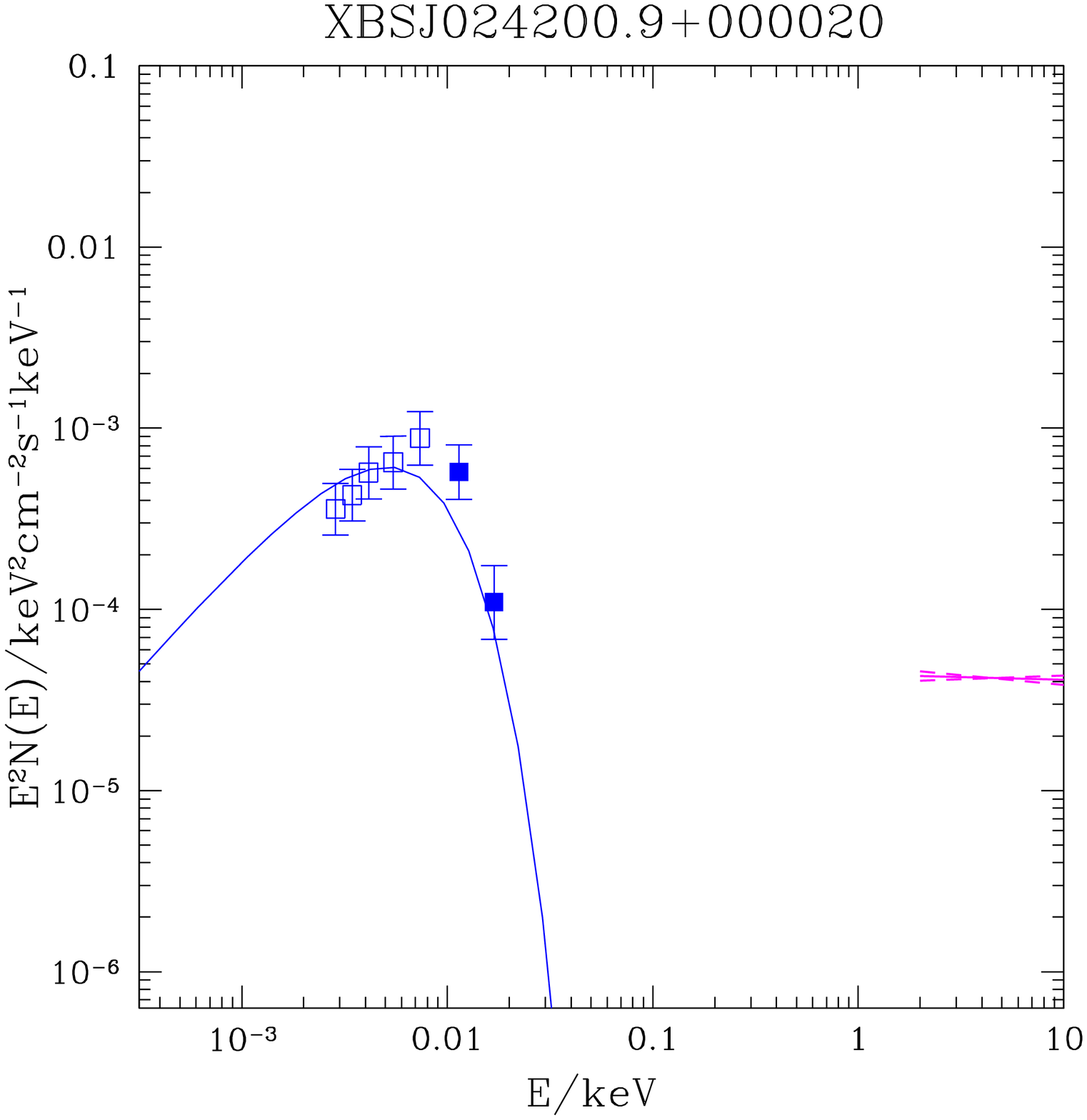}}   
  \subfigure{ 
  \includegraphics[height=5.6cm, width=6cm]{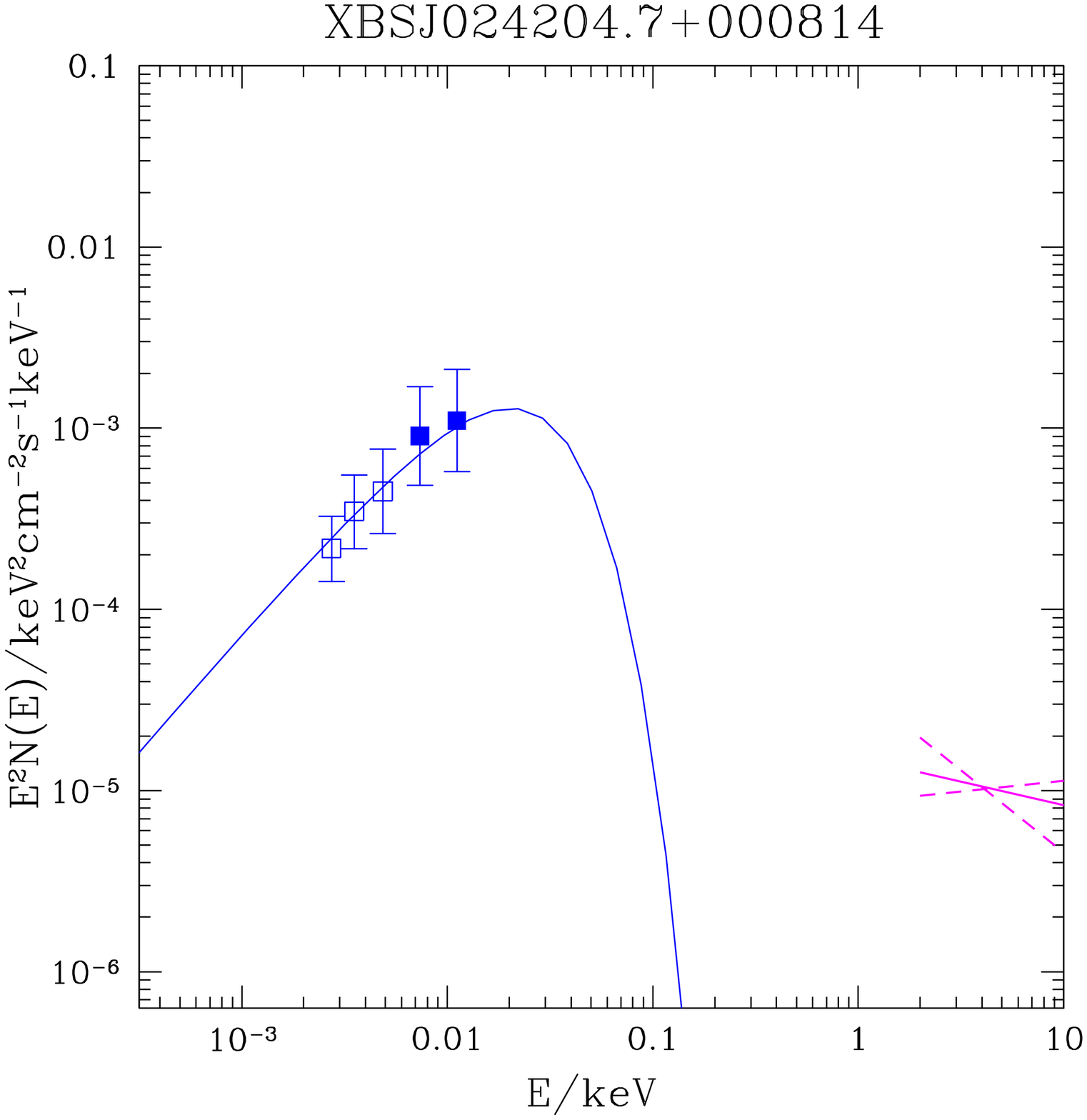}
  \includegraphics[height=5.6cm, width=6cm]{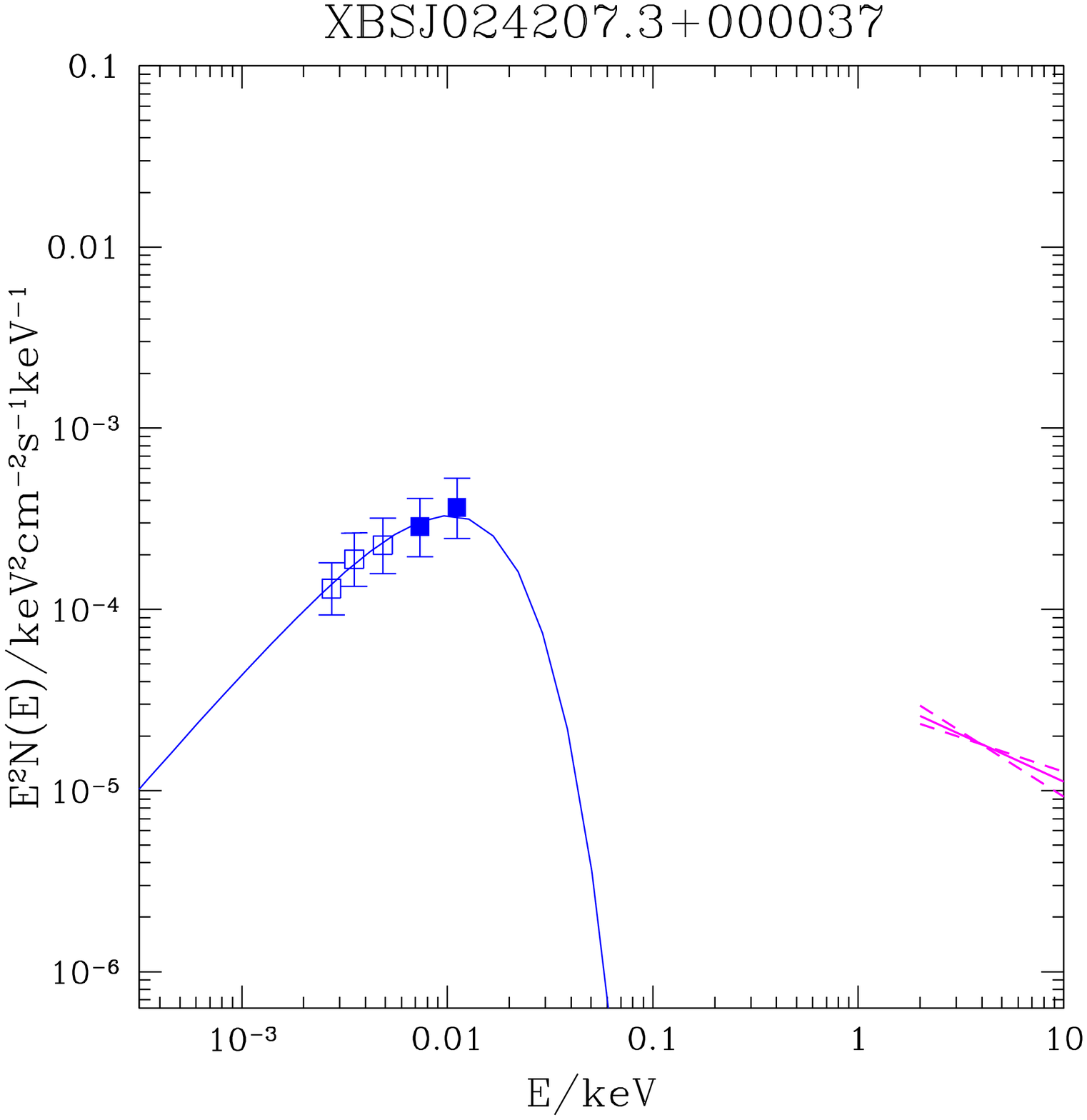}}    
  \subfigure{ 
  \includegraphics[height=5.6cm, width=6cm]{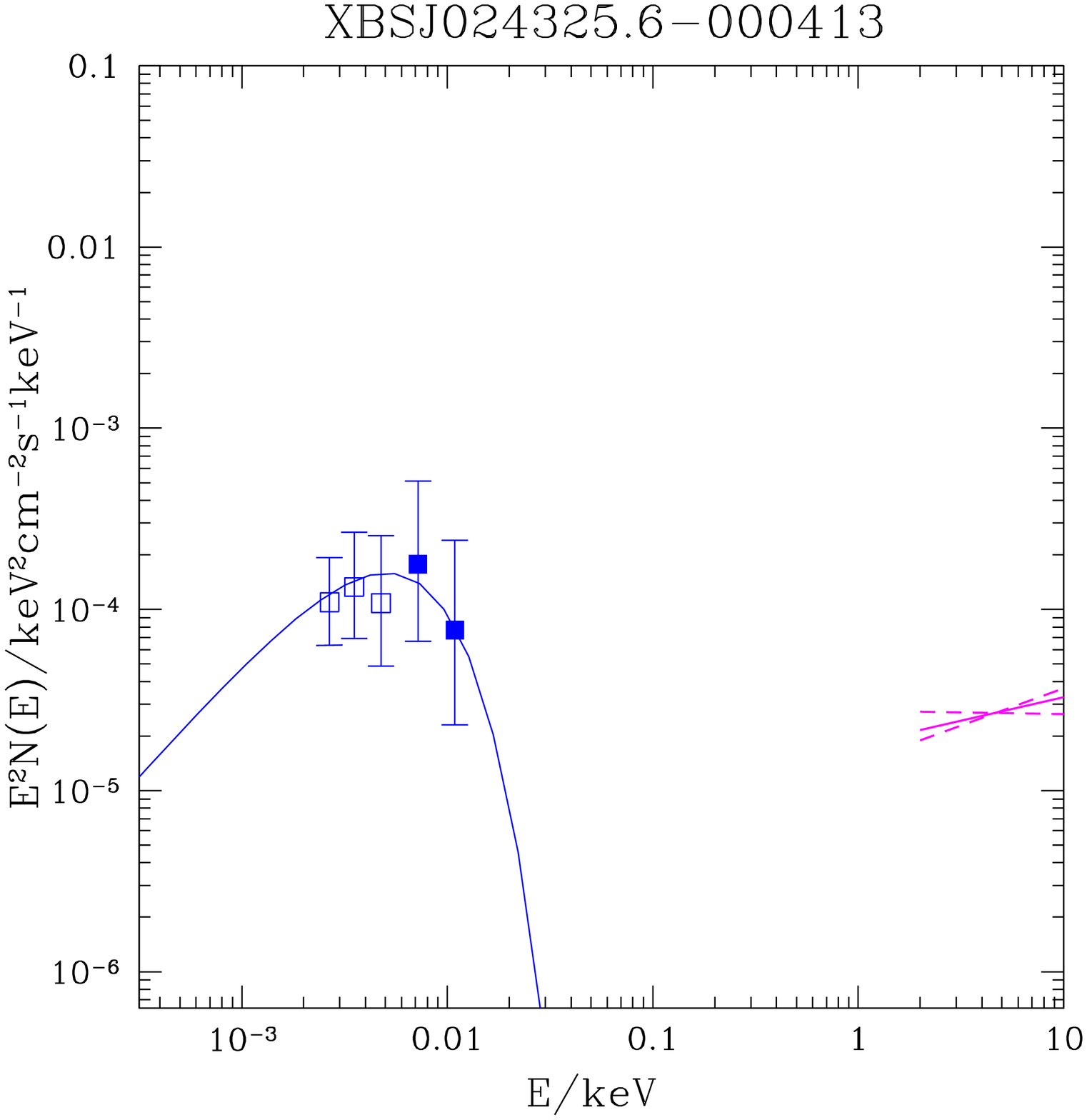}
  \includegraphics[height=5.6cm, width=6cm]{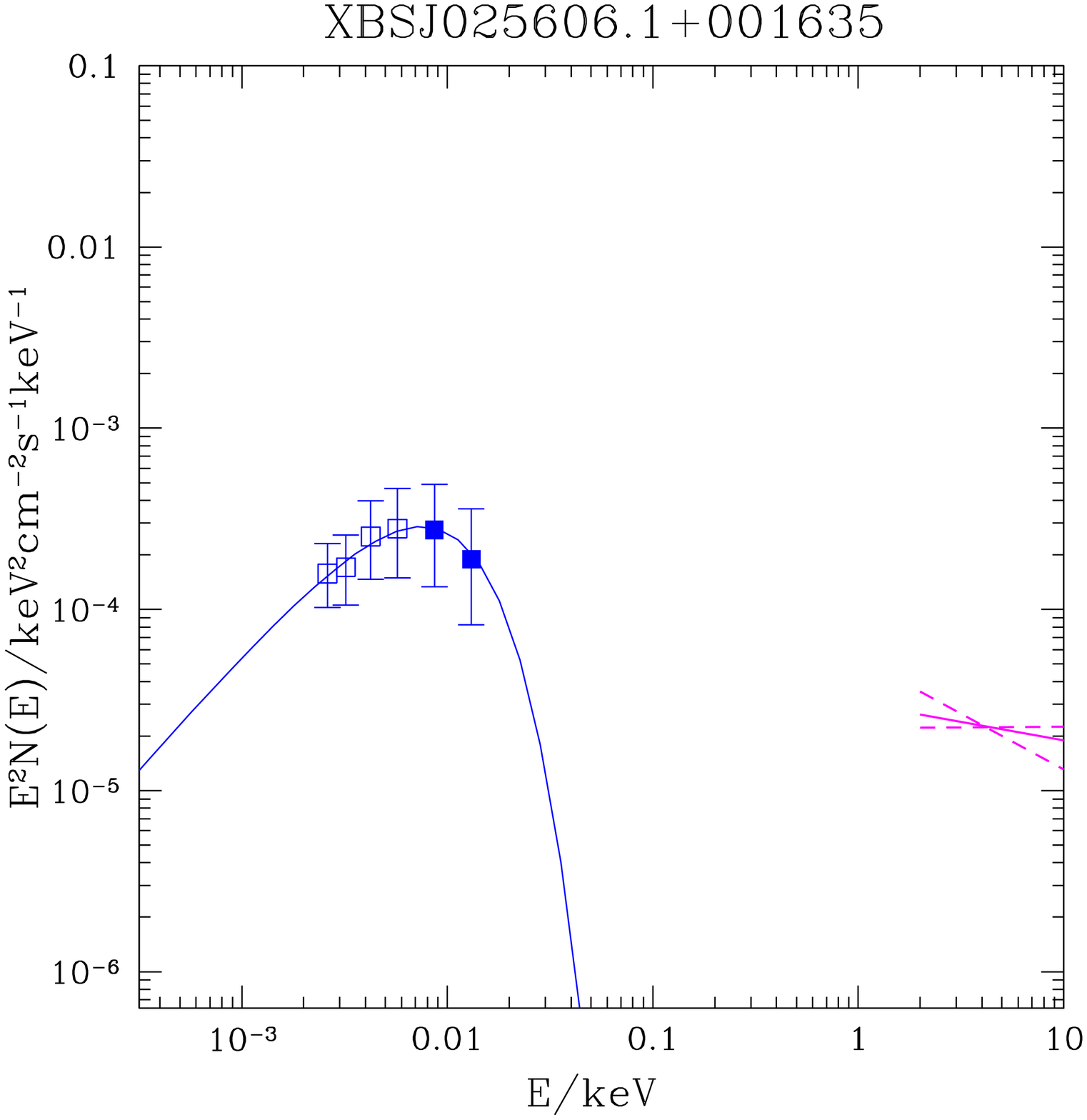}} 
  \end{figure*}
  
   \FloatBarrier 
     
    \begin{figure*}
\centering
  \subfigure{ 
  \includegraphics[height=5.6cm, width=6cm]{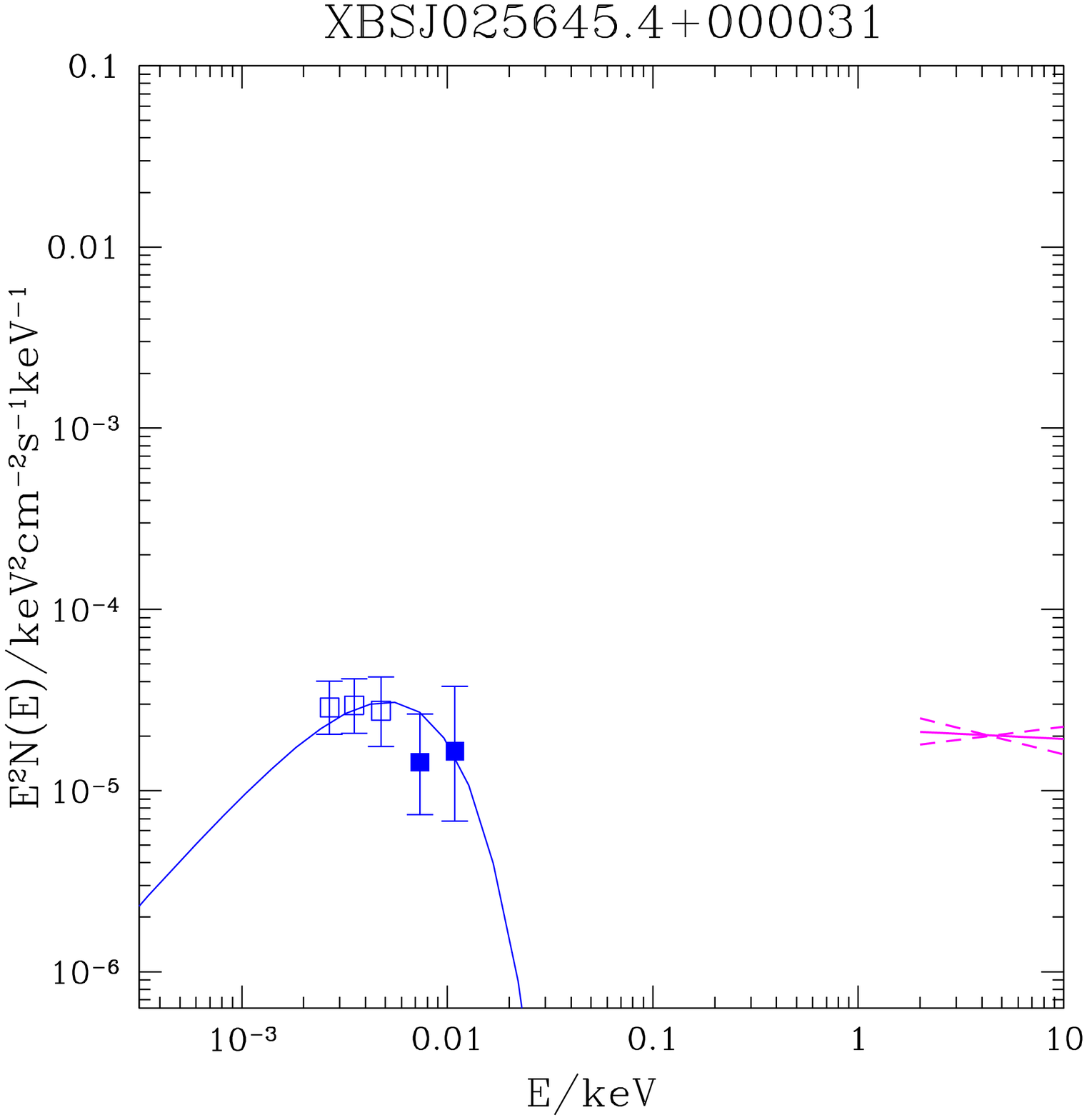}
  \includegraphics[height=5.6cm, width=6cm]{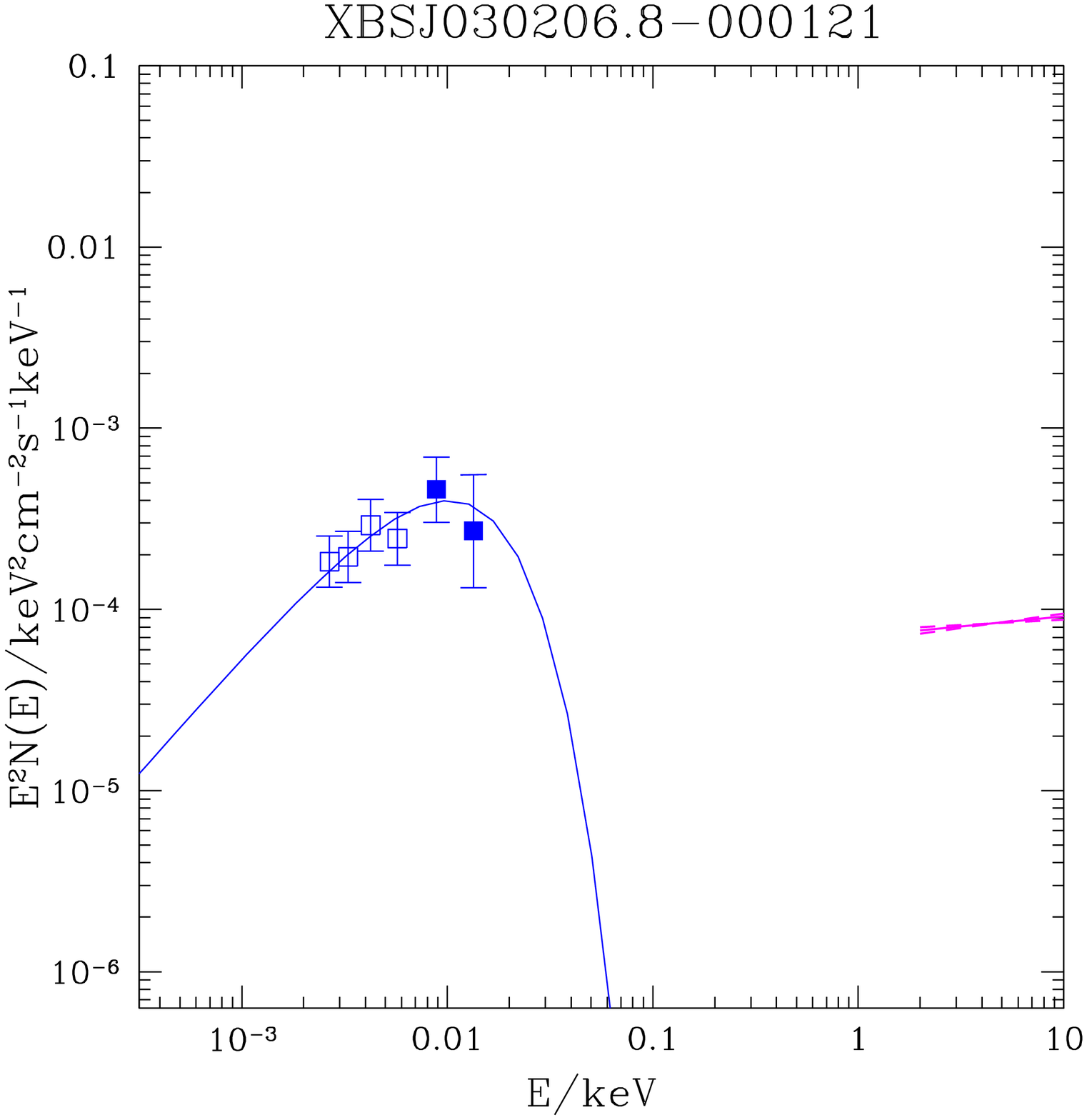}}   
   \subfigure{ 
  \includegraphics[height=5.6cm, width=6cm]{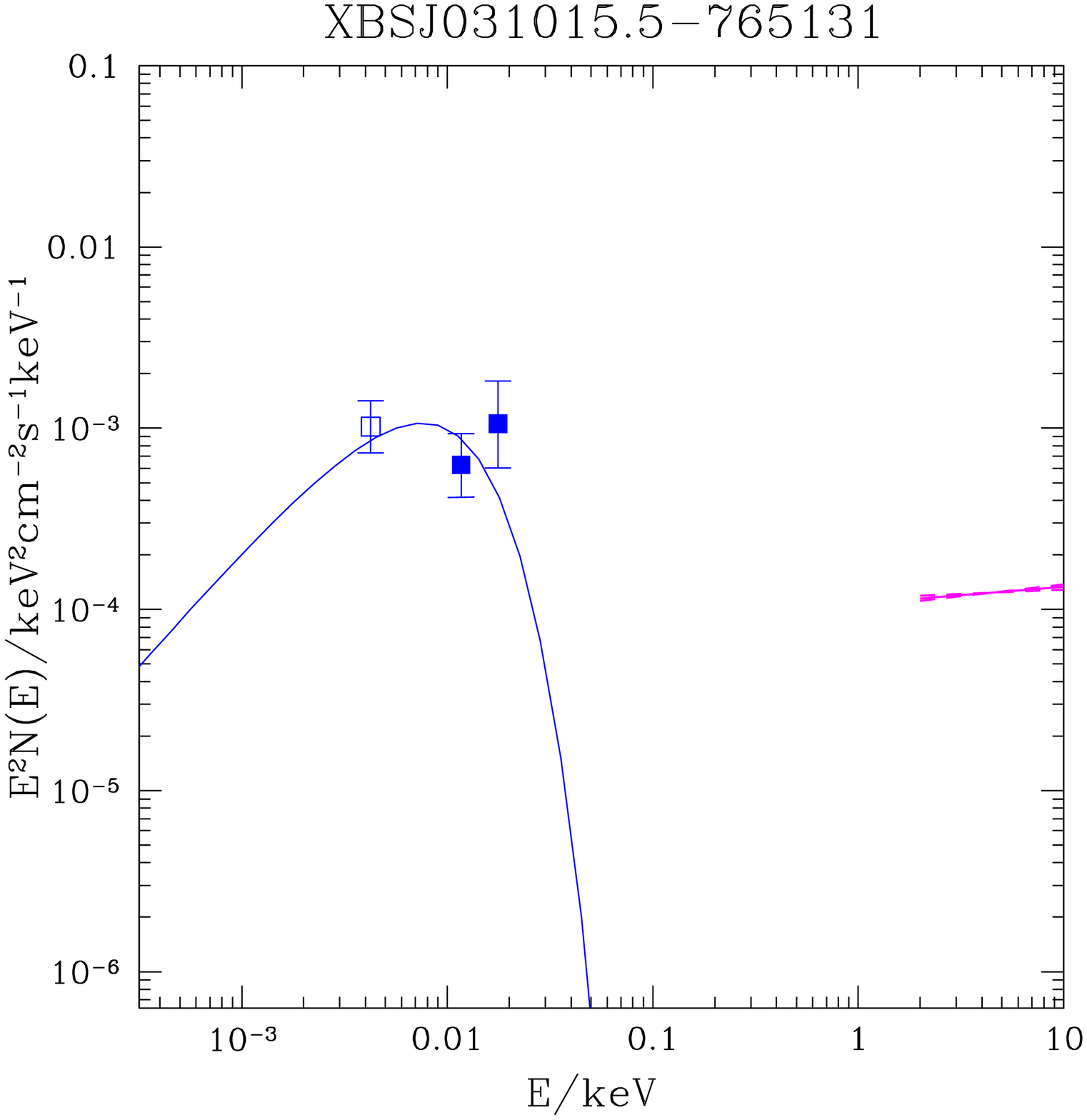}
  \includegraphics[height=5.6cm, width=6cm]{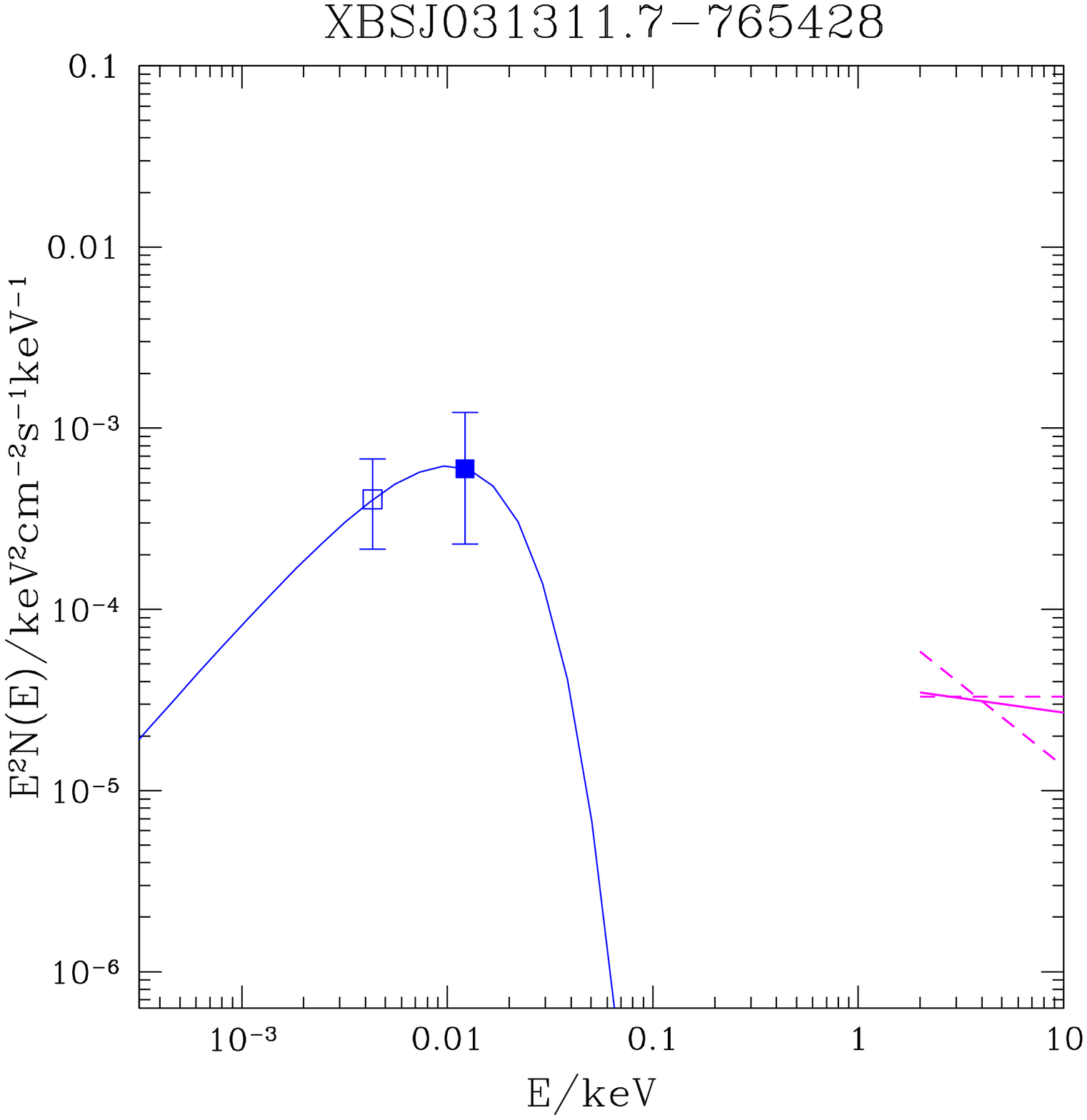}}  
  \subfigure{ 
  \includegraphics[height=5.6cm, width=6cm]{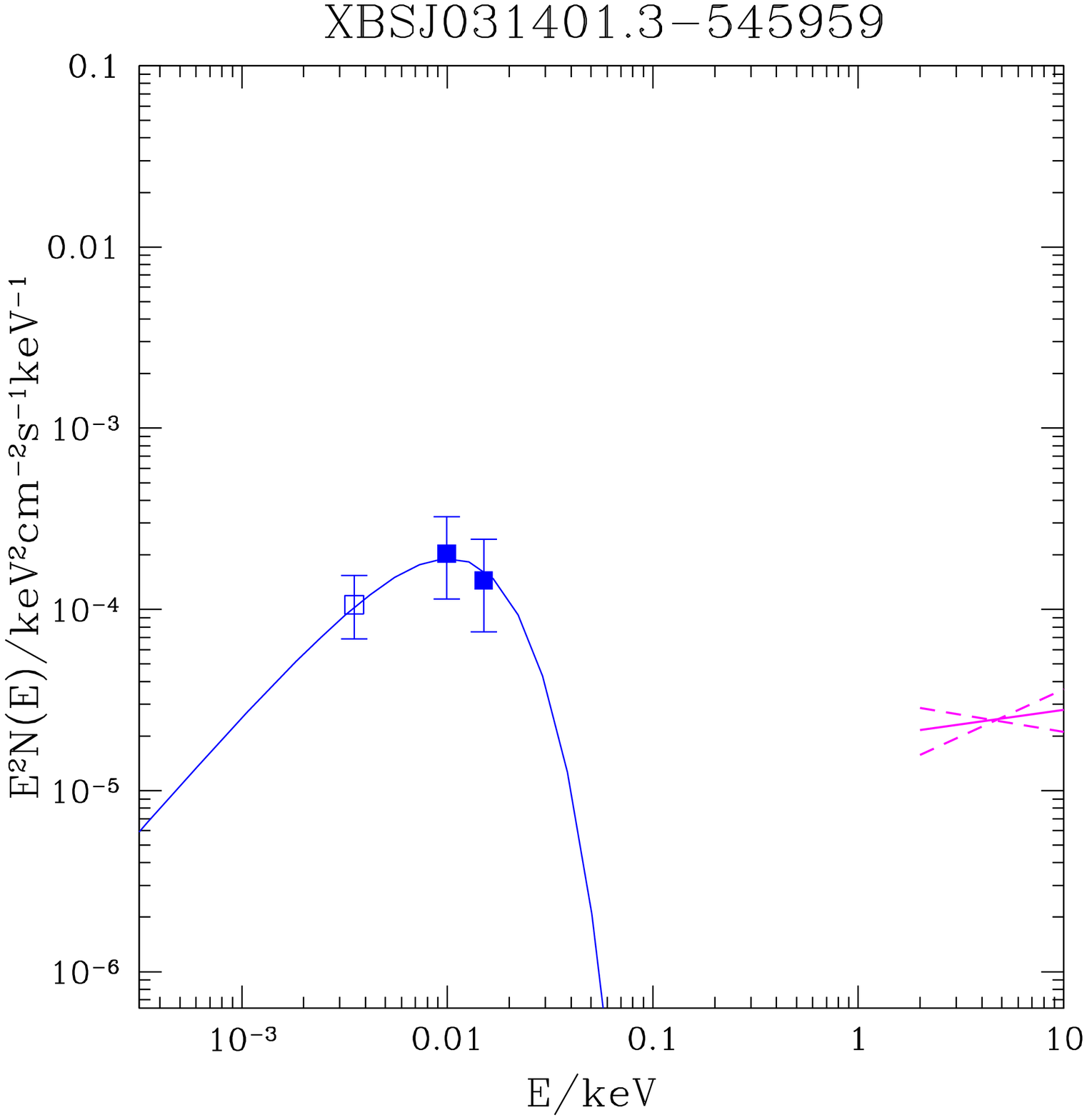}
  \includegraphics[height=5.6cm, width=6cm]{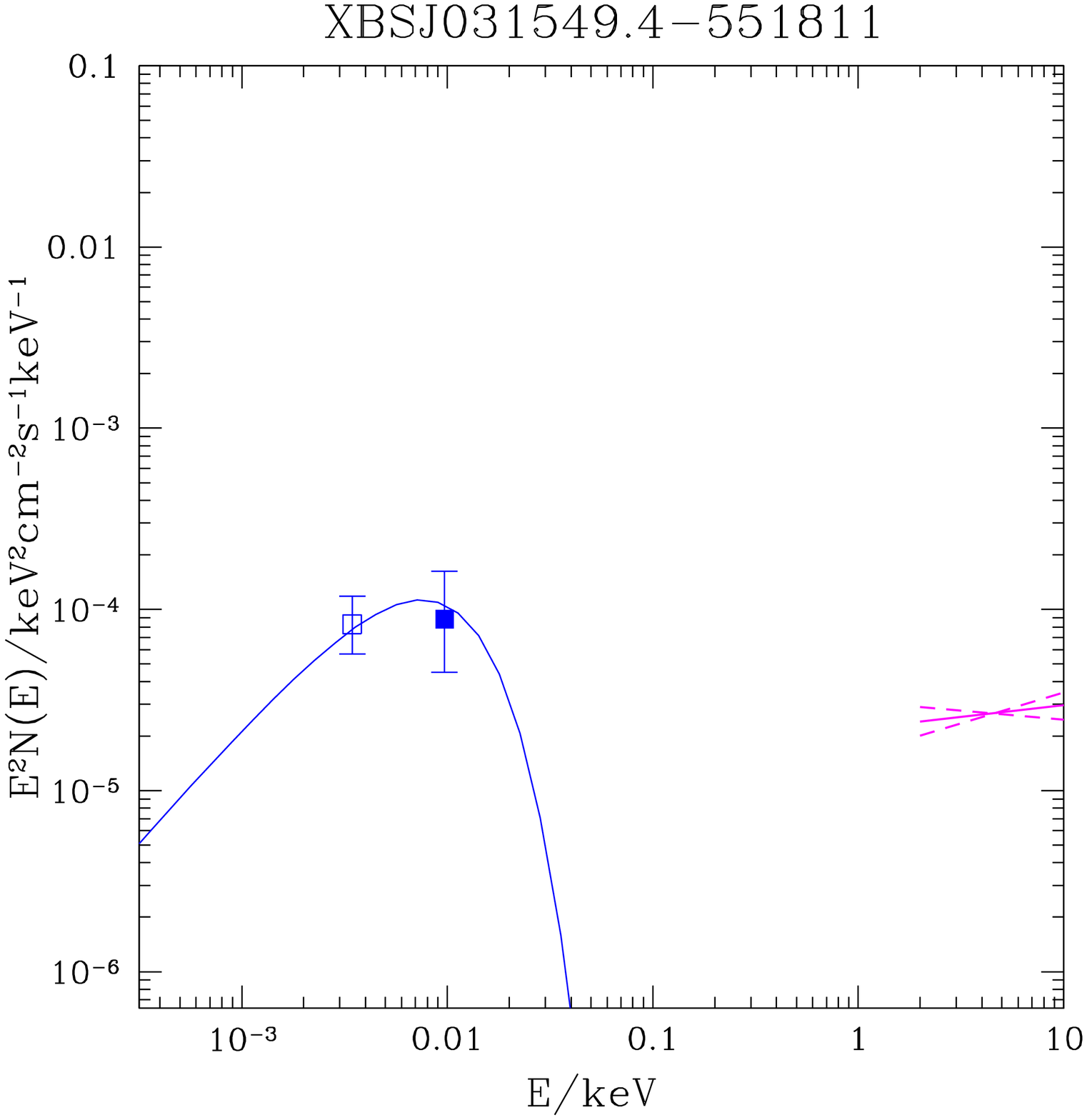}}  
    \subfigure{ 
  \includegraphics[height=5.6cm, width=6cm]{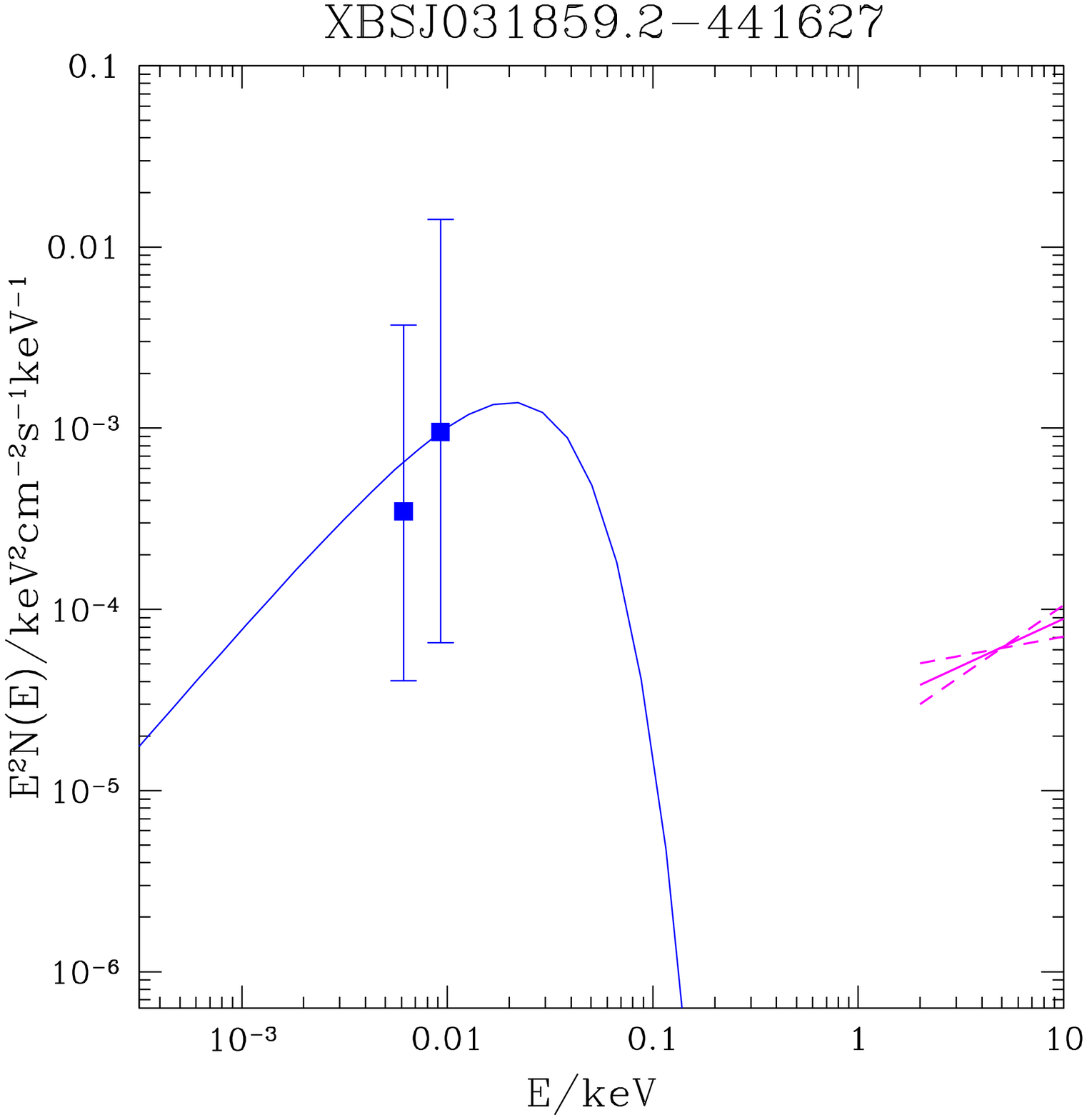}
  \includegraphics[height=5.6cm, width=6cm]{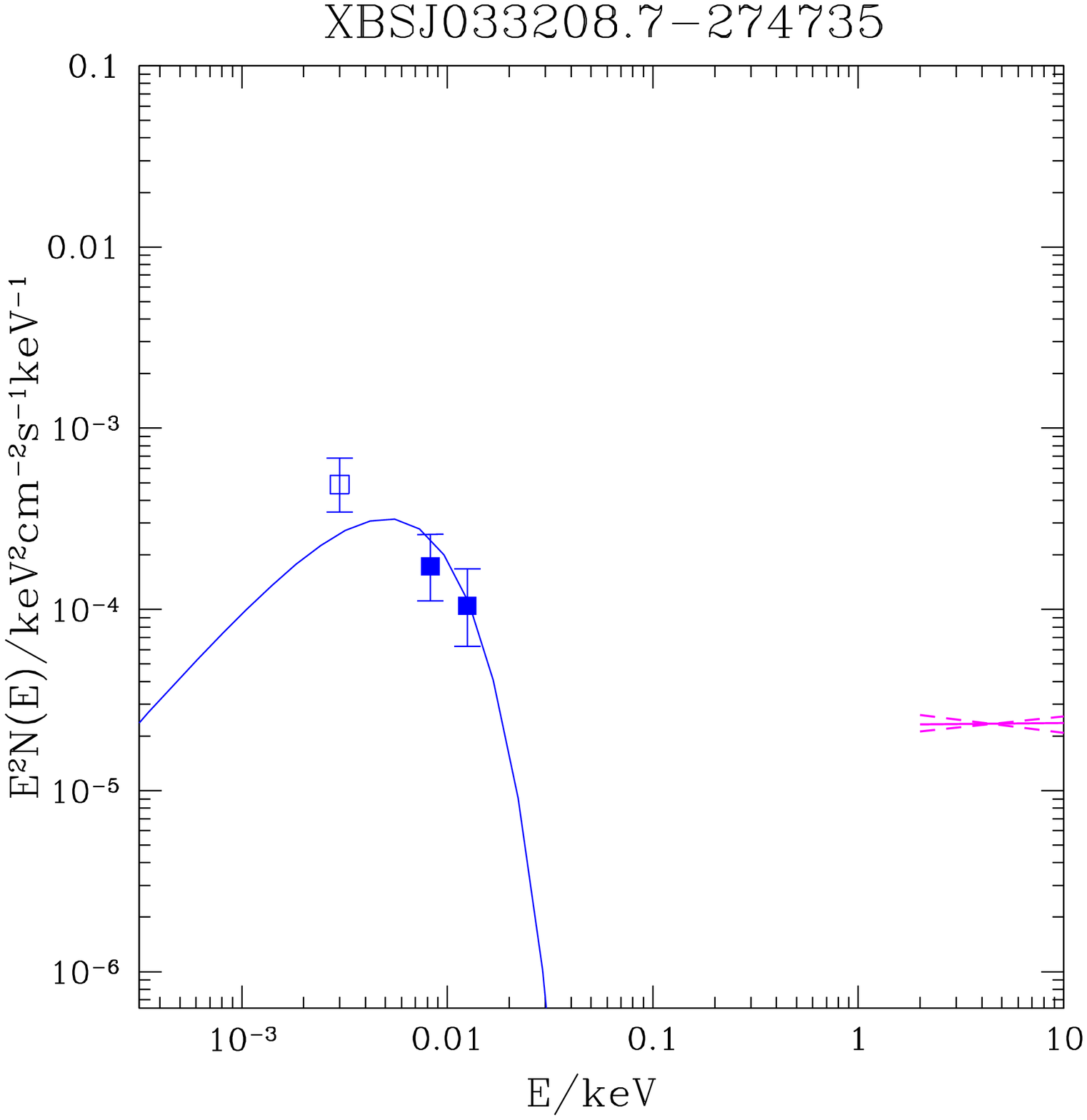}}   
  \end{figure*}
  
  \FloatBarrier
  
   \begin{figure*}
\centering
   \subfigure{ 
  \includegraphics[height=5.6cm, width=6cm]{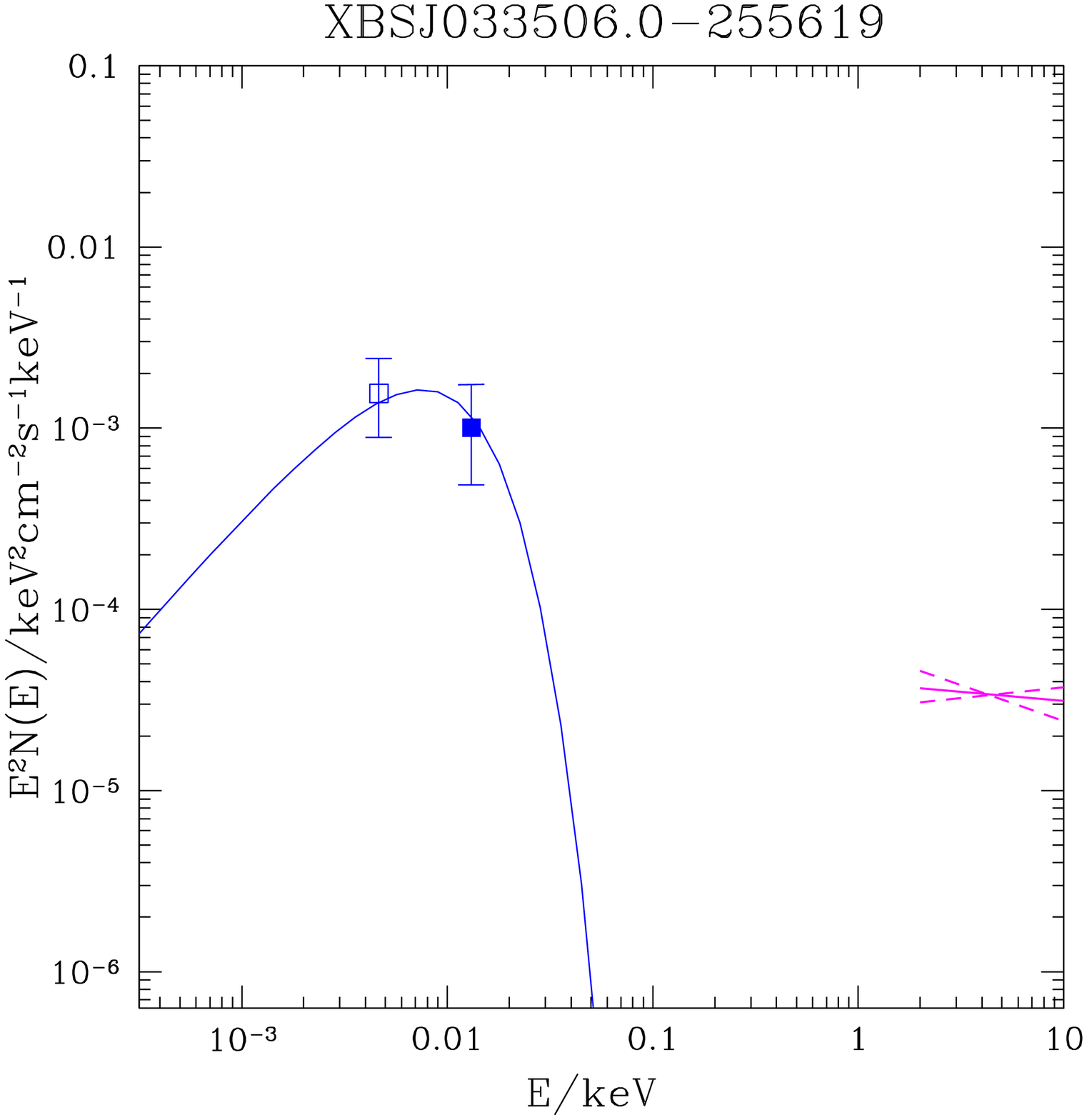}
  \includegraphics[height=5.6cm, width=6cm]{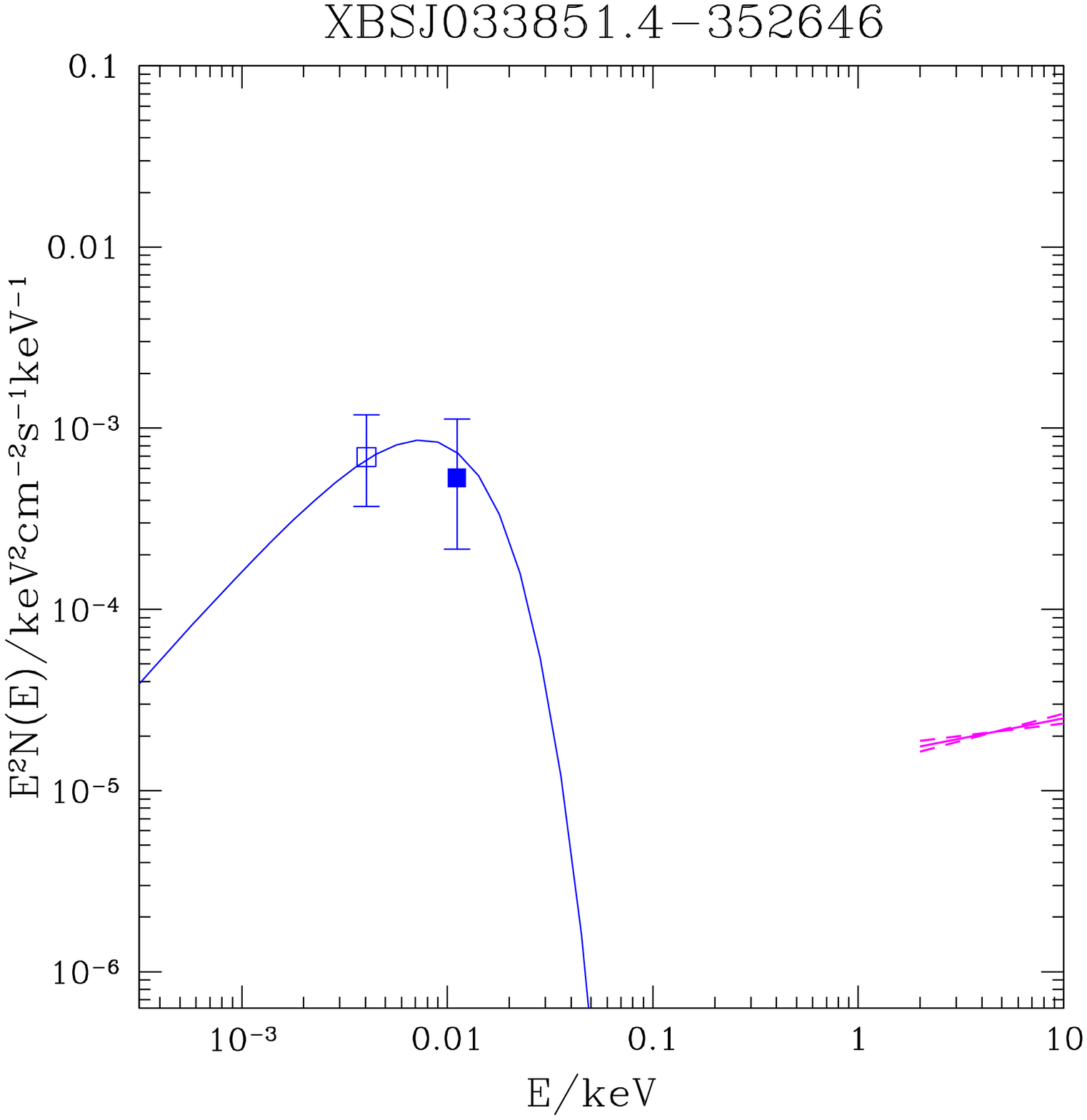}}    
  \subfigure{ 
  \includegraphics[height=5.6cm, width=6cm]{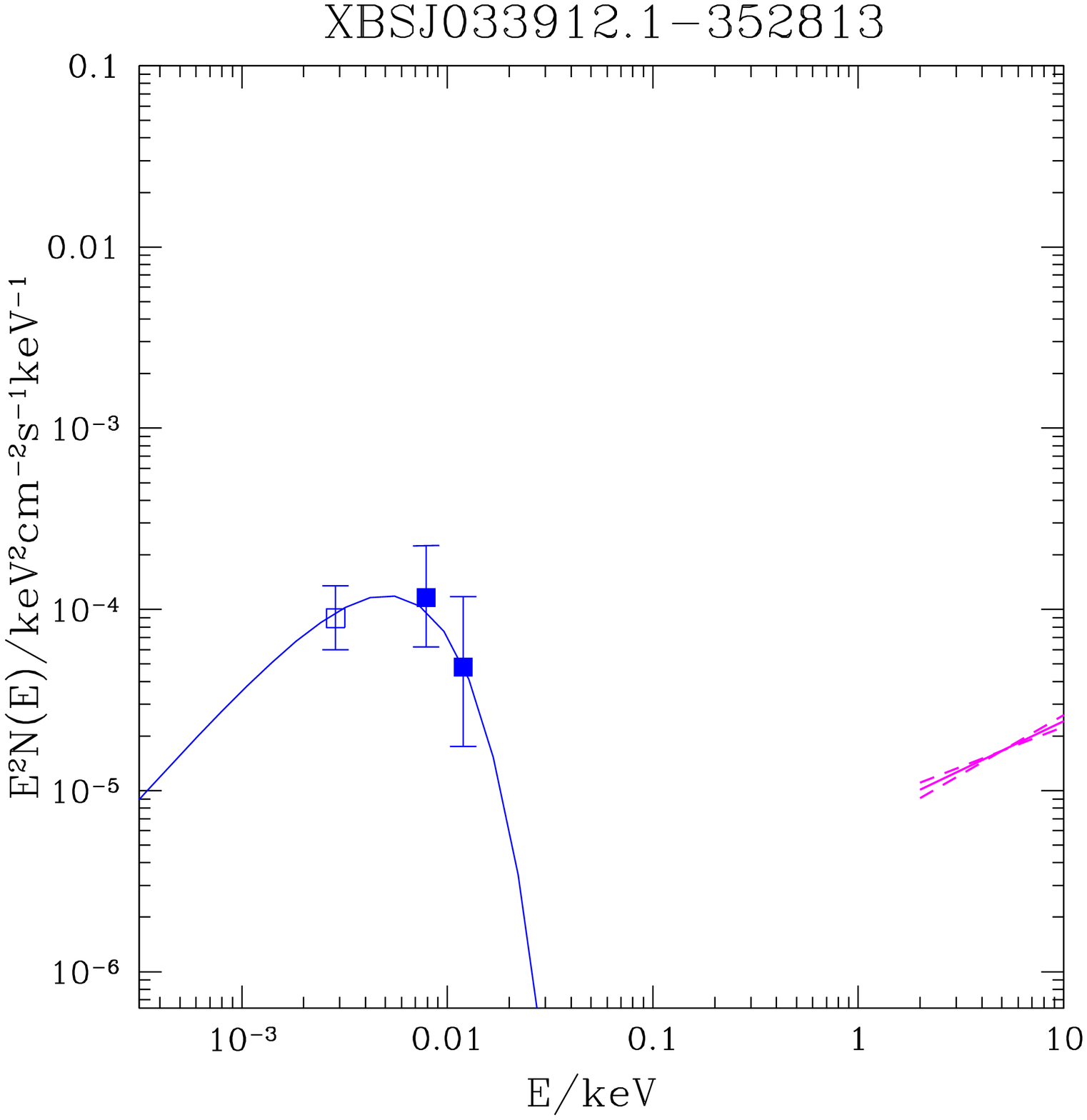}
  \includegraphics[height=5.6cm, width=6cm]{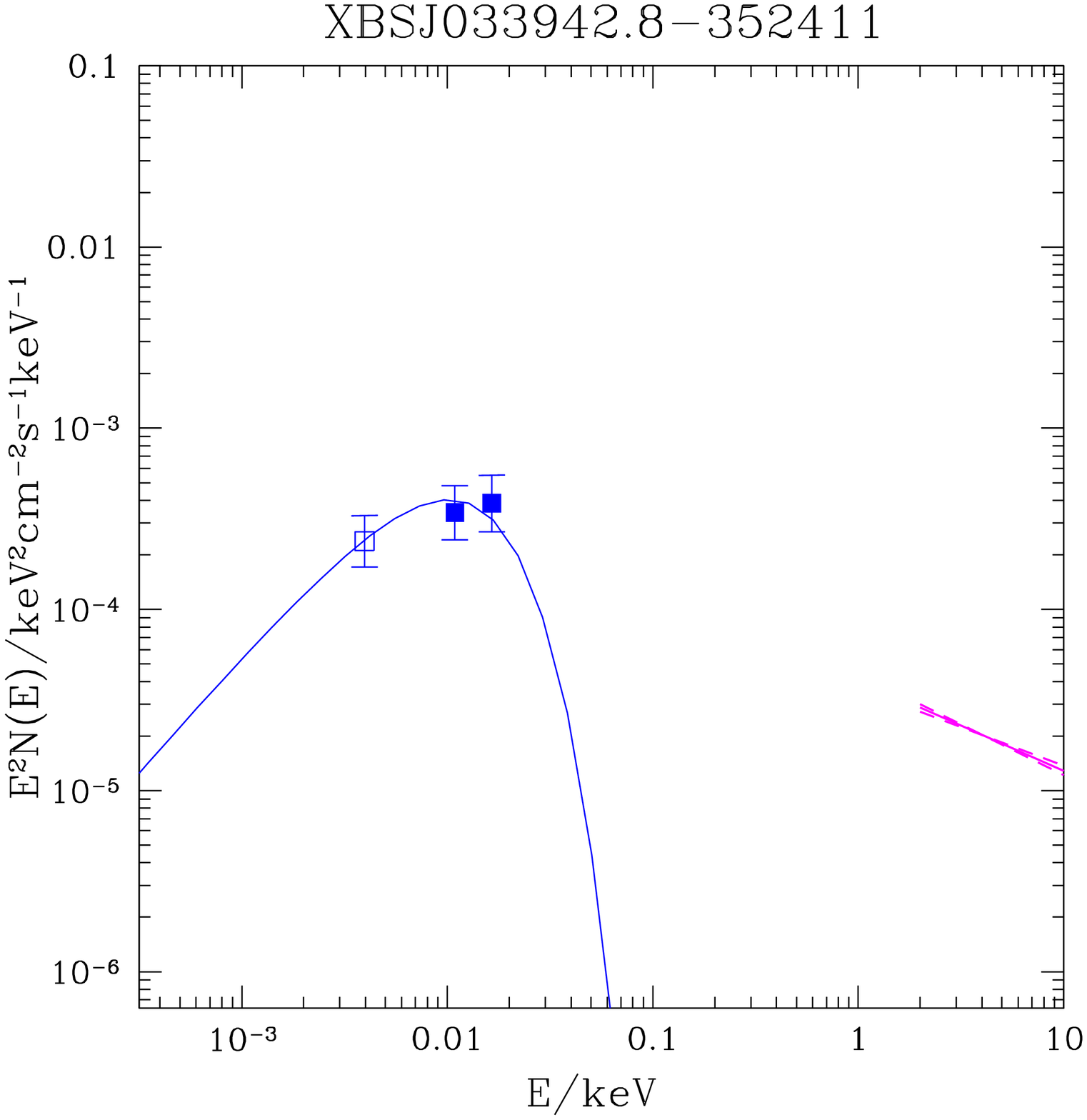}}  
  \subfigure{ 
  \includegraphics[height=5.6cm, width=6cm]{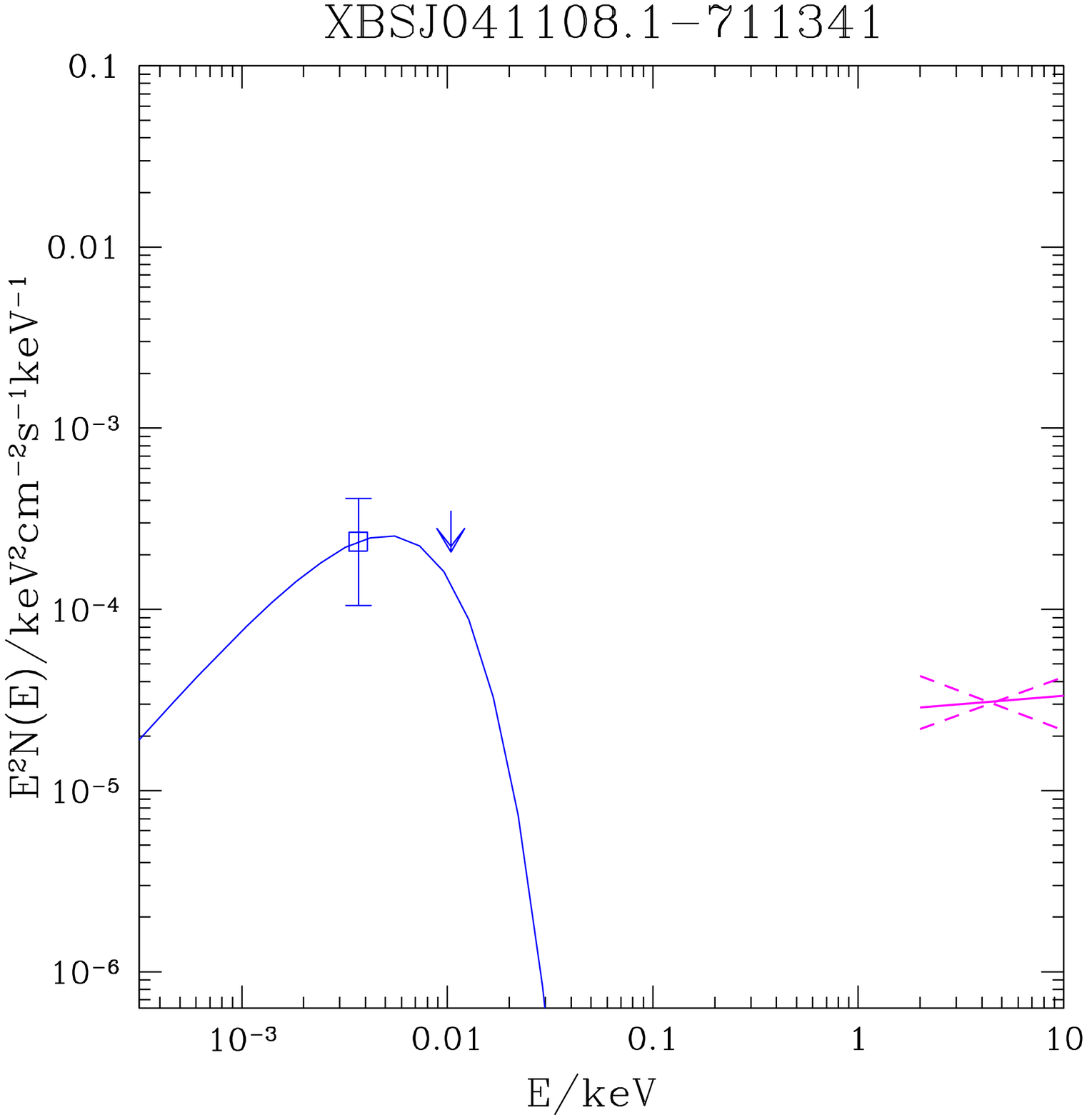}
  \includegraphics[height=5.6cm, width=6cm]{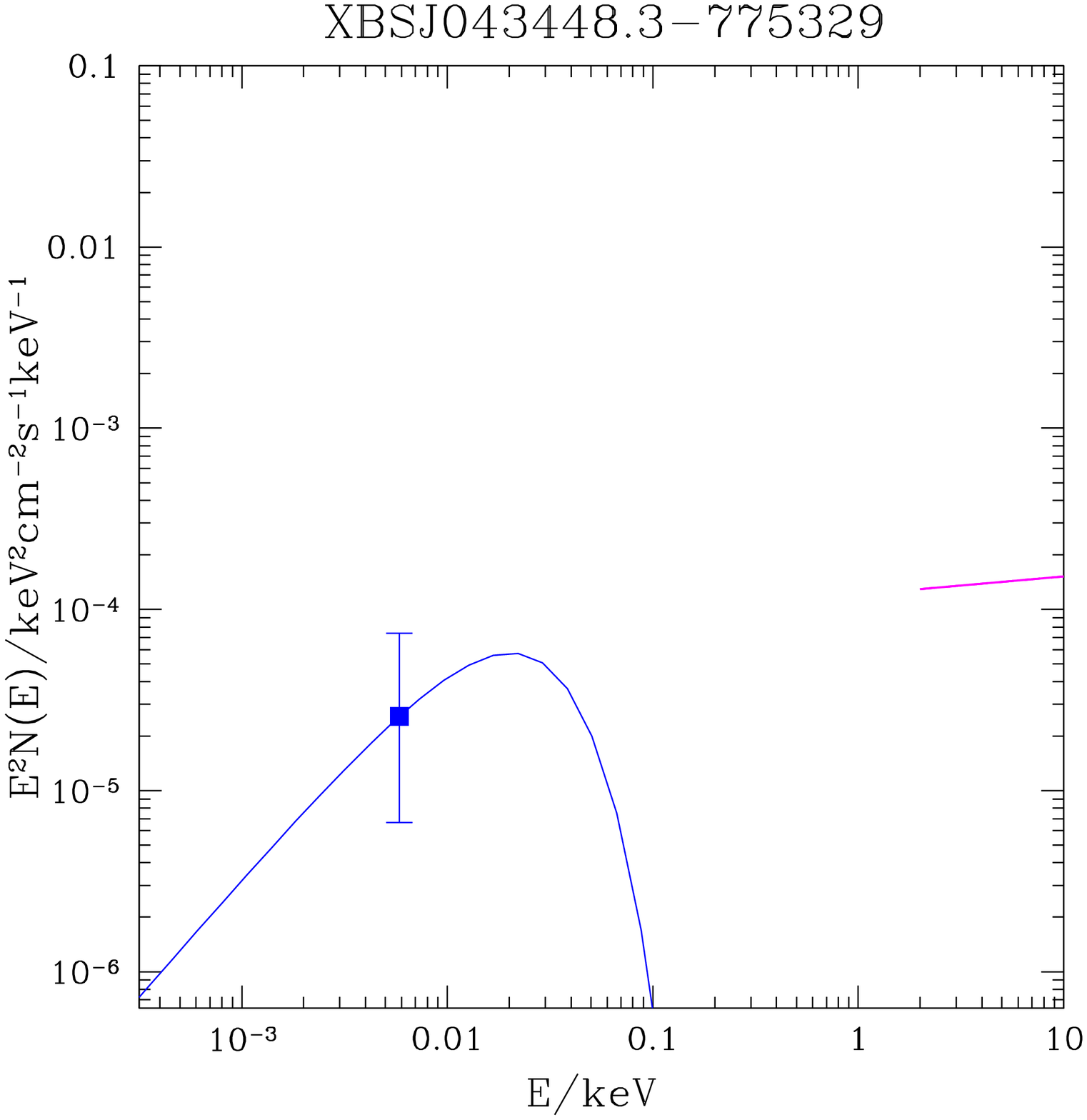}}      
\subfigure{ 
  \includegraphics[height=5.6cm, width=6cm]{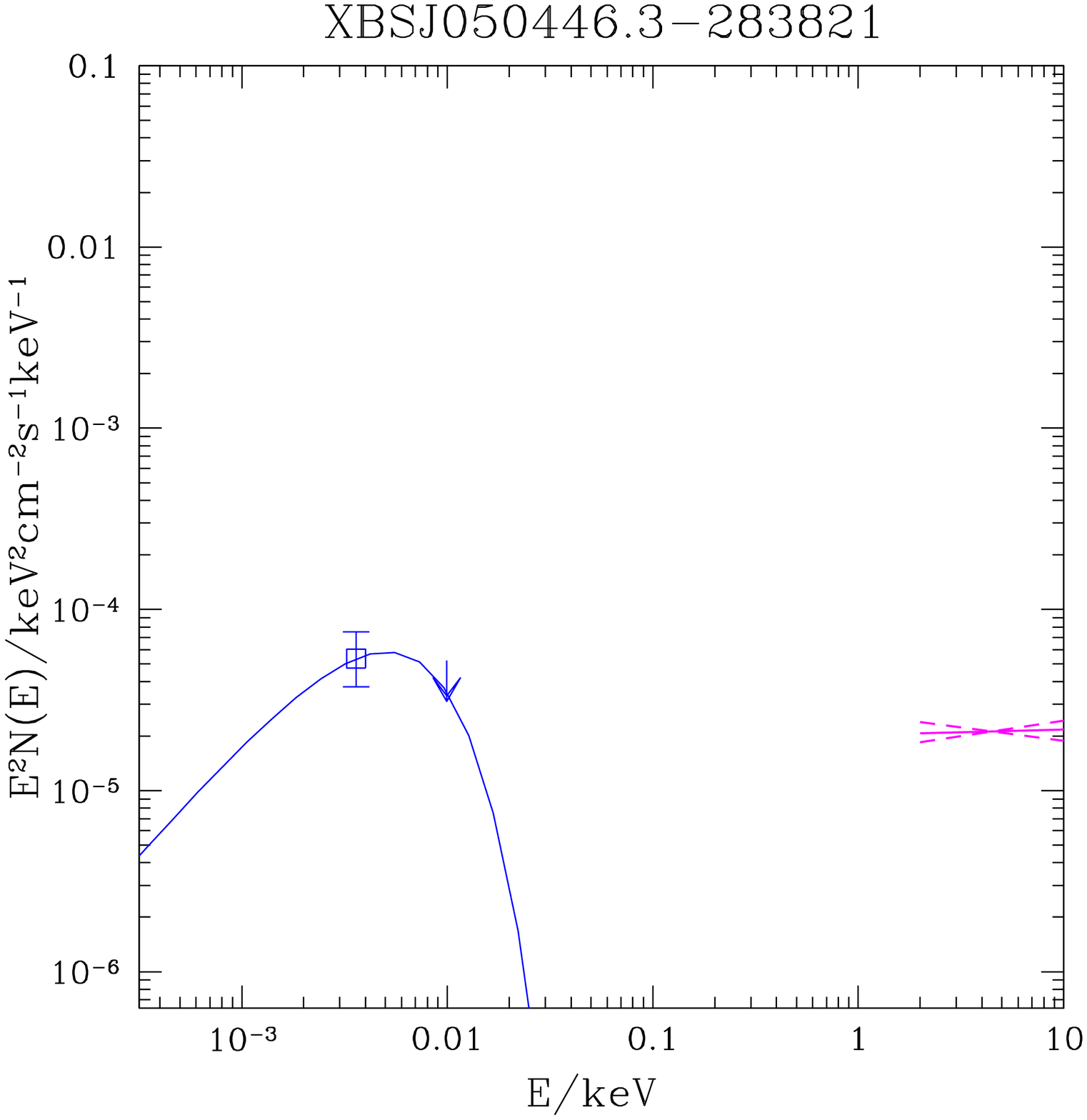}
  \includegraphics[height=5.6cm, width=6cm]{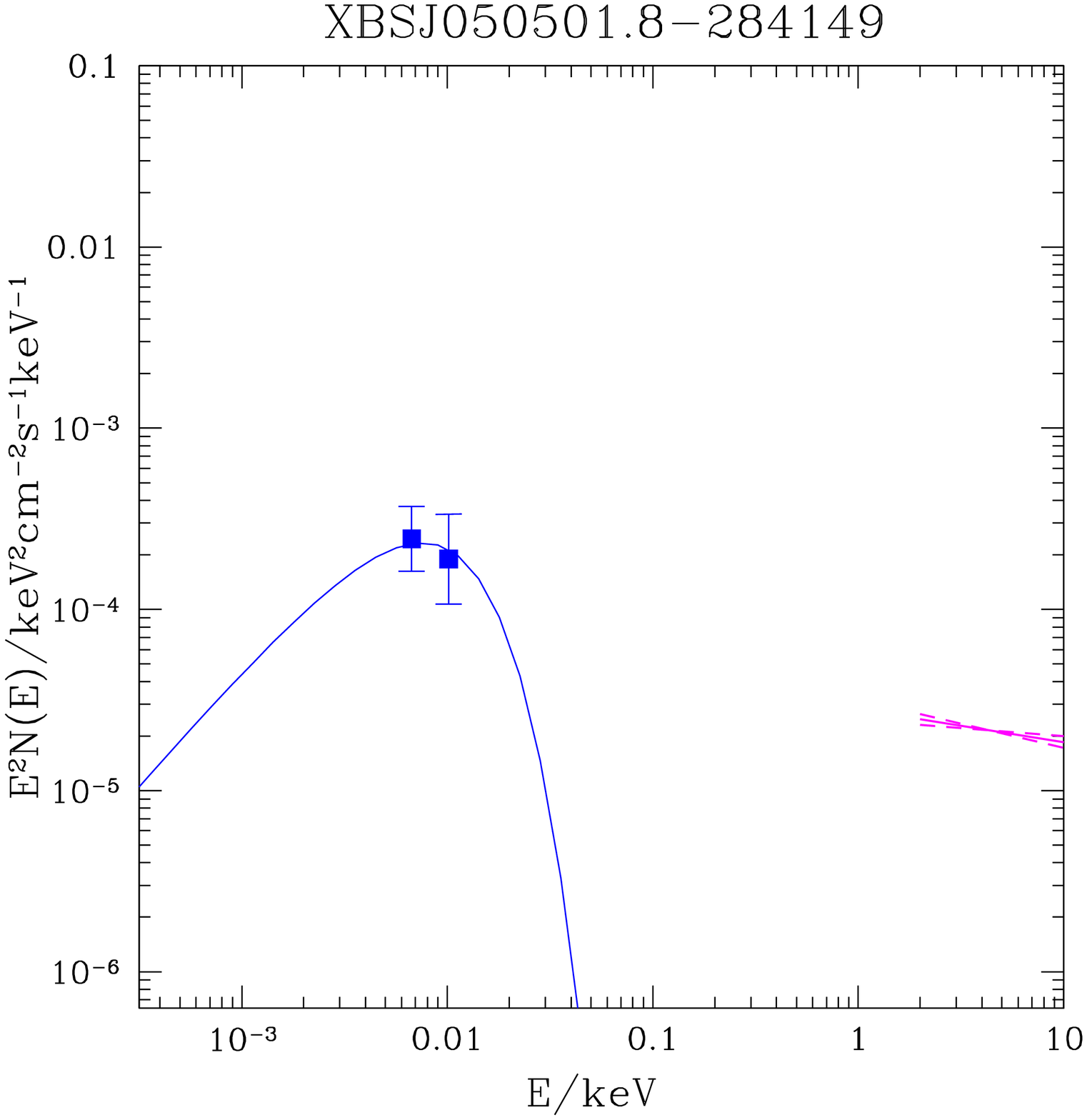}} 
  \end{figure*}
  
   \FloatBarrier
  
   \begin{figure*}
\centering     
\subfigure{ 
  \includegraphics[height=5.6cm, width=6cm]{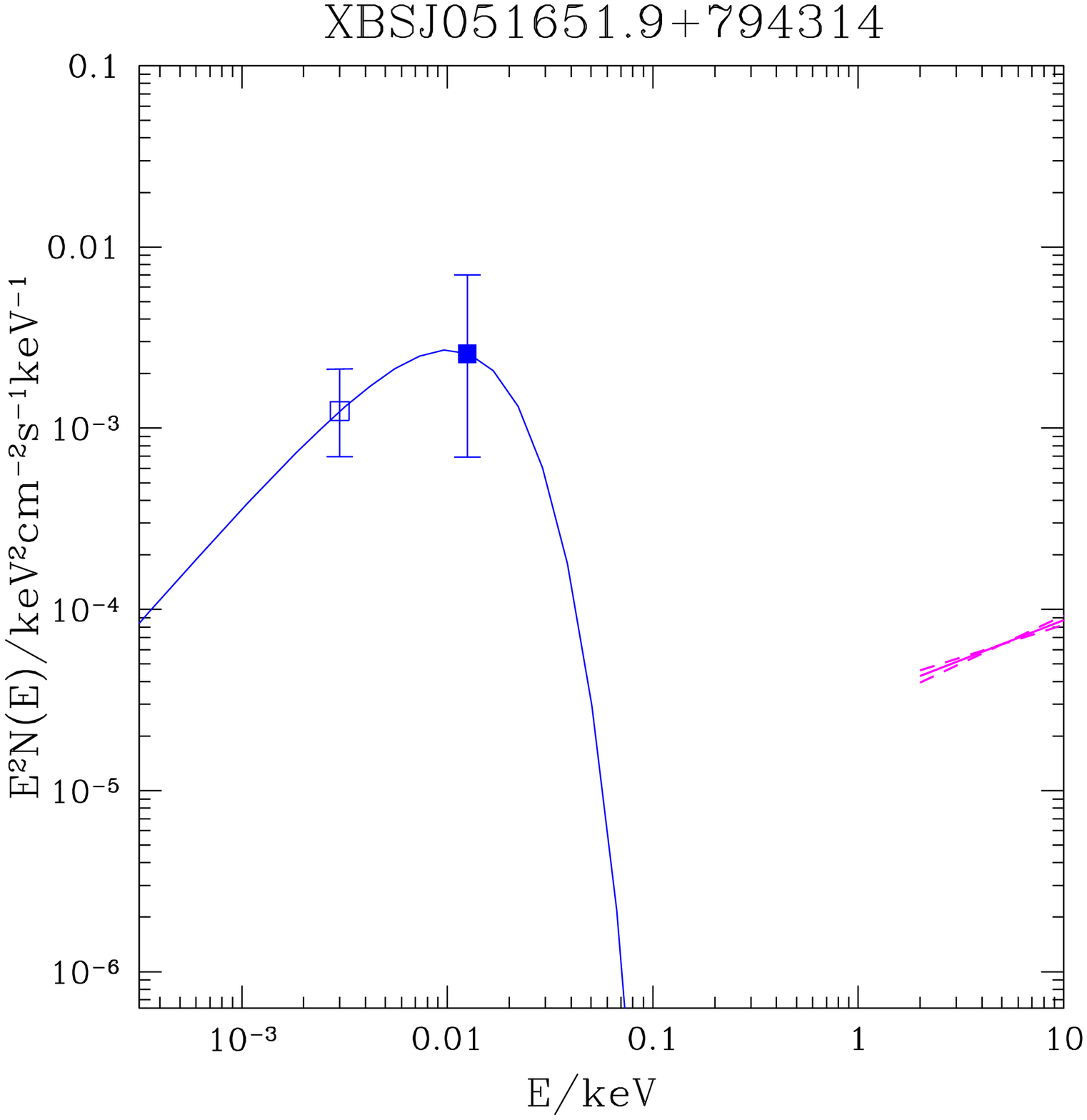}
  \includegraphics[height=5.6cm, width=6cm]{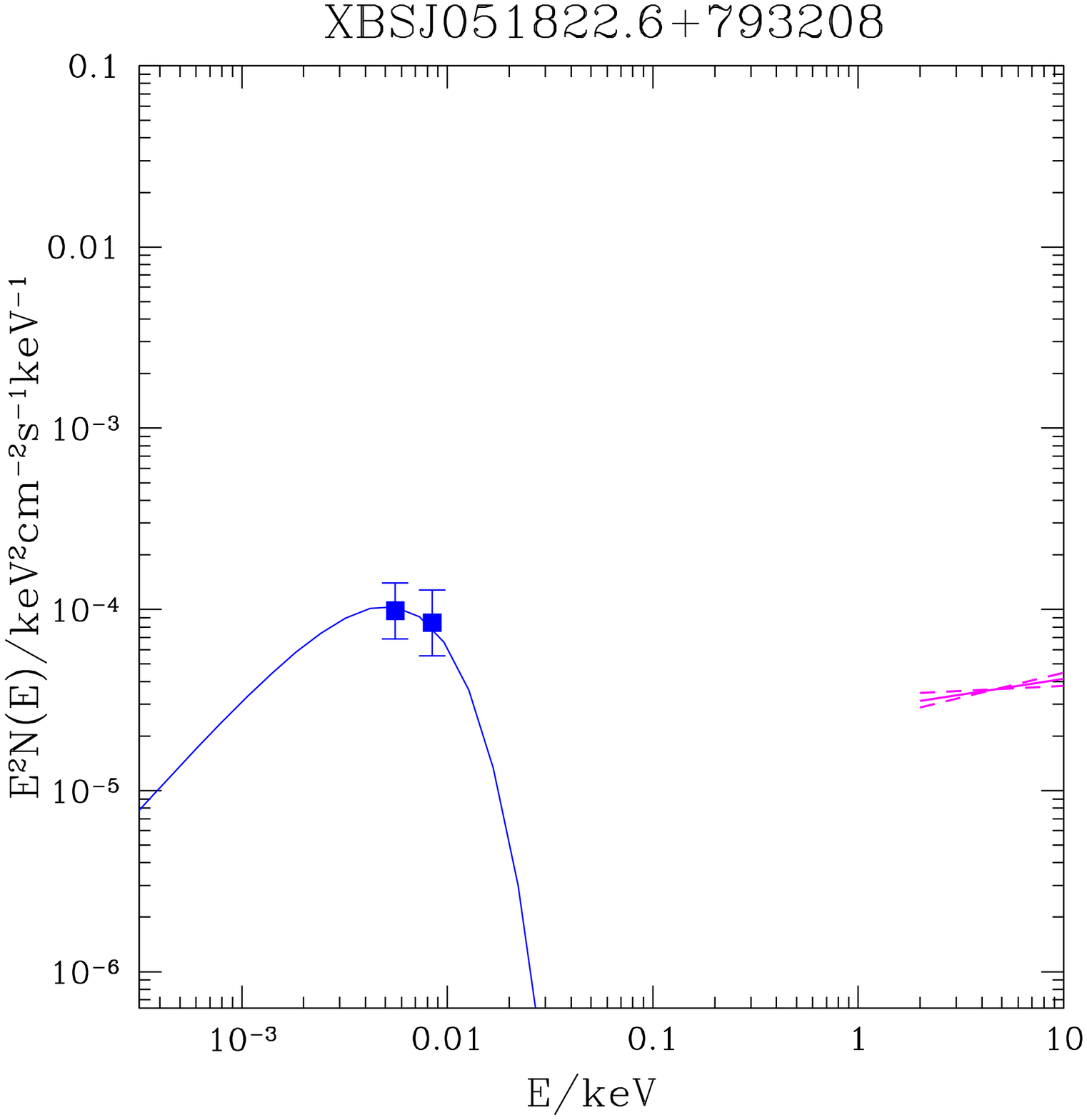}}     
 \subfigure{ 
  \includegraphics[height=5.6cm, width=6cm]{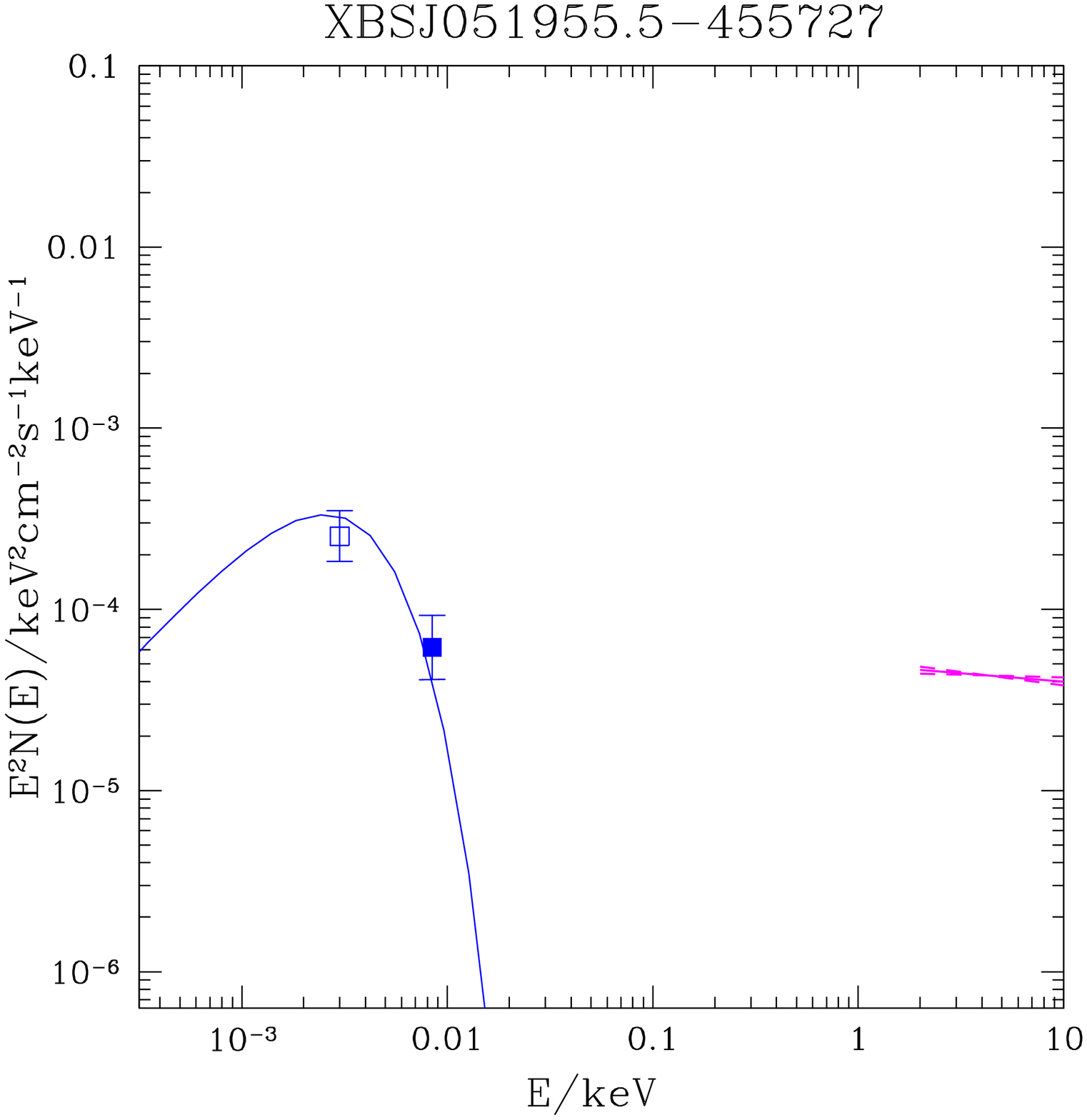}
  \includegraphics[height=5.6cm, width=6cm]{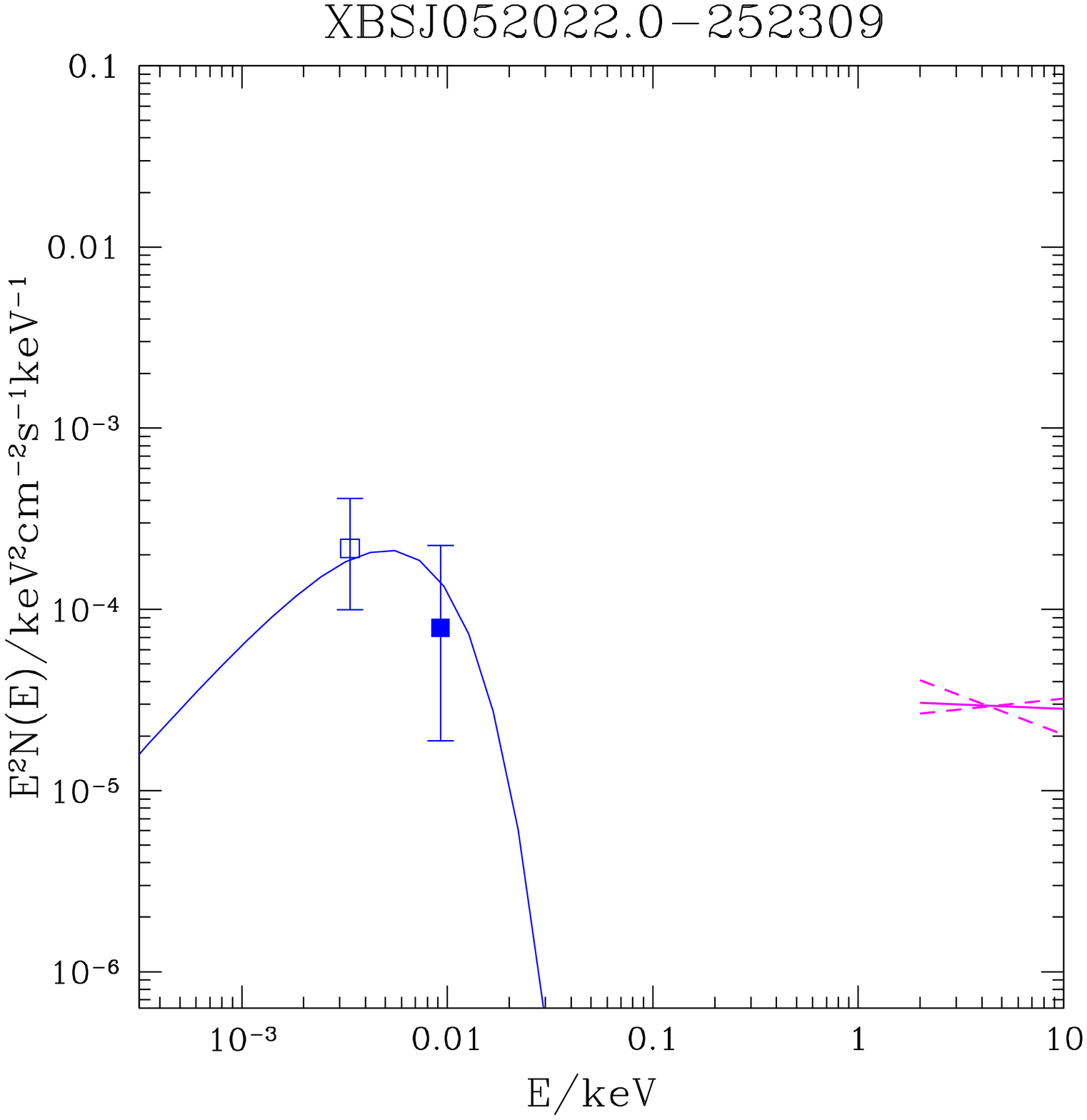}}     
\subfigure{ 
  \includegraphics[height=5.6cm, width=6cm]{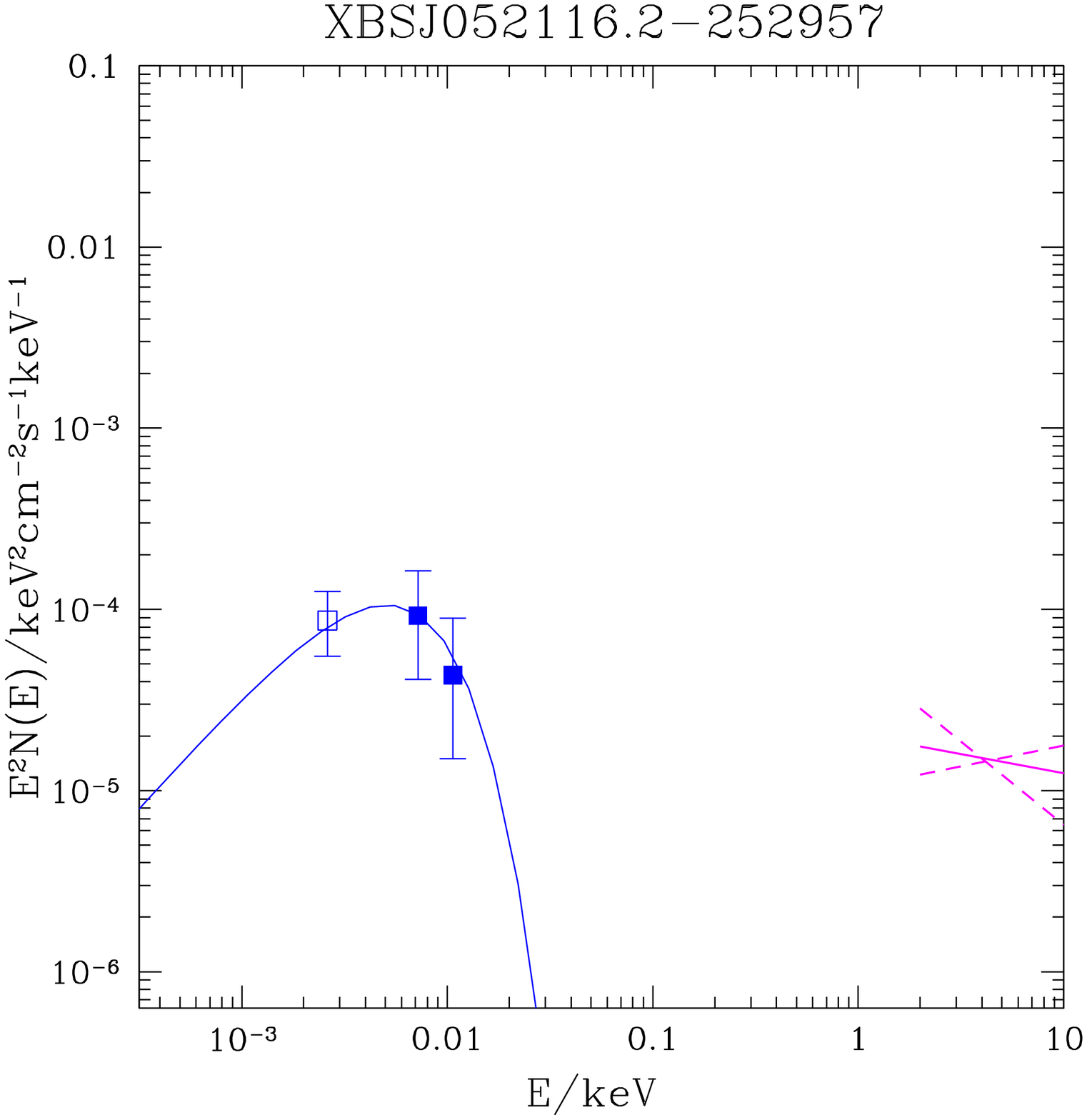}
  \includegraphics[height=5.6cm, width=6cm]{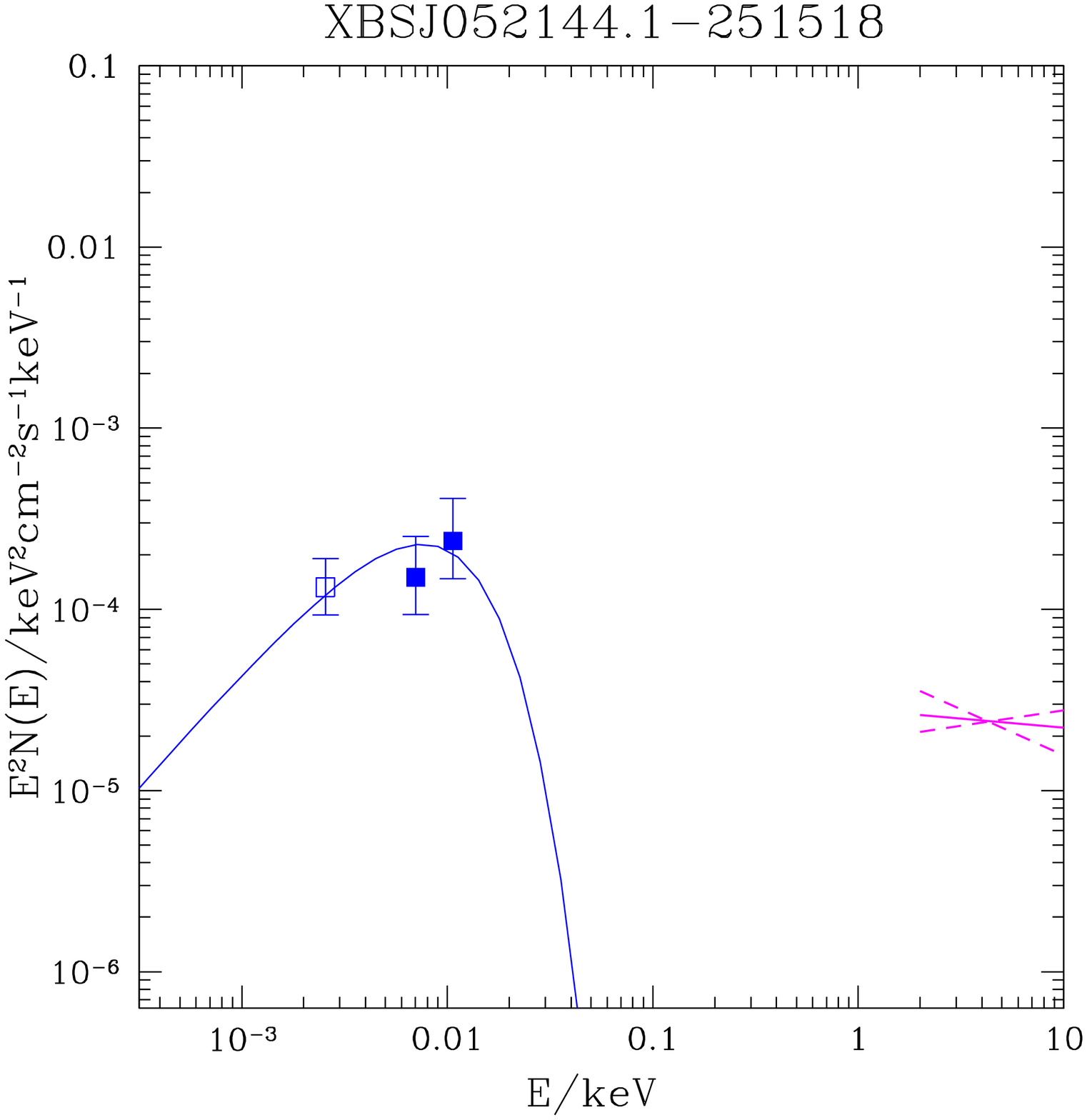}}      
\subfigure{ 
  \includegraphics[height=5.6cm, width=6cm]{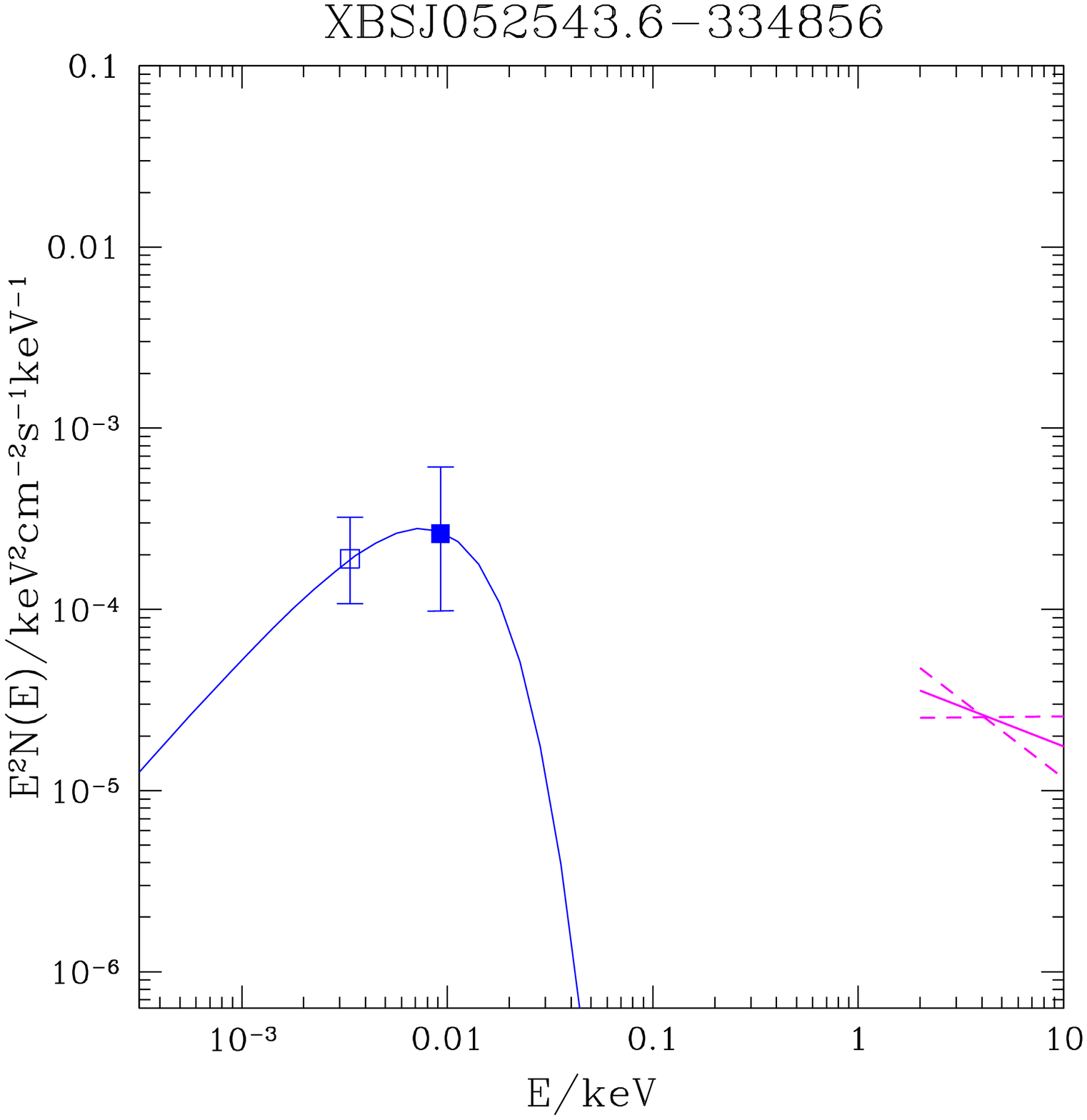}
  \includegraphics[height=5.6cm, width=6cm]{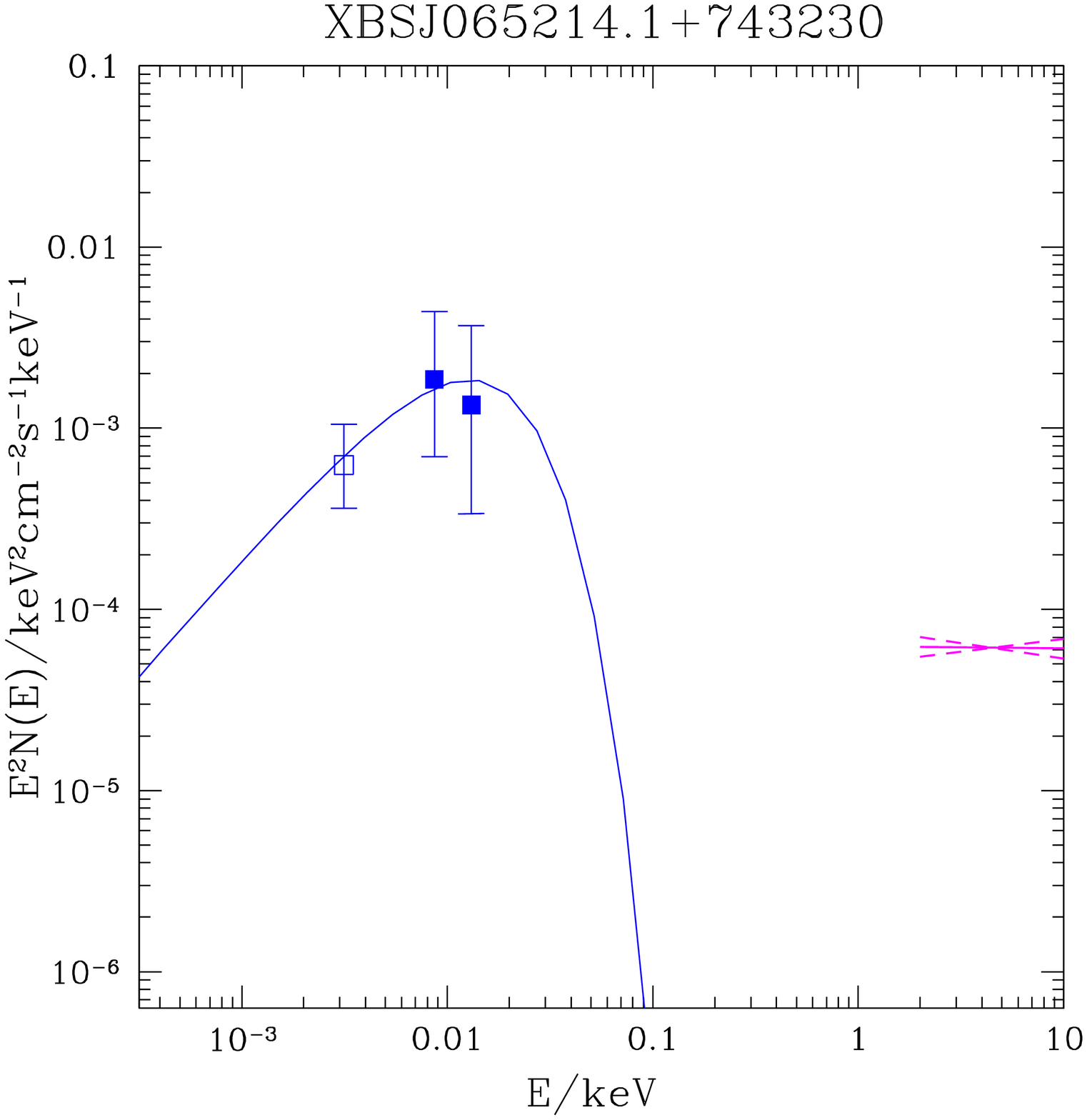}} 
  \end{figure*}

   \FloatBarrier
  
   \begin{figure*}
\centering     
\subfigure{ 
  \includegraphics[height=5.6cm, width=6cm]{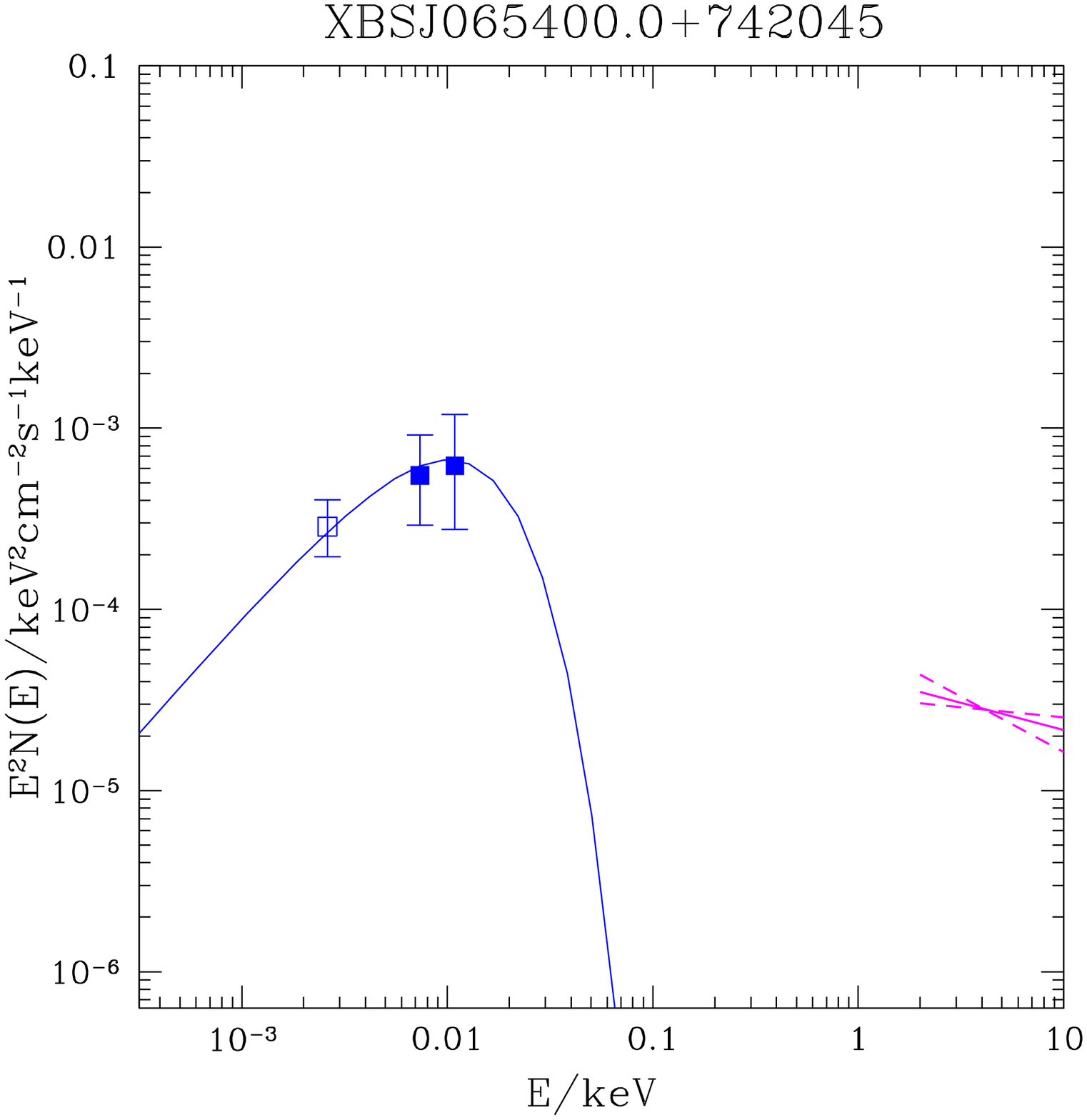}
  \includegraphics[height=5.6cm, width=6cm]{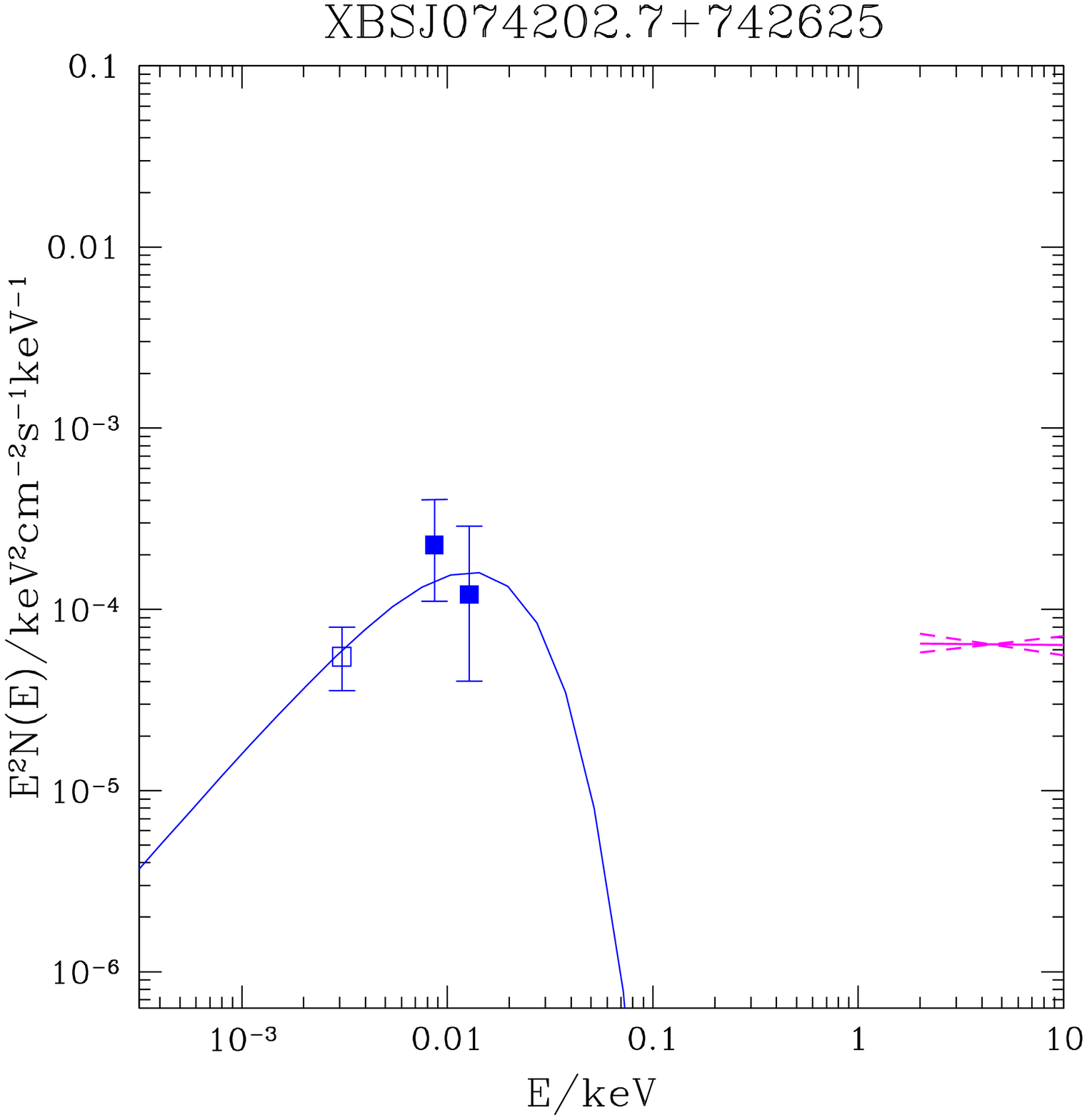}}     
 \subfigure{ 
  \includegraphics[height=5.6cm, width=6cm]{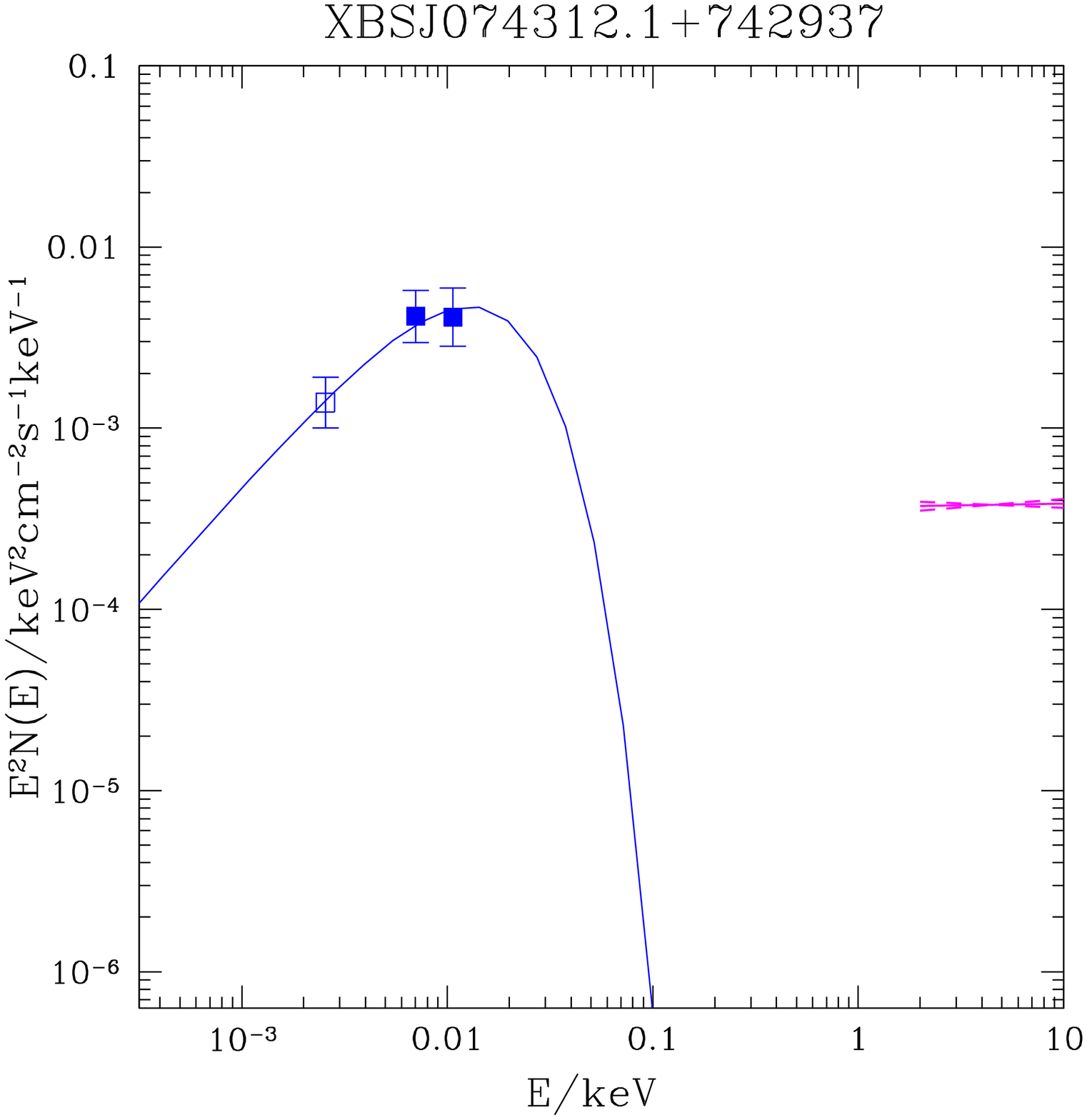}
  \includegraphics[height=5.6cm, width=6cm]{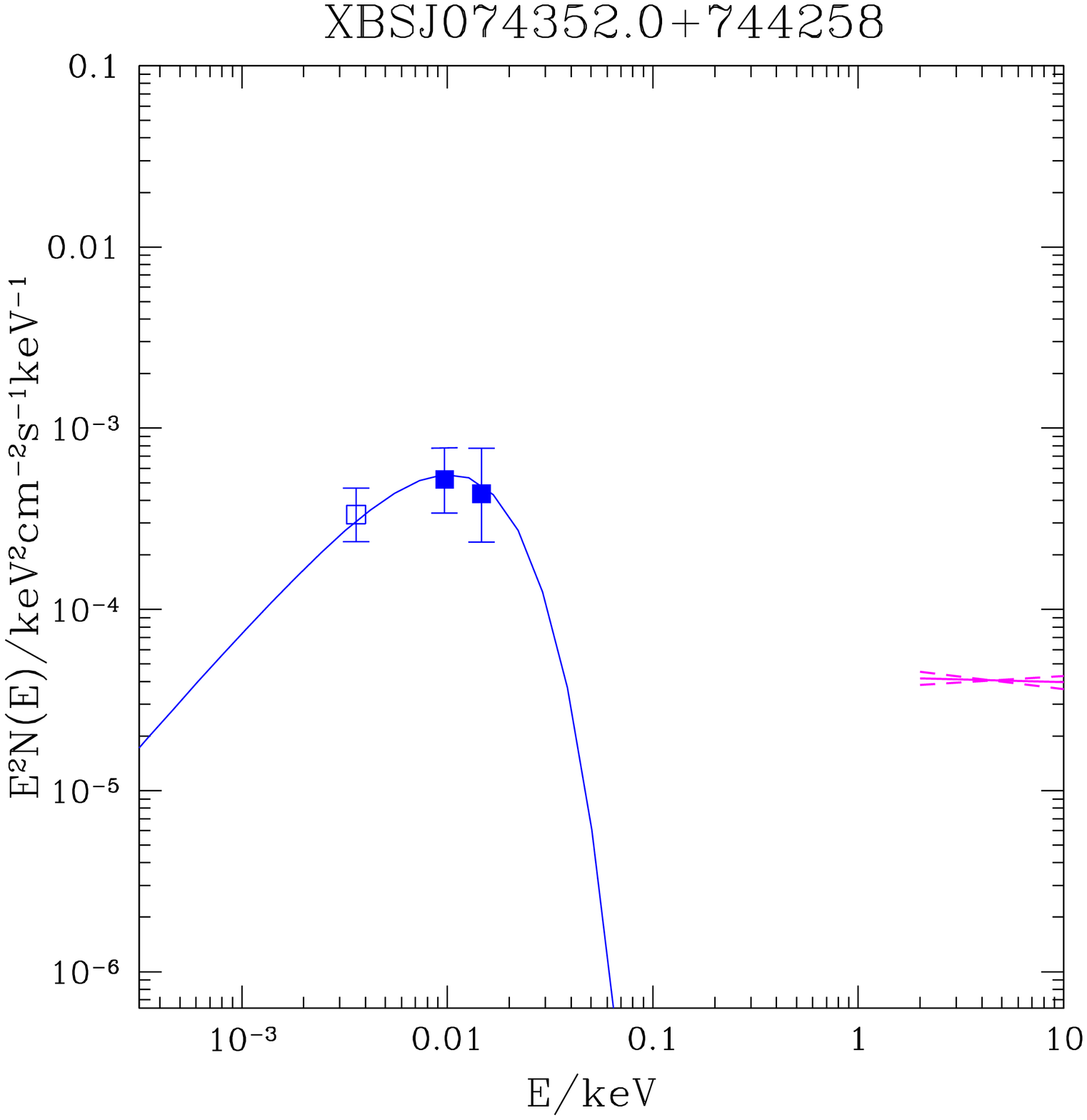}}  
\subfigure{ 
  \includegraphics[height=5.6cm, width=6cm]{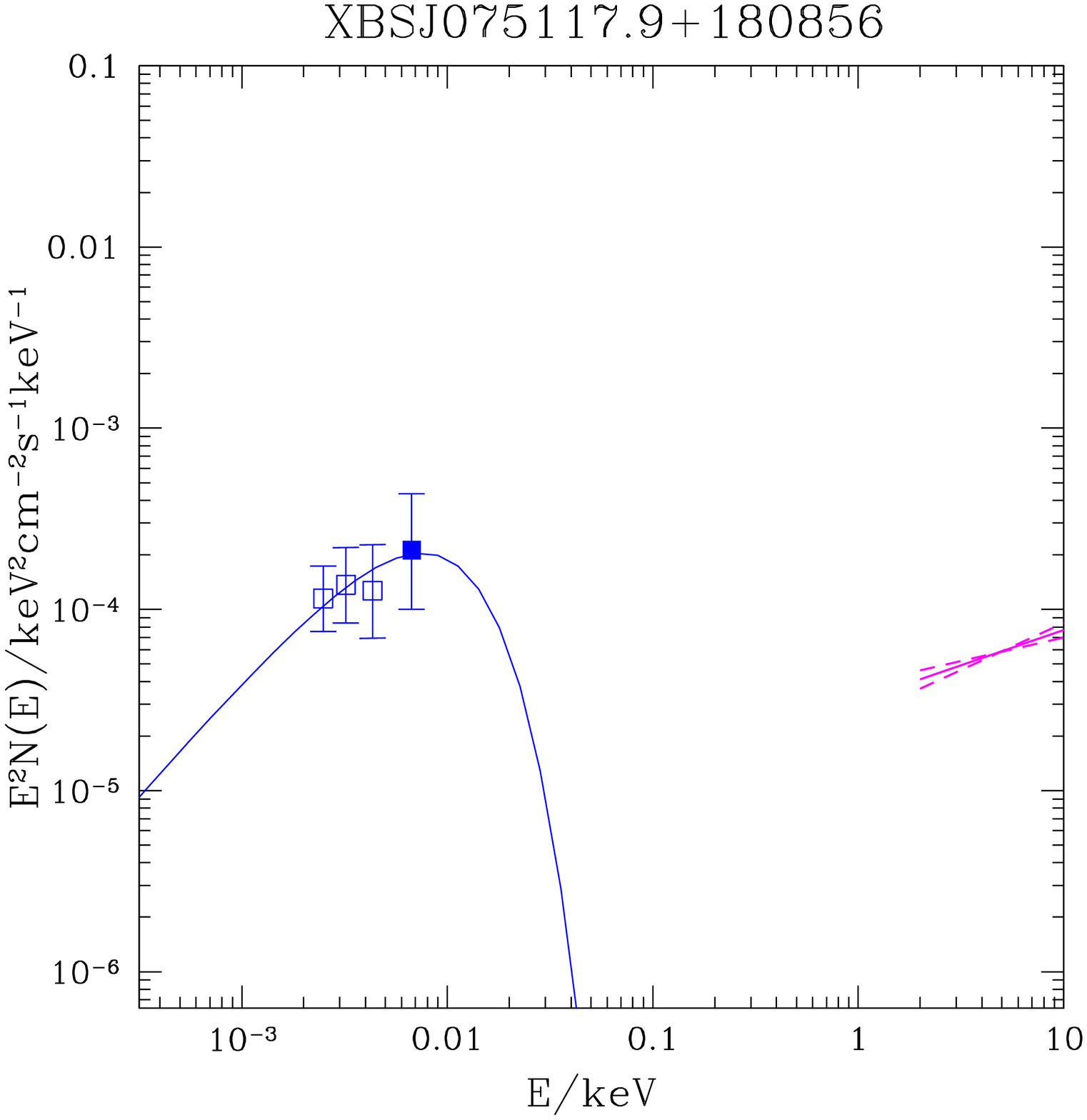}
  \includegraphics[height=5.6cm, width=6cm]{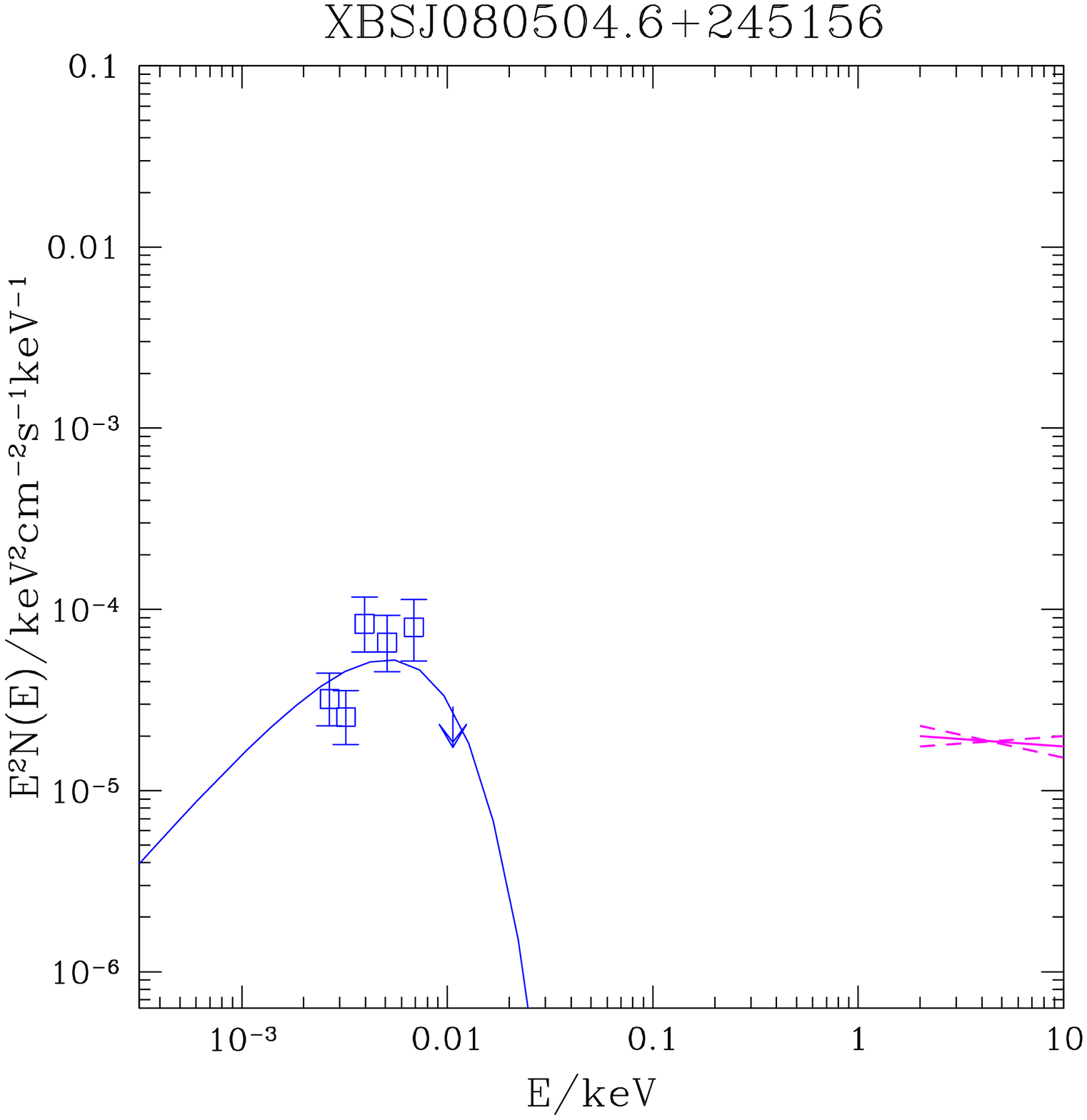}}     
 \subfigure{ 
  \includegraphics[height=5.6cm, width=6cm]{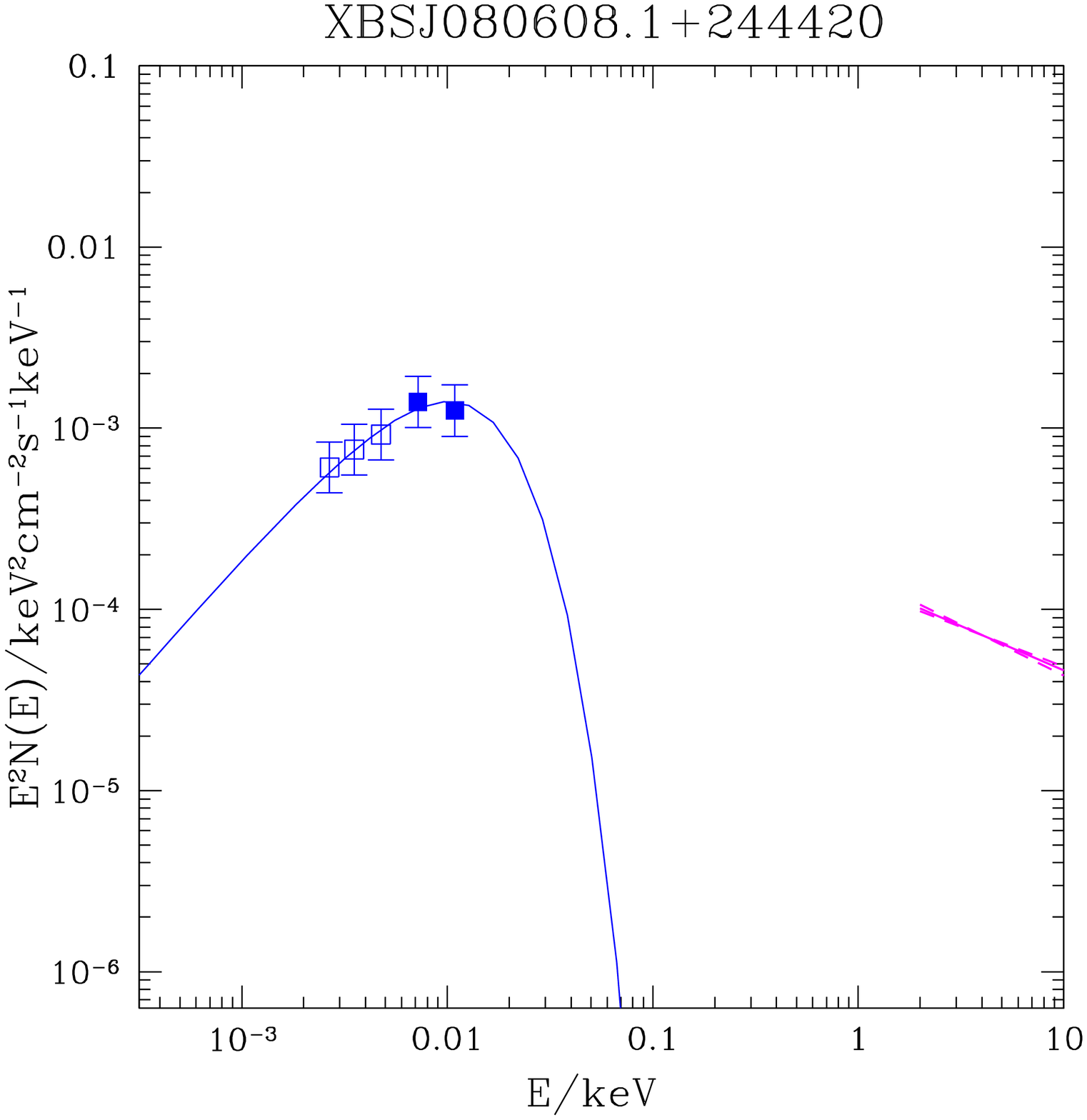}
  \includegraphics[height=5.6cm, width=6cm]{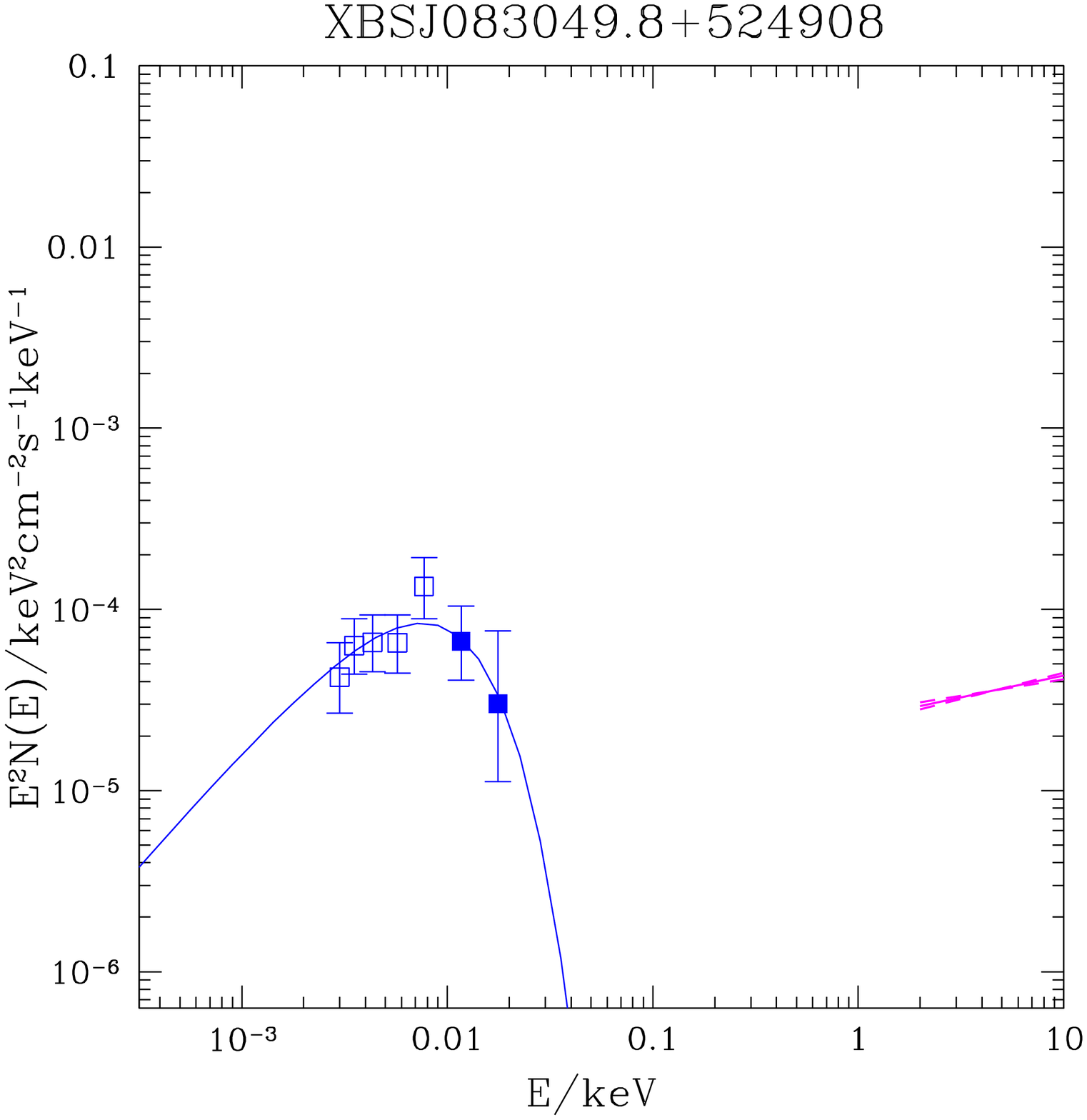}}  
  \end{figure*}
  
   \FloatBarrier
  
   \begin{figure*}
\centering    
\subfigure{ 
  \includegraphics[height=5.6cm, width=6cm]{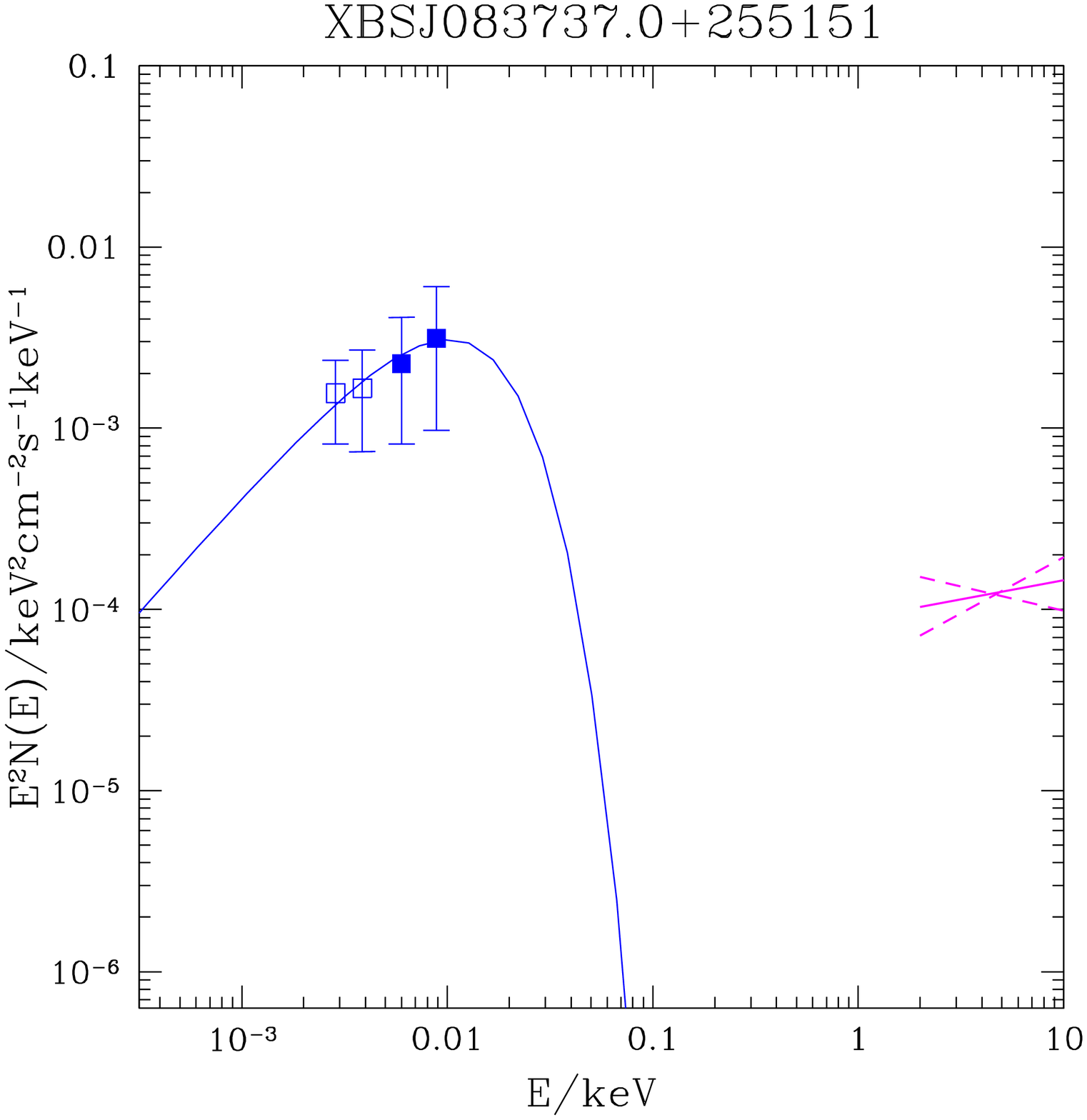}
  \includegraphics[height=5.6cm, width=6cm]{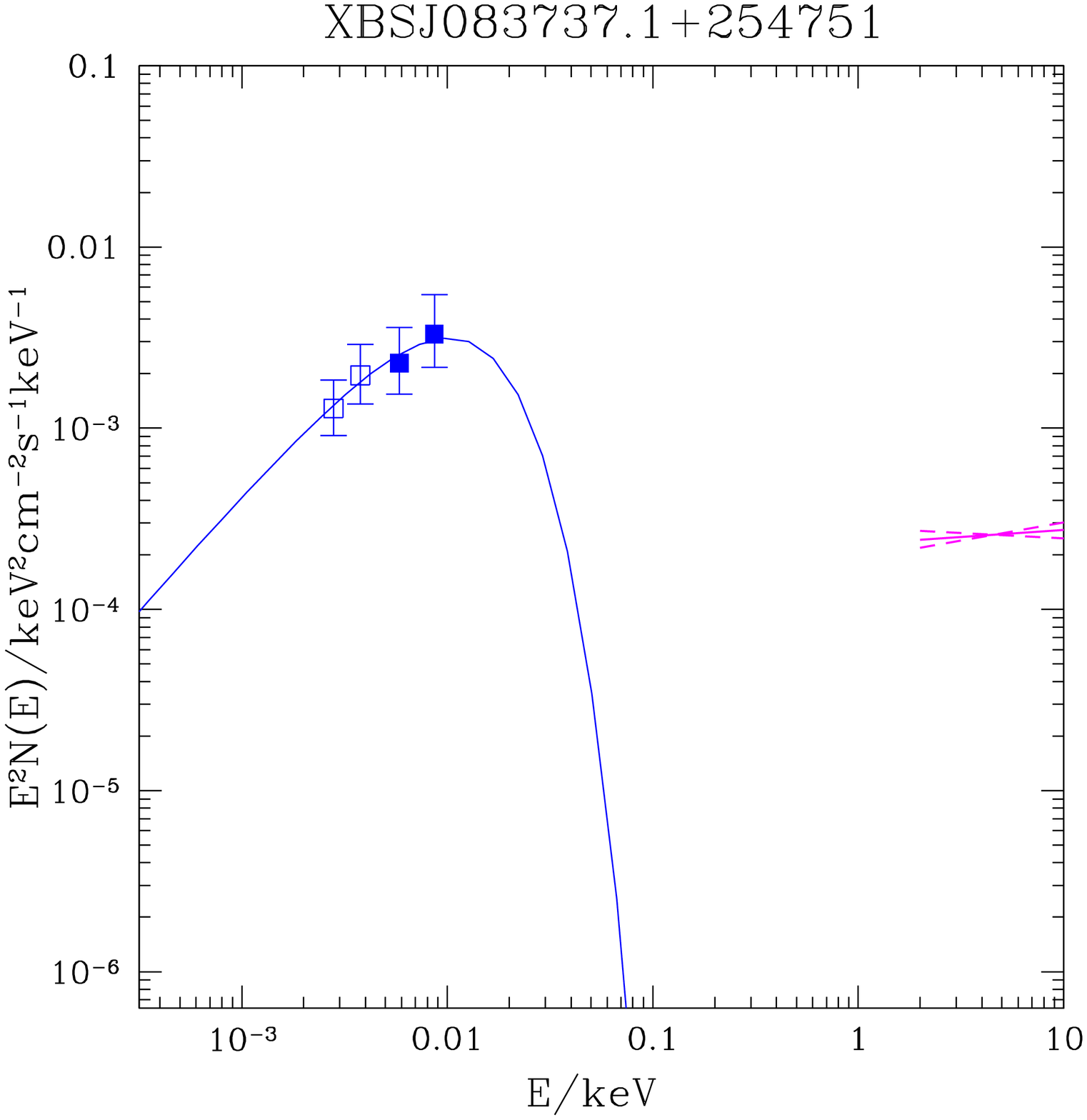}}      
\subfigure{ 
  \includegraphics[height=5.6cm, width=6cm]{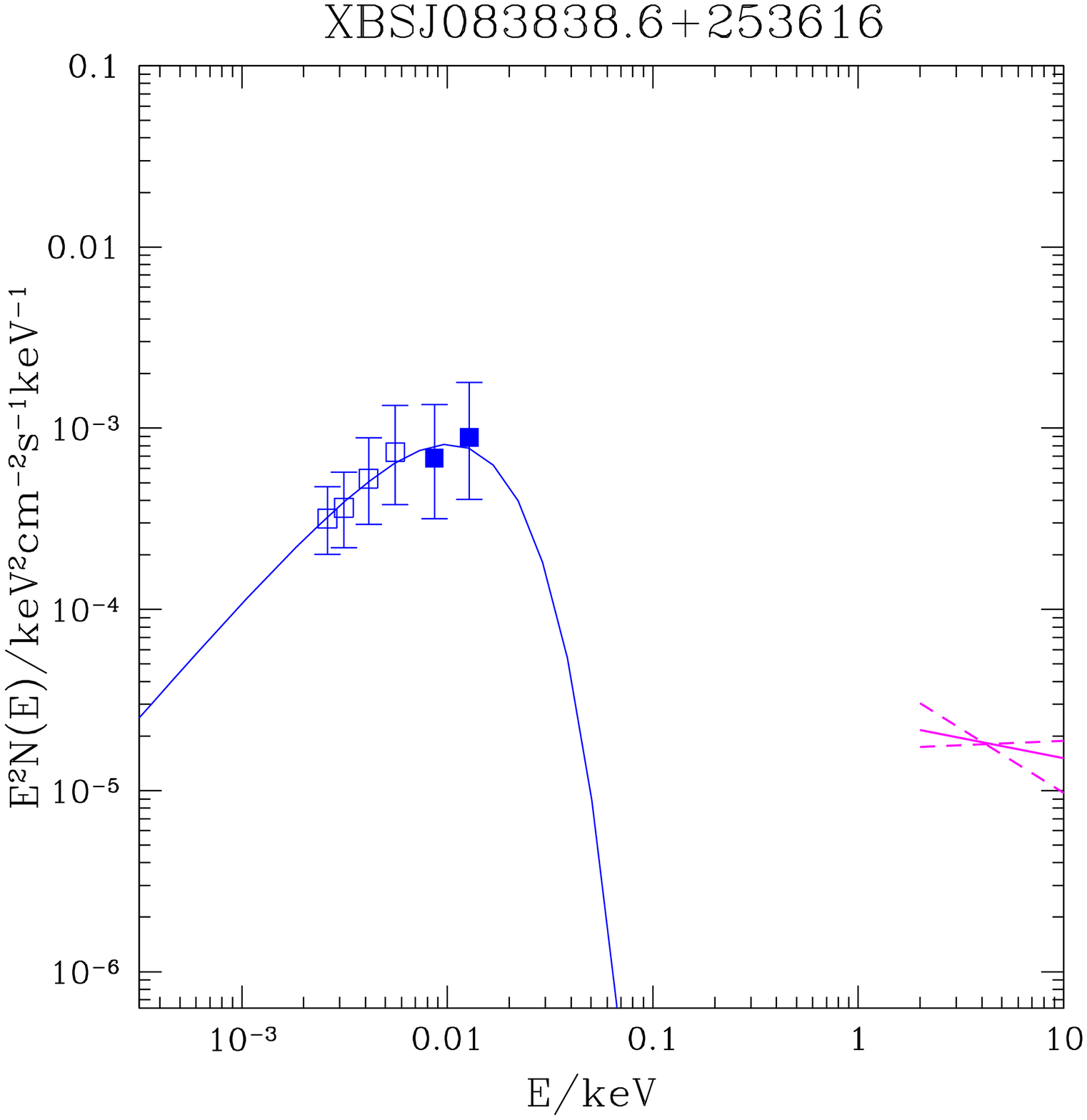}
  \includegraphics[height=5.6cm, width=6cm]{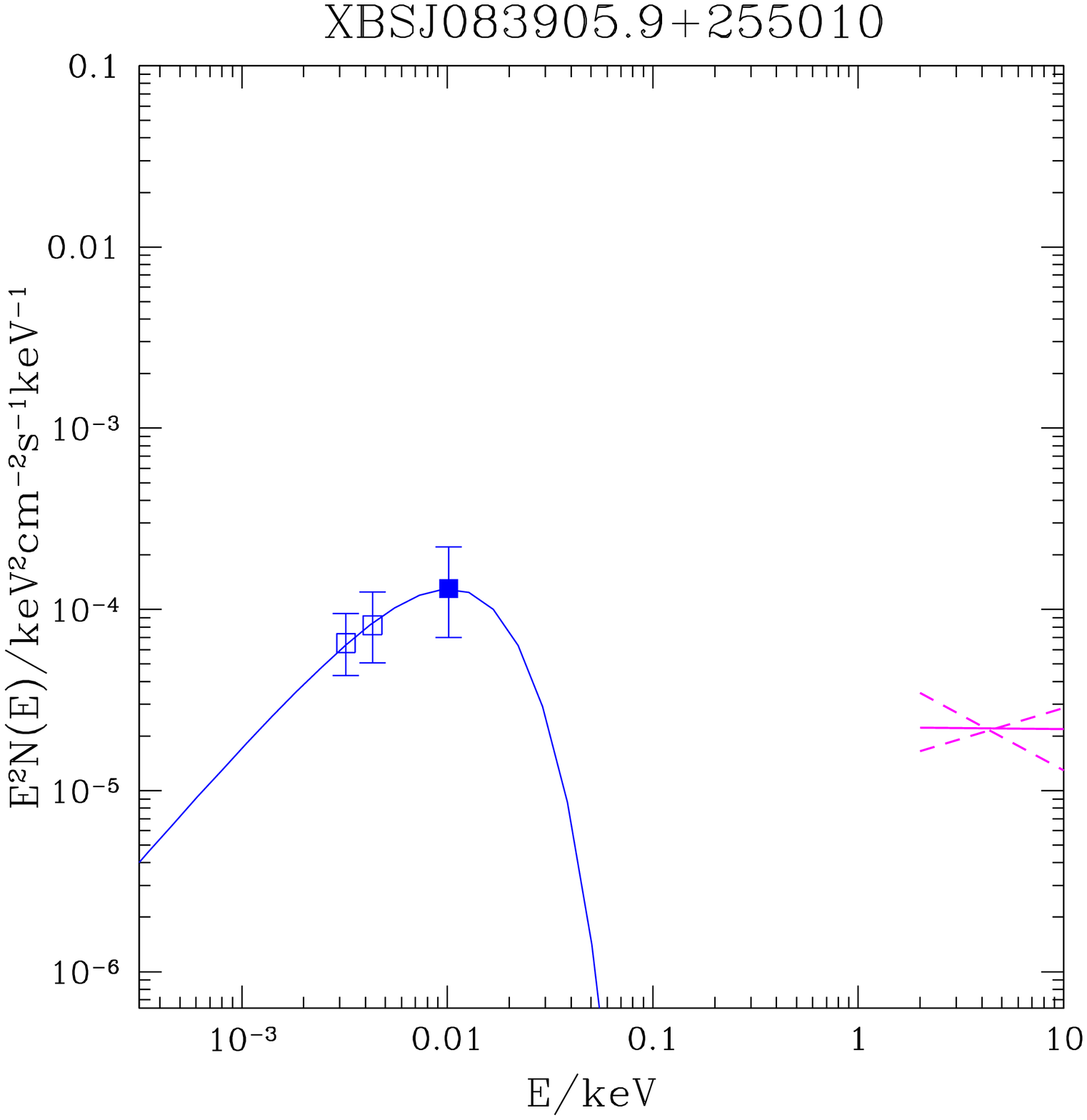}}     
\subfigure{ 
  \includegraphics[height=5.6cm, width=6cm]{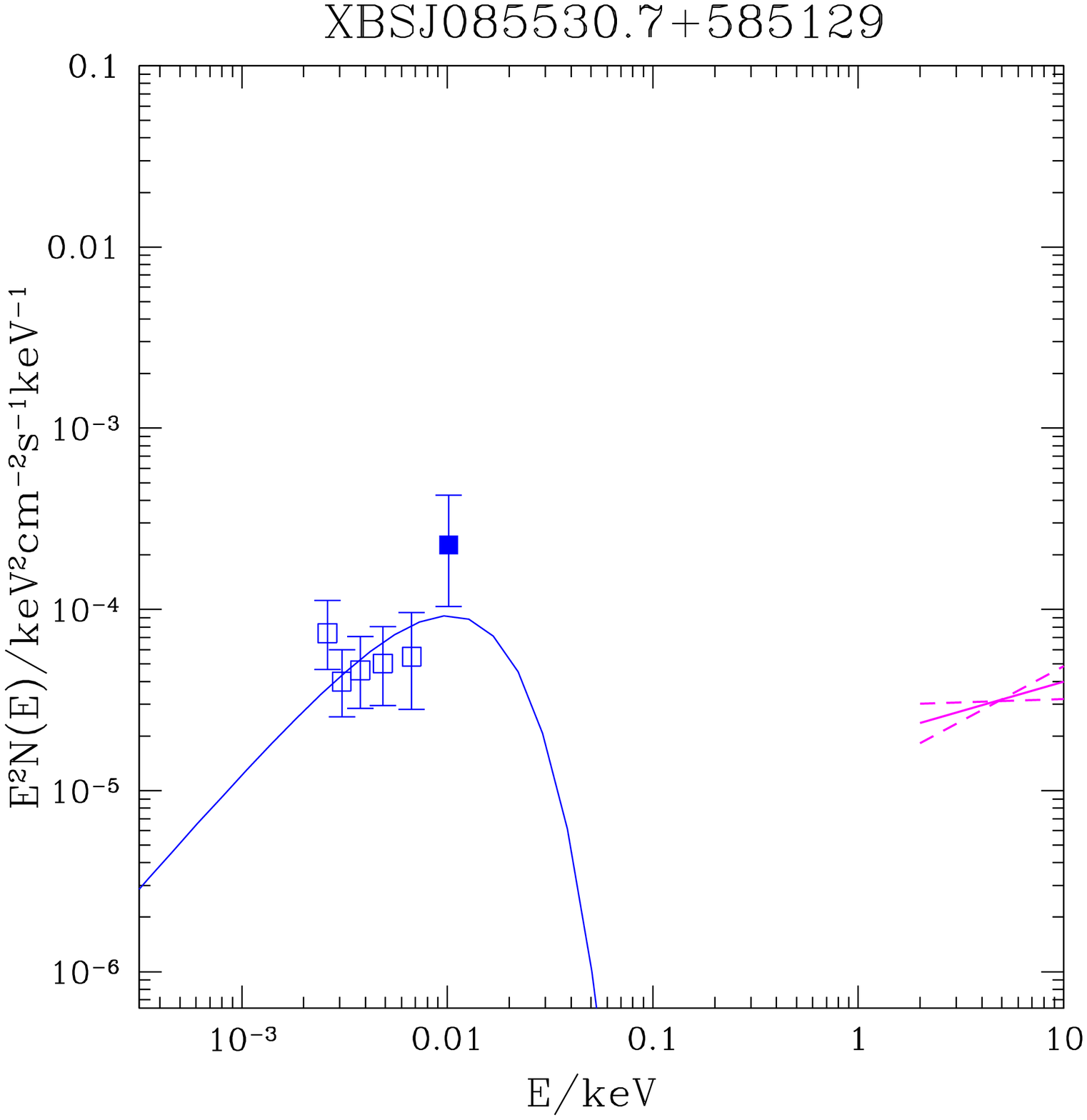}
  \includegraphics[height=5.6cm, width=6cm]{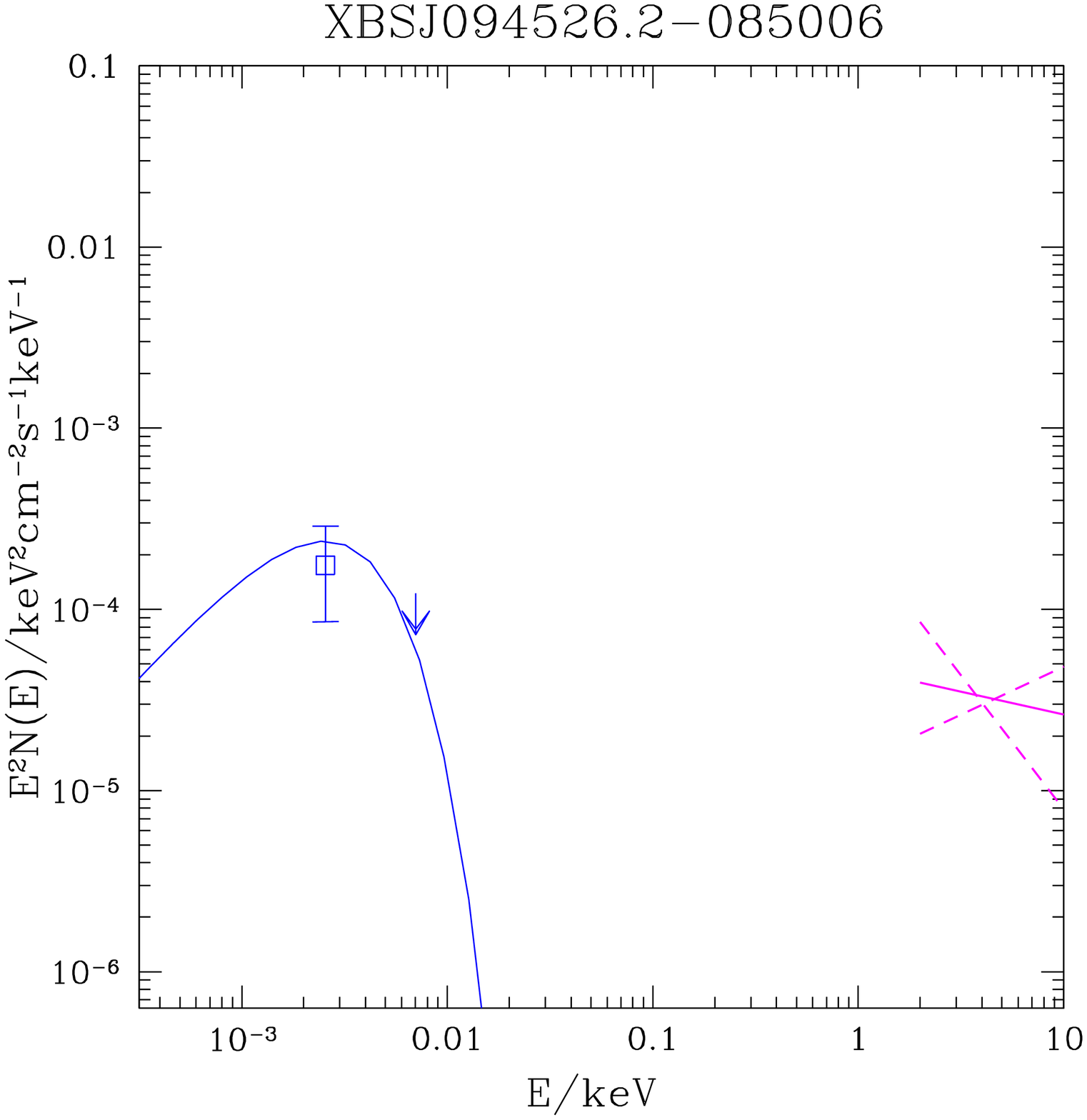}}      
\subfigure{ 
  \includegraphics[height=5.6cm, width=6cm]{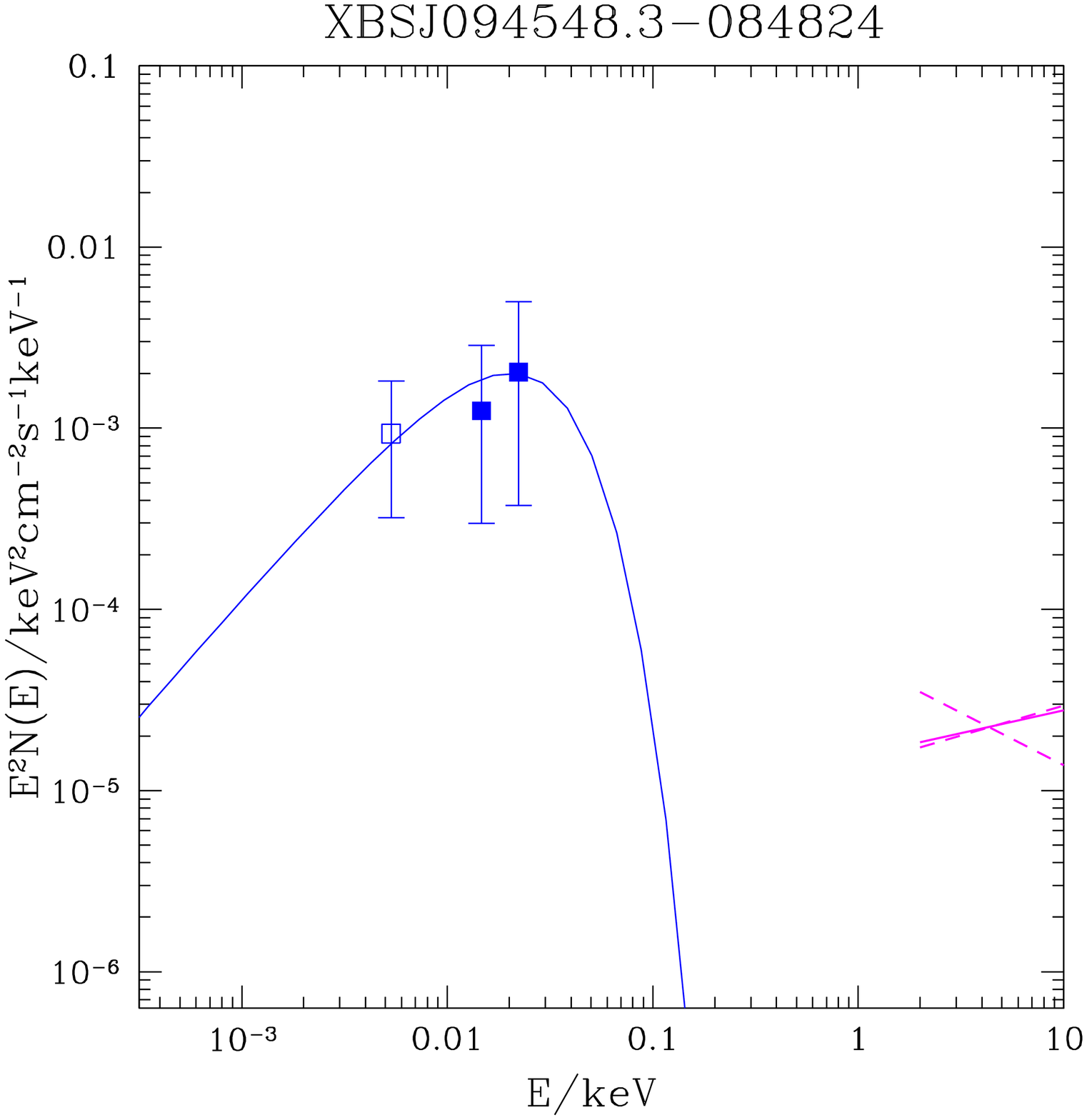}
  \includegraphics[height=5.6cm, width=6cm]{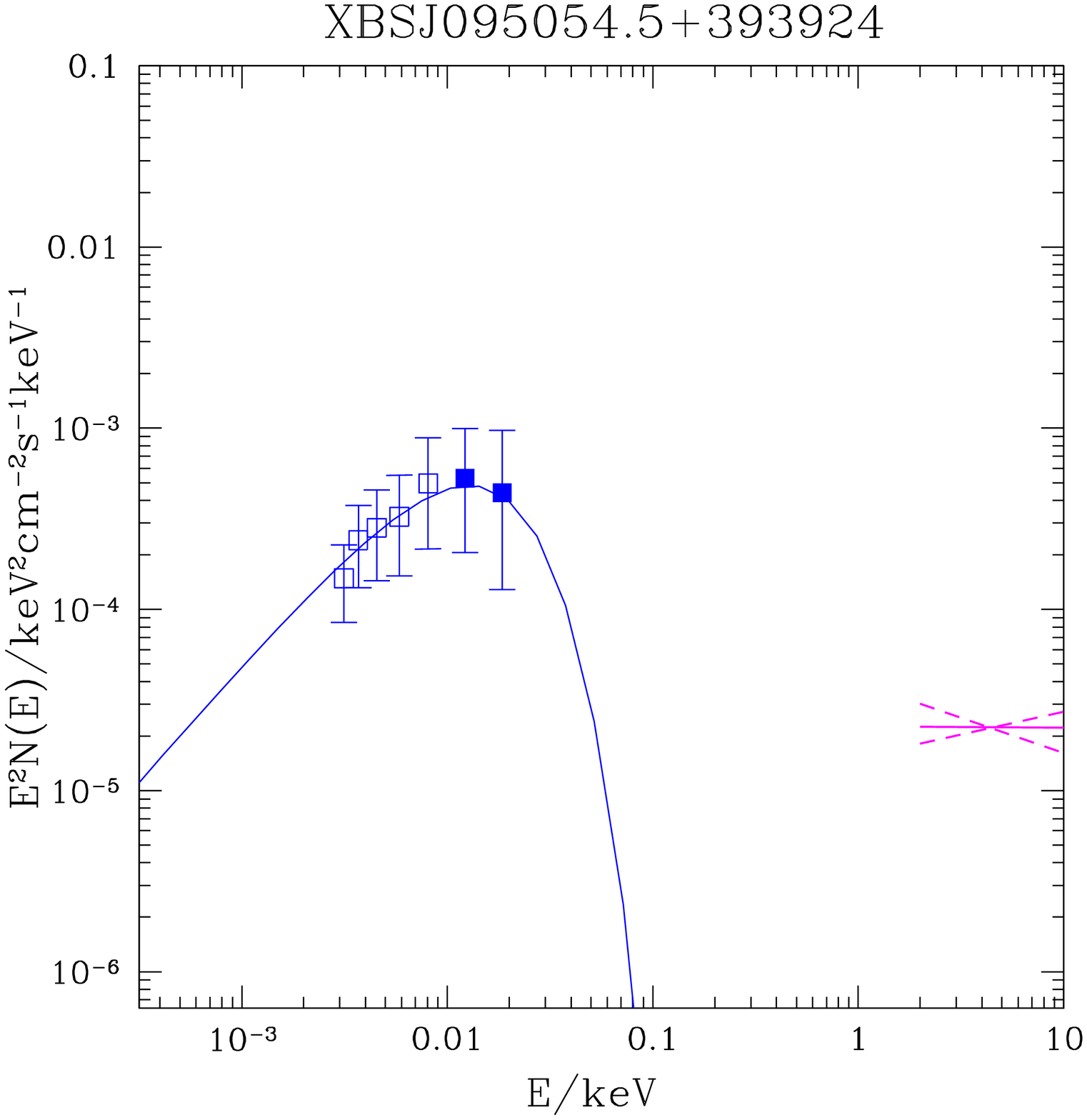}}  
  \end{figure*}
  
   \FloatBarrier

   \begin{figure*}
\centering    
\subfigure{ 
  \includegraphics[height=5.6cm, width=6cm]{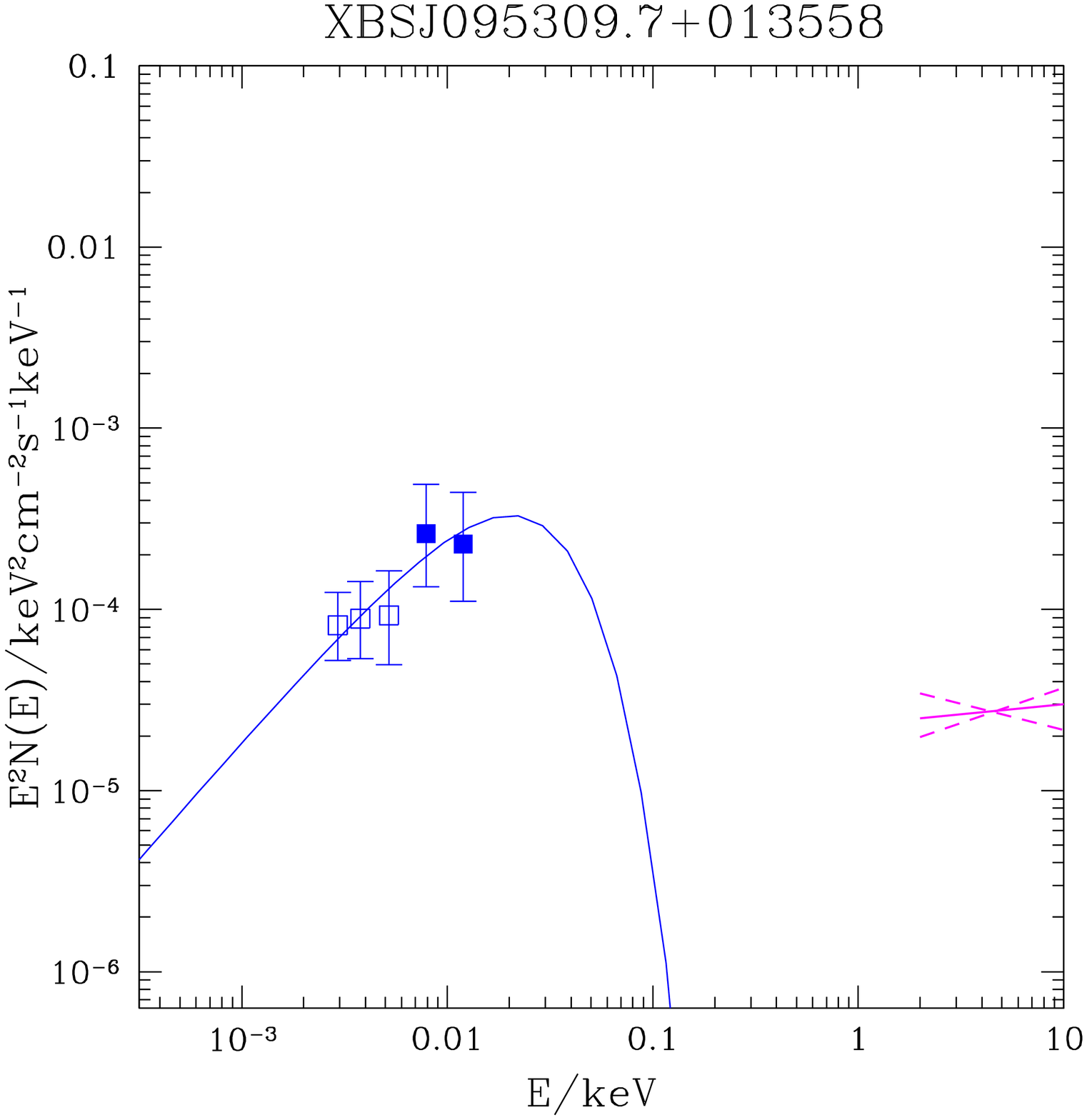}
  \includegraphics[height=5.6cm, width=6cm]{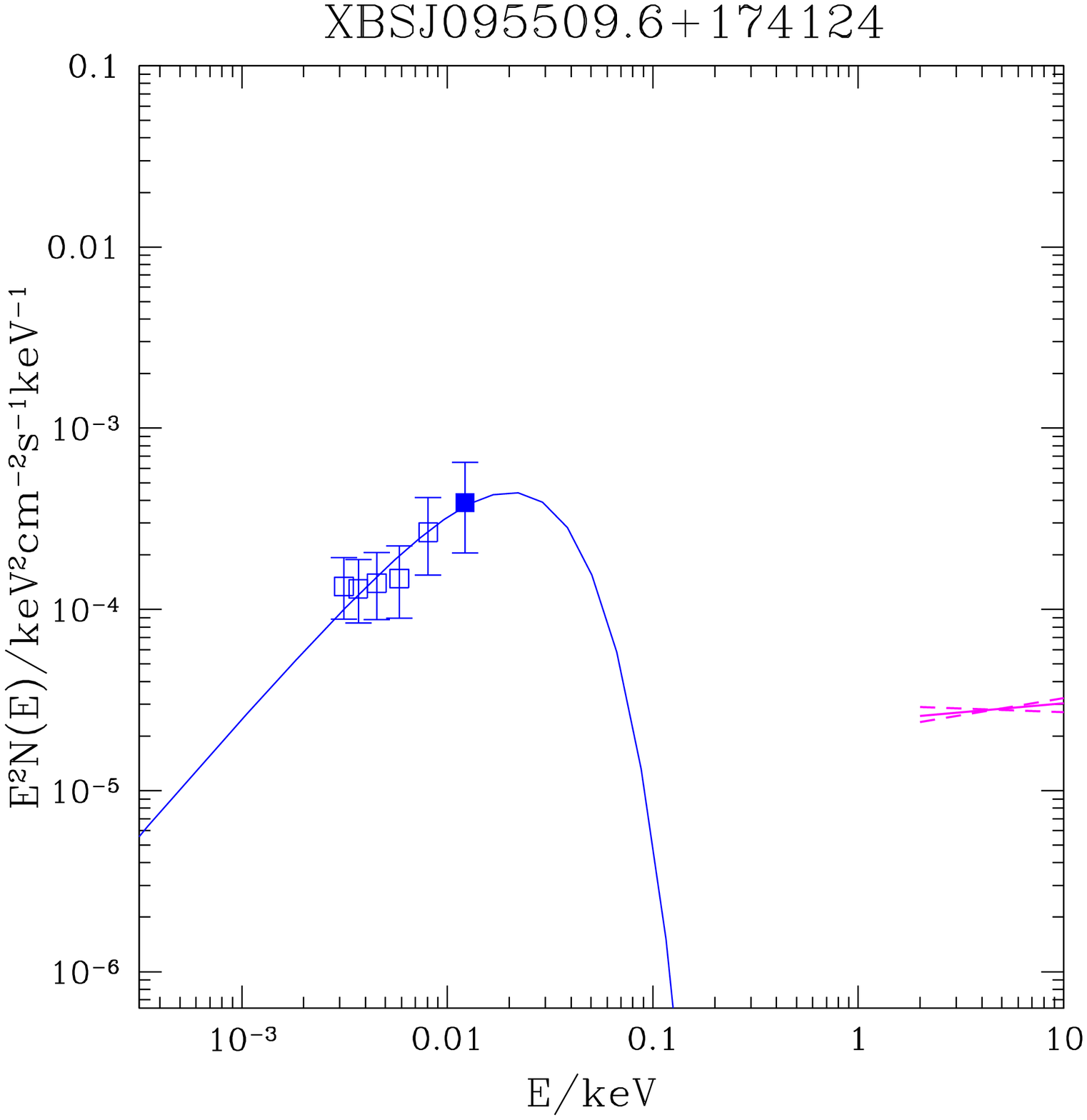}}      
\subfigure{ 
  \includegraphics[height=5.6cm, width=6cm]{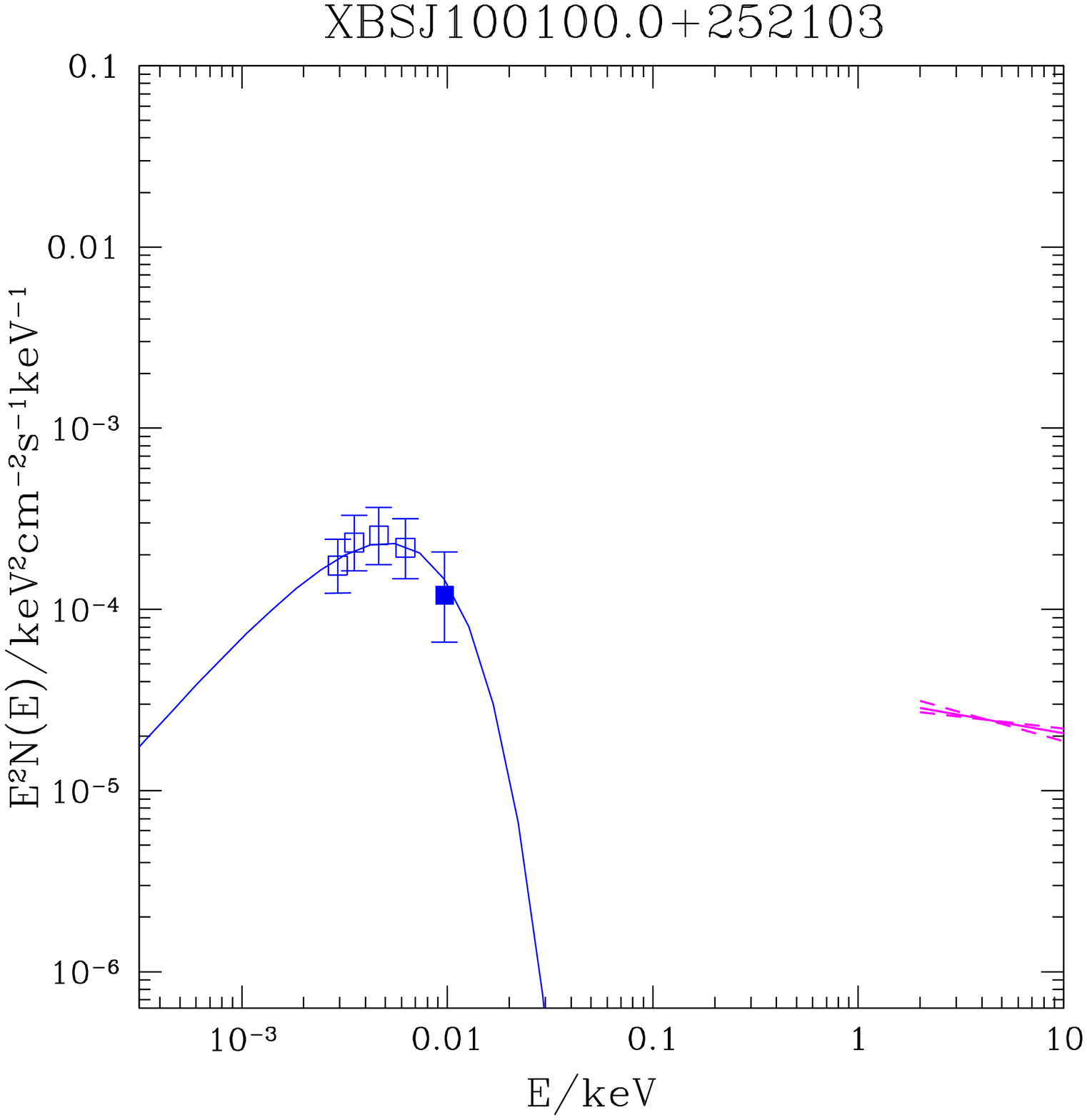}
  \includegraphics[height=5.6cm, width=6cm]{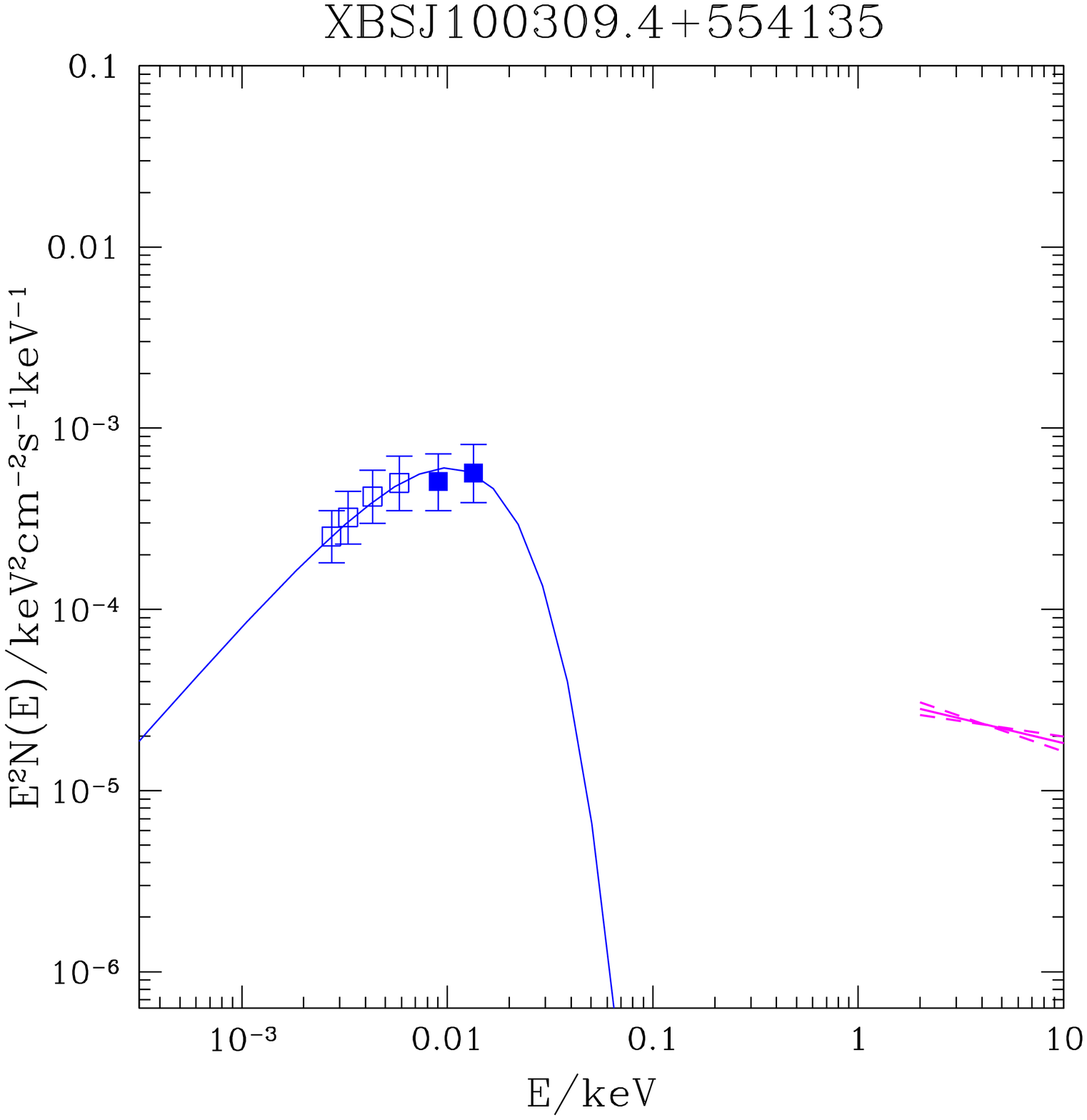}}    
\subfigure{ 
  \includegraphics[height=5.6cm, width=6cm]{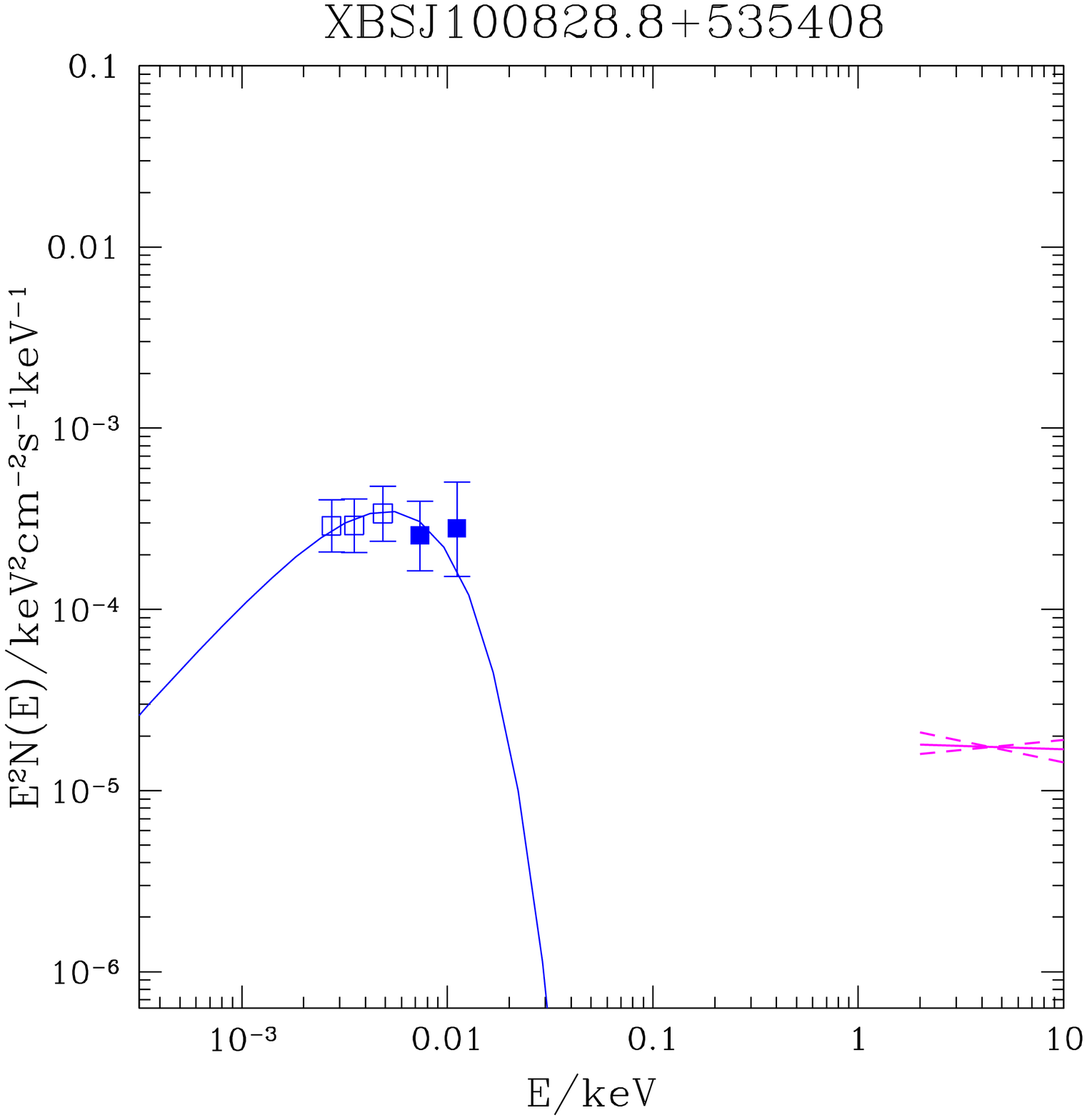}
  \includegraphics[height=5.6cm, width=6cm]{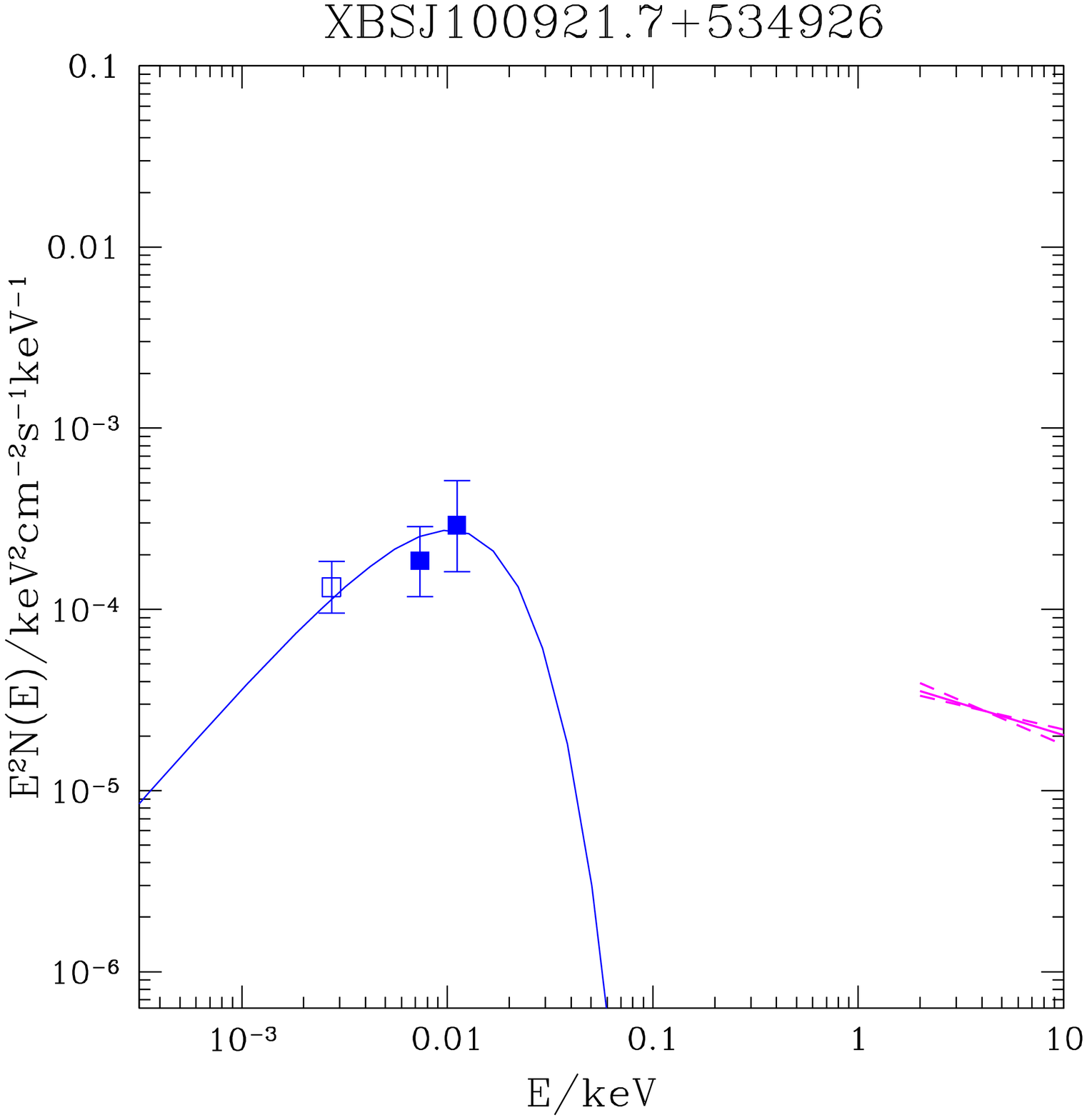}}      
\subfigure{ 
  \includegraphics[height=5.6cm, width=6cm]{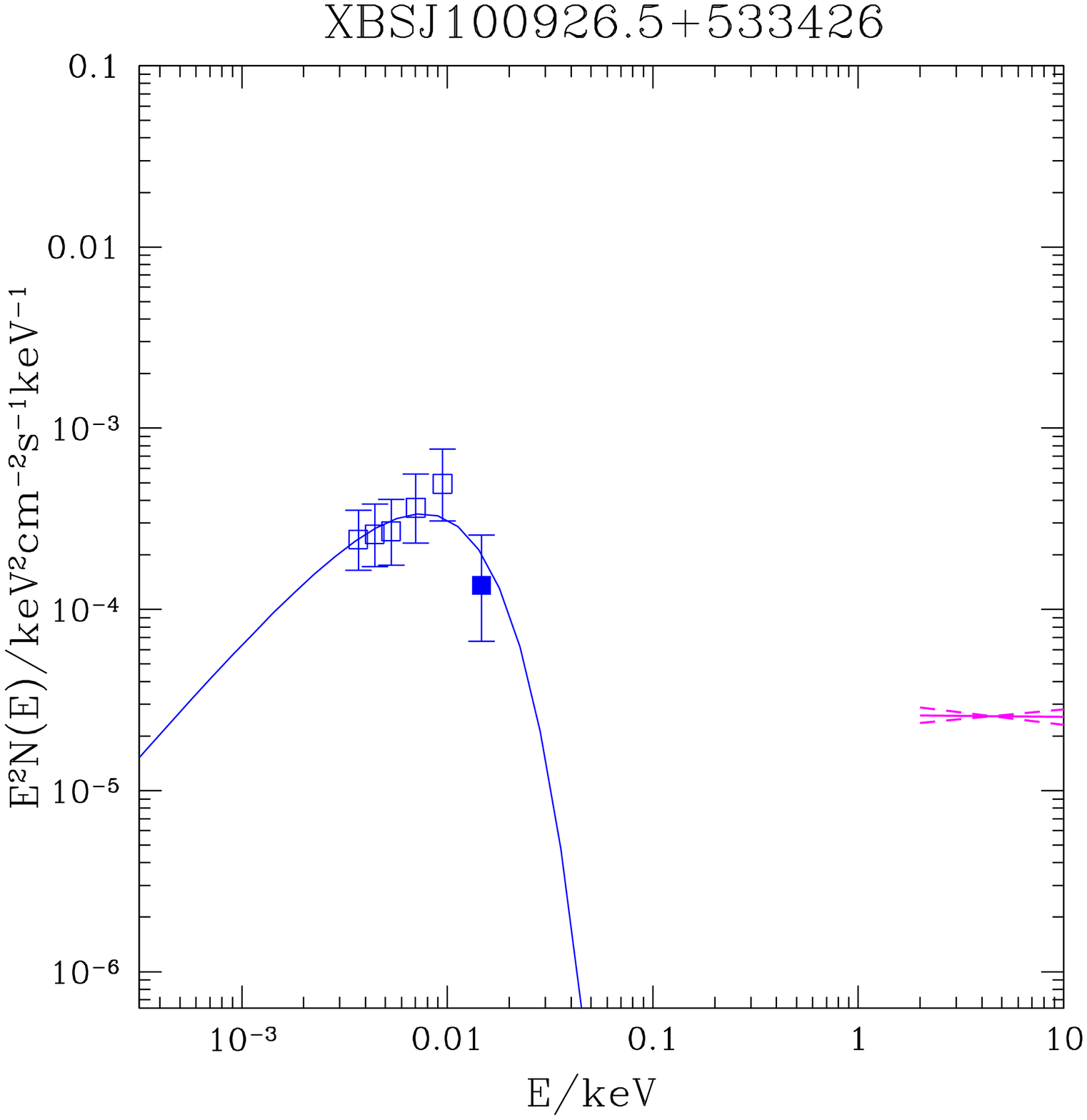}
  \includegraphics[height=5.6cm, width=6cm]{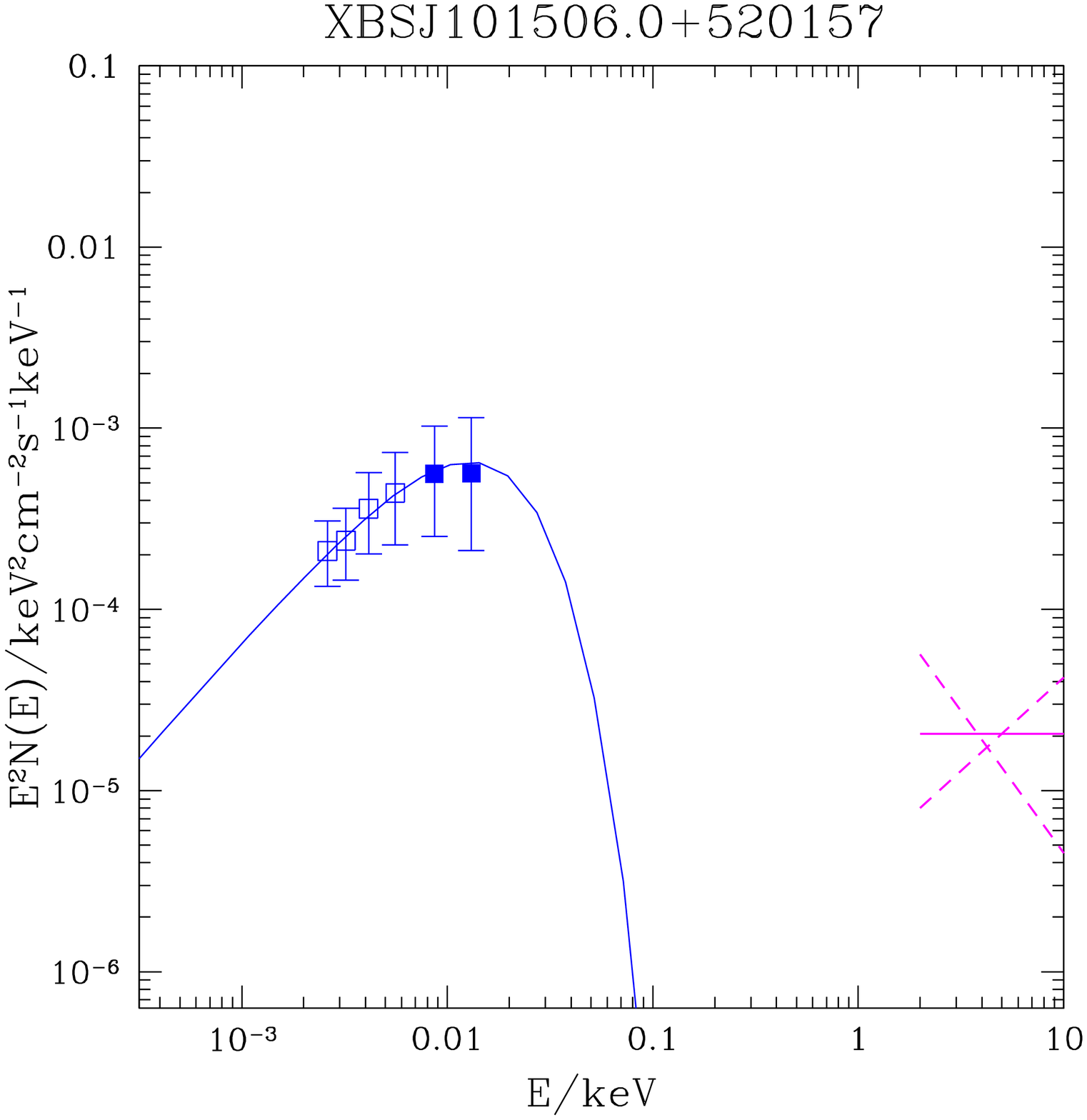}}    
  \end{figure*}
  
   \FloatBarrier
  
   \begin{figure*}
\centering  
\subfigure{ 
  \includegraphics[height=5.6cm, width=6cm]{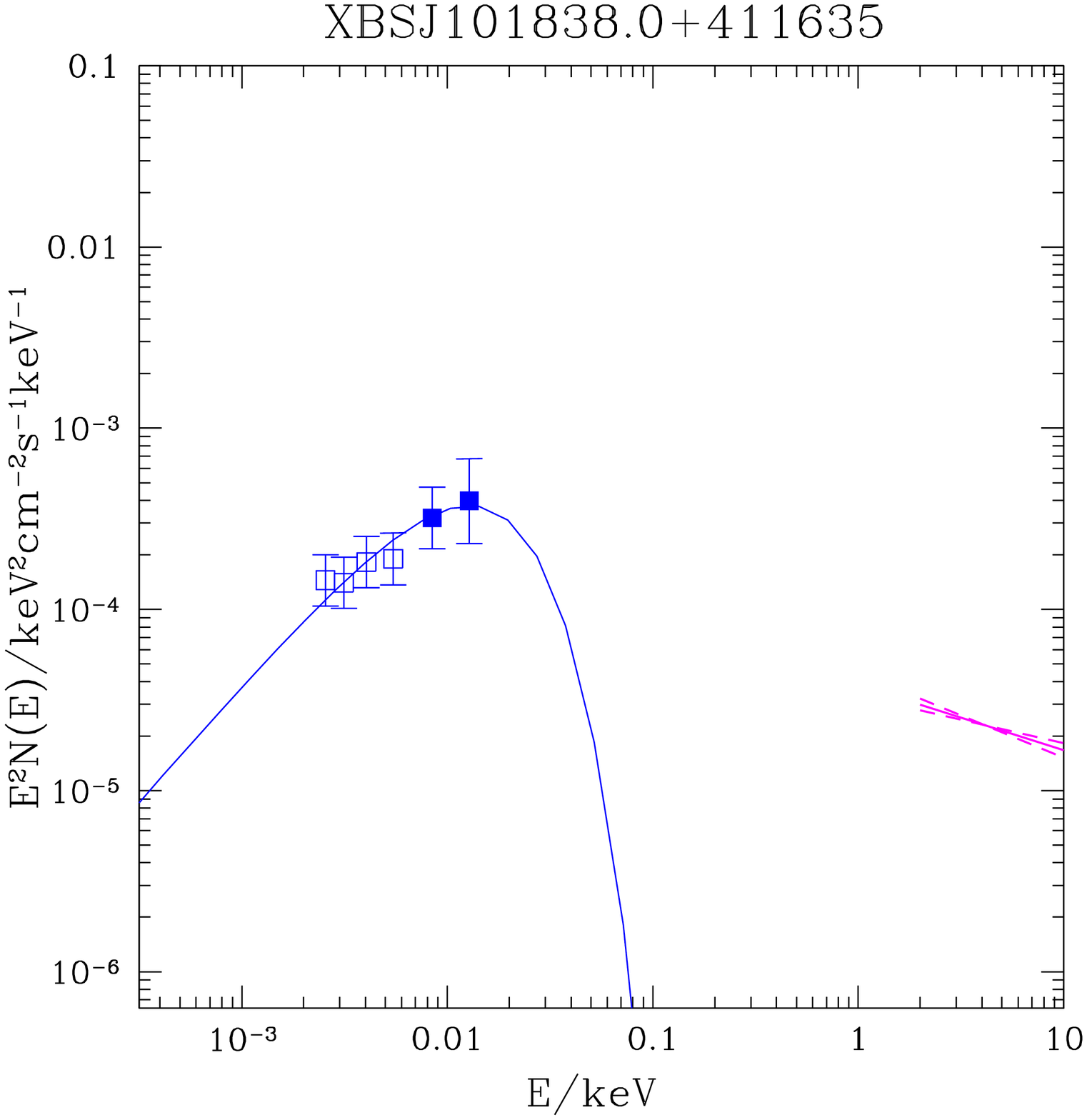}
  \includegraphics[height=5.6cm, width=6cm]{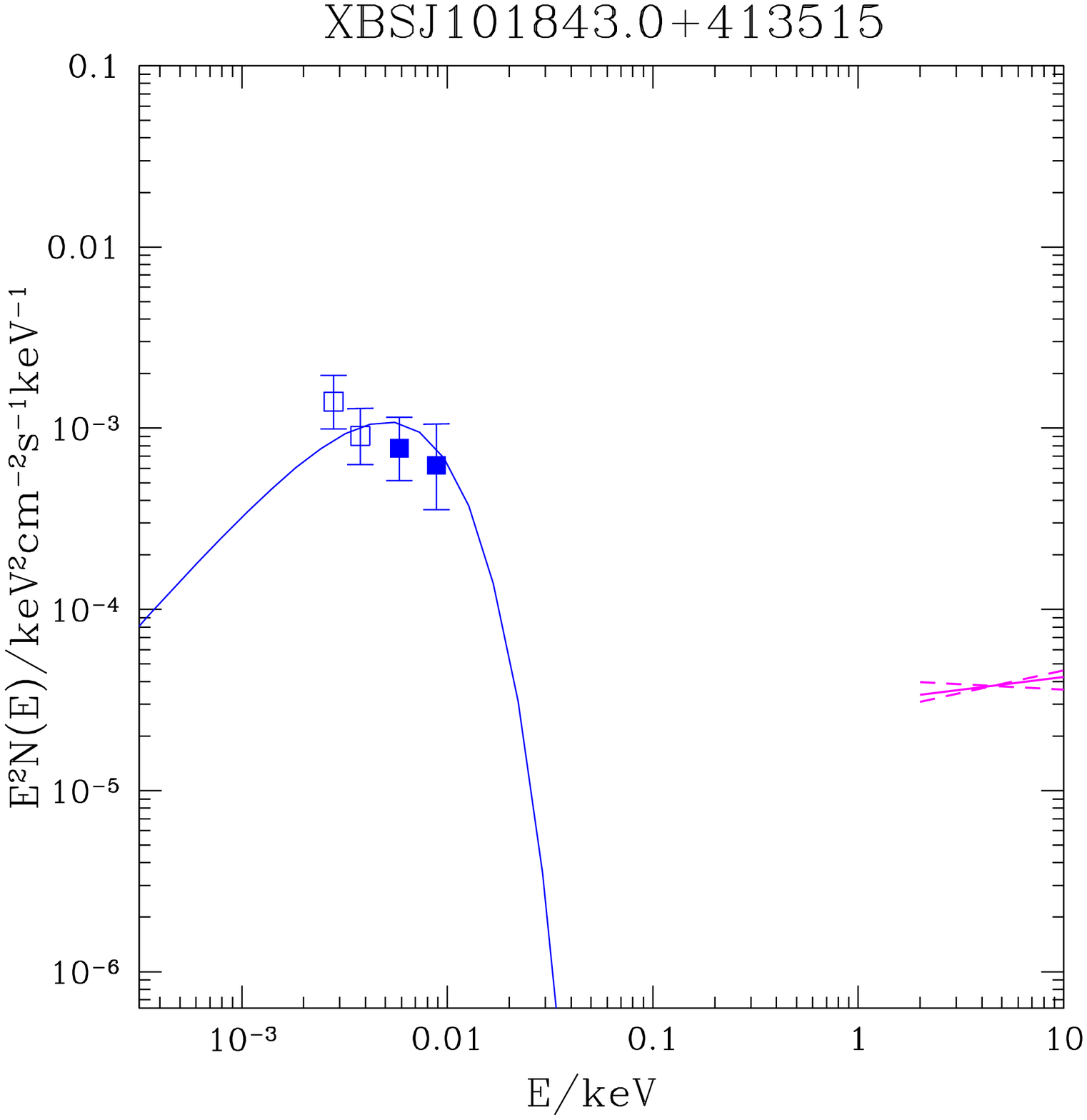}}      
\subfigure{ 
  \includegraphics[height=5.6cm, width=6cm]{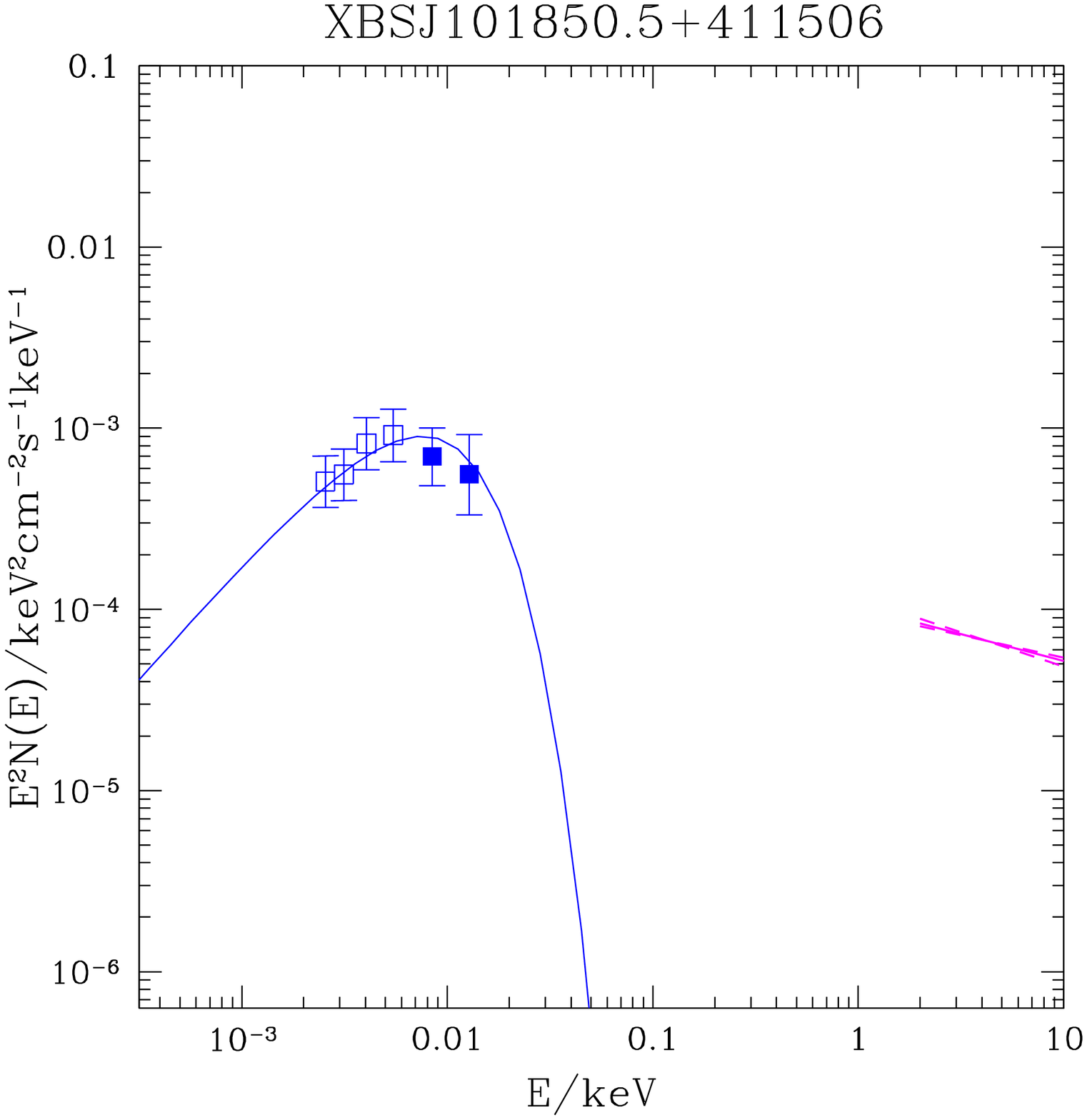}
  \includegraphics[height=5.6cm, width=6cm]{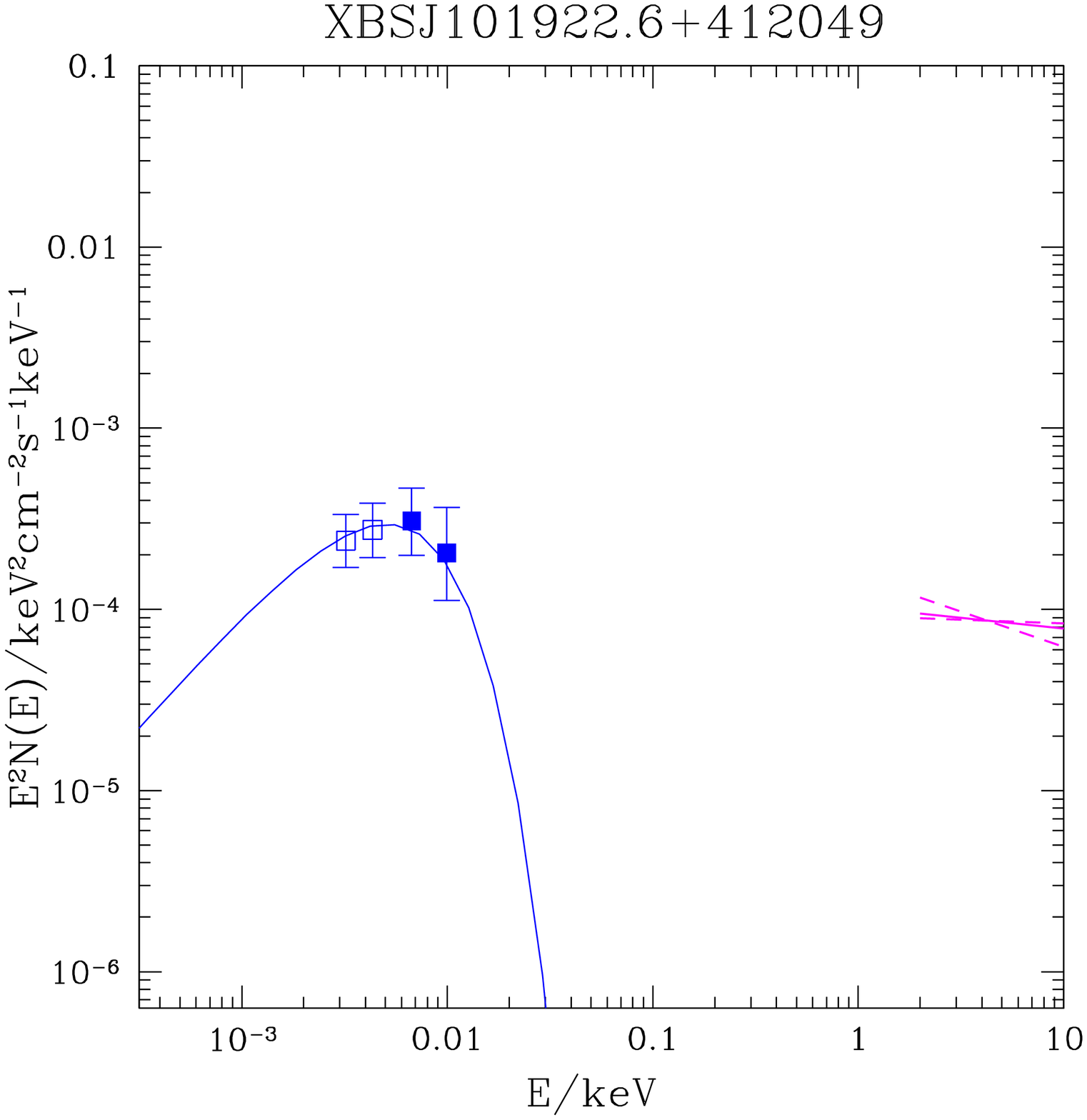}}   
\subfigure{ 
  \includegraphics[height=5.6cm, width=6cm]{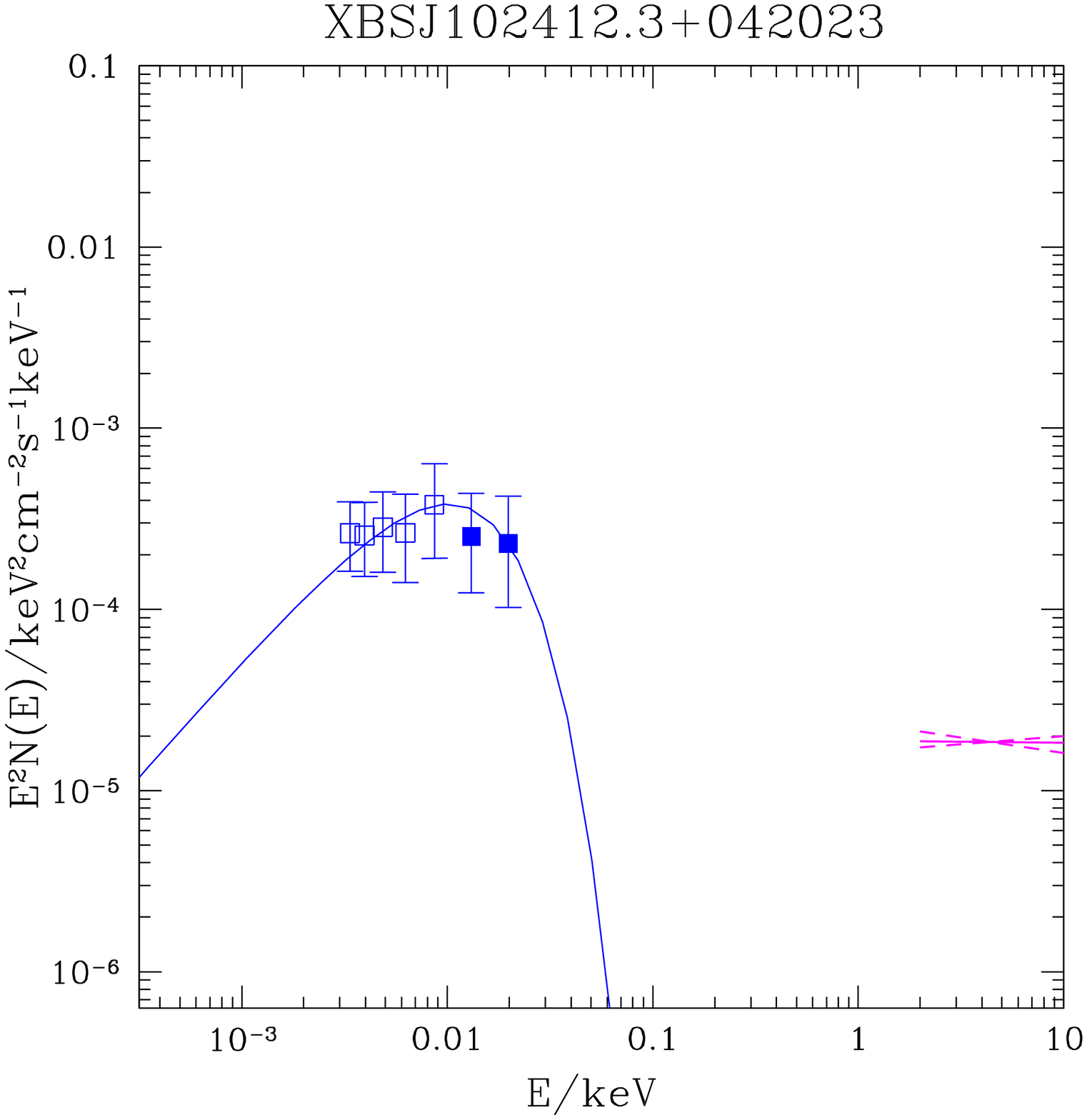}
  \includegraphics[height=5.6cm, width=6cm]{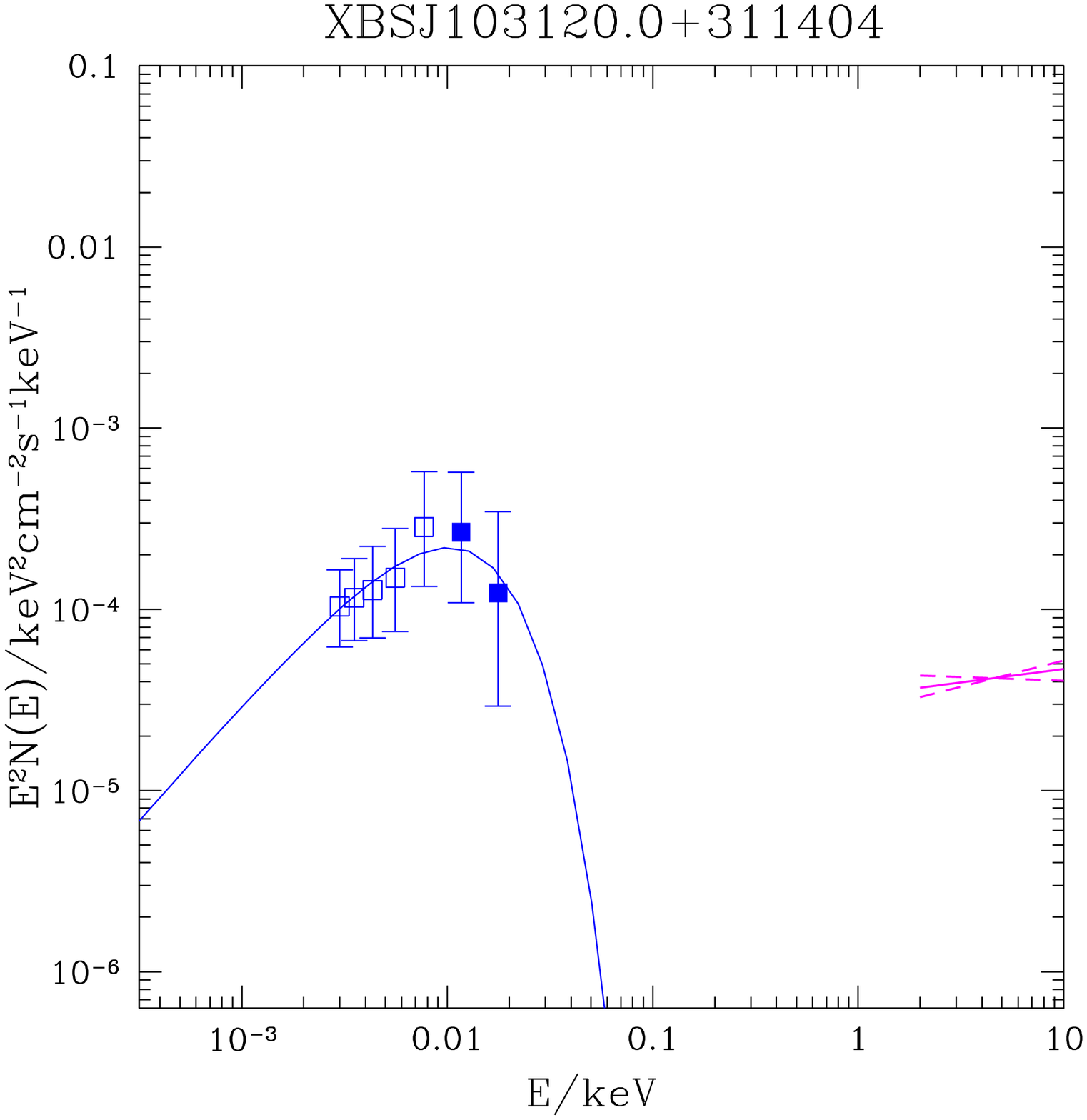}}      
\subfigure{ 
  \includegraphics[height=5.6cm, width=6cm]{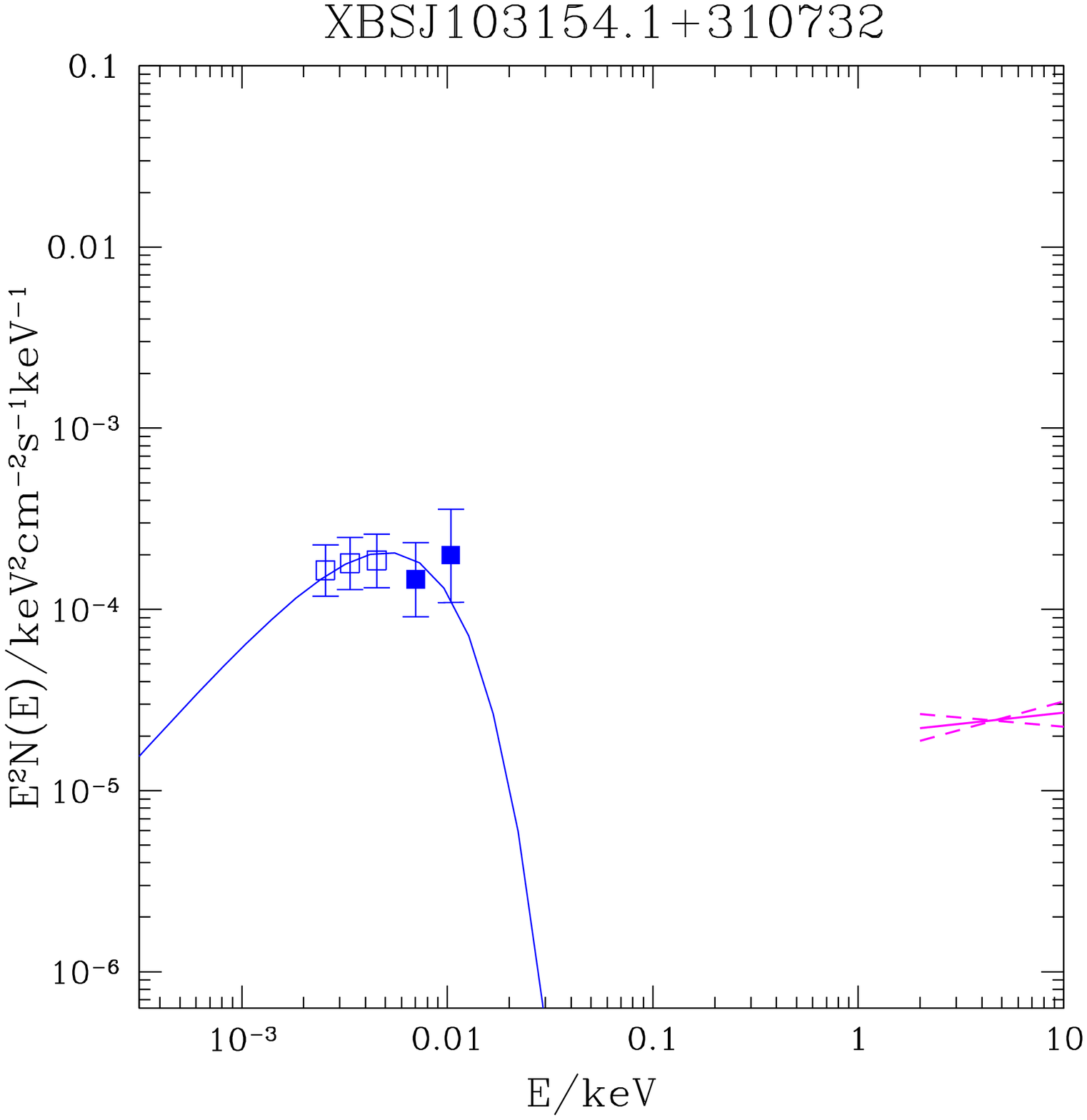}
  \includegraphics[height=5.6cm, width=6cm]{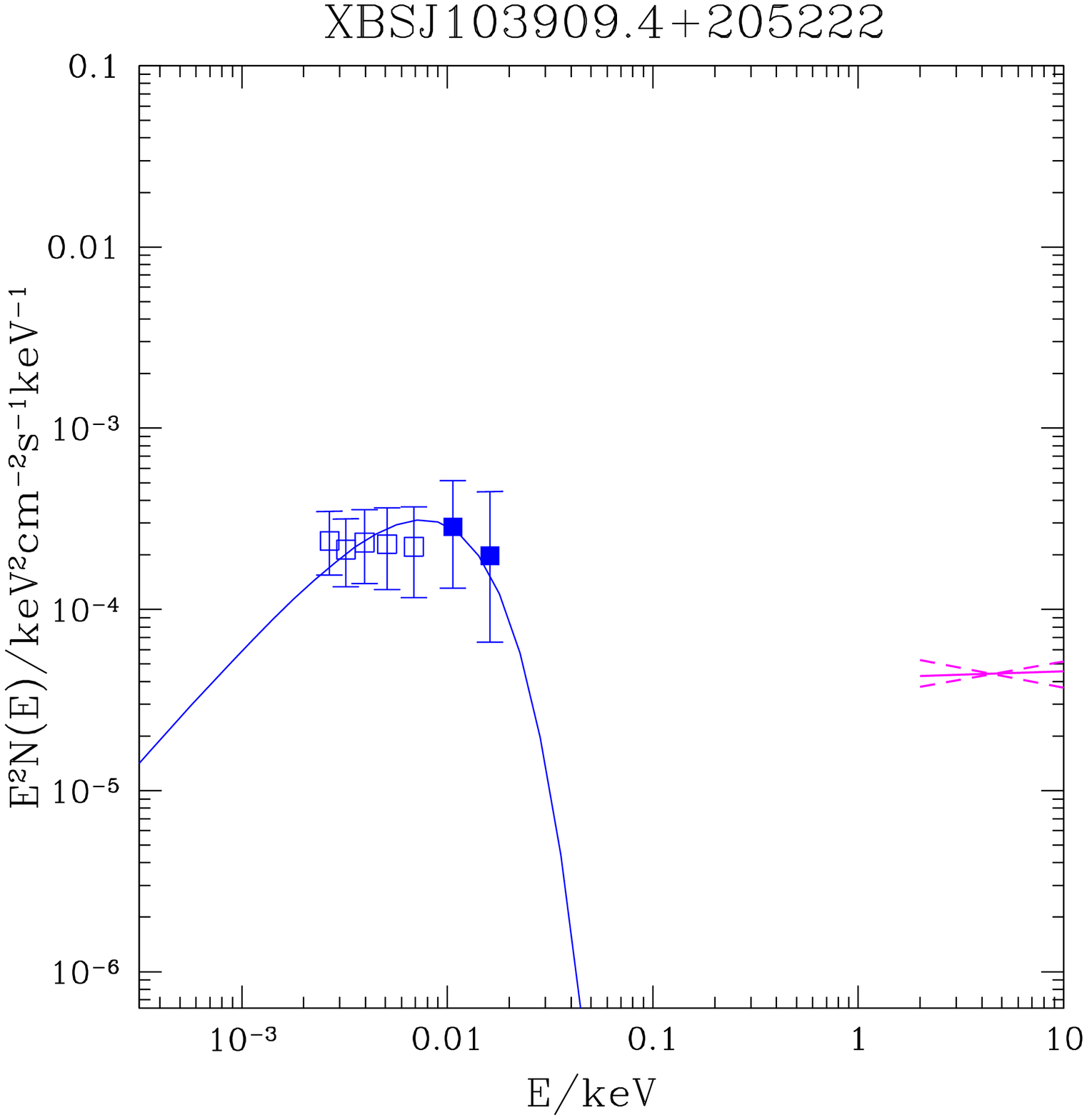}} 
  \end{figure*}
  
   \FloatBarrier
  
   \begin{figure*}
\centering     
\subfigure{ 
  \includegraphics[height=5.6cm, width=6cm]{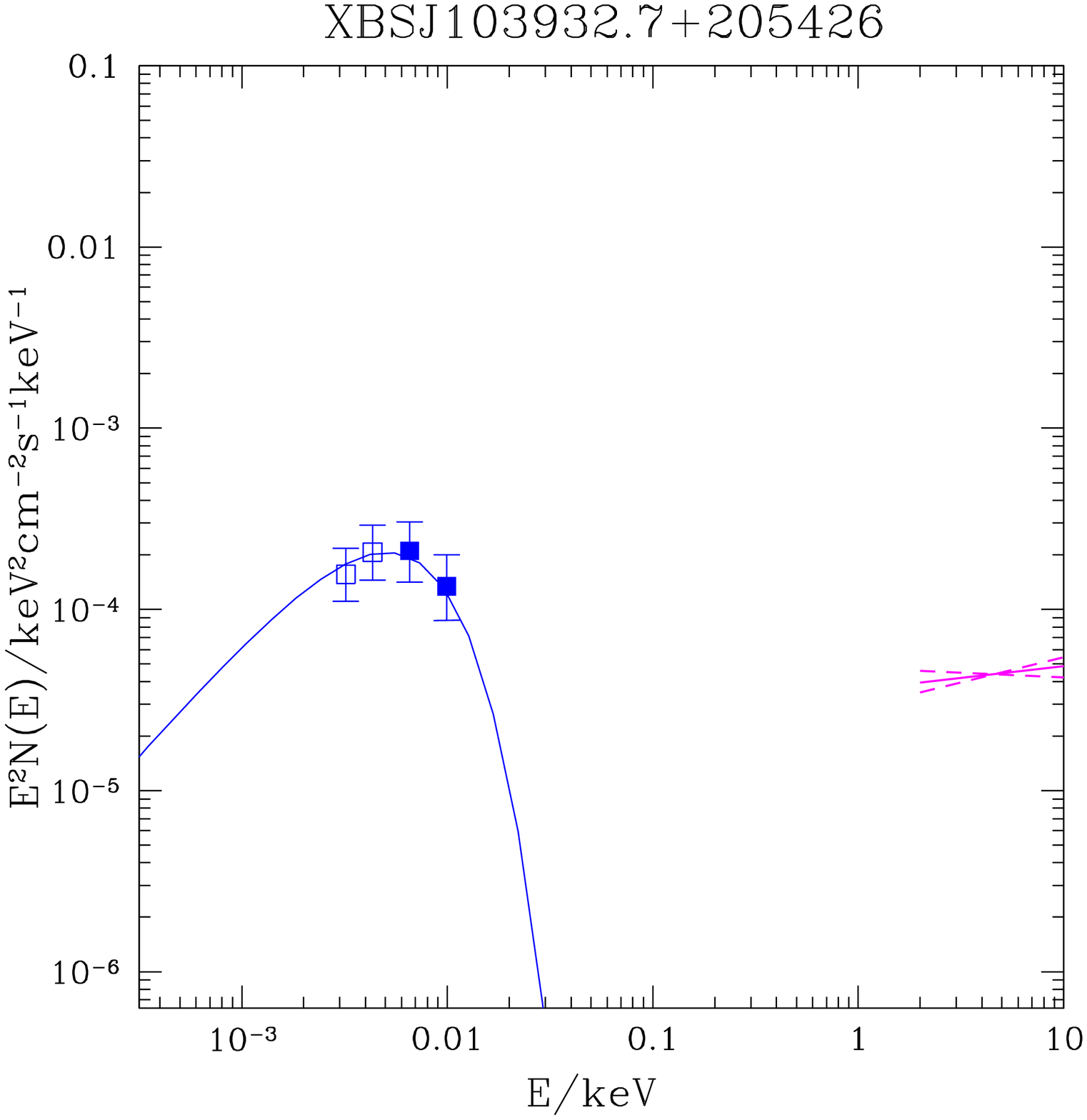}
  \includegraphics[height=5.6cm, width=6cm]{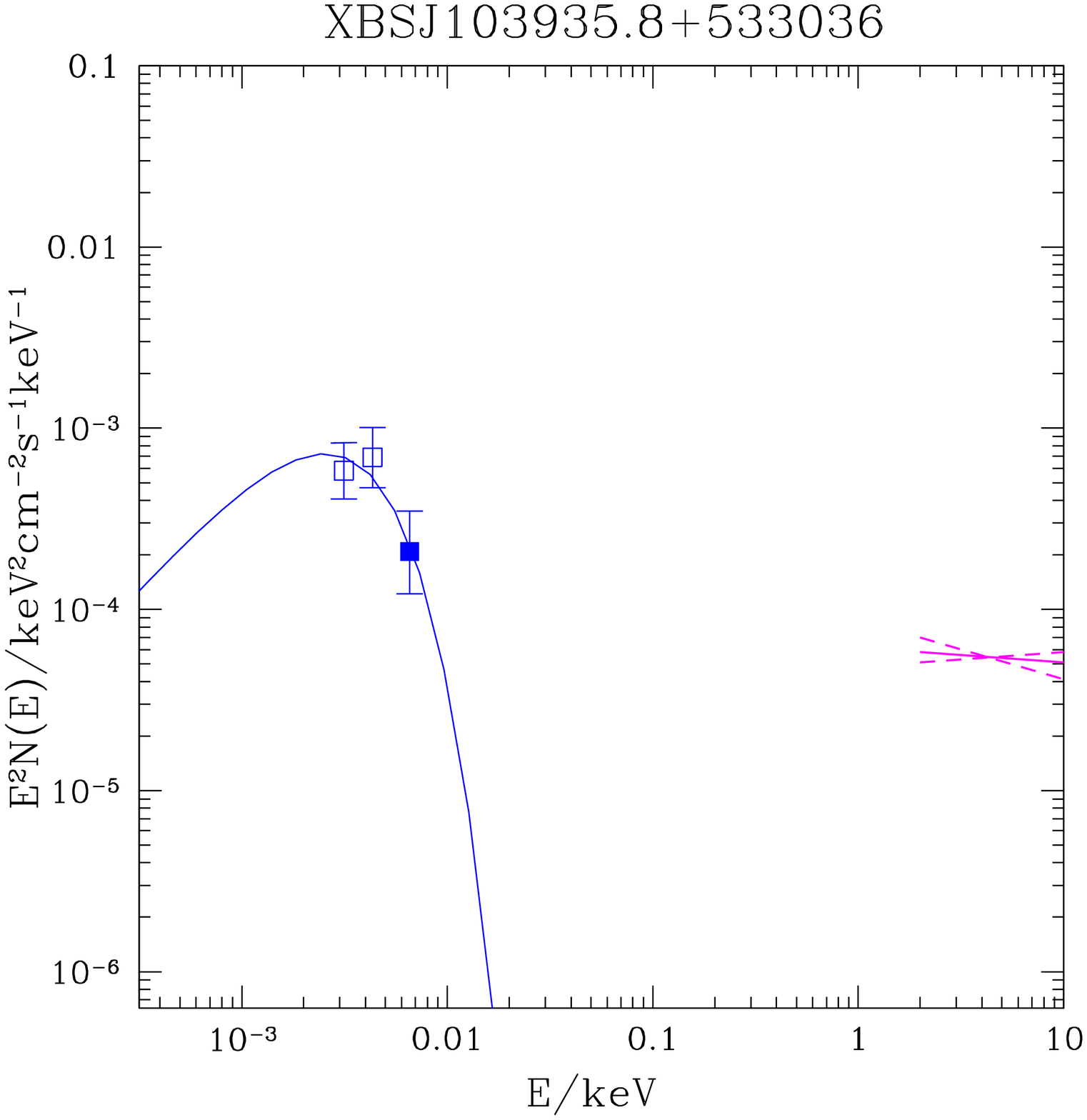}}      
\subfigure{ 
  \includegraphics[height=5.6cm, width=6cm]{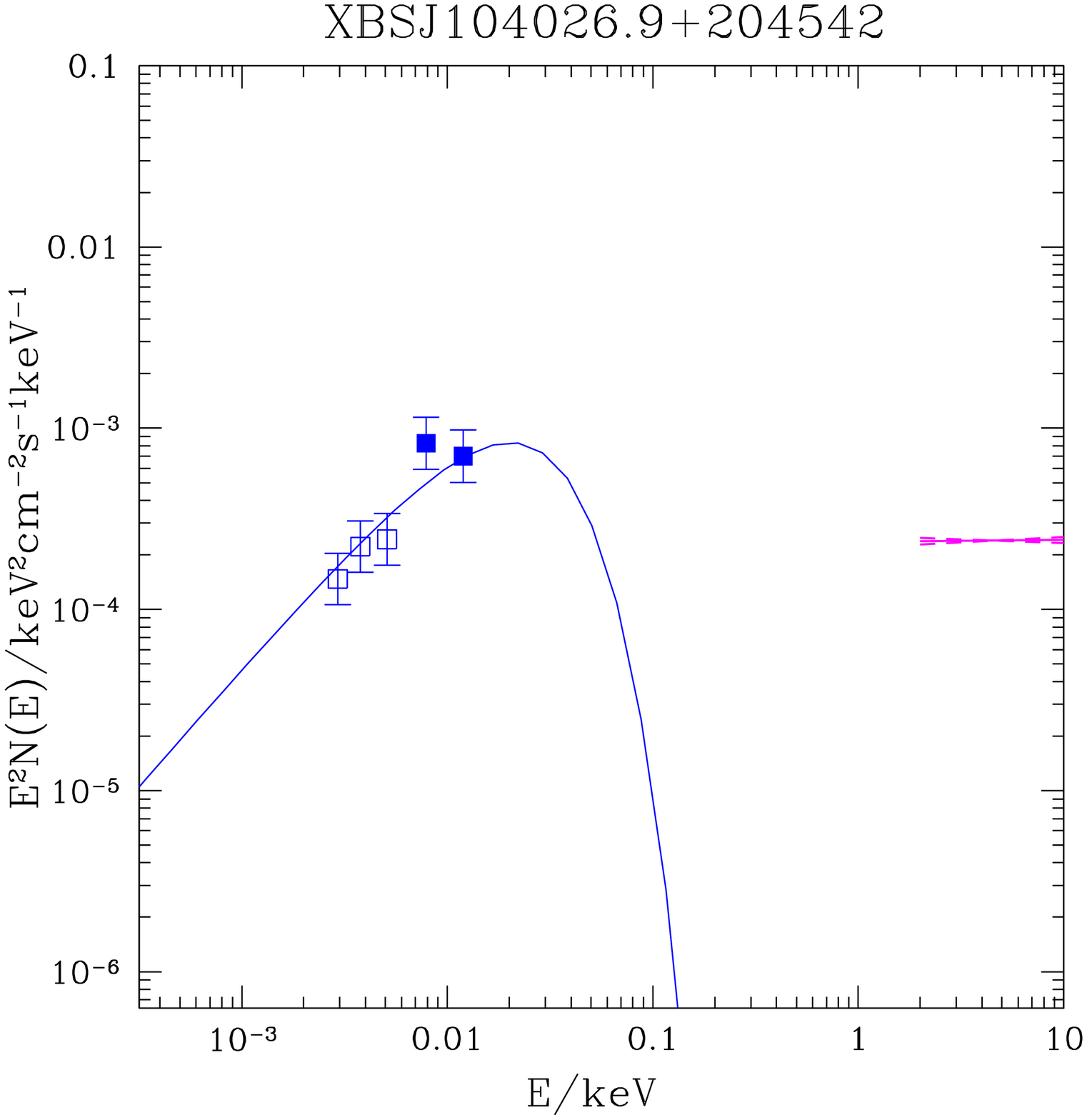}
  \includegraphics[height=5.6cm, width=6cm]{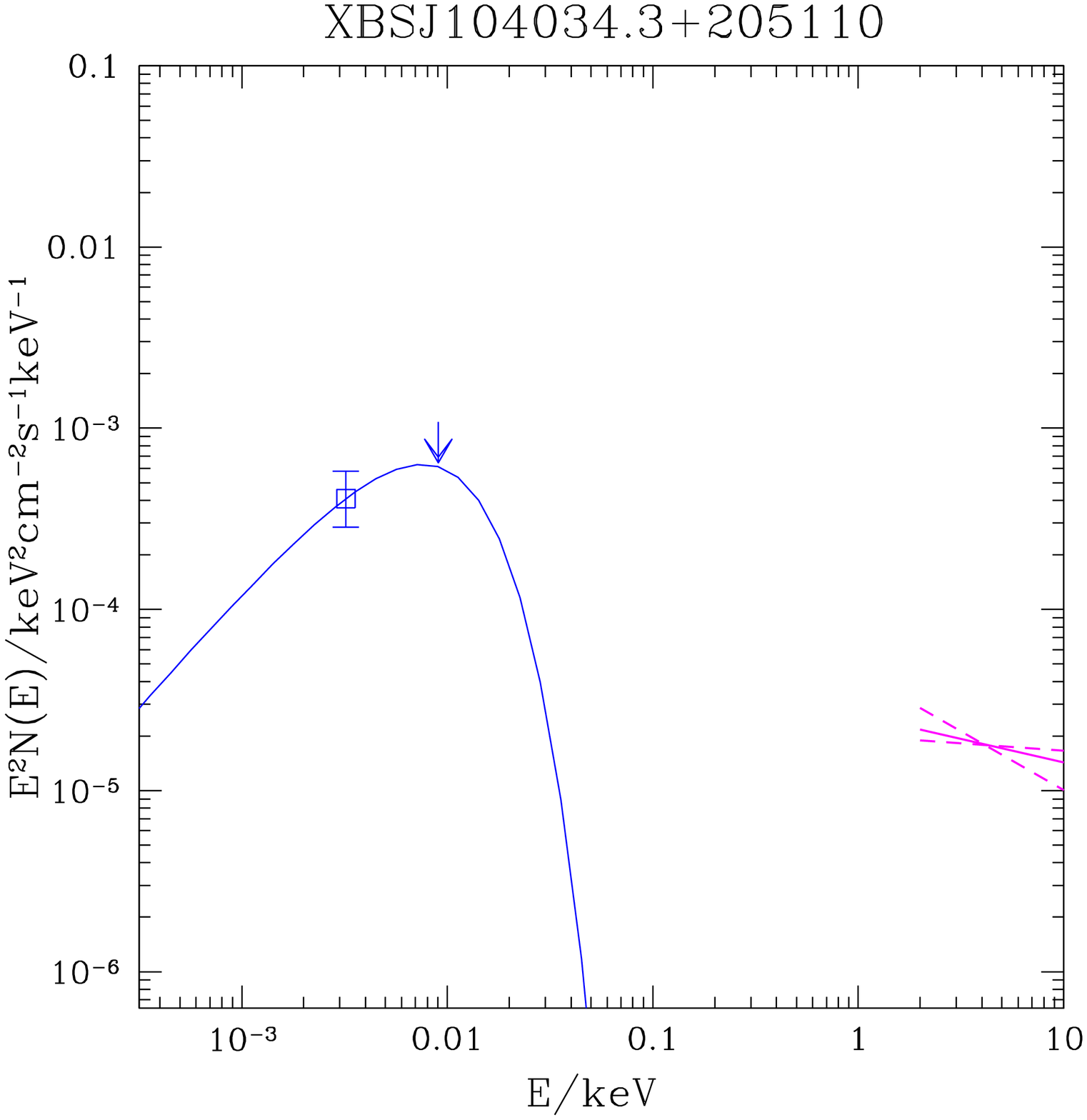}}    
\subfigure{ 
  \includegraphics[height=5.6cm, width=6cm]{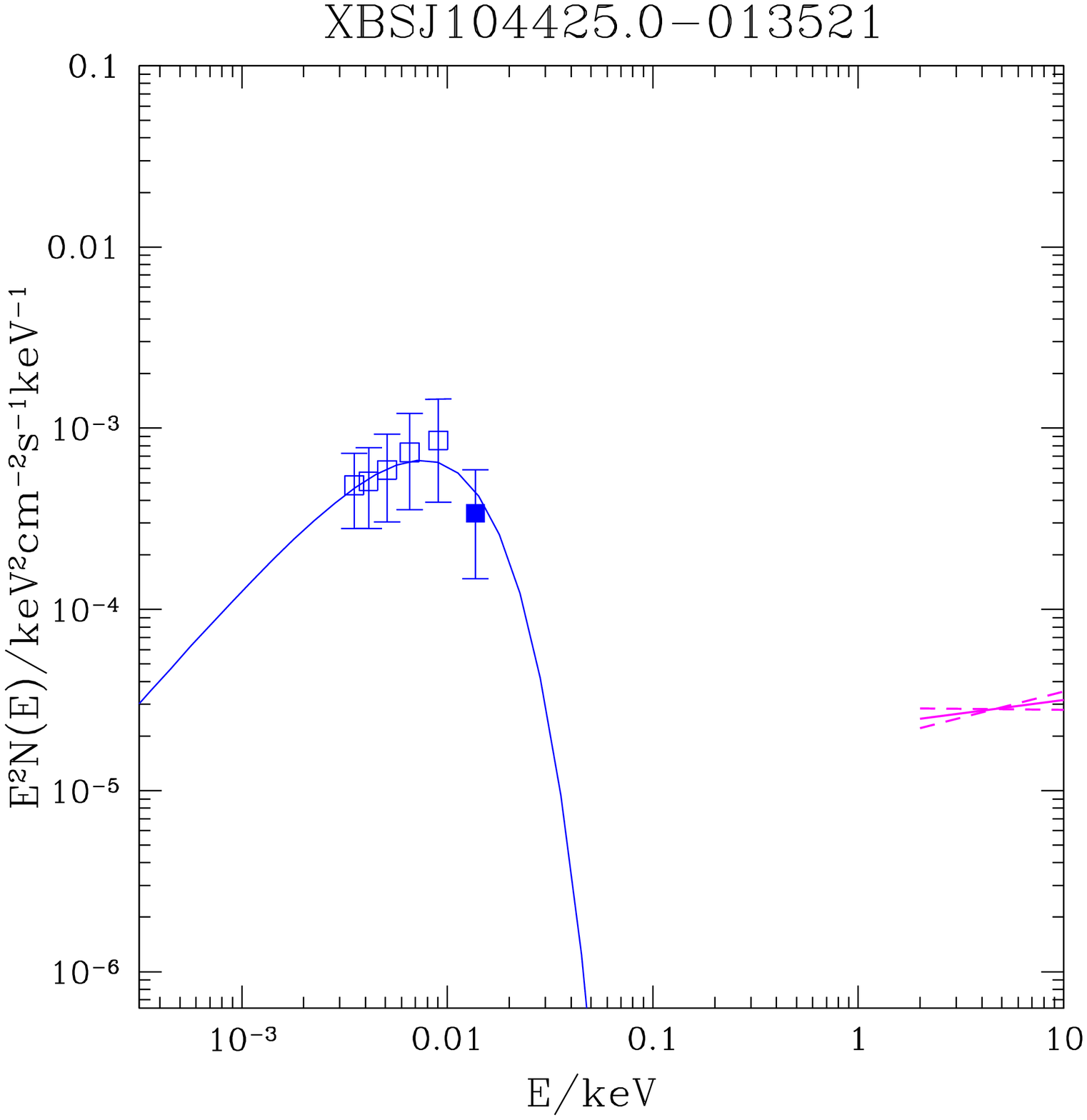}
  \includegraphics[height=5.6cm, width=6cm]{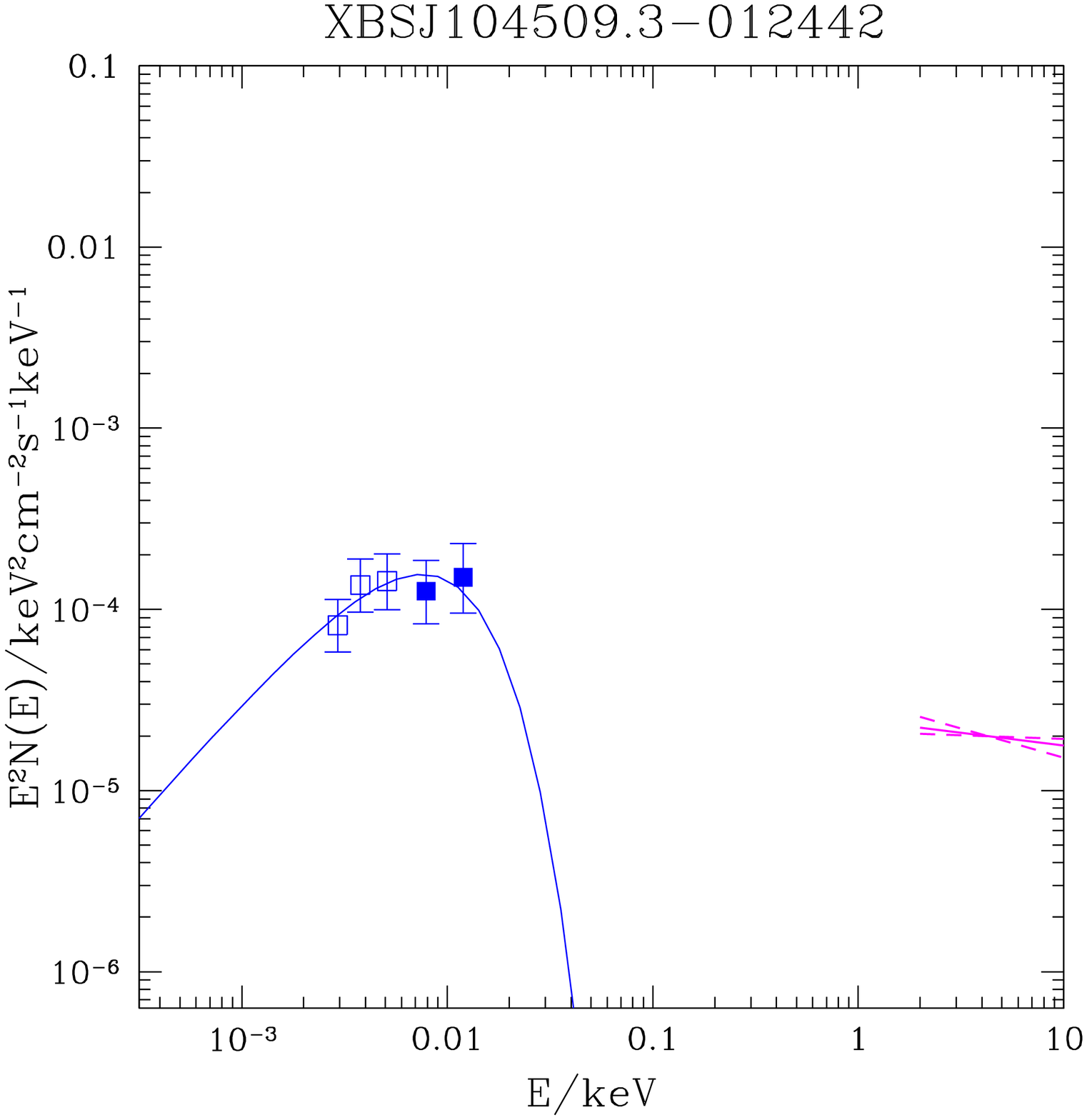}}      
\subfigure{ 
  \includegraphics[height=5.6cm, width=6cm]{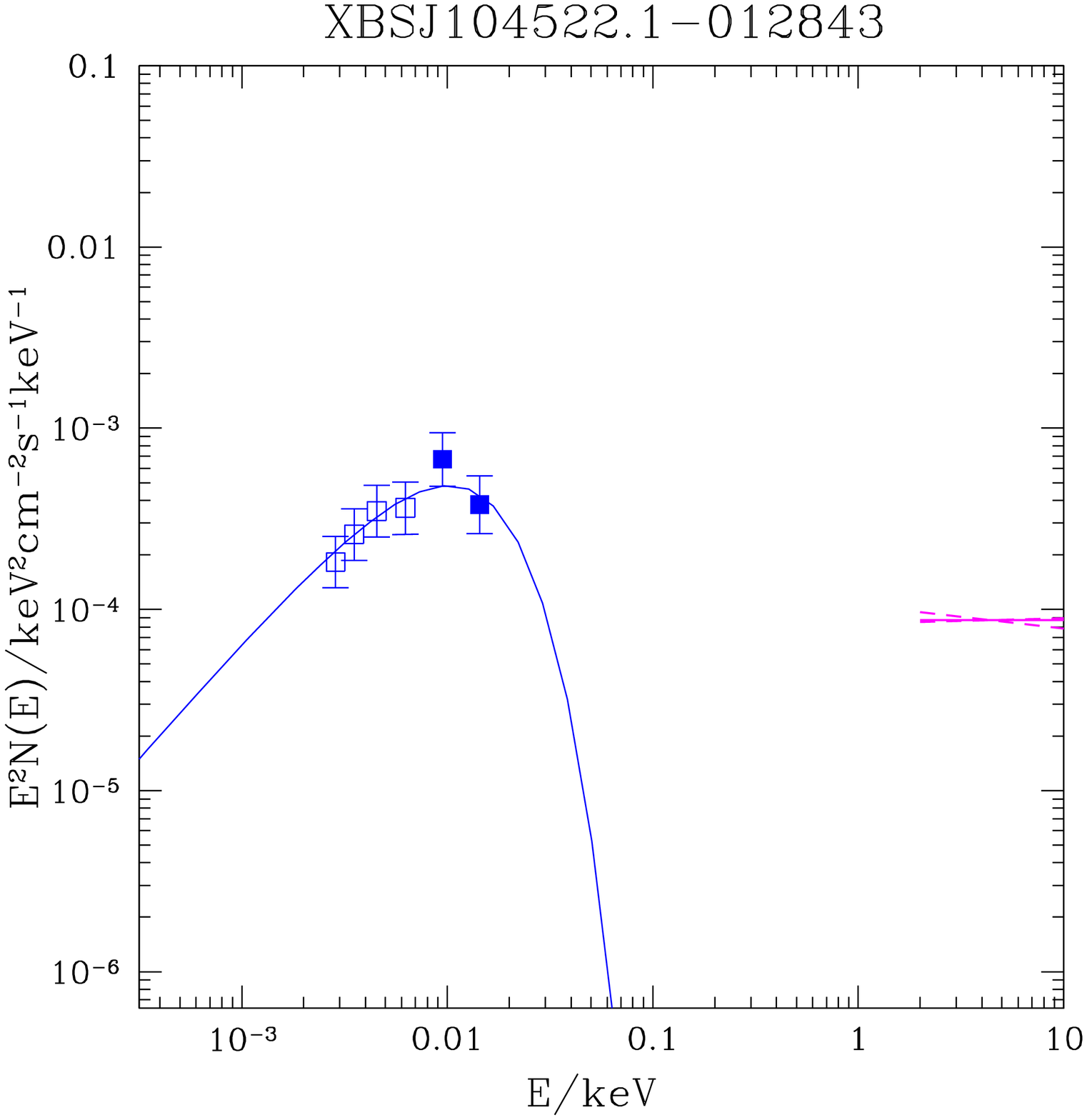}
  \includegraphics[height=5.6cm, width=6cm]{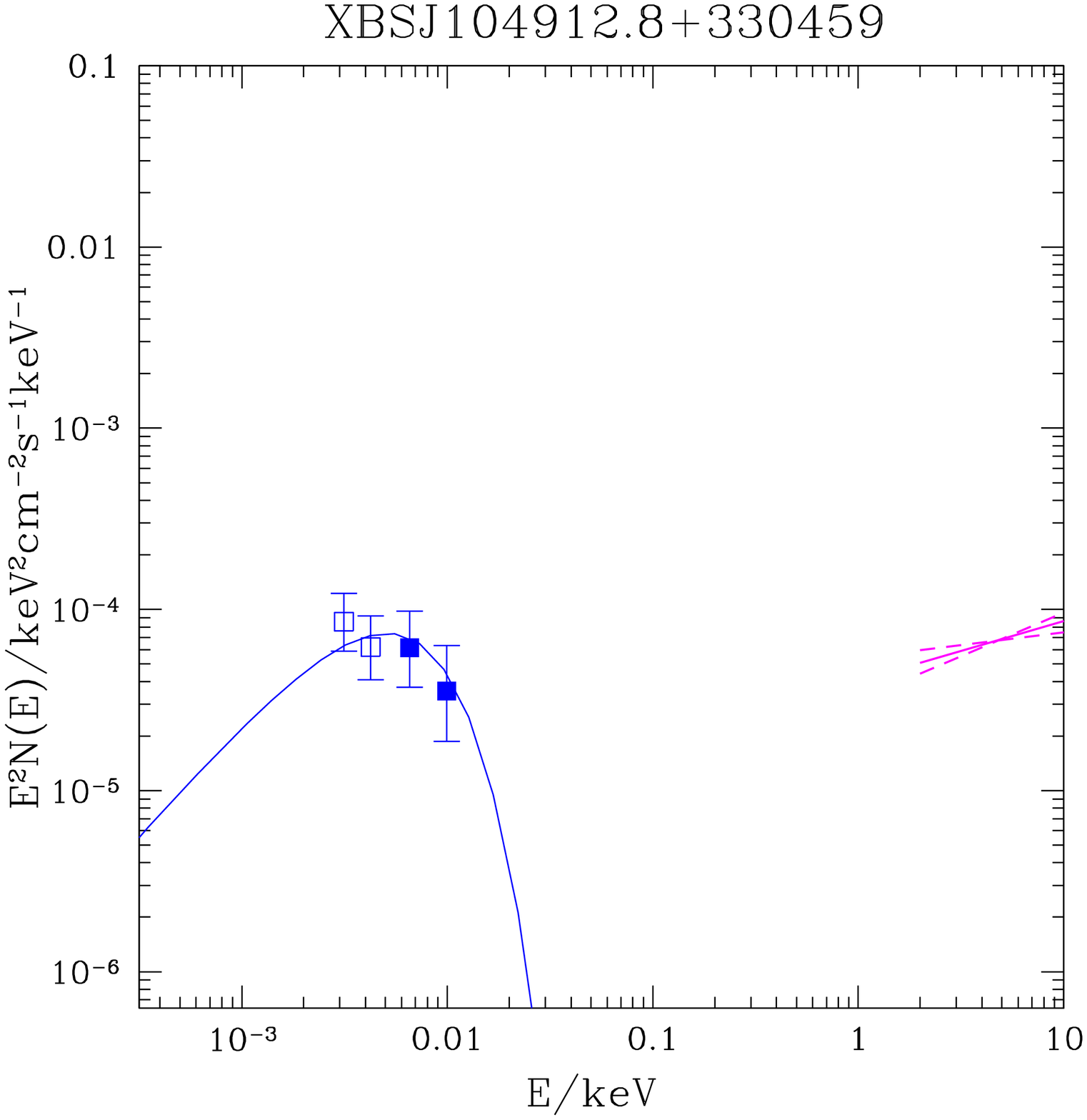}} 
  \end{figure*} 
  
   \FloatBarrier
  
   \begin{figure*}
\centering   
\subfigure{ 
  \includegraphics[height=5.6cm, width=6cm]{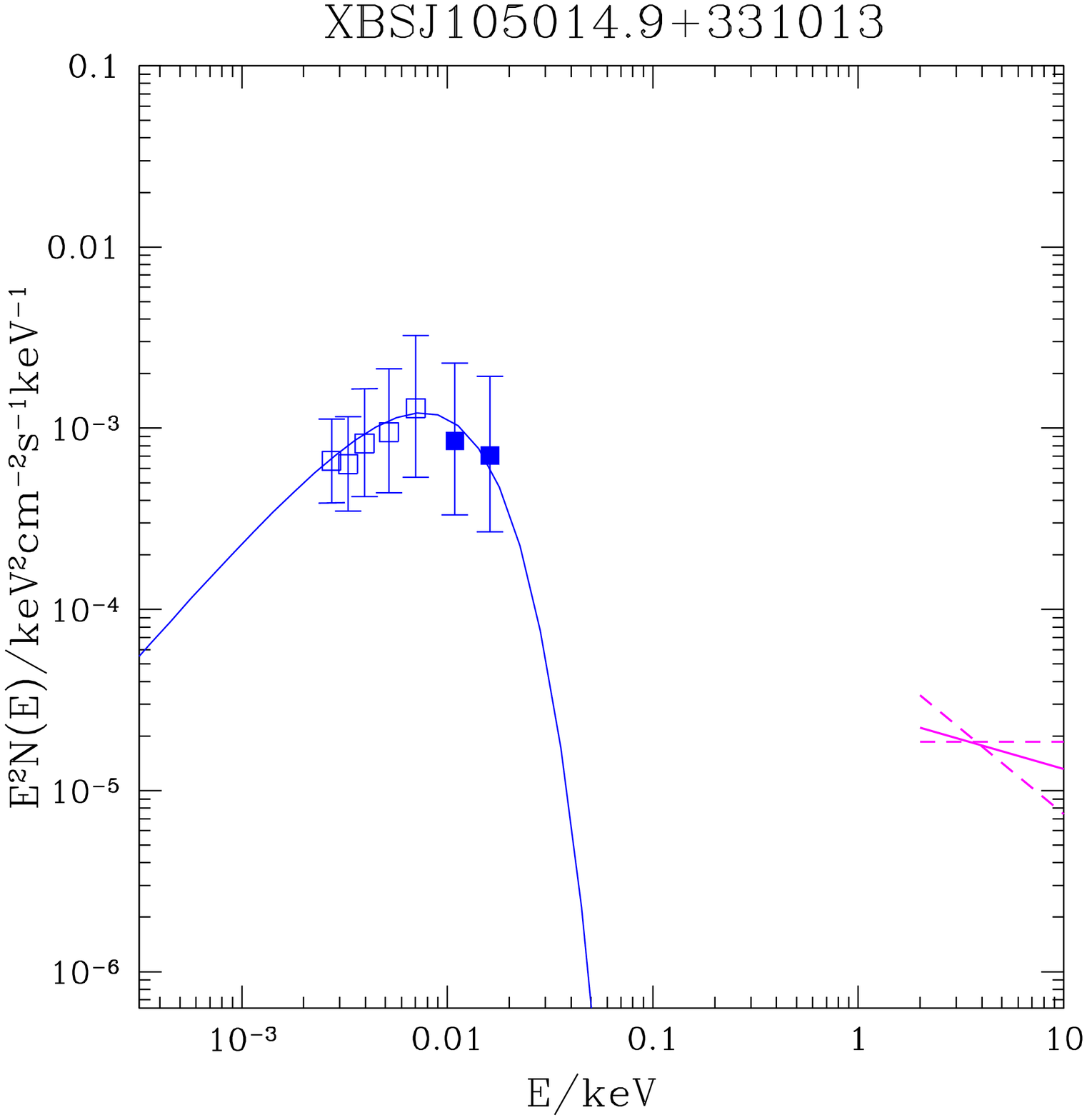}
  \includegraphics[height=5.6cm, width=6cm]{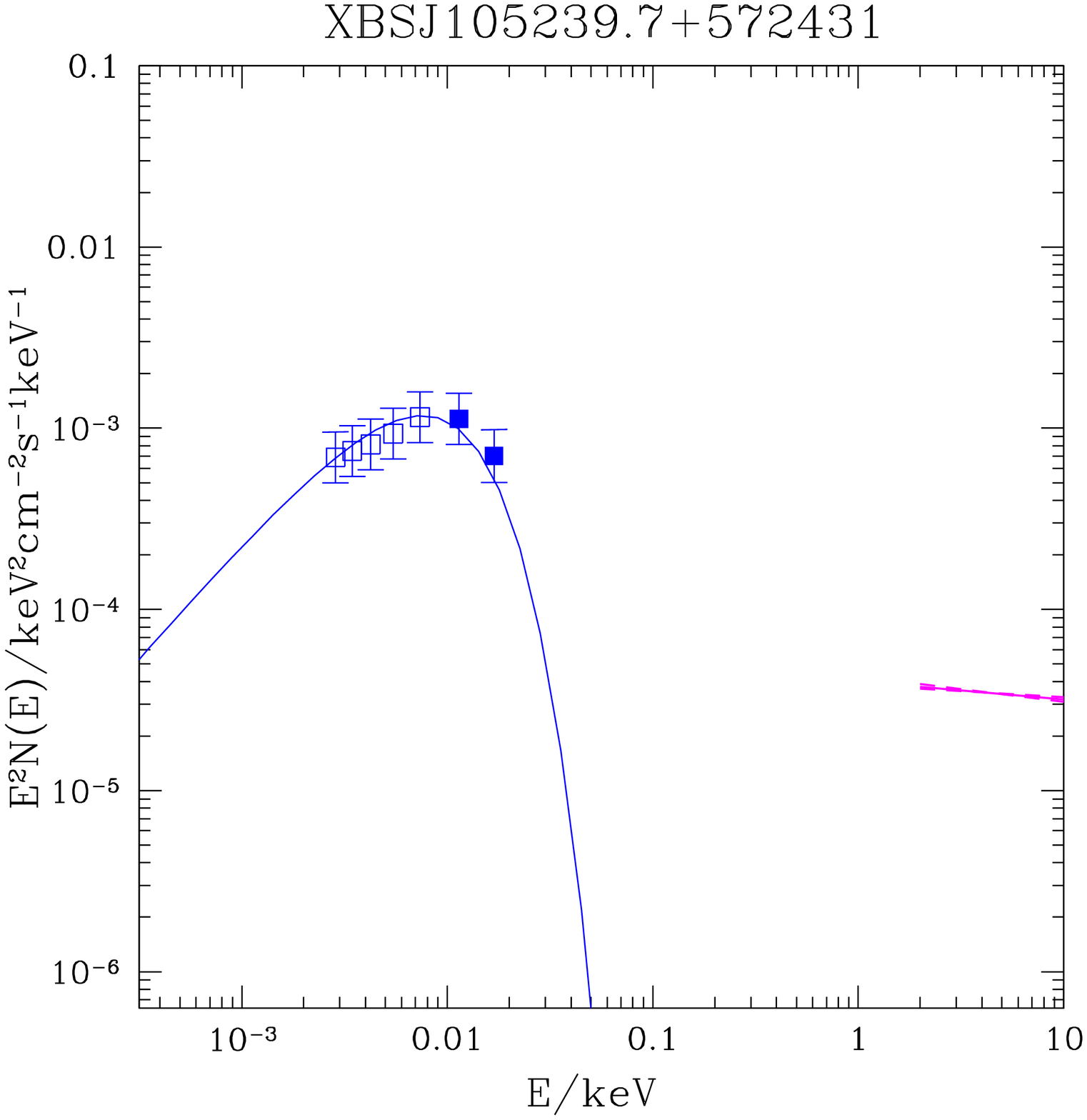}}      
\subfigure{ 
  \includegraphics[height=5.6cm, width=6cm]{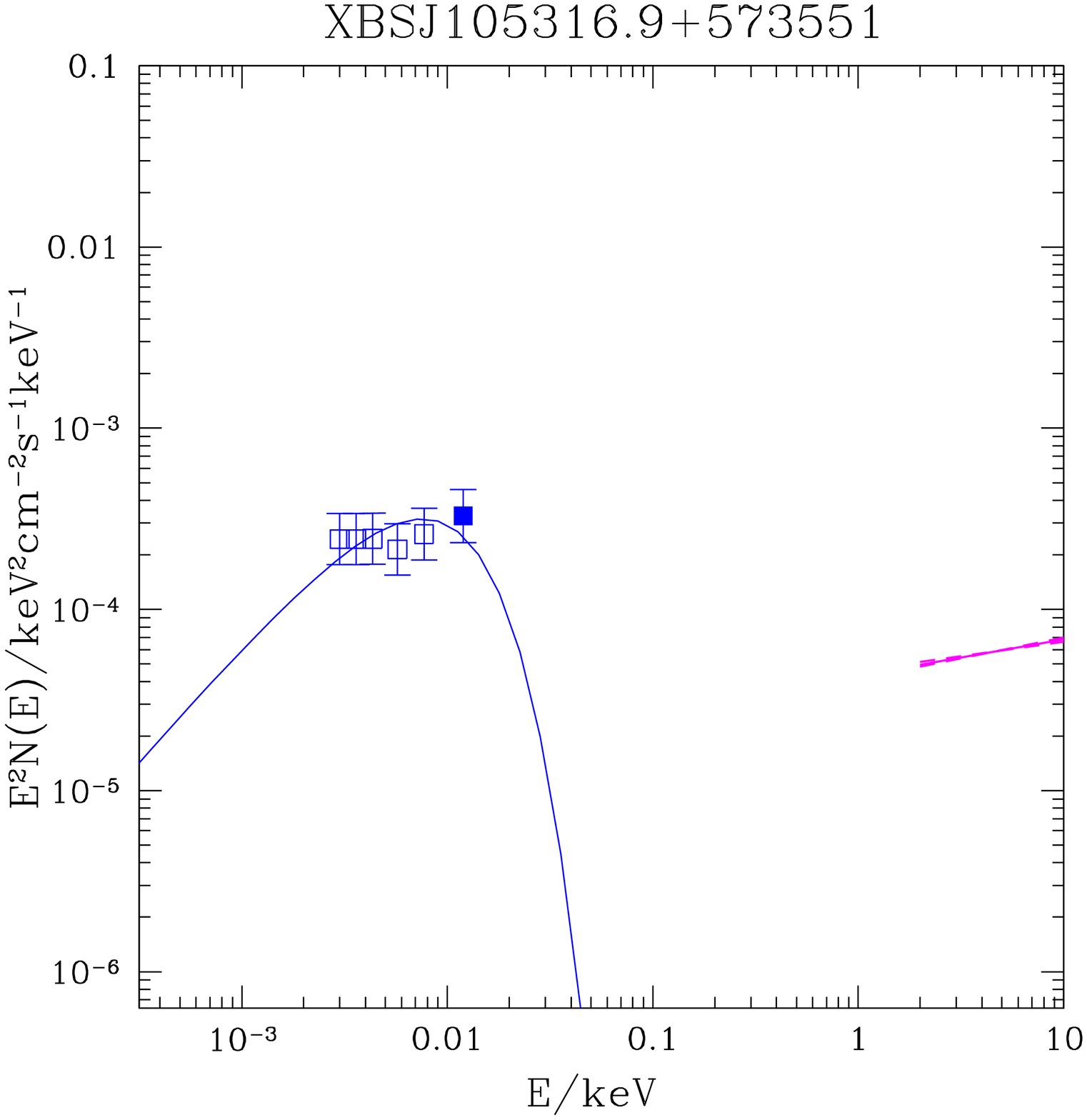}
  \includegraphics[height=5.6cm, width=6cm]{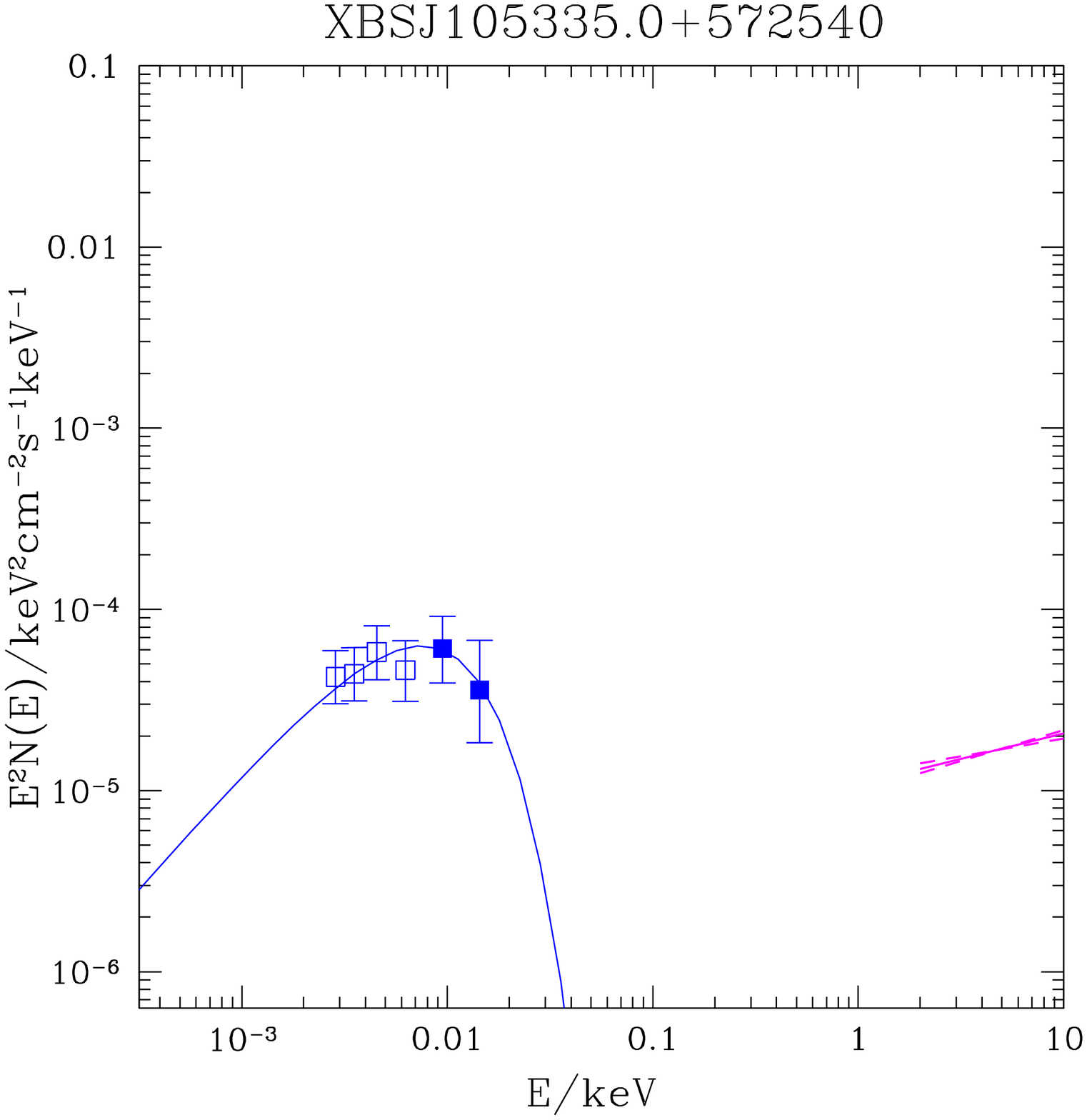}}     
 \subfigure{ 
  \includegraphics[height=5.6cm, width=6cm]{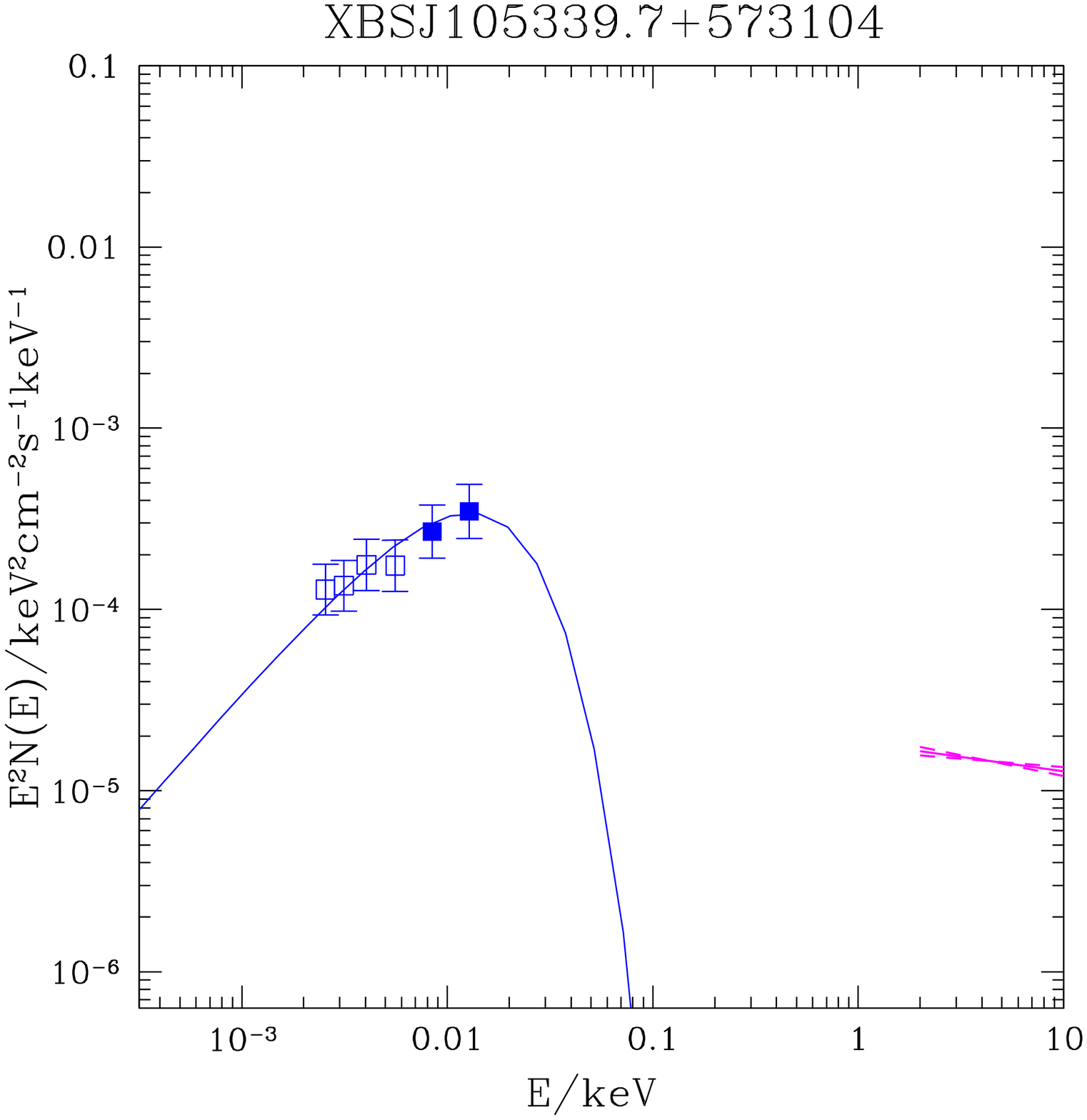}
  \includegraphics[height=5.6cm, width=6cm]{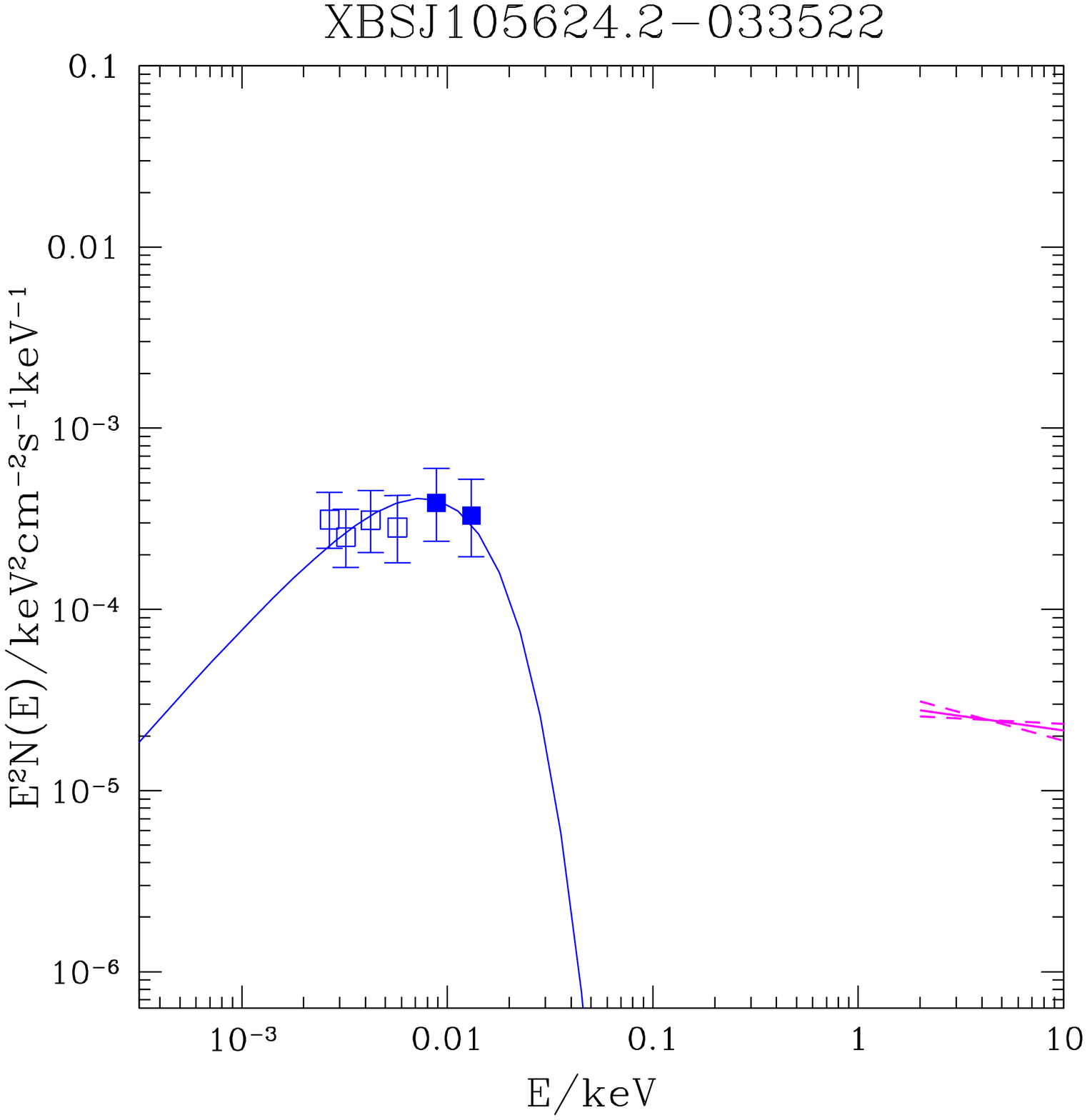}}      
\subfigure{ 
  \includegraphics[height=5.6cm, width=6cm]{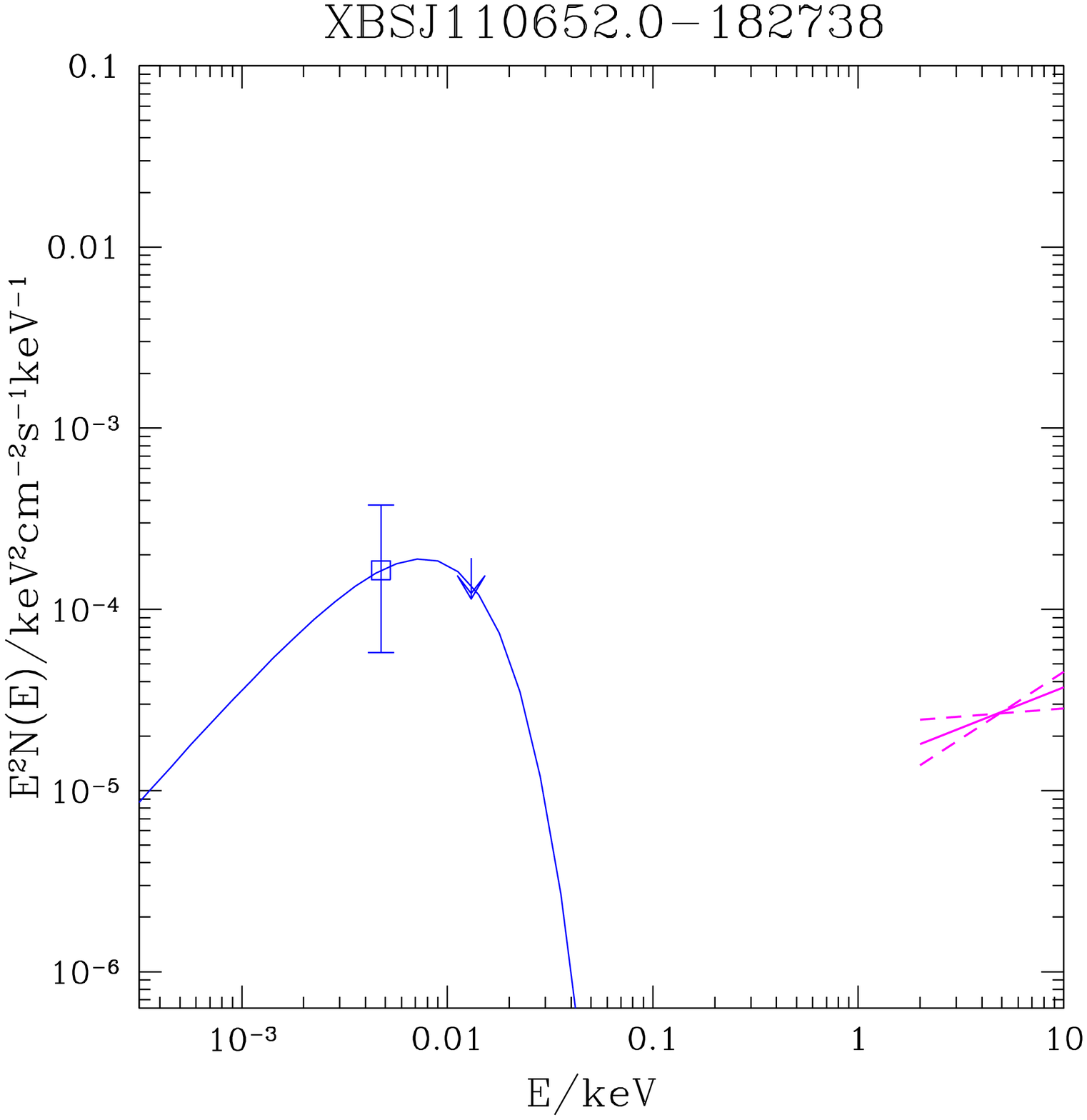}
  \includegraphics[height=5.6cm, width=6cm]{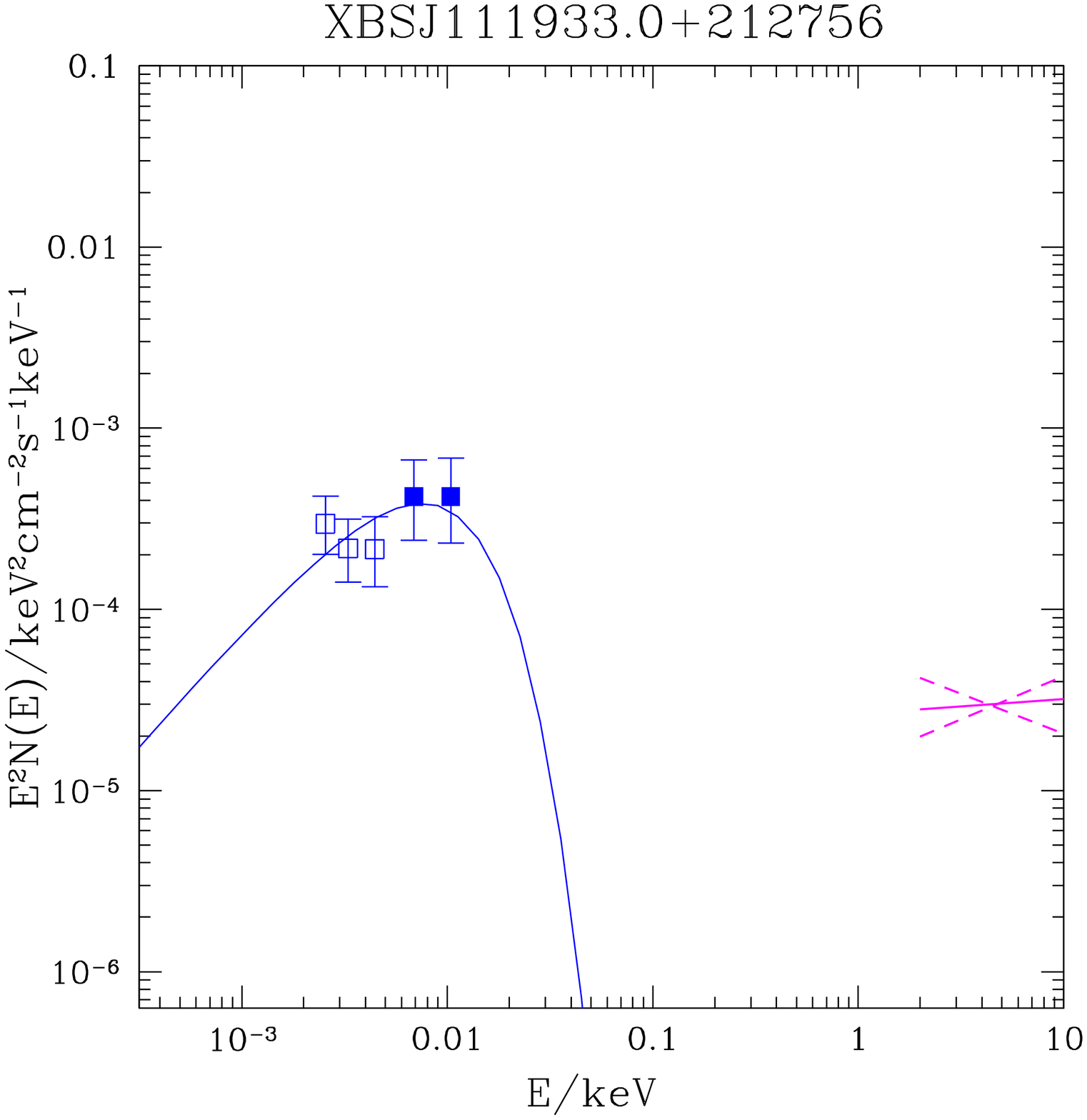}}  
  \end{figure*}
  
   \FloatBarrier
  
   \begin{figure*}
\centering    
\subfigure{ 
  \includegraphics[height=5.6cm, width=6cm]{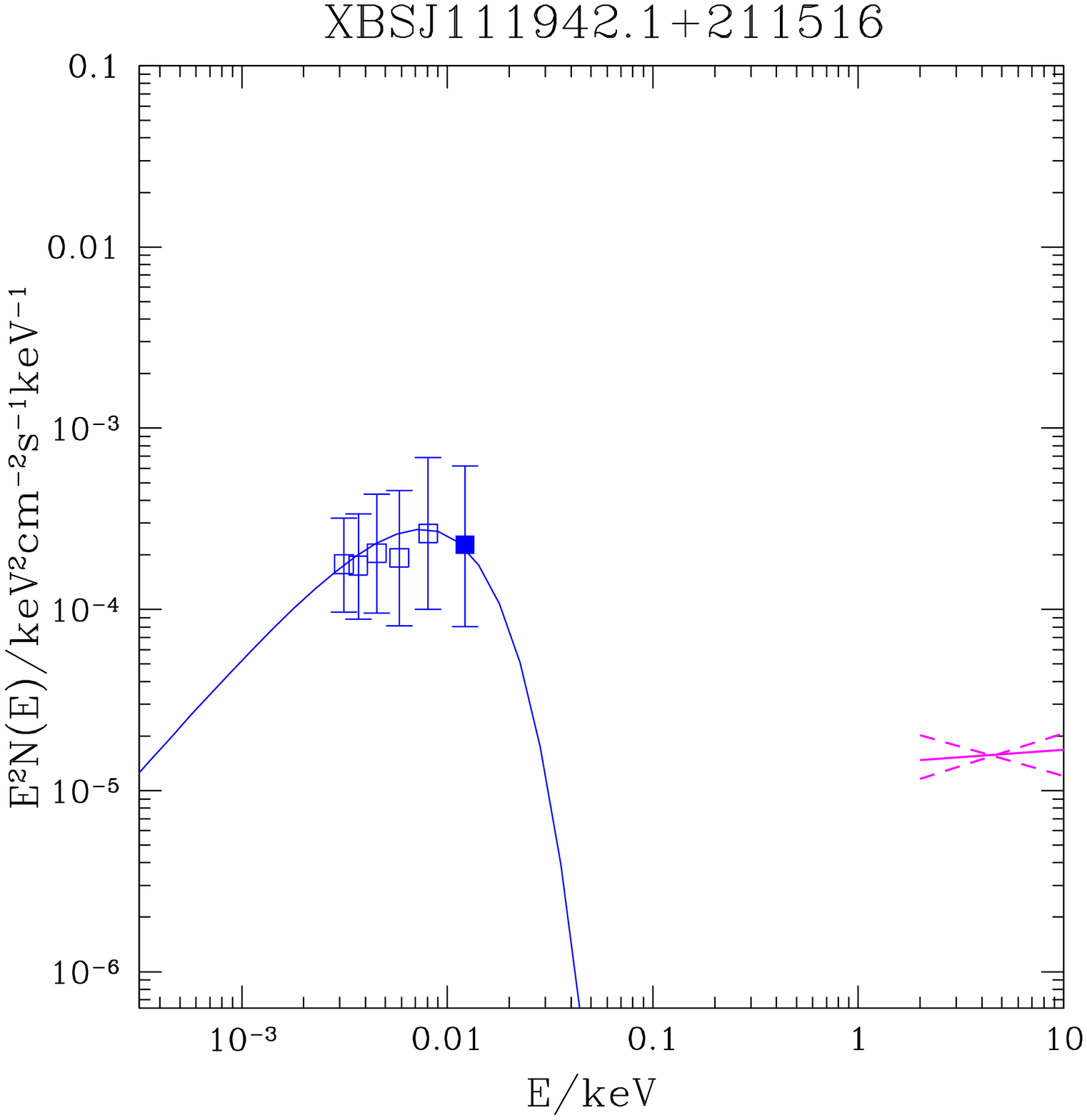}
  \includegraphics[height=5.6cm, width=6cm]{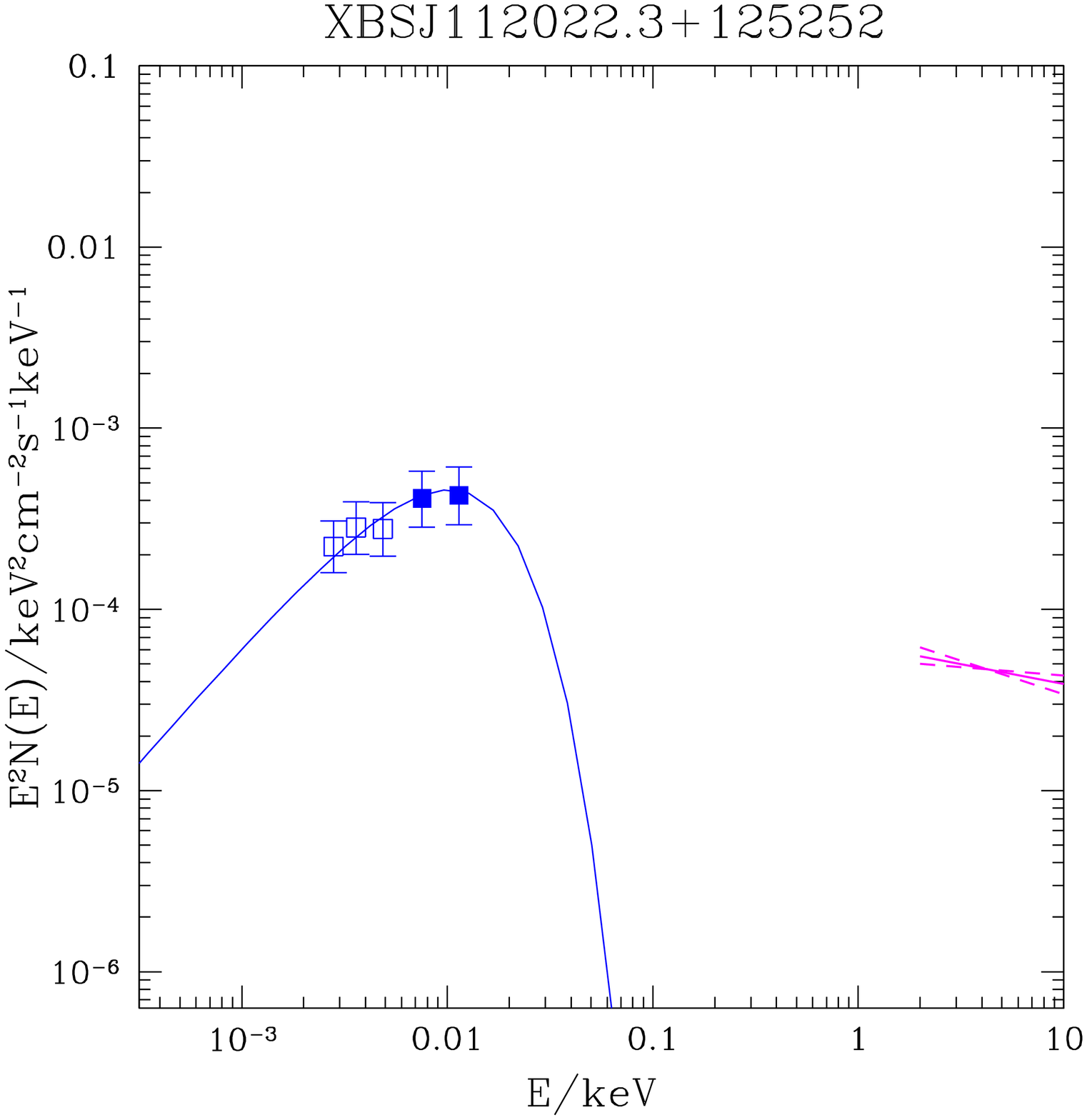}}      
\subfigure{ 
  \includegraphics[height=5.6cm, width=6cm]{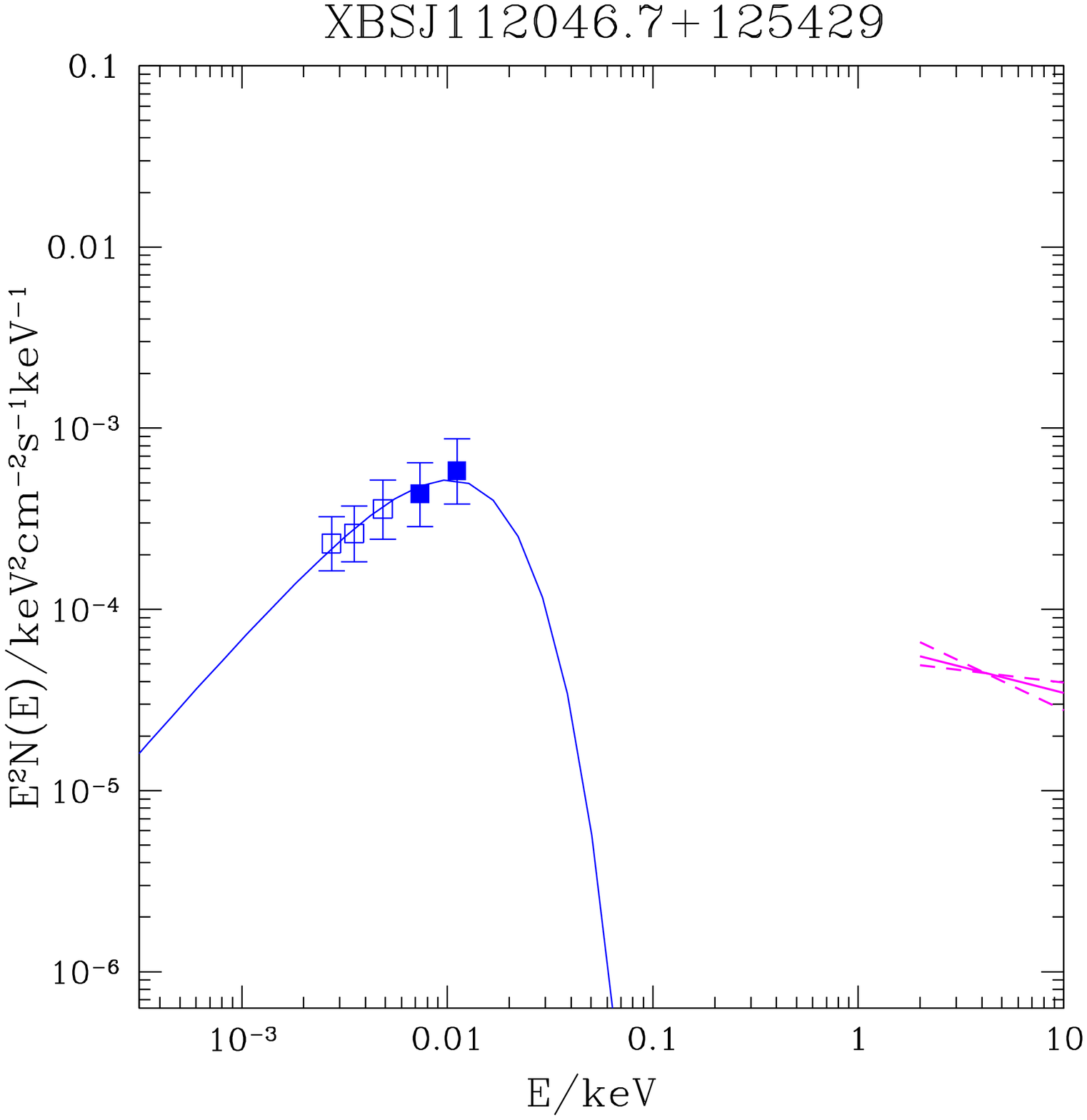}
  \includegraphics[height=5.6cm, width=6cm]{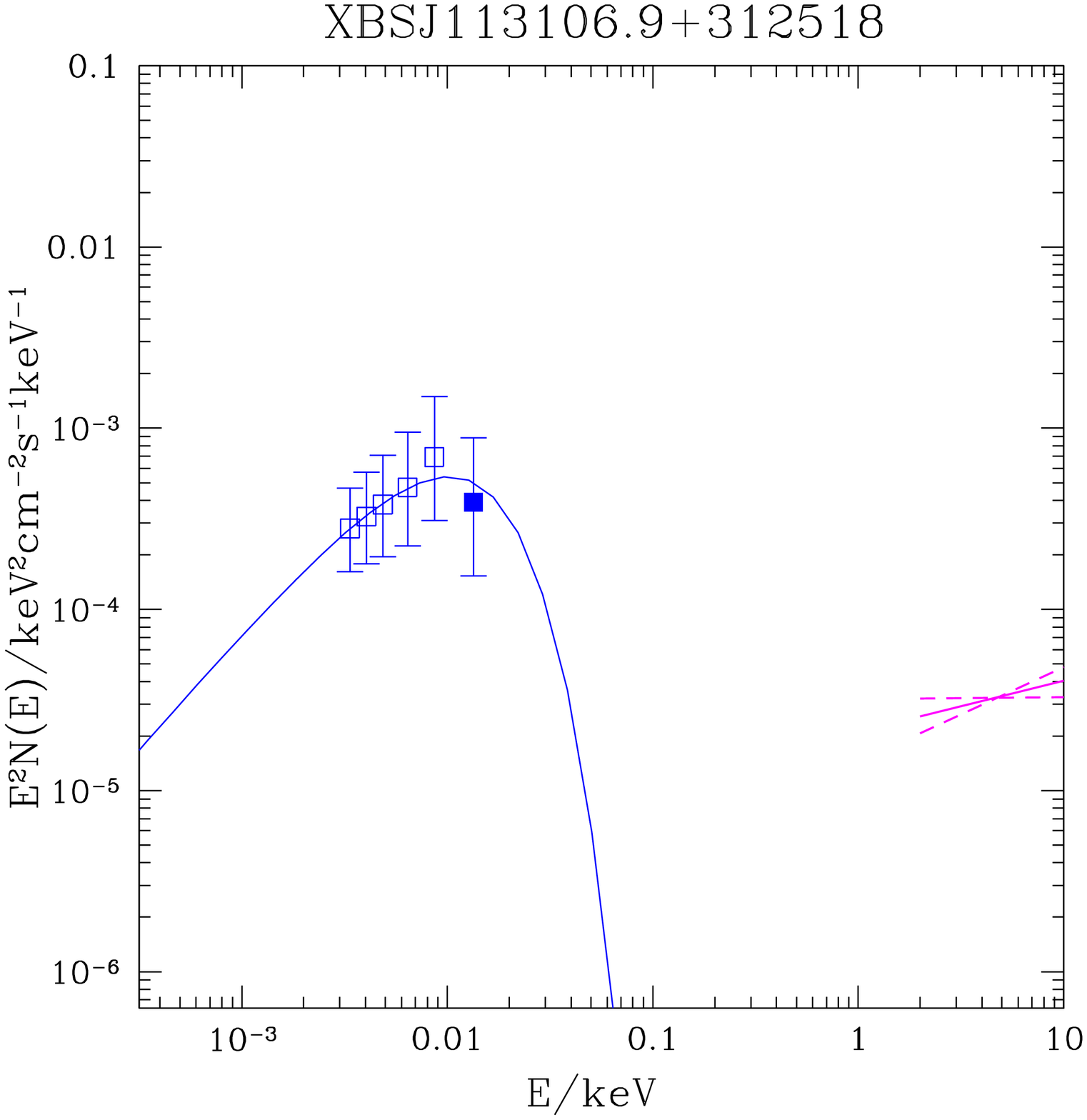}}      
\subfigure{ 
  \includegraphics[height=5.6cm, width=6cm]{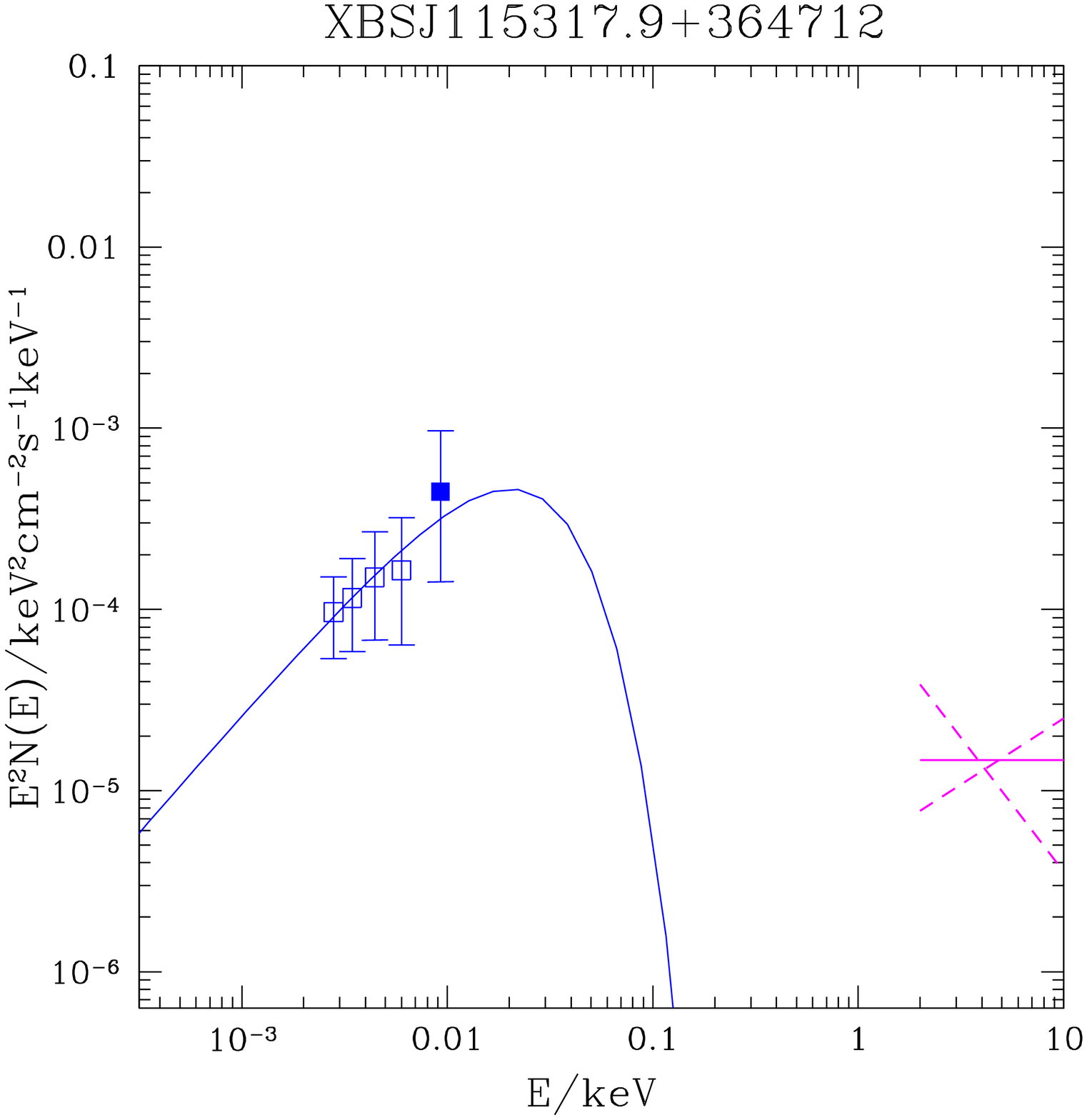}
  \includegraphics[height=5.6cm, width=6cm]{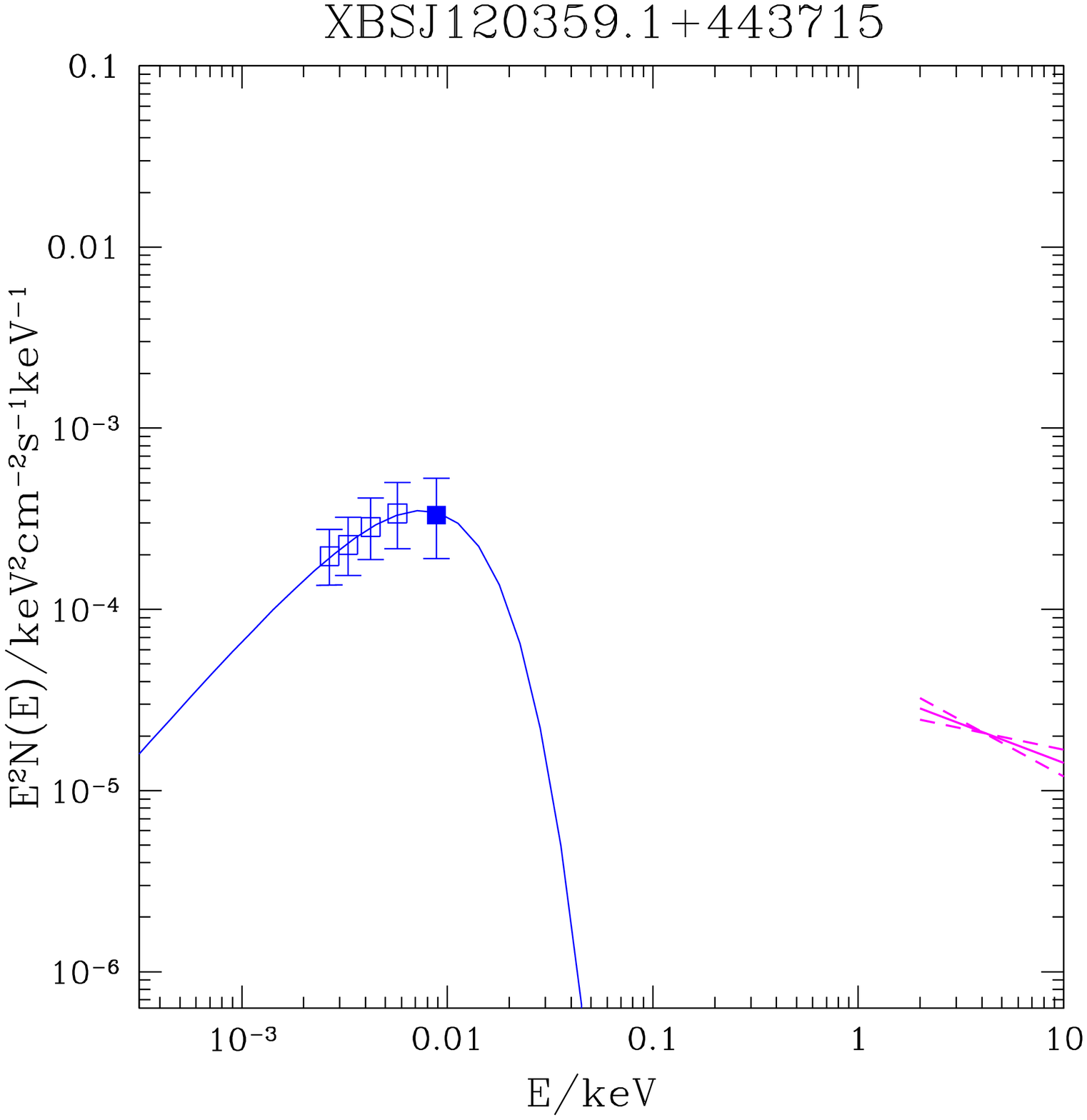}}      
\subfigure{ 
  \includegraphics[height=5.6cm, width=6cm]{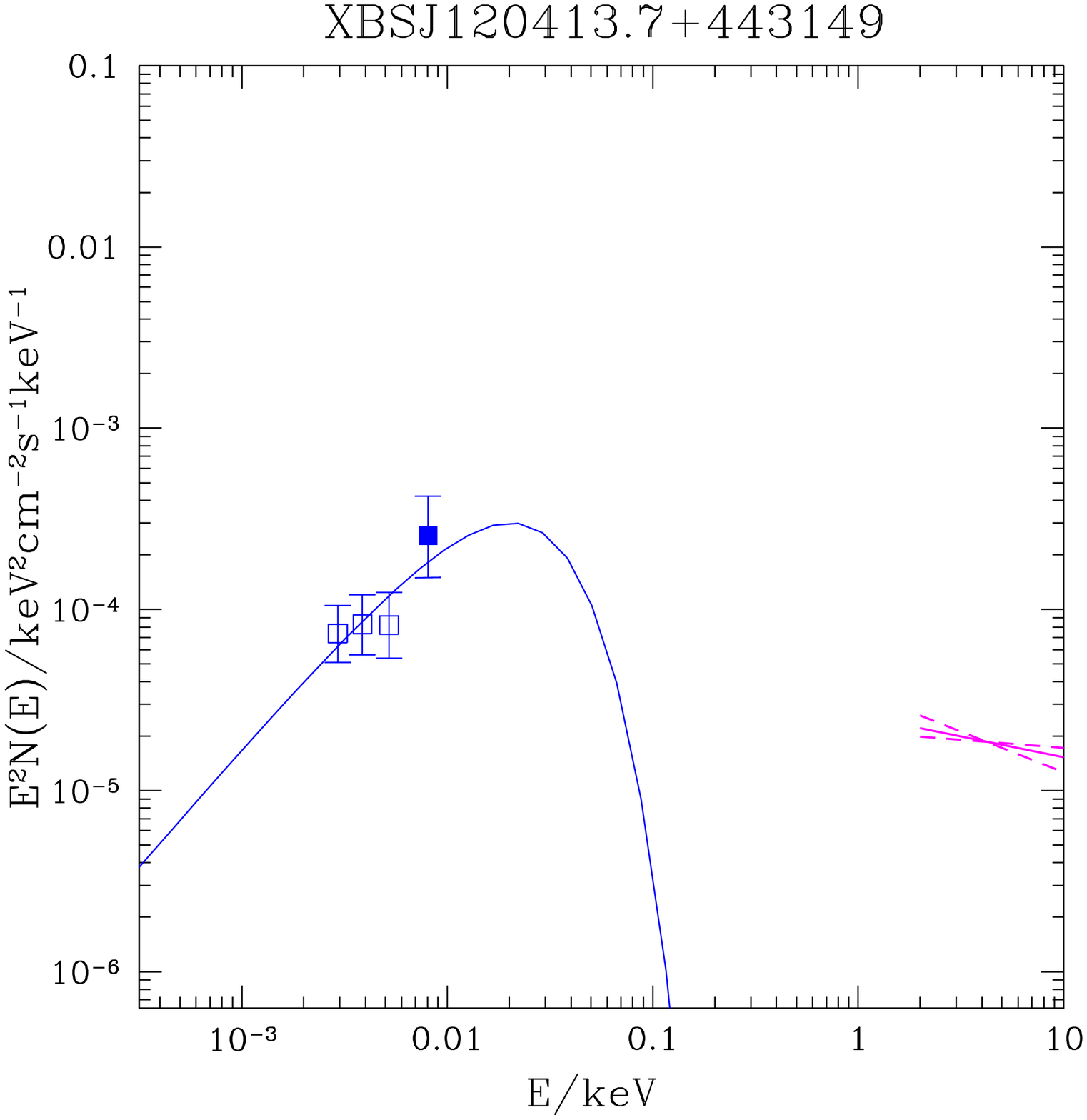}
  \includegraphics[height=5.6cm, width=6cm]{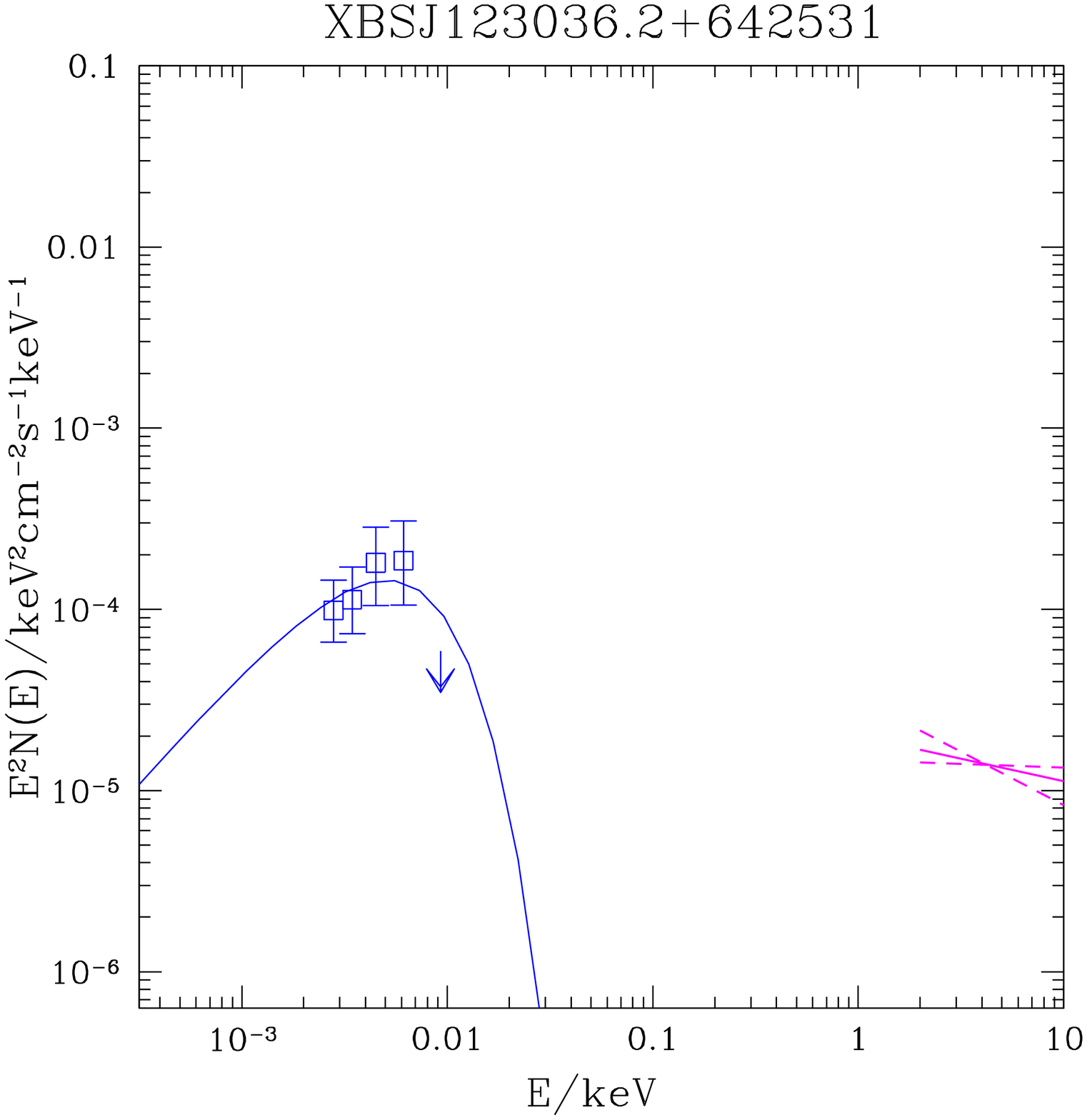}}  
  \end{figure*}
  
   \FloatBarrier
   
   \begin{figure*}
\centering    
\subfigure{ 
  \includegraphics[height=5.6cm, width=6cm]{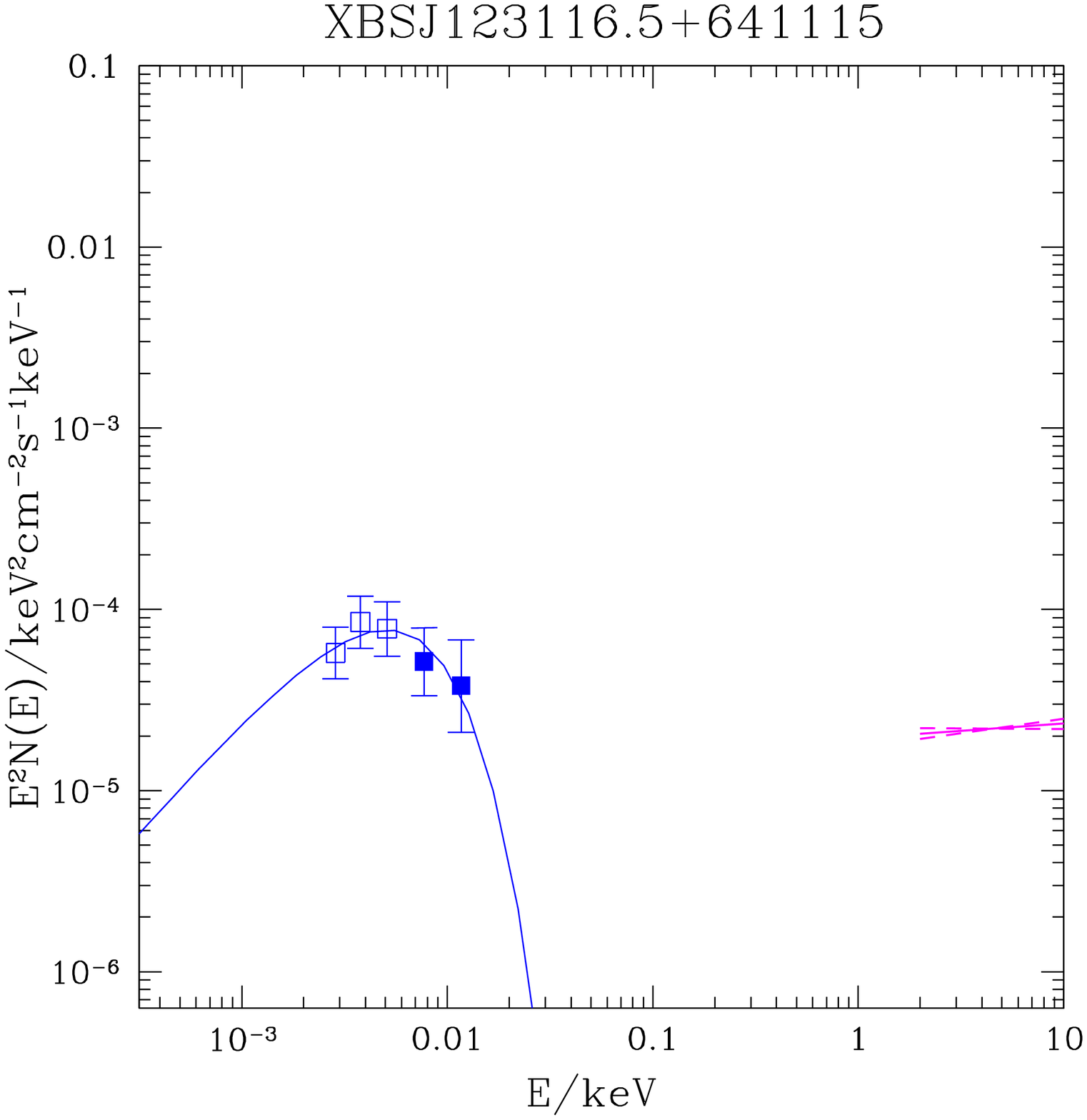}
  \includegraphics[height=5.6cm, width=6cm]{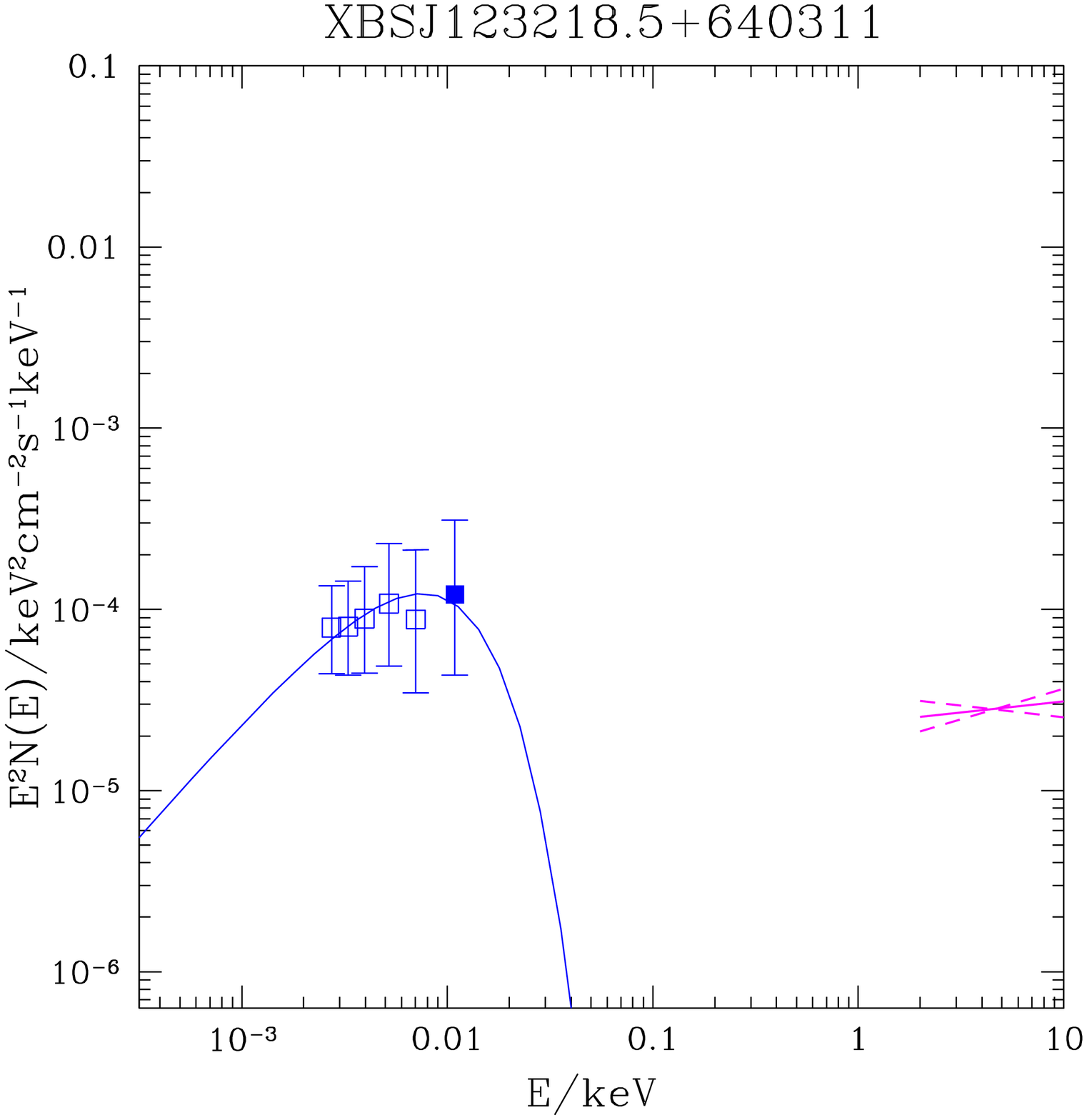}}      
\subfigure{ 
  \includegraphics[height=5.6cm, width=6cm]{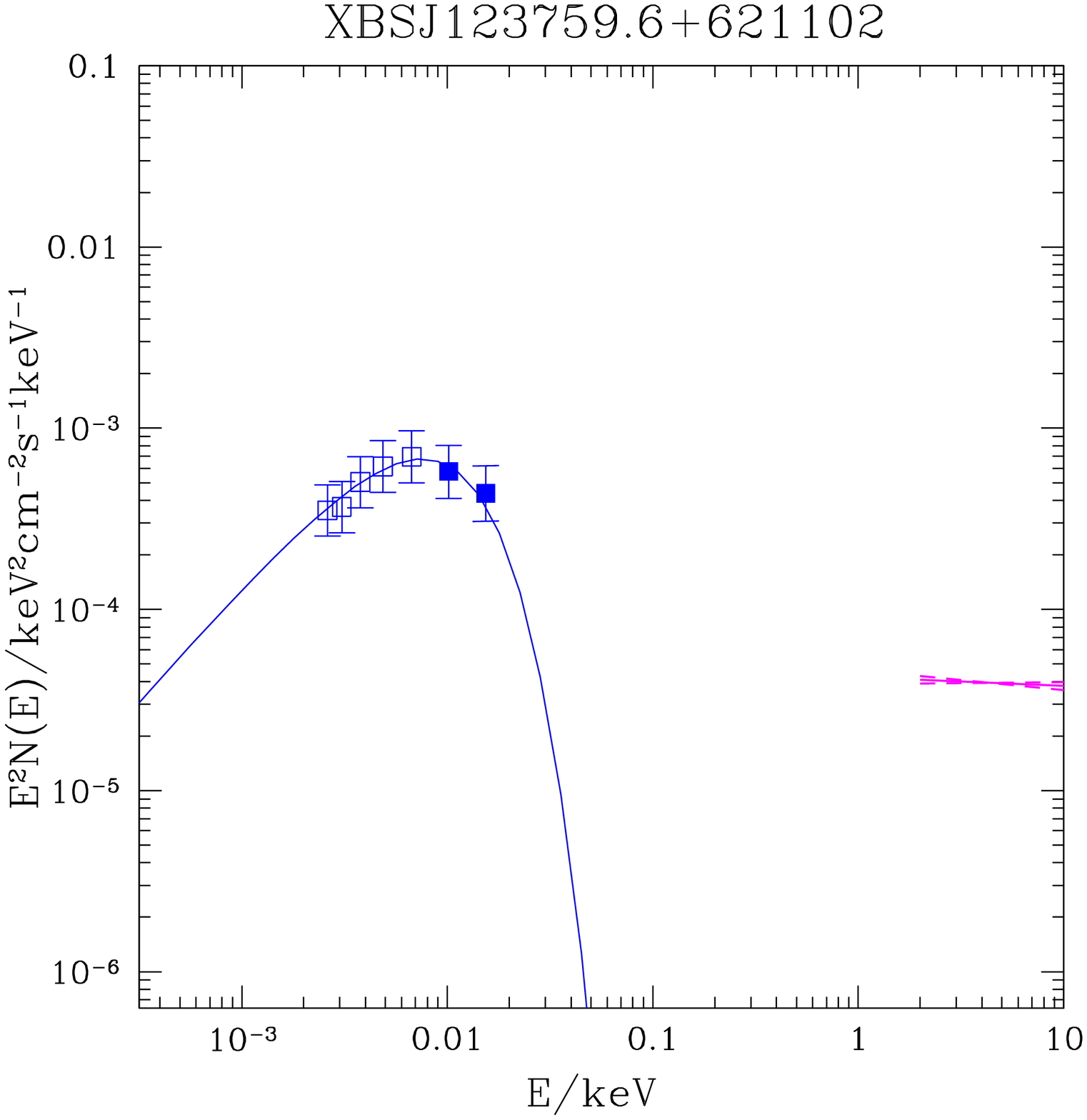}
  \includegraphics[height=5.6cm, width=6cm]{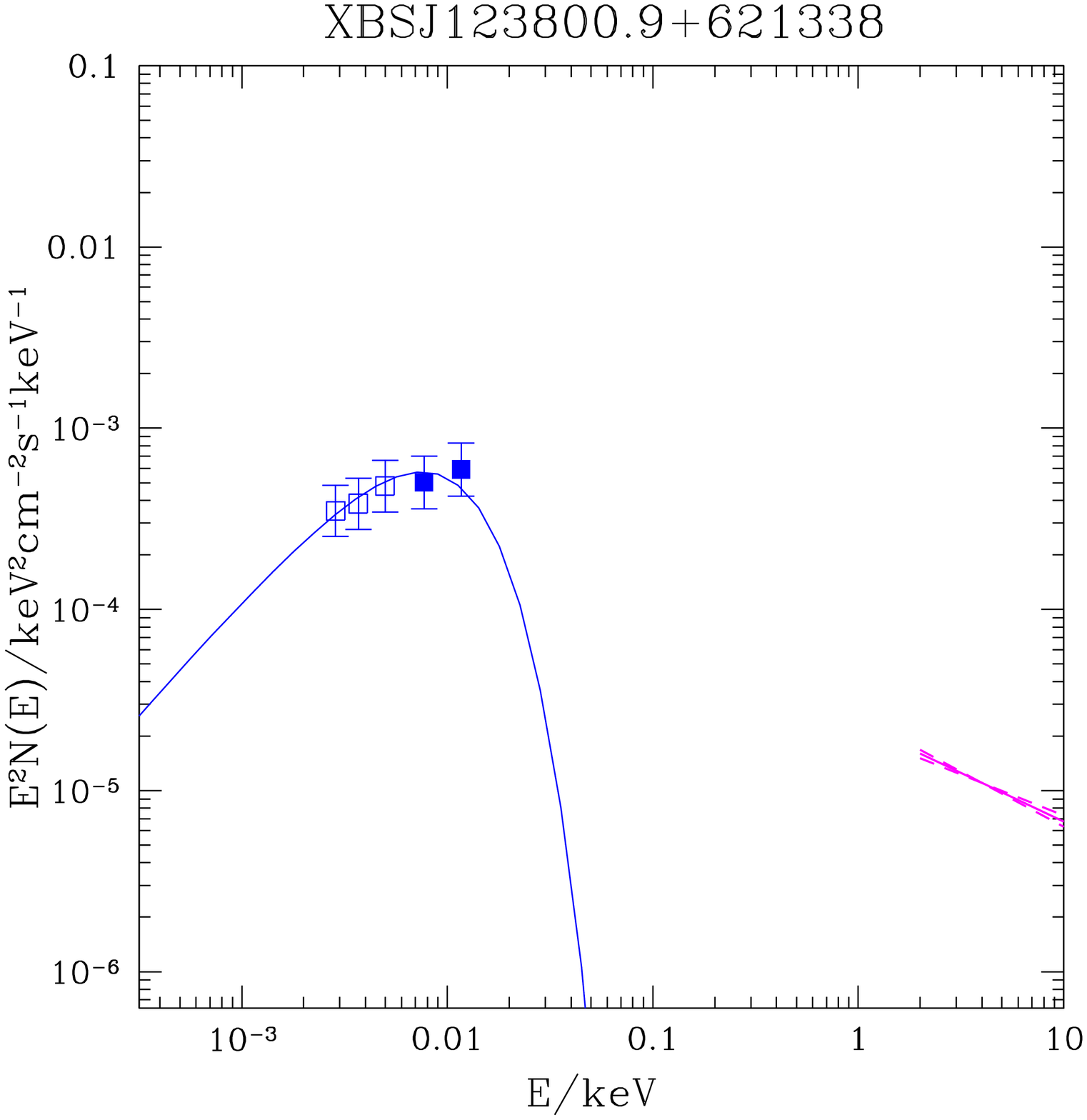}}   
\subfigure{ 
  \includegraphics[height=5.6cm, width=6cm]{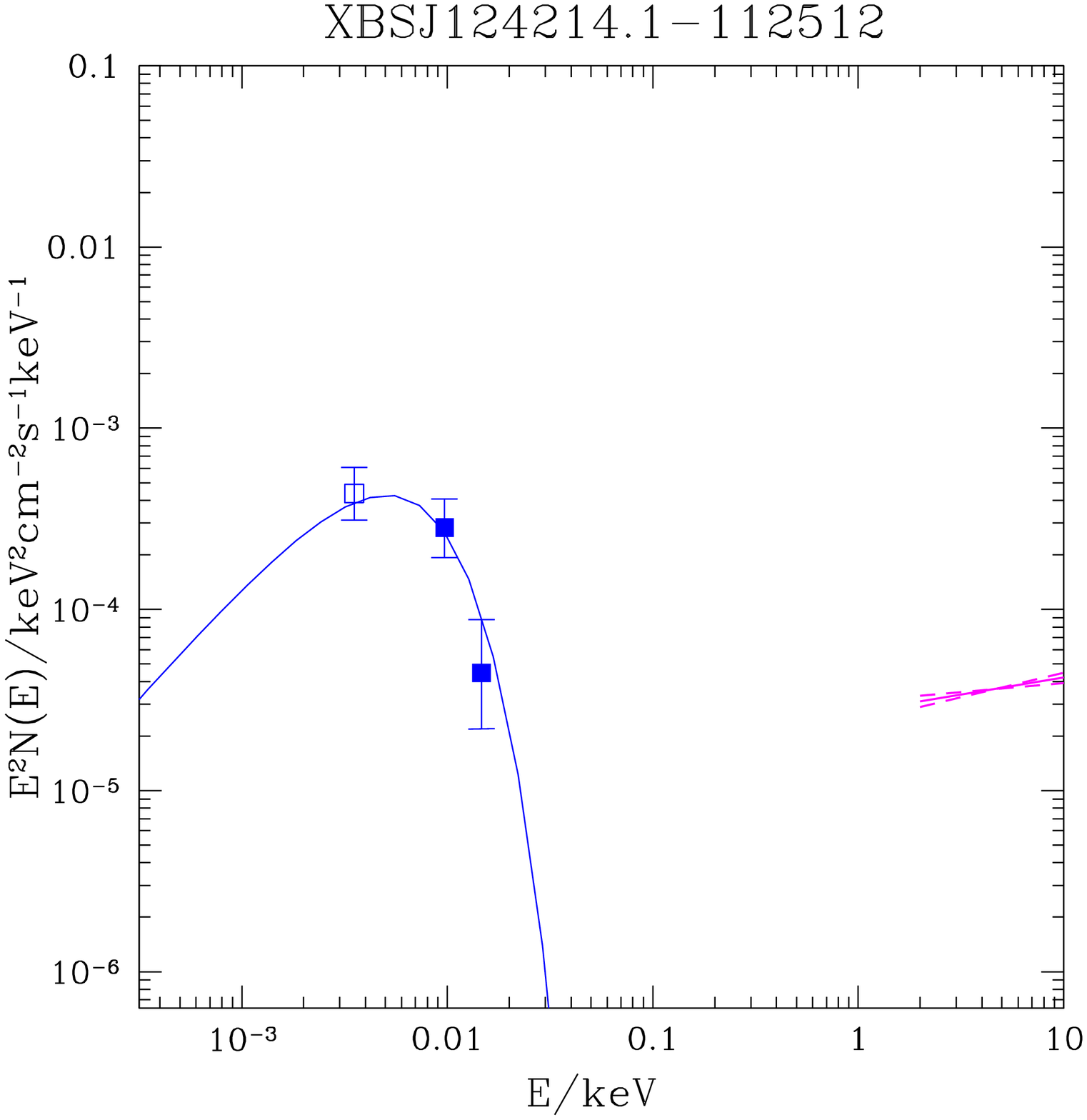}
  \includegraphics[height=5.6cm, width=6cm]{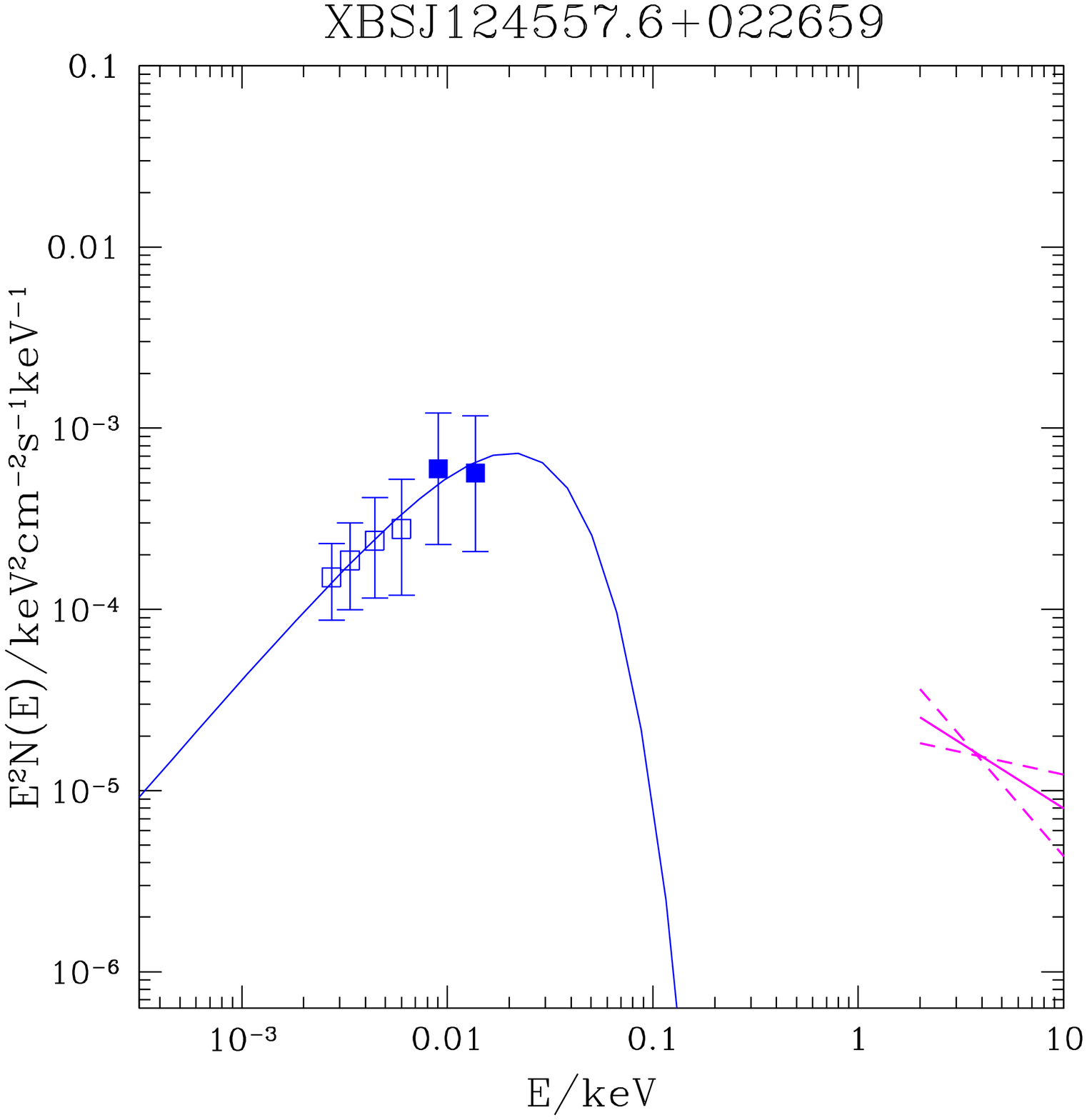}}      
\subfigure{ 
  \includegraphics[height=5.6cm, width=6cm]{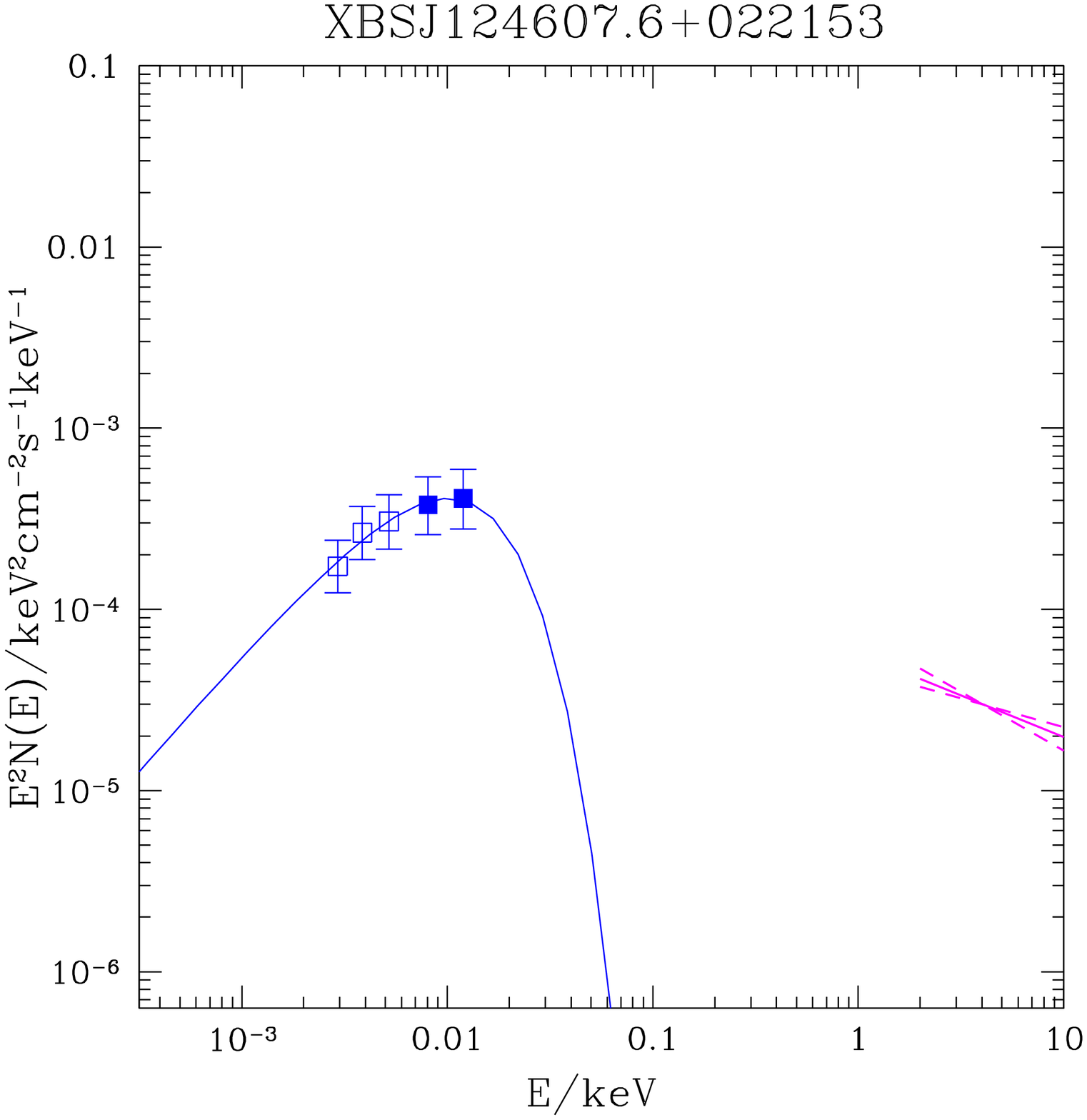}
  \includegraphics[height=5.6cm, width=6cm]{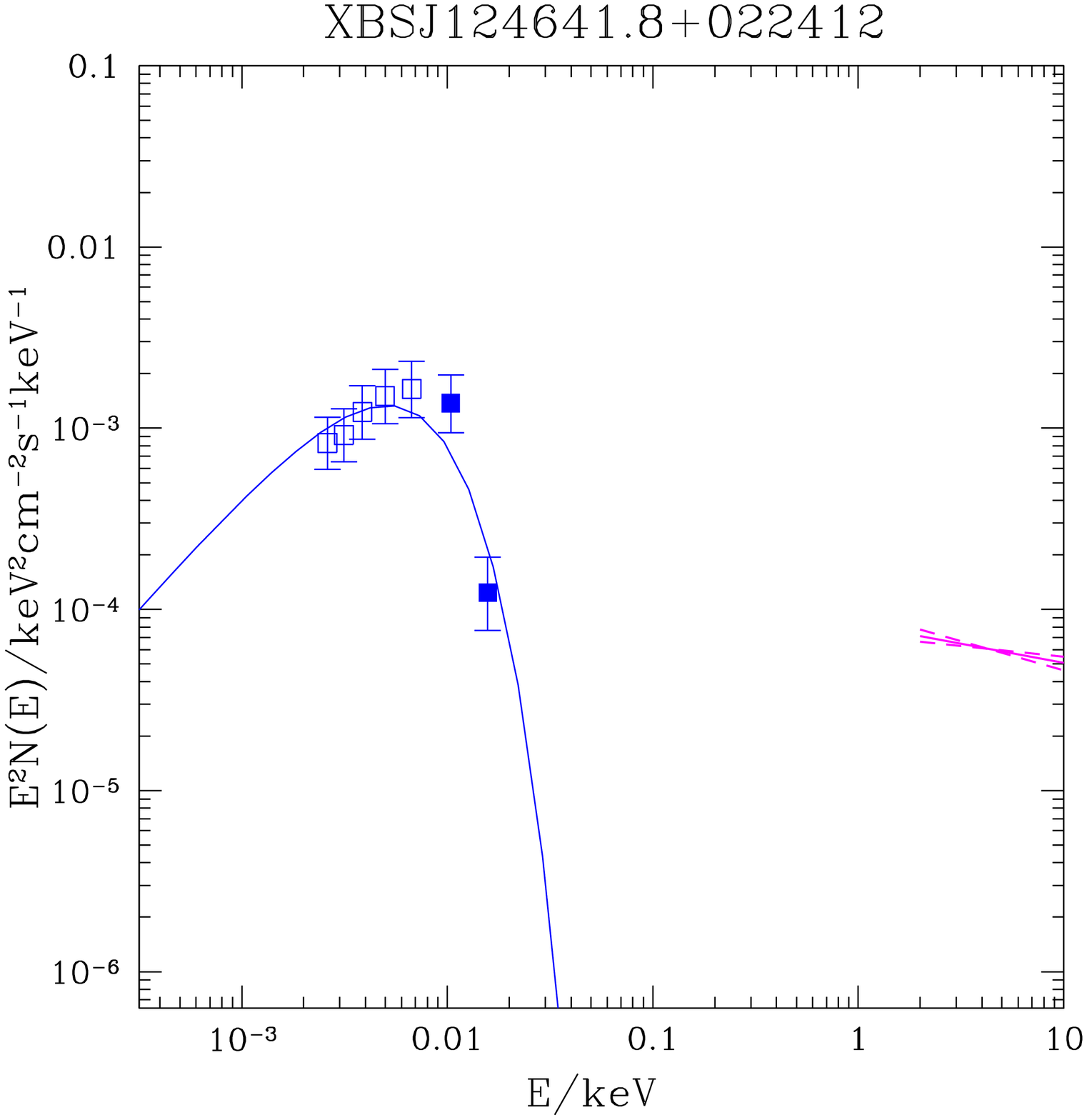}}  
  \end{figure*}
  
   \FloatBarrier
  
   \begin{figure*}
\centering    
\subfigure{ 
  \includegraphics[height=5.6cm, width=6cm]{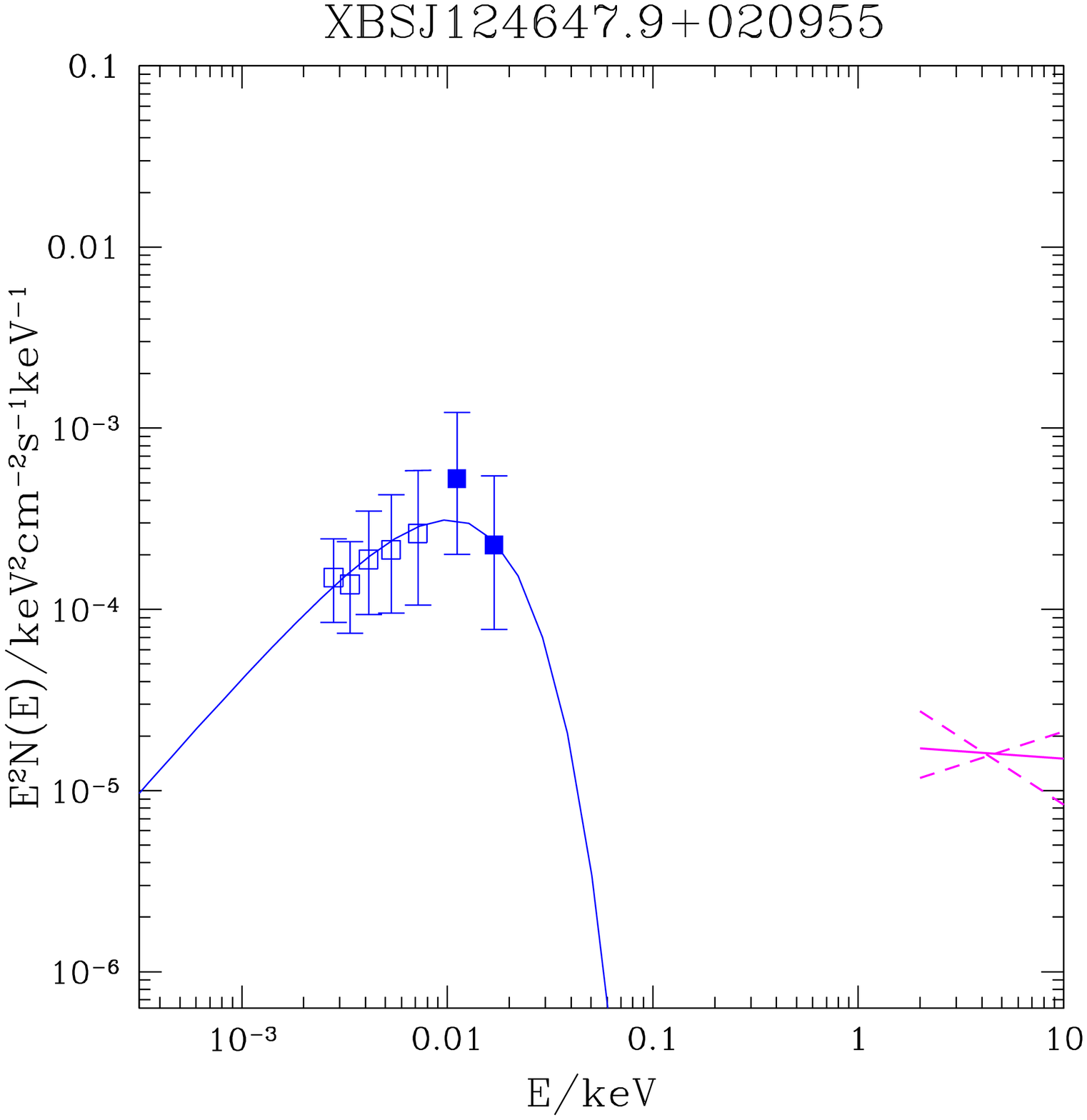}
  \includegraphics[height=5.6cm, width=6cm]{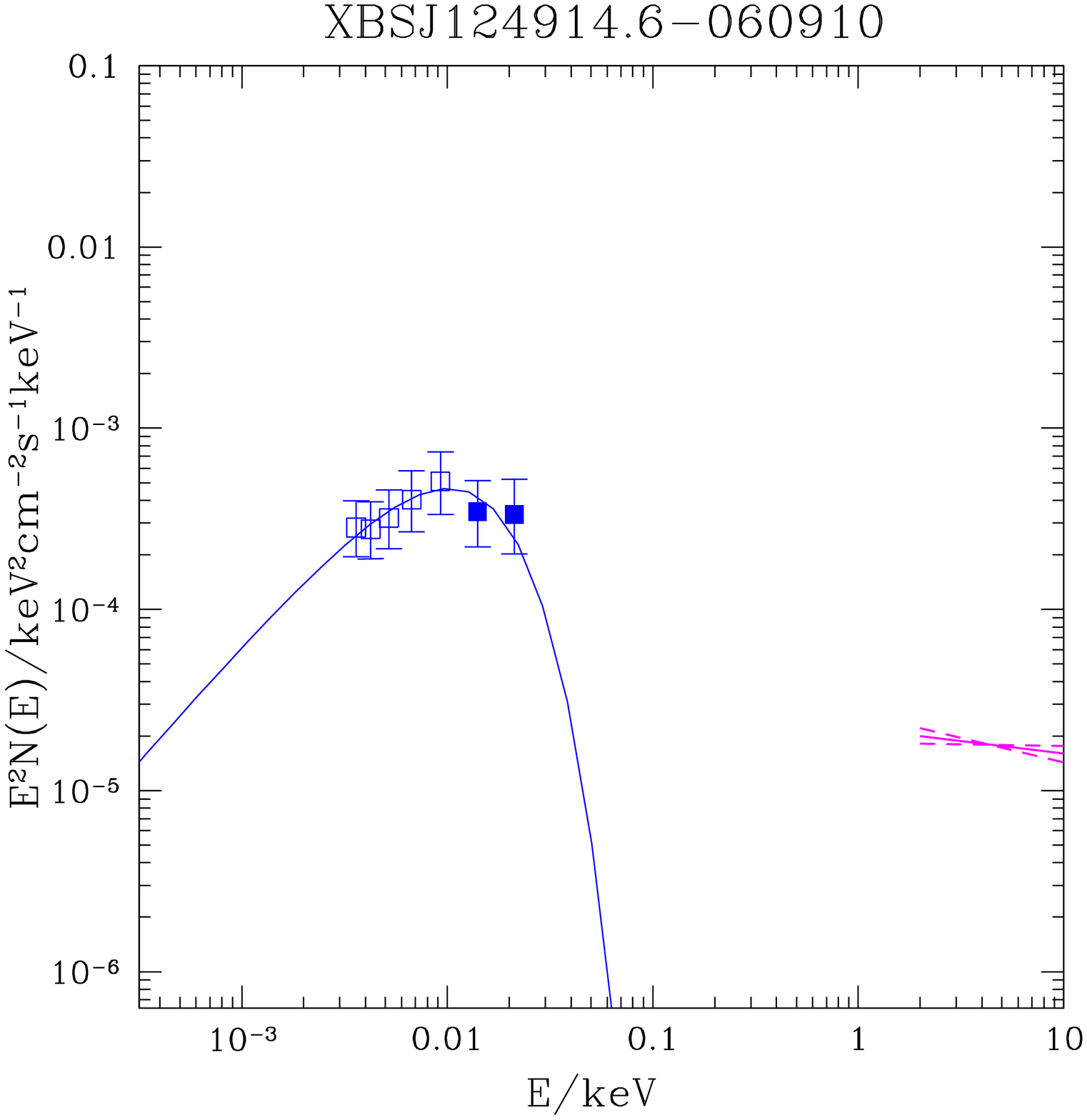}}      
\subfigure{ 
  \includegraphics[height=5.6cm, width=6cm]{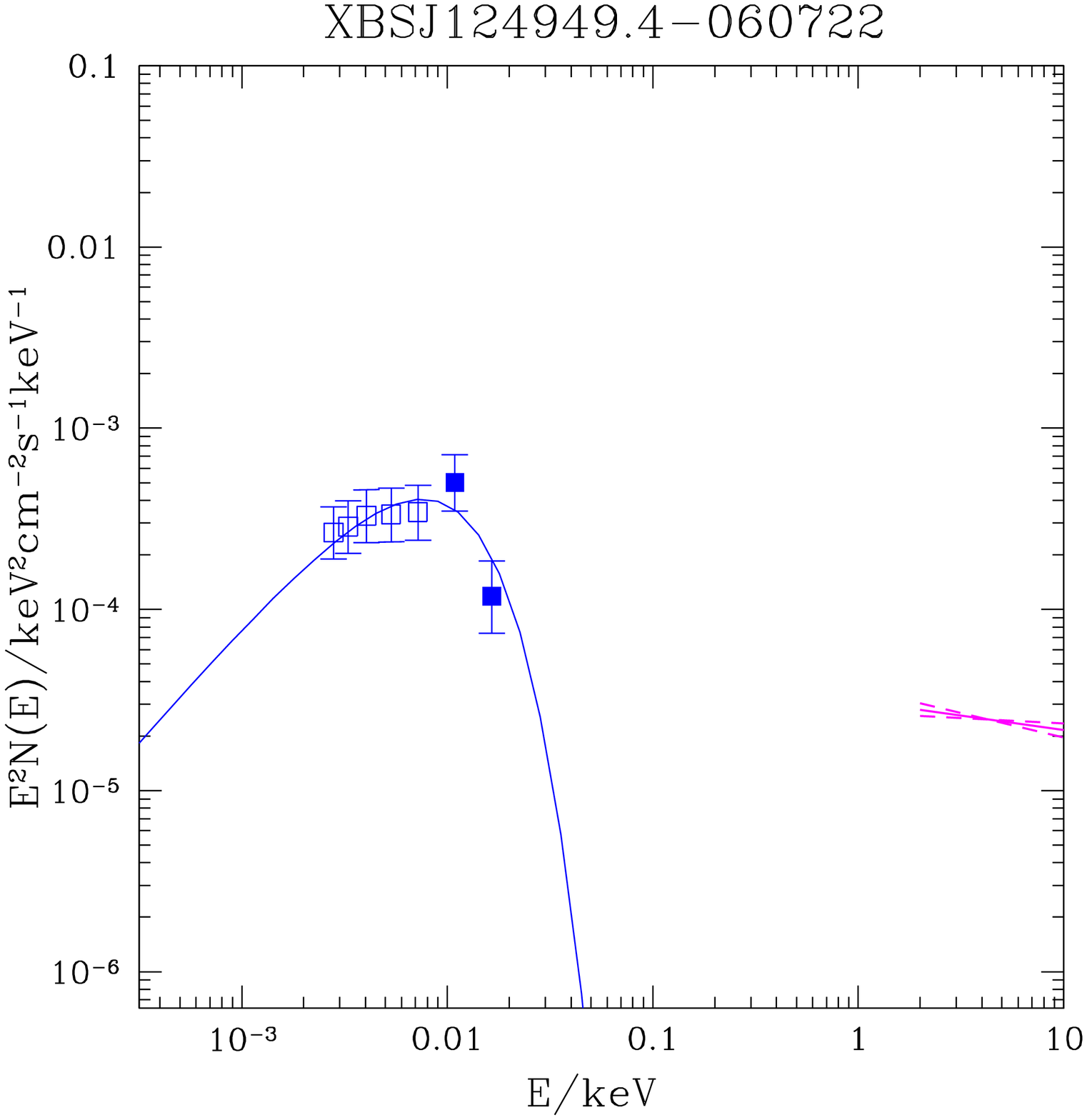}
  \includegraphics[height=5.6cm, width=6cm]{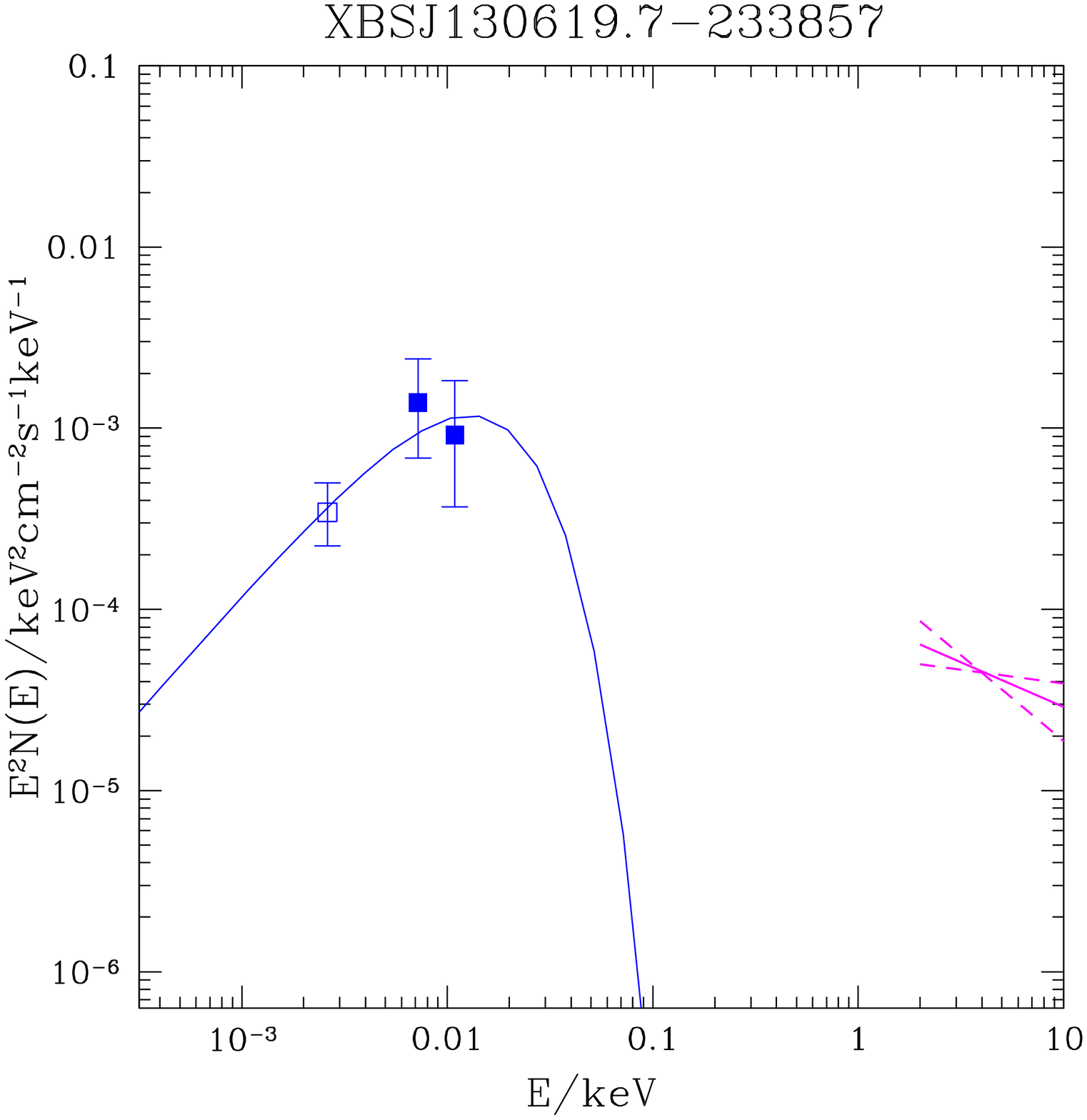}}      
\subfigure{ 
  \includegraphics[height=5.6cm, width=6cm]{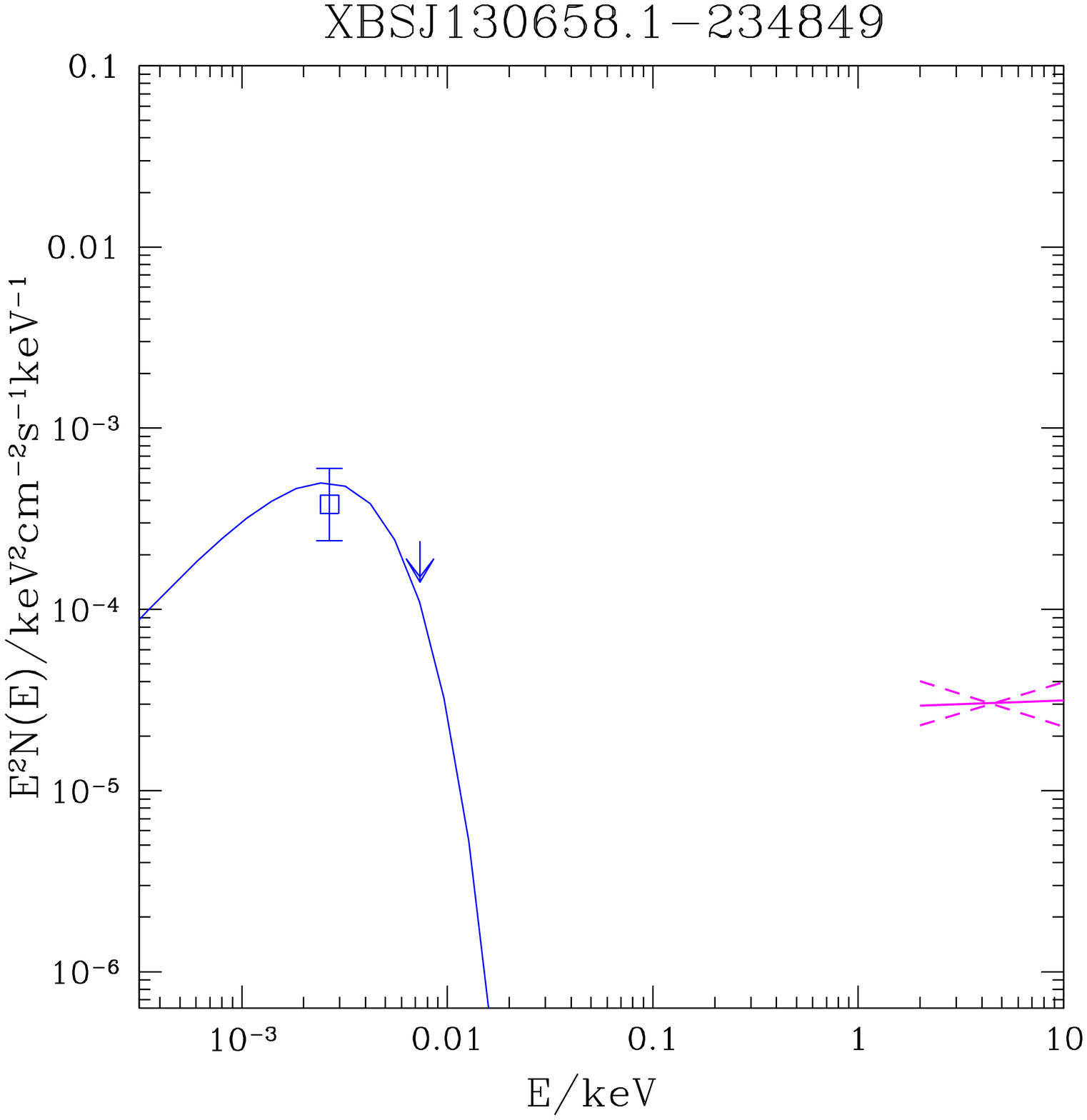}
  \includegraphics[height=5.6cm, width=6cm]{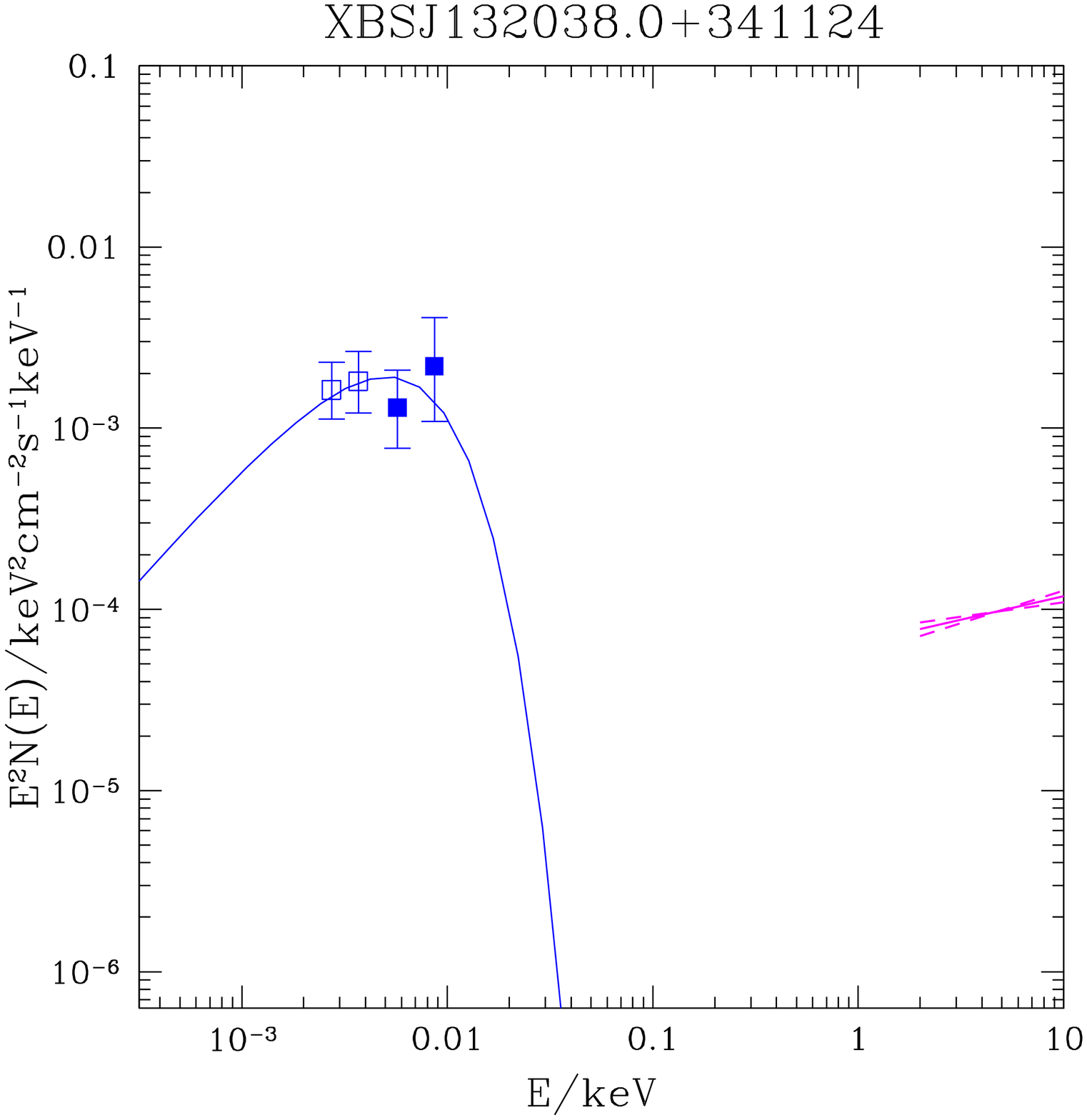}}      
\subfigure{ 
  \includegraphics[height=5.6cm, width=6cm]{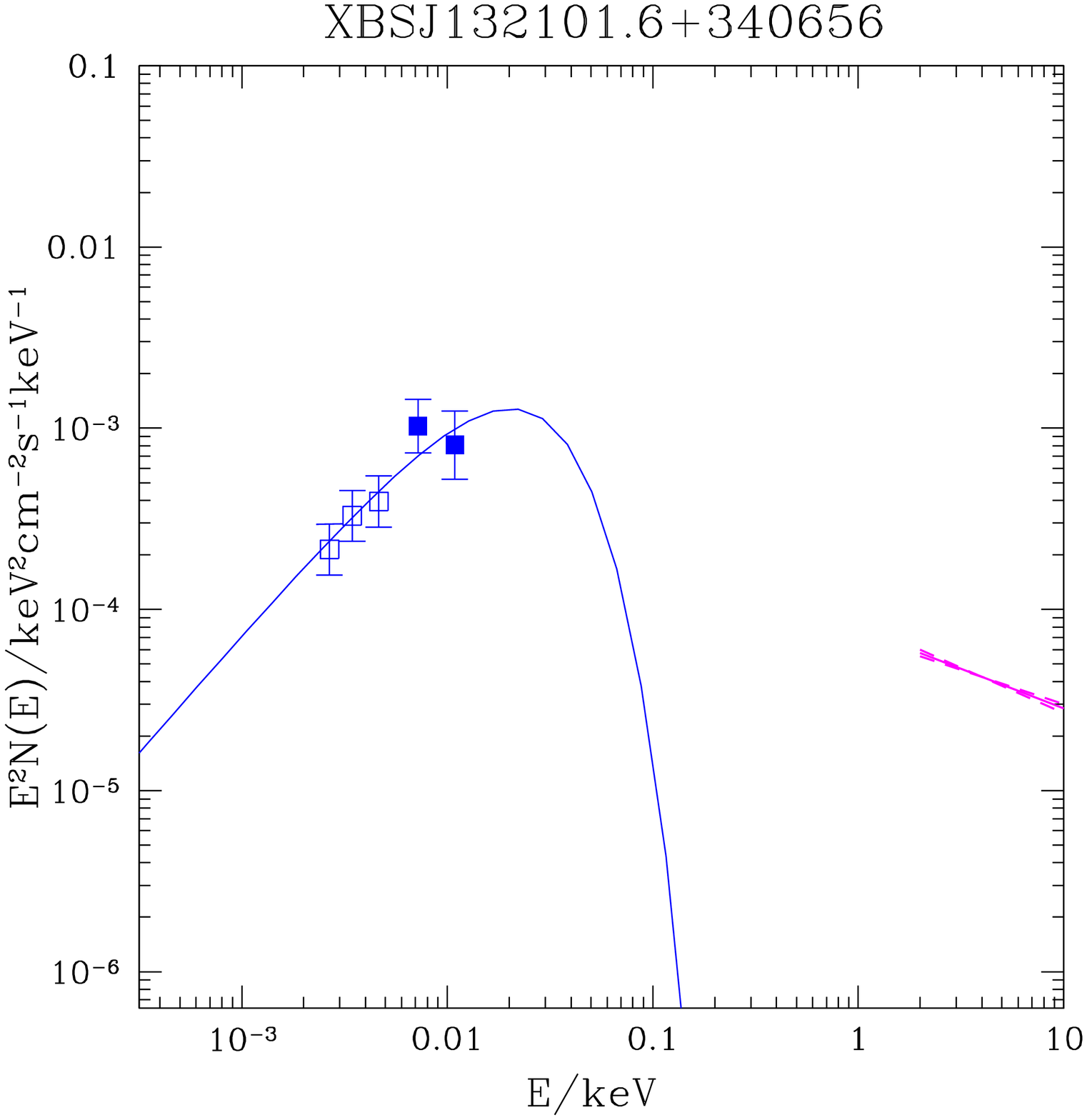}
  \includegraphics[height=5.6cm, width=6cm]{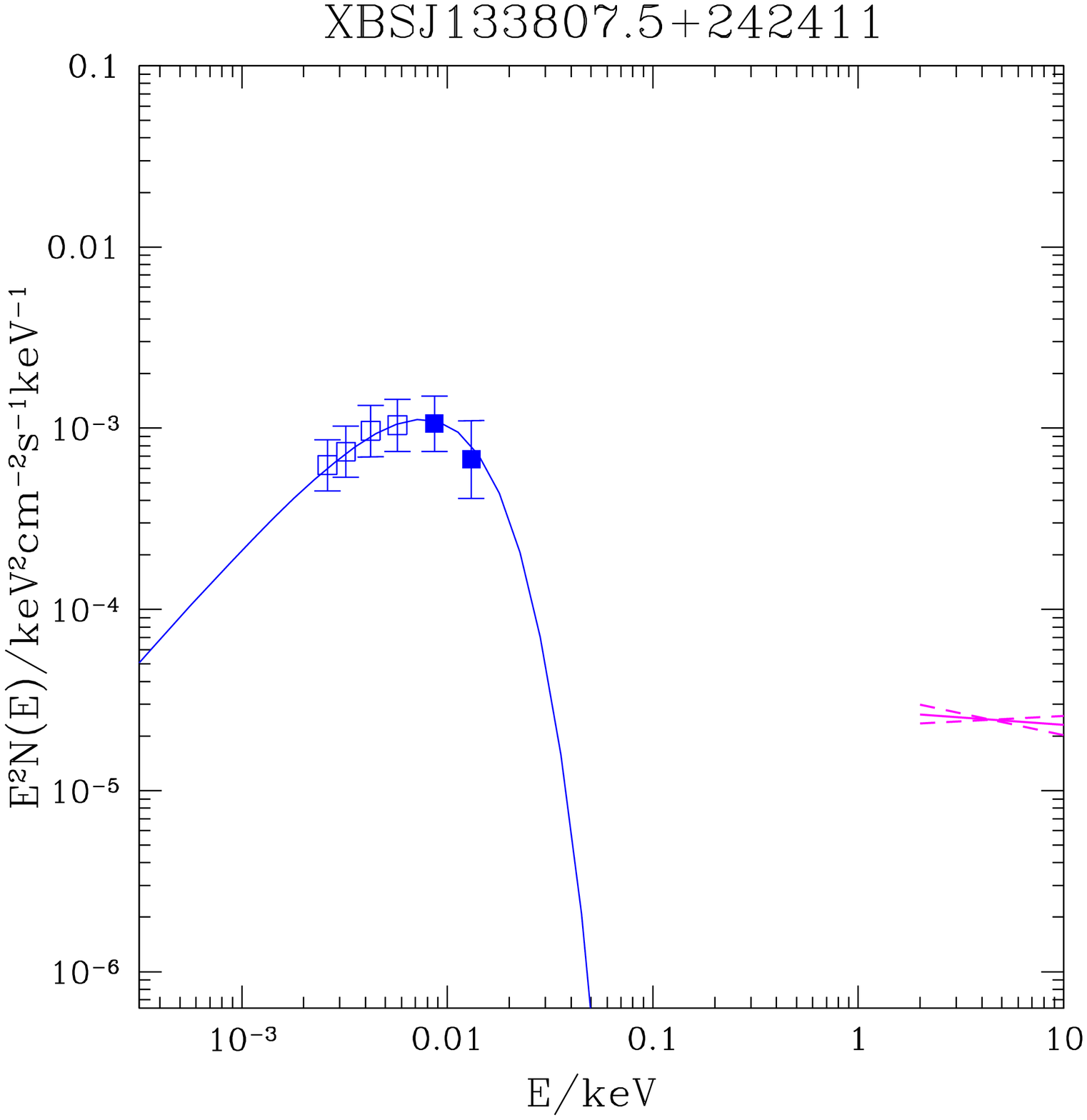}}   
  \end{figure*}
  
   \FloatBarrier
  
   \begin{figure*}
\centering   
\subfigure{ 
  \includegraphics[height=5.6cm, width=6cm]{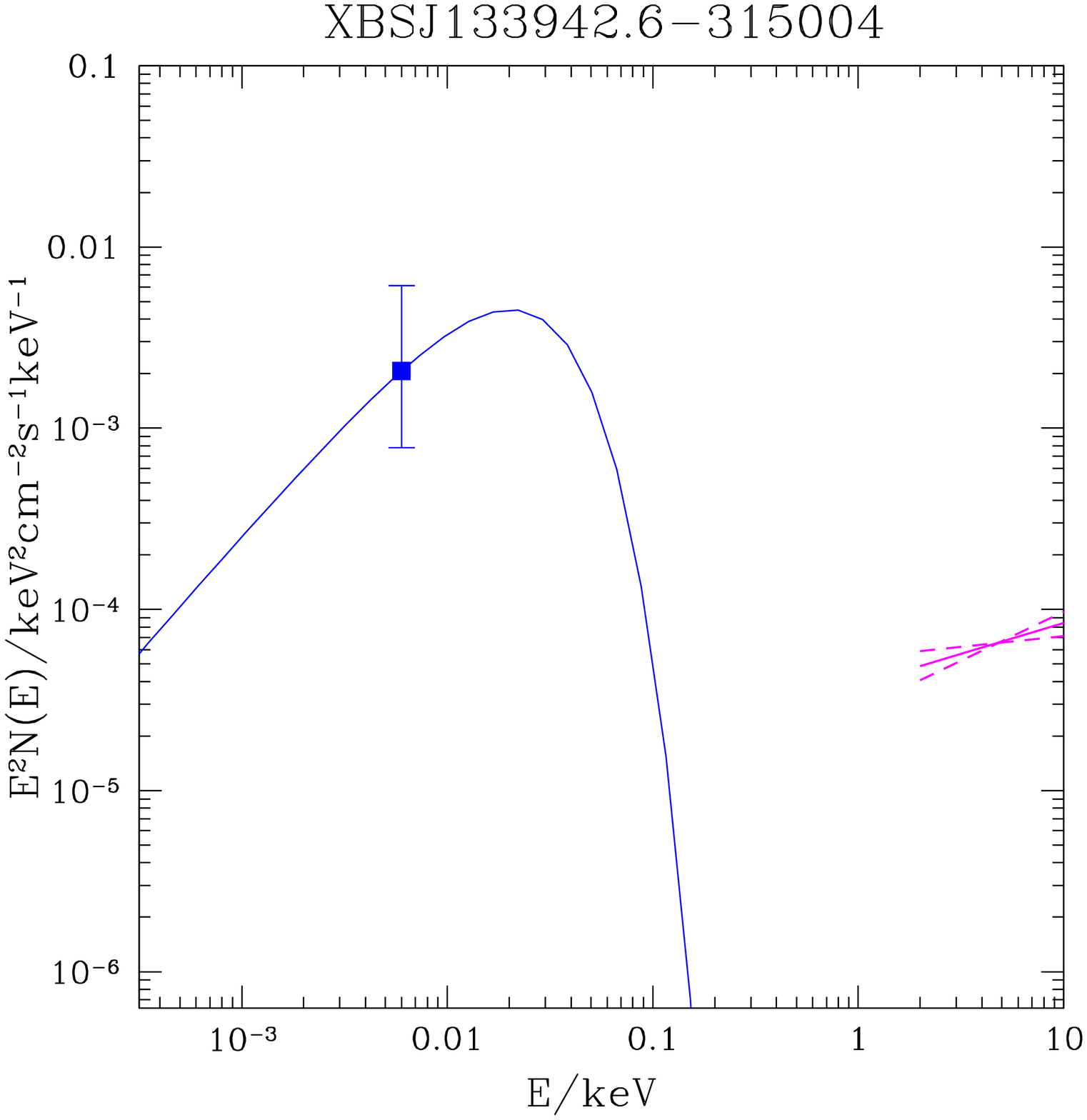}
  \includegraphics[height=5.6cm, width=6cm]{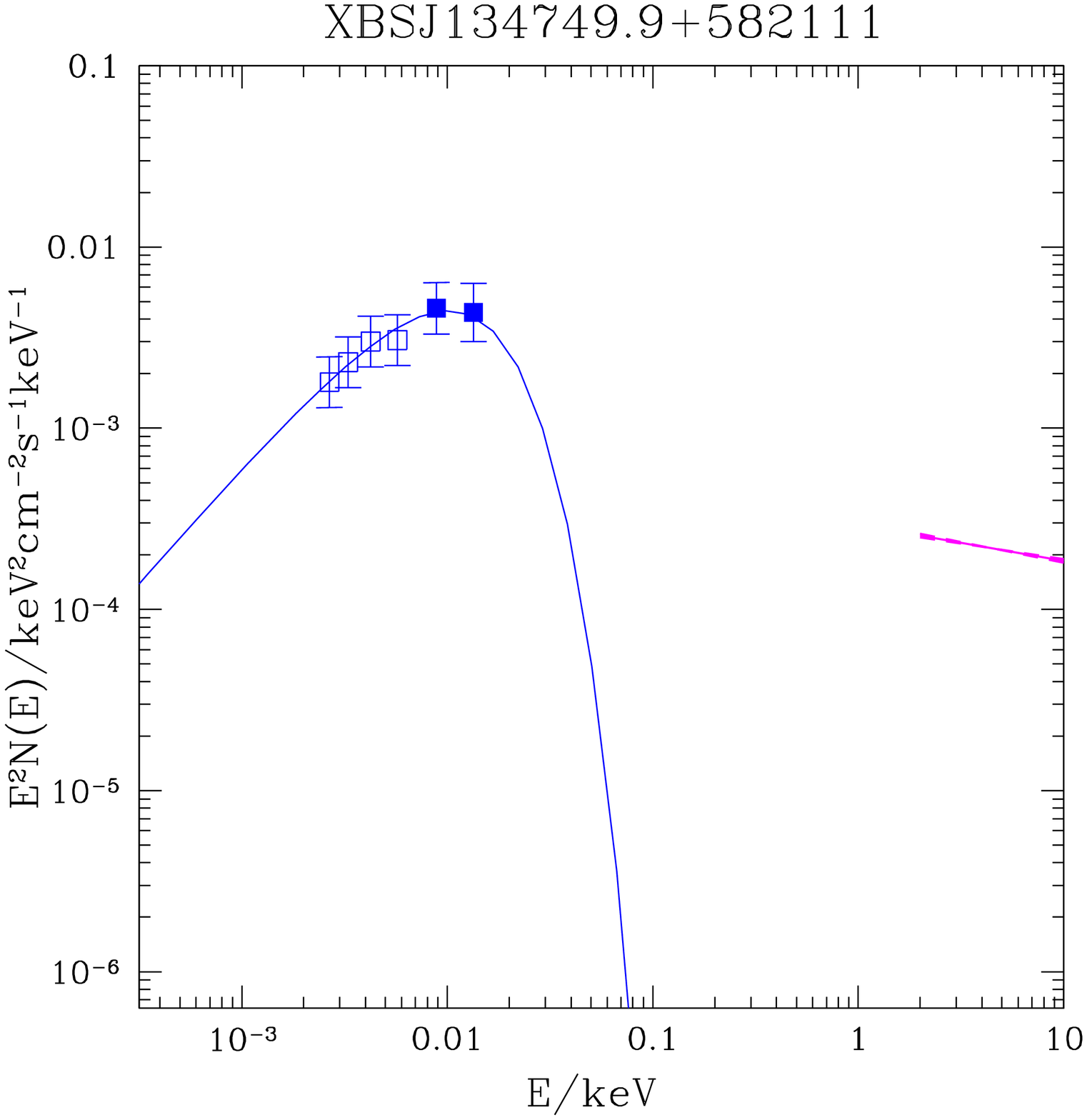}}      
\subfigure{ 
  \includegraphics[height=5.6cm, width=6cm]{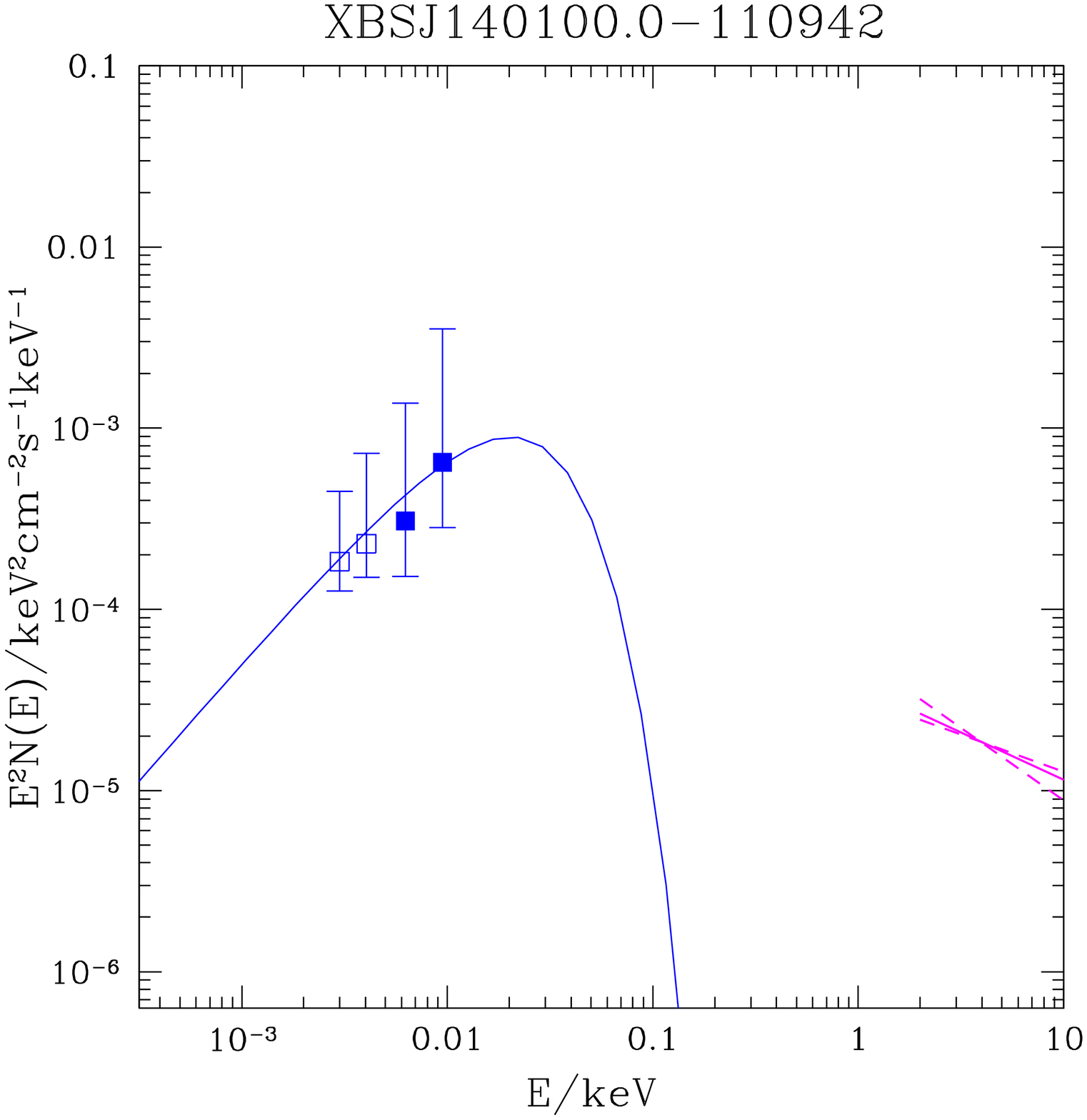}
  \includegraphics[height=5.6cm, width=6cm]{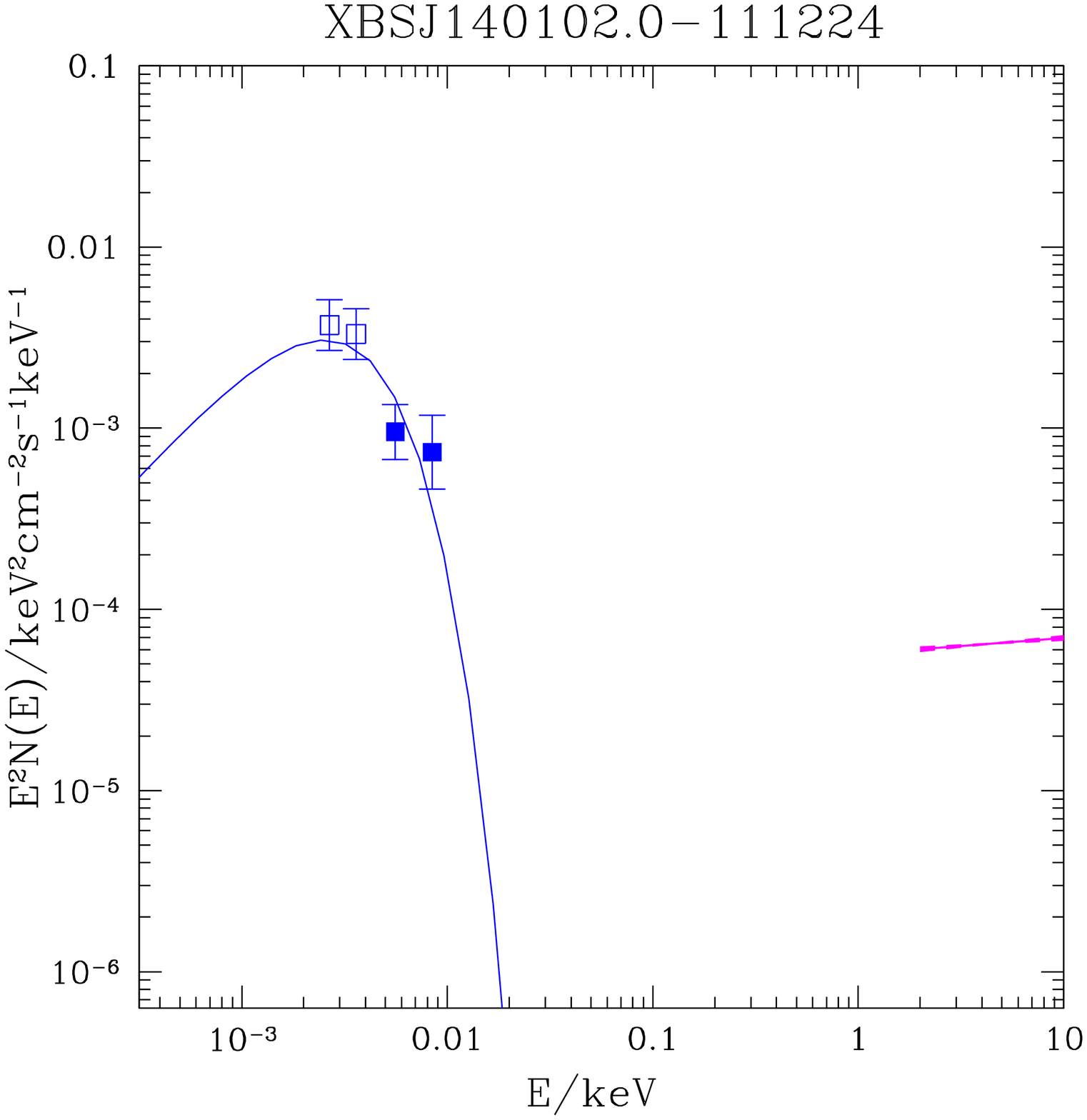}}      
\subfigure{ 
  \includegraphics[height=5.6cm, width=6cm]{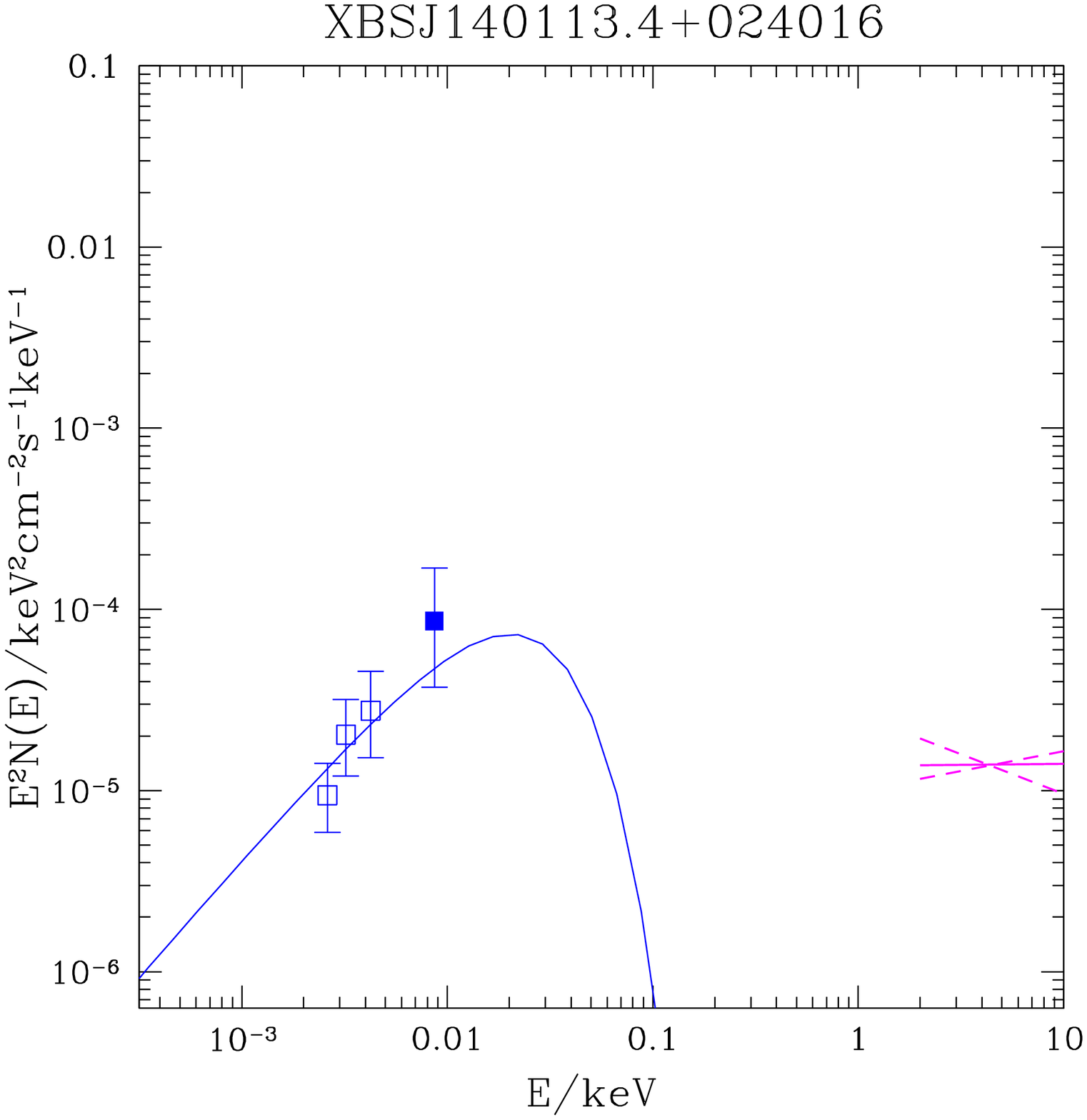}
  \includegraphics[height=5.6cm, width=6cm]{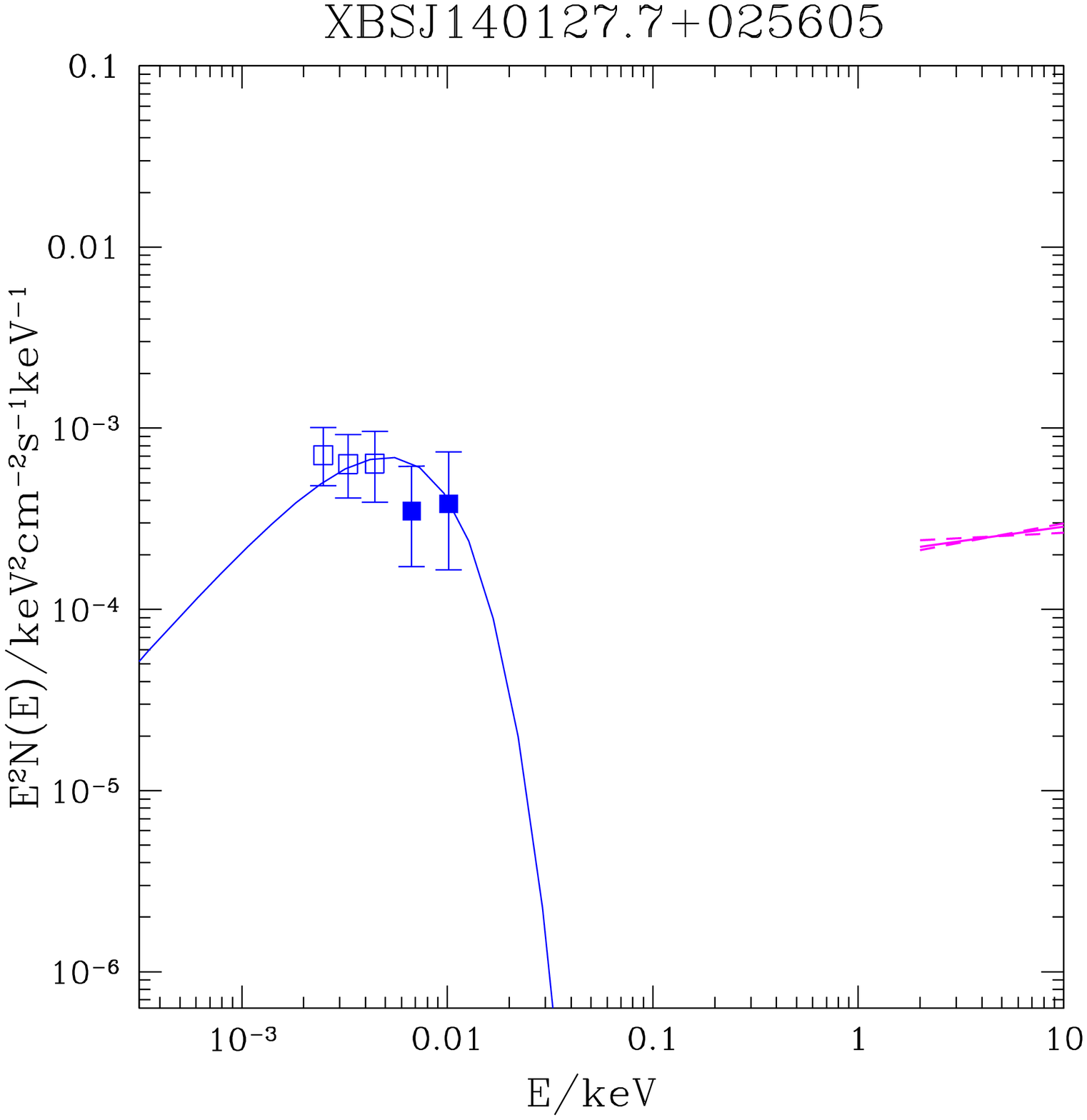}}      
\subfigure{ 
  \includegraphics[height=5.6cm, width=6cm]{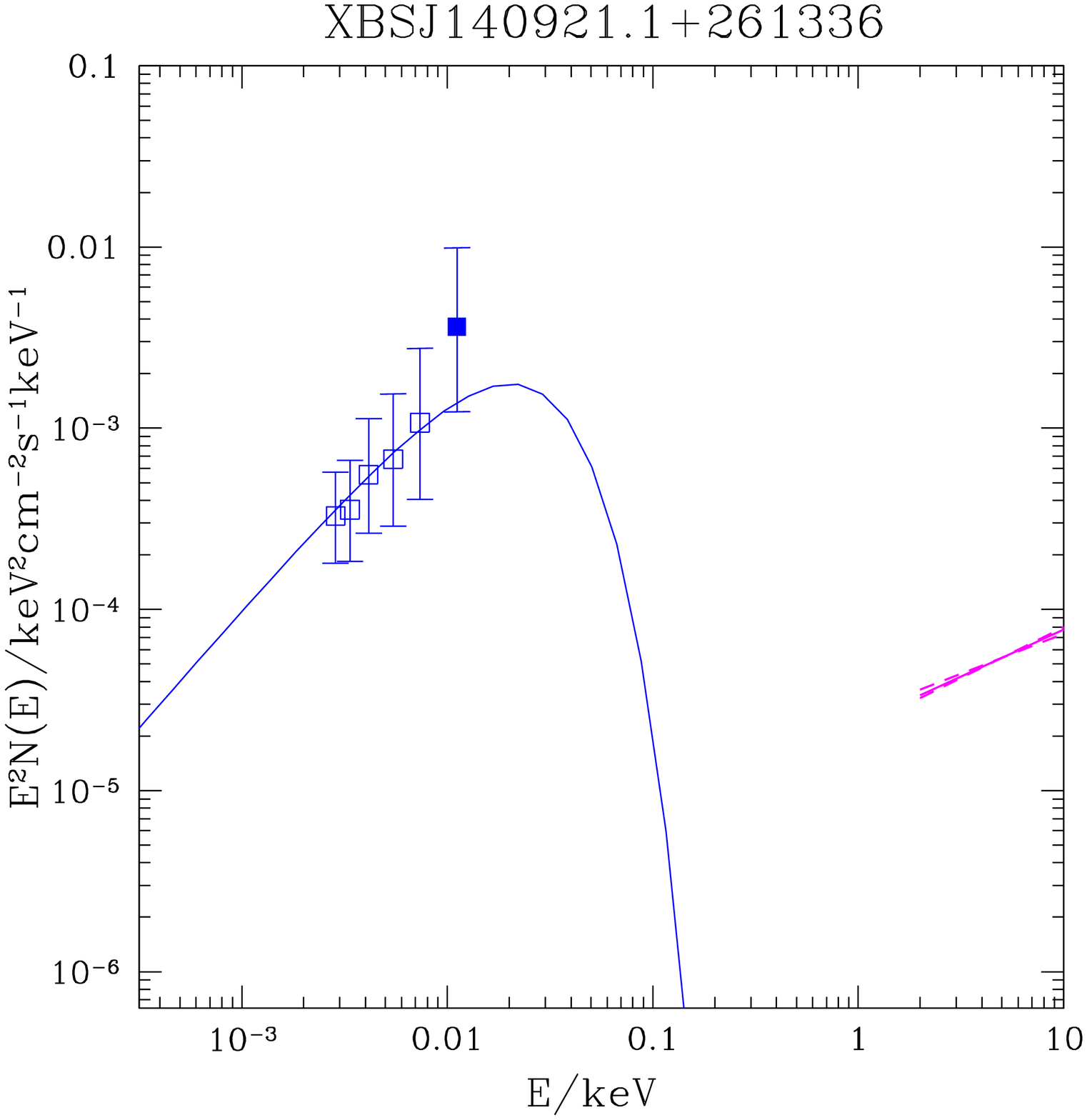}
  \includegraphics[height=5.6cm, width=6cm]{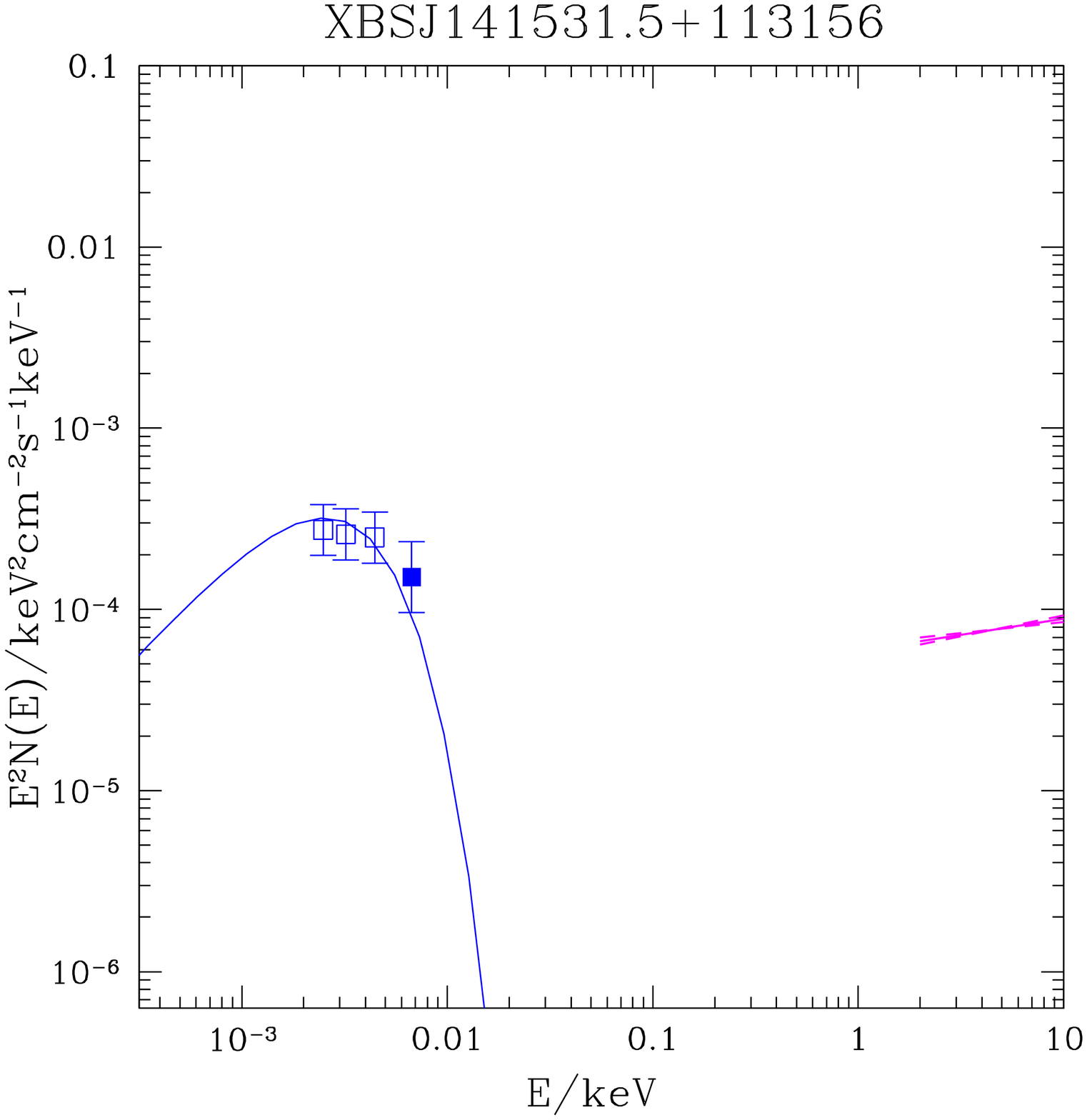}}  
  \end{figure*}
  
   \FloatBarrier
  
   \begin{figure*}
\centering    
\subfigure{ 
  \includegraphics[height=5.6cm, width=6cm]{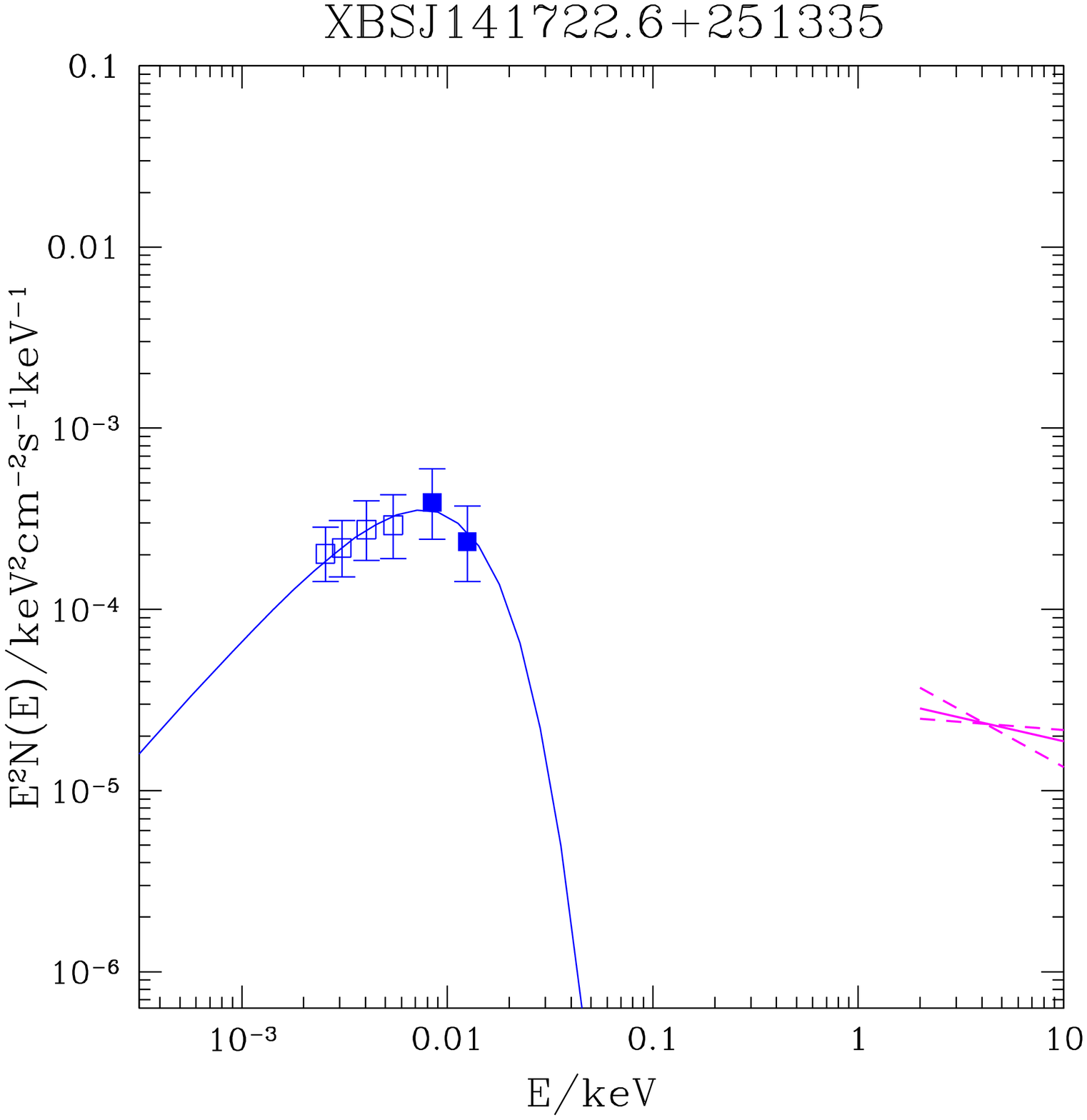}
  \includegraphics[height=5.6cm, width=6cm]{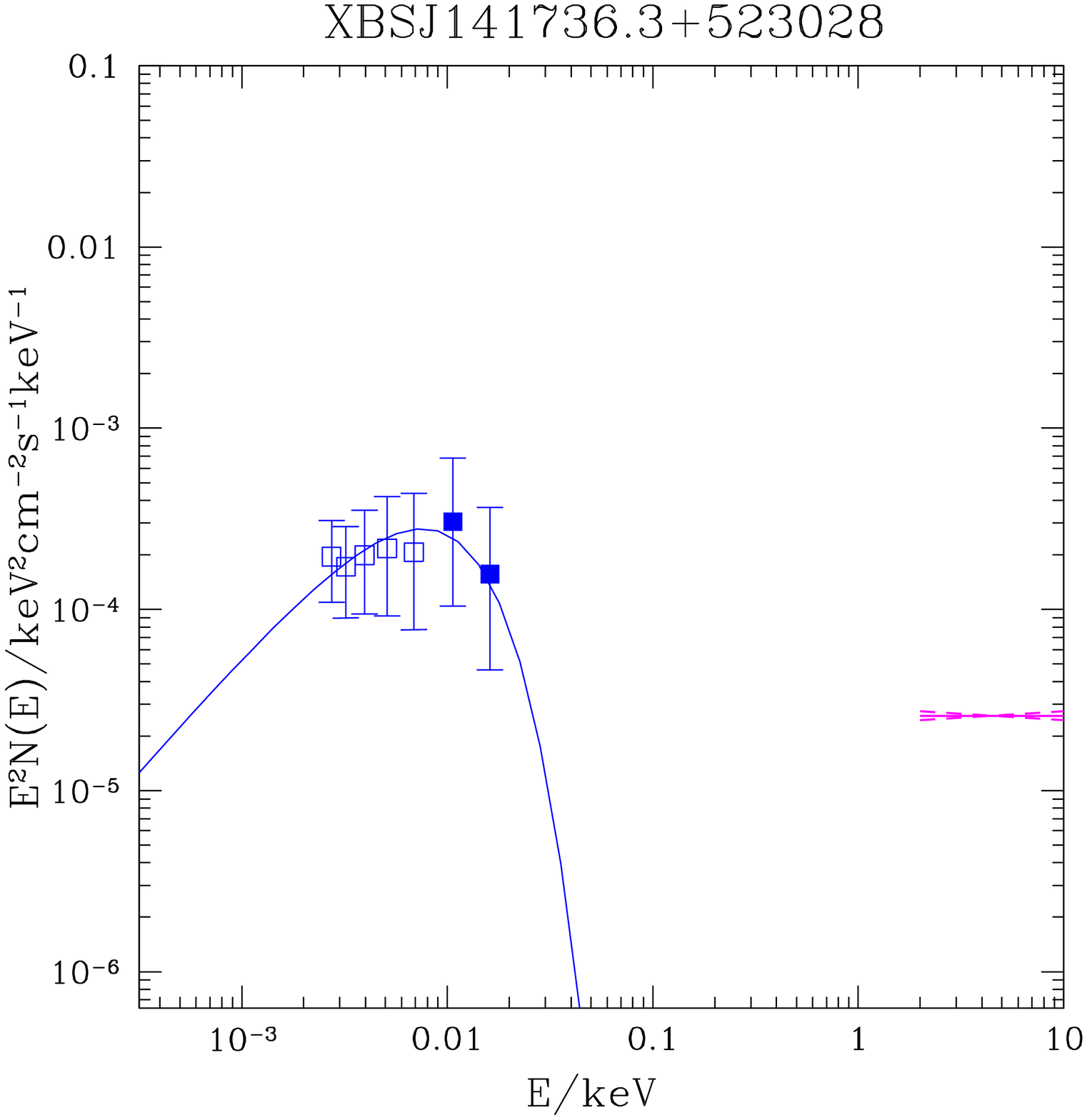}}      
\subfigure{ 
  \includegraphics[height=5.6cm, width=6cm]{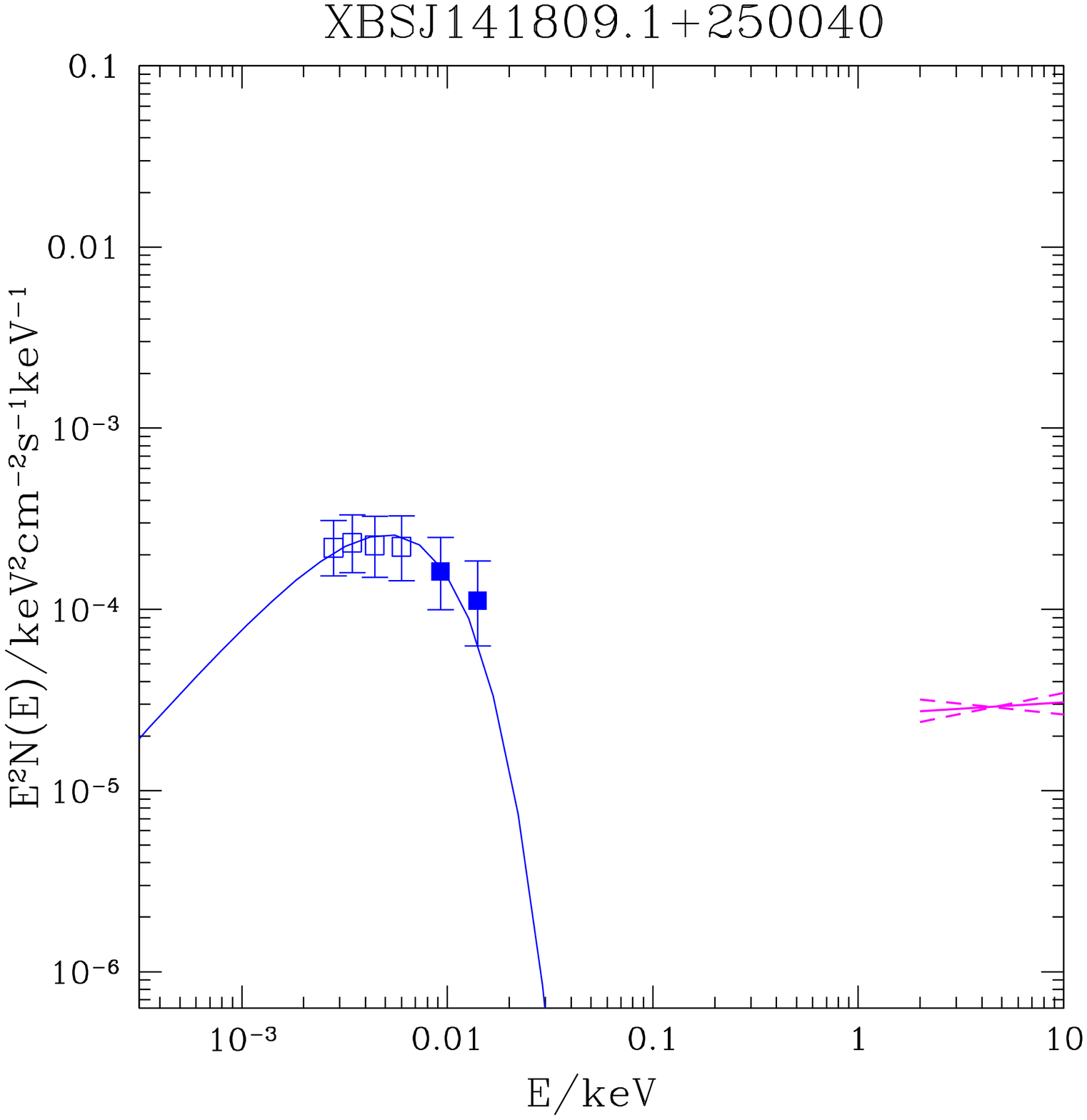}
  \includegraphics[height=5.6cm, width=6cm]{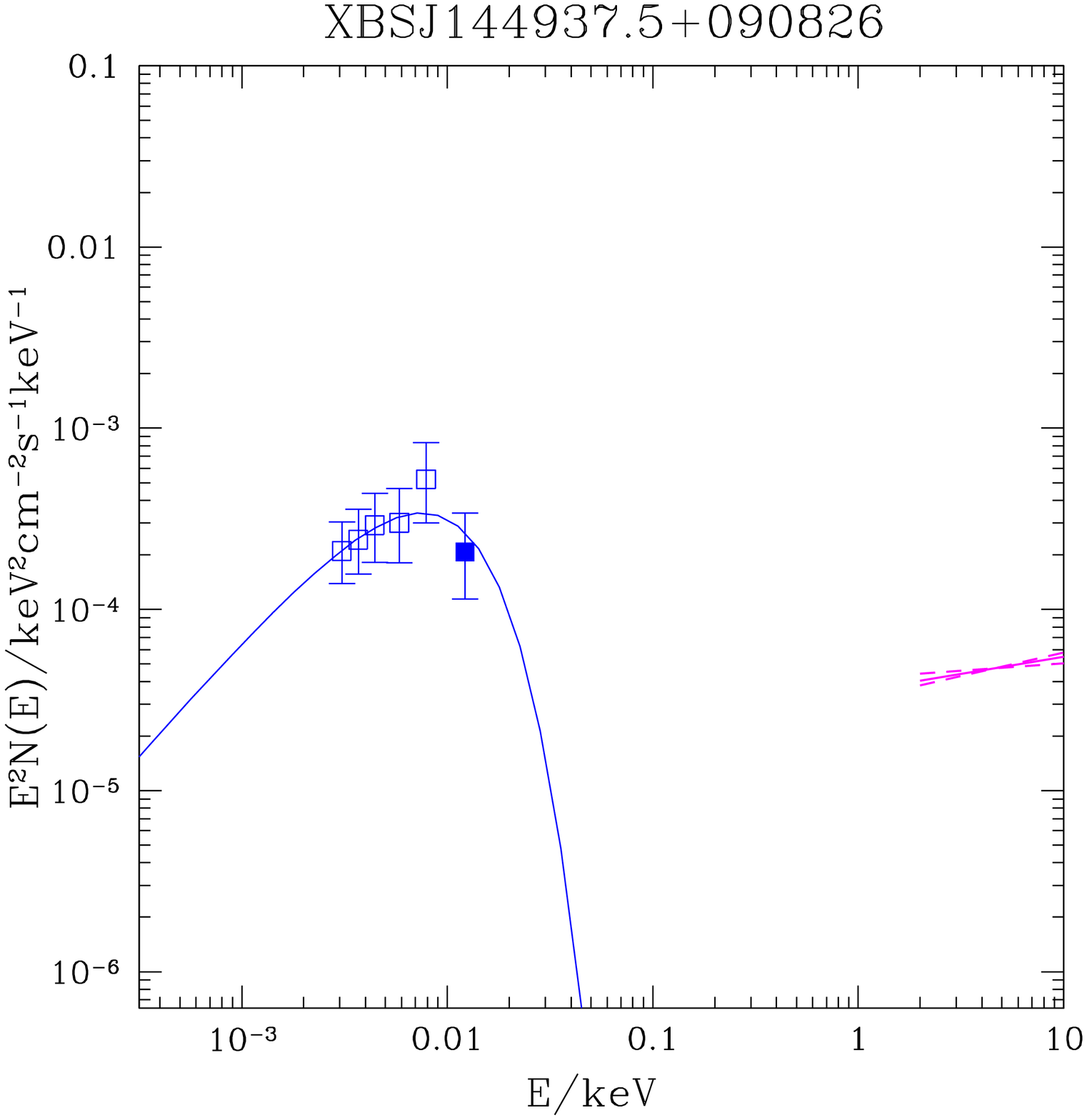}}     
\centering 
\subfigure{ 
  \includegraphics[height=5.6cm, width=6cm]{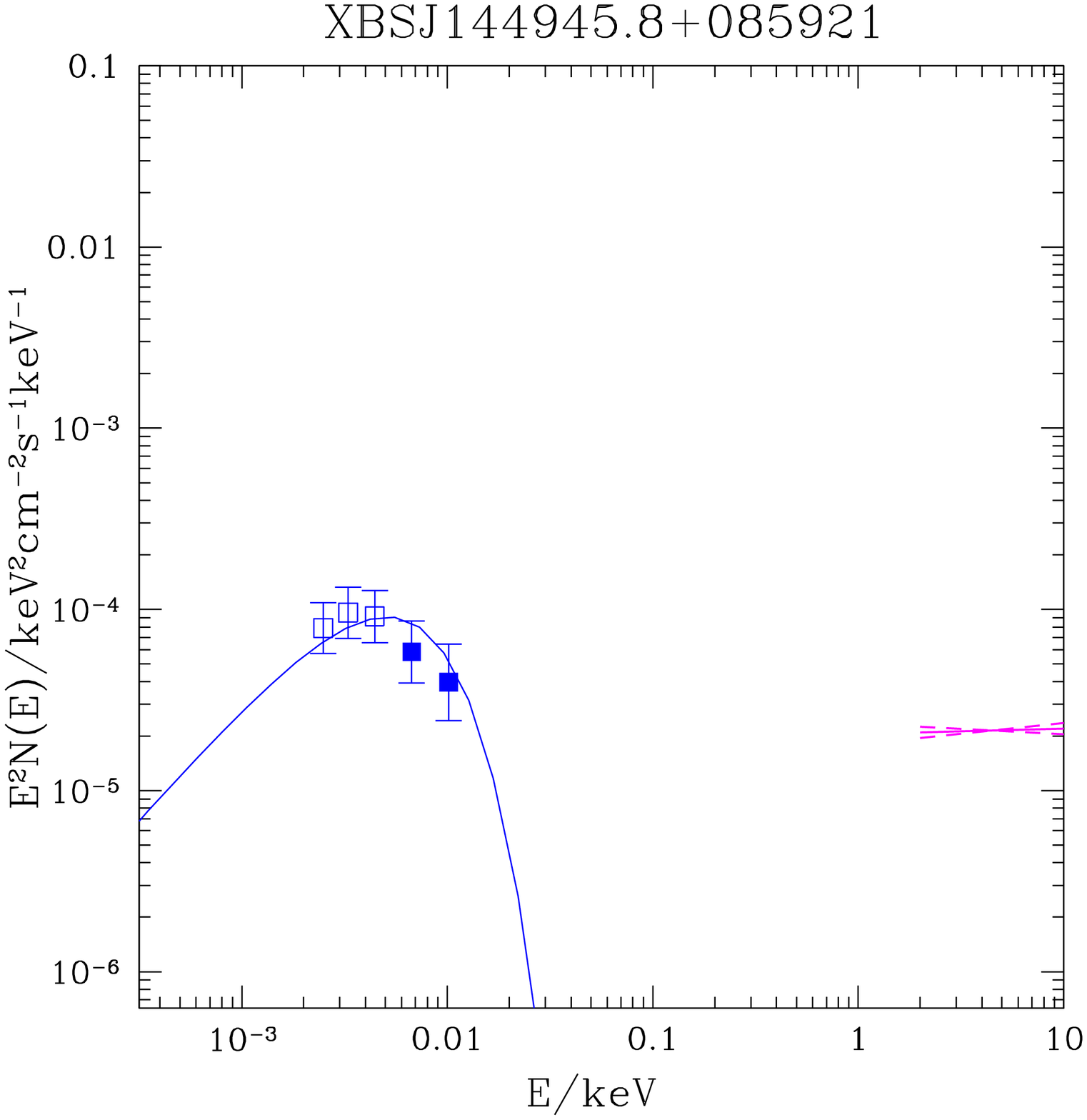}
  \includegraphics[height=5.6cm, width=6cm]{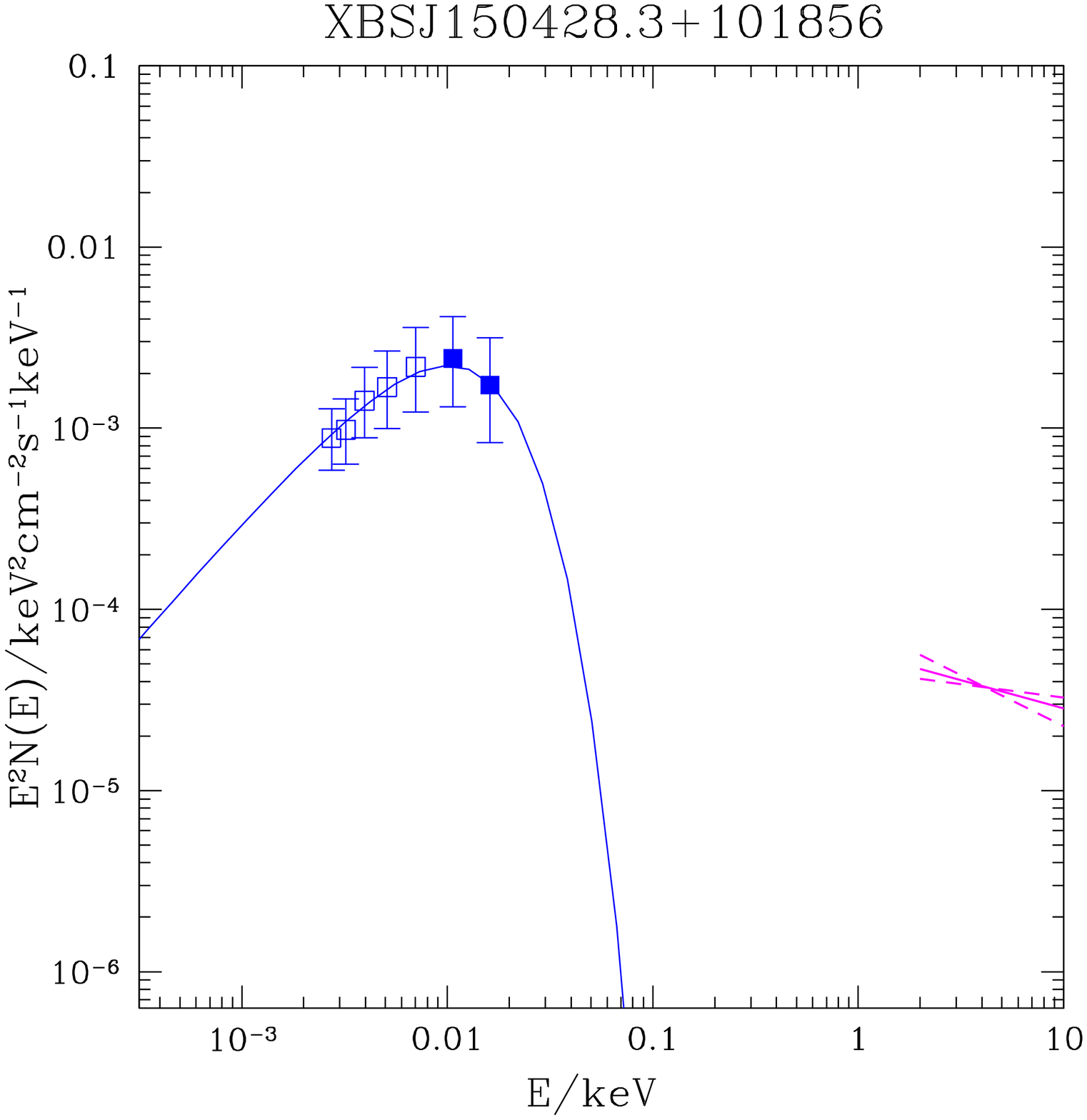}}      
\subfigure{ 
  \includegraphics[height=5.6cm, width=6cm]{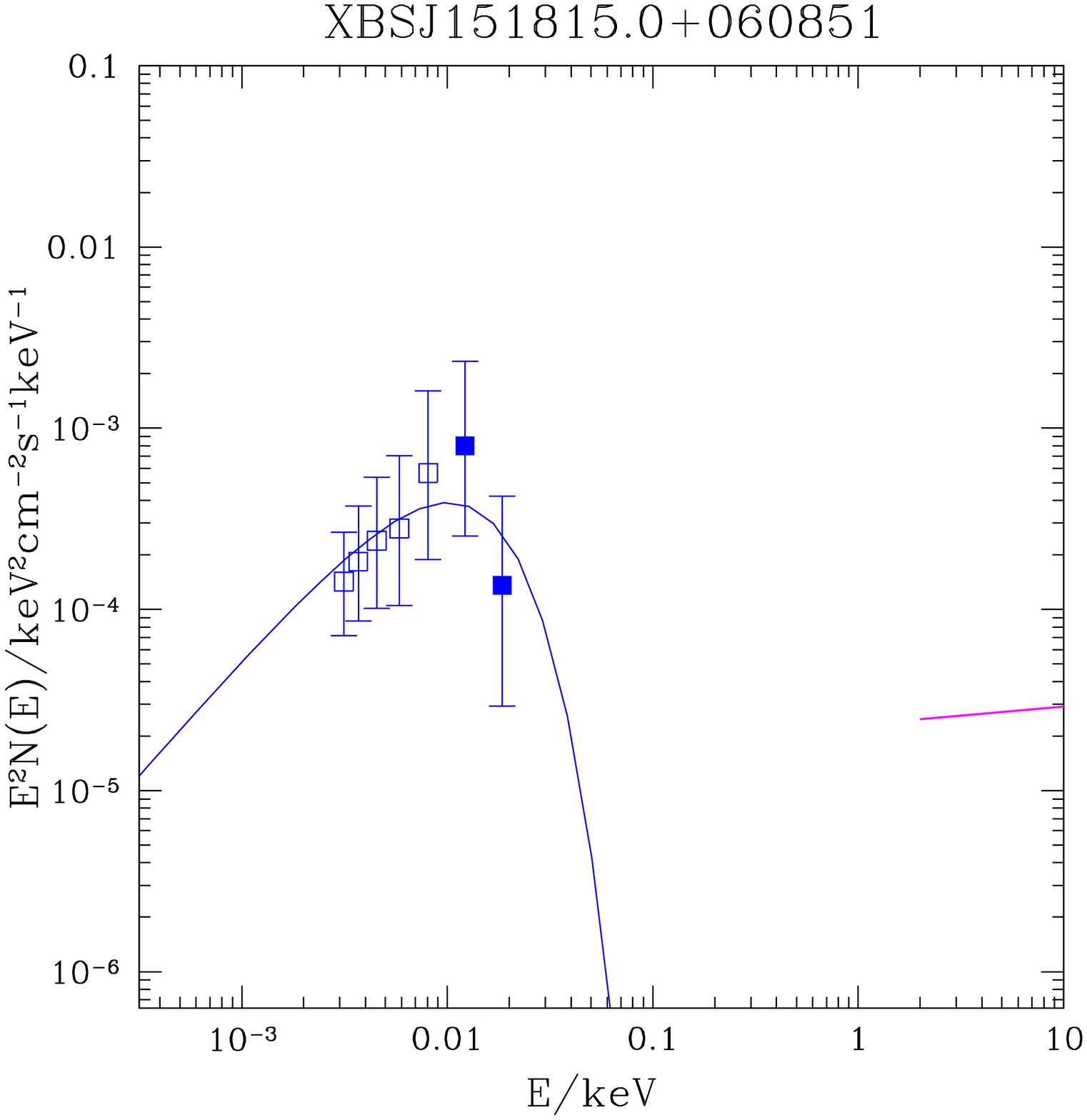}
  \includegraphics[height=5.6cm, width=6cm]{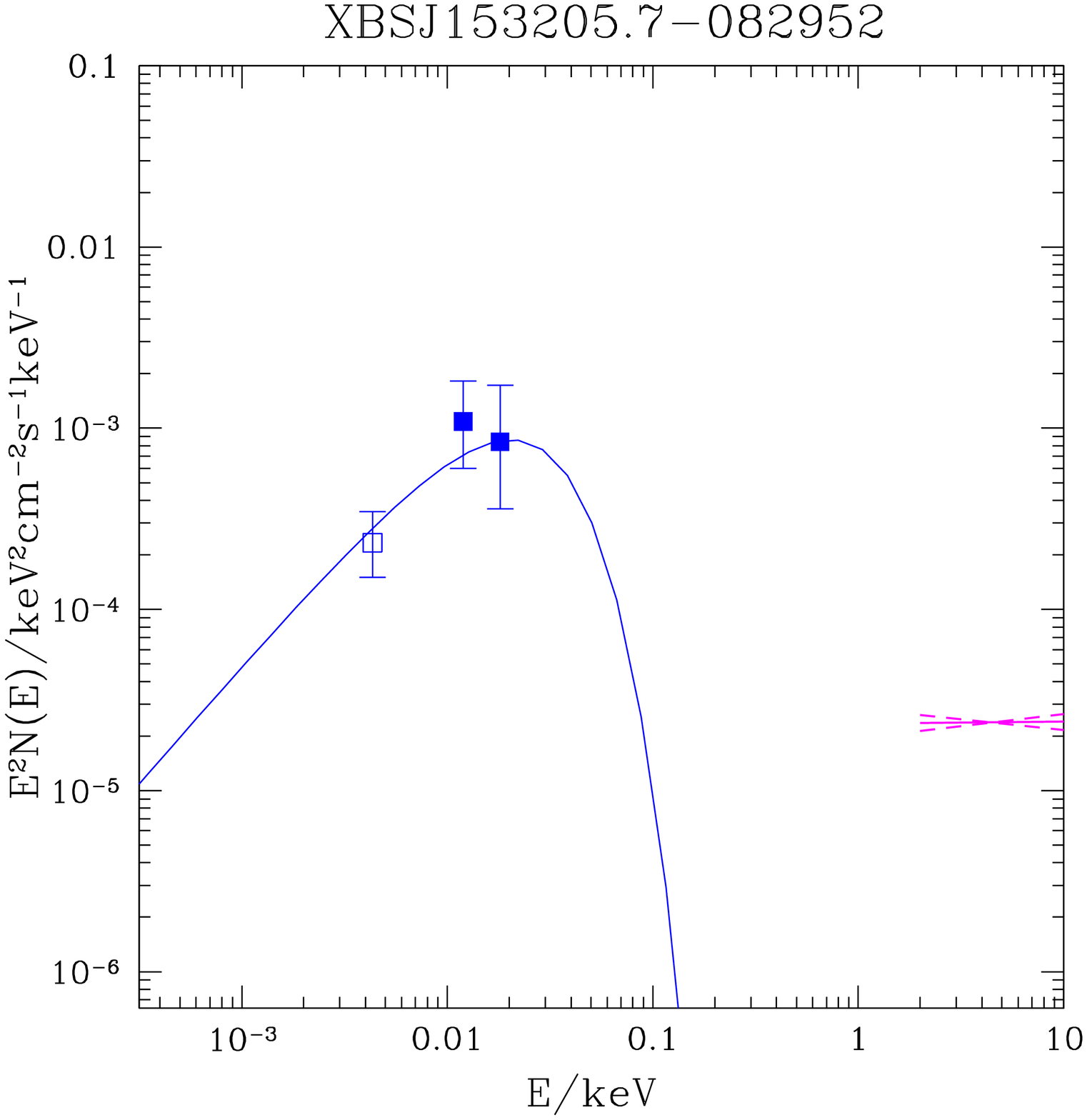}}  
\end{figure*}

 \FloatBarrier

 \begin{figure*}
\centering  
\subfigure{ 
  \includegraphics[height=5.6cm, width=6cm]{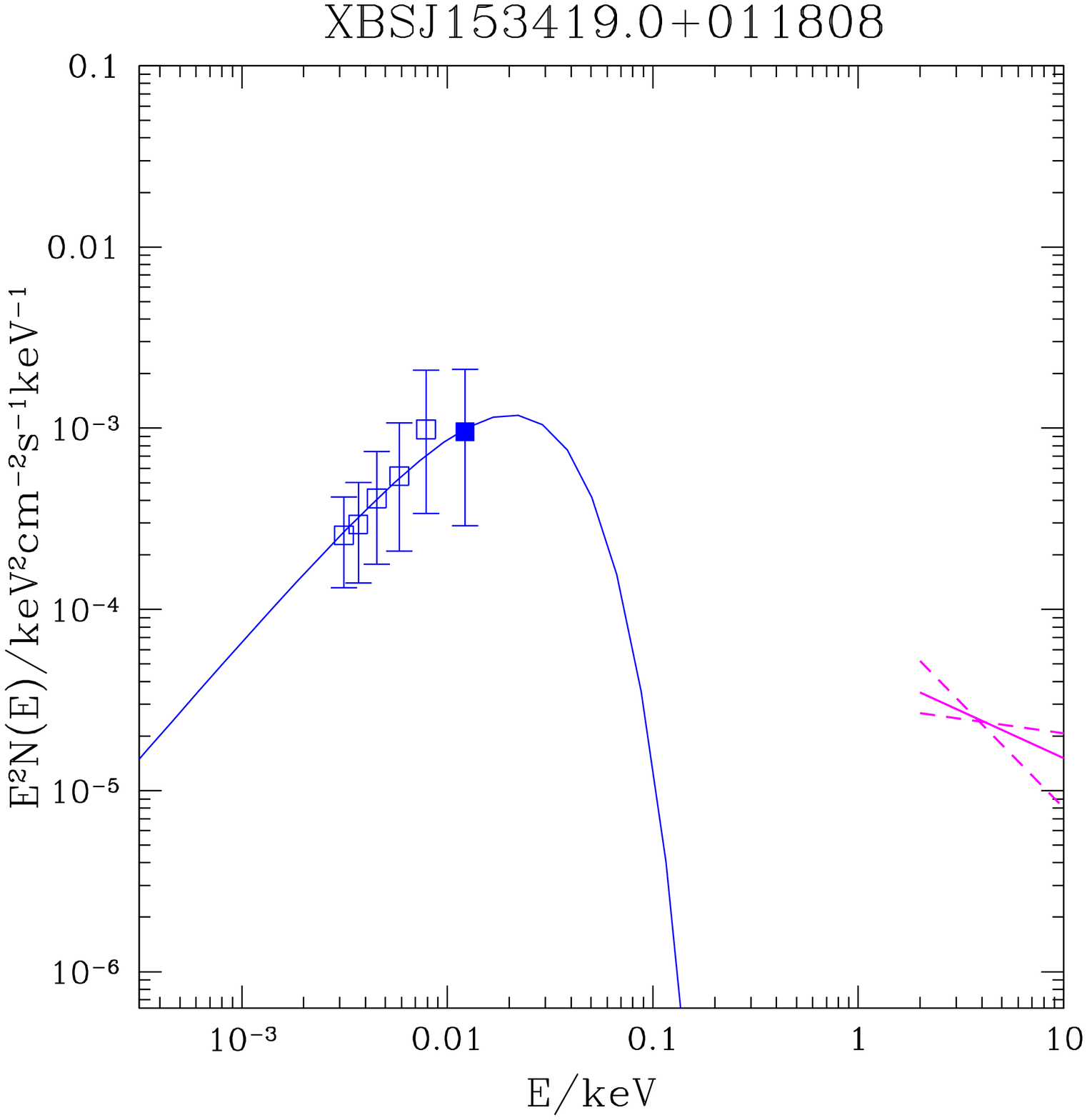}
  \includegraphics[height=5.6cm, width=6cm]{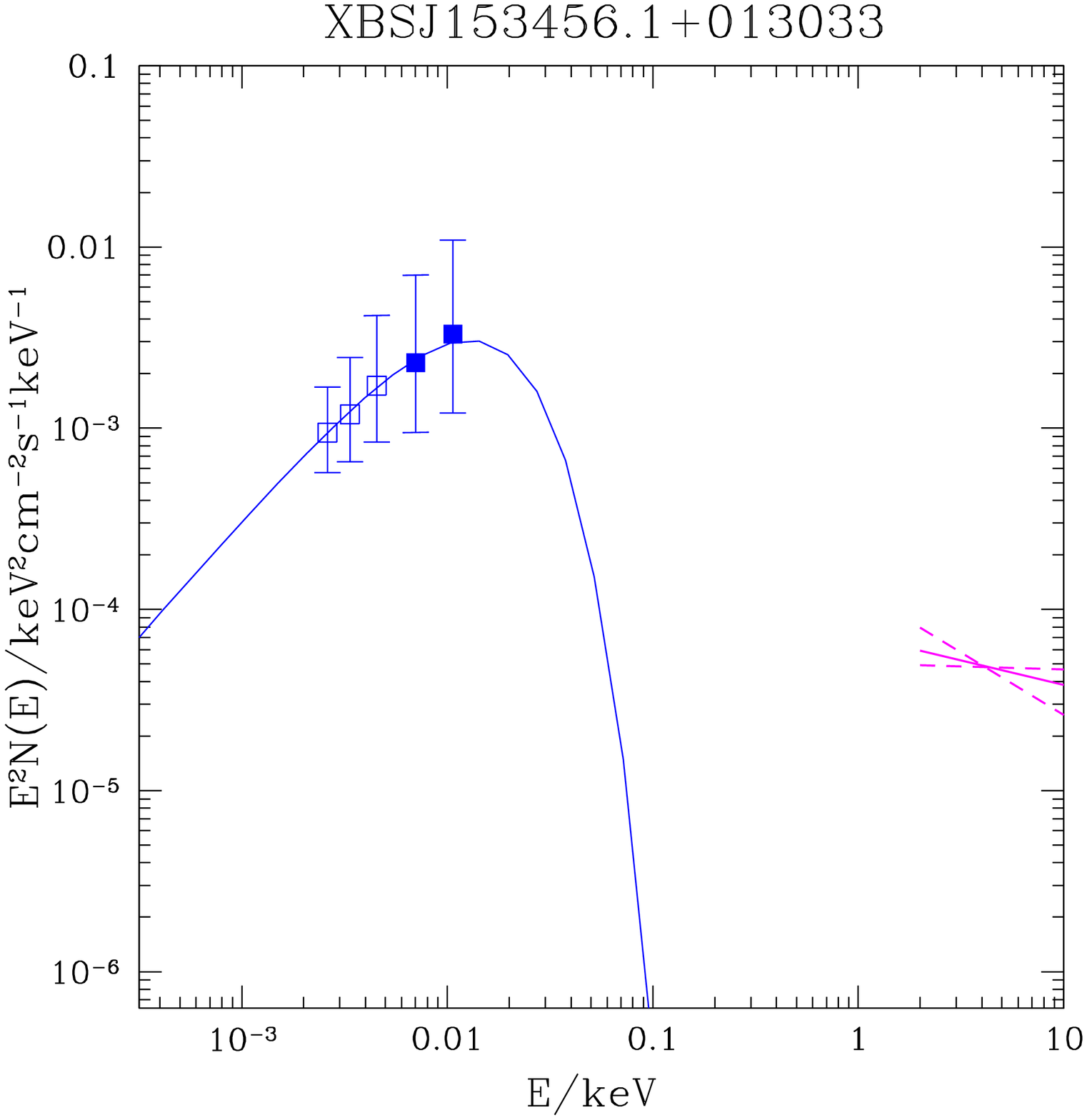}}     
 \subfigure{ 
  \includegraphics[height=5.6cm, width=6cm]{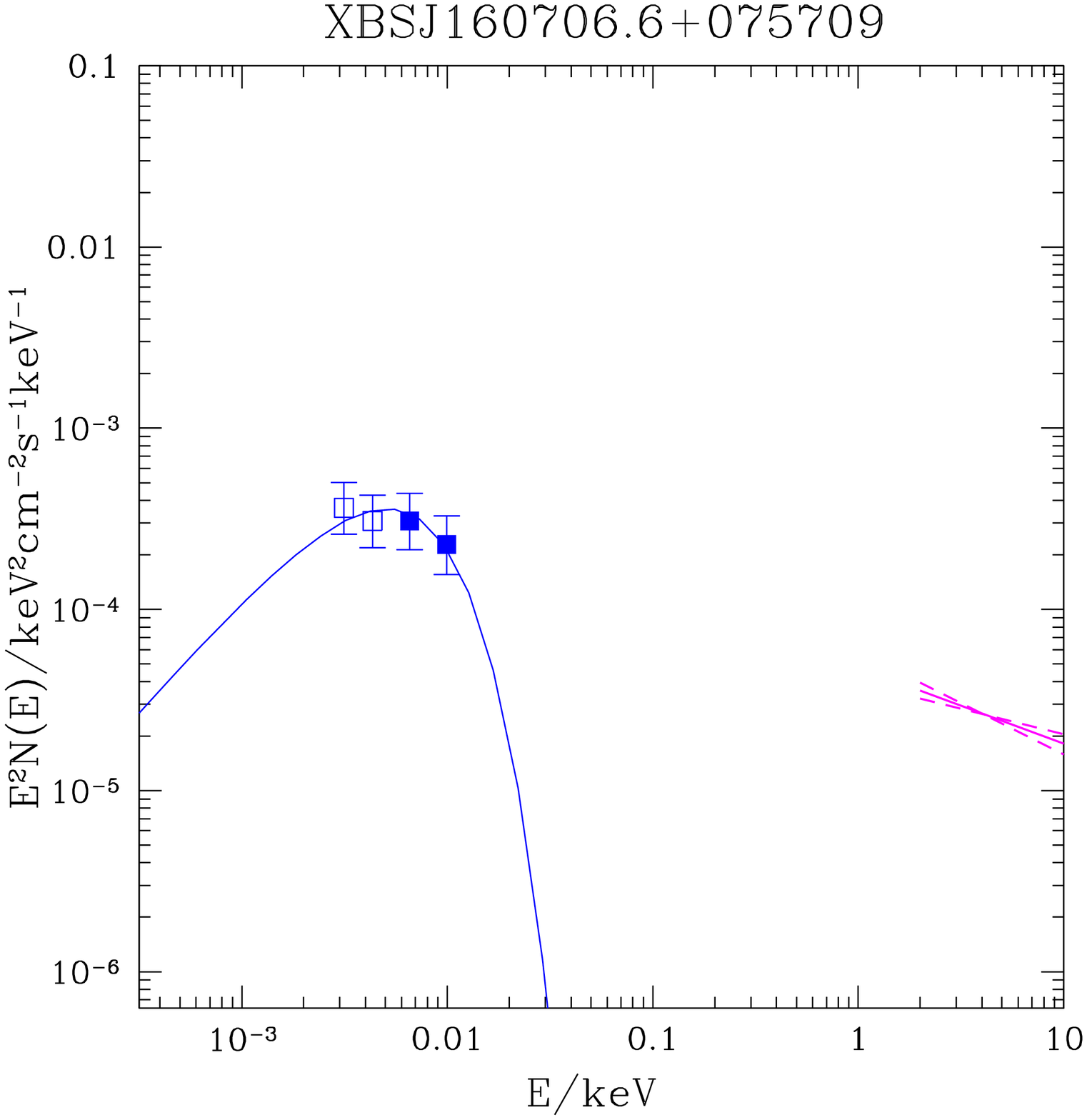}
  \includegraphics[height=5.6cm, width=6cm]{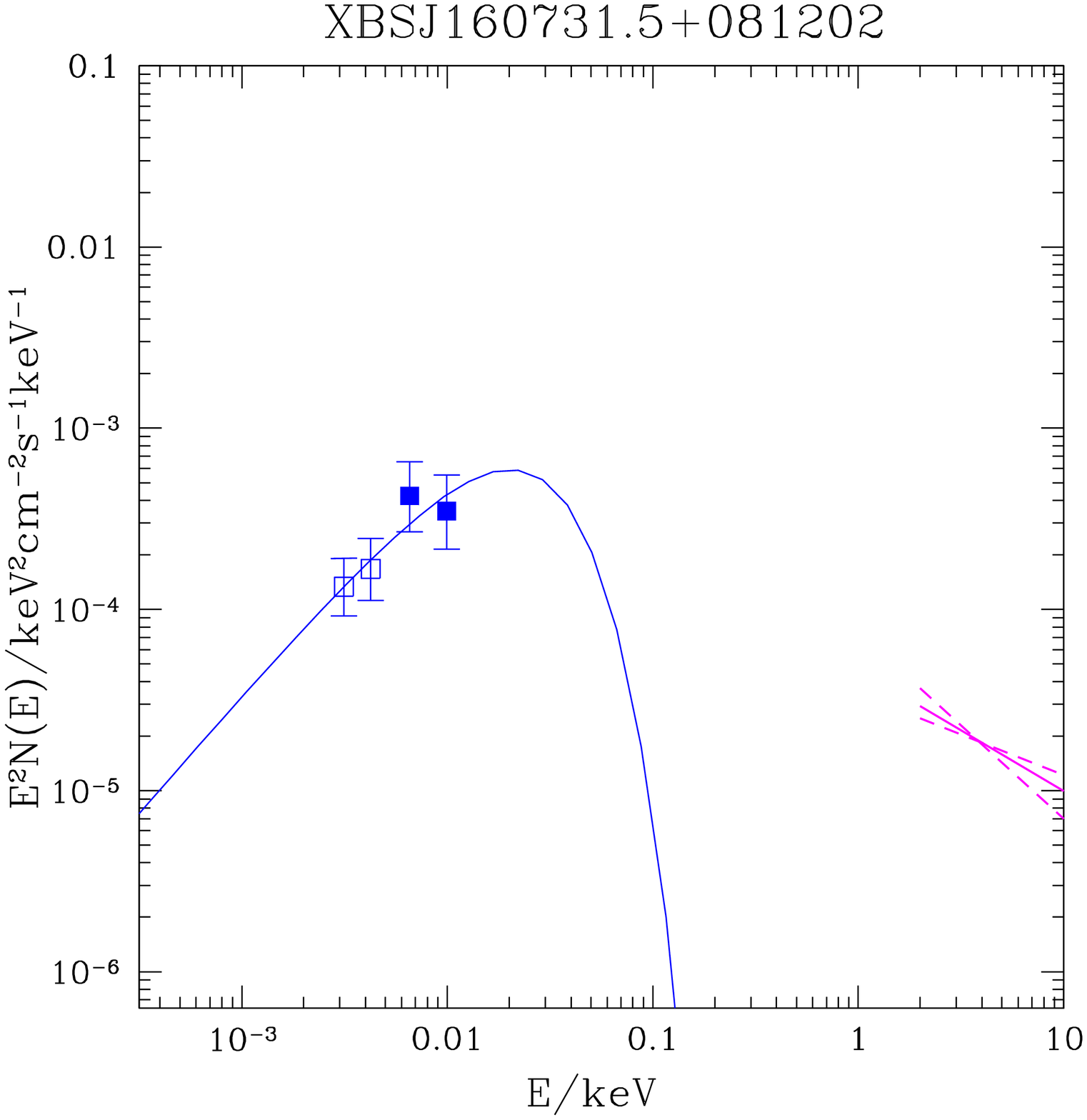}}    
 \subfigure{ 
  \includegraphics[height=5.6cm, width=6cm]{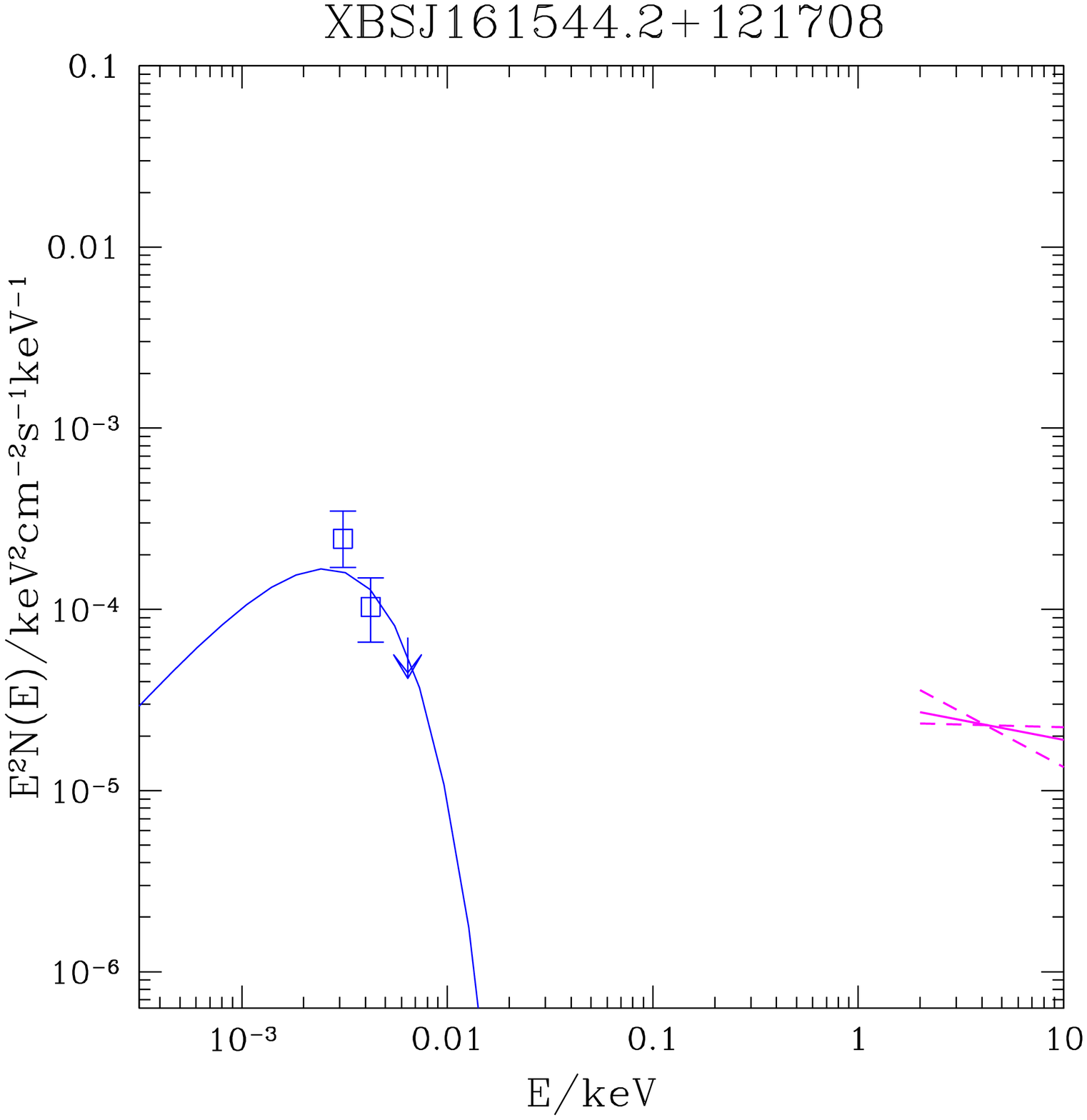}
  \includegraphics[height=5.6cm, width=6cm]{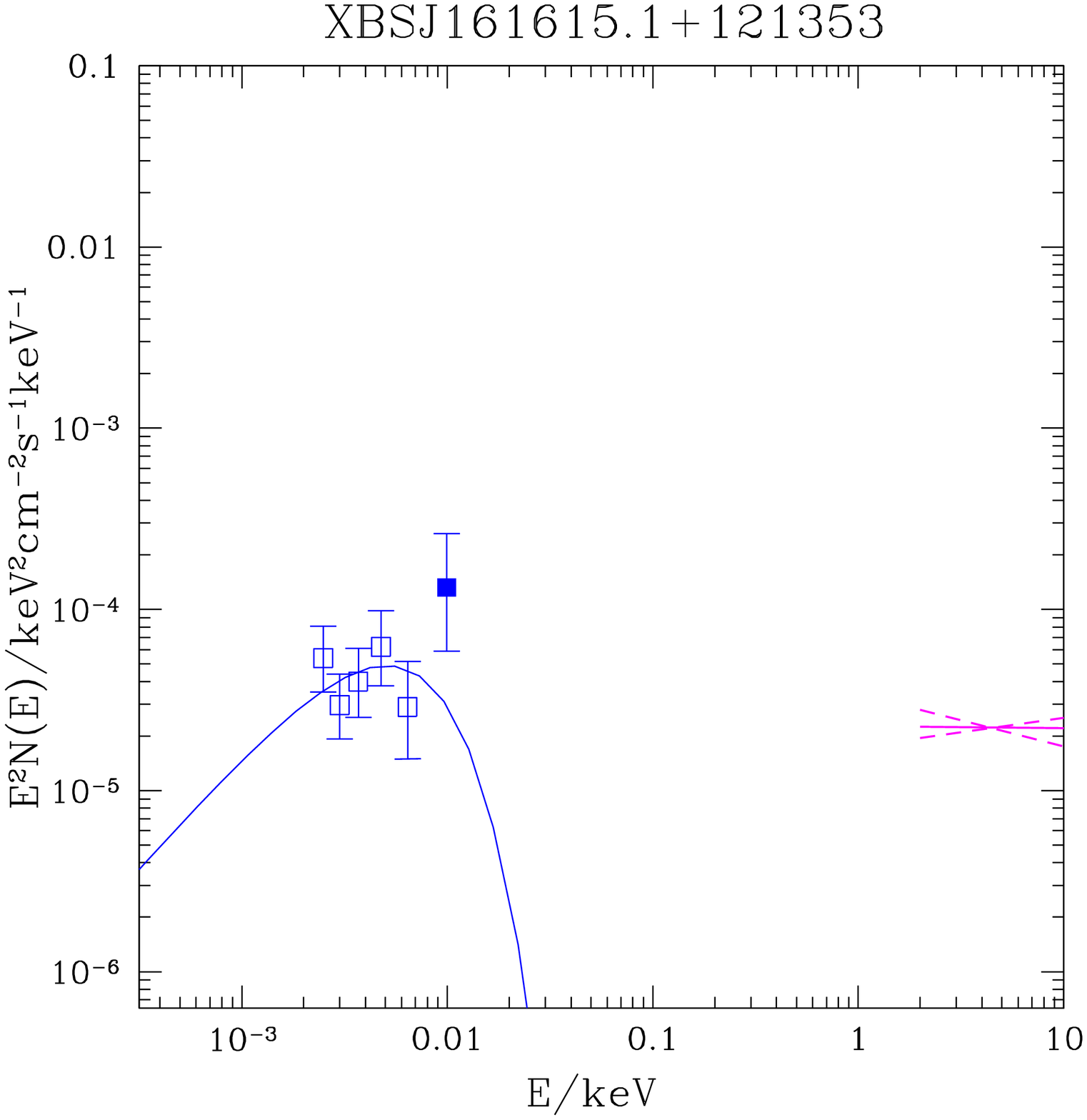}}      
\subfigure{ 
  \includegraphics[height=5.6cm, width=6cm]{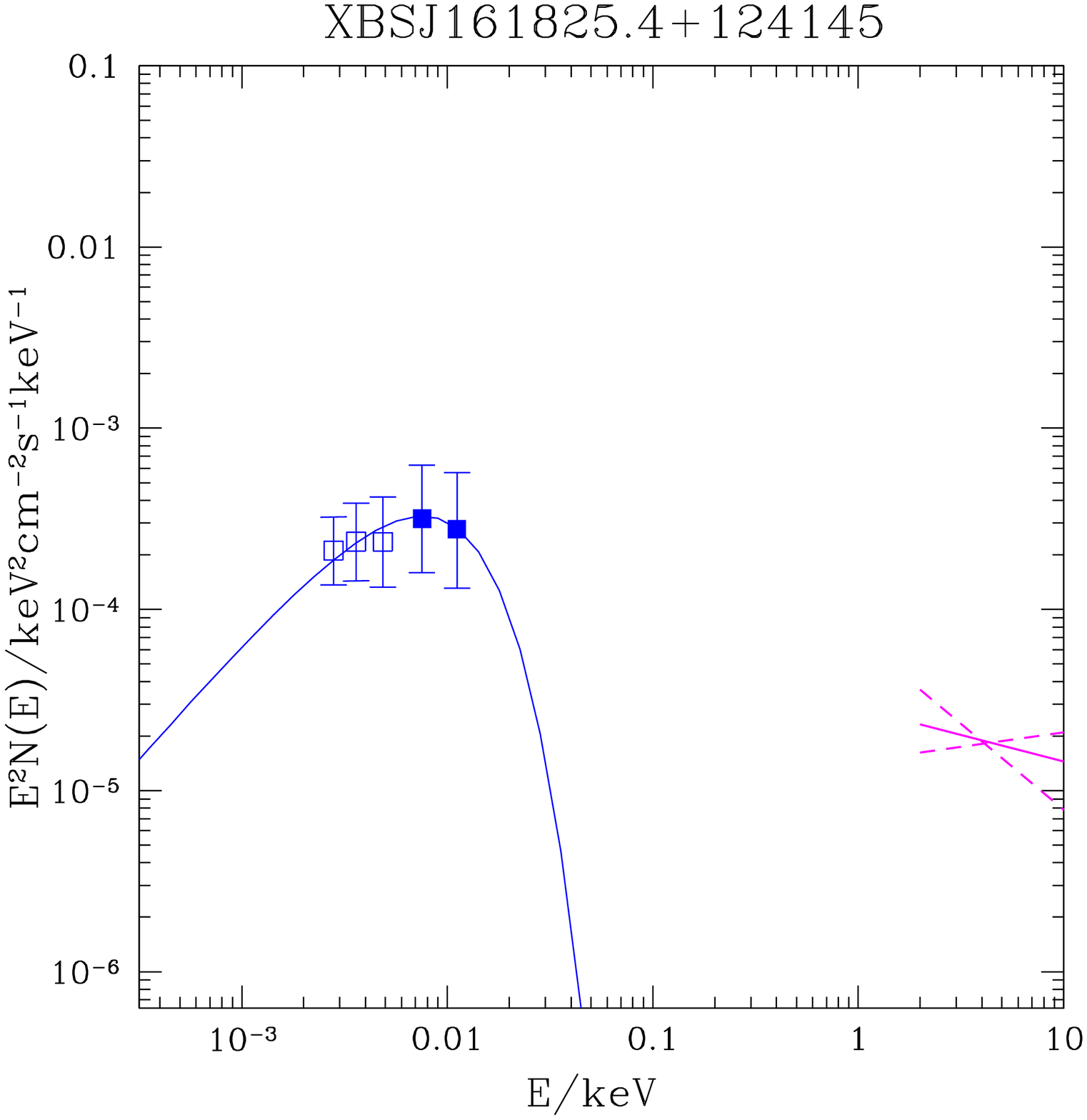}
  \includegraphics[height=5.6cm, width=6cm]{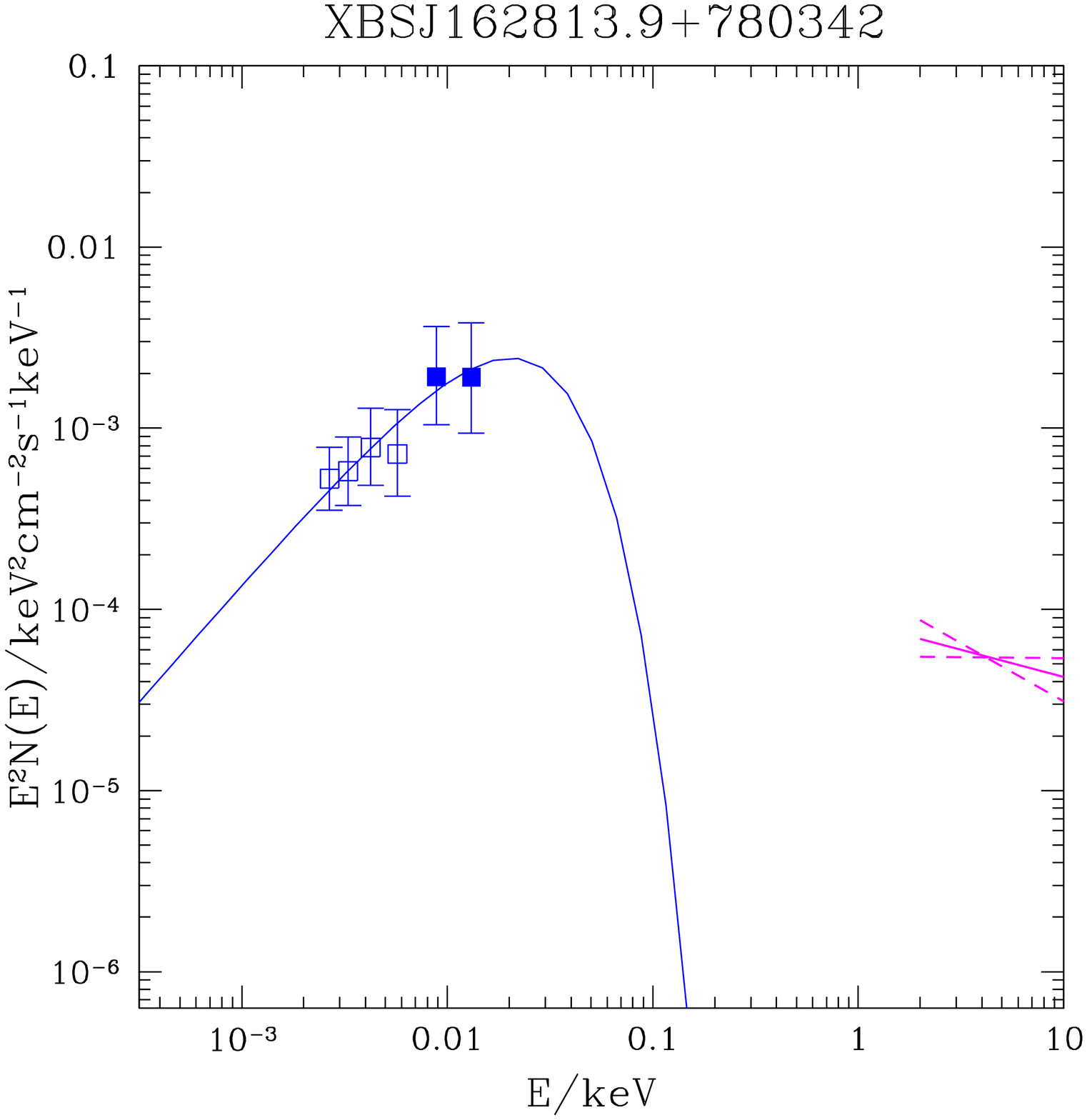}}
   \end{figure*}
   
    \FloatBarrier
   
    \begin{figure*}
\centering      
\subfigure{ 
  \includegraphics[height=5.6cm, width=6cm]{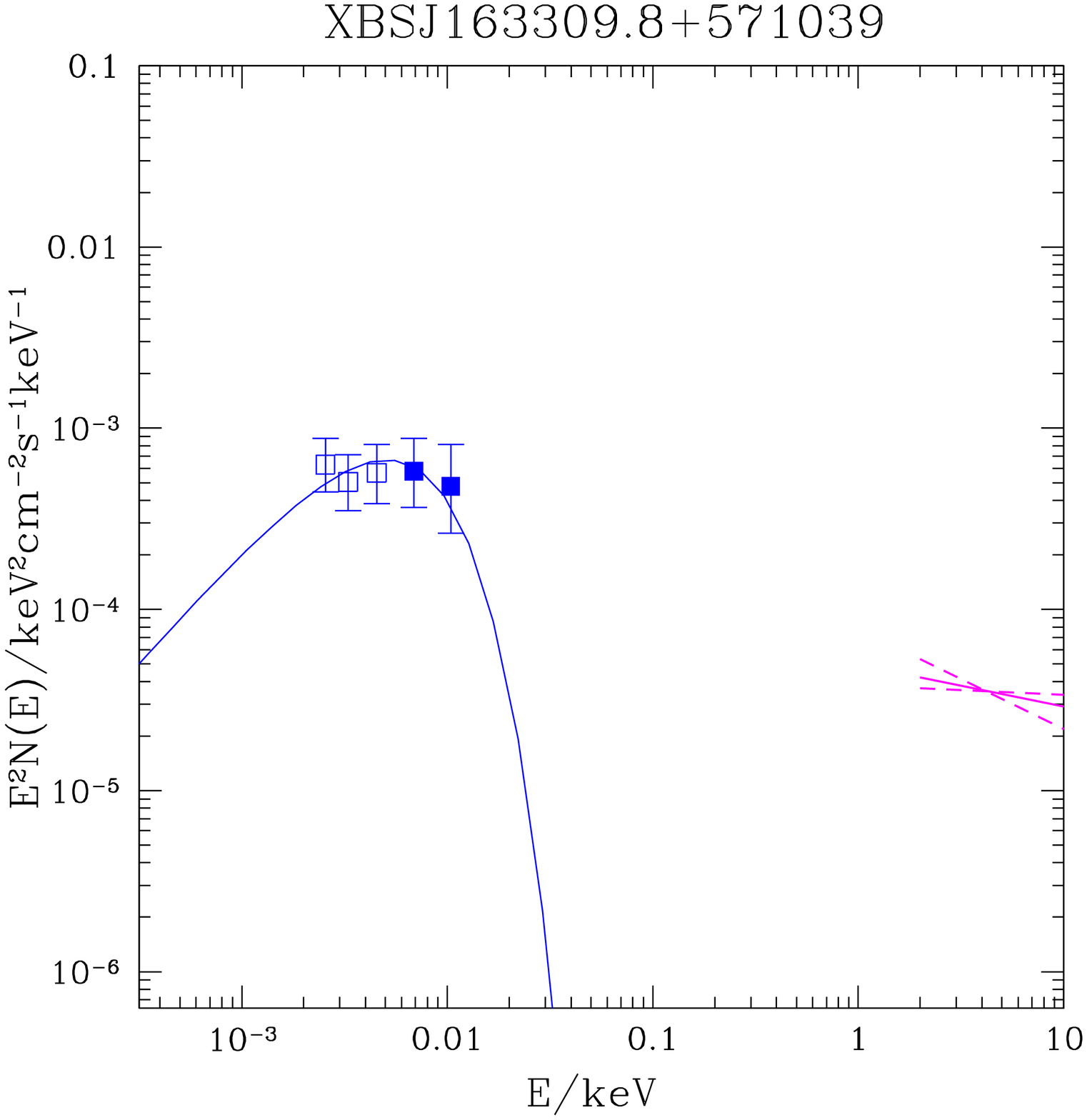}
  \includegraphics[height=5.6cm, width=6cm]{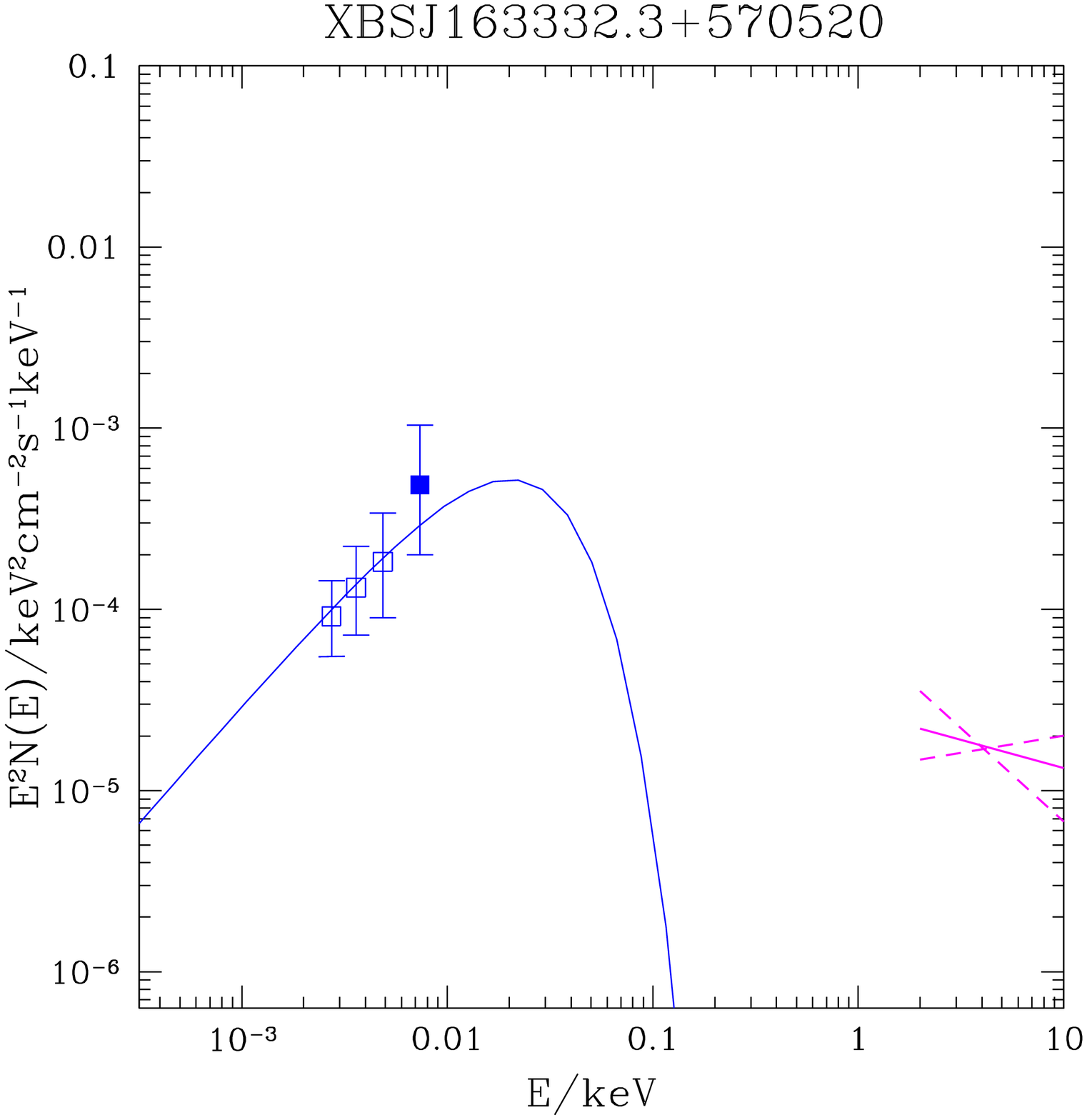}}     
 \subfigure{ 
  \includegraphics[height=5.6cm, width=6cm]{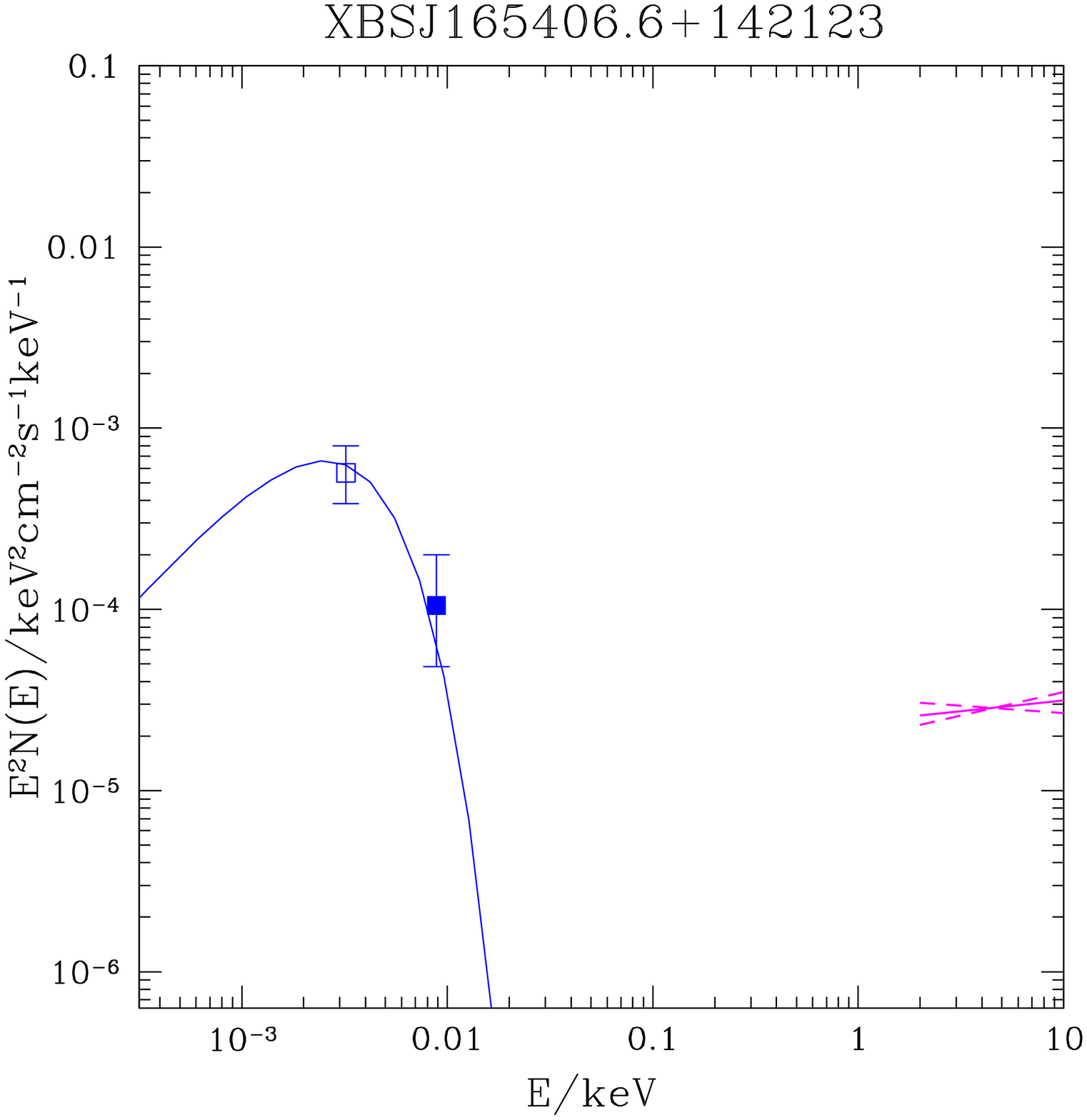}
  \includegraphics[height=5.6cm, width=6cm]{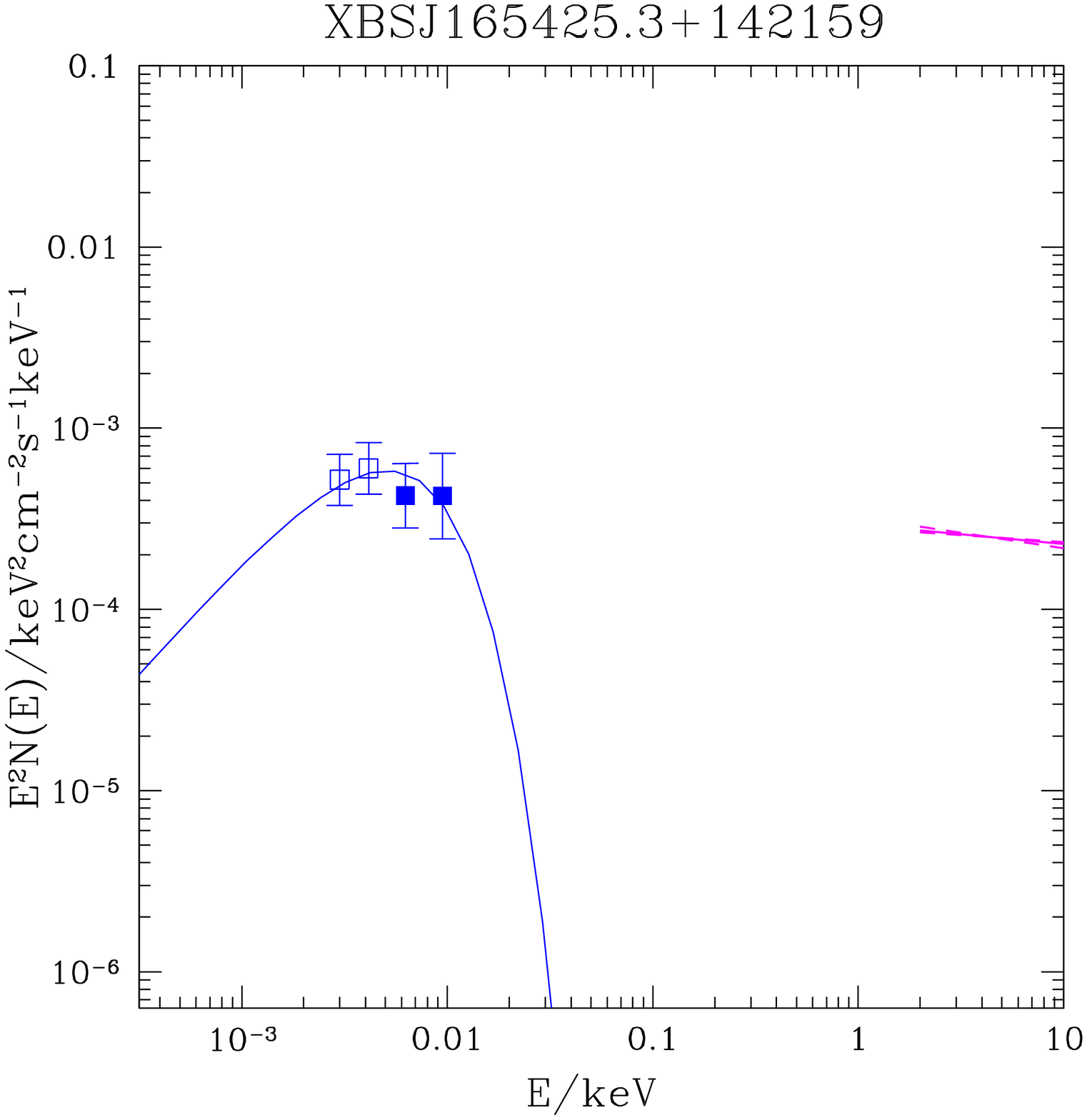}}  
\subfigure{ 
  \includegraphics[height=5.6cm, width=6cm]{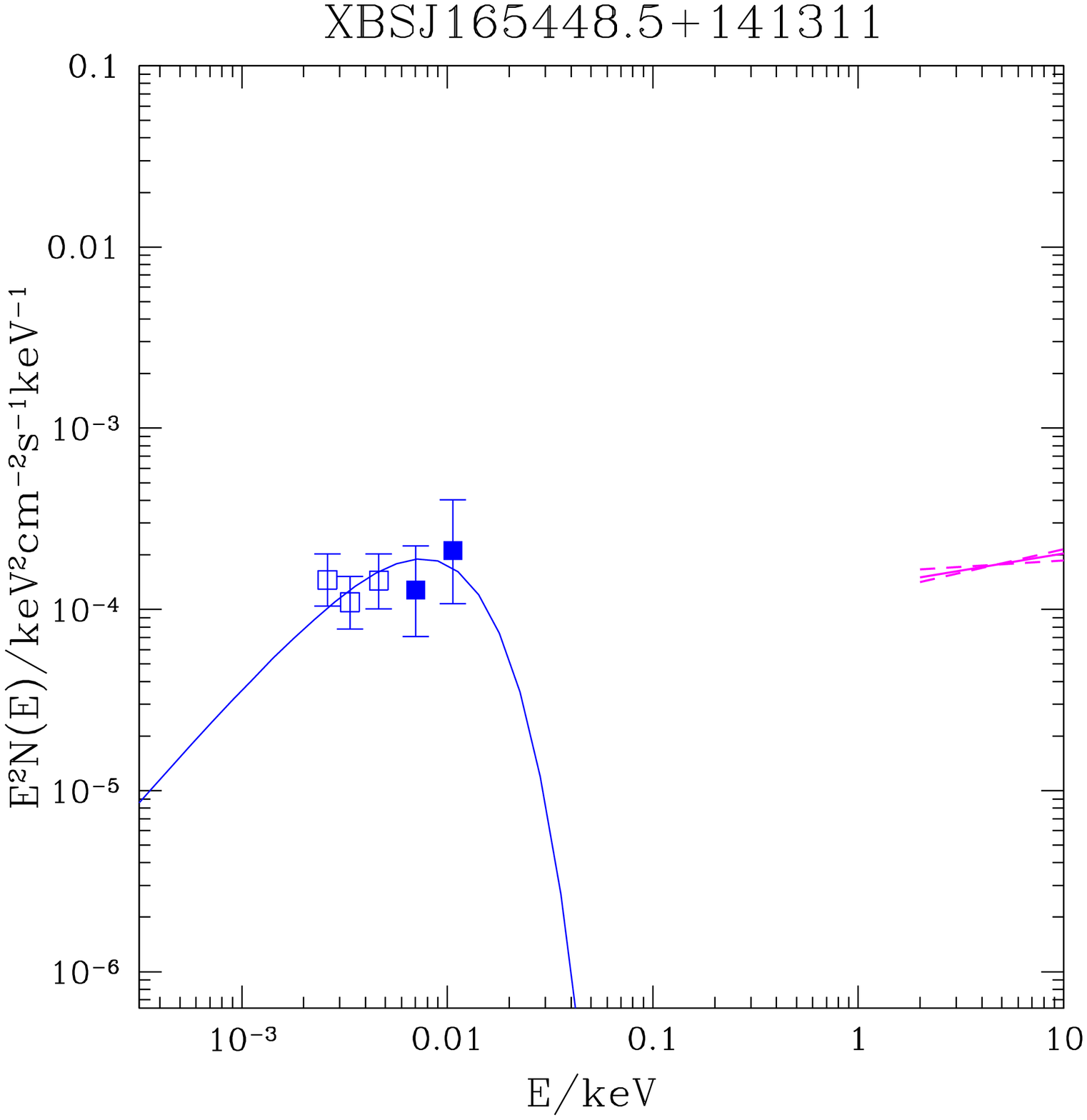}
  \includegraphics[height=5.6cm, width=6cm]{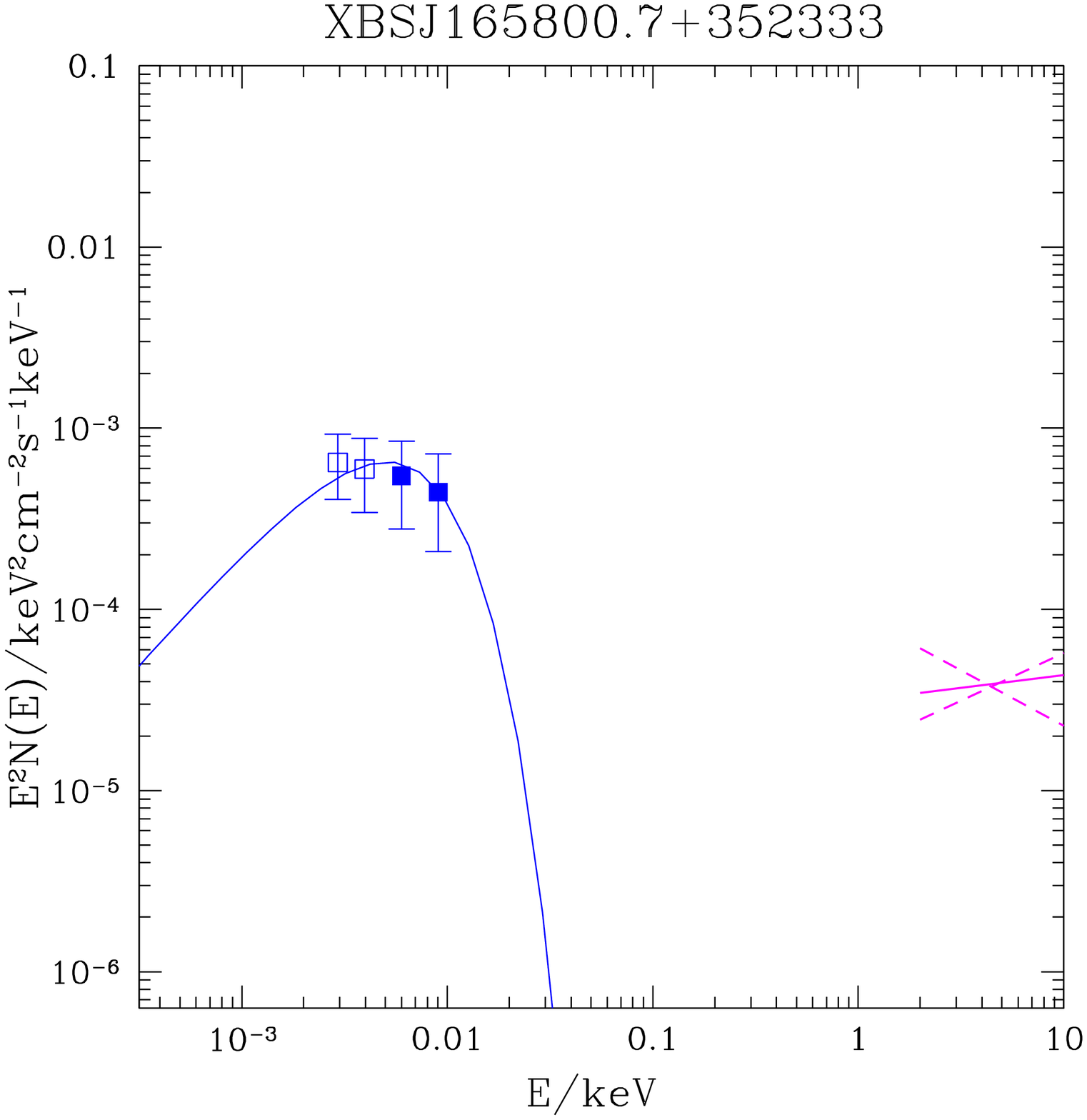}}      
\subfigure{ 
  \includegraphics[height=5.6cm, width=6cm]{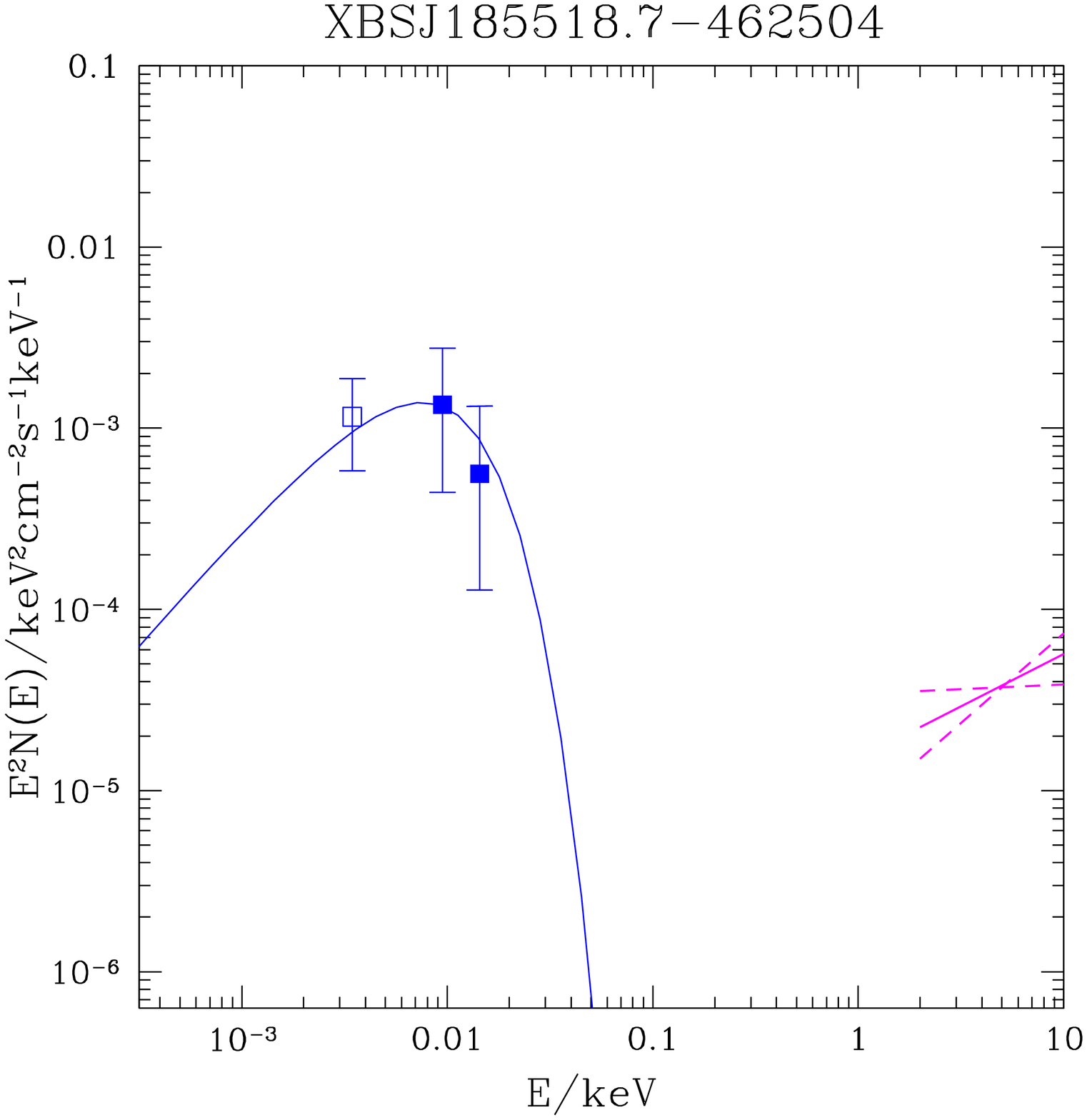}
  \includegraphics[height=5.6cm, width=6cm]{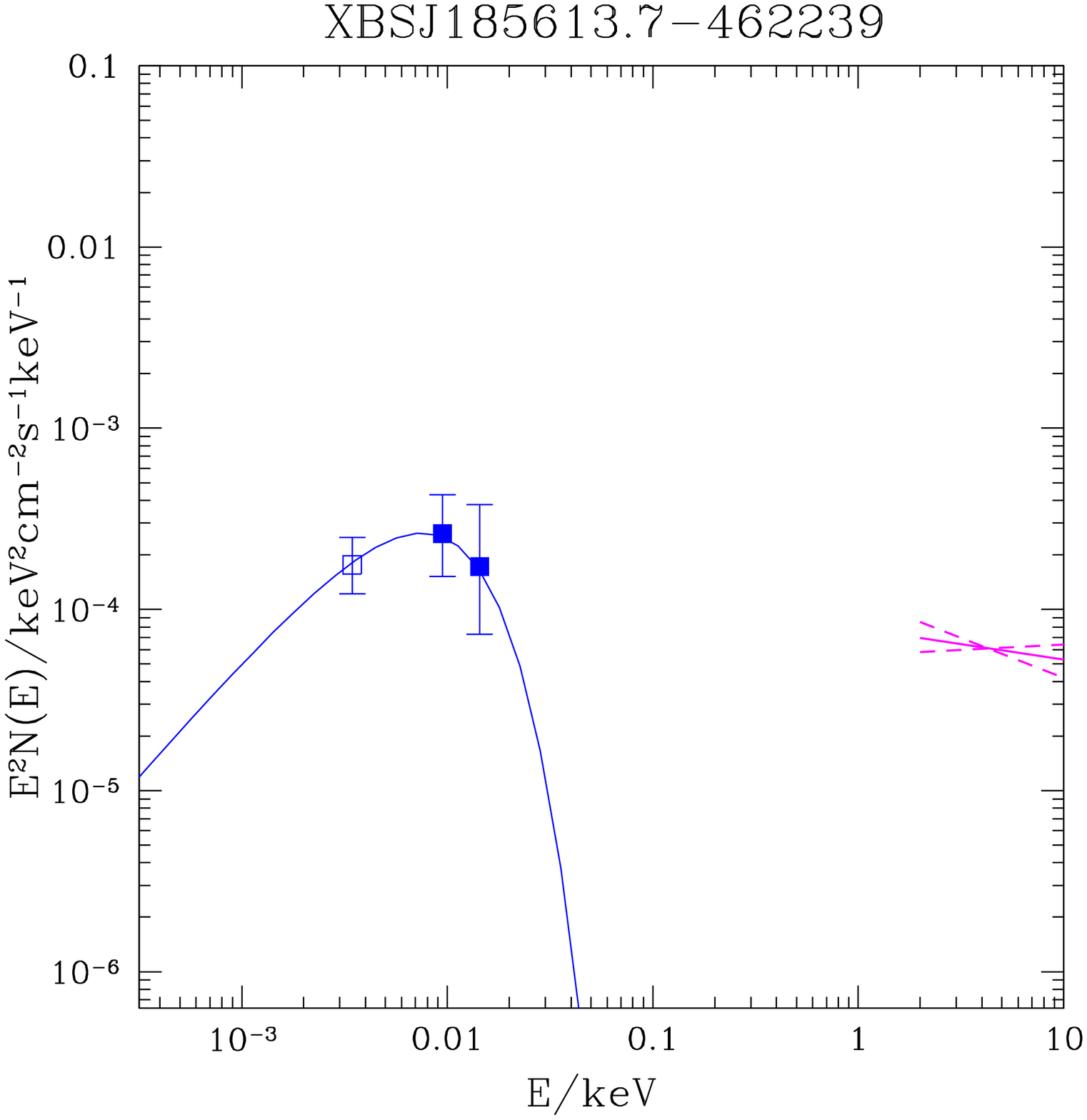}}    
  \end{figure*}
  
   \FloatBarrier
  
   \begin{figure*}
\centering  
\subfigure{ 
  \includegraphics[height=5.6cm, width=6cm]{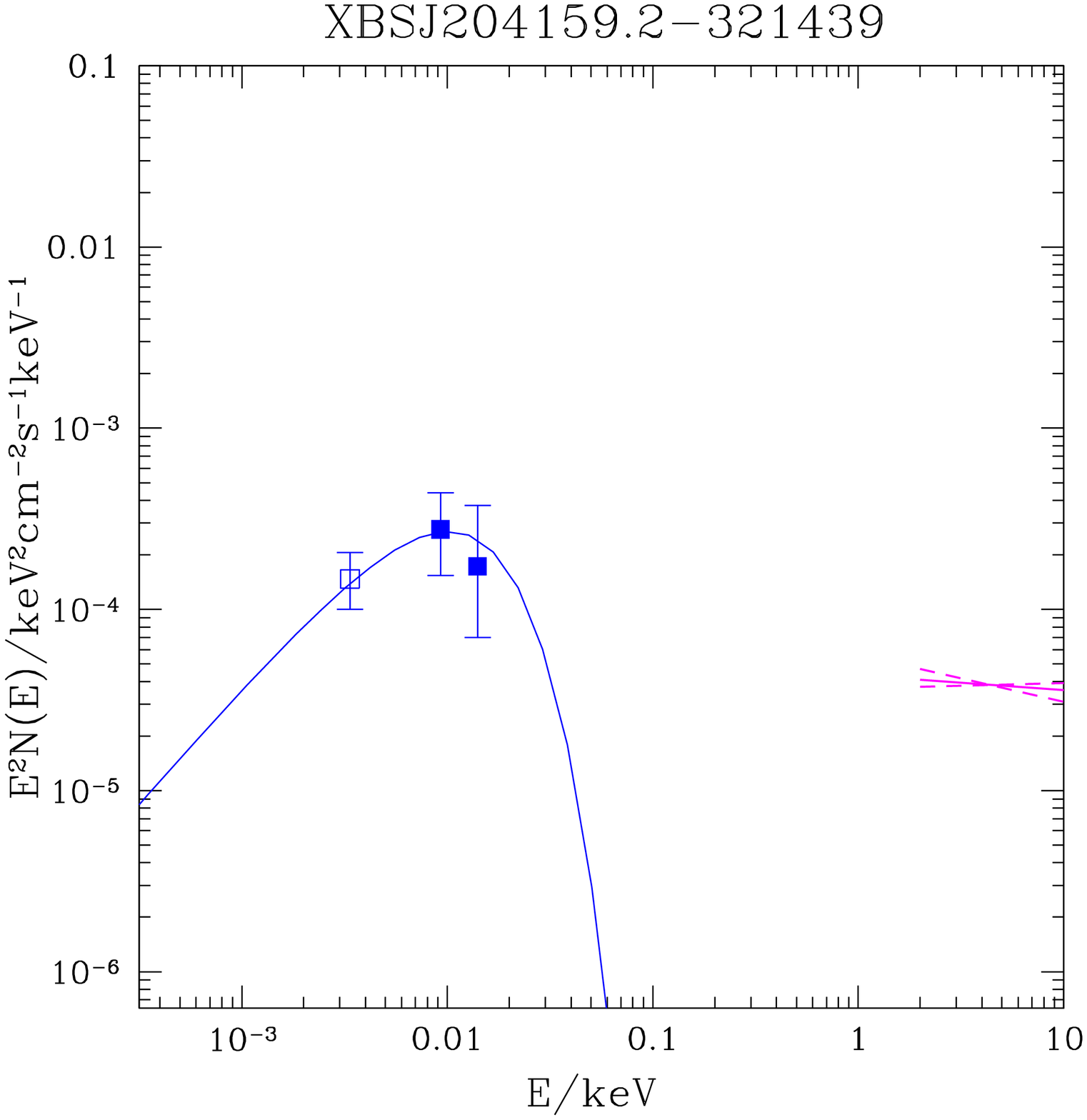}
  \includegraphics[height=5.6cm, width=6cm]{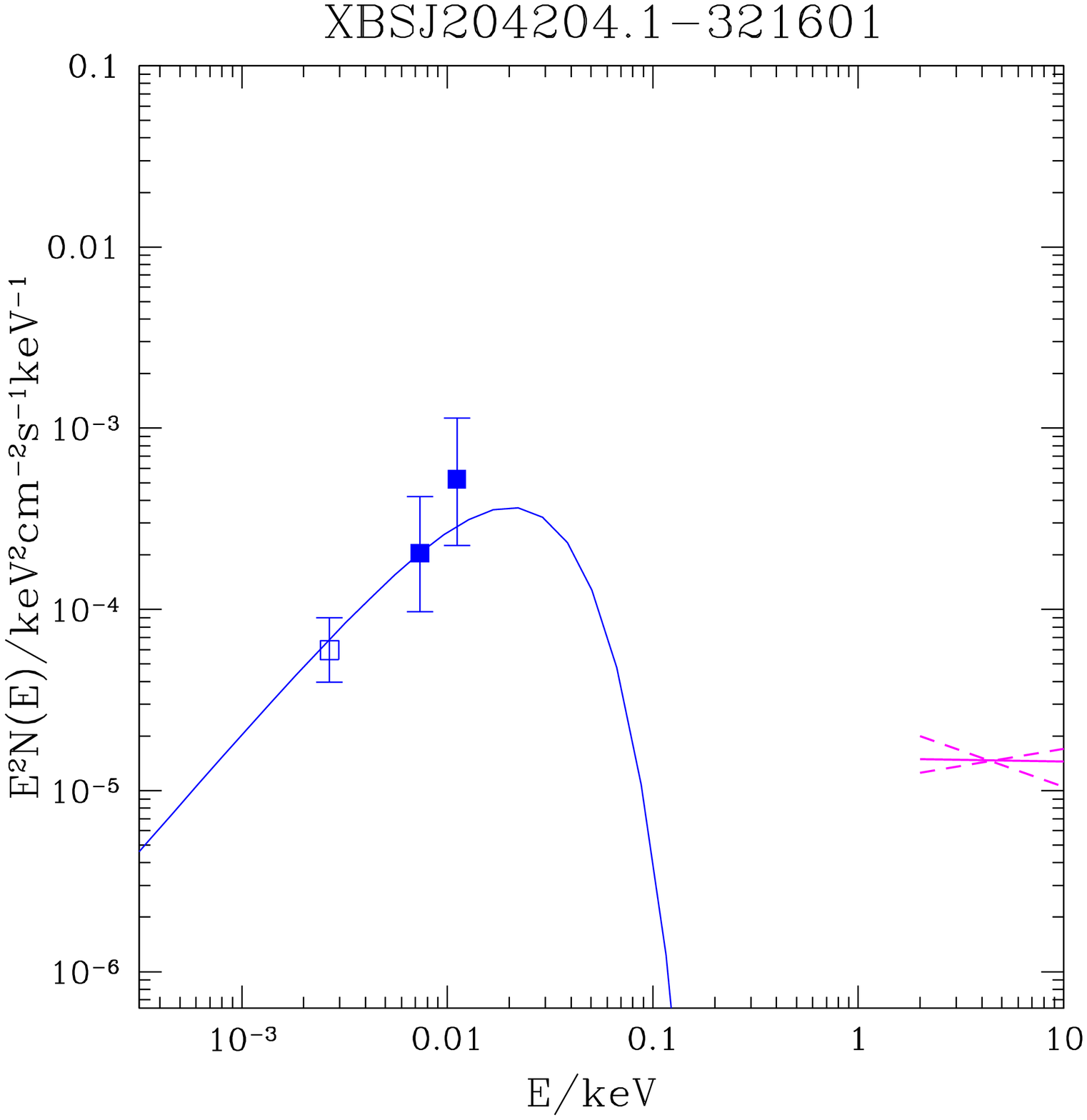}}      
\subfigure{ 
  \includegraphics[height=5.6cm, width=6cm]{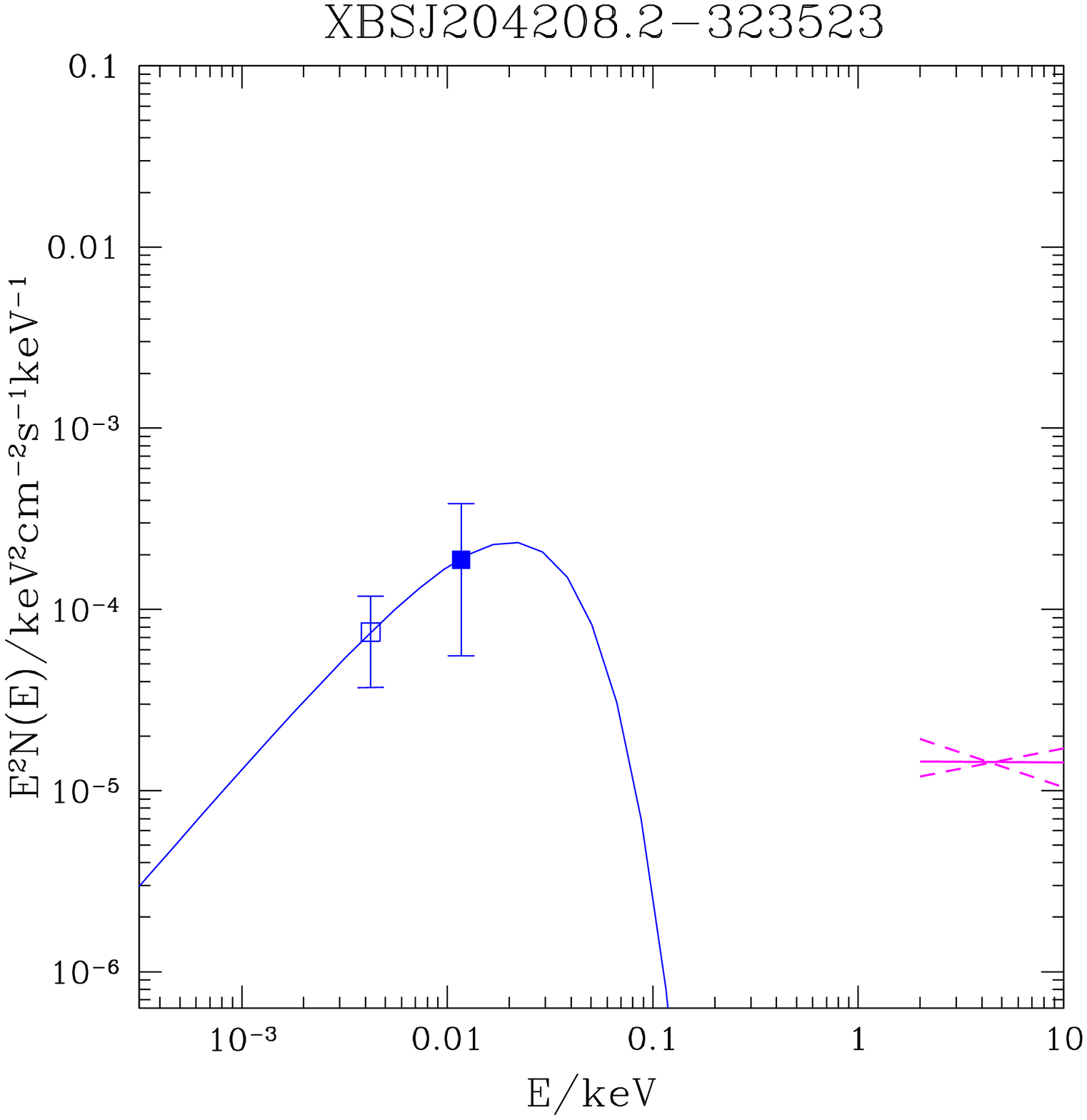}
  \includegraphics[height=5.6cm, width=6cm]{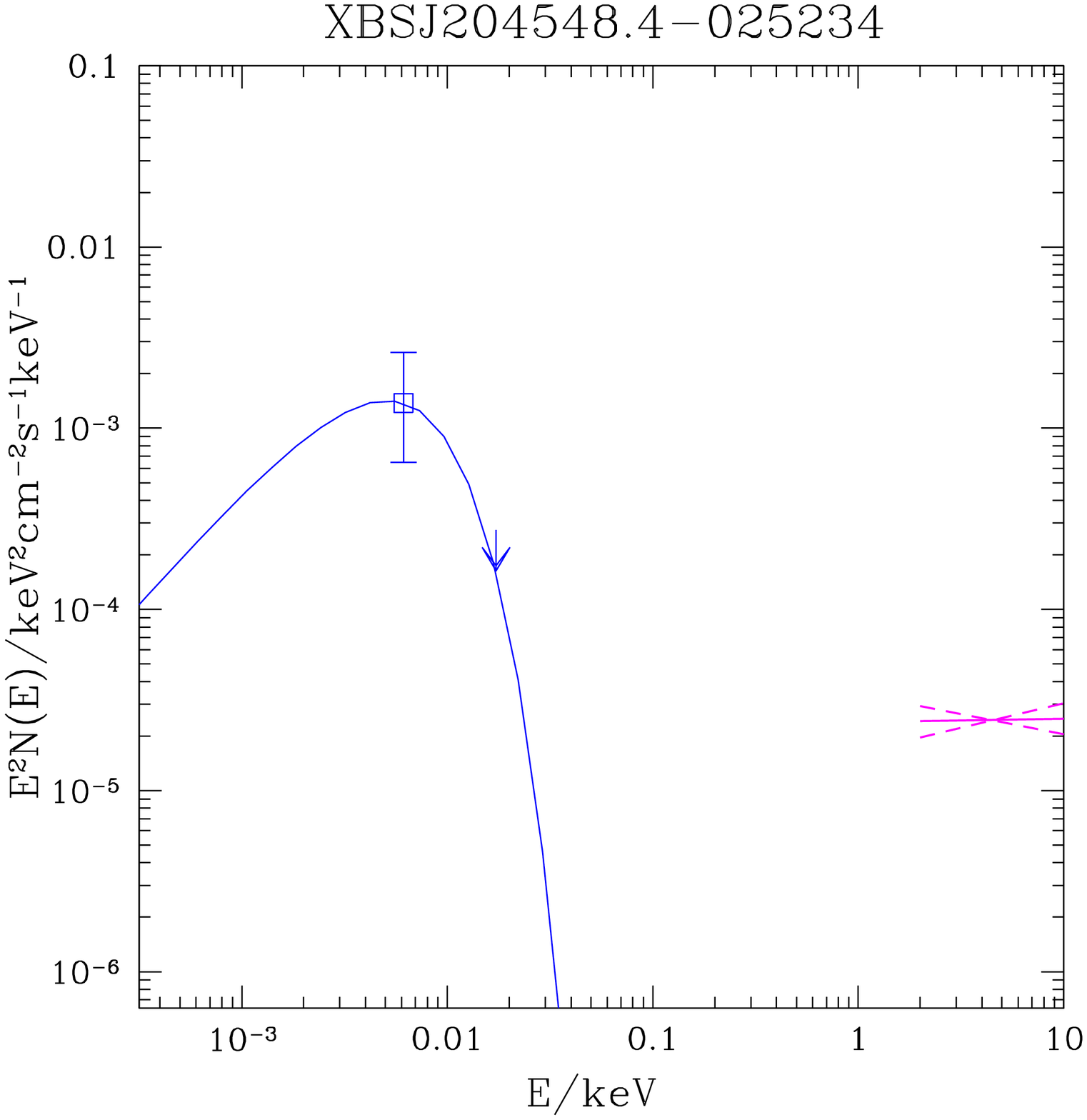}}    
\subfigure{ 
  \includegraphics[height=5.6cm, width=6cm]{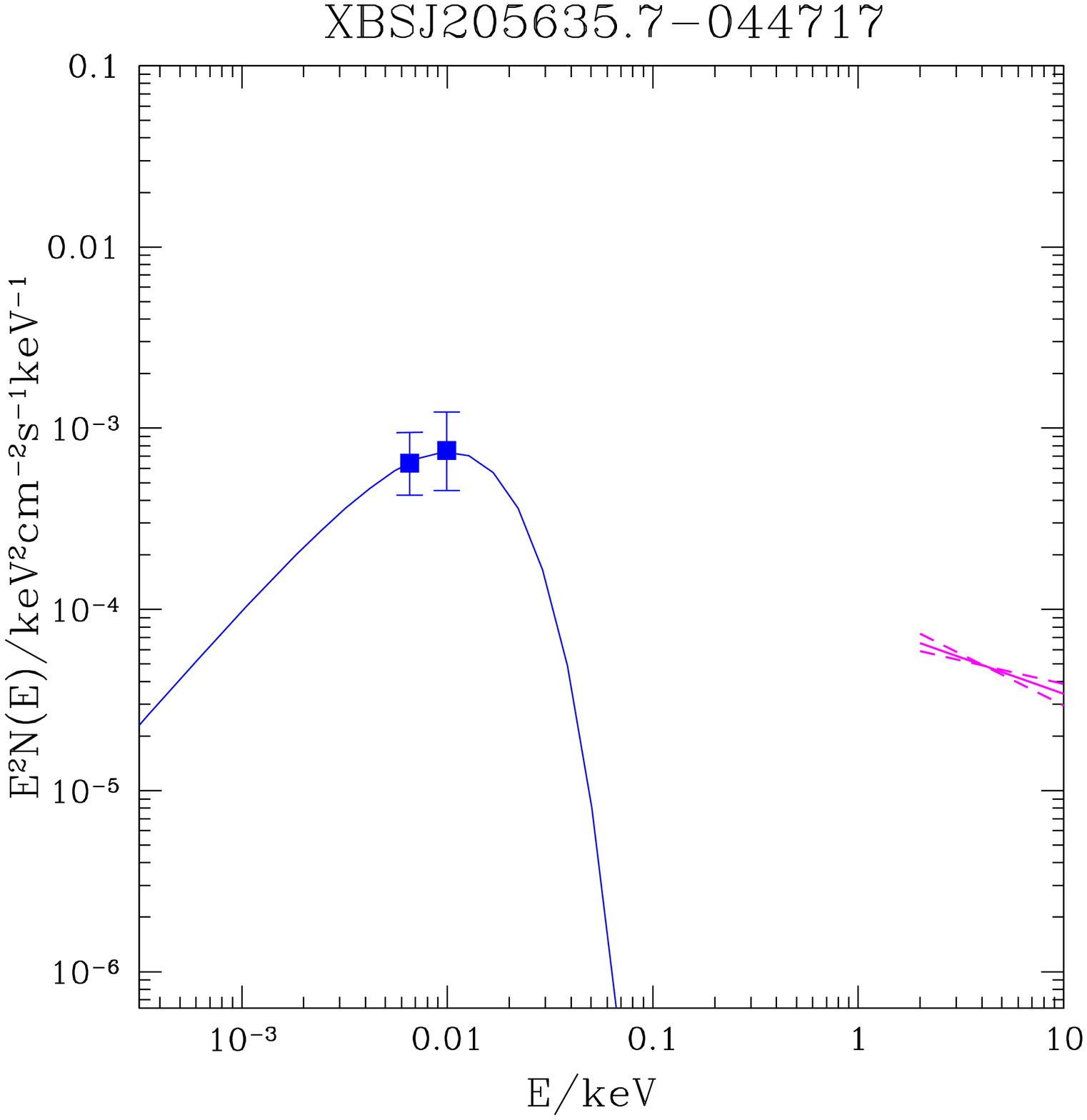}
  \includegraphics[height=5.6cm, width=6cm]{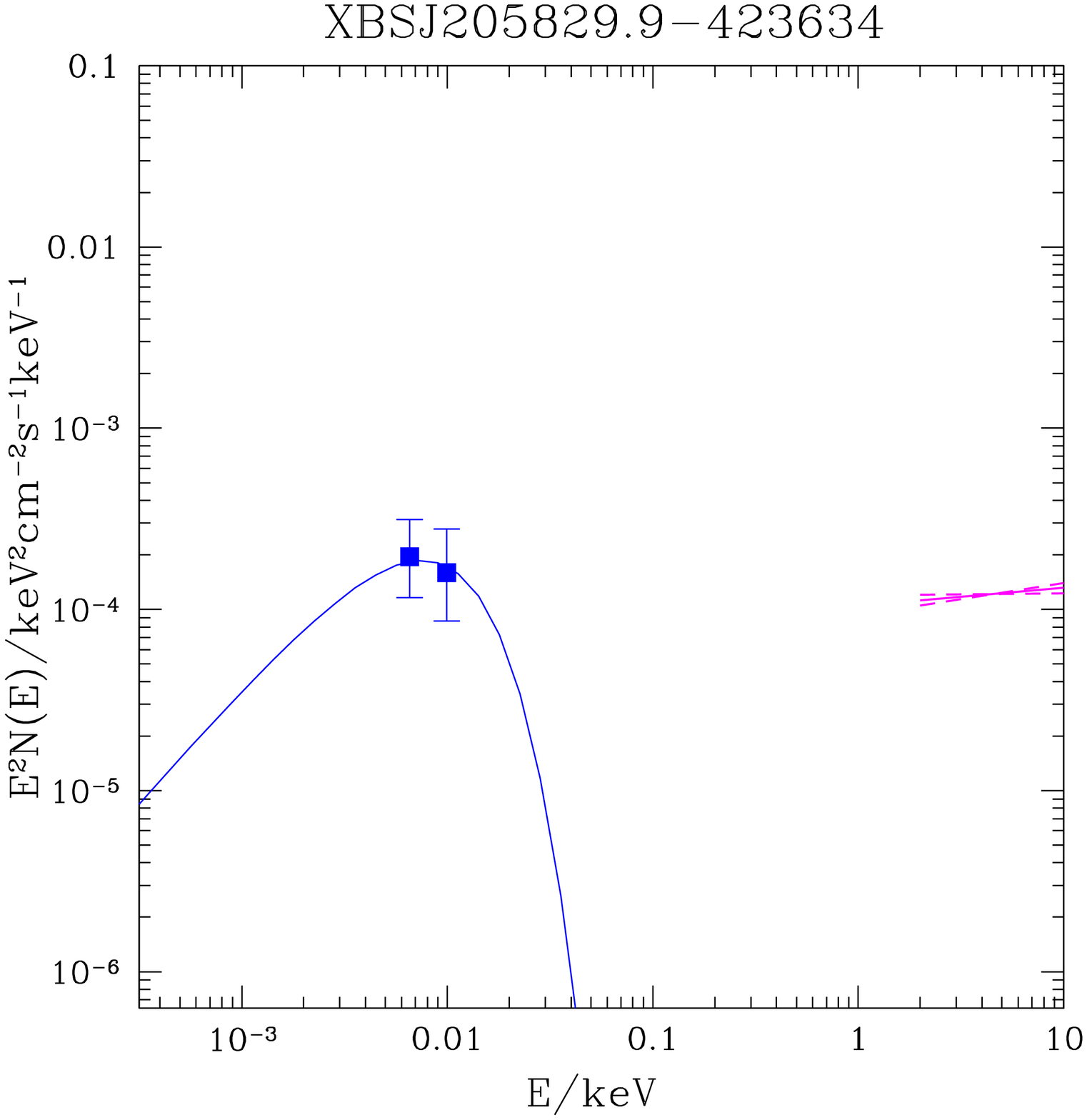}}      
\subfigure{ 
  \includegraphics[height=5.6cm, width=6cm]{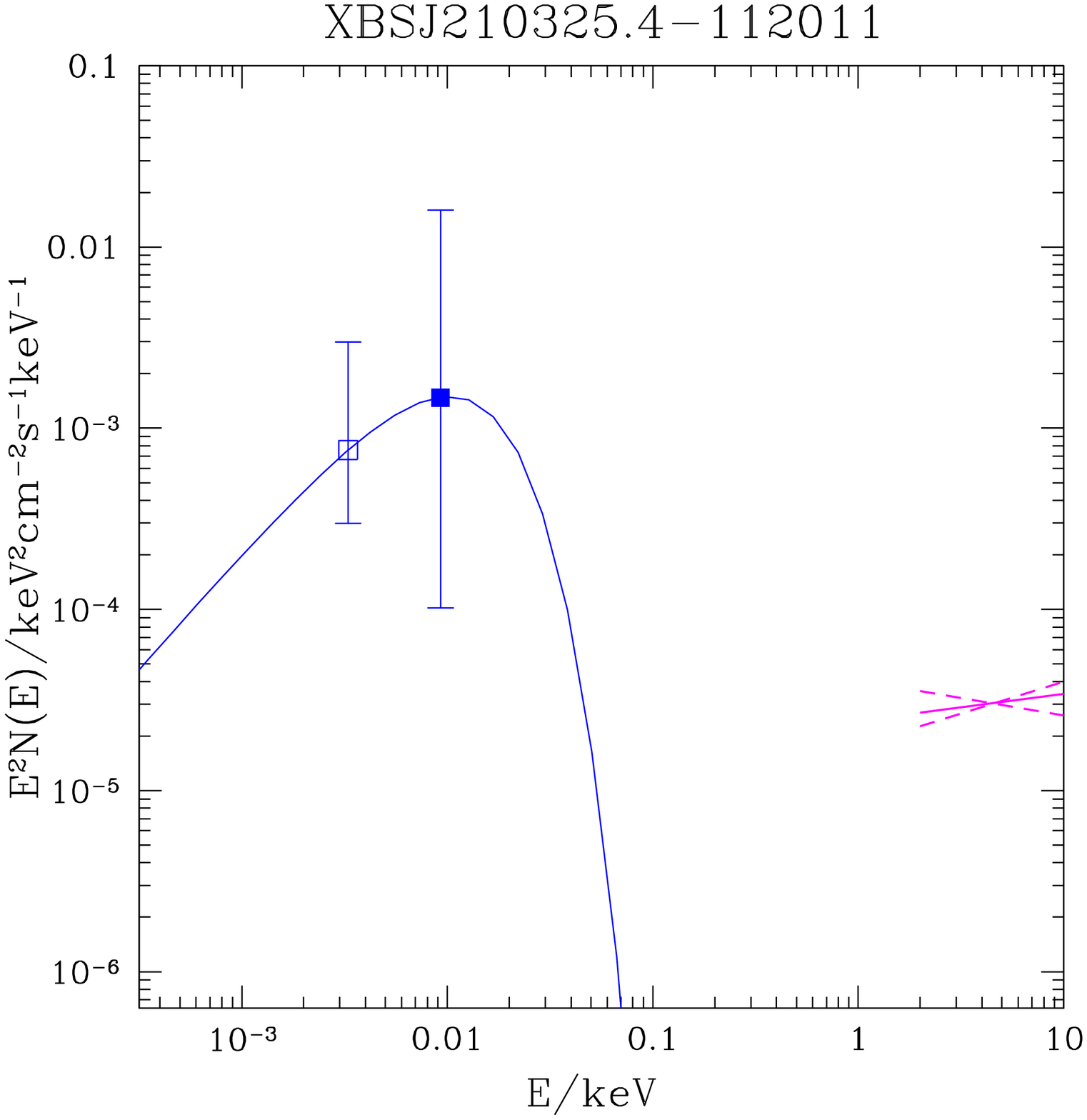}
  \includegraphics[height=5.6cm, width=6cm]{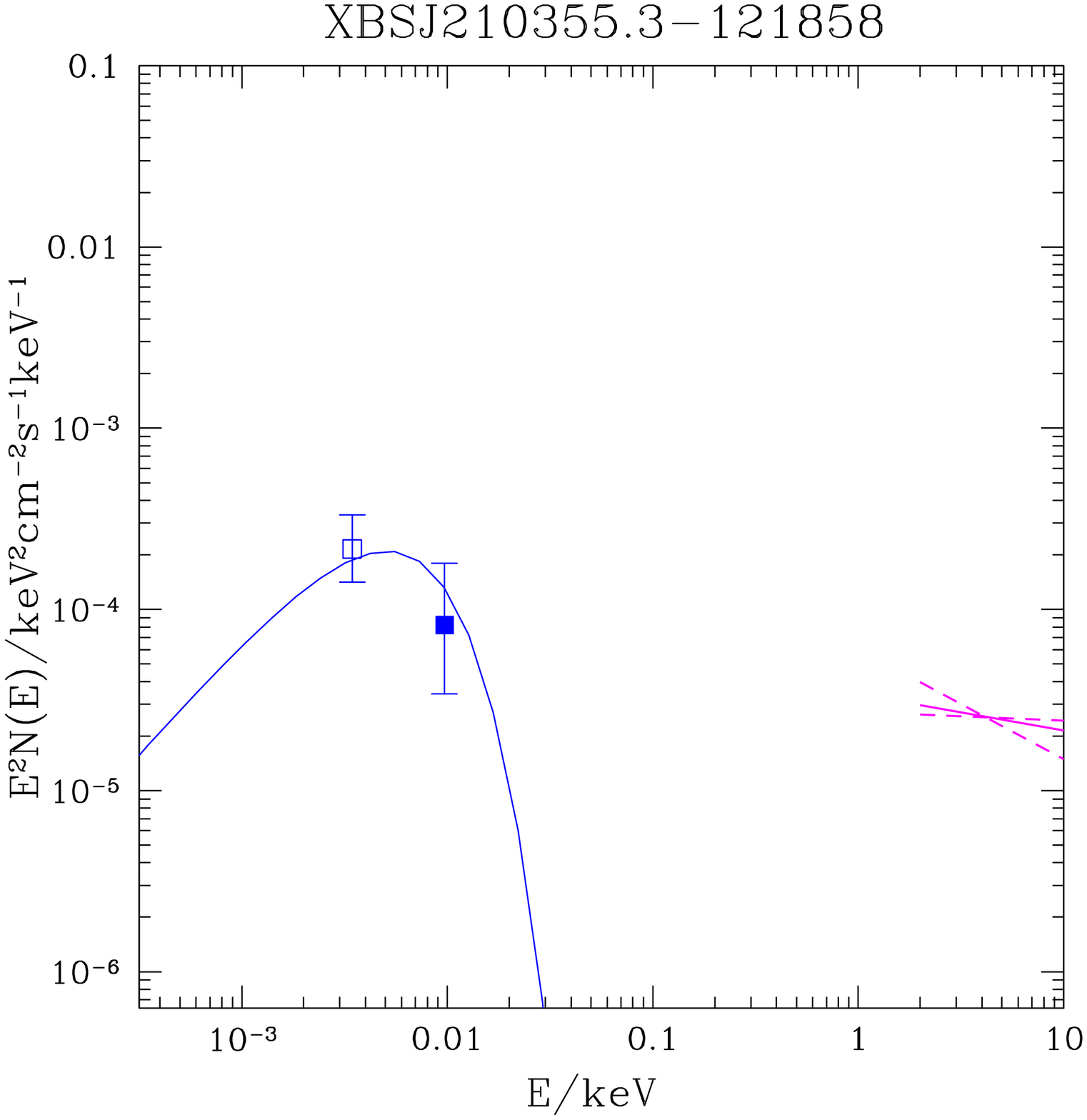}}      
  \end{figure*}
  
   \FloatBarrier
  
   \begin{figure*}
\centering
\subfigure{ 
  \includegraphics[height=5.6cm, width=6cm]{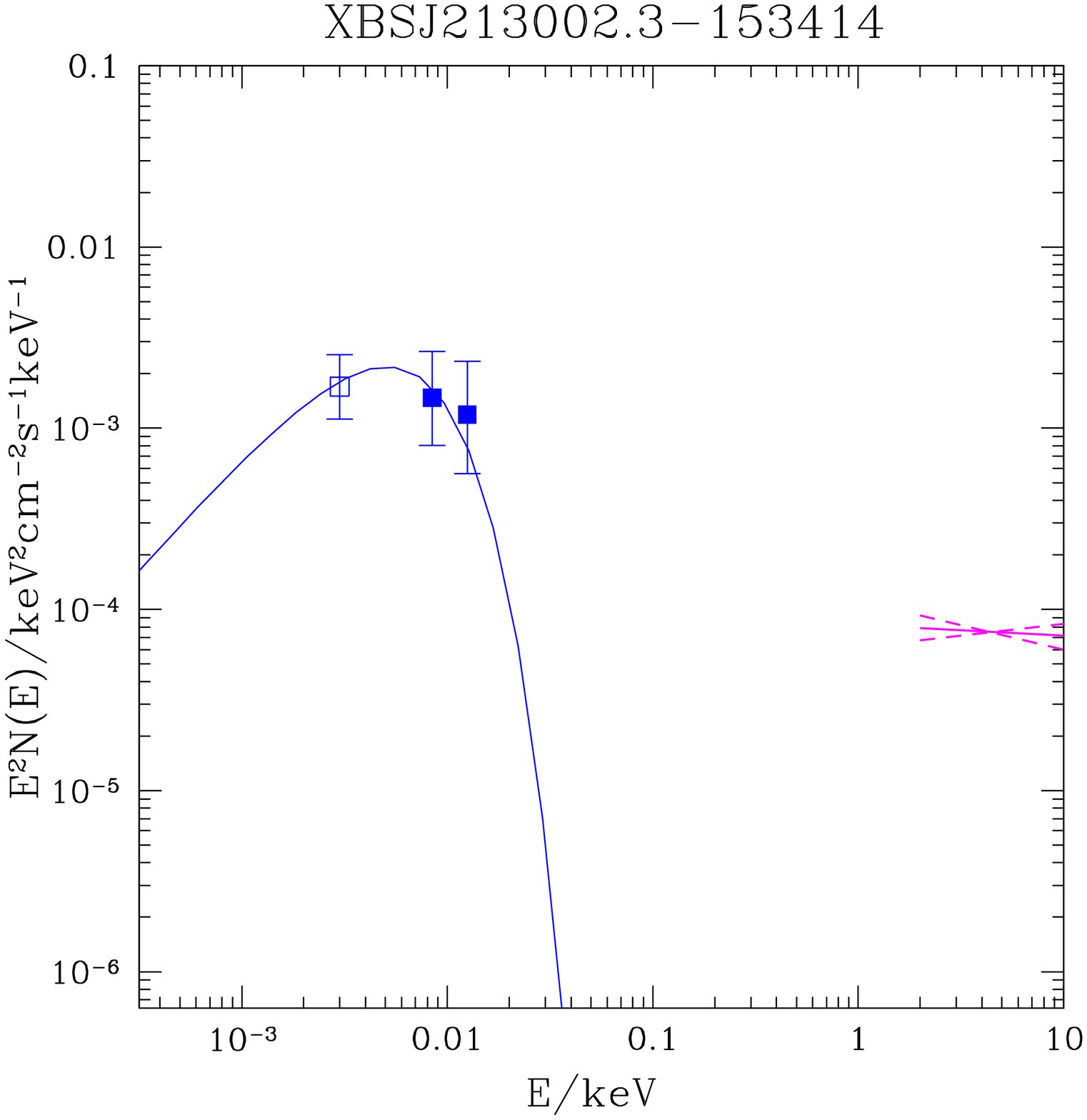}
  \includegraphics[height=5.6cm, width=6cm]{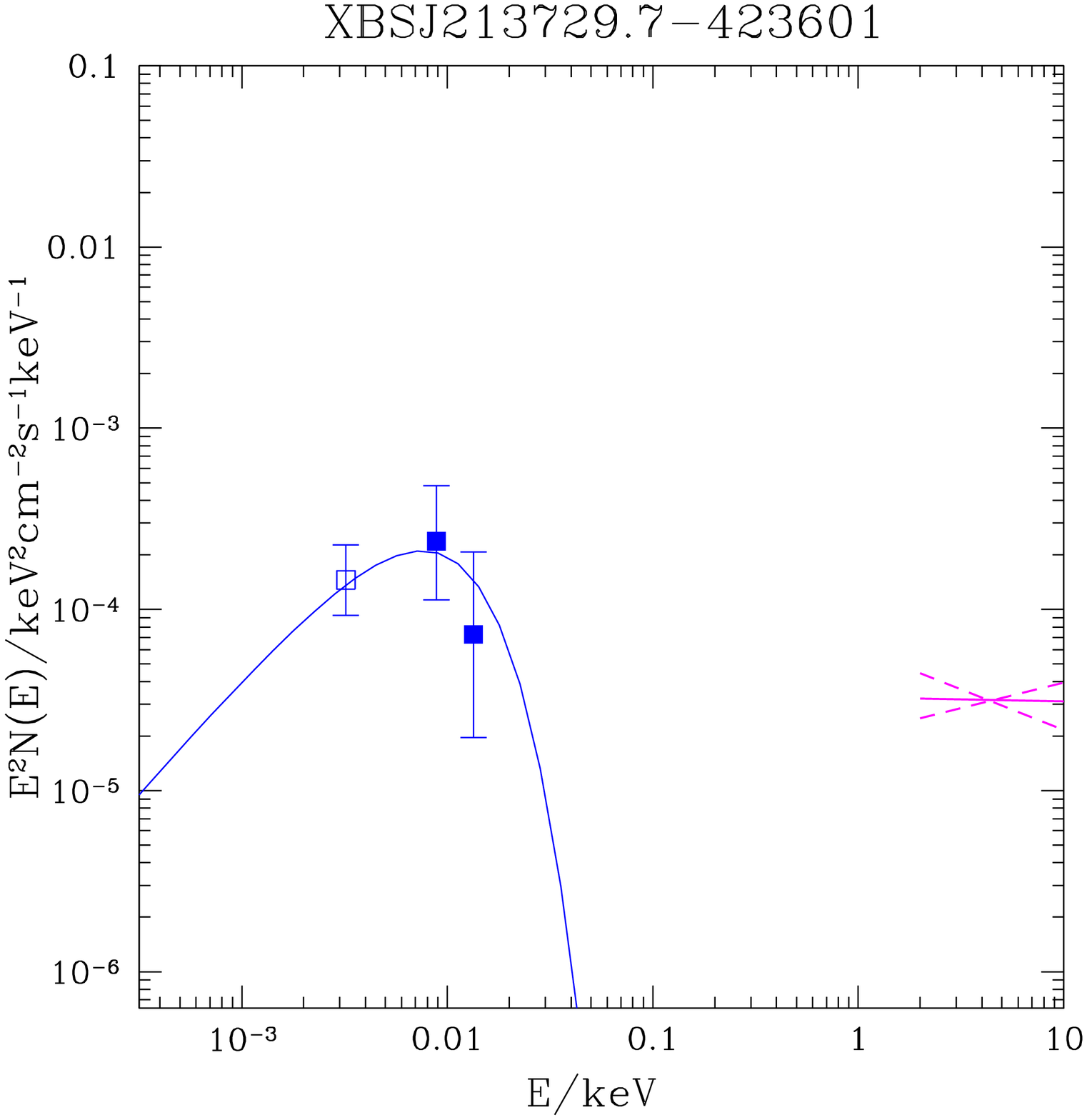}}      
\subfigure{ 
  \includegraphics[height=5.6cm, width=6cm]{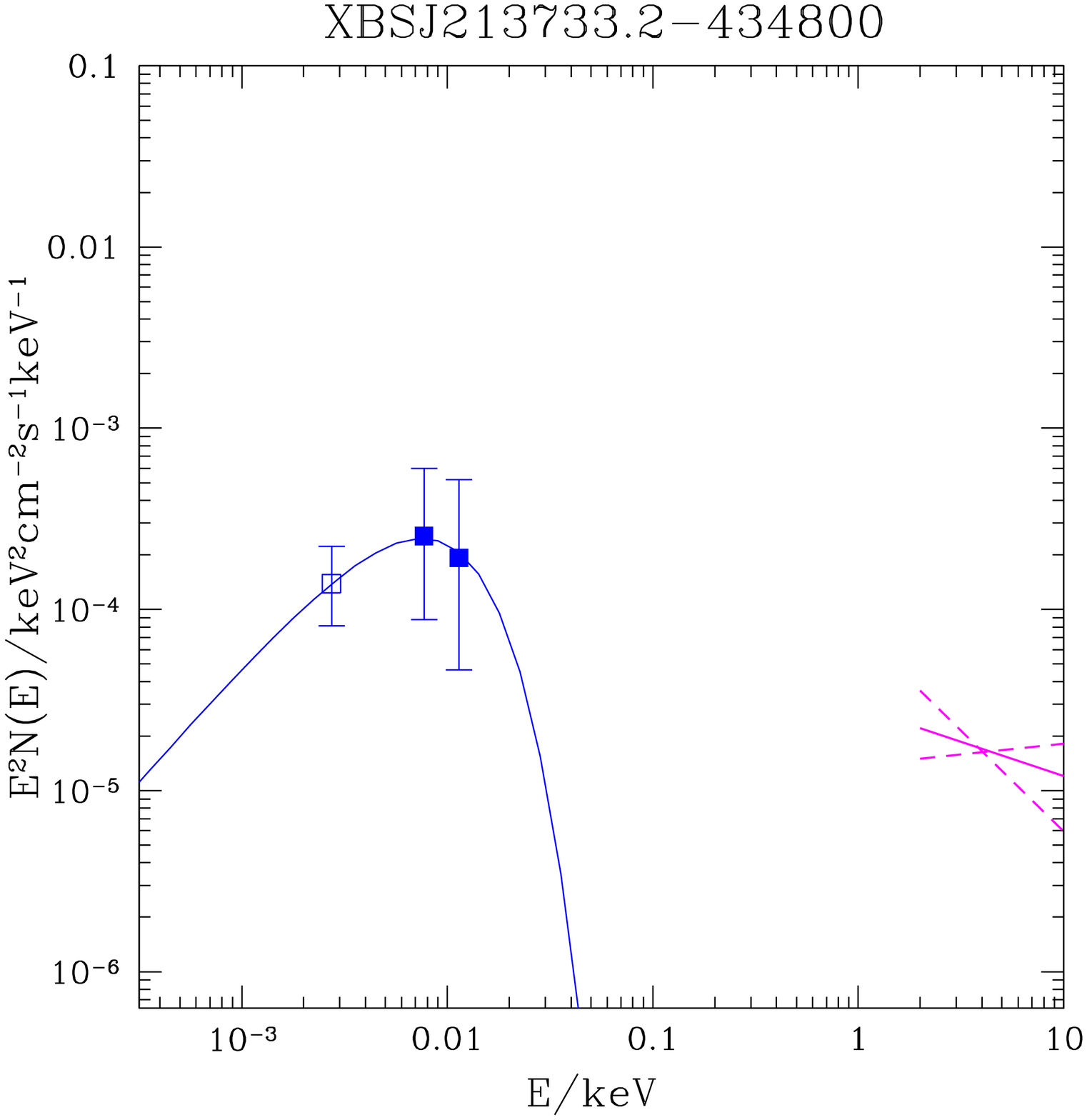}
  \includegraphics[height=5.6cm, width=6cm]{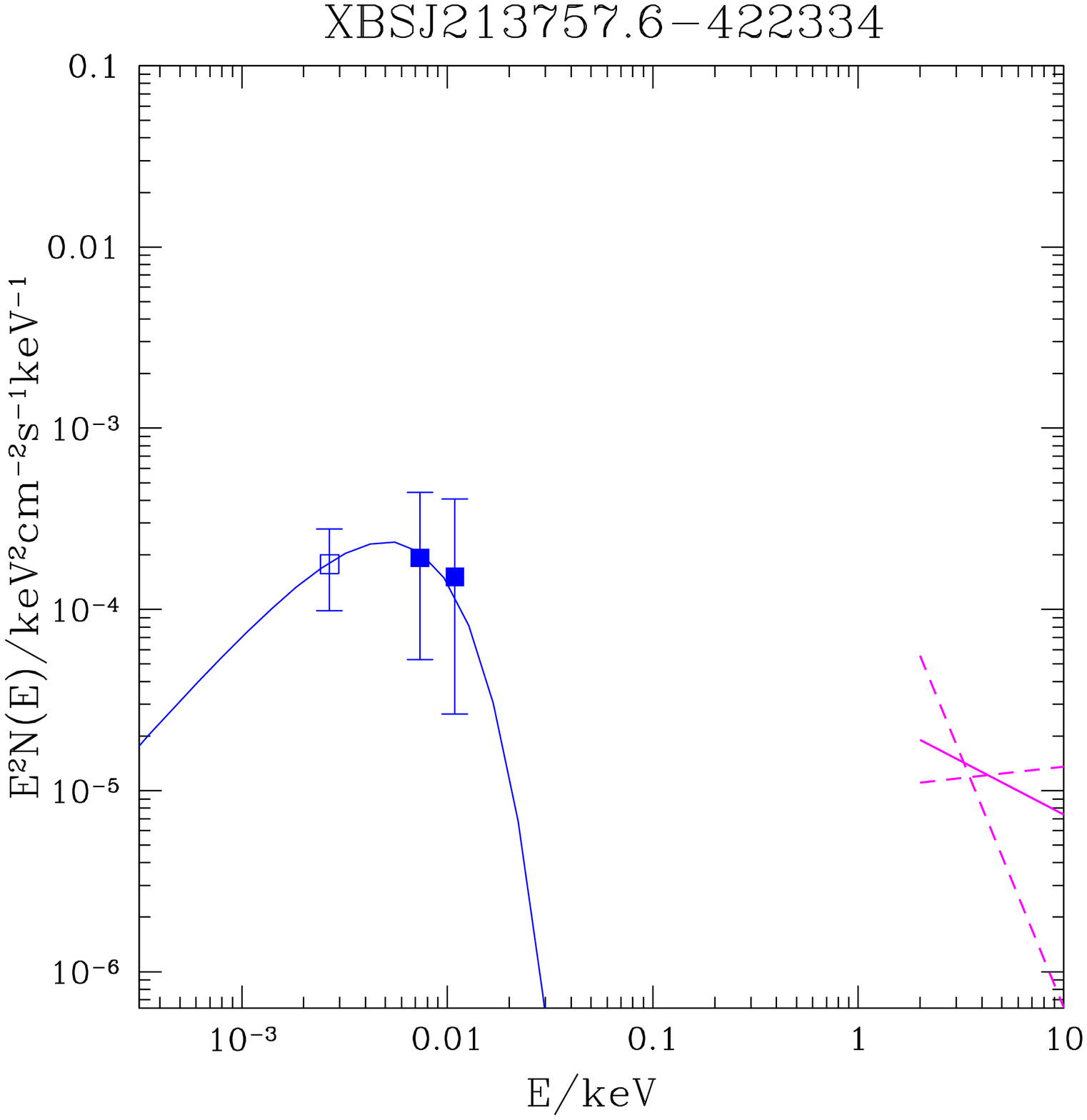}}    
\subfigure{ 
  \includegraphics[height=5.6cm, width=6cm]{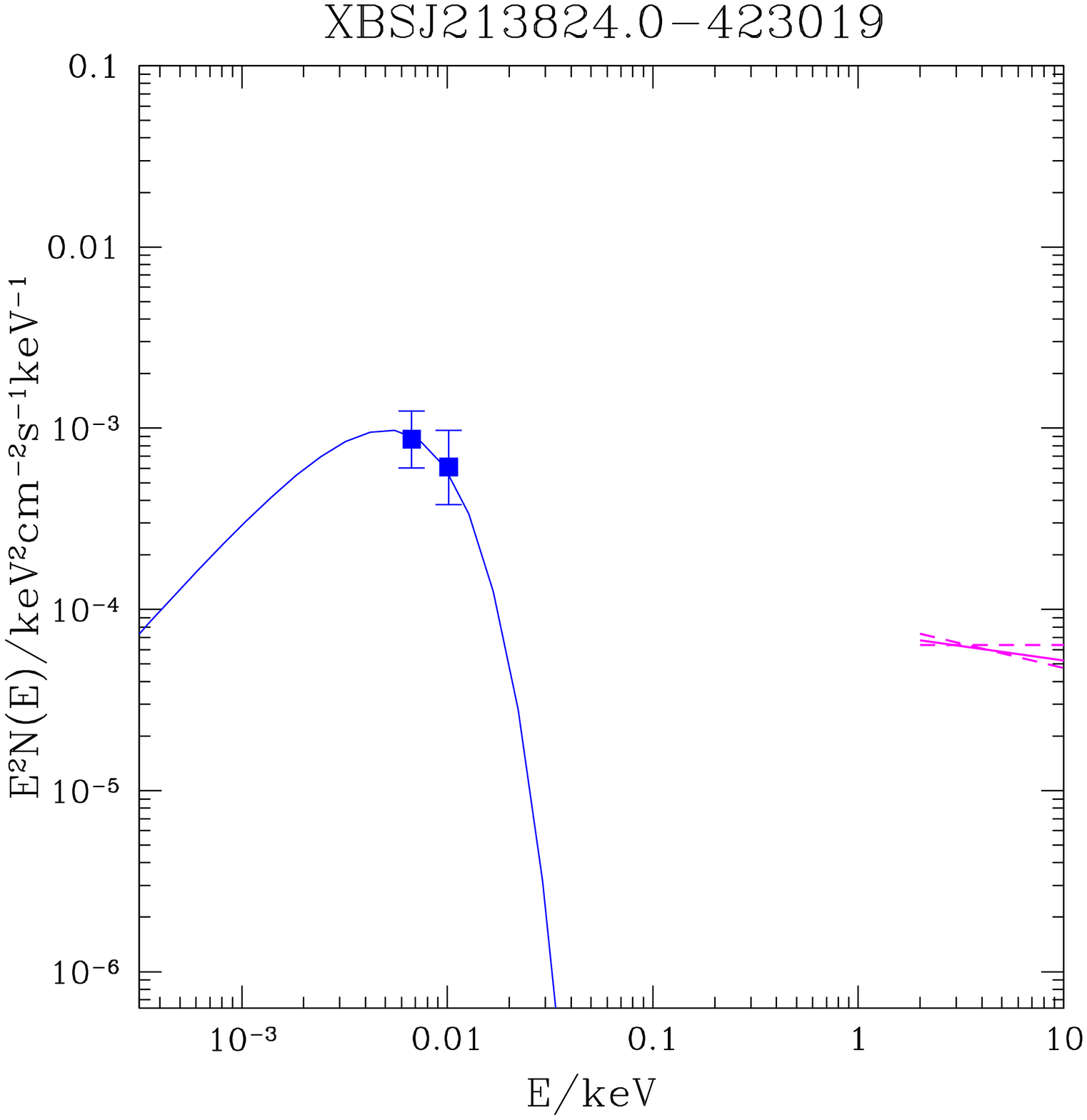}
  \includegraphics[height=5.6cm, width=6cm]{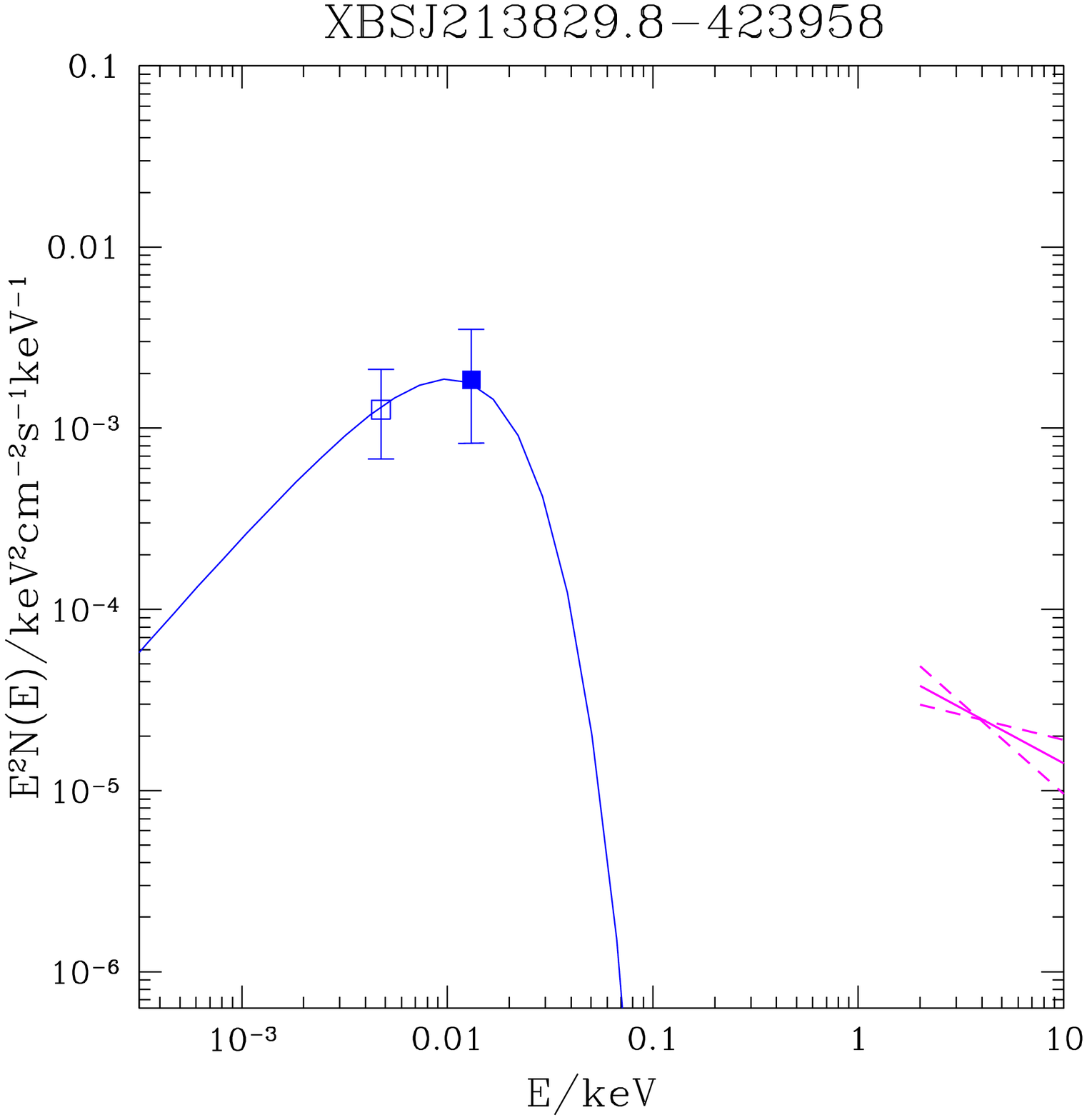}}      
\subfigure{ 
  \includegraphics[height=5.6cm, width=6cm]{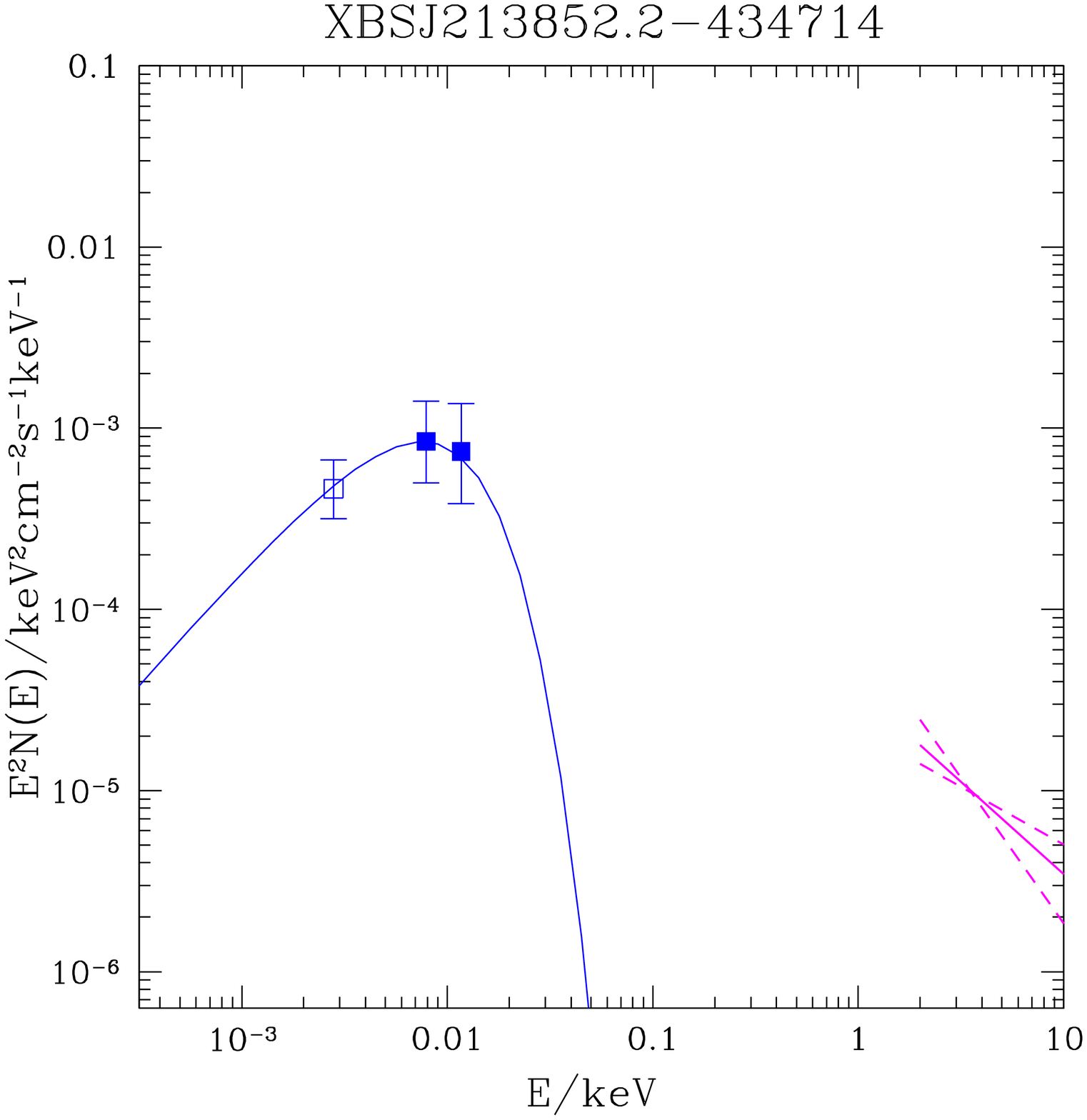}
  \includegraphics[height=5.6cm, width=6cm]{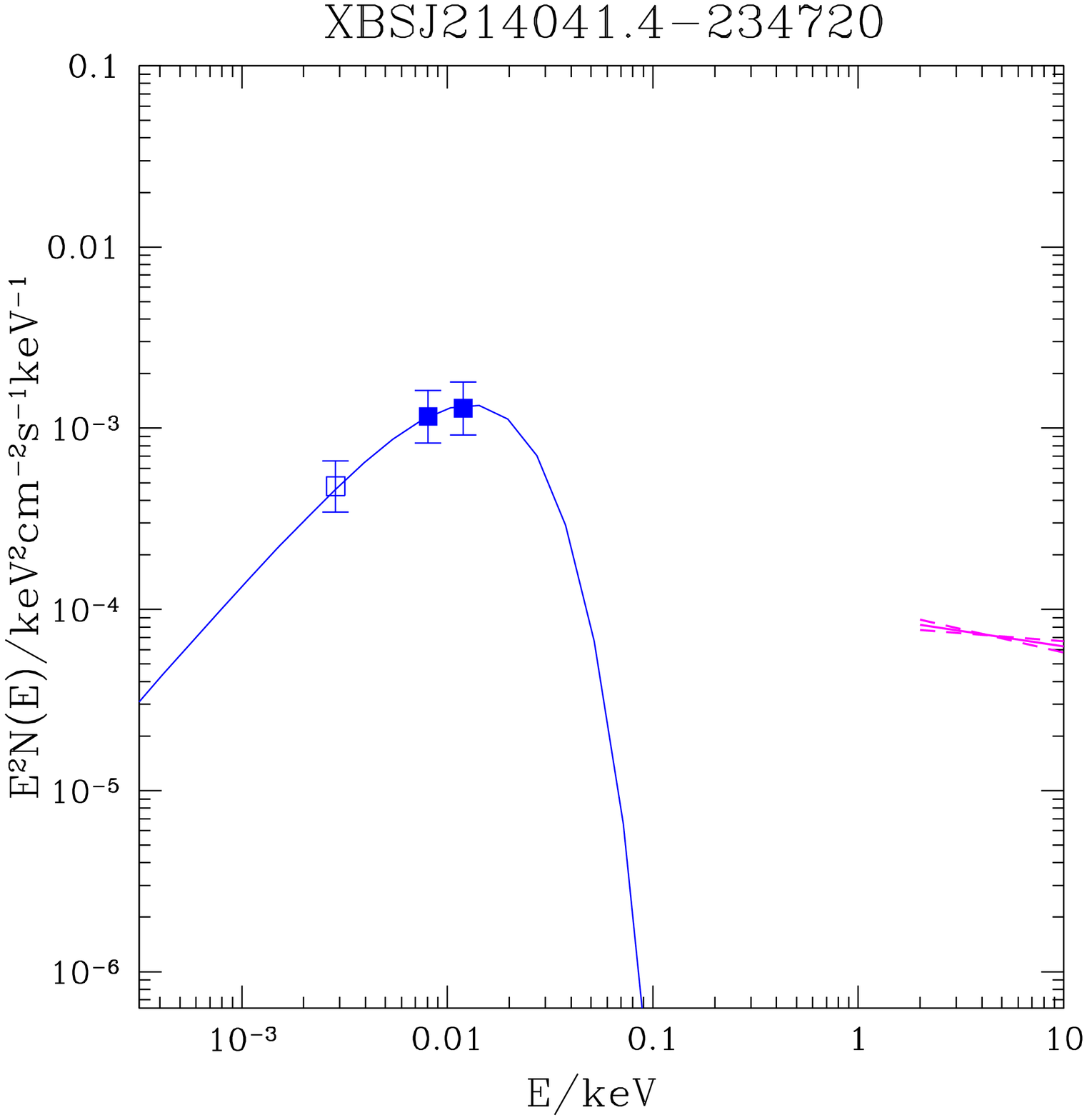}}      
  \end{figure*}
  
   \FloatBarrier
  
   \begin{figure*}
\centering
\subfigure{ 
  \includegraphics[height=5.6cm, width=6cm]{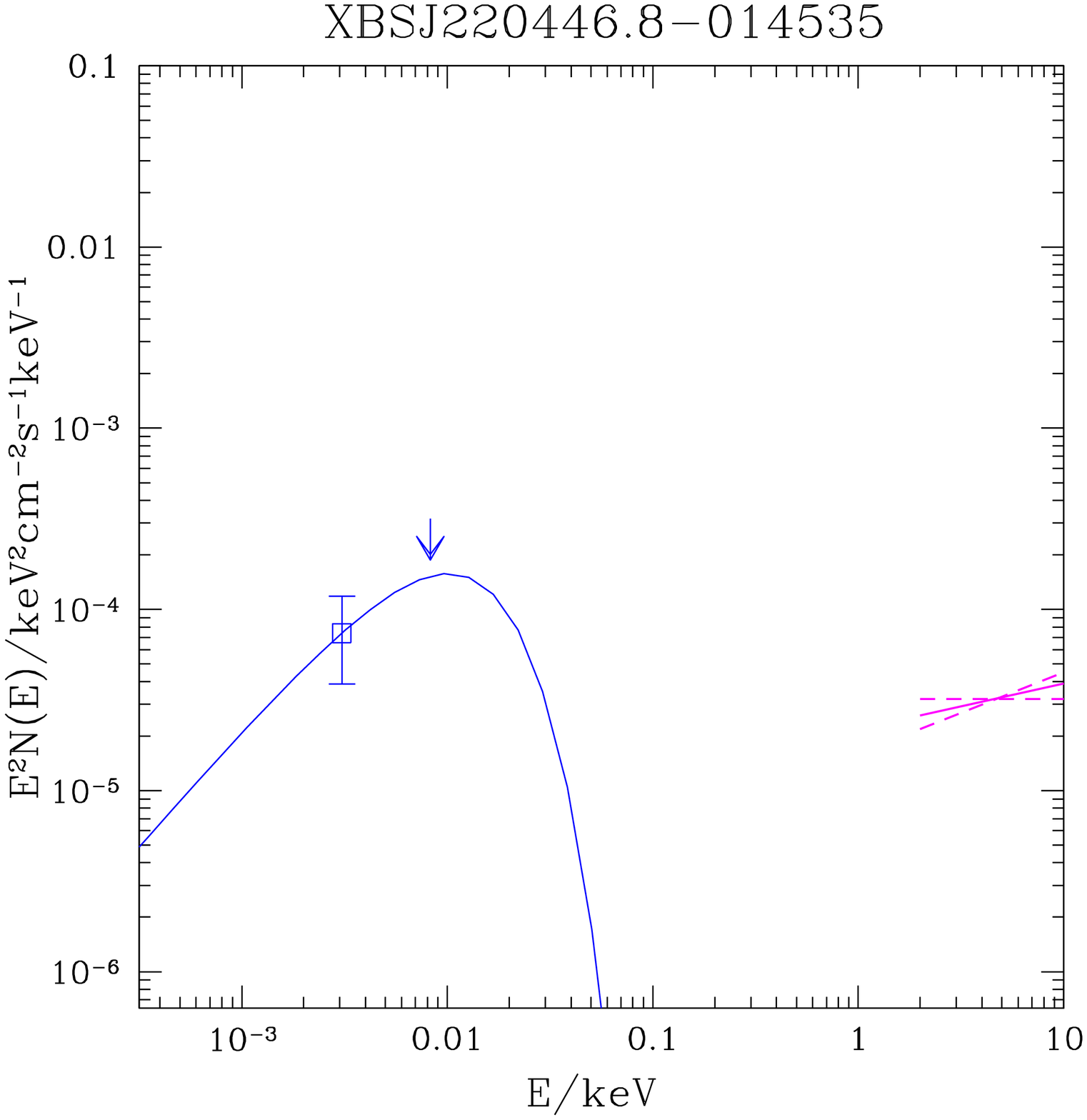}
  \includegraphics[height=5.6cm, width=6cm]{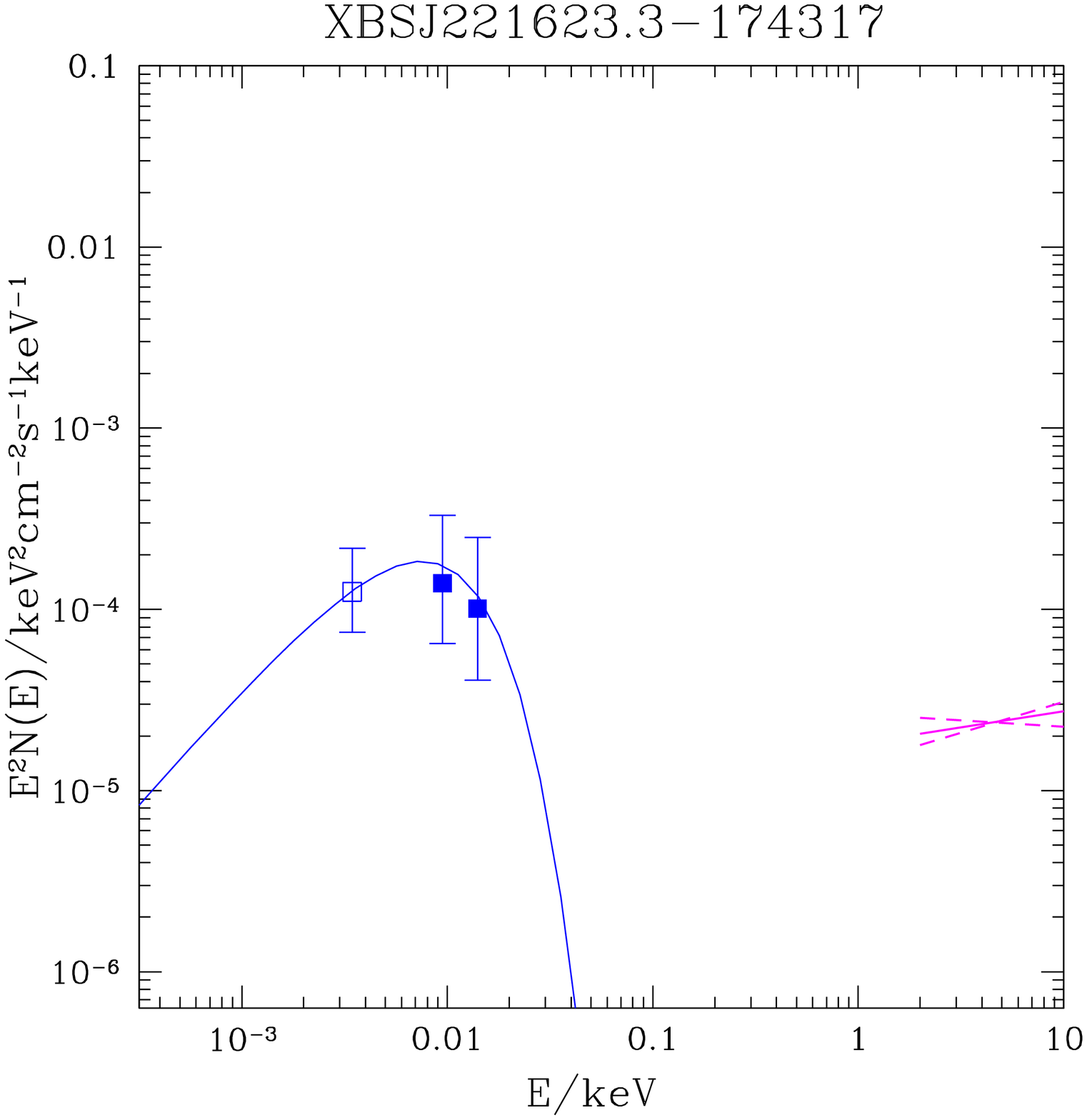}}      
\subfigure{ 
  \includegraphics[height=5.6cm, width=6cm]{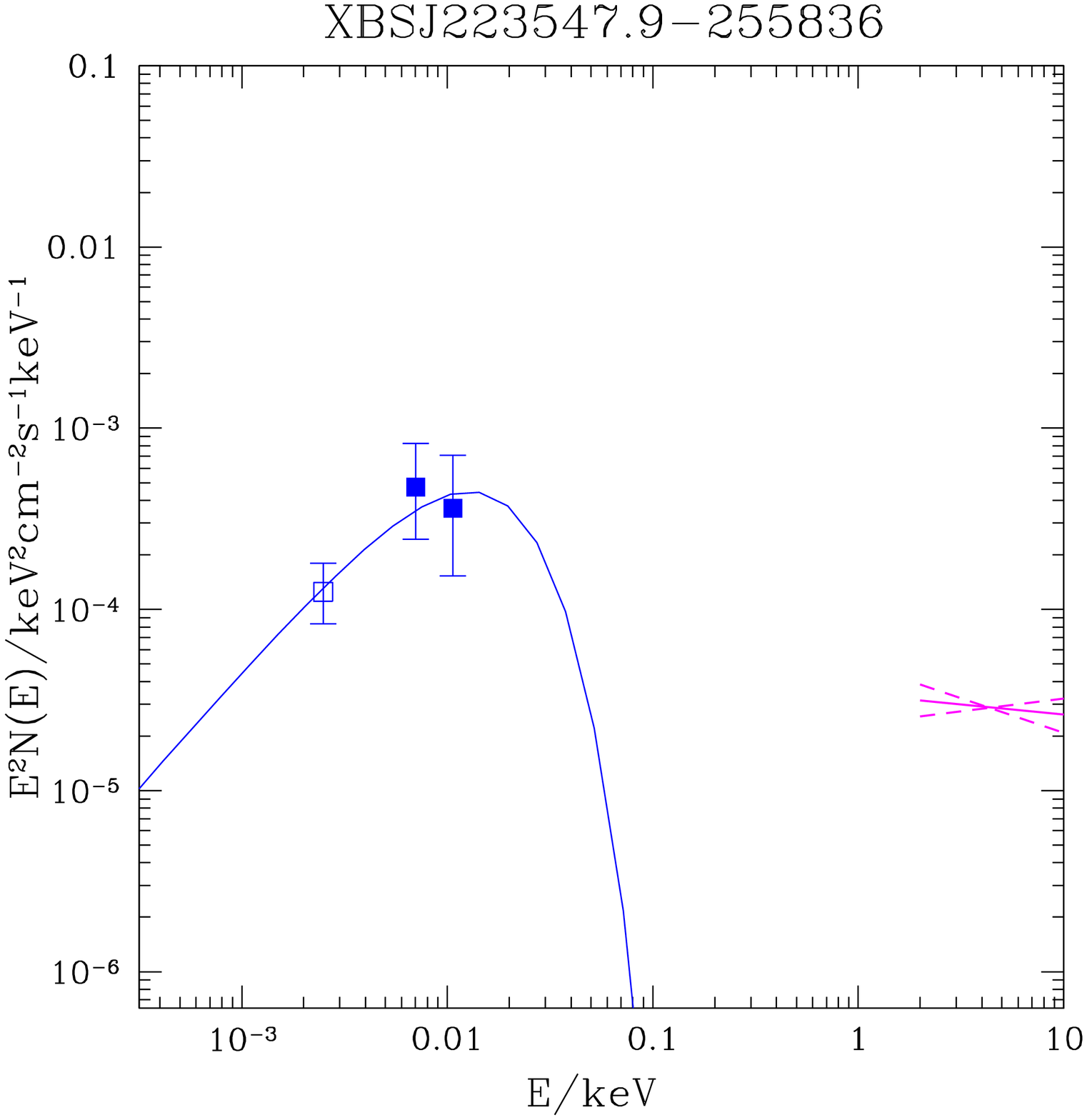}
  \includegraphics[height=5.6cm, width=6cm]{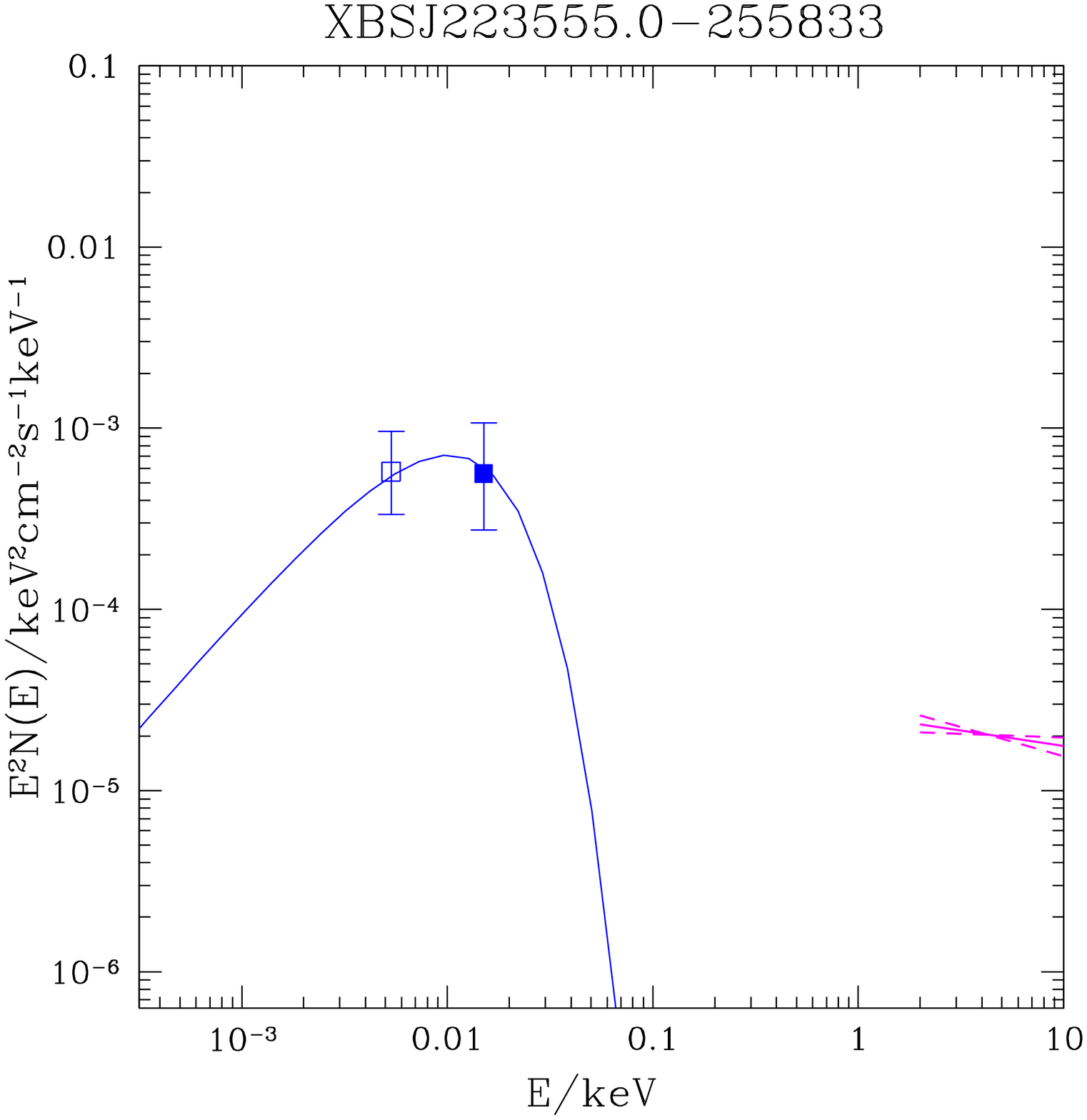}}     
 \subfigure{ 
  \includegraphics[height=5.6cm, width=6cm]{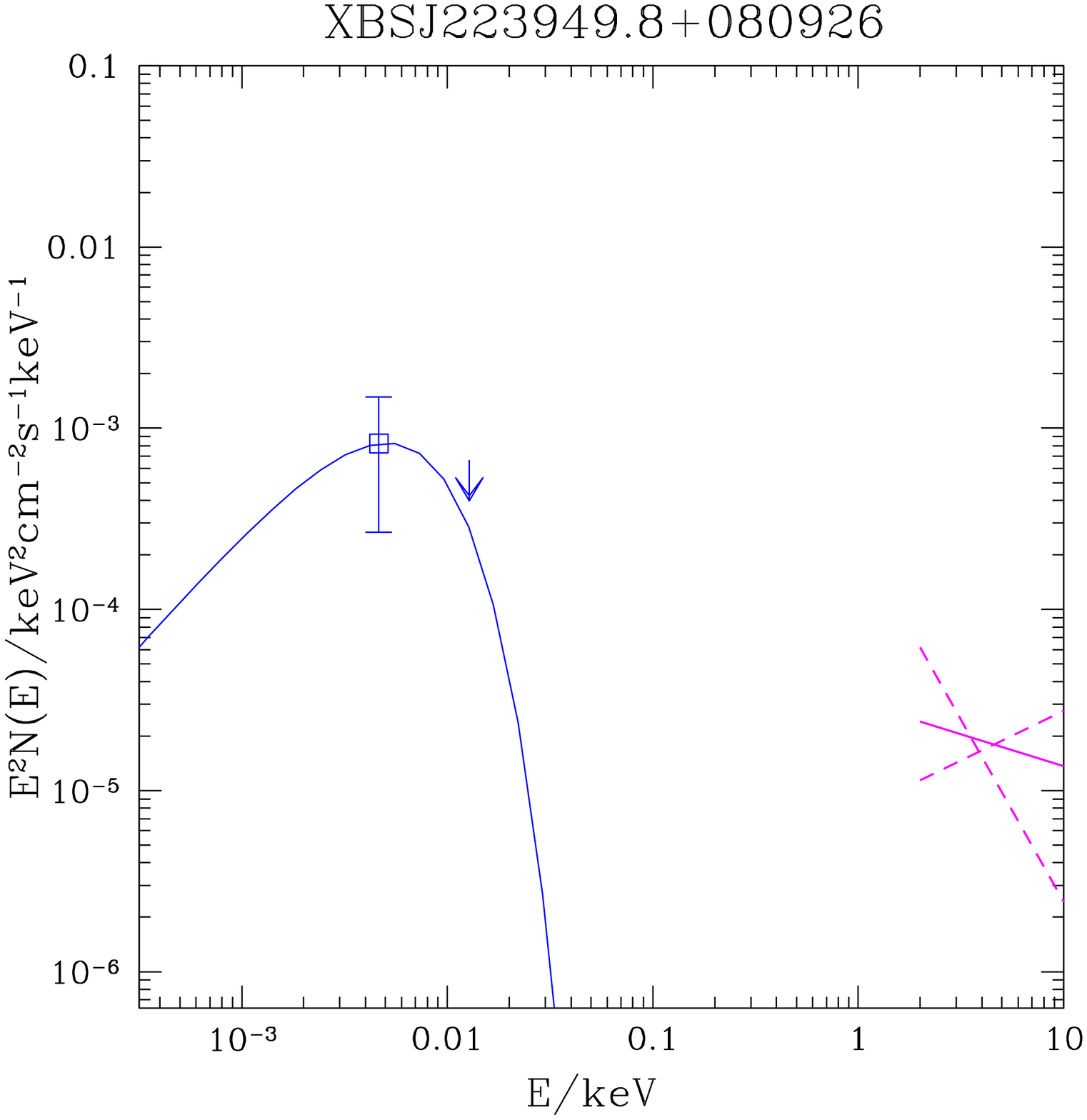}
  \includegraphics[height=5.6cm, width=6cm]{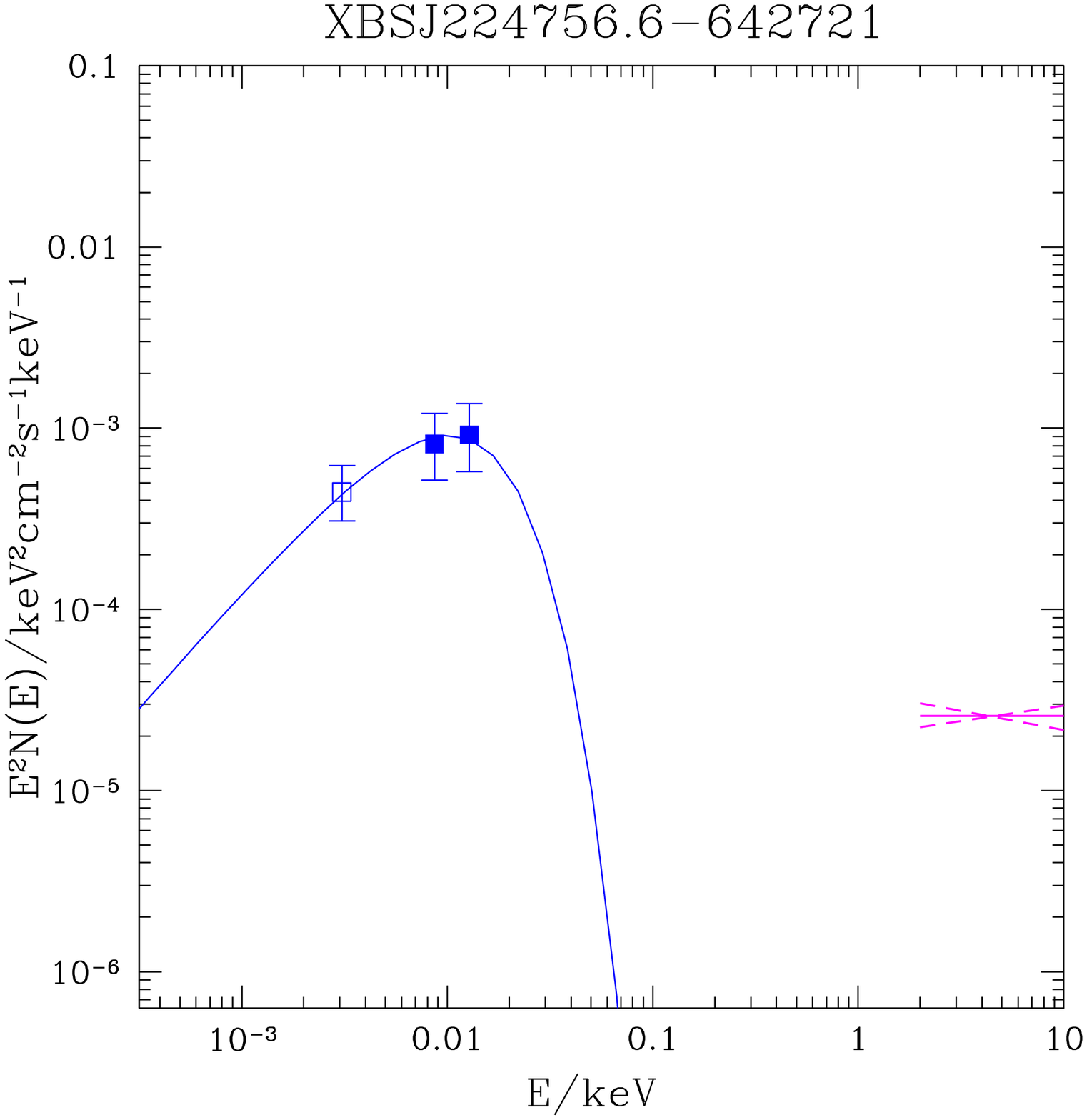}}     
 \subfigure{ 
  \includegraphics[height=5.6cm, width=6cm]{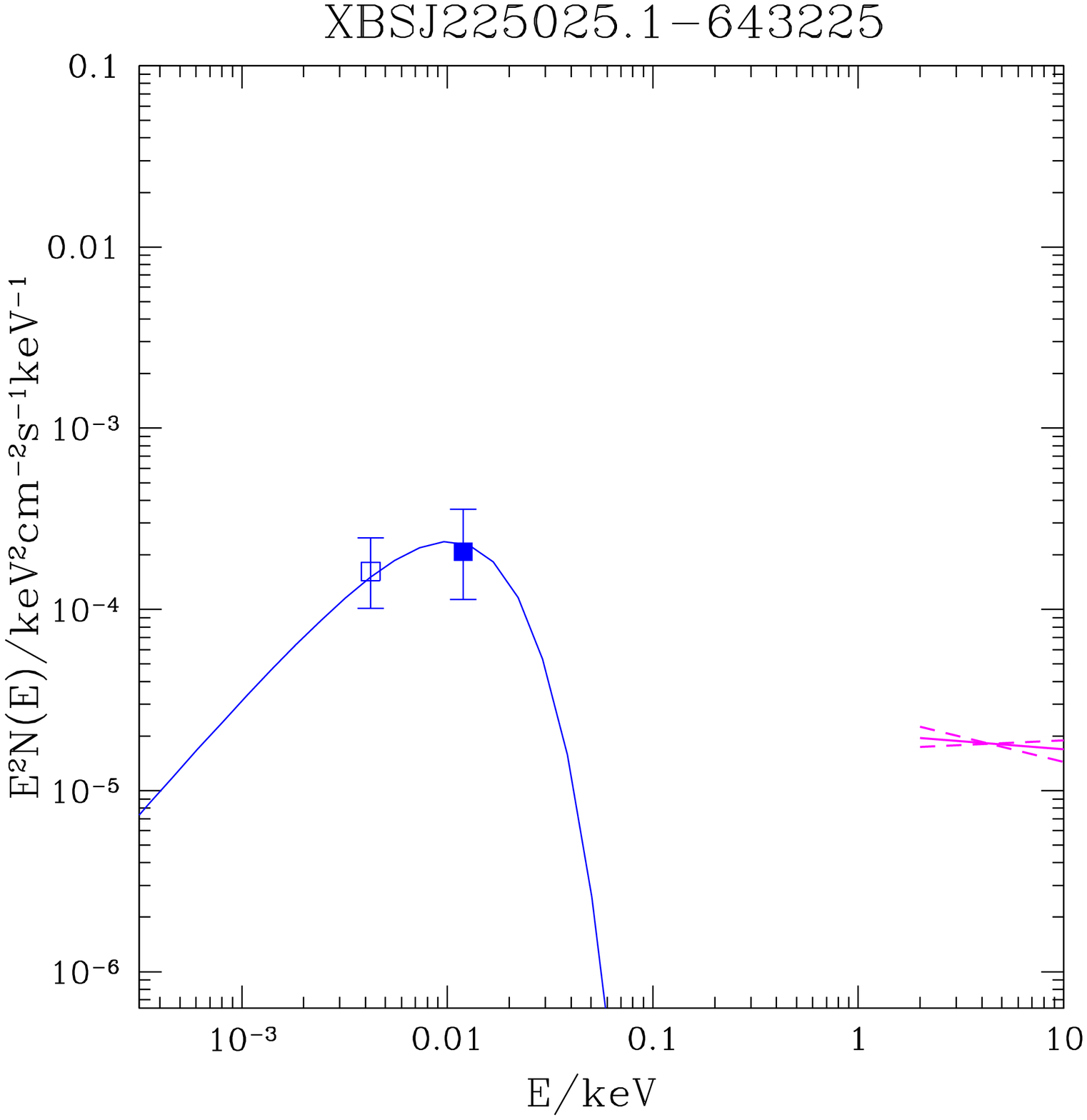}
  \includegraphics[height=5.6cm, width=6cm]{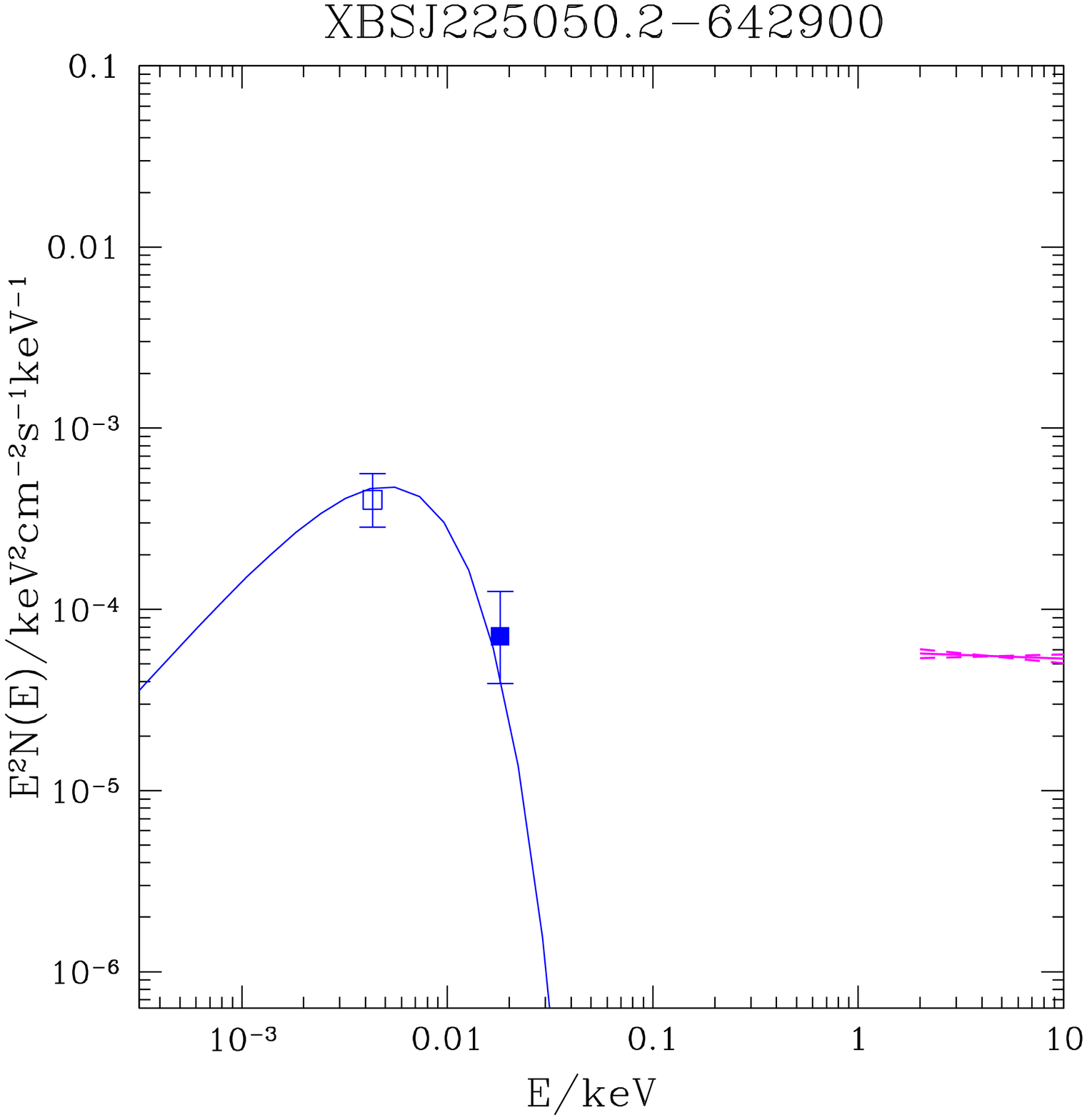}}      
  \end{figure*}
  
  \FloatBarrier
  
   \begin{figure*}
\centering
\subfigure{ 
  \includegraphics[height=5.6cm, width=6cm]{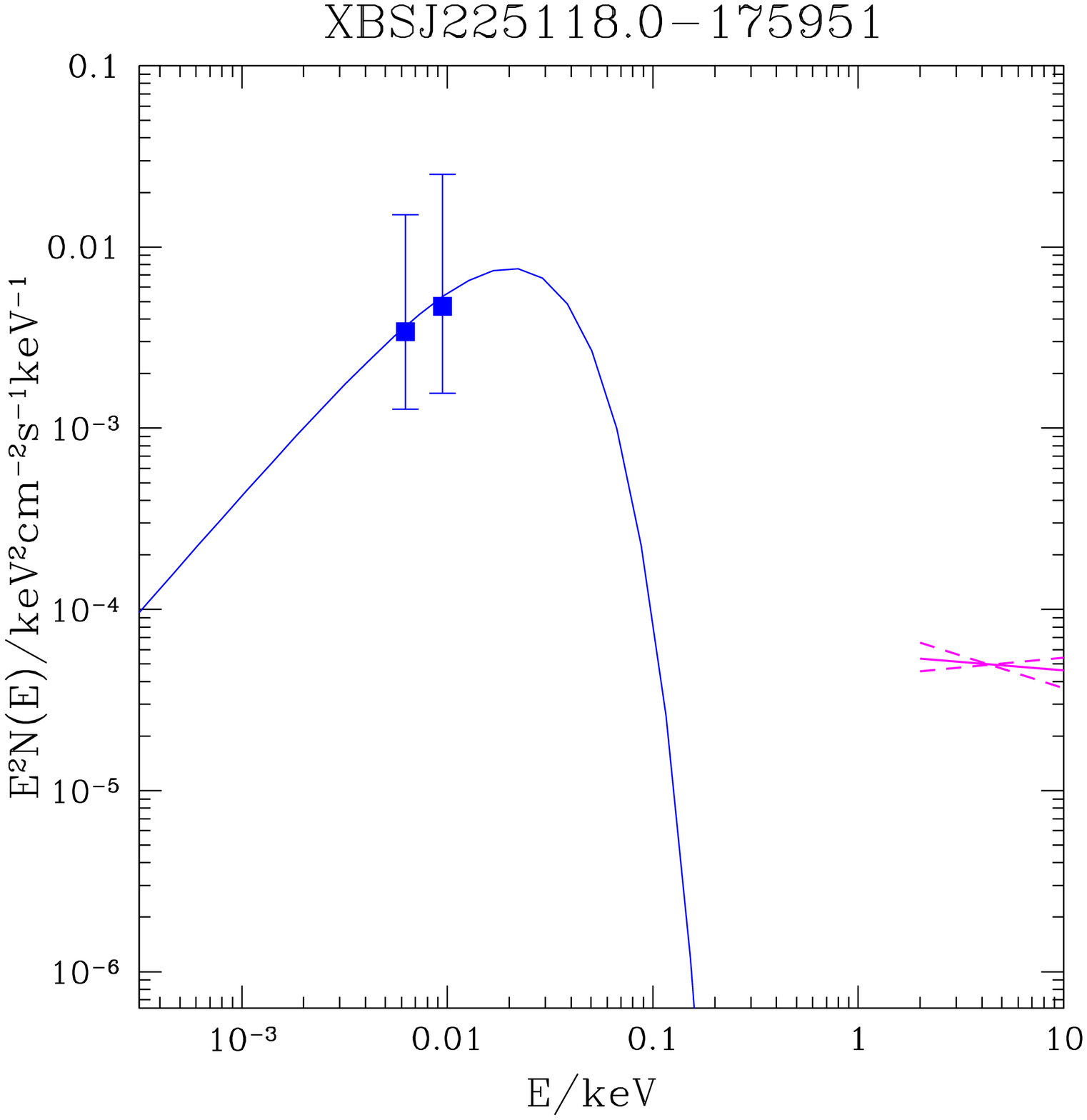}
  \includegraphics[height=5.6cm, width=6cm]{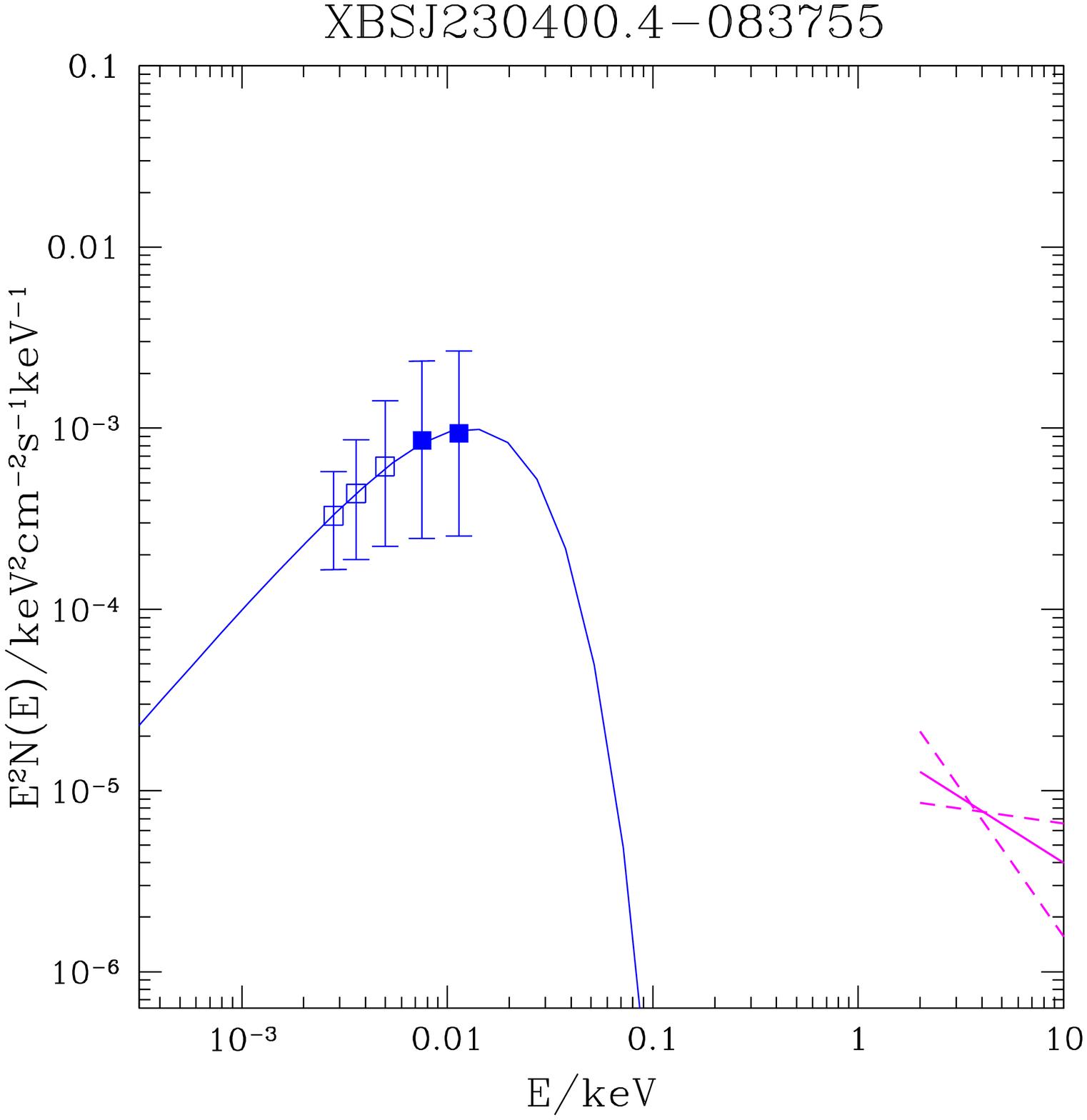}}     
 \subfigure{ 
  \includegraphics[height=5.6cm, width=6cm]{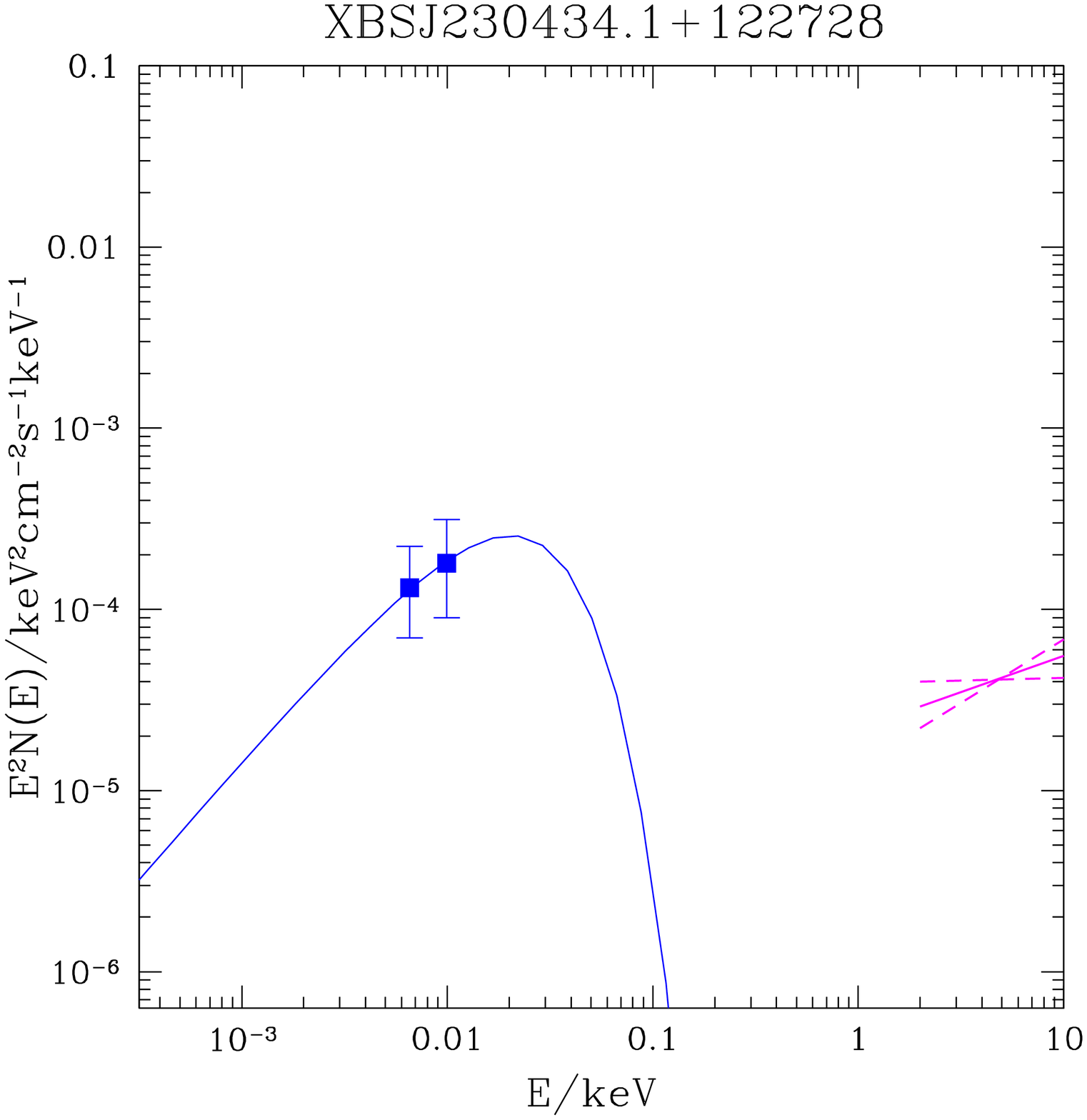}
  \includegraphics[height=5.6cm, width=6cm]{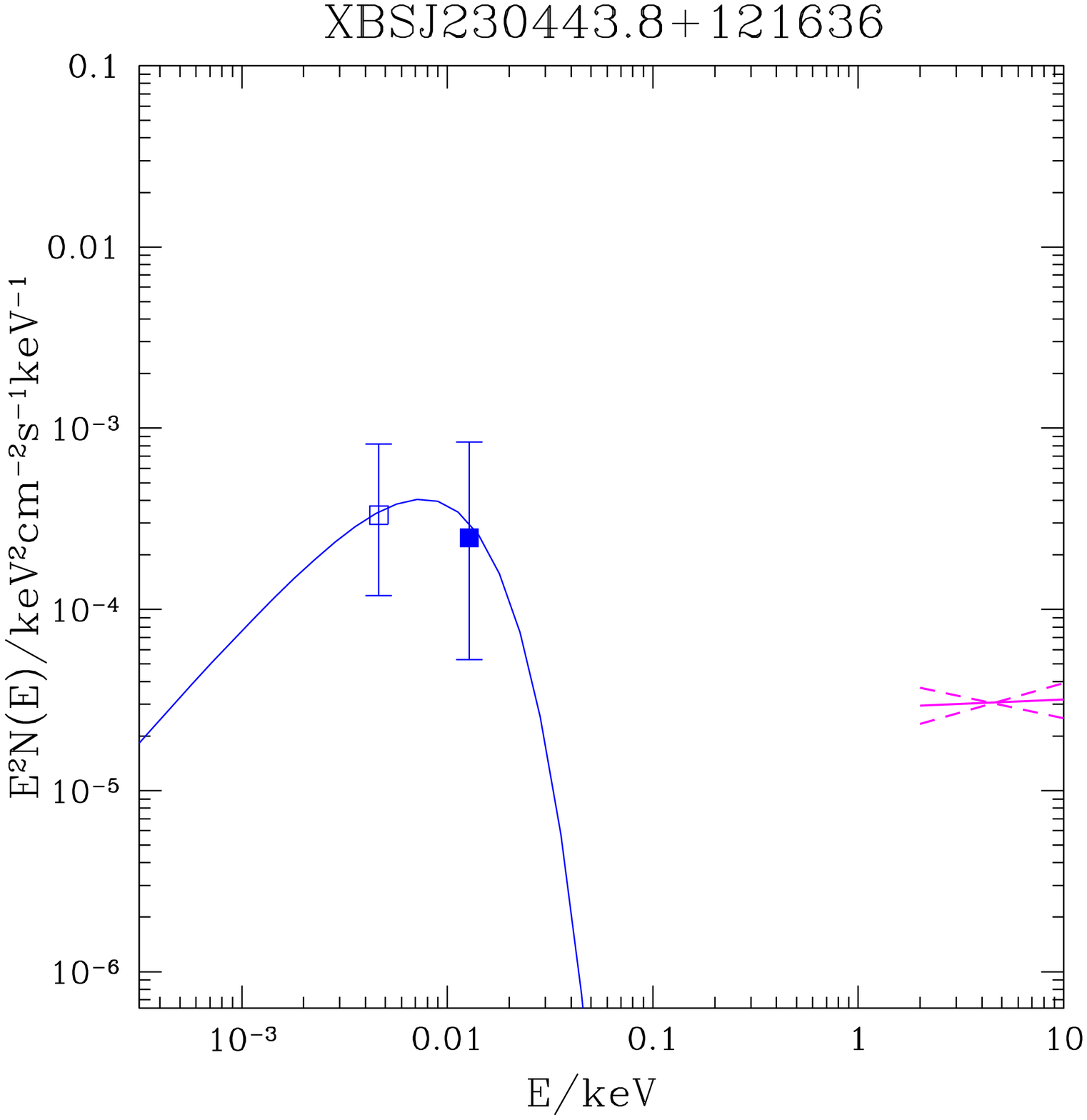}}    
 \subfigure{ 
  \includegraphics[height=5.6cm, width=6cm]{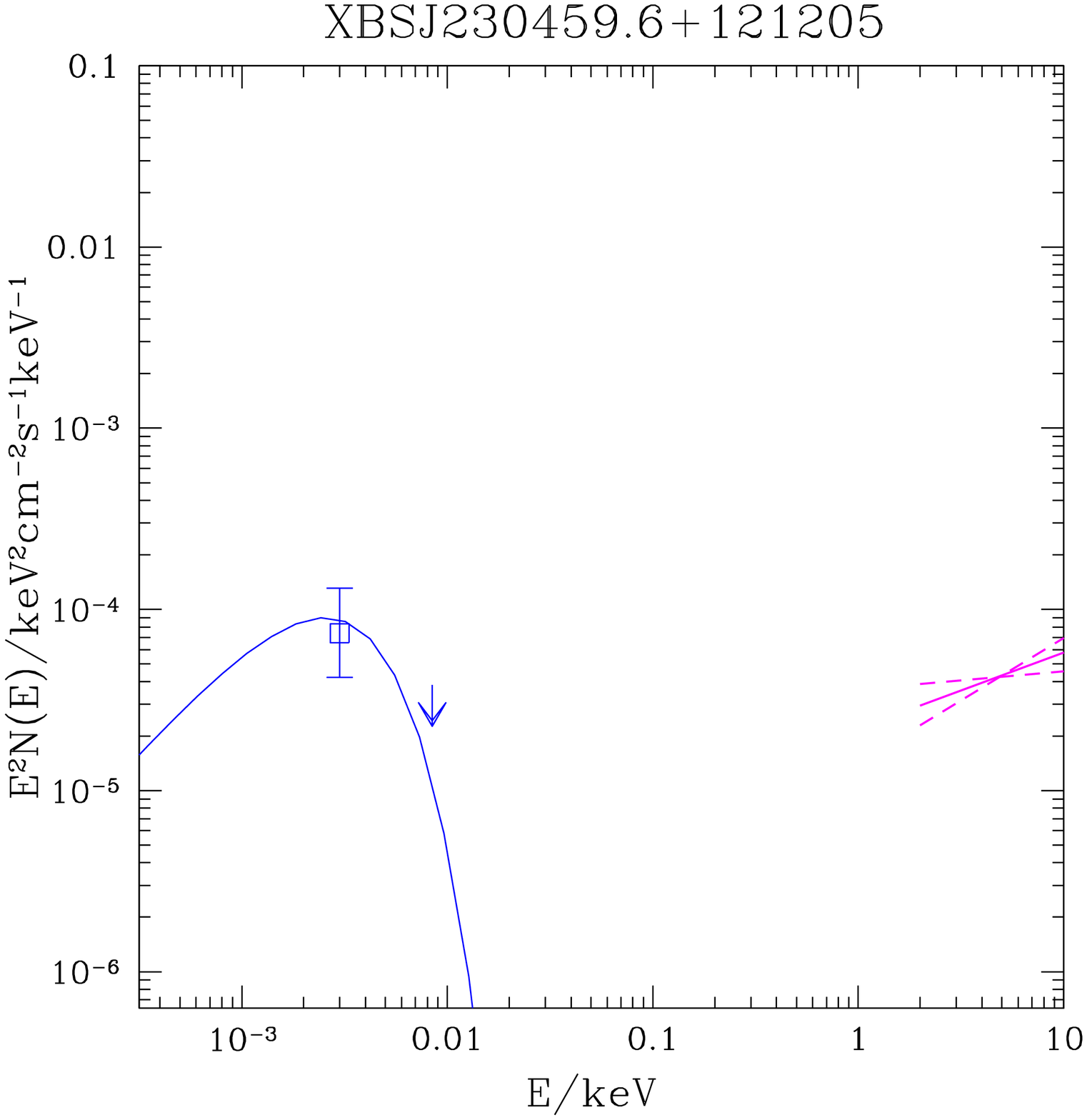}
  \includegraphics[height=5.6cm, width=6cm]{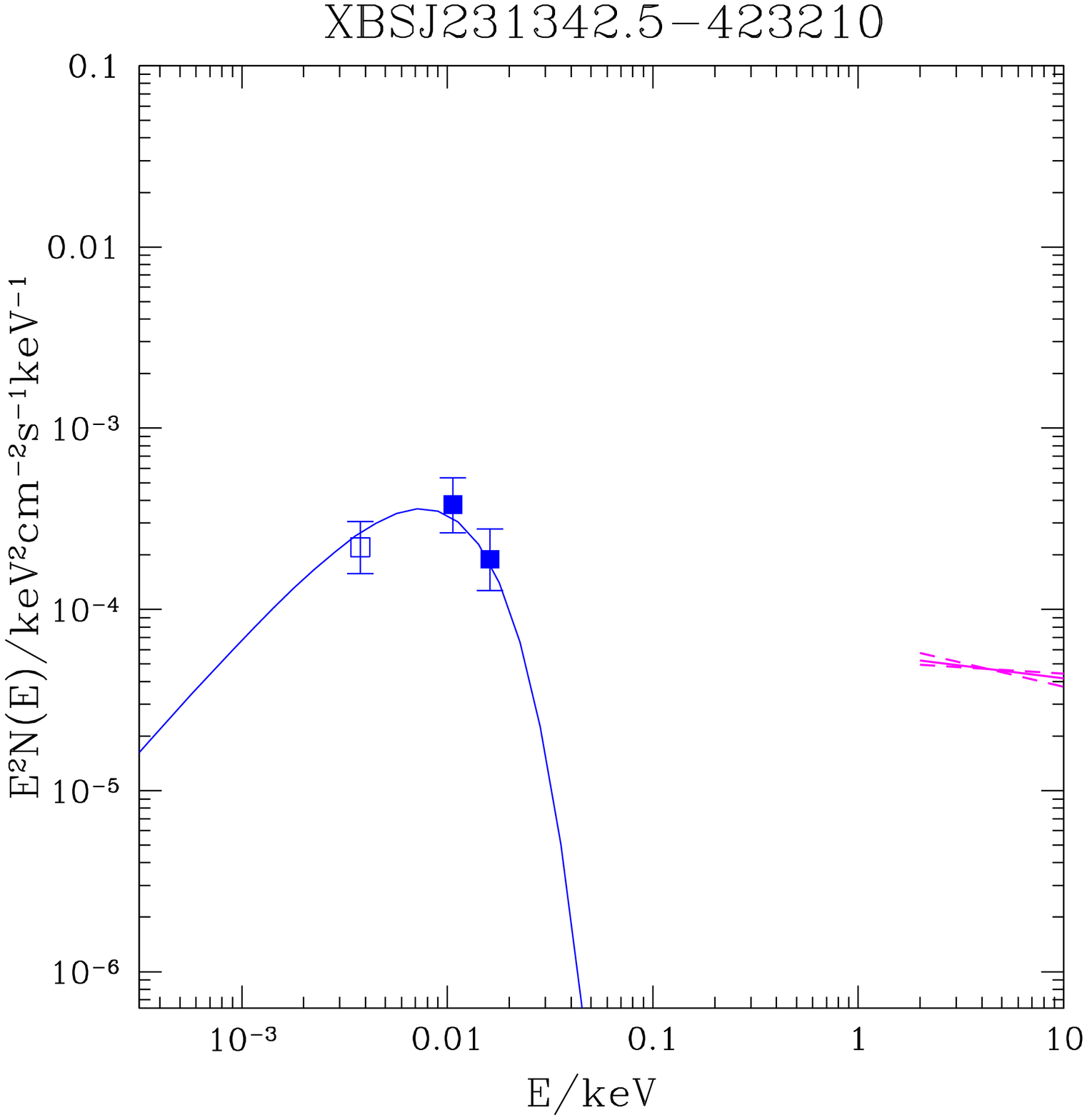}} 
  \end{figure*}
  
 \begin{figure*}
 \centering     
\subfigure{ 
  \includegraphics[height=5.6cm, width=6cm]{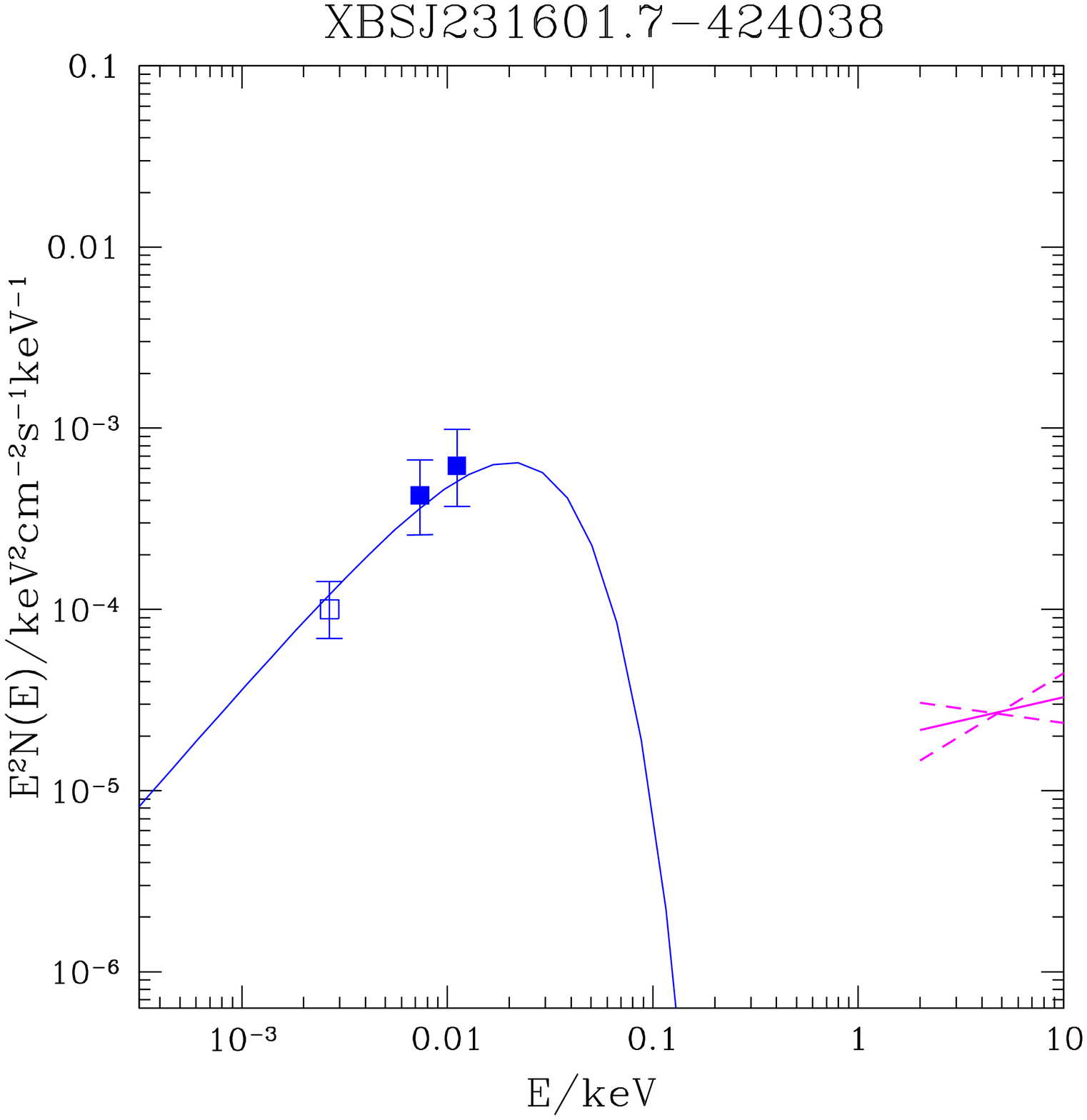}}
\end{figure*}

\end{appendix}

\end{document}